\documentclass[acmart, submit]{paper}

\begin{document}

\begin{abstract}
  Many effect systems for algebraic effect handlers are designed to guarantee that
all invoked effects are handled adequately.
However, respective researchers have developed their own effect systems that
differ in how to represent the collections of effects that may happen.
This situation results in blurring what is required for the representation and
manipulation of effect collections in a safe effect system.

In this work, we present a language {\lang} equipped with an effect system that
abstracts the existing effect systems for algebraic effect handlers.
The effect system of {\lang} is parameterized over \emph{effect algebras}, which
abstract the representation and manipulation of effect collections in safe
effect systems.
We prove the type-and-effect safety of {\lang} by assuming that a given effect
algebra meets certain properties called \emph{safety conditions}.
As a result, we can obtain the safety properties of a concrete effect system
by proving that an effect algebra corresponding to the concrete system meets the safety conditions.
We also show that effect algebras meeting the safety conditions are expressive
enough to accommodate some existing effect systems, each of which represents
effect collections in a different style.
Our framework can also differentiate the safety aspects of the effect
collections of the existing effect systems.
To this end, we extend {\lang} and the safety conditions to \emph{lift coercions} and \emph{type-erasure
semantics}, propose other effect algebras including ones for which no effect system has been studied in the literature, and
compare which effect algebra is safe and which is not for the extensions.

\end{abstract}

\maketitle

\section{Introduction}\label{sec:intro}

\subsection{Background: Effect Systems for Algebraic Effect Handlers}

Algebraic effect handlers~\cite{plotkin_handlers_2009, plotkin_handling_2013}
enable implementing user-defined computational effects, such as
mutable states, exceptions, backtracking, and generators, and structuring
programs with them in a modular way.
A significant aspect of algebraic effect handlers is compositionality.
Because of the algebraicity inherited from algebraic
effects~\cite{plotkin_algebraic_2003, kammar_handlers_2013}, they allow composing multiple effects
easily, unlike some other approaches to user-defined effects, such as
monads~\cite{moggi_notions_1991, wadler_marriage_1998}. 
Another benefit of algebraic effect handlers is to separate the interfaces and
implementations of effects.
For example, the manipulation of mutable states is expressed by two operations to
set a new state and get the current state.
While a program manipulates states via these operations, their implementation can
be determined dynamically by installing \emph{effect handlers}.
This separation of interfaces from implementations allows writing effectful
programs in a modular manner.

A key property expected in a statically typed language with algebraic effect handlers is \emph{type-and-effect safety}.
In the presence of effect handlers, type safety ensures that the type of an
operation is matched with that of its implementation provided by an effect handler.
\emph{Effect safety}~\cite{brachthauser_effects_2020}\footnote{The notion of
  effect safety itself and its importance have been recognized before the name
  was coined~\cite{kammar_handlers_2013}.} states that every operation call is
handled appropriately (i.e., it is performed under an effect
handler that provides the called operation with an implementation). Ensuring effect safety is crucial to guarantee the safety of programs
as an ``unhandled operation'' makes programs get stuck.

Several
researchers have proposed type-and-effect systems (effect systems for
short) to guarantee type-and-effect safety.
The effect systems in the literature are classified roughly into two groups
according to how they represent collections of effects that programs may invoke.
Certain effect systems adapt \emph{sets} to represent such
collections~\cite{kammar_handlers_2013,bauer_effect_2013,kammar_no_2017,forster_expressive_2017,saleh_explicit_2018,sekiyama_signature_2020}.
Another approach is
using \emph{rows}~\cite{hillerstrom_liberating_2016, leijen_type_2017, biernacki_abstracting_2019, xie_first-class_2022}, which allow
manipulating the collections of effects in a more structured manner.
For example, the effect system of \citet{hillerstrom_liberating_2016} can
represent the presence and absence of effects in rows, and that of
\citet{leijen_type_2017} allows the duplication of effects with the same name in one
row.
%

However, several issues are posed by the current situation that the effect systems in the different styles
have been studied independently.
First, it blurs what manipulation of effect collections are indispensable
to give an effect system.
Second, it is unclear what property an effect system requires for effect collections
and their manipulation to guarantee effect safety.
These unclarities cause the problem that designers of new effect systems grope in
the dark for the representations of effect collections, and even if they come up with an
appropriate representation, they need to prove the desired properties, such as
effect safety, from scratch.
The third issue is that, when extending languages with new features,
one needs to build the metatheory for each of the representations.
%

\subsection{Our Work}

This work aims to reveal the essence of safe effect systems for effect handlers.
Because we are interested in the shared nature of such effect systems, we avoid
choosing one concrete representation of effect collections.  Instead, we provide an effect
system that abstracts over the representations of effect collections and can derive concrete effect
systems by instantiating them.

More specifically, our effect system is parameterized over constructors
and manipulations of effect collections.
In general, effect systems for algebraic effect handlers require two kinds of
manipulation.
One is subeffecting, which overapproximates effects to
adjust the effects of different expressions.
The other is the removal of effects.
An installed effect handler removes the effect it handles and forwards the
remaining effects to outer effect handlers.
%
%
%
We formulate such manipulation of effect collections required by effect
systems as \emph{effect algebras}\footnote{The name ``effect algebra'' has
  been used to specify algebraic structures found in quantum mechanics~\cite{foulis_effect_1994},
  but we decided to use this name because the present work is far from quantum
  mechanics.} and ensure that our effect system relies only on the manipulations
allowed on them.
%
%

\OLD{
  Our language cannot have safety properties without some assumptions on effect signatures and ARE. For example, ARE with no assumptions can allow programs with unhandled operations to be well-typed. Thus, we formalize such assumptions as \emph{safety conditions}. For example, safety conditions include ``no operations cannot be treated as pure'' or ``different effects cannot be identified''. Safety conditions are essentially required to prove safety properties}

However, some effect algebras make the effect system unsound.
For instance, the effect system with an effect algebra that allows subeffecting
to remove some effects may typecheck unsafe programs (e.g., ones that cause
unhandled effects).
To prevent the use of such effect algebras, we formalize \emph{safety conditions}, which
are sufficient conditions on effect algebras to guarantee effect safety; we call effect algebras
meeting the conditions \emph{safe}.
We prove that the effect system instantiated with any safe effect algebra enjoys effect
safety as well as type safety---therefore, ones can ensure the safety of their
effect systems only by showing the safety of the corresponding effect algebras.
Furthermore, we also show what kind of unsafe programs each condition excludes.

\OLD{
  Our research aims to reveal the essential difference between effect systems and what is needed to prove safety properties. Towards this goal, there are mainly three difficulties. First, different effect systems have different syntax of effects. We must distinguish the essential difference we want to focus on with the difference caused by syntax. Secondly, some rules of type-and-effect systems have side conditions whose statements are involved with effects. These side conditions have different operations on syntactically different effects. Finally, there are syntactical or semantical extensions of languages with algebraic effects and handlers. We need to study what is essentially required for something like these kinds of extensions.

  Our crucial idea to discover the common core of effect systems is to introduce effect signatures and an appending relation on effects. Effect signatures abstract syntax of effects like $\Sigma$-algebra. An appending relation on effects, we call ARE, is used to represent some statements concerning effects. We discuss the properties of the effect system with these abstract components to achieve our goal. However, appending relations on effects must meet some conditions to prove safety properties. Then, we formalize such conditions as safety conditions. When considering extensions of language, we study what conditions need to be added to keep safety properties guaranteed.
}

\OLD{
  This paper presents a language and its type-and-effect system that are parameterized by effect signatures and an ARE; and proof of its safety properties within safety conditions. We can instantiate our language to concrete language proposed by several studies. This fact means that we can prove the safety properties of concrete languages via our language's safety properties. Furthermore, we must only check whether the instantiation is valid against safety conditions to prove the concrete type-and-effect system's safety properties. From the viewpoint of designing languages, we only need to verify that the new effects definition meets safety conditions. Even when extending languages, we must only verify that the expanded language meets extended safety conditions.
}

\OLD{
  \subsection{Contributions}
  \begin{itemize}
    \item We introduce the effect system with the relation over effects as a parameter.
    \item To prove the effect safety of the calculus, we define safety conditions that the relation over effects must satisfy.
    \item We also define type-erasure safety conditions that must be satisfied when we take type-erasure semantics.
    \item We give a sequence-based effect system as an instance to demonstrate the extensibility of our system.
  \end{itemize}

  \subsection{Organization}

  The organization of the rest of this paper is as follows. Section~\ref{sec:overview} reviews the syntax and behavior of algebraic effects and handlers, revisits effect systems presented in the past, and overviews our approach. Section~\ref{sec:effdef} introduces definitions of our abstract treatment of effects. Section~\ref{sec:calculus} presents our language's syntax, semantics, and type-and-effect system. Section~\ref{sec:safety} shows safety conditions that dealings with effects must meet to prove safety properties. Section~\ref{sec:erasure} defines extended conditions, type-erasure safety conditions, for parametric effects with type-erasure. Section~\ref{sec:effseq} illustrates one of the instances of our system. Section~\ref{sec:discussion} discusses the further extensions of our language. Section~\ref{sec:related} compare related works with our work. Finally, Section~\ref{sec:conclusion} concludes this paper and mentions the possible future directions of our work.
}

To show that our framework is expressive enough to capture the essence shared
among sound effect systems in the literature, we provide three instances of our
effect system.
The instances represent effect collections by sets and two styles of
rows---called \emph{simple rows}~\cite{hillerstrom_liberating_2016} and
\emph{scoped rows}~\cite{leijen_type_2017}. 
We define effect algebras for these three instances and prove their safety,
which means that all the instances satisfy type-and-effect safety.
We also show that these instances indeed model the existing effect systems~\cite{pretnar_introduction_2015,
  hillerstrom_liberating_2016, leijen_type_2017}.

Once it turns out that all the instances satisfy the desired property, what are
differences among them? How can they be compared?  To answer these questions, we
make two changes on the language: introduction of \emph{lift
  coercions}~\citep{biernacki_handle_2018,biernacki_abstracting_2019} and
employment of a \emph{type-erasure semantics}~\citep{biernacki_abstracting_2019}.

Lift coercions are a construct to prevent an operation call from being handled by
the closest effect handler, introduced to avoid \emph{accidental handling}, that
is, unintended handling of operation calls.
To reason about the effect of lift coercions soundly, effect collections
should be able to express how many effect handlers need to be installed on
effectful computation.
Effect collections represented by sets or simple rows cannot express it because
they collapse multiple occurrences of the same effect into one.
Thus, the instances with them result in being unsound.
By contrast, scoped rows can encode the number of necessary effect handlers due
to the ability to duplicate effects.
To enhance the importance of being able to represent the number of necessary
effect handlers in the presence of lift coercions, we propose a new instance
where effect collections are represented by \emph{multisets}.
Because multisets record the multiplicities of the elements they contain, it is
expected that the instance with multisets, as well as that with scoped rows,
satisfies type-and-effect safety even in the presence of lift coercions.
We show that it is the case by providing an additional safety condition for lift
coercions, proving that any instance of the effect system enjoys type-and-effect
safety if it meets the new safety condition as well as the original ones,
and showing that the effect algebra for scoped rows and the one for multisets meet both
the additional and original safety conditions.

The second change is to adopt a type-erasure semantics, which differs from the
original semantics in the effect comparison in the dynamic search for effect
handlers: the original semantics takes into account what type parameters effects
accompany to identify effects, while the type-erasure semantics does not.
\TS{What is a benefit of the type-erasure semantics?}
This nature of type-erasure semantics makes the instances with sets and
multisets unsound because it is in conflict with the nature of sets and multisets that the order of elements is ignored.
The row-based instances can be adapted to the type-erasure semantics by
restricting the commutativity in rows.
Even for sets and multisets, we can give type-and-effect safe instances based on
them if we admit restriction on swapping elements.

\OLD{

  \TY{[draft beginning]}


  To show that our framework is practical enough to capture the shared essence of sound effect systems, we provide five instances of our language.
  They represent effect collections by sets, multisets, and three styles of rows.
  The instances with sets and two styles of rows are inspired by the existing works~\cite{pretnar_introduction_2015, hillerstrom_liberating_2016, leijen_type_2017}.
  Concerning these three instances, we show that the translation from the cores of them to instances preserves well-typedness.
  \TY{May multisets style effect system new?}
  The instance with the last style of rows is new;
  it is a generalization of the existing row-based systems.
  We prove that the AREs of these instances are safe, and thus the
  instances satisfy type-and-effect safety.

  To formalize the difference between the instances,
  we adapt our language to two settings,
  \emph{type-erasure semantics} \citep{biernacki_abstracting_2019} and \emph{effect coercions} \citep{biernacki_handle_2018, biernacki_abstracting_2019},
  and define additional safety conditions for them.
  A type-erasure semantics ignores the type parameters in parametric effects in searching for effect handlers corresponding to invoked effects.
  Effect coercions explicitly manipulate the matching of operations and handlers.
  Whether we can soundly adapt an effect system to these settings or not splits the five instances into four categories, as summarized in Table~\ref{tbl:instance_soundness}.
  \begin{table}
    \caption{Soundness of instances in various settings.}
    \TY{The following tables are not final versions.}
    \label{tbl:instance_soundness}
    \begin{tabular}{c|c|c}
      \diagbox[width=3cm]{Effect                            \\coercions}{Type-erasure\\semantics} & Unsound   & Soundly extensible   \\
      \hline
      Unsound            & Sets      & Simple rows          \\
      \hline
      Soundly extensible & Multisets & Scoped and Free rows
    \end{tabular}

    \vspace{2ex}
    \begin{tabular}{c|c|c}
      \diagbox[height=3\line, width=5cm]{Effect coercions}{Type-erasure semantics} & Unsound   & Soundly extensible   \\
      \hline
      Unsound                                                                      & Sets      & Simple rows          \\
      \hline
      Soundly extensible                                                           & Multisets & Scoped and Free rows
    \end{tabular}

    \vspace{2ex}
    \begin{tabular}{c|c|c|c}
                  & Normal setting & Type-erasure semantics & Effect Coercions \\
      \hline
      Sets        & \cmark         & \xmark                 & \xmark           \\
      Multisets   & \cmark         & \xmark                 & \cmark           \\
      Simple rows & \cmark         & \cmark                 & \xmark           \\
      Scoped rows & \cmark         & \cmark                 & \cmark           \\
      Free rows   & \cmark         & \cmark                 & \cmark
    \end{tabular}
  \end{table}
  Sound extensibilities of these instances for type-erasure semantics and effect coercions correspond to their satisfiability for additional safety conditions.
  For example, while a multisets-based effect system is unsound with type-erasure semantics and sound with effect coercions, an ARE for multiset violates the safety condition for type-erasure semantics and meets the one for effect coercions.

  \TY{[draft end]}
}

The contributions of this work are summarized as follows.
\begin{itemize}
  \item We introduce an abstract effect system for algebraic effect handlers.
        It abstracts over effect algebras, which characterize the representation and manipulation of effect collections in the effect system.
  \item We define safety conditions that enforce the effect manipulation allowed by effect algebras
        to be safe.
  \item We prove that effect systems instantiated by safe effect algebras are type-and-effect safe.
  \item We extend the effect system to lift coercions and type-erasure semantics,
        define an additional safety condition for each of them, and
        prove that an instance of each extension is type-and-effect safe provided
        that the effect algebra in the instance meets the specified conditions.
  \item We give four examples of safe effect algebras and their
        variants for the type-erasure semantics.
\end{itemize}

The effect system presented in this paper supposes \emph{deep} effect handlers,
but we also have adapted the system to \emph{shallow} effect
handlers~\cite{kammar_handlers_2013}; readers interested in the formulation for
shallow effect handlers are referred to the supplementary material.

The rest of this paper is organized as follows.
Section~\ref{sec:overview} reviews algebraic effect handlers and the existing
effect systems, and overviews our approach.
Section~\ref{sec:effdef} introduces our type-and-effect language and effect algebras.  We also show the
instances based on sets and rows as their examples.
Section~\ref{sec:calculus} presents our calculus with the abstract effect
system.
Section~\ref{sec:safety} states safety conditions, explains their necessities,
and proves the type-and-effect safety of the calculus under the safe conditions.
\TY{The explanations of sections are changed.}
Section~\ref{sec:comparisons} shows that some existing effect systems can be modeled soundly by the corresponding instances of our calculus.
Section~\ref{sec:extensions} extends our language and the safety conditions to lift coercions and type-erasure semantics and
Section~\ref{sec:compare-effect-algebras} compares the effect algebras given in the paper.
%
%
%
%
%
Section~\ref{sec:related} describes additional related works and
Section~\ref{sec:conclusion} concludes this paper with future
works.
This paper only states certain key properties. All the auxiliary lemmas, proofs,
and full definition are given in the supplementary material.

\section{Overview}\label{sec:overview}

\TY{
  Changed $\mathsf{IO}$ to $\mathsf{Writer}$.
}

\OLD{
  This section reviews algebraic effects and handlers first, effect systems for them secondly, and introduces our approaches for abstracting effect systems lastly.
}

This section reviews algebraic effect handlers and the existing effect systems
for them, and provides an overview of our approach to abstracting the effect
systems.

\subsection{Review: Algebraic Effects and Handlers}\label{subsec:algeff}

\OLD{
  Algebraic effects and handlers consist of operations, which define the interface of effects, and handlers, which define the behavior of effects. Handlers consist of two parts: one return clause and some operation clauses. A return clause determines the action when a handled expression is a value. Operation clauses determine how operations work using delimited continuations.

  For example, we show the implementation of an exception handler via algebraic effects and handlers.
  \begin{flalign*}
     &  \mathsf{Exc}  ::  \ottsym{\{}  \mathsf{raise}  \ottsym{:}    \forall   \alpha  \ottsym{:}   \mathbf{Typ}    \ottsym{.}     \mathsf{Unit}    \Rightarrow   \alpha   \ottsym{\}}  & \\
     & \mathit{g} = \lambda  \mathit{x}  \ottsym{:}   \mathsf{Int}   \ottsym{.}  \lambda  \mathit{y}  \ottsym{:}   \mathsf{Int}   \ottsym{.}   \mathbf{handle}_{ \mathsf{Exc} }  \,      & \\ &    \quad    \quad    \quad   \mathbf{if} \, \mathit{y}  \ottsym{=}  0 \, \mathbf{then} \,  \mathsf{raise} _{ \mathsf{Exc} }  \,  \mathsf{Int}  \,  ()  \, \mathbf{else} \, \mathit{x}      \ottsym{/}  \mathit{y}    & \\ &    \quad    \quad  \, \mathbf{with} \,  \ottsym{\{} \, \mathbf{return} \, \mathit{z}  \mapsto   \textnormal{\ttfamily int\_to\_string}  \, \mathit{z}  \ottsym{\}}   \uplus   \ottsym{\{}  \mathsf{raise} \, \alpha  \ottsym{:}   \mathbf{Typ}  \, \mathit{p} \, \mathit{k}  \mapsto   \textnormal{\texttt{"divided by 0"} }   \ottsym{\}} 
  \end{flalign*}
  In this example, $\mathsf{Exc}$ is a label consisting of one operation $\mathsf{raise}$. $\mathsf{raise}$ takes an type $\alpha$ and a value of $ \mathsf{Unit} $, and returns an expression of $\alpha$. The handler
  \begin{flalign*}
     \ottsym{\{} \, \mathbf{return} \, \mathit{z}  \mapsto   \textnormal{\ttfamily int\_to\_string}  \, \mathit{z}  \ottsym{\}}   \uplus   \ottsym{\{}  \mathsf{raise} \, \alpha  \ottsym{:}   \mathbf{Typ}  \, \mathit{p} \, \mathit{k}  \mapsto   \textnormal{\texttt{"divided by 0"} }   \ottsym{\}} 
  \end{flalign*}
  means that: if a handled expression is a value, then the whole handling expression is $ \textnormal{\ttfamily int\_to\_string}  \, \mathit{z}$ binding the value to $\mathit{z}$; and if an operation $\mathsf{raise}$ arises in a handled expression, then the whole handling expression is $ \textnormal{\texttt{"divided by 0"} } $ discarding $\alpha$ and $\mathit{p}$, which bound to the parameters provided with $\mathsf{raise}$, and discarding $\mathit{k}$, which is a delimited continuation at the point where $\mathsf{raise}$ arises. Thus, $\mathit{g}$ takes two integers named as $\mathit{x}$ and $\mathit{y}$, and returns the string $ \textnormal{\texttt{"divided by 0"} } $ if $\mathit{y}  \ottsym{=}  0$, otherwise the sting of $\mathit{y}  \ottsym{/}  \mathit{x}$, e.g. $\mathit{g} \, 42 \, 2$ results in $\texttt{"21"}$ and $\mathit{g} \, 42 \, 0$ in $ \textnormal{\texttt{"divided by 0"} } $.

  We show another example using delimited continuations.
  \begin{flalign*}
     &  \mathsf{Choice}  ::  \ottsym{\{}  \mathsf{decide}  \ottsym{:}    \mathsf{Unit}    \Rightarrow    \mathsf{Bool}    \ottsym{\}}  & \\
     &  \mathbf{handle}_{ \mathsf{Choice} }  \,     & \\ &    \quad   \mathbf{let} \, \mathit{x}  \ottsym{=}  \mathbf{if} \,  \mathsf{decide} _{ \mathsf{Choice} }  \,  {}  \,  ()  \, \mathbf{then} \, 20 \, \mathbf{else} \, 10 \, \mathbf{in} \,  & \\ &    \quad   \mathbf{let} \, \mathit{y}  \ottsym{=}  \mathbf{if} \,  \mathsf{decide} _{ \mathsf{Choice} }  \,  {}  \,  ()  \, \mathbf{then} \, 5 \, \mathbf{else} \, 0 \, \mathbf{in} \,  & \\ &    \quad   \mathit{x}        -  \mathit{y}     & \\ &  \, \mathbf{with} \,  \ottsym{\{} \, \mathbf{return} \, \mathit{z}  \mapsto  \mathit{z}  \ottsym{\}}   \uplus   \ottsym{\{}  \mathsf{decide} \,  {}  \, \mathit{z} \, \mathit{k}  \mapsto   \textnormal{\texttt{max} } ( \mathit{k} \,  \mathsf{true}   ,  \mathit{k} \,  \mathsf{false}  )   \ottsym{\}} 
  \end{flalign*}
  The operation $\mathsf{decide}$ takes a value of $ \mathsf{Unit} $ and returns an expression of $ \mathsf{Bool} $. This handler
  \begin{flalign*}
     \ottsym{\{} \, \mathbf{return} \, \mathit{z}  \mapsto  \mathit{z}  \ottsym{\}}   \uplus   \ottsym{\{}  \mathsf{decide} \,  {}  \, \mathit{z} \, \mathit{k}  \mapsto   \textnormal{\texttt{max} } ( \mathit{k} \,  \mathsf{true}   ,  \mathit{k} \,  \mathsf{false}  )   \ottsym{\}} 
  \end{flalign*}
  has an operation clause returning the maximum value of all cases $ \mathsf{decide} _{ \mathsf{Choice} }  \,  {}  \,  () $ is true or false. The evaluation is as follows. At the first point where operation $\mathsf{decide}$ arises, the delimited continuation is
  \begin{flalign*}
     &  \mathbf{handle}_{ \mathsf{Choice} }  \,    & \\ &    \quad   \mathbf{let} \, \mathit{x}  \ottsym{=}  \mathbf{if} \, \Box \, \mathbf{then} \, 20 \, \mathbf{else} \, 10 \, \mathbf{in} \,   & \\ &    \quad   \mathbf{let} \, \mathit{y}  \ottsym{=}  \mathbf{if} \,  \mathsf{decide} _{ \mathsf{Choice} }  \,  {}  \,  ()  \, \mathbf{then} \, 5 \, \mathbf{else} \, 0 \, \mathbf{in} \,  & \\ &    \quad   \mathit{x}      -  \mathit{y}       & \\ &  \, \mathbf{with} \,  \ottsym{\{} \, \mathbf{return} \, \mathit{z}  \mapsto  \mathit{z}  \ottsym{\}}   \uplus   \ottsym{\{}  \mathsf{decide} \,  {}  \, \mathit{z} \, \mathit{k}  \mapsto   \textnormal{\texttt{max} } ( \mathit{k} \,  \mathsf{true}   ,  \mathit{k} \,  \mathsf{false}  )   \ottsym{\}} ,
  \end{flalign*}
  where $ \Box $ denotes a hole. This delimited continuation is bound to $\mathit{k}$ in the operation clause, and the evaluation result is $ \textnormal{\texttt{max} } ( \mathit{k} \,  \mathsf{true}   ,  \mathit{k} \,  \mathsf{false}  ) $. The function application $\mathit{k} \,  \mathsf{true} $ in this expression is the expression filling the delimited continuation with $ \mathsf{true} $. Then, the evaluation of $\mathit{k} \,  \mathsf{true} $ is
  \begin{flalign*}
     &  \mathbf{handle}_{ \mathsf{Choice} }  \,     & \\ &    \quad   \mathbf{let} \, \mathit{y}  \ottsym{=}  \mathbf{if} \,  \mathsf{decide} _{ \mathsf{Choice} }  \,  {}  \,  ()  \, \mathbf{then} \, 5 \, \mathbf{else} \, 0 \, \mathbf{in} \,  & \\ &    \quad   20      -  \mathit{y}     & \\ &  \, \mathbf{with} \,  \ottsym{\{} \, \mathbf{return} \, \mathit{z}  \mapsto  \mathit{z}  \ottsym{\}}   \uplus   \ottsym{\{}  \mathsf{decide} \,  {}  \, \mathit{z} \, \mathit{k}  \mapsto   \textnormal{\texttt{max} } ( \mathit{k} \,  \mathsf{true}   ,  \mathit{k} \,  \mathsf{false}  )   \ottsym{\}} .
  \end{flalign*}
  This expression evaluates to $ \textnormal{\texttt{max} } ( 15  ,  20 ) $ similarly. The function application $\mathit{k} \,  \mathsf{false} $ results in $ \textnormal{\texttt{max} } ( 5  ,  10 ) $ in the same way. Thus, the program evaluates to $20$.
}

Algebraic effect handlers are a means to implement user-defined effects in a modular way.
The interface of effects consists of operations, and their behavior is specified
by effect handlers.

For example, consider the following program that uses effect $\mathsf{Choice}$ (this
paper uses ML-like syntax to describe programs):
\begin{flalign*}
   & \effdecl{ \mathsf{Choice}  ::  \ottsym{\{}  \mathsf{decide}  \ottsym{:}    \mathsf{Unit}    \Rightarrow    \mathsf{Bool}    \ottsym{\}} } & \\
   &  \mathbf{handle}_{ \mathsf{Choice} }  \,     & \\ &    \quad   \mathbf{let} \, \mathit{x}  \ottsym{=}  \mathbf{if} \,  \mathsf{decide}  \,  ()  \, \mathbf{then} \, 20 \, \mathbf{else} \, 10 \, \mathbf{in} \, \mathbf{let} \, \mathit{y}  \ottsym{=}  \mathbf{if} \,  \mathsf{decide}  \,  ()  \, \mathbf{then} \, 5 \, \mathbf{else} \, 0 \, \mathbf{in} \, \mathit{x}    -  \mathit{y}     & \\ &  \, \mathbf{with} \,  \ottsym{\{} \, \mathbf{return} \, \mathit{z}  \mapsto  \mathit{z}  \ottsym{\}}   \uplus   \ottsym{\{}  \mathsf{decide} \, \mathit{z} \, \mathit{k}  \mapsto   \textnormal{\texttt{max} } ( \mathit{k} \,  \mathsf{true}   ,  \mathit{k} \,  \mathsf{false}  )   \ottsym{\}} 
\end{flalign*}
The first line declares \emph{effect label} $\mathsf{Choice}$ with only one operation $\mathsf{decide}$.
%
As indicated by its type, $\mathsf{decide}$ takes the unit value and returns a
Boolean.
The program invokes the operation in the third line, determines
numbers $\mathit{x}$ and $\mathit{y}$ depending on the results, and returns
$ \mathit{x}  -  \mathit{y} $ finally.
To install an effect handler, we use the handling construct $ \mathbf{handle}  \,  \!\textnormal{--}\!  \, \mathbf{with}$.

In general, an expression $ \mathbf{handle}_{ \mathit{l} }  \, \ottnt{e} \, \mathbf{with} \, \ottnt{h}$ means that an expression $\ottnt{e}$ is
executed under effect handler $\ottnt{h}$, which interprets the operations of effect
label $\mathit{l}$ invoked by $\ottnt{e}$; we call $\ottnt{e}$ a \emph{handled expression}.
An effect handler consists of one return clause and possibly several operation
clauses. A return clause $\ottsym{\{} \, \mathbf{return} \, \mathit{x}  \mapsto  \ottnt{e_{\ottmv{r}}}  \ottsym{\}}$,
which corresponds to $\ottsym{\{} \, \mathbf{return} \, \mathit{z}  \mapsto  \mathit{z}  \ottsym{\}}$ in the example,
is executed when a handled expression evaluates to a value,
which the body $\ottnt{e_{\ottmv{r}}}$ references by $\mathit{x}$.
An operation clause takes the form $\ottsym{\{}  \mathsf{op} \,  {}  \, \mathit{x} \, \mathit{k}  \mapsto  \ottnt{e}  \ottsym{\}}$, which
determines the implementation of operation $\mathsf{op}$.
%
When an operation $\mathsf{op}$ is called with an argument $\ottnt{v}$ under an effect
handler with operation clause $\ottsym{\{}  \mathsf{op} \,  {}  \, \mathit{x} \, \mathit{k}  \mapsto  \ottnt{e}  \ottsym{\}}$, the reduction proceeds
as follows.
First, the remaining computation from the point of the operation call up to the
$ \mathbf{handle}  \,  \!\textnormal{--}\!  \, \mathbf{with}$ construct installing the effect handler is
captured; such a computation is called a \emph{delimited continuation}.
Then, the body $\ottnt{e}$ of the corresponding operation clause is executed by passing
the argument $\ottnt{v}$ as $\mathit{x}$ and the delimited continuation as $\mathit{k}$.

In the example, the delimited continuation for the first call to $\mathsf{decide}$ is
\begin{flalign*}
   &  \mathbf{handle}_{ \mathsf{Choice} }  \,    & \\ &    \quad   \mathbf{let} \, \mathit{x}  \ottsym{=}  \mathbf{if} \, \Box \, \mathbf{then} \, 20 \, \mathbf{else} \, 10 \, \mathbf{in} \,  \mathbf{let} \, \mathit{y}  \ottsym{=}  \mathbf{if} \,  \mathsf{decide}  \,  ()  \, \mathbf{then} \, 5 \, \mathbf{else} \, 0 \, \mathbf{in} \, \mathit{x}  -  \mathit{y}       & \\ &  \, \mathbf{with} \,  \ottsym{\{} \, \mathbf{return} \, \mathit{z}  \mapsto  \mathit{z}  \ottsym{\}}   \uplus   \ottsym{\{}  \mathsf{decide} \, \mathit{z} \, \mathit{k}  \mapsto   \textnormal{\texttt{max} } ( \mathit{k} \,  \mathsf{true}   ,  \mathit{k} \,  \mathsf{false}  )   \ottsym{\}} ,
\end{flalign*}
where $ \Box $ denotes a hole.
The functional form $\ottnt{v_{{\mathrm{1}}}}$ of this delimited continuation is bound to variable $\mathit{k}$ in the operation clause
of $\mathsf{decide}$, and the program evaluates to $ \textnormal{\texttt{max} } ( \ottnt{v_{{\mathrm{1}}}} \,  \mathsf{true}   ,  \ottnt{v_{{\mathrm{1}}}} \,  \mathsf{false}  ) $. The
function application $\ottnt{v_{{\mathrm{1}}}} \,  \mathsf{true} $ fills the hole of the delimited continuation with argument $ \mathsf{true} $.
Thus, it reduces
\begin{flalign*}
   &  \mathbf{handle}_{ \mathsf{Choice} }  \,     & \\ &    \quad   \mathbf{let} \, \mathit{x}  \ottsym{=}  \mathbf{if} \,  \colorbox{syntaxhighlight}{$  \mathsf{true}  $}  \, \mathbf{then} \, 20 \, \mathbf{else} \, 10 \, \mathbf{in} \, \mathbf{let} \, \mathit{y}  \ottsym{=}  \mathbf{if} \,  \mathsf{decide}  \,  ()  \, \mathbf{then} \, 5 \, \mathbf{else} \, 0 \, \mathbf{in} \, \mathit{x}    -  \mathit{y}     & \\ &  \, \mathbf{with} \,  \ottsym{\{} \, \mathbf{return} \, \mathit{z}  \mapsto  \mathit{z}  \ottsym{\}}   \uplus   \ottsym{\{}  \mathsf{decide} \, \mathit{z} \, \mathit{k}  \mapsto   \textnormal{\texttt{max} } ( \mathit{k} \,  \mathsf{true}   ,  \mathit{k} \,  \mathsf{false}  )   \ottsym{\}} ,
\end{flalign*}
where $ \colorbox{syntaxhighlight}{$  \mathsf{true}  $} $ comes from the argument.
Then, it substitutes $20$ for $\mathit{x}$, and then calls $\mathsf{decide}$ again.
The operation clause invokes the delimited continuation $\ottnt{v_{{\mathrm{2}}}}$ captured by the
second call with arguments $ \mathsf{true} $ and $ \mathsf{false} $. The applications
$\ottnt{v_{{\mathrm{2}}}} \,  \mathsf{true} $ and $\ottnt{v_{{\mathrm{2}}}} \,  \mathsf{false} $ choose $5$ and $0$ as $\mathit{y}$ and return the
results of $ 20  -  5 $ and $ 20  -  0 $ (that is, $15$ and $20$), respectively.  Then, the operation clause
return $ \textnormal{\texttt{max} } ( 15  ,  20 ) $ as the result of $\ottnt{v_{{\mathrm{1}}}} \,  \mathsf{true} $.
Similarly, the function application $\ottnt{v_{{\mathrm{1}}}} \,  \mathsf{false} $ results in $ \textnormal{\texttt{max} } ( 5  ,  10 ) $.
Thus, the entire program evaluates to $ \textnormal{\texttt{max} } (  \textnormal{\texttt{max} } ( 15  ,  20 )   ,   \textnormal{\texttt{max} } ( 5  ,  10 )  ) $ and then to $20$ finally.

While the operation clause in the above example uses captured continuations,
effect handlers can also discard them.
Using this ability, we can implement exception handling, as the following program that divides $\mathit{x}$ by $\mathit{y}$ if $\mathit{y}$ is nonzero:
\begin{flalign*}
   & \effdecl{ \mathsf{Exc}  ::  \ottsym{\{}  \mathsf{raise}  \ottsym{:}    \mathsf{Unit}    \Rightarrow    \mathsf{Empty}    \ottsym{\}} }    & \\
   & \letdecl{\mathit{g}} \lambda  \mathit{x}  \ottsym{:}   \mathsf{Int}   \ottsym{.}  \lambda  \mathit{y}  \ottsym{:}   \mathsf{Int}   \ottsym{.}   \   \mathbf{handle}_{ \mathsf{Exc} }  \,                (  \mathbf{if} \, \mathit{y}  \ottsym{=}  0 \, \mathbf{then} \,  \mathsf{raise}  \,  ()  \, \mathbf{else} \, \mathit{x}  \ottsym{/}  \mathit{y}  )     & \\ &   \   \   \    \quad    \quad    \quad    \quad    \quad    \quad    \quad    \quad    \quad  \, \mathbf{with} \,  \ottsym{\{} \, \mathbf{return} \, \mathit{z}  \mapsto   \textnormal{\ttfamily int\_to\_string}  \, \mathit{z}  \ottsym{\}}   \uplus   \ottsym{\{}  \mathsf{raise} \, \mathit{p} \, \mathit{k}  \mapsto   \textnormal{\texttt{"divided by 0"} }   \ottsym{\}}  
\end{flalign*}
In this example, $\mathsf{Exc}$ is an effect label consisting of one operation
$\mathsf{raise}$ with type $  \mathsf{Unit}    \Rightarrow    \mathsf{Empty}  $.  Here, $ \mathsf{Empty} $ is a type having no
inhabitant, and we assume that an expression of this type can be regarded as that
of any type.
The return clause of the effect handler means that, when the handled expression evaluates to an
integer, the handling construct returns its string version.
Because the operation clause for $\mathsf{raise}$ discards the continuations, the
handling construct returns the string $ \textnormal{\texttt{"divided by 0"} } $ immediately once
$\mathsf{raise}$ is called.
Therefore, the operation call and effect handling in this example correspond to
excepting raising and handling, respectively.

\subsection{Effect Systems for Algebraic Effects and Handlers}\label{subsec:effsys}

This section briefly explains a role of effect systems for algebraic effect handlers and
summarizes the existing systems.

\subsubsection{A Role of Effect Systems}
\label{subsec:effsys:role}

\OLD{
  Effect systems statically guarantee that there are no unhandled operations. We call this property effect safety. One of the essential features of algebraic effects and effects handlers is that effect operators arise, and outer enclosing handlers handle them to determine the actions of these operators. Therefore, if an operator arises without being enclosed by appropriate handlers, the operator's behavior is unknown, and the whole program gets stuck.

  For example, consider the following program.
  \begin{flalign*}
     &  \mathsf{Exc}  ::  \ottsym{\{}  \mathsf{raise}  \ottsym{:}    \forall   \alpha  \ottsym{:}   \mathbf{Typ}    \ottsym{.}     \mathsf{Unit}    \Rightarrow   \alpha   \ottsym{\}}   \ottsym{,}   \mathsf{IO}  ::  \ottsym{\{}  \mathsf{print}  \ottsym{:}    \mathsf{String}    \Rightarrow    \mathsf{Unit}    \ottsym{\}}  & \\
     & \mathit{g} = \lambda  \mathit{x}  \ottsym{:}   \mathsf{Int}   \ottsym{.}  \lambda  \mathit{y}  \ottsym{:}   \mathsf{Int}   \ottsym{.}   \mathbf{handle}_{ \mathsf{Exc} }  \,      & \\ &    \quad    \quad    \quad   \mathbf{if} \, \mathit{y}  \ottsym{=}  0 \, \mathbf{then} \,  \mathsf{raise} _{ \mathsf{Exc} }  \,  \mathsf{Int}  \,  ()  \, \mathbf{else} \, \mathit{x}      \ottsym{/}  \mathit{y}    & \\ &    \quad    \quad  \, \mathbf{with} \,  \ottsym{\{} \, \mathbf{return} \, \mathit{z}  \mapsto   \mathsf{print} _{ \mathsf{IO} }  \,  {}  \,  (   \textnormal{\ttfamily int\_to\_string}  \, \mathit{z}  )   \ottsym{\}}    & \\ &    \quad    \quad    \quad    \quad   \uplus        \ottsym{\{}  \mathsf{raise} \, \alpha  \ottsym{:}   \mathbf{Typ}  \, \mathit{z} \, \mathit{k}  \mapsto   \mathsf{print} _{ \mathsf{IO} }  \,  {}  \,  \textnormal{\texttt{"divided by 0"} }   \ottsym{\}} 
  \end{flalign*}
  This function $\mathit{g}$ takes two integers named $x$ and $y$. If $y$ equals $0$, the raise operator arises; otherwise, the enclosing $\mathsf{Exc}$ handler takes the result of $x$ divided by $y$. When receiving the value, the handler gives the string of it to $ \mathsf{print} _{ \mathsf{IO} }  \,  {} $ as an argument. When the raise operator arises, the handler discards the delimited continuation and gives $ \textnormal{\texttt{"divided by 0"} } $ to $ \mathsf{print} _{ \mathsf{IO} }  \,  {} $. Thus, the function application like $\mathit{g} \, 42 \, 2$ and $\mathit{g} \, 42 \, 0$ is effect unsafe because no $\mathsf{IO}$ handlers enclose it. These applications get stuck indeed. In other words, the expression
  \begin{align*}
     \mathbf{handle}_{ \mathsf{IO} }  \, \mathit{g} \, 42 \, 0 \, \mathbf{with} \,  \ottsym{\{} \, \mathbf{return} \, \mathit{z}  \mapsto  \mathit{z}  \ottsym{\}}   \uplus   \ottsym{\{}  \mathsf{print} \,  {}  \, \mathit{p} \, \mathit{k}  \mapsto  \mathit{p}  \ottsym{\}} 
  \end{align*}
  is safe because no operations remain at the top level due to handling; it is reduced to $ \textnormal{\texttt{"divided by 0"} } $.

  There are mainly two styles of formalization of effect systems: the set-based approach \citep{kammar_handlers_2013, bauer_effect_2013} and the row-based approach \citep{leijen_type_2017,biernacki_abstracting_2019}.
}

A property ensured by many effect systems in the literature 
is \emph{effect safety}, which means that there is no unhandled operation.
A simple example that breaks effect safety is $ \mathsf{op}  \,  {}  \, \ottnt{v}$, which just
invokes operation $\mathsf{op}$.  Because no effect handler for $\mathsf{op}$ is
given---thus, there is no way to interpret it---the program gets stuck.
However, even if an operation call is enclosed by handling constructs,
effect safety can be broken.
For example, consider the following program:
\begin{flalign*}
   & \effdecl{ \mathsf{Exc}  ::  \ottsym{\{}  \mathsf{raise}  \ottsym{:}    \mathsf{Unit}    \Rightarrow    \mathsf{Empty}    \ottsym{\}} } & \\
   & \effdecl{ \mathsf{State}  ::  \ottsym{\{}  \mathsf{set}  \ottsym{:}    \mathsf{Int}    \Rightarrow    \mathsf{Unit}    \ottsym{,}  \mathsf{get}  \ottsym{:}    \mathsf{Unit}    \Rightarrow    \mathsf{Int}    \ottsym{\}} }   & \\
   & \letdecl{\mathit{g}} \lambda  \mathit{x}  \ottsym{:}   \mathsf{Int}   \ottsym{.}   \   \mathbf{handle}_{ \mathsf{Exc} }  \,           (  \mathbf{if} \, \mathit{x}  \ottsym{=}  0 \, \mathbf{then} \,  \mathsf{raise}  \,  ()  \, \mathbf{else} \,  (  \mathbf{let} \, \mathit{y}  \ottsym{=}   \mathsf{get}  \,  ()   \ottsym{/}  \mathit{x} \, \mathbf{in} \,   \mathsf{set}  \, \mathit{y}  \ottsym{;}  \mathit{y}   )   )     & \\ &   \    \quad    \quad    \quad    \quad    \quad    \quad  \, \mathbf{with} \,  \ottsym{\{} \, \mathbf{return} \, \mathit{z}  \mapsto   \textnormal{\ttfamily int\_to\_string}  \, \mathit{z}  \ottsym{\}}   \uplus   \ottsym{\{}  \mathsf{raise} \, \mathit{p} \, \mathit{k}  \mapsto   \textnormal{\texttt{"divided by 0"} }   \ottsym{\}}  
   &                                                   \\
   & \mathit{g} \, 42 \, 2
\end{flalign*}
The effect label $\mathsf{State}$ is for mutable state, providing two operations
$\mathsf{set}$ and $\mathsf{get}$ to update and get the current values in the state.
The function $\mathit{g}$ divides the current value of the state (returned by
$\mathsf{get}$) by $\mathit{x}$, sets the result to the state, and returns it if $\mathit{x}$
is nonzero.
All the operation calls in the application $\mathit{g} \, 42 \, 2$ at the last line are performed
under the effect handler, but the call to $\mathsf{get}$ is not handled.
Hence, this example is not effect safe.

In general, the effect systems enjoying effect safety need to track which effect
each expression may invoke and which effect an effect handler targets.
However, there are choices to represent the effects caused by expressions.
Thus far, mainly two styles of formalization of effect systems
have been studied: one is based on \emph{sets}~\cite{kammar_handlers_2013,bauer_effect_2013,kammar_no_2017,forster_expressive_2017,saleh_explicit_2018,sekiyama_signature_2020}, and the other is
based on \emph{rows}~\cite{hillerstrom_liberating_2016, leijen_type_2017, biernacki_abstracting_2019, xie_first-class_2022}.
%

\subsubsection{Set-Based Effect Systems}

Set-based effect systems assign to an expression a set of
effect labels that the expression may invoke.
%
%
For example, they assign to an operation call a set that includes the effect
label of the called operation.
This is formalized as follows, where typing judgment $\Gamma  \vdash  \ottnt{e}  \ottsym{:}  \ottnt{A}  \mid  {s}$
means that expression $\ottnt{e}$ is of type $\ottnt{A}$ under typing context $\Gamma$
and may invoke effects in set ${s}$:
\begin{mathpar}
  \inferrule
  {\text{Operation $\mathsf{op}  \ottsym{:}   \ottnt{A}   \Rightarrow   \ottnt{B} $ belongs to effect $\mathit{l}$} \\ \Gamma  \vdash  \ottnt{v}  \ottsym{:}  \ottnt{A}  \mid  \{  \}}
  {\Gamma  \vdash   \mathsf{op}  \,  {}  \, \ottnt{v}  \ottsym{:}  \ottnt{B}  \mid  \{  \mathit{l}  \}}
\end{mathpar}
Subeffecting, which is supported to unify the effects of multiple expressions (such as branches in conditional expressions),
is implemented by allowing the expansion of sets:
\begin{mathpar}
  \inferrule
  {\Gamma  \vdash  \ottnt{e}  \ottsym{:}  \ottnt{A}  \mid  {s} \\  {s}   \subseteq   {s}' }
  {\Gamma  \vdash  \ottnt{e}  \ottsym{:}  \ottnt{A}  \mid  {s}'}
\end{mathpar}
In the presence of algebraic effect handlers, sets not only expand but
also may shrink. Such manipulation is performed in handling constructs:
\begin{mathpar}
  \inferrule
  { {\Gamma  \vdash  \ottnt{e}  \ottsym{:}  \ottnt{A}  \mid  {s}} \and {\{  \mathit{l}  \} \cup {s}' = {s}} \and \cdots}
  {\Gamma  \vdash   \mathbf{handle}_{ \mathit{l} }  \, \ottnt{e} \, \mathbf{with} \, \ottnt{h}  \ottsym{:}  \ottnt{B}  \mid  {s}'}
\end{mathpar}
where the omitted premise states that $\ottnt{h}$ is a handler for effect $\mathit{l}$,
translating a computation of type $\ottnt{A}$ to type $\ottnt{B}$.
This inference rule is matched with the behavior of the handling constructs
because they can make handled effects $\mathit{l}$ ``unobservable.''
The set-based effect systems defined in such a way can soundly overapproximate
the observable effects of programs and guarantee the effect safety of
expressions to which the empty set can be assigned.

For instance, consider the example in Section~\ref{subsec:effsys:role}.
A set-based effect system would assign the set $\{  \mathsf{Exc}  \ottsym{,}  \mathsf{State}  \}$ to
the handled expression
$\mathbf{if} \, \mathit{x}  \ottsym{=}  0 \, \mathbf{then} \,  \mathsf{raise}  \,  ()  \, \mathbf{else} \,  (  \mathbf{let} \, \mathit{y}  \ottsym{=}   \mathsf{get}  \,  ()   \ottsym{/}  \mathit{x} \, \mathbf{in} \,   \mathsf{set}  \, \mathit{y}  \ottsym{;}  \mathit{y}   ) $
because it calls operation $\mathsf{raise}$ of $\mathsf{Exc}$ or $\mathsf{get}$ and $\mathsf{set}$ of $\mathsf{State}$.
Because this expression is only placed under the effect handler for $\mathsf{Exc}$, the
entire program $\mathit{g} \, 42 \, 2$ could have set $\{  \mathsf{State}  \}$.
As this set indicates that effect $\mathsf{State}$ may not be handled---and it \emph{is
  not} actually---the effect system would conclude that the program may not be effect
safe.
If the program were wrapped by a handling construct with an effect handler for
$\mathsf{State}$, the empty set could be assigned to it; then, we could conclude that the
program is effect safe.

\subsubsection{Row-Based Effect Systems}\label{subsec:effsys:row}

Rows express collections of effect labels in a more structured way.
In a monomorphic setting, they are just sequences of effect labels, as
$\langle  \mathit{l_{{\mathrm{1}}}}  \ottsym{,}  \ldots  \ottsym{,}  \mathit{l_{\ottmv{n}}}  \rangle$, which is the row consisting only of labels
$\mathit{l_{{\mathrm{1}}}}  \ottsym{,}  \ldots  \ottsym{,}  \mathit{l_{\ottmv{n}}}$.
Rows are identified up to the reordering of labels. For example, $\langle  \mathit{l_{{\mathrm{1}}}}  \ottsym{,}  \mathit{l_{{\mathrm{2}}}}  \rangle$
equals $\langle  \mathit{l_{{\mathrm{2}}}}  \ottsym{,}  \mathit{l_{{\mathrm{1}}}}  \rangle$.\footnote{The label reordering might need to be restricted
  if effect labels are parameterized over, e.g., types, as discussed in
  Section~\ref{sec:erasure}.}

Rows are often adapted in languages with effect
polymorphism~\cite{hillerstrom_liberating_2016, leijen_type_2017, biernacki_abstracting_2019}.
In such languages, rows are allowed to end with effect variables $\rho$, such as
$\langle  \mathit{l_{{\mathrm{1}}}}  \ottsym{,}  \ldots  \ottsym{,}  \mathit{l_{\ottmv{n}}}  \ottsym{,}  \rho  \rangle$, which means that an expression may invoke effects
$\mathit{l_{{\mathrm{1}}}}  \ottsym{,}  \ldots  \ottsym{,}  \mathit{l_{\ottmv{n}}}$ as well as those in an instance of effect variable
$\rho$.
This extension enables abstraction over rows by universally quantifying effect
variables.
For example, consider function \textsf{filtered\_set}, which, given an integer
list and a function $f$ from integers to Booleans, filters out the elements of
the list using function $f$ and then calls operation $\mathsf{set}$ of effect
$\mathsf{State}$ on the remaining elements.
Assume that the type of functions from type $\ottnt{A}$ to type $\ottnt{B}$ with effects in row ${r}$
is described as $ \ottnt{A}    \rightarrow_{ {r} }    \ottnt{B} $.
Then, \textsf{filtered\_set} can be given type
$  \forall   \rho   \ottsym{.}     \ottsym{(}    \mathsf{Int}  \,\mathsf{List}   \times  \ottsym{(}    \mathsf{Int}     \rightarrow_{ \rho }     \mathsf{Bool}    \ottsym{)}  \ottsym{)}    \rightarrow_{ \langle  \mathsf{State}  \ottsym{,}  \rho  \rangle }     \mathsf{Unit}   $.
By instantiating $\rho$ with $\langle  \mathit{l_{{\mathrm{1}}}}  \ottsym{,}  \ldots  \ottsym{,}  \mathit{l_{\ottmv{n}}}  \rangle$, this type can express that,
when passed a function $f$ that may cause effects $\mathit{l_{{\mathrm{1}}}},\cdots,\mathit{l_{\ottmv{n}}}$,
\textsf{filtered\_set} may also cause them via the application of $f$.

Inference rules of the row-based effect systems are similar to those of
set-based ones, except that subeffecting allows enlarging rows only when they do
not end with effect variables (such rows are called \emph{closed}, while rows ending
with effect variables are \emph{open}~\cite{hillerstrom_liberating_2016}):
\begin{mathpar}
  \inferrule
  {\Gamma  \vdash  \ottnt{e}  \ottsym{:}  \ottnt{A}  \mid  \langle  \mathit{l_{{\mathrm{1}}}}  \ottsym{,}  \ldots  \ottsym{,}  \mathit{l_{\ottmv{n}}}  \rangle}
  {\Gamma  \vdash  \ottnt{e}  \ottsym{:}  \ottnt{A}  \mid  \langle  \mathit{l_{{\mathrm{1}}}}  \ottsym{,}  \ldots  \ottsym{,}  \mathit{l_{\ottmv{n}}}  \ottsym{,}  {r}  \rangle}
\end{mathpar}
Rows shrink in handling constructs where handled effects are removed:
\begin{mathpar}
  \inferrule
  { {\Gamma  \vdash  \ottnt{e}  \ottsym{:}  \ottnt{A}  \mid  {r}} \and {\langle  \mathit{l}  \ottsym{,}  {r}'  \rangle = {r}} \and \cdots}
  {\Gamma  \vdash   \mathbf{handle}_{ \mathit{l} }  \, \ottnt{e} \, \mathbf{with} \, \ottnt{h}  \ottsym{:}  \ottnt{B}  \mid  {r}'}
\end{mathpar}
Similar to set-based ones, the row-based effect systems also ensure the effect
safety of expressions to which the empty row $\langle  \rangle$ can be assigned.
The reasoning about the example in Section~\ref{subsec:effsys:role} can be done
similarly to the case with simple rows.

These are the common core of the row-based effect systems, but they can be
further classified into two groups depending on the formalism of rows.
One is simple rows~\cite{hillerstrom_liberating_2016}, where each label can appear at most once in one row.
In this formalism, any $\mathit{l_{\ottmv{i}}}$ in row $\langle  \mathit{l_{{\mathrm{1}}}}  \ottsym{,}  \ldots  \ottsym{,}  \mathit{l_{\ottmv{n}}}  \rangle$ must be different
from $\mathit{l_{\ottmv{j}}}$ for any $\ottmv{j} \neq \ottmv{i}$.
The other is scoped rows~\cite{leijen_type_2017}, where the same label can appear in one row
multiple times.
Therefore, given a scoped row $\langle  \mathit{l_{{\mathrm{1}}}}  \ottsym{,}  \ldots  \ottsym{,}  \mathit{l_{\ottmv{n}}}  \rangle$,
any $\mathit{l_{\ottmv{i}}}$ is allowed to be equivalent to some $\mathit{l_{\ottmv{j}}}$, unlike simple rows.

\TY{I removed the remark about the difference between simple and scoped rows,
  because lift coercions make the difference.}

\OLD{
  \paragraph{Remark.}
  The effect systems with simple rows and those with scoped rows might look
  similar, but there are significant differences between them, as summarized as
  follows:
  \begin{itemize}
    \item Some effect systems with simple rows restrict rows substituted for
          effect variables to ensure the uniqueness of labels in a
          row~\cite{hillerstrom_liberating_2016}.
          For example, a row $\langle  \mathsf{Exc}  \ottsym{,}  \rho  \rangle$ with the effect variable $\rho$ becomes
          invalid if $\rho$ is allowed to be instantiated by a row including
          $\mathsf{Exc}$.
          The effect system of \citet{hillerstrom_liberating_2016} restricts
          instantiation of effect variables using a kind system, to support row
          polymorphism~\cite{remy_type_1993}.
          By contrast, the effect systems with scoped rows have no such restriction
          because they allow duplication of labels in the same row.
          This difference might influence the design and implementation of typechecking,
          type and effect inference, and unification for the effect systems.

    \item Given a handling construct $ \mathbf{handle}_{ \mathit{l} }  \, \ottnt{e} \, \mathbf{with} \, \ottnt{h}$, the effect systems
          with simple rows ensure that all the unhandled calls to operations of
          effect $\mathit{l}$ in $\ottnt{e}$ are interpreted by handler $\ottnt{h}$.
          However, this might not be true in the effect systems adapting scoped rows.
          In such effect systems, each call to an operation is matched with
          \emph{one} occurrence of the operation's effect.
          \TS{We need an example to make this more understandable.}
          Therefore, operation calls matched with the second or later occurrences
          of the effect may leak outside the closest handling construct.
          As a result, it is possible to interpret calls to the same operation by
          different effect handlers even when the calls are performed in the same
          context.
          This feature is useful in, for example, encoding mutable states by effect handlers
          because different operation calls may intend to manipulate different
          reference cells.~\footnote{To write such a program, we would need
            additional constructs as coercions.  Readers interested in more details
            are referred to the work of \citet{biernacki_handle_2018, biernacki_abstracting_2019}.}
  \end{itemize}
}

\OOLD{
Thus far, mainly two styles of formalization of effect systems for effect
handlers have been studied: one is based on simple
rows~\citep{kammar_handlers_2013, bauer_effect_2013,...} and the other is based on
scoped rows~\citep{leijen_type_2017,biernacki_abstracting_2019,...}.\footnote{\TS{Where do the words ``simple rows'' and ``scoped rows'' come from?}}

\subsubsection{Effect Systems with Simple Rows}

\OLD{
  A set-based effect system assigns a set of labels to an expression as an effect. It concludes that the expression is effect safe only when the effect can be the empty set; handlers remove a label from a label set. For example, an effect system of this style assigns the singleton $\{\mathsf{Exc}\}$ to $\mathbf{if} \, \mathit{y}  \ottsym{=}  0 \, \mathbf{then} \,  \mathsf{raise} _{ \mathsf{Exc} }  \,  \mathsf{Int}  \,  ()  \, \mathbf{else} \, \mathit{y}  \ottsym{/}  \mathit{x}$ and the singleton $\{\mathsf{IO}\}$ to $\mathit{g} \, 42 \, 0$. These label sets mean that the expressions may cause the effect $\mathsf{Exc}$ and the effect $\mathsf{IO}$, respectively. Note that the $\mathsf{Exc}$ handler, in the example above, removes the label $\mathsf{Exc}$ from the effect of the if-expression and adds the label $\mathsf{IO}$ to the entire label set. Thus, we need to enclose the expression by an $\mathsf{IO}$ handler to satisfy effect safety; in other words, to remove the $\mathsf{IO}$ label from $\{\mathsf{IO}\}$ resulting in the empty set.

  A set-based effect system has the advantage of simplicity in its definition. However, effect systems of this style has limited extensibility to parametric effects with type-erasure semantics.
}

Intuitively, effect systems with simple rows assign to an expression a row of
effect labels that the expression may invoke.  Rows adapted by these systems are
called \emph{simple} because they allow each label to appear at most once in the
same row.
Namely, given a simple row $\{ \mathit{l_{{\mathrm{1}}}}, \cdots, \mathit{l_{\ottmv{n}}} \}$ with effect labels
$\mathit{l_{{\mathrm{1}}}}, \cdots, \mathit{l_{\ottmv{n}}}$, any $\mathit{l_{\ottmv{i}}}$ must be different from $\mathit{l_{\ottmv{j}}}$ for any
$\ottmv{j} \neq \ottmv{i}$.
%

The effect systems manipulate rows according to each constructor.
For example, they assign to an operation call a row that includes the effect
label of the called operation.
This is formalized as follows, where typing judgment $\Gamma  \vdash  \ottnt{e}  \ottsym{:}  \ottnt{A}  \mid  {s}$
means that expression $\ottnt{e}$ is of type $\ottnt{A}$ under typing context $\Gamma$
and may invoke effects in simple row ${s}$:
\begin{mathpar}
  \inferrule
  {\text{Operation $\mathsf{op}  \ottsym{:}   \ottnt{A}   \Rightarrow   \ottnt{B} $ belongs to effect $\mathit{l}$} \\ \Gamma  \vdash  \ottnt{v}  \ottsym{:}  \ottnt{A}  \mid  \{  \}}
  {\Gamma  \vdash   \mathsf{op}  \,  {}  \, \ottnt{v}  \ottsym{:}  \ottnt{B}  \mid  \{  \mathit{l}  \}}
\end{mathpar}
Subeffecting corresponds to enlarging rows:
\begin{mathpar}
  \inferrule
  {\Gamma  \vdash  \ottnt{e}  \ottsym{:}  \ottnt{A}  \mid  {s} \\  {s}   \subseteq   {s}' }
  {\Gamma  \vdash  \ottnt{e}  \ottsym{:}  \ottnt{A}  \mid  {s}'}
\end{mathpar}
By supporting algebraic effect handlers, rows not only enlarge, but
also may shrink. Such manipulation is performed in handling constructs:
\begin{mathpar}
  \inferrule
  { {\Gamma  \vdash  \ottnt{e}  \ottsym{:}  \ottnt{A}  \mid  {s}} \and { \{  \mathit{l}  \}  \,\underline{ \cup }\,  {s}'  = {s}} \and \cdots}
  {\Gamma  \vdash   \mathbf{handle}_{ \mathit{l} }  \, \ottnt{e} \, \mathbf{with} \, \ottnt{h}  \ottsym{:}  \ottnt{B}  \mid  {s}'}
\end{mathpar}
where the omitted premise states that $\ottnt{h}$ is a handler for effect $\mathit{l}$,
translating computation of type $\ottnt{A}$ to type $\ottnt{B}$.
This inference rule is matched with the behavior of the handling constructs,
which can make handled effects $\mathit{l}$ ``unobservable.''
This manipulation of rows enables overapproximating the observable effects of programs.
Therefore, the effect systems guarantee the effect safety of expressions to
which the empty row can be assigned.

For instance, consider the example in Section~\ref{subsec:effsys:role}.
An effect system with simple rows would assign row $\{ \mathsf{Exc}, \mathsf{IO} \}$ to
the handled expression
$\mathbf{if} \, \mathit{y}  \ottsym{=}  0 \, \mathbf{then} \,  \mathsf{raise}  \,  {}  \,  ()  \, \mathbf{else} \,  (  \mathbf{let} \, \mathit{z}  \ottsym{=}  \mathit{x}  \ottsym{/}  \mathit{y} \, \mathbf{in} \,   \mathsf{print}  \,  {}  \, \mathit{z}  \ottsym{;}  \mathit{z}   ) $
because it may call operation $\mathsf{raise}$ of $\mathsf{Exc}$ and $\mathsf{print}$ of $\mathsf{IO}$.
Because this expression is placed under the effect handler for $\mathsf{Exc}$, the
entire program $\mathit{g} \, 42 \, 0$ would have row $\{  \mathsf{IO}  \}$.
As the row indicates that effect $\mathsf{IO}$ may not be handled---and it \emph{is
  not} actually---the effect system would conclude that the program may be effect
unsafe.
If the program were wrapped by a handling construct with an effect handler for
$\mathsf{IO}$, the empty row could be assigned; therefore, we could conclude that the
program effect safe.

\subsubsection{Effect Systems with Scoped Rows}

Scoped rows~\cite{} are another formalism of rows where the same label can
appear in the same row multiple times.
Namely, given a scoped row ${r} = \langle  \mathit{l_{{\mathrm{1}}}}  \ottsym{,}  \ldots  \ottsym{,}  \mathit{l_{\ottmv{n}}}  \rangle$,
any $\mathit{l_{\ottmv{i}}}$ is allowed to be equivalent to some $\mathit{l_{\ottmv{j}}}$, unlike simple rows.
The difference from simple rows can be found in the inference rules for
subeffecting and handling constructs.
We write $\langle  \mathit{l_{{\mathrm{1}}}}  \ottsym{,}  \ldots  \ottsym{,}  \mathit{l_{\ottmv{n}}}  \ottsym{,}  {r}  \rangle$ to express $\langle  \mathit{l_{{\mathrm{1}}}}  \ottsym{,}  \ldots  \ottsym{,}  \mathit{l_{\ottmv{n}}}  \ottsym{,}  \mathit{l'_{{\mathrm{1}}}}  \ottsym{,}  \ldots  \ottsym{,}  \mathit{l'_{\ottmv{m}}}  \rangle$ when
${r} = \langle  \mathit{l'_{{\mathrm{1}}}}  \ottsym{,}  \ldots  \ottsym{,}  \mathit{l'_{\ottmv{m}}}  \rangle$.
Then, the inference rules are given as follows:
\begin{mathpar}
  \inferrule
  {\Gamma  \vdash  \ottnt{e}  \ottsym{:}  \ottnt{A}  \mid  \langle  \mathit{l_{{\mathrm{1}}}}  \ottsym{,}  \ldots  \ottsym{,}  \mathit{l_{\ottmv{n}}}  \rangle}
  {\Gamma  \vdash  \ottnt{e}  \ottsym{:}  \ottnt{A}  \mid  \langle  \mathit{l_{{\mathrm{1}}}}  \ottsym{,}  \ldots  \ottsym{,}  \mathit{l_{\ottmv{n}}}  \ottsym{,}  {r}  \rangle}

  \inferrule
  { {\Gamma  \vdash  \ottnt{e}  \ottsym{:}  \ottnt{A}  \mid  {r}} \and {\langle  \mathit{l}  \ottsym{,}  {r}'  \rangle = {r}} \and \cdots}
  {\Gamma  \vdash   \mathbf{handle}_{ \mathit{l} }  \, \ottnt{e} \, \mathbf{with} \, \ottnt{h}  \ottsym{:}  \ottnt{B}  \mid  {r}'}
\end{mathpar}
Similar to those with simple rows, the effect systems with scoped rows also
ensure the effect safety of expressions to which the empty row $\langle  \rangle$ can be
assigned.
The reasoning about the example in Section~\ref{subsec:effsys:role} can be done
similarly to the case with simple rows.

\subsubsection{Comparison between Simple and Scoped Rows}
The effect systems with simple rows and those with scoped rows might look
similar, but there are significant differences between them, as summarized as
follows:
\begin{itemize}
  \item When extended with effect polymorphism, the effect systems with simple
        rows need a way to restrict simple rows passed to effect
        abstractions to ensure the uniqueness of labels in a row.
        For example, a row $ \{  \mathsf{Exc}  \}  \,\underline{ \cup }\,  \alpha $ with row variable $\alpha$
        become invalid if $\alpha$ is instantiated by a row including
        $\mathsf{Exc}$ (and if the axiom to collapse multiple occurrences of the same
        effect into one is not supported).
        The effect system of \citet{hillerstrom_liberating_2016} restricts instantiation
        of row variables using a kind system, to support row polymorphism in
        Remy's style~\cite{}.
        By contrast, the effect systems with scoped rows have no such restriction
        because they allow duplication of labels in the same row.
        This difference might influence the design and implementation of typechecking,
        type inference, and unification for the effect systems.

  \item Given a handling construct $ \mathbf{handle}_{ \mathit{l} }  \, \ottnt{e} \, \mathbf{with} \, \ottnt{h}$, the effect systems
        with simple rows ensures that all the unhandled calls to operations of
        effect $\mathit{l}$ in $\ottnt{e}$ are interpreted by handler $\ottnt{h}$.
        However, this might not be the case in the effect systems adapting scoped rows.
        In such effect systems, each call to an operation is matched with
        \emph{one} occurrence of the operation's effect.
        Therefore, operation calls matched with the second or later occurrences
        of the effect may leak outside the closest handling construct.
        As a result, it is possible to interpret calls to the same operation by
        different effect handlers even when the calls are performed in the same
        context.
        This feature is useful in, for example, encoding mutable state by effect handlers
        because different operation calls may intend to manipulate different
        reference cells.~\footnote{To write such a program, we would need
          additional constructs as coercions.  Readers interested in more details
          are referred to the work of \citet{...}.}
\end{itemize}
}

\OLD{
  \subsubsection{Parametric Effects and Type-Erasure Semantics}

  Parametric effects are the labels parameterised by typelike parameters. We write $\mathit{l} \,  \bm{ { S } } ^ {  \mathit{I}  } $ to denote parametric effects, where typelikes, written as $S$ and $T$, are composed of types, labels, and effects. When allowing the parametric effects, we call $\mathit{l}$ a label name and $\mathit{l} \,  \bm{ { S } } ^ {  \mathit{I}  } $ a label.

  For example, consider the following program.
  \begin{flalign*}
     &  \mathsf{Exc}  ::  \ottsym{\{}  \mathsf{raise}  \ottsym{:}    \forall   \alpha  \ottsym{:}   \mathbf{Typ}    \ottsym{.}     \mathsf{Unit}    \Rightarrow   \alpha   \ottsym{\}}   \ottsym{,}   \mathsf{IO}  ::    \forall   \alpha  \ottsym{:}   \mathbf{Typ}    \ottsym{.}    \ottsym{\{}  \mathsf{print}  \ottsym{:}   \alpha   \Rightarrow    \mathsf{Unit}    \ottsym{\}}   & \\
     & \mathit{g'} = \lambda  \mathit{x}  \ottsym{:}   \mathsf{Int}   \ottsym{.}  \lambda  \mathit{y}  \ottsym{:}   \mathsf{Int}   \ottsym{.}   \mathbf{handle}_{ \mathsf{Exc} }  \,      & \\ &    \quad    \quad    \quad   \mathbf{if} \, \mathit{y}  \ottsym{=}  0 \, \mathbf{then} \,  \mathsf{raise} _{ \mathsf{Exc} }  \,  \mathsf{Int}  \,  ()  \, \mathbf{else} \, \mathit{y}      \ottsym{/}  \mathit{x}    & \\ &    \quad    \quad  \, \mathbf{with} \,  \ottsym{\{} \, \mathbf{return} \, \mathit{z}  \mapsto   \mathsf{print} _{ \mathsf{IO} \,  \mathsf{String}  }  \,  {}  \,  (   \textnormal{\ttfamily int\_to\_string}  \, \mathit{z}  )   \ottsym{\}}    & \\ &    \quad    \quad    \quad    \quad   \uplus        \ottsym{\{}  \mathsf{raise} \, \alpha  \ottsym{:}   \mathbf{Typ}  \, \mathit{z} \, \mathit{k}  \mapsto   \mathsf{print} _{ \mathsf{IO} \,  \mathsf{String}  }  \,  {}  \,  \textnormal{\texttt{"divided by 0"} }   \ottsym{\}} 
  \end{flalign*}
  In this setting, the label name $\mathsf{IO}$ takes one parameter of types, which occurs in the signature of $\mathsf{print}$. Similarly as mentioned before, $\mathit{g'} \, 42 \, 0$ is unsafe, but the expression
  \begin{flalign*}
     \mathbf{handle}_{ \mathsf{IO} \,  \mathsf{Int}  }  \, \mathit{g'} \, 42 \, 0 \, \mathbf{with} \,  \ottsym{\{} \, \mathbf{return} \, \mathit{z}  \mapsto   \textnormal{\ttfamily int\_to\_string}  \, \mathit{z}  \ottsym{\}}   \uplus   \ottsym{\{}  \mathsf{print} \,  {}  \, \mathit{p} \, \mathit{k}  \mapsto   \textnormal{\ttfamily int\_to\_string}  \, \mathit{p}  \ottsym{\}} 
  \end{flalign*}
  is also unsafe. The outer $\mathsf{IO} \,  \mathsf{Int} $ handler handles the operation $ \mathsf{print} _{ \mathsf{IO} \,  \mathsf{Int}  }  \,  {} $ but not the operation $ \mathsf{print} _{ \mathsf{IO} \,  \mathsf{String}  }  \,  {} $. Thus, the expression gets stuck. Therefore, effect systems must treat $\mathit{l} \,  \bm{ { S } } ^ {  \mathit{I}  } $ as a different label from $\mathit{l} \,  \bm{ { T } } ^ {  \mathit{I}  } \,( \bm{ { S } } ^ {  \mathit{I}  }  \neq  \bm{ { T } } ^ {  \mathit{I}  } )$ when the parametric effects is allowed.

  Type-erasure semantics is the semantics that ignores the parameters of labels in handlings. Consider the function $\mathit{g'}$ written above. With type-erasure semantics,
  \begin{flalign*}
     \mathbf{handle}_{ \mathsf{IO} \,  \mathsf{Int}  }  \, \mathit{g'} \, 42 \, 0 \, \mathbf{with} \,  \ottsym{\{} \, \mathbf{return} \, \mathit{z}  \mapsto   \textnormal{\ttfamily int\_to\_string}  \, \mathit{z}  \ottsym{\}}   \uplus   \ottsym{\{}  \mathsf{print} \,  {}  \, \mathit{p} \, \mathit{k}  \mapsto   \textnormal{\ttfamily int\_to\_string}  \, \mathit{p}  \ottsym{\}} 
  \end{flalign*}
  is reduced to $ \textnormal{\ttfamily int\_to\_string}  \,  \textnormal{\texttt{"divided by 0"} } $. $\mathsf{IO} \,  \mathsf{Int} $ handler handles not only $ \mathsf{print} _{ \mathsf{IO} \,  \mathsf{Int}  }  \,  {} $ but also $ \mathsf{print} _{ \mathsf{IO} \,  \mathsf{String}  }  \,  {} $. Thus, effect systems must conclude that programs with mismatched handlings like this are effect unsafe.

  However, a set-based effect system introduced above cannot prevent mismatched handlings. For example,
  \begin{flalign*}
    \mathit{bad\_io} = \phantom{} &  \mathbf{handle}_{ \mathsf{IO} \,  \mathsf{String}  }  \,   \mathbf{handle}_{ \mathsf{IO} \,  \mathsf{Int}  }  \,     \mathsf{print} _{ \mathsf{IO} \,  \mathsf{String}  }  \,  {}  \,  \textnormal{\texttt{"foo"} }     & \\ &    \quad  \, \mathbf{with} \,  \ottsym{\{} \, \mathbf{return} \, \mathit{z}  \mapsto   \textnormal{\texttt{""} }   \ottsym{\}}   \uplus   \ottsym{\{}  \mathsf{print} \,  {}  \, \mathit{p} \, \mathit{k}  \mapsto   \textnormal{\ttfamily int\_to\_string}  \, \mathit{p}  \ottsym{\}}    & \\ &  \, \mathbf{with} \,  \ottsym{\{} \, \mathbf{return} \, \mathit{z}  \mapsto  \mathit{z}  \ottsym{\}}   \uplus   \ottsym{\{}  \mathsf{print} \,  {}  \, \mathit{p} \, \mathit{k}  \mapsto  \mathit{p}  \ottsym{\}} 
  \end{flalign*}
  is effect safe under a set-based effect system because the effect of $ \mathsf{print} _{ \mathsf{IO} \,  \mathsf{String}  }  \,  {}  \,  \textnormal{\texttt{"foo"} } $ is $\{\mathsf{IO} \,  \mathsf{String} \}$ and handlers remove their responsible labels from it to results in $\emptyset$. On the other hand, this program is reduced to a stuck expression
  \begin{flalign*}
     \mathbf{handle}_{ \mathsf{IO} \,  \mathsf{String}  }  \,   \textnormal{\ttfamily int\_to\_string}  \,  \textnormal{\texttt{"foo"} }   \, \mathbf{with} \,  \ottsym{\{} \, \mathbf{return} \, \mathit{z}  \mapsto  \mathit{z}  \ottsym{\}}   \uplus   \ottsym{\{}  \mathsf{print} \,  {}  \, \mathit{p} \, \mathit{k}  \mapsto  \mathit{p}  \ottsym{\}} 
  \end{flalign*}
  with type-erasure semantics. In that sense, a set-based effect system is limited.

  \subsubsection{A Row-Based Effect System}

  A row-based effect system is more powerful than a set-based one, sacrificing simplicity. In this style, the row and its equivalence relation are usually defined as follows:
  \begin{align*}
    \varepsilon \Coloneqq \langle  \rangle  \mid  \rho  \mid  \langle  \mathit{l}  \mid  \varepsilon  \rangle \qquad
    \inferrule{\varepsilon_{{\mathrm{1}}} \simeq \varepsilon_{{\mathrm{2}}}}{\langle  \mathit{l}  \mid  \varepsilon_{{\mathrm{1}}}  \rangle \simeq \langle  \mathit{l}  \mid  \varepsilon_{{\mathrm{2}}}  \rangle} \qquad
    \inferrule{\mathit{l_{{\mathrm{1}}}} \neq \mathit{l_{{\mathrm{2}}}}}{\langle  \mathit{l_{{\mathrm{1}}}}  \mid  \langle  \mathit{l_{{\mathrm{2}}}}  \mid  \varepsilon  \rangle  \rangle \simeq \langle  \mathit{l_{{\mathrm{2}}}}  \mid  \langle  \mathit{l_{{\mathrm{1}}}}  \mid  \varepsilon  \rangle  \rangle}
  \end{align*}
  where $\varepsilon$ and $\mathit{l}$ and $\rho$ denote effects, labels, and effect variables, respectively. If the effect can be $\langle  \rangle$, this system concludes that the expression is effect safe. Handlers remove their responsible label from the top of the row. An effect variable at the end of an effect can be replaced with an effect: effect polymorphism.

  One of the powerful features of a row-based effect system is the extensibility to the parametric effects with type-erasure semantics. Allowing the parametric effects changes the equivalence relation on rows to
  \begin{gather*}
    \inferrule{\varepsilon_{{\mathrm{1}}} \simeq \varepsilon_{{\mathrm{2}}}}{\langle  \mathit{l} \,  \bm{ { S } } ^ {  \mathit{I}  }   \mid  \varepsilon_{{\mathrm{1}}}  \rangle \simeq \langle  \mathit{l} \,  \bm{ { S } } ^ {  \mathit{I}  }   \mid  \varepsilon_{{\mathrm{2}}}  \rangle} \qquad
    \inferrule{\mathit{l_{{\mathrm{1}}}} \neq \mathit{l_{{\mathrm{2}}}}}{\langle  \mathit{l_{{\mathrm{1}}}} \,  \bm{ { S_{{\mathrm{1}}} } } ^ {  \mathit{I_{{\mathrm{1}}}}  }   \mid  \langle  \mathit{l_{{\mathrm{2}}}} \,  \bm{ { S_{{\mathrm{2}}} } } ^ {  \mathit{I_{{\mathrm{2}}}}  }   \mid  \varepsilon  \rangle  \rangle \simeq \langle  \mathit{l_{{\mathrm{2}}}} \,  \bm{ { S_{{\mathrm{2}}} } } ^ {  \mathit{I_{{\mathrm{2}}}}  }   \mid  \langle  \mathit{l_{{\mathrm{1}}}} \,  \bm{ { S_{{\mathrm{1}}} } } ^ {  \mathit{I_{{\mathrm{1}}}}  }   \mid  \varepsilon  \rangle  \rangle}.
  \end{gather*}
  This change means that if the expression $\ottnt{e}$ has the effect $\langle  \mathsf{IO} \,  \mathsf{Int}   \mid  \langle  \mathsf{IO} \,  \mathsf{String}   \mid  \langle  \rangle  \rangle  \rangle$, we must enclose it by the $\mathsf{IO} \,  \mathsf{Int} $ handler and next by the $\mathsf{IO} \,  \mathsf{String} $ handler for effect safety. Recall $\mathit{bad\_io}$. This program is ill-typed under a row-based effect system. The effect of $ \mathsf{print} _{ \mathsf{IO} \,  \mathsf{String}  }  \,  {}  \,  \textnormal{\texttt{"foo"} } $ must be $\langle  \mathsf{IO} \,  \mathsf{Int}   \mid  \langle  \mathsf{IO} \,  \mathsf{String}   \mid  \langle  \rangle  \rangle  \rangle$ for effect safety because the $\mathsf{IO} \,  \mathsf{Int} $ handler and $\mathsf{IO} \,  \mathsf{String} $ handler enclose it in this order. However, the effect of $ \mathsf{print} _{ \mathsf{IO} \,  \mathsf{String}  }  \,  {}  \,  \textnormal{\texttt{"foo"} } $ cannot be because there are no $\varepsilon$ such that $\langle  \mathsf{IO} \,  \mathsf{String}   \mid  \varepsilon  \rangle \simeq \langle  \mathsf{IO} \,  \mathsf{Int}   \mid  \langle  \mathsf{IO} \,  \mathsf{String}   \mid  \langle  \rangle  \rangle  \rangle$. Thus, a row-based effect system concludes that $\mathit{bad\_io}$ is effect unsafe.
}

\subsection{Our Work: Abstracting Effect Systems}

All effect systems based on sets, simple rows, or scoped rows exploit the
structures of the respective representations to augment and shrink the
information about effects.
However, it is not clear which part of these structures essentially contributes
to type-and-effect safety.
To reveal it, we provide an abstract model of effect collections and their
manipulation and give an effect system relying only on the abstract model.
We also state sufficient conditions on the abstract model to guarantee the safety of our effect system.
%
%
With the effect system depending only on the abstract nature of effect
collections, we reveal the essence of safe effect systems for algebraic effect
handlers.

\TY{I changed the explanation about AREs to the one about effect algebras.}
We abstract the effect collections and manipulation in the effect systems for algebraic effect handlers
by an \emph{effect algebra}, which consists of an equivalence relation $ \sim $ and
a partial binary operation $ \odot $,
which mean the equivalence over effects and effect concatenation, respectively
(these notations come from \citet{morris_abstracting_2019}).
For example, $  \varepsilon_{{\mathrm{1}}}  \mathop{ \odot }  \varepsilon_{{\mathrm{2}}}    \sim   \varepsilon_{{\mathrm{3}}} $ intends to state that
the concatenation of effects $\varepsilon_{{\mathrm{1}}}$ and $\varepsilon_{{\mathrm{2}}}$ is equal to $\varepsilon_{{\mathrm{3}}}$.
\OLD{
  We abstract the effect manipulation in the effect systems for algebraic effect
  handlers by a relation $  \varepsilon_{{\mathrm{1}}}  \mathop{ \odot }  \varepsilon_{{\mathrm{2}}}    \sim   \varepsilon_{{\mathrm{3}}} $ over
  abstract effect collections $\varepsilon_{{\mathrm{1}}}$, $\varepsilon_{{\mathrm{2}}}$, and $\varepsilon_{{\mathrm{3}}}$,
  inspired by \citet{morris_abstracting_2019}. \AI{need to cite ROSE}
  Intuitively, this relation means that the appending of $\varepsilon_{{\mathrm{2}}}$ to
  $\varepsilon_{{\mathrm{1}}}$ is equivalent to $\varepsilon_{{\mathrm{3}}}$; thus, we call it an
  \emph{appending relation on effects}, ARE for short.
}
Our effect system is parameterized by effect algebras and manipulate effect
collections only through the operation $ \odot $ of a given effect algebra;
hence, it does not suppose any concrete effect manipulation.
%

%
%
To abstract over the representations of effect collections,
our effect system assumes two effect constructors.
One is $ \bbZero $, which
represents the empty collection and corresponds to the empty set and row in the set- and row-based effect systems, respectively.
The other constructor is $ \lift{ \mathit{l} } $, which constructs the effect collection
composed only of effect label $\mathit{l}$.

With these abstractions, the inference rules that manipulate effect collections---i.e., those for
operation calls, subeffecting, and handling constructs---are given as follows:
\begin{gather*}
  \inferrule
  {\text{Operation $\mathsf{op}  \ottsym{:}   \ottnt{A}   \Rightarrow   \ottnt{B} $ belongs to effect $\mathit{l}$} \\ \Gamma  \vdash  \ottnt{v}  \ottsym{:}  \ottnt{A}  \mid   \bbZero }
  {\Gamma  \vdash   \mathsf{op}  \,  {}  \, \ottnt{v}  \ottsym{:}  \ottnt{B}  \mid   \lift{ \mathit{l} } }
  \\
  \inferrule
  {\Gamma  \vdash  \ottnt{e}  \ottsym{:}  \ottnt{A}  \mid  \varepsilon \and   \varepsilon  \mathop{ \odot }  \varepsilon_{{\mathrm{0}}}    \sim   \varepsilon' }
  {\Gamma  \vdash  \ottnt{e}  \ottsym{:}  \ottnt{A}  \mid  \varepsilon'}
  \qquad
  \inferrule
  { {\Gamma  \vdash  \ottnt{e}  \ottsym{:}  \ottnt{A}  \mid  \varepsilon} \and {   \lift{ \mathit{l} }   \mathop{ \odot }  \varepsilon'    \sim   \varepsilon } \and \cdots}
  {\Gamma  \vdash   \mathbf{handle}_{ \mathit{l} }  \, \ottnt{e} \, \mathbf{with} \, \ottnt{h}  \ottsym{:}  \ottnt{B}  \mid  \varepsilon'}
\end{gather*}

%
The rule for operation call $ \mathsf{op}  \,  {}  \, \ottnt{v}$ simply injects the
corresponding effect label into the effect collection.
The subsumption rule with subeffecting means that the effect $\varepsilon$ of an expression can be expanded to
$\varepsilon'$ by appending some effects $\varepsilon_{{\mathrm{0}}}$.
The rule for handling constructs means that, if a handled expression may invoke
effects in $\varepsilon$, only the remaining $\varepsilon'$ of excluding the handled
effect $\mathit{l}$ from $\varepsilon$ is observable from the outer context.

It is noteworthy that the above usage of effect algebras pays attention to the order of effects appearing in effect collections.
Specifically, the subsumption rule only allows
appending extra effects $\varepsilon_{{\mathrm{0}}}$ and does not allow prepending them, and
the rule for handling constructs removes only the handled effect label that
occurs first in $\varepsilon$.
This mirrors the nature of the effect handling that an operation call is handled
by the effect handler closest to the call.
%
%
%
The importance of considering the order of effects is confirmed in, e.g.,
adopting a type-erasure semantics: as discussed in
Section~\ref{sec:erasure}, our effect system becomes unsound under the
type-erasure semantics if a given effect algebra is equipped with commutative
$ \odot $, which makes the effect system \emph{in}sensitive to the order of
effects.  \TS{Check that Section~\ref{sec:erasure} discusses that point.}

While effect algebras are expressive enough to represent the manipulation of effect collections, some
effect algebras make the effect system unsafe.
For example, consider an effect algebra where
$   \lift{ \mathit{l} }   \mathop{ \odot }  \varepsilon    \sim    \bbZero  $ holds.
Given an operation $\mathsf{op}$ of the effect label $\mathit{l}$,
the subsumption rule allows coercing the effect $ \lift{ \mathit{l} } $ of an operation call $ \mathsf{op}  \, \ottnt{v}$ to $ \bbZero $.
It means that the effect system can state that $ \mathsf{op}  \, \ottnt{v}$ invokes no
unhandled operation, so the effect system with such an effect algebra is unsafe.

To prevent the use of such effect algebras, we establish conditions on effect algebras;
we call them \emph{safety conditions} and also call effect algebras meeting them \emph{safe}.
We prove that, given a safe effect algebra, our effect system satisfies type and effect
safety.
%
%
We also demonstrate the expressibility of our framework by providing effect
algebras for sets, simple rows, and scoped rows from the literature, as well as
one for multisets, which are a new representation of effect collections.
\TY{I removed the mention of free
rows.}

\OLD{
  \subsection{Sketch of Abstracting Effect Systems}

  \subsubsection{Motivation}
  Both set-based effect system and row-based effect system have representativity to the following two features:
  \begin{itemize}
    \item an effect is a subeffect of another effect and
    \item removing a label from an effect results in another effect.
  \end{itemize}
  First, subeffecting is needed for let-bindings. The typing rule for let-bindings requires that a bound and succeeding expression have the same effect. For example, the effect $\varepsilon$ of
  \begin{align*}
    \mathbf{let} \, \_  \ottsym{=}   \mathsf{op_{{\mathrm{1}}}} _{ \mathit{l_{{\mathrm{1}}}} }  \,  {}  \, \ottnt{v_{{\mathrm{1}}}} \, \mathbf{in} \,  \mathsf{op_{{\mathrm{2}}}} _{ \mathit{l_{{\mathrm{2}}}} }  \,  {}  \, \ottnt{v_{{\mathrm{2}}}}
  \end{align*}
  must be an supereffect of both $\mathit{l_{{\mathrm{1}}}}$ and $\mathit{l_{{\mathrm{2}}}}$. Secondly, the removing relation is needed for handlings. In a set-based effect system, a handling expression removes its responsible label from the effect set of a handled expression. In a row-based effect system, a handling expression removes its responsible label from the top of the effect row of a handled expression. Not only accumulating but removing effects is one of the features of effect handlers.

  Unlike a set-based one, a row-based effect system can reason about handling orders. This feature enables us to extend effect systems to parametric effects with type-erasure semantics.

  These observations motivate us to formalize these features and describe the crucial difference between set and row structures. To achieve this goal, we create a language that can be instantiated to some concrete effect systems, such as a set-based and row-based one. By discovering what is needed to prove type and effect safety of this language, we reveal the essence of effect systems. Furthermore, by extending this language to parametric effects with type-erasure semantics, we also reveal the essence of some kinds of effect systems that can adapt to type-erasure semantics.

  \subsubsection{Our Approach}
  We use only one relation $\Delta  \vdash    \varepsilon_{{\mathrm{1}}}  \mathop{ \odot }  \varepsilon_{{\mathrm{2}}}    \sim   \varepsilon_{{\mathrm{3}}} $ to represent the features that effect systems must have. This relation has extensibility to parametric effects with type-erasure semantics. Our language takes $\Delta  \vdash    \varepsilon_{{\mathrm{1}}}  \mathop{ \odot }  \varepsilon_{{\mathrm{2}}}    \sim   \varepsilon_{{\mathrm{3}}} $ as a parameter.

  The relation $\Delta  \vdash    \varepsilon_{{\mathrm{1}}}  \mathop{ \odot }  \varepsilon_{{\mathrm{2}}}    \sim   \varepsilon_{{\mathrm{3}}} $ intuitively means that the appending of $\varepsilon_{{\mathrm{2}}}$ to $\varepsilon_{{\mathrm{1}}}$ is equivalent to $\varepsilon_{{\mathrm{3}}}$ under the context $ \Delta $. This relation represents the three features above. First, labels contained by $\varepsilon_{{\mathrm{1}}}$ are handled before labels in $\varepsilon_{{\mathrm{2}}}$. Secondly, $\varepsilon_{{\mathrm{1}}}$ is an subeffect of $\varepsilon_{{\mathrm{2}}}$ if there exists $\varepsilon$ such that $\Delta  \vdash    \varepsilon_{{\mathrm{1}}}  \mathop{ \odot }  \varepsilon    \sim   \varepsilon_{{\mathrm{2}}} $. The fact that an effect $\varepsilon_{{\mathrm{1}}}$ is a subeffect of another effect $\varepsilon_{{\mathrm{2}}}$ means that $\varepsilon_{{\mathrm{1}}}$ is more precise than $\varepsilon_{{\mathrm{2}}}$. Because the effect $\varepsilon$ assigned to the expression $\ottnt{e}$ means that the evaluation of $\ottnt{e}$ may cause the effect $\varepsilon$, the more labels the effect has, the less precise the effect is. Thus, if $\varepsilon_{{\mathrm{1}}}$ is assigned to the expression $\ottnt{e}$, then the supereffect of $\varepsilon_{{\mathrm{1}}}$ has more labels handled after $\varepsilon_{{\mathrm{1}}}$. At last, for label-removing relations, we use $\Delta  \vdash     \lift{ \mathit{l} }   \mathop{ \odot }  \varepsilon_{{\mathrm{1}}}    \sim   \varepsilon_{{\mathrm{2}}} $ to state that removing the label at the top position of $\varepsilon_{{\mathrm{2}}}$ results in $\varepsilon_{{\mathrm{1}}}$. We denote $ \lift{ \mathit{l} } $ to distinguish the effect consisting of a single label $\mathit{l}$ from a single label $\mathit{l}$.

  However, any relation does not be accepted. There exist some relations allowing us to write the programs violating effect safety. For example, consider the relation on rows that treats the empty effect as the bottom and all the other effects as the top. In other words, $\Delta  \vdash    \langle  \rangle  \mathop{ \odot }  \langle  \rangle    \sim   \langle  \rangle $, $\Delta  \vdash    \langle  \rangle  \mathop{ \odot }  \varepsilon    \sim   \varepsilon $, $\Delta  \vdash    \varepsilon  \mathop{ \odot }  \langle  \rangle    \sim   \varepsilon $, and $\Delta  \vdash    \varepsilon_{{\mathrm{1}}}  \mathop{ \odot }  \varepsilon_{{\mathrm{2}}}    \sim   \varepsilon_{{\mathrm{3}}} $ (where $\varepsilon_{{\mathrm{1}}}, \varepsilon_{{\mathrm{2}}}, \varepsilon_{{\mathrm{3}}} \neq \langle  \rangle$). In this case, $ \mathbf{handle}_{ \mathsf{Exc} }  \,  \mathsf{print} _{ \mathsf{IO} }  \,  {}  \, 42 \, \mathbf{with} \, \ottnt{h}$ is treated as a safe program, though it gets stuck. Thus, this instance of $\Delta  \vdash    \varepsilon_{{\mathrm{1}}}  \mathop{ \odot }  \varepsilon_{{\mathrm{2}}}    \sim   \varepsilon_{{\mathrm{3}}} $ does not satisfy type and effect safety.

  Therefore, we define safety conditions. If the relation $\Delta  \vdash    \varepsilon_{{\mathrm{1}}}  \mathop{ \odot }  \varepsilon_{{\mathrm{2}}}    \sim   \varepsilon_{{\mathrm{3}}} $ meets safety conditions, then type-and-effect safety is guaranteed. Furthermore, if the relation violates any safety condition, then there exist some programs that the type-and-effect system regards as safe ones, but the evaluation of them gets stuck.
}

\section{Abstracting Effects}\label{sec:effdef}

This section introduces the core notions of our effect system: effect algebras, an abstract model
of effect collections and their manipulations.
Because we aim at a formal effect system, we need to decide the syntactic
representation of effect collections manipulated by the effect system.  However,
relying on specific representations prevents accommodating a variety of effect
systems in the literature.
To address this problem, we parameterize our effect system over the
representations of effect collections and assume that the interface of their
constructs is given by an \emph{effect signature}.

Throughout this paper, we use the notation $ \bm{ { \alpha } } ^ {  \mathit{I}  } $ for a finite sequence
$\alpha_{{\mathrm{0}}}  \ottsym{,}  \ldots  \ottsym{,}  \alpha_{\ottmv{n}}$ with an index set $\mathit{I} = \{0, \ldots, n\}$,
where $\alpha$ is any metavariable.
We also write $\{  \bm{ { \alpha } } ^ {  \mathit{I}  }  \}$ for the set consisting of the elements of
$ \bm{ { \alpha } } ^ {  \mathit{I}  } $.
Index sets are designated by $\mathit{I}$, $\mathit{J}$, and $\mathit{N}$.
We omit index sets and write $ \bm{ { \alpha } } $ simply when they are not important
(e.g., all the sequences of interest have the same length).

\subsection{Syntax}

\begin{figure}[t]
  \[\begin{array}{@{}r@{\ \ }c@{\ \ }llr@{\ \ }c@{\ \ }ll}
    \multicolumn{8}{@{}c@{}}{
     \mathit{f}, \mathit{g}, \mathit{x}, \mathit{y}, \mathit{z}, \mathit{p}, \mathit{k} \ \ \text{(variables)} \qquad
     \alpha, \beta, \gamma, \tau, \iota, \rho \ \ \text{(typelike variables)} \qquad
     \mathsf{op} \ \ \text{(operation names)}
    } \\[.5ex]
    \multicolumn{8}{@{}c@{}}{
     \mathit{l} \in \labels \ \ \text{(label names)} \quad%
     \mathcal{F} \in \EC \ \ \text{(effect constructors)} \quad%
     \mathcal{C} \in \labels \cup \EC
    } \\[1ex]
    \ottnt{K}               & \Coloneqq &  \mathbf{Typ}   \mid   \mathbf{Lab}   \mid   \mathbf{Eff}                                        & \text{(kinds)}                &
    S, T        & \Coloneqq & \ottnt{A}  \mid  \ottnt{L}  \mid  \varepsilon                                       & \text{(typelikes)}            \\
    \ottnt{A}, \ottnt{B}, \ottnt{C} & \Coloneqq & \tau  \mid   \ottnt{A}    \rightarrow_{ \varepsilon }    \ottnt{B}   \mid    \forall   \alpha  \ottsym{:}  \ottnt{K}   \ottsym{.}    \ottnt{A}    ^{ \varepsilon }   & \text{(types)}                &
    \ottnt{L}               & \Coloneqq & \iota  \mid  \mathit{l} \,  \bm{ { S } } ^ {  \mathit{I}  }                                                 & \text{(labels)}               \\
    \varepsilon         & \Coloneqq & \rho  \mid  \mathcal{F} \,  \bm{ { S } } ^ {  \mathit{I}  }                                                  & \text{(effects)}              &
    \Xi               & \Coloneqq &  \emptyset   \mid  \Xi  \ottsym{,}   \mathit{l}  ::    \forall    {\bm{ \alpha } }^{ \mathit{I} } : {\bm{ \ottnt{K} } }^{ \mathit{I} }    \ottsym{.}    \sigma                       & \text{(effect contexts)}      \\
    \sigma             & \Coloneqq & \multicolumn{2}{l}{ \{\}   \mid   \sigma   \uplus   \ottsym{\{}  \mathsf{op}  \ottsym{:}    \forall    {\bm{ \beta } }^{ \mathit{J} } : {\bm{ \ottnt{K} } }^{ \mathit{J} }    \ottsym{.}    \ottnt{A}   \Rightarrow   \ottnt{B}   \ottsym{\}} } & \multicolumn{4}{l}{\text{(operation signatures)}} \\
    \Gamma               & \Coloneqq & \multicolumn{2}{l}{ \emptyset   \mid  \Gamma  \ottsym{,}  \mathit{x}  \ottsym{:}  \ottnt{A}  \mid  \Gamma  \ottsym{,}  \alpha  \ottsym{:}  \ottnt{K}}                   & \multicolumn{4}{l}{\text{(typing contexts)}}
    \end{array}
  \]
  \caption{Typelike syntax over an label signature $ \Slabel $ and an effect signature $ \Sbase $.}
  \label{fig:typelike}
\end{figure}

We start by defining \emph{label} and \emph{effect signatures}, which specify
available \emph{label names} (the names of effects) and effect collection
constructors as well as their kinds, respectively.
We then introduce the syntax of types, effect labels, and effect collections
using a given label and effect signature.
Kinds, ranged by $\ottnt{K}$, are $ \mathbf{Typ} $ for types, $ \mathbf{Lab} $ for effect labels,
or $ \mathbf{Eff} $ for effect collections.
\begin{definition}[Signatures]\label{def:effsig}
  Given a set $S$ of label names, a label signature $ \Slabel $ is a functional relation whose domain $\labels$ is $S$.
  The codomain of $ \Slabel $ is the set of functional kinds of the form $ \Pi   _{ \ottmv{i}  \in  \mathit{I} }    \ottnt{K_{\ottmv{i}}}   \rightarrow   \mathbf{Lab} $ for some $\mathit{I}$ and $\ottnt{K}_i^{i \in \mathit{I}}$
  (if $\mathit{I} = \emptyset$, it means $ \mathbf{Lab} $ simply).
  Similarly, given a set $S$ of effect constructors, an effect signature $ \Sbase $ is a functional relation whose domain $\EC$ is $S$ and
  its codomain is the set of functional kinds of the form $ \Pi   _{ \ottmv{i}  \in  \mathit{I} }    \ottnt{K_{\ottmv{i}}}   \rightarrow   \mathbf{Eff} $ for some $\mathit{I}$ and $\ottnt{K}_i^{i \in \mathit{I}}$.
  A signature $\Sigma$ is the disjoint union of a label signature and an effect signature.
  We write $ \Pi {\bm{ { \ottnt{K} } } }^{ \mathit{I} }   \rightarrow  \ottnt{K}$, and more simply, $ \Pi {\bm{ { \ottnt{K} } } }   \rightarrow  \ottnt{K}$
  as an abbreviation for $ \Pi   _{ \ottmv{i}  \in  \mathit{I} }    \ottnt{K_{\ottmv{i}}}   \rightarrow  \ottnt{K}$.
  %
\end{definition}

We write $\mathcal{C} :  \Pi {\bm{ { \ottnt{K} } } }   \rightarrow  \ottnt{K}$ to denote the pair $\langle \mathcal{C},  \Pi {\bm{ { \ottnt{K} } } }   \rightarrow  \ottnt{K} \rangle$
for label name or effect constructor $\ottnt{C}$.
\begin{example}[Label Signatures of $\mathsf{Exc}$ and $\mathsf{State}$]
 \label{exa:label-sig}
 The label signature for label names $\mathsf{Exc}$ and $\mathsf{State}$ used in Section~\ref{subsec:effsys:role} are given as
 $\{ \mathsf{Exc}  \ottsym{:}   \mathbf{Lab} , \ \mathsf{State}  \ottsym{:}   \mathbf{Lab}  \}$.
 The label $\mathsf{State}$ in Section~\ref{subsec:effsys:role} assumes the values of state to be integers, but,
 if one wants to parameterize label $\mathsf{State}$ over the types of the state values,
 the signature of $\mathsf{State}$ changes to $\mathsf{State}  \ottsym{:}   \mathbf{Typ}   \rightarrow   \mathbf{Lab} $.
 This signature indicates that $\mathsf{State}$ can take a type argument $\ottnt{A}$ that
 represents the type of the state values.
 We call parameterized label names, as $\mathsf{State}$ of kind $ \mathbf{Typ}   \rightarrow   \mathbf{Lab} $,
 \emph{parametric effects}, which facilitate the reuse of program components as
 explained later.

\AI{explain the difference between parametric operations and parametric effects.}
\end{example}
The following is an effect signature for \emph{effect sets}, effect collections implemented by sets.
\begin{example}[Effect Signature of Effect Sets]\label{exa:effset-effsig}
  The effect signature $\SbaseSet$ of effect sets consists of
  %
  %
    the pairs
    $\{  \}  \ottsym{:}   \mathbf{Eff} $ (for the empty set),
    $\{  \ottsym{-}  \}  \ottsym{:}   \mathbf{Lab}   \rightarrow   \mathbf{Eff} $ (for singleton sets), and
    $ \ottsym{-}  \,\underline{ \cup }\,  \ottsym{-}   \ottsym{:}   \mathbf{Eff}   \times   \mathbf{Eff}   \rightarrow   \mathbf{Eff} $ (for set unions).\footnote{We use ``$ \!\textnormal{--}\! $'' for unnamed arguments.  Multiple
    occurrences of ``$ \!\textnormal{--}\! $'' are distinguished from each other; the $i$-th
    occurrence from the left represents the $i$-th argument.}
\end{example}

Given a signature $\Sigma =  \Slabel   \uplus   \Sbase $, the
syntax of types, ranged over by $\ottnt{A}$, $\ottnt{B}$, and $\ottnt{C}$, effect labels (or
labels for short), ranged over by $\ottnt{L}$, and effect collections (or effects
for short), ranged over by $\varepsilon$, is defined as in Figure~\ref{fig:typelike}.
This work allows three kinds of polymorphism, that is, type, label, and effect
polymorphism.
To simplify their presentation, we introduce a syntactic category that unifies
types, labels, and effects; we call its entities
\emph{typelikes}~\cite{biernacki_abstracting_2019}, which are ranged over by $S$
and $T$.  Typelikes are classified into types, labels, and effects using the
kind system presented in Section~\ref{subsec:kind-system}.
%
%
We use $\tau$, $\iota$, and $\rho$ to designate type, label, and effect
variables (i.e., typelike variables with kind $ \mathbf{Typ} $, $ \mathbf{Eff} $, and
$ \mathbf{Lab} $), respectively, and $\alpha$, $\beta$, and $\gamma$ in a
general context.

Types consist of: type variables; function types $ \ottnt{A}    \rightarrow_{ \varepsilon }    \ottnt{B} $, which
represent functions from type $\ottnt{A}$ to $\ottnt{B}$ with effect
$\varepsilon$; and polymorphic types $  \forall   \alpha  \ottsym{:}  \ottnt{K}   \ottsym{.}    \ottnt{A}    ^{ \varepsilon }  $, which
represent (suspended) computation with effect $\varepsilon$ abstracting over typelikes of
kind $\ottnt{K}$.
We omit base types such as $ \mathsf{Int} $ for simplification, but assume them and
some operations on them (such as $+$ for integers) in giving examples.

A label is a label variable or a label name, ranged over by $\mathit{l}$, possibly
with type arguments.
%
%
For example, consider $\mathsf{State}  \ottsym{:}   \mathbf{Typ}   \rightarrow   \mathbf{Lab} $ given in Example~\ref{exa:label-sig}.
A label $\mathsf{State} \, \ottnt{A}$ represents
mutable state possessing the values of the type $\ottnt{A}$.
%
%
We can implement $\mathsf{State} \, \ottnt{A}$ using a state-passing effect handler, which
abstracts over type arguments $\ottnt{A}$~\cite{leijen_type_2017}.
Thus, the effect handler can be reused for different type arguments.

Effects are composed of effect variables and effect constructors, ranged
over by $\mathcal{F}$, given by $ \Sbase $.
As label names, effect constructors can take typelikes as arguments.
For example, effect set $\{  \mathsf{Exc}  \}$ is represented by $\mathcal{F} \, \mathsf{Exc}$ where
$\mathcal{F}$ is the constructor $\{  \ottsym{-}  \}$ for singleton sets.
\OLD{
  We assume that the set $\EC$ of effect constructors contains the constructor
  $ \bbZero $, which is an effect assigned to pure computation, and
  $ \lift{ \ottsym{-} } $,\footnote{We use ``$ \!\textnormal{--}\! $'' for unnamed arguments.  Multiple
    occurrences of ``$ \!\textnormal{--}\! $'' are distinguished from each other; the $i$-th
    occurrence from the left represents the $i$-th argument.} which constructs an
  effect composed only of a given effect label.
}
%

Effect contexts, ranged over by $\Xi$, are finite sequences of declarations of
effect label names.
Each label name $\mathit{l}$ is associated with a type scheme of the form $  \forall    \bm{ \alpha } : \bm{ \ottnt{K} }    \ottsym{.}    \sigma $,
where $\sigma$ is an operation signature parameterized
over typelike variables $ \bm{ { \alpha } } $ of kinds $ {\bm{ { \ottnt{K} } } } $.
In general, the functional kind $ \Pi {\bm{ { \ottnt{K'} } } }   \rightarrow   \mathbf{Lab} $ of $\mathit{l}$ in $ \Slabel $ needs to be consistent with
the kind of the type scheme, that is, $ {\bm{ { \ottnt{K'} } } }  =  {\bm{ { \ottnt{K} } } } $; we will formalize this requirement in Section~\ref{subsec:type-and-effect-safety}.
An \emph{operation signature} is a set of pairs of an operation name $\mathsf{op}$ and its type $  \forall    \bm{ \beta } : \bm{ \ottnt{K} }    \ottsym{.}    \ottnt{A}   \Rightarrow   \ottnt{B} $.
%
Here, $\ottnt{A}$ and $\ottnt{B}$ are the argument and return types of the operation, respectively,
and they are parameterized over $ \bm{ { \beta } } $ of kinds $ {\bm{ { \ottnt{K} } } } $.
Namely, not only effect labels but also operations can be parametric.
For example, the effect context for nonparametric effect labels $\mathsf{Exc}$ and $\mathsf{State}$ in Section~\ref{subsec:algeff} is given as
\begin{align*}
   \mathsf{Exc}  ::  \ottsym{\{}  \mathsf{raise}  \ottsym{:}    \mathsf{Unit}    \Rightarrow    \mathsf{Empty}    \ottsym{\}}   \ottsym{,}   \mathsf{State}  ::  \ottsym{\{}  \mathsf{set}  \ottsym{:}    \mathsf{Int}    \Rightarrow    \mathsf{Unit}    \ottsym{,}  \mathsf{get}  \ottsym{:}    \mathsf{Unit}    \Rightarrow    \mathsf{Int}    \ottsym{\}}  ~.
\end{align*}
If one wants to parameterize label $\mathsf{State}$ over the types of the state values,
and operation $\mathsf{raise}$ of label $\mathsf{Exc}$ over return types (because it returns no value actually),
the effect context can change to \AI{explain the difference between parametric operations and parametric effects.}
\begin{align*}
   \mathsf{Exc}  ::  \ottsym{\{}  \mathsf{raise}  \ottsym{:}    \forall   \alpha  \ottsym{:}   \mathbf{Typ}    \ottsym{.}     \mathsf{Unit}    \Rightarrow   \alpha   \ottsym{\}}   \ottsym{,}   \mathsf{State}  ::    \forall   \alpha  \ottsym{:}   \mathbf{Typ}    \ottsym{.}    \ottsym{\{}  \mathsf{set}  \ottsym{:}   \alpha   \Rightarrow    \mathsf{Unit}    \ottsym{,}  \mathsf{get}  \ottsym{:}    \mathsf{Unit}    \Rightarrow   \alpha   \ottsym{\}}   ~.
\end{align*}
A difference between parametric effects and operations is that, while effect
handlers for parametric effects can be typechecked depending on given type
arguments, ones for parametric operations must abstract over type
arguments. See \citet{Sekiyama/Igarashi_2019_ESOP} for detail.

Typing contexts, ranged over by $\Gamma$, are finite sequences of
bindings of the form $\mathit{x}  \ottsym{:}  \ottnt{A}$ or $\alpha  \ottsym{:}  \ottnt{K}$.
%

\OLD{
  This section presents the core notions of {\lang}: effect signatures and ARE. Effect signatures and ARE are the parameters of the type-and-effect system of {\lang}. In order to get an instance of {\lang}, what is required is to give instances of them.

  Before introducing effect signatures and ARE, we show the notation used in this paper and the syntax of type language.
  \begin{remark}[Notation]
    We write $ \bm{ { \alpha } } ^ {  \mathit{I}  } $ for a finite sequence $\alpha_{{\mathrm{0}}}, \ldots, \alpha_{\ottmv{n}}$ with an index set $\mathit{I} = \{0, \ldots, n\}$, where $\alpha$ is any metavariable.
    We also write $\{  \bm{ { \alpha } } ^ {  \mathit{I}  }  \}$ for the set consisting of the elements of $ \bm{ { \alpha } } ^ {  \mathit{I}  } $.
  \end{remark}

  \begin{figure}[t]
    \begin{flalign*}
      \mathrm{Index}   & \ni \mathit{I}, \mathit{J}, \ottnt{K},                                       & \text{(index sets)}         &  &
      \mathrm{Var}     & \ni \mathit{g}, \mathit{x}, \mathit{y}, \mathit{z}, \mathit{p}, \mathit{k}                   & \text{(variables)}               \\
      \mathrm{TVar}    & \ni \alpha, \beta, \gamma, \tau, \iota, \rho & \text{(typelike variables)} &  &
      \mathrm{OpNames} & \ni \mathsf{op}                                                     & \text{(operation names)}         \\
      \labels          & \ni \mathit{l}                                                      & \text{(label names)}        &  &
      \EC              & \ni \mathit{f}                                                      & \text{(function symbols)}
    \end{flalign*}
    \begin{flalign*}
      \ottnt{K}               & \Coloneqq  \mathbf{Typ}   \mid   \mathbf{Lab}   \mid   \mathbf{Eff}                                        & \text{(kinds)}                &  &
      S, T        & \Coloneqq \ottnt{A}  \mid  \ottnt{L}  \mid  \varepsilon                                       & \text{(typelikes)}                 \\
      \ottnt{A}, \ottnt{B}, \ottnt{C} & \Coloneqq \tau  \mid   \ottnt{A}    \rightarrow_{ \varepsilon }    \ottnt{B}   \mid    \forall   \alpha  \ottsym{:}  \ottnt{K}   \ottsym{.}    \ottnt{A}    ^{ \varepsilon }   & \text{(types)}                &  &
      \ottnt{L}               & \Coloneqq \iota  \mid  \mathit{l} \,  \bm{ { S } } ^ {  \mathit{I}  }                                                 & \text{(labels)}                    \\
      \varepsilon         & \Coloneqq \rho  \mid  \mathcal{F} \,  \bm{ { S } } ^ {  \mathit{I}  }                                                  & \text{(effects)}                   \\
      \Gamma               & \Coloneqq  \emptyset   \mid  \Gamma  \ottsym{,}  \mathit{x}  \ottsym{:}  \ottnt{A}  \mid  \Gamma  \ottsym{,}  \alpha  \ottsym{:}  \ottnt{K}                       &                               &  &
                          &                                                                                     & \text{(typing contexts)}           \\
      \Delta               & \Coloneqq  \emptyset   \mid  \Gamma  \ottsym{,}  \alpha  \ottsym{:}  \ottnt{K}                                          &                               &  &
                          &                                                                                     & \text{(typelike contexts)}         \\
      \sigma             & \Coloneqq \ottsym{\{}  \mathsf{op_{\ottmv{i}}}  \ottsym{:}    \forall    {\bm{ \beta_{\ottmv{i}} } }^{ \mathit{J_{\ottmv{i}}} } : {\bm{ \ottnt{K_{\ottmv{i}}} } }^{ \mathit{J_{\ottmv{i}}} }    \ottsym{.}    \ottnt{A_{\ottmv{i}}}   \Rightarrow   \ottnt{B_{\ottmv{i}}}   \ottsym{\}}   ^{ \ottmv{i}  \in  \mathit{I} }                      &                               &  &
                          &                                                                                     & \text{(operation signatures)}      \\
      \Xi               & \Coloneqq  \emptyset   \mid  \Xi  \ottsym{,}   \mathit{l}  ::    \forall    {\bm{ \alpha } }^{ \mathit{I} } : {\bm{ \ottnt{K} } }^{ \mathit{I} }    \ottsym{.}    \sigma                       &                               &  &
                          &                                                                                     & \text{(effect contexts)}
    \end{flalign*}
    \caption{Typelike syntax.}
    \label{fig:typelike}
  \end{figure}

  We show the syntax of typelikes (including effects) in Fig~\ref{fig:typelike}. We use metavariables $\mathit{I}$, $\mathit{J}$, and $\mathit{N}$ to denote index sets; $\mathit{g}$, $\mathit{x}$, $\mathit{y}$, $\mathit{z}$, $\mathit{p}$, $\mathit{k}$ to variables; $\alpha$, $\beta$, $\gamma$, $\tau$, $\iota$, $\rho$ to typelike variables; and $\mathsf{op}$ to operation names of algebraic effects. As a convention, We use $\alpha$, $\beta$, and $\gamma$ to denote whole typelikes; $\tau$ to types; $\iota$ to labels; and $\rho$ to effects. $\labels$ is a set of label names and $\EC$ is a set of function symbols such that $\EC \supseteq \labels$.

  Kinds, ranged by $\ottnt{K}$, are composed of $ \mathbf{Typ} $ means types; $ \mathbf{Lab} $ means labels; and $ \mathbf{Eff} $ means effects. Kinds are not only for classifying typelikes but also for sorts of $\Sigma$-algebra. Typelikes ranged over by $S$ and $T$, are composed of: types ranged over by $\ottnt{A}$, $\ottnt{B}$, and $\ottnt{C}$; labels; and effects ranged over by $\varepsilon$. Types are composed of: variables $\tau$; function type $ \ottnt{A}    \rightarrow_{ \varepsilon }    \ottnt{B} $; and polymorphic type $  \forall   \alpha  \ottsym{:}  \ottnt{K}   \ottsym{.}    \ottnt{A}    ^{ \varepsilon }  $. An effect $\varepsilon$ occurs in all the types but variables. This annotated effect intuitively means an application may cause an effect $\varepsilon$. Labels and effects are defined using functional symbols.

  A typing context $\Gamma$ is usually defined as a finite sequence of variable bindings $\mathit{x}  \ottsym{:}  \ottnt{A}$ and typelike variable bindings $\alpha  \ottsym{:}  \ottnt{K}$. $\Delta$ is a sequence including only typelike variable bindings.

  Effect contexts ranged over by $\Xi$, are finite sequences of bindings of a label name to its typelike parameters and operation signatures $ \mathit{l}  ::    \forall    {\bm{ \alpha } }^{ \mathit{I} } : {\bm{ \ottnt{K} } }^{ \mathit{I} }    \ottsym{.}    \sigma  $. Operation signatures, ranged over by $\sigma$, are finite sets of tuples, including an operation name and signature. For example, only $\mathsf{IO}$ and $\mathsf{Exc}$ can be used in programs when $\Xi$ is
  \begin{align*}
     \mathsf{Exc}  ::  \ottsym{\{}  \mathsf{raise}  \ottsym{:}    \forall   \alpha  \ottsym{:}   \mathbf{Typ}    \ottsym{.}     \mathsf{Unit}    \Rightarrow   \alpha   \ottsym{\}}   \ottsym{,}   \mathsf{IO}  ::    \forall   \alpha  \ottsym{:}   \mathbf{Typ}    \ottsym{.}    \ottsym{\{}  \mathsf{print}  \ottsym{:}   \alpha   \Rightarrow    \mathsf{Unit}    \ottsym{\}}  .
  \end{align*}
}

\subsection{Kind System}
\label{subsec:kind-system}

\begin{figure}
  \textbf{Kinding}\tquad\fbox{$\Gamma  \vdash  S  \ottsym{:}  \ottnt{K}$}\tqquad\fbox{$\Gamma  \vdash   \bm{ { S } }^{ \mathit{I} } : \bm{ \ottnt{K} }^{ \mathit{I} } $} $\phantom{} \iff \forall \ottmv{i} \in \mathit{I} . (\Gamma  \vdash  S_{\ottmv{i}}  \ottsym{:}  \ottnt{K_{\ottmv{i}}})$ \hfill\phantom{}
  \begin{mathpar}
    \inferrule{\vdash  \Gamma \quad  \alpha   \ottsym{:}   \ottnt{K}   \in   \Gamma }{\Gamma  \vdash  \alpha  \ottsym{:}  \ottnt{K}}\ \rname{K}{Var}

    \inferrule{\vdash  \Gamma \quad  \mathcal{C}   \ottsym{:}    \Pi {\bm{ { \ottnt{K} } } }   \rightarrow  \ottnt{K_{{\mathrm{0}}}}   \in   \Sigma  \quad \Gamma  \vdash   \bm{ { S } } : \bm{ \ottnt{K} } }{\Gamma  \vdash  \mathcal{C} \,  \bm{ { S } }   \ottsym{:}  \ottnt{K_{{\mathrm{0}}}}}\ \rname{K}{Cons}

    \inferrule{\Gamma  \vdash  \ottnt{A}  \ottsym{:}   \mathbf{Typ}  \quad \Gamma  \vdash  \varepsilon  \ottsym{:}   \mathbf{Eff}  \quad \Gamma  \vdash  \ottnt{B}  \ottsym{:}   \mathbf{Typ} }{\Gamma  \vdash   \ottnt{A}    \rightarrow_{ \varepsilon }    \ottnt{B}   \ottsym{:}   \mathbf{Typ} }\ \rname{K}{Fun}
    \hfil
    \inferrule{\Gamma  \ottsym{,}  \alpha  \ottsym{:}  \ottnt{K}  \vdash  \ottnt{A}  \ottsym{:}   \mathbf{Typ}  \quad \Gamma  \ottsym{,}  \alpha  \ottsym{:}  \ottnt{K}  \vdash  \varepsilon  \ottsym{:}   \mathbf{Eff} }{\Gamma  \vdash    \forall   \alpha  \ottsym{:}  \ottnt{K}   \ottsym{.}    \ottnt{A}    ^{ \varepsilon }    \ottsym{:}   \mathbf{Typ} }\ \rname{K}{Poly}
  \end{mathpar}
  \caption{Kinding rules.}
  \label{fig:kinding}
\end{figure}

We show our kind system in Figure~\ref{fig:kinding}. We omit the rules for well-formedness of typing contexts because they are defined as usual~\cite{sekiyama_signature_2020, kawamata_answer_2024}.
The rules other than \rname{K}{Cons} are standard or straightforward.
When signature $\Sigma$ assigns $ \Pi {\bm{ { \ottnt{K} } } }   \rightarrow  \ottnt{K_{{\mathrm{0}}}}$ to label name or effect constructor $\mathcal{C}$, and typelike arguments $ \bm{ { S } } $ are of the kinds $ {\bm{ { \ottnt{K} } } } $, respectively,
the rule \rname{K}{Cons} assigns kind $\ottnt{K_{{\mathrm{0}}}}$ to the typelike $\mathcal{C} \,  \bm{ { S } } $.
%

\OLD{
  \subsection{Effect Signatures}
  We introduce the definition of arities before defining effect signatures.
  \begin{definition}[Arities]
    An arity takes the form $\Pi_{1 \leq i \leq n} K_i \rightarrow K$ for some $n$, and $\ari$ is the set of arities. We write $\Pi \bm{K}^I \rightarrow K$ as an abbreviation to $\Pi_{i \in I} K^i \rightarrow K$. $\labari$ is the function from $\labels$ to $\ari$ such that
    \begin{align*}
      \forall l \in \labels . \exists \bm{K}^I . (\labari(l) = \Pi \bm{K}^I \rightarrow  \mathbf{Lab}  ).
    \end{align*}
  \end{definition}
  A label name is a functional symbol whose codomain sort is $ \mathbf{Lab} $. This definition reflects the parametric effects in that label names can take some typelikes as arguments. For example, the sort of $\mathsf{IO}$ is $ \mathbf{Typ}  \rightarrow  \mathbf{Lab} $.

  \begin{definition}[Effect Signatures]\label{def:effsig}
    An effect signature $\Sbase$ is a functional relation that is a subset of $\EC \times \ari$ and satisfies the following conditions:
    \begin{itemize}
      \item $ \bbZero  :  \mathbf{Eff}  \in \Sbase$,
      \item $ \lift{ \ottsym{-} }  :  \mathbf{Lab}   \rightarrow   \mathbf{Eff}  \in \Sbase$,
      \item $\{ l : \labari(l) \mid l \in \labels \} \subseteq \Sbase$, and
      \item $\forall \mathit{f} : \Pi \bm{\ottnt{K}}^I \rightarrow \ottnt{K} \in \Sbase . (\mathit{f} \notin \labels \imply \ottnt{K} =  \mathbf{Eff} )$,
    \end{itemize}
    where we write $f : \Pi_{1 \leq i \leq n} K_i \rightarrow K$ for the pair $\langle f, \Pi_{1 \leq i \leq n} K_i \rightarrow K \rangle$.
  \end{definition}
  $ \bbZero $ intuitively means the empty effect representing the purity of expressions. $ \lift{ \ottsym{-} } $ can be understood as the function: it takes a label and returns the effect, which only includes that label. Other conditions say that the effect signature must include label names; all function symbols other than label names have $ \mathbf{Eff} $ as codomain sort.

  \begin{remark}[Proper Effect Contexts]\label{rem:proper_effctx}
    For any effect context $\Xi$, and any $ \mathit{l}  ::    \forall    {\bm{ \alpha } }^{ \mathit{I} } : {\bm{ \ottnt{K} } }^{ \mathit{I} }    \ottsym{.}    \ottsym{\{}  \mathsf{op_{\ottmv{j}}}  \ottsym{:}    \forall    {\bm{ \beta_{\ottmv{j}} } }^{ \mathit{N_{\ottmv{j}}} } : {\bm{ \ottnt{K_{\ottmv{j}}} } }^{ \mathit{N_{\ottmv{j}}} }    \ottsym{.}    \ottnt{A_{\ottmv{j}}}   \Rightarrow   \ottnt{B_{\ottmv{j}}}   \ottsym{\}}   ^{ \ottmv{j}  \in  \mathit{J} }     \in   \Xi $,
    we assume the following:
    \begin{itemize}
      \item $ \mathit{l}   \ottsym{:}    {\bm{ { \ottnt{K} } } }^{ \mathit{I} }   \rightarrow   \mathbf{Lab}    \in   \Sbase $,
      \item if $ \mathit{l}  ::    \forall    {\bm{ \alpha' } }^{ \mathit{I'} } : {\bm{ \ottnt{K'} } }^{ \mathit{I'} }    \ottsym{.}    \sigma    \in   \Xi $, then $ {\bm{ \alpha } }^{ \mathit{I} } : {\bm{ \ottnt{K} } }^{ \mathit{I} }  =  {\bm{ \alpha' } }^{ \mathit{I'} } : {\bm{ \ottnt{K'} } }^{ \mathit{I'} } $ and $\sigma = \ottsym{\{}  \mathsf{op_{\ottmv{j}}}  \ottsym{:}    \forall    {\bm{ \beta_{\ottmv{j}} } }^{ \mathit{N_{\ottmv{j}}} } : {\bm{ \ottnt{K_{\ottmv{j}}} } }^{ \mathit{N_{\ottmv{j}}} }    \ottsym{.}    \ottnt{A_{\ottmv{j}}}   \Rightarrow   \ottnt{B_{\ottmv{j}}}   \ottsym{\}}   ^{ \ottmv{j}  \in  \mathit{J} } $; and
      \item for any $\ottmv{j} \in \mathit{J}$, the well-formedness judgments $ {\bm{ \alpha } }^{ \mathit{I} } : {\bm{ \ottnt{K} } }^{ \mathit{I} }   \ottsym{,}   {\bm{ \beta_{\ottmv{j}} } }^{ \mathit{N_{\ottmv{j}}} } : {\bm{ \ottnt{K_{\ottmv{j}}} } }^{ \mathit{N_{\ottmv{j}}} }   \vdash  \ottnt{A_{\ottmv{j}}}  \ottsym{:}   \mathbf{Typ} $ and $ {\bm{ \alpha } }^{ \mathit{I} } : {\bm{ \ottnt{K} } }^{ \mathit{I} }   \ottsym{,}   {\bm{ \beta_{\ottmv{j}} } }^{ \mathit{N_{\ottmv{j}}} } : {\bm{ \ottnt{K_{\ottmv{j}}} } }^{ \mathit{N_{\ottmv{j}}} }   \vdash  \ottnt{B_{\ottmv{j}}}  \ottsym{:}   \mathbf{Typ} $
            are derivable.
    \end{itemize}
  \end{remark}
}

\subsection{Effect Algebras}

Now, we define effect algebras.
In short, an effect algebra provides an effect signature $ \Sbase $, a partial monoid
on effects defined over $ \Sbase $, and a function $ \lift{ \ottsym{-} } $
that injects labels to effects, but more formally, it also
requires that each involved operation preserve well-formedness and kind-aware
typelike substitution make a homomorphism.
In what follows, we denote the sets of types, effect labels, and effect
collections over a signature $\Sigma$ by
$\TypSet{\Sigma}$, $\LabelSet{\Sigma}$, and $\EffSet{\Sigma}$, respectively
(we refer to the set of entities at kind $\ottnt{K}$ by $\KSet{\ottnt{K}}{\Sigma}$).

\begin{definition}[Well-Formedness-Preserving Functions]
 Given a signature $\Sigma$, a (possibly partial) function $f \in \KSet{K_i}{\Sigma}^{i \in \{ 1, \ldots, n\}} \rightharpoonup \KSet{\ottnt{K}}{\Sigma}$
 preserves well-formedness if
 \[
  \forall \Gamma, S_{{\mathrm{1}}}, \ldots, S_{\ottmv{n}} . \, \Gamma  \vdash  S_{{\mathrm{1}}}  \ottsym{:}  \ottnt{K_{{\mathrm{1}}}} \land \cdots \land \Gamma  \vdash  S_{\ottmv{n}}  \ottsym{:}  \ottnt{K_{\ottmv{n}}} \land f(S_{{\mathrm{1}}}, \ldots, S_{\ottmv{n}}) \in \KSet{\ottnt{K}}{\Sigma} \implies \Gamma \vdash f(S_{{\mathrm{1}}}, \ldots, S_{\ottmv{n}}) : \ottnt{K} ~.
 \]
 Similarly, $f \in \KSet{\ottnt{K}}{\Sigma}$ preserves well-formedness if $\Gamma \vdash f : \ottnt{K}$ for any $\Gamma$.
\end{definition}
In what follows, we write $ \alpha   \mapsto   T   \vdash    \bm{ { S } }   :  \ottnt{K_{{\mathrm{0}}}} $
for a quadruple $\langle \alpha, T,  \bm{ { S } } , \ottnt{K_{{\mathrm{0}}}} \rangle$ such that
$\exists \Gamma_{{\mathrm{1}}}, \ottnt{K}, \Gamma_{{\mathrm{2}}} . \, (\forall S_{{\mathrm{0}}} \in  \bm{ { S } } . \, \Gamma_{{\mathrm{1}}}  \ottsym{,}  \alpha  \ottsym{:}  \ottnt{K}  \ottsym{,}  \Gamma_{{\mathrm{2}}}  \vdash  S_{{\mathrm{0}}}  \ottsym{:}  \ottnt{K_{{\mathrm{0}}}}) \land \Gamma_{{\mathrm{1}}}  \vdash  T  \ottsym{:}  \ottnt{K}$;
it means that typelikes $ \bm{ { S } } $ are well formed at kind $\ottnt{K_{{\mathrm{0}}}}$ and substituting typelike $T$ for typelike variable $\alpha$ in $ \bm{ { S } } $ preserves their well-formedness.
\begin{definition}[Effect algebras]\label{def:relation}
  {\sloppy{
  Given a label signature $ \Slabel $, an effect algebra is a quintuple $\langle  \Sbase ,  \odot ,  \bbZero ,  \lift{ \ottsym{-} } ,  \sim  \rangle$
  satisfying the following, where we let $\Sigma =  \Slabel   \uplus   \Sbase $.
 }}
  \begin{itemize}
   \OLD{
   \item $ \odot  \in \EffSet{ \Sbase } \times \EffSet{ \Sbase } \rightharpoonup \EffSet{ \Sbase }$ and
         it preserves well-formedness, that is,
         \[
          \forall \Delta, \varepsilon_{{\mathrm{1}}}, \varepsilon_{{\mathrm{2}}} . \, \Delta  \vdash  \varepsilon_{{\mathrm{1}}}  \ottsym{:}   \mathbf{Eff}  \land \Delta  \vdash  \varepsilon_{{\mathrm{2}}}  \ottsym{:}   \mathbf{Eff}  \land  \varepsilon_{{\mathrm{1}}}  \mathop{ \odot }  \varepsilon_{{\mathrm{2}}}  \in \EffSet{ \Sbase } \implies \Delta  \vdash    \varepsilon_{{\mathrm{1}}}  \mathop{ \odot }  \varepsilon_{{\mathrm{2}}}    \ottsym{:}   \mathbf{Eff}  ~.
         \]
   \item $ \bbZero  \in \EffSet{ \Sbase }$ and
         it is closed and well formed, that is,
         $\emptyset  \vdash   \bbZero   \ottsym{:}   \mathbf{Eff} $.
   \item $ \lift{ \ottsym{-} }  \in \LabelSet{ \Sbase } \rightarrow \EffSet{ \Sbase }$ and it
         preserves well-formedness, that is,
         \[
          \forall \Delta, \ottnt{L}. \, \Delta  \vdash  \ottnt{L}  \ottsym{:}   \mathbf{Lab}  \implies \Delta  \vdash   \lift{ \ottnt{L} }   \ottsym{:}   \mathbf{Eff}  ~.
         \]
   \item $ \sim $ is an equivalence relation on $\EffSet{ \Sbase }$ and preserves well-formedness, that is,
         \[
          \forall \varepsilon_{{\mathrm{1}}}, \varepsilon_{{\mathrm{2}}} . \,  \varepsilon_{{\mathrm{1}}}   \sim   \varepsilon_{{\mathrm{2}}}  \implies (\forall \Delta . \, \Delta  \vdash  \varepsilon_{{\mathrm{1}}}  \ottsym{:}   \mathbf{Eff}  \iff \Delta  \vdash  \varepsilon_{{\mathrm{2}}}  \ottsym{:}   \mathbf{Eff} ) ~.
         \]
   }

   \item $ \odot  \in \EffSet{\Sigma} \times \EffSet{\Sigma} \rightharpoonup \EffSet{\Sigma}$,
         $ \bbZero  \in \EffSet{\Sigma}$, and
         $ \lift{ \ottsym{-} }  \in \LabelSet{\Sigma} \rightarrow \EffSet{\Sigma}$ preserve well-formedness.
         Furthermore,
         $ \sim $ is an equivalence relation on $\EffSet{\Sigma}$ and preserves well-formedness, that is,
         $\forall \varepsilon_{{\mathrm{1}}}, \varepsilon_{{\mathrm{2}}} . \,  \varepsilon_{{\mathrm{1}}}   \sim   \varepsilon_{{\mathrm{2}}}  \implies (\forall \Gamma . \, \Gamma  \vdash  \varepsilon_{{\mathrm{1}}}  \ottsym{:}   \mathbf{Eff}  \iff \Gamma  \vdash  \varepsilon_{{\mathrm{2}}}  \ottsym{:}   \mathbf{Eff} )$.

   \item $\langle \EffSet{\Sigma},  \odot ,  \bbZero  \rangle$ is a partial
         monoid under $ \sim $, that is, the following holds:
           \begin{itemize}


            \item $\forall \varepsilon \in \EffSet{\Sigma} . \,
                    \varepsilon  \mathop{ \odot }   \bbZero     \sim   \varepsilon  \land    \bbZero   \mathop{ \odot }  \varepsilon    \sim   \varepsilon $; and

            \item $\forall \varepsilon_{{\mathrm{1}}}, \varepsilon_{{\mathrm{2}}}, \varepsilon_{{\mathrm{3}}} \in \EffSet{\Sigma} . \,$ \\
                    $ \ottsym{(}   \varepsilon_{{\mathrm{1}}}  \mathop{ \odot }  \varepsilon_{{\mathrm{2}}}   \ottsym{)}  \mathop{ \odot }  \varepsilon_{{\mathrm{3}}}  \in \EffSet{\Sigma} \lor
                       \varepsilon_{{\mathrm{1}}}  \mathop{ \odot }  \ottsym{(}   \varepsilon_{{\mathrm{2}}}  \mathop{ \odot }  \varepsilon_{{\mathrm{3}}}   \ottsym{)}  \in \EffSet{\Sigma}
                     \implies
                        \ottsym{(}   \varepsilon_{{\mathrm{1}}}  \mathop{ \odot }  \varepsilon_{{\mathrm{2}}}   \ottsym{)}  \mathop{ \odot }  \varepsilon_{{\mathrm{3}}}    \sim    \varepsilon_{{\mathrm{1}}}  \mathop{ \odot }  \ottsym{(}   \varepsilon_{{\mathrm{2}}}  \mathop{ \odot }  \varepsilon_{{\mathrm{3}}}   \ottsym{)}  $.
           \end{itemize}

   \OLD{
   \item $ \odot $, $ \bbZero $, $ \lift{ \ottsym{-} } $, and $ \sim $ preserve well-formedness, that is, the following holds:
         \begin{itemize}
          \item $\forall \Delta, \varepsilon_{{\mathrm{1}}}, \varepsilon_{{\mathrm{2}}} . \, \Delta  \vdash  \varepsilon_{{\mathrm{1}}}  \ottsym{:}   \mathbf{Eff}  \land \Delta  \vdash  \varepsilon_{{\mathrm{2}}}  \ottsym{:}   \mathbf{Eff}  \land  \varepsilon_{{\mathrm{1}}}  \mathop{ \odot }  \varepsilon_{{\mathrm{2}}}  \in \EffSet{\Sigma} \implies \Delta  \vdash    \varepsilon_{{\mathrm{1}}}  \mathop{ \odot }  \varepsilon_{{\mathrm{2}}}    \ottsym{:}   \mathbf{Eff} $;
          \item $\forall \Delta . \, \Delta  \vdash   \bbZero   \ottsym{:}   \mathbf{Eff} $;
          \item $\forall \Delta, \ottnt{L}. \, \Delta  \vdash  \ottnt{L}  \ottsym{:}   \mathbf{Lab}  \implies \Delta  \vdash   \lift{ \ottnt{L} }   \ottsym{:}   \mathbf{Eff} $;
          \item $\forall \varepsilon_{{\mathrm{1}}}, \varepsilon_{{\mathrm{2}}} . \,  \varepsilon_{{\mathrm{1}}}   \sim   \varepsilon_{{\mathrm{2}}}  \implies (\forall \Delta . \, \Delta  \vdash  \varepsilon_{{\mathrm{1}}}  \ottsym{:}   \mathbf{Eff}  \iff \Delta  \vdash  \varepsilon_{{\mathrm{2}}}  \ottsym{:}   \mathbf{Eff} )$.
         \end{itemize}
   }

   \item Typelike substitution respecting well-formedness is a homomorphism
         for $ \odot $, $ \lift{ \ottsym{-} } $, and $ \sim $, that is,
         the following holds:
         \begin{itemize}
          \item $\forall \alpha, S, \varepsilon_{{\mathrm{1}}}, \varepsilon_{{\mathrm{2}}} . \,
                 \alpha   \mapsto   S   \vdash   \varepsilon_{{\mathrm{1}}}  \ottsym{,}  \varepsilon_{{\mathrm{2}}}  :   \mathbf{Eff}   {\,\land\,}  \varepsilon_{{\mathrm{1}}}  \mathop{ \odot }  \varepsilon_{{\mathrm{2}}}  \in \EffSet{\Sigma} \implies \ottsym{(}   \varepsilon_{{\mathrm{1}}}  \mathop{ \odot }  \varepsilon_{{\mathrm{2}}}   \ottsym{)} \,  \! [  S  /  \alpha   ]  =  \varepsilon_{{\mathrm{1}}} \,  \! [  S  /  \alpha   ]   \mathop{ \odot }  \varepsilon_{{\mathrm{2}}}  \,  \! [  S  /  \alpha   ] $;
          \item $\forall \alpha, S, \ottnt{L} . \,
                 \alpha   \mapsto   S   \vdash   \ottnt{L}  :   \mathbf{Lab}   \implies   \lift{ \ottnt{L} }   \,  \! [  S  /  \alpha   ]  =  \lift{  \ottnt{L} \,  \! [  S  /  \alpha   ]   } $; and
          \item $\forall \alpha, S, \varepsilon_{{\mathrm{1}}}, \varepsilon_{{\mathrm{2}}} . \,
                 \alpha   \mapsto   S   \vdash   \varepsilon_{{\mathrm{1}}}  \ottsym{,}  \varepsilon_{{\mathrm{2}}}  :   \mathbf{Eff}   \land  \varepsilon_{{\mathrm{1}}}   \sim   \varepsilon_{{\mathrm{2}}}  \implies  \varepsilon_{{\mathrm{1}}} \,  \! [  S  /  \alpha   ]    \sim   \varepsilon \,  \! [  S  /  \alpha   ]  $.
         \end{itemize}
  \end{itemize}
\end{definition}
\OLD{
Note that the above definition ensures $ \bbZero  \,  \! [  S  /  \alpha   ]  =  \bbZero $ for any $S$ and $\alpha$
because $ \bbZero $ should be closed by $\emptyset  \vdash   \bbZero   \ottsym{:}   \mathbf{Eff} $.
}

\OLD{
  For readability, we introduce the following abbreviations.
  \begin{itemize}
    \item $ \varepsilon_{{\mathrm{1}}}  \olessthan  \varepsilon_{{\mathrm{2}}}  \defeq \exists \varepsilon'_{{\mathrm{1}}} . (  \varepsilon_{{\mathrm{1}}}  \mathop{ \odot }  \varepsilon'_{{\mathrm{1}}}    \sim   \varepsilon_{{\mathrm{2}}} )$

    \item
          $\begin{aligned}[t]
               & \Delta  \vdash    \varepsilon_{{\mathrm{0}}}  \mathop{ \odot } \cdots \mathop{ \odot }  \varepsilon_{\ottmv{n}}    \sim    \varepsilon'_{{\mathrm{0}}}  \mathop{ \odot } \cdots \mathop{ \odot }  \varepsilon'_{\ottmv{m}}   \defeq                                                                              \\[-1ex]
               &   \varepsilon_{{\mathrm{0}}}  \mathop{ \odot } \cdots \mathop{ \odot }  \varepsilon_{\ottmv{n}}    \sim    \varepsilon'_{{\mathrm{0}}}  \mathop{ \odot } \cdots \mathop{ \odot }  \varepsilon'_{\ottmv{m}}   \tand \forall i . (\Delta  \vdash  \varepsilon_{\ottmv{i}}  \ottsym{:}   \mathbf{Eff} ) \tand \forall j . (\Delta  \vdash  \varepsilon'_{\ottmv{j}}  \ottsym{:}   \mathbf{Eff} )
            \end{aligned}$

  \end{itemize}
  \TS{Check $ \nsim $ is not used anywhere.}
  We use $\Delta  \vdash   \varepsilon_{{\mathrm{1}}}  \olessthan  \varepsilon_{{\mathrm{2}}} $ to state that $\varepsilon_{{\mathrm{1}}}$ is a subeffect of $\varepsilon_{{\mathrm{2}}}$.
}

For example, an effect algebra for effect sets can be given as follows.
\begin{example}[Effect Sets]\label{exa:effset}
  An effect algebra {\eaSet} for effect sets is a tuple
  $\langle \SbaseSet,  \ottsym{-}  \,\underline{ \cup }\,  \ottsym{-} , \{  \}, \{  \ottsym{-}  \},  \sim_{\eanameSet}  \rangle$
  where $ \sim_{\eanameSet} $ is the least equivalence relation satisfying the following rules:
  \begin{mathpar}
    \inferrule{ }{  \varepsilon  \,\underline{ \cup }\,  \{  \}     \sim_{\eanameSet}    \varepsilon }

    \inferrule{ }{  \varepsilon_{{\mathrm{1}}}  \,\underline{ \cup }\,  \varepsilon_{{\mathrm{2}}}     \sim_{\eanameSet}     \varepsilon_{{\mathrm{2}}}  \,\underline{ \cup }\,  \varepsilon_{{\mathrm{1}}}  }

    \inferrule{ }{  \varepsilon  \,\underline{ \cup }\,  \varepsilon     \sim_{\eanameSet}    \varepsilon } \\

    \inferrule{ }{  \ottsym{(}   \varepsilon_{{\mathrm{1}}}  \,\underline{ \cup }\,  \varepsilon_{{\mathrm{2}}}   \ottsym{)}  \,\underline{ \cup }\,  \varepsilon_{{\mathrm{3}}}     \sim_{\eanameSet}     \varepsilon_{{\mathrm{1}}}  \,\underline{ \cup }\,  \ottsym{(}   \varepsilon_{{\mathrm{2}}}  \,\underline{ \cup }\,  \varepsilon_{{\mathrm{3}}}   \ottsym{)}  }

    \inferrule{
     \varepsilon_{{\mathrm{1}}}    \sim_{\eanameSet}    \varepsilon_{{\mathrm{2}}}  \\  \varepsilon_{{\mathrm{3}}}    \sim_{\eanameSet}    \varepsilon_{{\mathrm{4}}} 
    }{
      \varepsilon_{{\mathrm{1}}}  \,\underline{ \cup }\,  \varepsilon_{{\mathrm{3}}}     \sim_{\eanameSet}     \varepsilon_{{\mathrm{2}}}  \,\underline{ \cup }\,  \varepsilon_{{\mathrm{4}}}  
    }
  \end{mathpar}
  These rules reflect that the union operator in sets has the identity element
  $\{  \}$ and satisfies commutativity, idempotence, associativity, and
  compatibility.
\end{example}

We also show an instance for simple rows and scoped rows.
\begin{example}[Simple Rows]\label{exa:eff_simple_row}
  The effect signature $\SbaseRow$ for simple rows is the set of
  $\langle  \rangle  \ottsym{:}   \mathbf{Eff} $ and
  $\langle  \ottsym{-}  \mid  \ottsym{-}  \rangle  \ottsym{:}   \mathbf{Lab}   \times   \mathbf{Eff}   \rightarrow   \mathbf{Eff} $.
  \OLD{
    \begin{itemize}
      \item $\langle  \rangle  \ottsym{:}   \mathbf{Eff} $ and
      \item $\langle  \ottsym{-}  \mid  \langle  \rangle  \rangle  \ottsym{:}   \mathbf{Lab}   \rightarrow   \mathbf{Eff} $, and
      \item $\langle  \ottsym{-}  \mid  \ottsym{-}  \rangle  \ottsym{:}   \mathbf{Lab}   \times   \mathbf{Eff}   \rightarrow   \mathbf{Eff} $.
    \end{itemize}
    The constructors $ \bbZero $ and $ \lift{ \ottsym{-} } $ are defined to
    be $\langle  \rangle$ and $\langle  \ottsym{-}  \mid  \langle  \rangle  \rangle$, respectively.
  }
  An effect algebra {\eaSimpleRow} for them is
  $\langle \SbaseRow,  \odot_\mathrm{SimpR} , \langle  \rangle, \langle  \ottsym{-}  \mid  \langle  \rangle  \rangle,  \sim_{\eanameSimpleRow}  \rangle$
  where
  \begin{align*}
     \varepsilon_{{\mathrm{1}}}  \mathop{  \odot_\mathrm{SimpR}  }  \varepsilon_{{\mathrm{2}}}  \defeq
    \begin{dcases}
      \, \langle \ottnt{L_{{\mathrm{1}}}} \mid \langle \cdots \langle  \ottnt{L_{\ottmv{n}}}  \mid  \varepsilon_{{\mathrm{2}}}  \rangle \rangle \rangle
       & (\tif \varepsilon_{{\mathrm{1}}} = \langle \ottnt{L_{{\mathrm{1}}}} \mid \langle \cdots \langle  \ottnt{L_{\ottmv{n}}}  \mid  \langle  \rangle  \rangle \rangle \rangle )                           \\
      \, \varepsilon_{{\mathrm{1}}}
       & (\tif \varepsilon_{{\mathrm{1}}} = \langle \ottnt{L_{{\mathrm{1}}}} \mid \langle \cdots \langle  \ottnt{L_{\ottmv{n}}}  \mid  \rho  \rangle \rangle \rangle \tand \varepsilon_{{\mathrm{2}}} = \langle  \rangle ) \\
    \end{dcases}
  \end{align*}
  and $ \sim_{\eanameSimpleRow} $ is the least equivalence relation satisfying the following.
  \begin{mathpar}
    \inferrule{ \varepsilon_{{\mathrm{1}}}    \sim_{\eanameSimpleRow}    \varepsilon_{{\mathrm{2}}} }{ \langle  \ottnt{L}  \mid  \varepsilon_{{\mathrm{1}}}  \rangle    \sim_{\eanameSimpleRow}    \langle  \ottnt{L}  \mid  \varepsilon_{{\mathrm{2}}}  \rangle }

    \inferrule{ \ottnt{L_{{\mathrm{1}}}}   \neq   \ottnt{L_{{\mathrm{2}}}} }{ \langle  \ottnt{L_{{\mathrm{1}}}}  \mid  \langle  \ottnt{L_{{\mathrm{2}}}}  \mid  \varepsilon  \rangle  \rangle    \sim_{\eanameSimpleRow}    \langle  \ottnt{L_{{\mathrm{2}}}}  \mid  \langle  \ottnt{L_{{\mathrm{1}}}}  \mid  \varepsilon  \rangle  \rangle }

    \inferrule{ }{ \langle  \ottnt{L}  \mid  \varepsilon  \rangle    \sim_{\eanameSimpleRow}    \langle  \ottnt{L}  \mid  \langle  \ottnt{L}  \mid  \varepsilon  \rangle  \rangle }
  \end{mathpar}

  Note that the definition of $ \varepsilon_{{\mathrm{1}}}  \mathop{  \odot_\mathrm{SimpR}  }  \varepsilon_{{\mathrm{2}}} $
  depends on whether effect $\varepsilon_{{\mathrm{1}}}$ ends with an effect variable.
  If it does, $\varepsilon_{{\mathrm{2}}}$ must be empty because simple rows ending with effect variables cannot be
  extended.
  Otherwise, $ \varepsilon_{{\mathrm{1}}}  \mathop{  \odot_\mathrm{SimpR}  }  \varepsilon_{{\mathrm{2}}} $ simply concatenates $\varepsilon_{{\mathrm{1}}}$ and $\varepsilon_{{\mathrm{2}}}$.

  The first rule of $ \sim_{\eanameSimpleRow} $ means that the results of adding the same label to equivalent effects are also equivalent.
  The remaining two rules allow reordering different labels
  and collapsing multiple occurrences of the same label into one, respectively.
  %
  %
  The collapsing of multiple occurrences reflects the characteristic of simple
  rows that the same label appears at most once in a row because it means that two
  or more occurrences of a label cannot be distinguished from one occurrence of
  it.
\end{example}
\begin{example}[Scoped Rows]\label{exa:effrow}
  An effect algebra {\eaScopedRow} for scoped rows is defined in a way similar to that for simple rows.
  The only difference is in the definition of equivalence $ \sim_{\eanameScopedRow} $.
  The equivalence $ \sim_{\eanameScopedRow} $ for scoped rows is defined as the least equivalence relation satisfying the following rules:
  \begin{mathpar}
    \inferrule{ \varepsilon_{{\mathrm{1}}}    \sim_{\eanameScopedRow}    \varepsilon_{{\mathrm{2}}} }{ \langle  \ottnt{L}  \mid  \varepsilon_{{\mathrm{1}}}  \rangle    \sim_{\eanameScopedRow}    \langle  \ottnt{L}  \mid  \varepsilon_{{\mathrm{2}}}  \rangle }

    \inferrule{ \ottnt{L_{{\mathrm{1}}}}   \neq   \ottnt{L_{{\mathrm{2}}}} }{ \langle  \ottnt{L_{{\mathrm{1}}}}  \mid  \langle  \ottnt{L_{{\mathrm{2}}}}  \mid  \varepsilon  \rangle  \rangle    \sim_{\eanameScopedRow}    \langle  \ottnt{L_{{\mathrm{2}}}}  \mid  \langle  \ottnt{L_{{\mathrm{1}}}}  \mid  \varepsilon  \rangle  \rangle }
  \end{mathpar}
  Unlike simple rows, scoped rows are distinguished if they have different numbers of occurrences of some label.
\end{example}

\OLD{
  \subsection{Appending Relation on Effects}

  We show the definition of ARE as follows.
  \begin{definition}[Appending Relation on Effects; ARE]\label{def:relation}
    An appending relation on effects (ARE) is a relation on tuples of one typelike
    context $\Delta$ and three effects well-formed under $\Delta$ (i.e., for each
    effect $\varepsilon$, $\Delta  \vdash  \varepsilon  \ottsym{:}   \mathbf{Eff} $ holds).
    We write $\Delta  \vdash    \varepsilon_{{\mathrm{1}}}  \mathop{ \odot }  \varepsilon_{{\mathrm{2}}}    \sim   \varepsilon_{{\mathrm{3}}} $ when a tuple
    $(\Delta, \varepsilon_{{\mathrm{1}}}, \varepsilon_{{\mathrm{2}}}, \varepsilon_{{\mathrm{3}}})$ is contained in the ARE.
    For readability, we introduce the following abbreviations.
    \begin{align*}
      \Delta  \not\vdash    \varepsilon_{{\mathrm{1}}}  \mathop{ \odot }  \varepsilon_{{\mathrm{2}}}    \sim   \varepsilon_{{\mathrm{3}}}  & \iff \lnot \Delta  \vdash    \varepsilon_{{\mathrm{1}}}  \mathop{ \odot }  \varepsilon_{{\mathrm{2}}}    \sim   \varepsilon_{{\mathrm{3}}}                                                                                                            \\
      \Delta  \vdash   \varepsilon_{{\mathrm{1}}}  \olessthan  \varepsilon_{{\mathrm{2}}}                   & \iff \exists \varepsilon . ( \Delta  \vdash    \varepsilon_{{\mathrm{1}}}  \mathop{ \odot }  \varepsilon    \sim   \varepsilon_{{\mathrm{2}}} )                                                                                         \\
      \Delta  \vdash   \varepsilon_{{\mathrm{1}}}  \olessthan_M  \varepsilon_{{\mathrm{2}}}                  & \iff \exists \varepsilon', \varepsilon'', \varepsilon''' . ( \Delta  \vdash    \varepsilon'  \mathop{ \odot }  \varepsilon_{{\mathrm{1}}}    \sim   \varepsilon''  \tand \Delta  \vdash    \varepsilon''  \mathop{ \odot }  \varepsilon'''    \sim   \varepsilon_{{\mathrm{2}}} ) \\
      \Delta  \vdash   \varepsilon_{{\mathrm{1}}}   \cong   \varepsilon_{{\mathrm{2}}}                & \iff (\Delta  \vdash     \bbZero   \mathop{ \odot }  \varepsilon_{{\mathrm{1}}}    \sim   \varepsilon_{{\mathrm{2}}}  \tand \Delta  \vdash     \bbZero   \mathop{ \odot }  \varepsilon_{{\mathrm{2}}}    \sim   \varepsilon_{{\mathrm{1}}}                                                            \\
                                                    & \hphantom{\iff} \quad  \tand \Delta  \vdash    \varepsilon_{{\mathrm{1}}}  \mathop{ \odot }   \bbZero     \sim   \varepsilon_{{\mathrm{2}}}  \tand \Delta  \vdash    \varepsilon_{{\mathrm{2}}}  \mathop{ \odot }   \bbZero     \sim   \varepsilon_{{\mathrm{1}}} )
    \end{align*}
  \end{definition}
  The intuitive meaning of $\Delta  \vdash    \varepsilon_{{\mathrm{1}}}  \mathop{ \odot }  \varepsilon_{{\mathrm{2}}}    \sim   \varepsilon_{{\mathrm{3}}} $ is that the appending of $\varepsilon_{{\mathrm{2}}}$ to $\varepsilon_{{\mathrm{1}}}$ equals  $\varepsilon_{{\mathrm{3}}}$. We use $\Delta  \vdash   \varepsilon_{{\mathrm{1}}}  \olessthan  \varepsilon_{{\mathrm{2}}} $ to state that $\varepsilon_{{\mathrm{1}}}$ is a subeffect of $\varepsilon_{{\mathrm{2}}}$. $\Delta  \vdash   \varepsilon_{{\mathrm{1}}}  \olessthan_M  \varepsilon_{{\mathrm{2}}} $ shows that $\varepsilon_{{\mathrm{2}}}$ contains $\varepsilon_{{\mathrm{1}}}$. This relation is more general than $\Delta  \vdash   \varepsilon_{{\mathrm{1}}}  \olessthan  \varepsilon_{{\mathrm{2}}} $. At last, $\Delta  \vdash   \varepsilon_{{\mathrm{1}}}   \cong   \varepsilon_{{\mathrm{2}}} $ means $\varepsilon_{{\mathrm{1}}}$ and $\varepsilon_{{\mathrm{2}}}$ are regarded as the same effects.
}

\OLD{
  \subsection{Instance Examples: Set and Row}
  In the following, we give set and row instances of ARE. These examples show that our abstraction via effect signatures and ARE accommodates a set and a row. First, we show the instance giving a set-based effect system.
  \begin{example}[Effect Set]\label{exa:effset}
    \addtolength{\jot}{0.5ex}
    \begin{gather*}
      \varepsilon \Coloneqq \{\}  \mid  \rho  \mid  \{\ottnt{L}\}  \mid  \varepsilon_{{\mathrm{1}}} \cup \varepsilon_{{\mathrm{2}}} \qquad
       \bbZero  = \{\} \qquad  \lift{ \ottsym{-} }  = \{ - \} \\
      \Sbase = \{\{\} :  \mathbf{Eff} ,  \lift{ \ottsym{-} }  :  \mathbf{Lab}   \rightarrow   \mathbf{Eff} , \cup :  \mathbf{Eff}  \times  \mathbf{Eff}   \rightarrow   \mathbf{Eff}  \} \cup \{ l : \labari(l) \mid l \in \labels \}
    \end{gather*}
    A relation $\simeq$ is the least equivalence relation satisfying the following.
    \begin{gather*}
      \inferrule{ }{\varepsilon_{{\mathrm{1}}} \cup \varepsilon_{{\mathrm{2}}} \simeq \varepsilon_{{\mathrm{2}}} \cup \varepsilon_{{\mathrm{1}}}} \qquad
      \inferrule{ }{\varepsilon \cup \{\} \simeq \varepsilon} \qquad
      \inferrule{ }{\varepsilon \cup \varepsilon \simeq \varepsilon} \\
      \inferrule{ }{(\varepsilon_{{\mathrm{1}}} \cup \varepsilon_{{\mathrm{2}}}) \cup \varepsilon_{{\mathrm{3}}} \simeq \varepsilon_{{\mathrm{1}}} \cup (\varepsilon_{{\mathrm{2}}} \cup \varepsilon_{{\mathrm{3}}}) } \qquad
      \inferrule{\varepsilon_{{\mathrm{1}}} \simeq \varepsilon_{{\mathrm{2}}} \\ \varepsilon_{{\mathrm{3}}} \simeq \varepsilon_{{\mathrm{4}}}}{\varepsilon_{{\mathrm{1}}} \cup \varepsilon_{{\mathrm{3}}} \simeq \varepsilon_{{\mathrm{2}}} \cup \varepsilon_{{\mathrm{4}}}} \\
      \Delta  \vdash    \varepsilon_{{\mathrm{1}}}  \mathop{ \odot }  \varepsilon_{{\mathrm{2}}}    \sim   \varepsilon_{{\mathrm{3}}}  \iff \varepsilon_{{\mathrm{1}}} \cup \varepsilon_{{\mathrm{2}}} \simeq \varepsilon_{{\mathrm{3}}}
    \end{gather*}
  \end{example}
  Secondly, we show the instance providing a row-based effect system.
  \begin{example}[Effect Row]\label{exa:effrow}
    \addtolength{\jot}{0.5ex}
    \begin{gather*}
      \varepsilon \Coloneqq \langle  \rangle  \mid  \rho  \mid  \langle  \ottnt{L}  \mid  \varepsilon  \rangle \qquad
       \bbZero  = \langle  \rangle \qquad  \lift{ \ottsym{-} }  = \langle - \mid \langle  \rangle \rangle \\
      \Sbase = \{\langle  \rangle :  \mathbf{Eff} ,  \lift{ \ottsym{-} }  :  \mathbf{Lab}   \rightarrow   \mathbf{Eff} , \langle - \mid - \rangle :  \mathbf{Lab}  \times  \mathbf{Eff}   \rightarrow   \mathbf{Eff}  \} \cup \{ l : \labari(l) \mid l \in \labels \}
    \end{gather*}
    A relation $\simeq$ is the least reflexive, transitive relation satisfying the following.
    \begin{gather*}
      \inferrule{\varepsilon_{{\mathrm{1}}} \simeq \varepsilon_{{\mathrm{2}}}}{\langle  \ottnt{L}  \mid  \varepsilon_{{\mathrm{1}}}  \rangle \simeq \langle  \ottnt{L}  \mid  \varepsilon_{{\mathrm{2}}}  \rangle} \qquad
      \inferrule{\mathit{l_{{\mathrm{1}}}} \neq \mathit{l_{{\mathrm{2}}}}}{\langle  \mathit{l_{{\mathrm{1}}}} \,  \bm{ { S_{{\mathrm{1}}} } } ^ {  \mathit{I_{{\mathrm{1}}}}  }   \mid  \langle  \mathit{l_{{\mathrm{2}}}} \,  \bm{ { S_{{\mathrm{2}}} } } ^ {  \mathit{I_{{\mathrm{2}}}}  }   \mid  \varepsilon  \rangle  \rangle \simeq \langle  \mathit{l_{{\mathrm{2}}}} \,  \bm{ { S_{{\mathrm{2}}} } } ^ {  \mathit{I_{{\mathrm{2}}}}  }   \mid  \langle  \mathit{l_{{\mathrm{1}}}} \,  \bm{ { S_{{\mathrm{1}}} } } ^ {  \mathit{I_{{\mathrm{1}}}}  }   \mid  \varepsilon  \rangle  \rangle} \\
      \Delta  \vdash    \varepsilon_{{\mathrm{1}}}  \mathop{ \odot }  \varepsilon_{{\mathrm{2}}}    \sim   \varepsilon_{{\mathrm{3}}}  \iff
      \begin{dcases}
         & \varepsilon_{{\mathrm{3}}} \simeq \langle \mathit{l_{{\mathrm{1}}}} \,  \bm{ { S_{{\mathrm{1}}} } } ^ {  \mathit{I_{{\mathrm{1}}}}  }  \mid \langle \cdots \langle \mathit{l_{\ottmv{n}}} \,  \bm{ { S_{\ottmv{n}} } } ^ {  \mathit{I_{\ottmv{n}}}  }  \mid \varepsilon_{{\mathrm{2}}} \rangle \rangle \rangle            \\
         & \quad (\tif \varepsilon_{{\mathrm{1}}} = \langle \mathit{l_{{\mathrm{1}}}} \,  \bm{ { S_{{\mathrm{1}}} } } ^ {  \mathit{I_{{\mathrm{1}}}}  }  \mid \langle \cdots \langle \mathit{l_{\ottmv{n}}} \,  \bm{ { S_{\ottmv{n}} } } ^ {  \mathit{I_{\ottmv{n}}}  }  \mid \langle\rangle \rangle \rangle \rangle ) \\
         & \varepsilon_{{\mathrm{2}}} = \langle\rangle \tand \varepsilon_{{\mathrm{3}}} \simeq \varepsilon_{{\mathrm{1}}}                                                                       \\
         & \quad (\tif \varepsilon_{{\mathrm{1}}} = \langle \mathit{l_{{\mathrm{1}}}} \,  \bm{ { S_{{\mathrm{1}}} } } ^ {  \mathit{I_{{\mathrm{1}}}}  }  \mid \langle \cdots \langle \mathit{l_{\ottmv{n}}} \,  \bm{ { S_{\ottmv{n}} } } ^ {  \mathit{I_{\ottmv{n}}}  }  \mid \rho \rangle \rangle \rangle )
      \end{dcases}
    \end{gather*}
  \end{example}
  The fact that the definition of $\Delta  \vdash    \varepsilon_{{\mathrm{1}}}  \mathop{ \odot }  \varepsilon_{{\mathrm{2}}}    \sim   \varepsilon_{{\mathrm{3}}} $ is divided by cases means that the row ending at an effect variable cannot be extended with non-empty effects. This case dividing is because the definition of effects shows that any effect does not include more than one effect variable.
}

\section{{\lang}: A Calculus with Abstract Effect System}\label{sec:calculus}
\label{sec:lang}

\TS{
  The following will be inserted somewhere.

  For readability, we introduce the following abbreviations.
  \begin{itemize}
    \item $ \varepsilon_{{\mathrm{1}}}  \olessthan  \varepsilon_{{\mathrm{2}}}  \defeq \exists \varepsilon'_{{\mathrm{1}}} . (  \varepsilon_{{\mathrm{1}}}  \mathop{ \odot }  \varepsilon'_{{\mathrm{1}}}    \sim   \varepsilon_{{\mathrm{2}}} )$

    \item
          $\begin{aligned}[t]
               & \Delta  \vdash    \varepsilon_{{\mathrm{0}}}  \mathop{ \odot } \cdots \mathop{ \odot }  \varepsilon_{\ottmv{n}}    \sim    \varepsilon'_{{\mathrm{0}}}  \mathop{ \odot } \cdots \mathop{ \odot }  \varepsilon'_{\ottmv{m}}   \defeq                                                                              \\[-1ex]
               &   \varepsilon_{{\mathrm{0}}}  \mathop{ \odot } \cdots \mathop{ \odot }  \varepsilon_{\ottmv{n}}    \sim    \varepsilon'_{{\mathrm{0}}}  \mathop{ \odot } \cdots \mathop{ \odot }  \varepsilon'_{\ottmv{m}}   \tand \forall i . (\Delta  \vdash  \varepsilon_{\ottmv{i}}  \ottsym{:}   \mathbf{Eff} ) \tand \forall j . (\Delta  \vdash  \varepsilon'_{\ottmv{j}}  \ottsym{:}   \mathbf{Eff} )
            \end{aligned}$

  \end{itemize}
}

This section shows the syntax, semantics, and type-and-effect system of our
language {\lang}. It is similar to the call-by-value polymorphic $\lambda$-calculi with
algebraic effect handlers in the literature~\cite{leijen_type_2017, biernacki_handle_2018, sekiyama_signature_2020} except that it is
parameterized over effect algebras.
Throughout this and the next sections, we fix a label signature $ \Slabel $,
effect algebra $\langle  \Sbase ,  \odot ,  \bbZero ,  \lift{ \ottsym{-} } ,  \sim  \rangle$ over $ \Slabel $, and
effect context $\Xi$, which are given as parameters.

\subsection{Syntax}
\begin{figure}[t]
  \[
   \begin{array}{rcll}
    \ottnt{e} &\Coloneqq& \ottnt{v}  \mid  \ottnt{v_{{\mathrm{1}}}} \, \ottnt{v_{{\mathrm{2}}}}  \mid  \ottnt{v} \, S  \mid  \mathbf{let} \, \mathit{x}  \ottsym{=}  \ottnt{e_{{\mathrm{1}}}} \, \mathbf{in} \, \ottnt{e_{{\mathrm{2}}}}  \mid   \mathbf{handle}_{ \mathit{l} \,  \bm{ { S } } ^ {  \mathit{I}  }  }  \, \ottnt{e} \, \mathbf{with} \, \ottnt{h}                                 & \text{(expressions)}                          \\
    \ottnt{v} &\Coloneqq& \mathit{x}  \mid  \ottkw{fun} \, \ottsym{(}  \mathit{f}  \ottsym{,}  \mathit{x}  \ottsym{,}  \ottnt{e}  \ottsym{)}  \mid  \Lambda  \alpha  \ottsym{:}  \ottnt{K}  \ottsym{.}  \ottnt{e}  \mid   \mathsf{op} _{ \mathit{l} \,  \bm{ { S } } ^ {  \mathit{I}  }  }  \,  \bm{ { T } } ^ {  \mathit{J}  }                                        & \text{(values)}                               \\
    \ottnt{h} &\Coloneqq& \ottsym{\{} \, \mathbf{return} \, \mathit{x}  \mapsto  \ottnt{e}  \ottsym{\}}  \mid   \ottnt{h}   \uplus   \ottsym{\{}  \mathsf{op} \,  {\bm{ \beta } }^{ \mathit{J} } : {\bm{ \ottnt{K} } }^{ \mathit{J} }  \, \mathit{p} \, \mathit{k}  \mapsto  \ottnt{e}  \ottsym{\}}                                                                & \text{(handlers)}                             \\
    \ottnt{E} &\Coloneqq&  \Box   \mid  \mathbf{let} \, \mathit{x}  \ottsym{=}  \ottnt{E} \, \mathbf{in} \, \ottnt{e}  \mid   \mathbf{handle}_{ \mathit{l} \,  \bm{ { S } } ^ {  \mathit{I}  }  }  \, \ottnt{E} \, \mathbf{with} \, \ottnt{h}                                                              & \text{(evaluation contexts)}                  \\
   \end{array}
  \]
  \caption{Program syntax of {\lang}.}
  \label{fig:syntax}
\end{figure}

We show the program syntax of {\lang} in Figure~\ref{fig:syntax}.

Expressions, ranged over $\ottnt{e}$, are composed of: values; function applications $\ottnt{v_{{\mathrm{1}}}} \, \ottnt{v_{{\mathrm{2}}}}$; typelike applications $\ottnt{v} \, S$; let-bindings $\mathbf{let} \, \mathit{x}  \ottsym{=}  \ottnt{e_{{\mathrm{1}}}} \, \mathbf{in} \, \ottnt{e_{{\mathrm{2}}}}$; and handling expressions $ \mathbf{handle}_{ \mathit{l} \,  \bm{ { S } } ^ {  \mathit{I}  }  }  \, \ottnt{e} \, \mathbf{with} \, \ottnt{h}$.
%
%
Values are: variables $\mathit{x}$; recursive functions $\ottkw{fun} \, \ottsym{(}  \mathit{f}  \ottsym{,}  \mathit{x}  \ottsym{,}  \ottnt{e}  \ottsym{)}$; typelike abstractions $\Lambda  \alpha  \ottsym{:}  \ottnt{K}  \ottsym{.}  \ottnt{e}$; or operations $ \mathsf{op} _{ \mathit{l} \,  \bm{ { S } } ^ {  \mathit{I}  }  }  \,  \bm{ { T } } ^ {  \mathit{J}  } $.
An operation $ \mathsf{op} _{ \mathit{l} \,  \bm{ { S } } ^ {  \mathit{I}  }  }  \,  \bm{ { T } } ^ {  \mathit{J}  } $ accompanies two typelike sequences
$ \bm{ { S } } ^ {  \mathit{I}  } $ and $ \bm{ { T } } ^ {  \mathit{J}  } $, which are parameters of effect label $\mathit{l}$ and
operation $\mathsf{op}$, respectively.
We write $\lambda  \mathit{x}  \ottsym{.}  \ottnt{e}$ for $\ottkw{fun} \, \ottsym{(}  \mathit{f}  \ottsym{,}  \mathit{x}  \ottsym{,}  \ottnt{e}  \ottsym{)}$ when variable $\mathit{f}$ does not occur free in expression $\ottnt{e}$.

An effect handler for label name $\mathit{l}$ possesses one return clause and clauses for the operations of $\mathit{l}$.
For a return clause $\ottsym{\{} \, \mathbf{return} \, \mathit{x}  \mapsto  \ottnt{e}  \ottsym{\}}$, the body $\ottnt{e}$ is executed once a handled expression evaluates to a value $\ottnt{v}$;
$\mathit{x}$ is used to refer to the value $\ottnt{v}$.
For an operation clause $\ottsym{\{}  \mathsf{op} \,  \bm{ \beta } : \bm{ \ottnt{K} }  \, \mathit{p} \, \mathit{k}  \mapsto  \ottnt{e}  \ottsym{\}}$,
the body $\ottnt{e}$ is executed once a handled expression calls operation $\mathsf{op}$.
Typelike variables $ \bm{ { \beta } } $, variable $\mathit{p}$, and variable
$\mathit{k}$ are replaced by typelike parameters attached to the operation call, the
argument of the call, and the delimited continuation from the call up to the
handling expression installing the effect handler, respectively.

Evaluation contexts, ranged over by $\ottnt{E}$, are defined in a standard manner.
They may wrap a hole $ \Box $ by let-constructs and handling constructs.

\subsection{Operational Semantics}

\TY{Freeness of labels is introduced.}

\begin{figure}[t]
  \textbf{Freeness of labels}\tquad\fbox{$ \mathit{n}  \mathrm{-free} ( \ottnt{L} ,  \ottnt{E} ) $} \hfill \phantom{}
  \begin{mathpar}
    \inferrule{
    }{
       0  \mathrm{-free} ( \ottnt{L} ,  \Box ) 
    }

    \inferrule{
       \mathit{n}  \mathrm{-free} ( \ottnt{L} ,  \ottnt{E} ) 
    }{
       \mathit{n}  \mathrm{-free} ( \ottnt{L} ,  \mathbf{let} \, \mathit{x}  \ottsym{=}  \ottnt{E} \, \mathbf{in} \, \ottnt{e} ) 
    }

\OLD{
    \inferrule{
       \ottsym{(}  \mathit{n}  \ottsym{+}  1  \ottsym{)}  \mathrm{-free} ( \ottnt{L} ,  \ottnt{E} ) 
    }{
       \mathit{n}  \mathrm{-free} ( \ottnt{L} ,   \mathbf{handle}_{ \ottnt{L} }  \, \ottnt{E} \, \mathbf{with} \, \ottnt{h} ) 
    }

}
    \inferrule{
     \mathit{n}  \mathrm{-free} ( \ottnt{L} ,  \ottnt{E} )  \\ \ottnt{L} \neq \ottnt{L'}
    }{
     \mathit{n}  \mathrm{-free} ( \ottnt{L} ,   \mathbf{handle}_{ \ottnt{L'} }  \, \ottnt{E} \, \mathbf{with} \, \ottnt{h} ) 
    }
  \end{mathpar}

  \phantom{}\\
  \textbf{Reduction}\tquad\fbox{$\ottnt{e}  \longmapsto  \ottnt{e'}$} \hfill\phantom{} \\[-1ex]

  \begin{mathpar}
    \inferrule{ }{\ottkw{fun} \, \ottsym{(}  \mathit{f}  \ottsym{,}  \mathit{x}  \ottsym{,}  \ottnt{e}  \ottsym{)} \, \ottnt{v}  \longmapsto  \ottnt{e} \,  \! [  \ottkw{fun} \, \ottsym{(}  \mathit{f}  \ottsym{,}  \mathit{x}  \ottsym{,}  \ottnt{e}  \ottsym{)}  /  \mathit{f}  ]  \,  \! [  \ottnt{v}  /  \mathit{x}  ] } \ \rname{R}{App}

    \inferrule{ }{\ottsym{(}  \Lambda  \alpha  \ottsym{:}  \ottnt{K}  \ottsym{.}  \ottnt{e}  \ottsym{)} \, S  \longmapsto  \ottnt{e} \,  \! [  S  /  \alpha   ] } \ \rname{R}{TApp}

    \inferrule{ }{\mathbf{let} \, \mathit{x}  \ottsym{=}  \ottnt{v} \, \mathbf{in} \, \ottnt{e}  \longmapsto  \ottnt{e} \,  \! [  \ottnt{v}  /  \mathit{x}  ] } \ \rname{R}{Let}

    \inferrule{ \mathbf{return} \, \mathit{x}  \mapsto  \ottnt{e_{\ottmv{r}}}   \in   \ottnt{h} }{ \mathbf{handle}_{ \mathit{l} \,  \bm{ { S } } ^ {  \mathit{I}  }  }  \, \ottnt{v} \, \mathbf{with} \, \ottnt{h}  \longmapsto  \ottnt{e_{\ottmv{r}}} \,  \! [  \ottnt{v}  /  \mathit{x}  ] } \ \rname{R}{Handle1}
    %
    %

    \inferrule{
   \OLD{
     \mathit{l}  ::    \forall    {\bm{ \alpha } }^{ \mathit{I} } : {\bm{ \ottnt{K} } }^{ \mathit{I} }    \ottsym{.}    \sigma    \in   \Xi  \\  \mathsf{op}  \ottsym{:}    \forall    {\bm{ \beta } }^{ \mathit{J} } : {\bm{ \ottnt{K} } }^{ \mathit{J} }    \ottsym{.}    \ottnt{A}   \Rightarrow   \ottnt{B}    \in   \sigma \,  \! [ {\bm{ { S } } }^{ \mathit{I} } / {\bm{ \alpha } }^{ \mathit{I} } ]   \\  \mathsf{op} \,  {\bm{ \beta } }^{ \mathit{J} } : {\bm{ \ottnt{K} } }^{ \mathit{J} }  \, \mathit{p} \, \mathit{k}  \mapsto  \ottnt{e}   \in   \ottnt{h}  \\\\
    \ottnt{v_{\ottmv{cont}}} = \lambda  \mathit{z}  \ottsym{.}   \mathbf{handle}_{ \mathit{l} \,  \bm{ { S } } ^ {  \mathit{I}  }  }  \, \ottnt{E}  \ottsym{[}  \mathit{z}  \ottsym{]} \, \mathbf{with} \, \ottnt{h} \\  0  \mathrm{-free} ( \mathit{l} \,  \bm{ { S } } ^ {  \mathit{I}  }  ,  \ottnt{E} ) 
   }
     \mathsf{op} \,  {\bm{ \beta } }^{ \mathit{J} } : {\bm{ \ottnt{K} } }^{ \mathit{J} }  \, \mathit{p} \, \mathit{k}  \mapsto  \ottnt{e}   \in   \ottnt{h}  \\ \ottnt{v_{\ottmv{cont}}} = \lambda  \mathit{z}  \ottsym{.}   \mathbf{handle}_{ \mathit{l} \,  \bm{ { S } } ^ {  \mathit{I}  }  }  \, \ottnt{E}  \ottsym{[}  \mathit{z}  \ottsym{]} \, \mathbf{with} \, \ottnt{h} \\  0  \mathrm{-free} ( \mathit{l} \,  \bm{ { S } } ^ {  \mathit{I}  }  ,  \ottnt{E} ) 
    }{
     \mathbf{handle}_{ \mathit{l} \,  \bm{ { S } } ^ {  \mathit{I}  }  }  \, \ottnt{E}  \ottsym{[}   \mathsf{op} _{ \mathit{l} \,  \bm{ { S } } ^ {  \mathit{I}  }  }  \,  \bm{ { T } } ^ {  \mathit{J}  }  \, \ottnt{v}  \ottsym{]} \, \mathbf{with} \, \ottnt{h}  \longmapsto  \ottnt{e} \,  \! [ {\bm{ { T } } }^{ \mathit{J} } / {\bm{ \beta } }^{ \mathit{J} } ]  \,  \! [  \ottnt{v}  /  \mathit{p}  ]  \,  \! [  \ottnt{v_{\ottmv{cont}}}  /  \mathit{k}  ] 
    } \ \rname{R}{Handle2}
  \end{mathpar}

  \textbf{Evaluation}\tquad\fbox{$\ottnt{e}  \longrightarrow  \ottnt{e'}$} \hfill\phantom{} \\[-3ex]
  \begin{mathpar}
    \inferrule{\ottnt{e_{{\mathrm{1}}}}  \longmapsto  \ottnt{e_{{\mathrm{2}}}}}{\ottnt{E}  \ottsym{[}  \ottnt{e_{{\mathrm{1}}}}  \ottsym{]}  \longrightarrow  \ottnt{E}  \ottsym{[}  \ottnt{e_{{\mathrm{2}}}}  \ottsym{]}}\ \rname{E}{Eval}
  \end{mathpar}
  \caption{Operational semantics of {\lang}.}
  \label{fig:semantics}
\end{figure}

The operational semantics of {\lang} is defined in Figure~\ref{fig:semantics}.
Following \citet{biernacki_handle_2018}, it uses
the notion of \emph{freeness}, which helps define the operational semantics of lift coercions in Section~\ref{sec:coercions}.
Figure~\ref{fig:semantics} defines $0$-\emph{freeness} of
labels~\citep{biernacki_handle_2018}.
The judgment $ 0  \mathrm{-free} ( \ottnt{L} ,  \ottnt{E} ) $, which is read as ``an label $\ottnt{L}$ is $0$-free
in an evaluation context $\ottnt{E}$,'' means that any operation of $\ottnt{L}$ called
under $\ottnt{E}$ is not handled.
The operational semantics of {\lang} uses this notion to ensure that every call
to an operation of effect label $\ottnt{L}$ is handled by the innermost $\ottnt{L}$'s
effect handler enclosing the operation call.
We generalize $0$-freeness to $n$-freeness for an arbitrary natural number $n$ in
introducing lift coercions (see Section~\ref{sec:coercions} for detail).

We show the operational semantics of {\lang} in Figure~\ref{fig:semantics}. The
semantics comprises two binary relations: the reduction relation $ \longmapsto $ and
the evaluation relation $ \longrightarrow $. The reduction relation defines the basic
computation; in contrast, the evaluation relation gives a way of reducing
subexpressions.

The reduction relation is defined by five rules. Function applications, typelike
applications, let-bindings are reduced as usual.
The remaining are the standard rules to reduce handling expressions.
Consider an expression $ \mathbf{handle}_{ \mathit{l} \,  \bm{ { S } } ^ {  \mathit{I}  }  }  \, \ottnt{e} \, \mathbf{with} \, \ottnt{h}$.
If the handled expression $\ottnt{e}$ is a value $\ottnt{v}$, the rule {\rname{R}{Handle1}}
reduces the handling expression to the body $\ottnt{e_{\ottmv{r}}}$ of the return clause $\ottsym{\{} \, \mathbf{return} \, \mathit{x}  \mapsto  \ottnt{e_{\ottmv{r}}}  \ottsym{\}}$ of $\ottnt{h}$ by substituting $\ottnt{v}$ for $\mathit{x}$ in $\ottnt{e_{\ottmv{r}}}$.
\TY{The explanation of \rname{R}{Handle2} is changed.}
The other rule \rname{R}{Handle2} is used when $\ottnt{e}$ calls an operation $\mathsf{op}$ of label name $\mathit{l}$, that is, $\ottnt{e}$ takes the form $\ottnt{E}  \ottsym{[}   \mathsf{op} _{ \mathit{l} \,  \bm{ { S } } ^ {  \mathit{I}  }  }  \,  \bm{ { T } } ^ {  \mathit{J}  }  \, \ottnt{v}  \ottsym{]}$
for some $\ottnt{E}$, $ \bm{ { T } } ^ {  \mathit{J}  } $, and $\ottnt{v}$ (it is guaranteed by the type-and-effect system that the typelike arguments to $\mathit{l}$ in the operation call are $ \bm{ { S } } ^ {  \mathit{I}  } $).
The reduction rule \rname{R}{Handle2} assumes $ 0  \mathrm{-free} ( \mathit{l} \,  \bm{ { S } } ^ {  \mathit{I}  }  ,  \ottnt{E} ) $, which
ensures that $\ottnt{h}$ is the effect handler closest to the operation call among
the ones for $\mathit{l} \,  \bm{ { S } } ^ {  \mathit{I}  } $.
After substituting the argument typelikes $ \bm{ { T } } ^ {  \mathit{J}  } $, the argument value
$\ottnt{v}$, and the captured delimited continuation $\ottnt{v_{\ottmv{cont}}}$ (which installs the
effect handler $\ottnt{h}$ on the captured evaluation context $\ottnt{E}$ because effect
handlers in {\lang} are \emph{deep}) for the corresponding variables
of $\mathsf{op}$'s operation clause in $\ottnt{h}$, the evaluation proceeds
to reducing the clause's body.

The evaluation relation only has one rule {\rname{E}{Eval}}. It means that the evaluation of an entire program proceeds by decomposing it into a redex $\ottnt{e}$ and an evaluation context $\ottnt{E}$, reducing $\ottnt{e}$ to an expression $\ottnt{e'}$, and then filling the hole of $\ottnt{E}$ with the reduction result $\ottnt{e'}$.

\subsection{Type-and-Effect System}
\begin{figure}[t]
  \phantom{}\\
  \textbf{Typing}\tquad\fbox{$\Gamma  \vdash  \ottnt{e}  \ottsym{:}  \ottnt{A}  \mid  \varepsilon$}\hfill\phantom{}
  \begin{mathpar}
    \inferrule{\vdash  \Gamma \\  \mathit{x}   \ottsym{:}   \ottnt{A}   \in   \Gamma }{\Gamma  \vdash  \mathit{x}  \ottsym{:}  \ottnt{A}  \mid   \bbZero }\ \rname{T}{Var}

    \inferrule{\Gamma  \ottsym{,}  \mathit{f}  \ottsym{:}   \ottnt{A}    \rightarrow_{ \varepsilon }    \ottnt{B}   \ottsym{,}  \mathit{x}  \ottsym{:}  \ottnt{A}  \vdash  \ottnt{e}  \ottsym{:}  \ottnt{B}  \mid  \varepsilon}{\Gamma  \vdash  \ottkw{fun} \, \ottsym{(}  \mathit{f}  \ottsym{,}  \mathit{x}  \ottsym{,}  \ottnt{e}  \ottsym{)}  \ottsym{:}   \ottnt{A}    \rightarrow_{ \varepsilon }    \ottnt{B}   \mid   \bbZero }\ \rname{T}{Abs}

    \inferrule{\Gamma  \vdash  \ottnt{v_{{\mathrm{1}}}}  \ottsym{:}   \ottnt{A}    \rightarrow_{ \varepsilon }    \ottnt{B}   \mid   \bbZero  \\ \Gamma  \vdash  \ottnt{v_{{\mathrm{2}}}}  \ottsym{:}  \ottnt{A}  \mid   \bbZero }{\Gamma  \vdash  \ottnt{v_{{\mathrm{1}}}} \, \ottnt{v_{{\mathrm{2}}}}  \ottsym{:}  \ottnt{B}  \mid  \varepsilon}\ \rname{T}{App}

    \inferrule{\Gamma  \ottsym{,}  \alpha  \ottsym{:}  \ottnt{K}  \vdash  \ottnt{e}  \ottsym{:}  \ottnt{A}  \mid  \varepsilon}{\Gamma  \vdash  \Lambda  \alpha  \ottsym{:}  \ottnt{K}  \ottsym{.}  \ottnt{e}  \ottsym{:}    \forall   \alpha  \ottsym{:}  \ottnt{K}   \ottsym{.}    \ottnt{A}    ^{ \varepsilon }    \mid   \bbZero }\ \rname{T}{TAbs}

    \inferrule{\Gamma  \vdash  \ottnt{v}  \ottsym{:}    \forall   \alpha  \ottsym{:}  \ottnt{K}   \ottsym{.}    \ottnt{A}    ^{ \varepsilon }    \mid   \bbZero  \\ \Gamma  \vdash  S  \ottsym{:}  \ottnt{K}}{\Gamma  \vdash  \ottnt{v} \, S  \ottsym{:}  \ottnt{A} \,  \! [  S  /  \alpha   ]   \mid  \varepsilon \,  \! [  S  /  \alpha   ] }\ \rname{T}{TApp}

    \inferrule{\Gamma  \vdash  \ottnt{e_{{\mathrm{1}}}}  \ottsym{:}  \ottnt{A}  \mid  \varepsilon \\ \Gamma  \ottsym{,}  \mathit{x}  \ottsym{:}  \ottnt{A}  \vdash  \ottnt{e_{{\mathrm{2}}}}  \ottsym{:}  \ottnt{B}  \mid  \varepsilon}{\Gamma  \vdash  \mathbf{let} \, \mathit{x}  \ottsym{=}  \ottnt{e_{{\mathrm{1}}}} \, \mathbf{in} \, \ottnt{e_{{\mathrm{2}}}}  \ottsym{:}  \ottnt{B}  \mid  \varepsilon}\ \rname{T}{Let}

    \inferrule{\Gamma  \vdash  \ottnt{e}  \ottsym{:}  \ottnt{A}  \mid  \varepsilon \\ \Gamma  \vdash  \ottnt{A}  \mid  \varepsilon  <:  \ottnt{A'}  \mid  \varepsilon'}{\Gamma  \vdash  \ottnt{e}  \ottsym{:}  \ottnt{A'}  \mid  \varepsilon'}\ \rname{T}{Sub}

    \inferrule{
     \mathit{l}  ::    \forall    {\bm{ \alpha } }^{ \mathit{I} } : {\bm{ \ottnt{K} } }^{ \mathit{I} }    \ottsym{.}    \sigma    \in   \Xi  \\  \mathsf{op}  \ottsym{:}    \forall    {\bm{ \beta } }^{ \mathit{J} } : {\bm{ \ottnt{K'} } }^{ \mathit{J} }    \ottsym{.}    \ottnt{A}   \Rightarrow   \ottnt{B}    \in   \sigma \,  \! [ {\bm{ { S } } }^{ \mathit{I} } / {\bm{ \alpha } }^{ \mathit{I} } ]   \\\\
    \vdash  \Gamma \\ \Gamma  \vdash   \bm{ { S } }^{ \mathit{I} } : \bm{ \ottnt{K} }^{ \mathit{I} }  \\ \Gamma  \vdash   \bm{ { T } }^{ \mathit{J} } : \bm{ \ottnt{K'} }^{ \mathit{J} } 
    }{
    \Gamma  \vdash   \mathsf{op} _{ \mathit{l} \,  \bm{ { S } } ^ {  \mathit{I}  }  }  \,  \bm{ { T } } ^ {  \mathit{J}  }   \ottsym{:}   \ottsym{(}  \ottnt{A} \,  \! [ {\bm{ { T } } }^{ \mathit{J} } / {\bm{ \beta } }^{ \mathit{J} } ]   \ottsym{)}    \rightarrow_{  \lift{ \mathit{l} \,  \bm{ { S } } ^ {  \mathit{I}  }  }  }    \ottsym{(}  \ottnt{B} \,  \! [ {\bm{ { T } } }^{ \mathit{I} } / {\bm{ \beta } }^{ \mathit{I} } ]   \ottsym{)}   \mid   \bbZero 
    }\ \rname{T}{Op}

    \inferrule{
    \\ \Gamma  \vdash  \ottnt{e}  \ottsym{:}  \ottnt{A}  \mid  \varepsilon' \\ \\  \mathit{l}  ::    \forall    {\bm{ \alpha } }^{ \mathit{I} } : {\bm{ \ottnt{K} } }^{ \mathit{I} }    \ottsym{.}    \sigma    \in   \Xi  \\ \\ \Gamma  \vdash   \bm{ { S } }^{ \mathit{I} } : \bm{ \ottnt{K} }^{ \mathit{I} }  \\\\
     \Gamma  \vdash _{ \sigma \,  \! [ {\bm{ { S } } }^{ \mathit{I} } / {\bm{ \alpha } }^{ \mathit{I} } ]  }  \ottnt{h}  :  \ottnt{A}   \Rightarrow  ^ { \varepsilon }  \ottnt{B}  \\    \lift{ \mathit{l} \,  \bm{ { S } } ^ {  \mathit{I}  }  }   \mathop{ \odot }  \varepsilon    \sim   \varepsilon' 
    }{
    \Gamma  \vdash   \mathbf{handle}_{ \mathit{l} \,  \bm{ { S } } ^ {  \mathit{I}  }  }  \, \ottnt{e} \, \mathbf{with} \, \ottnt{h}  \ottsym{:}  \ottnt{B}  \mid  \varepsilon
    }\ \rname{T}{Handling}
  \end{mathpar}

  \textbf{Handler Typing}\tquad\fbox{$ \Gamma  \vdash _{ \sigma }  \ottnt{h}  :  \ottnt{A}   \Rightarrow  ^ { \varepsilon }  \ottnt{B} $}\hfill\phantom{}
  \begin{mathpar}
    \inferrule{
    \Gamma  \ottsym{,}  \mathit{x}  \ottsym{:}  \ottnt{A}  \vdash  \ottnt{e_{\ottmv{r}}}  \ottsym{:}  \ottnt{B}  \mid  \varepsilon
    }{
     \Gamma  \vdash _{  \{\}  }  \ottsym{\{} \, \mathbf{return} \, \mathit{x}  \mapsto  \ottnt{e_{\ottmv{r}}}  \ottsym{\}}  :  \ottnt{A}   \Rightarrow  ^ { \varepsilon }  \ottnt{B} 
    }\ \rname{H}{Return}

    \inferrule{
    \sigma  \ottsym{=}   \sigma'   \uplus   \ottsym{\{}  \mathsf{op}  \ottsym{:}    \forall    {\bm{ \beta } }^{ \mathit{J} } : {\bm{ \ottnt{K} } }^{ \mathit{J} }    \ottsym{.}    \ottnt{A'}   \Rightarrow   \ottnt{B'}   \ottsym{\}}  \\\\
     \Gamma  \vdash _{ \sigma' }  \ottnt{h}  :  \ottnt{A}   \Rightarrow  ^ { \varepsilon }  \ottnt{B}  \\ \\ \Gamma  \ottsym{,}   {\bm{ \beta } }^{ \mathit{J} } : {\bm{ \ottnt{K} } }^{ \mathit{J} }   \ottsym{,}  \mathit{p}  \ottsym{:}  \ottnt{A'}  \ottsym{,}  \mathit{k}  \ottsym{:}   \ottnt{B'}    \rightarrow_{ \varepsilon }    \ottnt{B}   \vdash  \ottnt{e}  \ottsym{:}  \ottnt{B}  \mid  \varepsilon
    }{
     \Gamma  \vdash _{ \sigma }   \ottnt{h}   \uplus   \ottsym{\{}  \mathsf{op} \,  {\bm{ \beta } }^{ \mathit{J} } : {\bm{ \ottnt{K} } }^{ \mathit{J} }  \, \mathit{p} \, \mathit{k}  \mapsto  \ottnt{e}  \ottsym{\}}   :  \ottnt{A}   \Rightarrow  ^ { \varepsilon }  \ottnt{B} 
    }\ \rname{H}{Op}
  \end{mathpar}

  \textbf{Subtyping}\tquad\fbox{$\Gamma  \vdash  \ottnt{A}  <:  \ottnt{B}$} \quad \fbox{$\Gamma  \vdash  \ottnt{A}  \mid  \varepsilon_{{\mathrm{1}}}  <:  \ottnt{B}  \mid  \varepsilon_{{\mathrm{2}}}$} \hfill\phantom{}
  \begin{mathpar}
    \inferrule{\Gamma  \vdash  \ottnt{A}  \ottsym{:}   \mathbf{Typ} }{\Gamma  \vdash  \ottnt{A}  <:  \ottnt{A}} \ \rname{ST}{Refl}

    \inferrule
    {\Gamma  \vdash  \ottnt{A_{{\mathrm{2}}}}  <:  \ottnt{A_{{\mathrm{1}}}} \\ \Gamma  \vdash  \ottnt{B_{{\mathrm{1}}}}  \mid  \varepsilon_{{\mathrm{1}}}  <:  \ottnt{B_{{\mathrm{2}}}}  \mid  \varepsilon_{{\mathrm{2}}}}
    {\Gamma  \vdash   \ottnt{A_{{\mathrm{1}}}}    \rightarrow_{ \varepsilon_{{\mathrm{1}}} }    \ottnt{B_{{\mathrm{1}}}}   <:   \ottnt{A_{{\mathrm{2}}}}    \rightarrow_{ \varepsilon_{{\mathrm{2}}} }    \ottnt{B_{{\mathrm{2}}}} }
    \ \rname{ST}{Fun}

    \inferrule
    {\Gamma  \ottsym{,}  \alpha  \ottsym{:}  \ottnt{K}  \vdash  \ottnt{A_{{\mathrm{1}}}}  \mid  \varepsilon_{{\mathrm{1}}}  <:  \ottnt{A_{{\mathrm{2}}}}  \mid  \varepsilon_{{\mathrm{2}}}}
    {\Gamma  \vdash    \forall   \alpha  \ottsym{:}  \ottnt{K}   \ottsym{.}    \ottnt{A_{{\mathrm{1}}}}    ^{ \varepsilon_{{\mathrm{1}}} }    <:    \forall   \alpha  \ottsym{:}  \ottnt{K}   \ottsym{.}    \ottnt{A_{{\mathrm{2}}}}    ^{ \varepsilon_{{\mathrm{2}}} }  }
    \  \rname{ST}{Poly}

    \inferrule
    {\Gamma  \vdash  \ottnt{A_{{\mathrm{1}}}}  <:  \ottnt{B} \\ \Gamma  \vdash   \varepsilon_{{\mathrm{1}}}  \olessthan  \varepsilon_{{\mathrm{2}}} }
    {\Gamma  \vdash  \ottnt{A}  \mid  \varepsilon_{{\mathrm{1}}}  <:  \ottnt{B}  \mid  \varepsilon_{{\mathrm{2}}}}
    \  \rname{ST}{Comp}
  \end{mathpar}

  \caption{Type-and-effect system of {\lang}.}
  \label{fig:typing}
\end{figure}

\TY{I placed abbreviations here.}

We show the type-and-effect system of {\lang} in Figure~\ref{fig:typing}. Typing judgments are of the form $\Gamma  \vdash  \ottnt{e}  \ottsym{:}  \ottnt{A}  \mid  \varepsilon$, meaning that an expression $\ottnt{e}$ is typed at $\ottnt{A}$ under a typing context $\Gamma$ and the evaluation of $\ottnt{e}$ may cause effect $\varepsilon$. The rules for variables, function abstractions, function applications, typelike abstractions, typelike applications, and let-bindings are standard.
%

The rule \rname{T}{Sub} allows subsumption by subtyping.
We show the subtyping relation $\Gamma  \vdash  \ottnt{A}  <:  \ottnt{B}$ for values and the one $\Gamma  \vdash  \ottnt{A}  \mid  \varepsilon_{{\mathrm{1}}}  <:  \ottnt{B}  \mid  \varepsilon_{{\mathrm{2}}}$ for computations at the bottom of Figure~\ref{fig:typing}. The subtyping rules
are standard except for the subeffecting $\Gamma  \vdash   \varepsilon_{{\mathrm{1}}}  \olessthan  \varepsilon_{{\mathrm{2}}} $, which is used in the rule \rname{ST}{Comp} for the second subtyping relation.
The subeffecting is defined via the given effect algebra:
\[
 \Gamma  \vdash   \varepsilon_{{\mathrm{1}}}  \olessthan  \varepsilon_{{\mathrm{2}}}  \defeq \exists \varepsilon . \,   \varepsilon_{{\mathrm{1}}}  \mathop{ \odot }  \varepsilon    \sim   \varepsilon_{{\mathrm{2}}}  \land (\forall \varepsilon' \in \{ \varepsilon_{{\mathrm{1}}}, \varepsilon_{{\mathrm{2}}}, \varepsilon \} . \, \Gamma  \vdash  \varepsilon'  \ottsym{:}   \mathbf{Eff} ) ~.
\]

The rule \rname{T}{Op} typecheckes operation $ \mathsf{op} _{ \mathit{l} \,  \bm{ { S } } ^ {  \mathit{I}  }  }  \,  \bm{ { T } } ^ {  \mathit{J}  } $ if
$\mathsf{op}$ belongs to effect label $\mathit{l}$, and if the kinds of typelike
arguments $ \bm{ { S } } ^ {  \mathit{I}  } $ and $ \bm{ { T } } ^ {  \mathit{J}  } $ are matched with those of parameters of
$\mathit{l}$ in the effect context $\Xi$.
The operation is given a function type determined by the argument and return
type of $\mathsf{op}$ in $\Xi$ and typelike arguments $ \bm{ { S } } ^ {  \mathit{I}  } $ and $ \bm{ { T } } ^ {  \mathit{J}  } $.
Because every call to the operation only invokes effect label $\mathit{l} \,  \bm{ { S } } ^ {  \mathit{I}  } $,
the latent effect of the function type is given by injecting $\mathit{l} \,  \bm{ { S } } ^ {  \mathit{I}  } $ via $ \lift{ \ottsym{-} } $.

The rule \rname{T}{Handling} is for handling expressions.
Assume that a handled expression $\ottnt{e}$ is of type $\ottnt{A}$ and has effect $\varepsilon'$.
If it is handled by an effect handler for effect label $\mathit{l} \,  \bm{ { S } } ^ {  \mathit{I}  } $,
the operations of $\mathit{l} \,  \bm{ { S } } ^ {  \mathit{I}  } $ become unobservable from the outer context.
Thus, the effect $\varepsilon$ of the handling expression is the result of removing label
$\mathit{l} \,  \bm{ { S } } ^ {  \mathit{I}  } $ from effect $\varepsilon'$.
This ``label-removing manipulation'' is represented as $\Gamma  \vdash     \lift{ \mathit{l} \,  \bm{ { S } } ^ {  \mathit{I}  }  }   \mathop{ \odot }  \varepsilon    \sim   \varepsilon' $.
Therefore, the result $\varepsilon$ of the label-removing manipulation depends on the given effect algebra.
For example, if the effect algebra {\eaSimpleRow} for simple rows is given,
the result of removing the label $\mathsf{Exc}$ from the effect $\langle  \mathsf{Exc}  \mid  \langle  \mathsf{Exc}  \mid  \langle  \mathsf{Choice}  \mid  \langle  \rangle  \rangle  \rangle  \rangle$ can be
$\langle  \mathsf{Choice}  \mid  \langle  \rangle  \rangle$ because $  \langle  \mathsf{Exc}  \mid  \langle  \rangle  \rangle  \mathop{  \odot_\mathrm{SimpR}  }  \langle  \mathsf{Choice}  \mid  \langle  \rangle  \rangle     \sim_{\eanameSimpleRow}    \langle  \mathsf{Exc}  \mid  \langle  \mathsf{Exc}  \mid  \langle  \mathsf{Choice}  \mid  \langle  \rangle  \rangle  \rangle  \rangle $
holds (recall that simple rows can collapse multiple occurrences of the same label into one).
On the contrary, the removing result in the algebra {\eaScopedRow} for scoped rows can be
$\langle  \mathsf{Exc}  \mid  \langle  \mathsf{Choice}  \mid  \langle  \rangle  \rangle  \rangle$ but cannot be $\langle  \mathsf{Choice}  \mid  \langle  \rangle  \rangle$.

The type $\ottnt{B}$ of the handling expression is determined by handler $\ottnt{h}$:
typing judgments for handlers take the form $ \Gamma  \vdash _{ \sigma }  \ottnt{h}  :  \ottnt{A}   \Rightarrow  ^ { \varepsilon }  \ottnt{B} $,
which means that handler $\ottnt{h}$ transforms computation of type $\ottnt{A}$
involving an effect label with operation signature $\sigma$ to that of type
$\ottnt{B}$ with effect $\varepsilon$.
The rules \rname{H}{Return} and \rname{H}{Op} are for return and
operation clauses and reflect the reduction rules \rname{R}{Handle1} and
\rname{R}{Handle2}, respectively.
Note that the return type of a continuation variable $\mathit{k}$ equals the type
$\ottnt{B}$ of the handling expression as the effect handlers in {\lang} are deep~\cite{kammar_handlers_2013}.

\OLD{
This section shows the syntax, semantics, and type-and-effect system of {\lang}. It is based on a call-by-value $\lambda$-calculus with algebraic effects and handlers \cite{leijen_type_2017, biernacki_abstracting_2019, sekiyama_signature_2020}, except for having effect signatures and ARE as parameters.

\subsection{Syntax}

\begin{figure}[t]
  \begin{flalign*}
    \ottnt{e}               \Coloneqq\phantom{} & \ottnt{v}  \mid  \ottnt{v_{{\mathrm{1}}}} \, \ottnt{v_{{\mathrm{2}}}}  \mid  \ottnt{v} \, S  \mid  \mathbf{let} \, \mathit{x}  \ottsym{=}  \ottnt{e_{{\mathrm{1}}}} \, \mathbf{in} \, \ottnt{e_{{\mathrm{2}}}}  \mid   \mathbf{handle}_{ \mathit{l} \,  \bm{ { S } } ^ {  \mathit{I}  }  }  \, \ottnt{e} \, \mathbf{with} \, \ottnt{h}                                 & \text{(expressions)}                          \\
    \ottnt{v}               \Coloneqq\phantom{} & \mathit{x}  \mid  \ottkw{fun} \, \ottsym{(}  \mathit{g}  \ottsym{,}  \mathit{x}  \ottsym{,}  \ottnt{e}  \ottsym{)}  \mid  \Lambda  \alpha  \ottsym{:}  \ottnt{K}  \ottsym{.}  \ottnt{e}  \mid   \mathsf{op} _{ \mathit{l} \,  \bm{ { S } } ^ {  \mathit{I}  }  }  \,  \bm{ { T } } ^ {  \mathit{J}  }                                        & \text{(values)}                               \\
    \ottnt{h}               \Coloneqq\phantom{} & \ottsym{\{} \, \mathbf{return} \, \mathit{x}  \mapsto  \ottnt{e}  \ottsym{\}}  \mid   \ottnt{h}   \uplus   \ottsym{\{}  \mathsf{op} \,  {\bm{ \beta } }^{ \mathit{J} } : {\bm{ \ottnt{K} } }^{ \mathit{J} }  \, \mathit{p} \, \mathit{k}  \mapsto  \ottnt{e}  \ottsym{\}}                                                                & \text{(handlers)}                             \\
    \ottnt{E}               \Coloneqq\phantom{} &  \Box   \mid  \mathbf{let} \, \mathit{x}  \ottsym{=}  \ottnt{E} \, \mathbf{in} \, \ottnt{e}  \mid   \mathbf{handle}_{ \mathit{l} \,  \bm{ { S } } ^ {  \mathit{I}  }  }  \, \ottnt{E} \, \mathbf{with} \, \ottnt{h}                                                              & \text{(evaluation contexts)}                  \\
    \texttt{\textcolor{red}{<<no parses (char 10): E(l \{U I\})*** >>}}      \Coloneqq\phantom{} &  \Box   \mid  \texttt{\textcolor{red}{<<no parses (char 23): let x = E(l \{U I\}) in e*** >>}}  \mid  \texttt{\textcolor{red}{<<no parses (char 35): handle l' \{U' I'\} E(l \{U I\}) with h*** >>}} \quad (\mathit{l} \,  \bm{ { S } } ^ {  \mathit{I}  }  \neq \mathit{l'} \,  \bm{ { S' } } ^ {  \mathit{I'}  } ) & \text{(evaluation context for $\mathit{l} \,  \bm{ { S } } ^ {  \mathit{I}  } $)} \\[2ex]
    \sigma             \Coloneqq\phantom{} & \ottsym{\{}  \mathsf{op_{\ottmv{i}}}  \ottsym{:}    \forall    {\bm{ \beta_{\ottmv{i}} } }^{ \mathit{J} } : {\bm{ \ottnt{K_{\ottmv{i}}} } }^{ \mathit{J} }    \ottsym{.}    \ottnt{A_{\ottmv{i}}}   \Rightarrow   \ottnt{B_{\ottmv{i}}}   \ottsym{\}}   ^{ \ottmv{i}  \in  \mathit{I} }                                                                              & \text{(operation signatures)}                 \\
    \Xi               \Coloneqq\phantom{} &  \emptyset   \mid  \Xi  \ottsym{,}   \mathit{l}  ::    \forall    \bm{ \alpha } : \bm{ \ottnt{K} }    \ottsym{.}    \sigma                                                                                & \text{(effect contexts)}
  \end{flalign*}
  \caption{Syntax of {\lang}.}
  \label{fig:syntax}
\end{figure}

We show the syntax of {\lang} in Fig~\ref{fig:syntax}. The syntax of kinds or typelikes is already mentioned by Section~\ref{sec:effdef}.

Expressions are ranged over by $\ottnt{e}$, and values by $\ottnt{v}$. Expressions are composed of: values $\ottnt{v}$; function applications $\ottnt{v_{{\mathrm{1}}}} \, \ottnt{v_{{\mathrm{2}}}}$; typelike applications $\ottnt{v} \, S$; let-bindings $\mathbf{let} \, \mathit{x}  \ottsym{=}  \ottnt{e_{{\mathrm{1}}}} \, \mathbf{in} \, \ottnt{e_{{\mathrm{2}}}}$; and handlings $ \mathbf{handle}_{ \mathit{l} \,  \bm{ { S } } ^ {  \mathit{I}  }  }  \, \ottnt{e} \, \mathbf{with} \, \ottnt{h}$. A label $\mathit{l}$ in a handling expression $ \mathbf{handle}_{ \mathit{l} \,  \bm{ { S } } ^ {  \mathit{I}  }  }  \, \ottnt{e} \, \mathbf{with} \, \ottnt{h}$ means the handler $\ottnt{h}$ handles operations of $\mathit{l}$. Values are composed of: variables $\mathit{x}$; possibly recursive functions $\ottkw{fun} \, \ottsym{(}  \mathit{g}  \ottsym{,}  \mathit{x}  \ottsym{,}  \ottnt{e}  \ottsym{)}$; typelike abstractions $\Lambda  \alpha  \ottsym{:}  \ottnt{K}  \ottsym{.}  \ottnt{e}$; and operation calls $ \mathsf{op} _{ \mathit{l} \,  \bm{ { S } } ^ {  \mathit{I}  }  }  \,  \bm{ { T } } ^ {  \mathit{J}  } $.

An effect handler is a finite set comprising a return clause and operation clauses. A return clause $\ottsym{\{} \, \mathbf{return} \, \mathit{x}  \mapsto  \ottnt{e}  \ottsym{\}}$ binds a value to $\mathit{x}$, and an operation clause $\{ \mathsf{op}\, \bm{\beta}^{\mathit{J}} : \bm{\ottnt{K}}^{\mathit{J}} \, \mathit{p} \, \mathit{k} \mapsto \ottnt{e} \}$ binds typelikes to $\bm{\beta}^{\mathit{J}}$, the argument of the operation $\mathsf{op}$ to $\mathit{p}$, and the delimited continuation of the operation to $\mathit{k}$.

Evaluation contexts are composed of two variations. The one, ranged over by $\ottnt{E}$, is standard evaluation contexts: a hole $ \Box $, let-bindings $\mathbf{let} \, \mathit{x}  \ottsym{=}  \ottnt{E} \, \mathbf{in} \, \ottnt{e}$, and handler $ \mathbf{handle}_{ \mathit{l} \,  \bm{ { S } } ^ {  \mathit{I}  }  }  \, \ottnt{E} \, \mathbf{with} \, \ottnt{h}$. Another, ranged over by $\texttt{\textcolor{red}{<<no parses (char 10): E(l \{U I\})*** >>}}$, is parameterized by a label $\mathit{l} \,  \bm{ { S } } ^ {  \mathit{I}  } $ and omits handlers handling operations of $\mathit{l} \,  \bm{ { S } } ^ {  \mathit{I}  } $.

Operation signatures, ranged over by $\sigma$, are finite sets of tuples, including an operation name and signature. Effect contexts ranged over by $\Xi$, are finite sequences of bindings of a label name to its typelike parameters and operation signatures $ \mathit{l}  ::    \forall    {\bm{ \alpha } }^{ \mathit{I} } : {\bm{ \ottnt{K} } }^{ \mathit{I} }    \ottsym{.}    \sigma  $. For example, only $\mathsf{Exc}$ and $\mathsf{IO}$ can be used in programs when $\Xi$ is
\begin{align*}
   \mathsf{Exc}  ::  \ottsym{\{}  \mathsf{raise}  \ottsym{:}    \forall   \alpha  \ottsym{:}   \mathbf{Typ}    \ottsym{.}     \mathsf{Unit}    \Rightarrow   \alpha   \ottsym{\}}   \ottsym{,}   \mathsf{IO}  ::    \forall   \alpha  \ottsym{:}   \mathbf{Typ}    \ottsym{.}    \ottsym{\{}  \mathsf{print}  \ottsym{:}   \alpha   \Rightarrow    \mathsf{Unit}    \ottsym{\}}  .
\end{align*}
We suppose that any $\Xi$ is \emph{proper effect contexts} defined as follows.
\begin{definition}[Proper Effect Contexts]\label{rem:proper_effctx}
  For any effect context $\Xi$, and any $ \mathit{l}  ::    \forall    {\bm{ \alpha } }^{ \mathit{I} } : {\bm{ \ottnt{K} } }^{ \mathit{I} }    \ottsym{.}    \sigma    \in   \Xi $,
  we assume the following:
  \begin{itemize}
    \item $ \mathit{l}   \ottsym{:}    {\bm{ { \ottnt{K} } } }^{ \mathit{I} }   \rightarrow   \mathbf{Lab}    \in   \Sbase $,
    \item if $ \mathit{l}  ::    \forall    {\bm{ \alpha' } }^{ \mathit{I'} } : {\bm{ \ottnt{K'} } }^{ \mathit{I'} }    \ottsym{.}    \sigma'    \in   \Xi $, then $ {\bm{ \alpha } }^{ \mathit{I} } : {\bm{ \ottnt{K} } }^{ \mathit{I} }  =  {\bm{ \alpha' } }^{ \mathit{I'} } : {\bm{ \ottnt{K'} } }^{ \mathit{I'} } $ and $\sigma = \sigma'$; and
    \item for any $ \mathsf{op}  \ottsym{:}    \forall    {\bm{ \beta } }^{ \mathit{J} } : {\bm{ \ottnt{K'} } }^{ \mathit{J} }    \ottsym{.}    \ottnt{A}   \Rightarrow   \ottnt{B}    \in   \sigma $, the well-formedness judgments
          \[
             {\bm{ \alpha } }^{ \mathit{I} } : {\bm{ \ottnt{K} } }^{ \mathit{I} }   \ottsym{,}   {\bm{ \beta } }^{ \mathit{J} } : {\bm{ \ottnt{K'} } }^{ \mathit{J} }   \vdash  \ottnt{A}  \ottsym{:}   \mathbf{Typ}  \quad \tand \quad  {\bm{ \alpha } }^{ \mathit{I} } : {\bm{ \ottnt{K} } }^{ \mathit{I} }   \ottsym{,}   {\bm{ \beta } }^{ \mathit{J} } : {\bm{ \ottnt{K'} } }^{ \mathit{J} }   \vdash  \ottnt{B}  \ottsym{:}   \mathbf{Typ} 
          \] are derivable.
  \end{itemize}
\end{definition}

\subsection{Operational Semantics}

\begin{figure}[t]
  \phantom{}\\
  \textbf{Reduction}\tquad\fbox{$\ottnt{e}  \longmapsto  \ottnt{e'}$} \hfill\phantom{}
  \begin{mathpar}
    \inferrule{ }{\ottkw{fun} \, \ottsym{(}  \mathit{g}  \ottsym{,}  \mathit{x}  \ottsym{,}  \ottnt{e}  \ottsym{)} \, \ottnt{v}  \longmapsto  \ottnt{e} \,  \! [  \ottkw{fun} \, \ottsym{(}  \mathit{g}  \ottsym{,}  \mathit{x}  \ottsym{,}  \ottnt{e}  \ottsym{)}  /  \mathit{g}  ]  \,  \! [  \ottnt{v}  /  \mathit{x}  ] } \quad \rname{R}{App}

    \inferrule{ }{\ottsym{(}  \Lambda  \alpha  \ottsym{:}  \ottnt{K}  \ottsym{.}  \ottnt{e}  \ottsym{)} \, S  \longmapsto  \ottnt{e} \,  \! [  S  /  \alpha   ] } \quad \rname{R}{TApp}

    \inferrule{ }{\mathbf{let} \, \mathit{x}  \ottsym{=}  \ottnt{v} \, \mathbf{in} \, \ottnt{e}  \longmapsto  \ottnt{e} \,  \! [  \ottnt{v}  /  \mathit{x}  ] } \quad \rname{R}{Let}

    \inferrule{ \mathbf{return} \, \mathit{x}  \mapsto  \ottnt{e_{\ottmv{r}}}   \in   \ottnt{h} }{ \mathbf{handle}_{ \mathit{l} \,  \bm{ { S } } ^ {  \mathit{I}  }  }  \, \ottnt{v} \, \mathbf{with} \, \ottnt{h}  \longmapsto  \ottnt{e_{\ottmv{r}}} \,  \! [  \ottnt{v}  /  \mathit{x}  ] } \quad \rname{R}{Handle1}

    \inferrule
    { \mathit{l}  ::    \forall    {\bm{ \alpha } }^{ \mathit{I} } : {\bm{ \ottnt{K} } }^{ \mathit{I} }    \ottsym{.}    \sigma    \in   \Xi  \\  \mathsf{op}  \ottsym{:}    \forall    {\bm{ \beta } }^{ \mathit{J} } : {\bm{ \ottnt{K} } }^{ \mathit{J} }    \ottsym{.}    \ottnt{A}   \Rightarrow   \ottnt{B}    \in   \sigma \,  \! [ {\bm{ { S } } }^{ \mathit{I} } / {\bm{ \alpha } }^{ \mathit{I} } ]   \\  \mathsf{op} \,  {\bm{ \beta } }^{ \mathit{J} } : {\bm{ \ottnt{K} } }^{ \mathit{J} }  \, \mathit{p} \, \mathit{k}  \mapsto  \ottnt{e}   \in   \ottnt{h}  \\\\ \ottnt{v_{\ottmv{cont}}} = \texttt{\textcolor{red}{<<no parses (char 57): \mbox{$\backslash{}$}z : B[\{V/beta J\}] . handle l \{U I\} E(l \{U I\}) [z] with h***  >>}}}
    {\texttt{\textcolor{red}{<<no parses (char 51): handle l \{U I\} E(l \{U I\}) [op \{l \{U I\} \} \{V J\} v] w***ith h \mbox{$\mid$}--> e [\{V/beta J\}] [v/p] [vcont / k]  >>}}} \quad \rname{R}{Handle2}
  \end{mathpar}

  \textbf{Evaluation}\tquad\fbox{$\ottnt{e}  \longrightarrow  \ottnt{e'}$} \hfill\phantom{}
  \begin{mathpar}
    \inferrule{\ottnt{e_{{\mathrm{1}}}}  \longmapsto  \ottnt{e_{{\mathrm{2}}}}}{\ottnt{E}  \ottsym{[}  \ottnt{e_{{\mathrm{1}}}}  \ottsym{]}  \longrightarrow  \ottnt{E}  \ottsym{[}  \ottnt{e_{{\mathrm{2}}}}  \ottsym{]}}\quad\rname{E}{Eval}
  \end{mathpar}
  \caption{Operational semantics of {\lang}.}
  \label{fig:semantics}
\end{figure}

We show the operational semantics of {\lang} in Fig~\ref{fig:semantics}. The semantics comprises two binary relations: the reduction relation $ \longmapsto $ and the evaluation relation $ \longrightarrow $. The reduction relation gives a way of reducing a redex; in contrast, the evaluation relation gives a way of reducing a whole program.

The reduction relation comprises six rules. A function application $\ottkw{fun} \, \ottsym{(}  \mathit{g}  \ottsym{,}  \mathit{x}  \ottsym{,}  \ottnt{e}  \ottsym{)} \, \ottnt{v}$ is reduced to $\ottnt{e} \,  \! [  \ottkw{fun} \, \ottsym{(}  \mathit{g}  \ottsym{,}  \mathit{x}  \ottsym{,}  \ottnt{e}  \ottsym{)}  /  \mathit{g}  ]  \,  \! [  \ottnt{v}  /  \mathit{x}  ] $ by {\rname{R}{App}}, a typelike application $\ottsym{(}  \Lambda  \alpha  \ottsym{:}  \ottnt{K}  \ottsym{.}  \ottnt{e}  \ottsym{)} \, S$ is reduced to $\ottnt{e} \,  \! [  S  /  \alpha   ] $ by {\rname{R}{TApp}}, and a let binding $\mathbf{let} \, \mathit{x}  \ottsym{=}  \ottnt{v} \, \mathbf{in} \, \ottnt{e}$ is reduced to $\ottnt{e} \,  \! [  \ottnt{v}  /  \mathit{x}  ] $ by {\rname{R}{Let}} as usual. The remaining two rules are about handling expressions. If a handled expression is a value $\ottnt{v}$, namely, a handle expression is $ \mathbf{handle}_{ \mathit{l} \,  \bm{ { S } } ^ {  \mathit{I}  }  }  \, \ottnt{v} \, \mathbf{with} \, \ottnt{h}$, then the rule {\rname{R}{Handle1}} reduces it using the return clause $\ottsym{\{} \, \mathbf{return} \, \mathit{x}  \mapsto  \ottnt{e_{\ottmv{r}}}  \ottsym{\}}$ of $\ottnt{h}$. In this case, binding $\ottnt{v}$ to $\mathit{x}$ in $\ottnt{e_{\ottmv{r}}}$ gives a reduced result $\ottnt{e_{\ottmv{r}}} \,  \! [  \ottnt{v}  /  \mathit{x}  ] $. On the other hand, if a handled expression contains calling an operation $\mathsf{op}$ belonging to a label $\mathit{l}$, then a redex is $\texttt{\textcolor{red}{<<no parses (char 48): handle l \{U I\} E(l \{U I\}) [op \{l \{U I\}\} \{V\} v] w***ith h >>}}$. The reason why we use $\texttt{\textcolor{red}{<<no parses (char 10): E(l \{U I\})*** >>}}$ instead of $\ottnt{E}$ is to require that only the innermost handler can handle operations. For example, the redex of
\begin{align*}
   \mathbf{handle}_{ \mathit{l_{{\mathrm{1}}}} \,  {}  }  \,  \mathbf{handle}_{ \mathit{l_{{\mathrm{1}}}} \,  {}  }  \,  \mathsf{op} _{ \mathit{l_{{\mathrm{1}}}} \,  {}  }  \,  \bm{ { T } }  \, \ottnt{v} \, \mathbf{with} \, \ottnt{h_{{\mathrm{1}}}} \, \mathbf{with} \, \ottnt{h_{{\mathrm{1}}}}
\end{align*}
is $ \mathbf{handle}_{ \mathit{l_{{\mathrm{1}}}} \,  {}  }  \,  \mathsf{op_{{\mathrm{1}}}} _{ \mathit{l_{{\mathrm{1}}}} \,  {}  }  \,  \bm{ { T } }  \, \ottnt{v} \, \mathbf{with} \, \ottnt{h_{{\mathrm{1}}}}$, not the whole expression. Moreover, the redex of
\begin{align*}
   \mathbf{handle}_{ \mathit{l_{{\mathrm{1}}}} \,  {}  }  \,  \mathbf{handle}_{ \mathit{l_{{\mathrm{2}}}} \,  {}  }  \,  \mathsf{op} _{ \mathit{l_{{\mathrm{1}}}} \,  {}  }  \,  \bm{ { T } }  \, \ottnt{v} \, \mathbf{with} \, \ottnt{h_{{\mathrm{2}}}} \, \mathbf{with} \, \ottnt{h_{{\mathrm{1}}}} \quad (\text{where } \mathit{l_{{\mathrm{1}}}} \neq \mathit{l_{{\mathrm{2}}}})
\end{align*}
is the whole expression. Then, the reduced result of {\rname{R}{Handle2}} is given by binding arguments $ \bm{ { T } } ^ {  \mathit{J}  } $ and $\ottnt{v}$ of operation call $ \mathsf{op} _{ \mathit{l} \,  \bm{ { S } } ^ {  \mathit{I}  }  }  \,  \bm{ { T } } ^ {  \mathit{J}  }  \, \ottnt{v}$ to $ \bm{ { \beta } } ^ {  \mathit{J}  } $ and $\mathit{p}$ and binding a current continuation
\begin{align*}
  \texttt{\textcolor{red}{<<no parses (char 71):  \mbox{$\backslash{}$}z : B[\{U/alpha I\}][\{V/beta J\}] . handle l \{U I\} E(l \{U I\}) [z] with h***  >>}}
\end{align*}
to $\mathit{k}$ in $\ottnt{e}$, where $\ottnt{h}$ includes an operation clause $\ottsym{\{}  \mathsf{op} \,  {\bm{ \beta } }^{ \mathit{J} } : {\bm{ \ottnt{K} } }^{ \mathit{J} }  \, \mathit{p} \, \mathit{k}  \mapsto  \ottnt{e}  \ottsym{\}}$.

The evaluation relation has one rule {\rname{E}{Eval}}. A whole program is decomposed to a redex $\ottnt{e}$ and an evaluation context $\ottnt{E}$, and then, a hole of $\ottnt{E}$ is filled by a reduced result of $\ottnt{e}$.

\subsection{Type-and-Effect System}

\begin{figure}[t]
  \phantom{}\\
  \textbf{Typing}\tquad\fbox{$\Gamma  \vdash  \ottnt{e}  \ottsym{:}  \ottnt{A}  \mid  \varepsilon$}\hfill\phantom{}
  \begin{mathpar}
    \inferrule{\vdash  \Gamma \\  \mathit{x}   \ottsym{:}   \ottnt{A}   \in   \Gamma }{\Gamma  \vdash  \mathit{x}  \ottsym{:}  \ottnt{A}  \mid   \bbZero }\quad\rname{T}{Var}

    \inferrule{\Gamma  \ottsym{,}  \mathit{g}  \ottsym{:}   \ottnt{A}    \rightarrow_{ \varepsilon }    \ottnt{B}   \ottsym{,}  \mathit{x}  \ottsym{:}  \ottnt{A}  \vdash  \ottnt{e}  \ottsym{:}  \ottnt{B}  \mid  \varepsilon}{\Gamma  \vdash  \ottkw{fun} \, \ottsym{(}  \mathit{g}  \ottsym{,}  \mathit{x}  \ottsym{,}  \ottnt{e}  \ottsym{)}  \ottsym{:}   \ottnt{A}    \rightarrow_{ \varepsilon }    \ottnt{B}   \mid   \bbZero }\quad\rname{T}{Abs}

    \inferrule{\Gamma  \vdash  \ottnt{v_{{\mathrm{1}}}}  \ottsym{:}   \ottnt{A}    \rightarrow_{ \varepsilon }    \ottnt{B}   \mid   \bbZero  \\ \Gamma  \vdash  \ottnt{v_{{\mathrm{2}}}}  \ottsym{:}  \ottnt{A}  \mid   \bbZero }{\Gamma  \vdash  \ottnt{v_{{\mathrm{1}}}} \, \ottnt{v_{{\mathrm{2}}}}  \ottsym{:}  \ottnt{B}  \mid  \varepsilon}\quad\rname{T}{App}

    \inferrule{\Gamma  \ottsym{,}  \alpha  \ottsym{:}  \ottnt{K}  \vdash  \ottnt{e}  \ottsym{:}  \ottnt{A}  \mid  \varepsilon}{\Gamma  \vdash  \Lambda  \alpha  \ottsym{:}  \ottnt{K}  \ottsym{.}  \ottnt{e}  \ottsym{:}    \forall   \alpha  \ottsym{:}  \ottnt{K}   \ottsym{.}    \ottnt{A}    ^{ \varepsilon }    \mid   \bbZero }\quad\rname{T}{TAbs}

    \inferrule{\Gamma  \vdash  \ottnt{v}  \ottsym{:}    \forall   \alpha  \ottsym{:}  \ottnt{K}   \ottsym{.}    \ottnt{A}    ^{ \varepsilon }    \mid   \bbZero  \\ \Gamma  \vdash  S  \ottsym{:}  \ottnt{K}}{\Gamma  \vdash  \ottnt{v} \, S  \ottsym{:}  \ottnt{A} \,  \! [  S  /  \alpha   ]   \mid  \varepsilon \,  \! [  S  /  \alpha   ] }\quad\rname{T}{TApp}

    \inferrule{\Gamma  \vdash  \ottnt{e_{{\mathrm{1}}}}  \ottsym{:}  \ottnt{A}  \mid  \varepsilon \\ \Gamma  \ottsym{,}  \mathit{x}  \ottsym{:}  \ottnt{A}  \vdash  \ottnt{e_{{\mathrm{2}}}}  \ottsym{:}  \ottnt{B}  \mid  \varepsilon}{\Gamma  \vdash  \mathbf{let} \, \mathit{x}  \ottsym{=}  \ottnt{e_{{\mathrm{1}}}} \, \mathbf{in} \, \ottnt{e_{{\mathrm{2}}}}  \ottsym{:}  \ottnt{B}  \mid  \varepsilon}\quad\rname{T}{Let}

    \inferrule{\Gamma  \vdash  \ottnt{e}  \ottsym{:}  \ottnt{A}  \mid  \varepsilon \\ \Gamma  \vdash  \ottnt{A}  <:  \ottnt{A'} \\  \Delta   \ottsym{(}   \Gamma   \ottsym{)}   \vdash   \varepsilon  \olessthan  \varepsilon' }{\Gamma  \vdash  \ottnt{e}  \ottsym{:}  \ottnt{A'}  \mid  \varepsilon'}\quad\rname{T}{Sub}

    \inferrule{
     \mathit{l}  ::    \forall    {\bm{ \alpha } }^{ \mathit{I} } : {\bm{ \ottnt{K} } }^{ \mathit{I} }    \ottsym{.}    \sigma    \in   \Xi  \\  \mathsf{op}  \ottsym{:}    \forall    {\bm{ \beta } }^{ \mathit{J} } : {\bm{ \ottnt{K'} } }^{ \mathit{J} }    \ottsym{.}    \ottnt{A}   \Rightarrow   \ottnt{B}    \in   \sigma \,  \! [ {\bm{ { S } } }^{ \mathit{I} } / {\bm{ \alpha } }^{ \mathit{I} } ]   \\\\
    \vdash  \Gamma \\ \Gamma  \vdash   \bm{ { S } }^{ \mathit{I} } : \bm{ \ottnt{K} }^{ \mathit{I} }  \\ \Gamma  \vdash   \bm{ { T } }^{ \mathit{J} } : \bm{ \ottnt{K'} }^{ \mathit{J} } 
    }{
    \Gamma  \vdash   \mathsf{op} _{ \mathit{l} \,  \bm{ { S } } ^ {  \mathit{I}  }  }  \,  \bm{ { T } } ^ {  \mathit{J}  }   \ottsym{:}   \ottsym{(}  \ottnt{A} \,  \! [ {\bm{ { T } } }^{ \mathit{J} } / {\bm{ \beta } }^{ \mathit{J} } ]   \ottsym{)}    \rightarrow_{  \lift{ \mathit{l} \,  \bm{ { S } } ^ {  \mathit{I}  }  }  }    \ottsym{(}  \ottnt{B} \,  \! [ {\bm{ { T } } }^{ \mathit{I} } / {\bm{ \beta } }^{ \mathit{I} } ]   \ottsym{)}   \mid   \bbZero 
    }\quad\rname{T}{Op}

    \inferrule{
    \\ \Gamma  \vdash  \ottnt{e}  \ottsym{:}  \ottnt{A}  \mid  \varepsilon' \\ \\  \mathit{l}  ::    \forall    {\bm{ \alpha } }^{ \mathit{N} } : {\bm{ \ottnt{K} } }^{ \mathit{N} }    \ottsym{.}    \sigma    \in   \Xi  \\\\
     \Gamma  \vdash _{ \sigma \,  \! [ {\bm{ { S } } }^{ \mathit{N} } / {\bm{ \alpha } }^{ \mathit{N} } ]  }  \ottnt{h}  :  \ottnt{A}   \Rightarrow  ^ { \varepsilon }  \ottnt{B}  \\  \Delta   \ottsym{(}   \Gamma   \ottsym{)}   \vdash     \lift{ \mathit{l} \,  \bm{ { S } } ^ {  \mathit{N}  }  }   \mathop{ \odot }  \varepsilon    \sim   \varepsilon' 
    }{
    \Gamma  \vdash   \mathbf{handle}_{ \mathit{l} \,  \bm{ { S } } ^ {  \mathit{N}  }  }  \, \ottnt{e} \, \mathbf{with} \, \ottnt{h}  \ottsym{:}  \ottnt{B}  \mid  \varepsilon
    }\quad\rname{T}{Handling}
  \end{mathpar}

  \textbf{Handler Typing}\tquad\fbox{$ \Gamma  \vdash _{ \sigma }  \ottnt{h}  :  \ottnt{A}   \Rightarrow  ^ { \varepsilon }  \ottnt{B} $}\hfill\phantom{}
  \begin{mathpar}
    \inferrule{
    \Gamma  \ottsym{,}  \mathit{x}  \ottsym{:}  \ottnt{A}  \vdash  \ottnt{e_{\ottmv{r}}}  \ottsym{:}  \ottnt{B}  \mid  \varepsilon
    }{
     \Gamma  \vdash _{  \{\}  }  \ottsym{\{} \, \mathbf{return} \, \mathit{x}  \mapsto  \ottnt{e_{\ottmv{r}}}  \ottsym{\}}  :  \ottnt{A}   \Rightarrow  ^ { \varepsilon }  \ottnt{B} 
    }\quad\rname{H}{Return}

    \inferrule{
    \sigma  \ottsym{=}   \sigma'   \uplus   \ottsym{\{}  \mathsf{op}  \ottsym{:}    \forall    {\bm{ \beta } }^{ \mathit{J} } : {\bm{ \ottnt{K} } }^{ \mathit{J} }    \ottsym{.}    \ottnt{A'}   \Rightarrow   \ottnt{B'}   \ottsym{\}}  \\ \\ \\  \Gamma  \vdash _{ \sigma' }  \ottnt{h}  :  \ottnt{A}   \Rightarrow  ^ { \varepsilon }  \ottnt{B}  \\\\
    \Gamma  \ottsym{,}   {\bm{ \beta } }^{ \mathit{J} } : {\bm{ \ottnt{K} } }^{ \mathit{J} }   \ottsym{,}  \mathit{p}  \ottsym{:}  \ottnt{A'}  \ottsym{,}  \mathit{k}  \ottsym{:}   \ottnt{B'}    \rightarrow_{ \varepsilon }    \ottnt{B}   \vdash  \ottnt{e}  \ottsym{:}  \ottnt{B}  \mid  \varepsilon \\  \{   \bm{ { \beta } } ^ {  \mathit{J}  }   \}   \cap   \ottsym{(}    \mathrm{FTV}   \ottsym{(}   \ottnt{B}   \ottsym{)}    \cup    \mathrm{FTV}   \ottsym{(}   \varepsilon   \ottsym{)}    \ottsym{)}   \ottsym{=}  \emptyset
    }{
     \Gamma  \vdash _{ \sigma }   \ottnt{h}   \uplus   \ottsym{\{}  \mathsf{op} \,  {\bm{ \beta } }^{ \mathit{J} } : {\bm{ \ottnt{K} } }^{ \mathit{J} }  \, \mathit{p} \, \mathit{k}  \mapsto  \ottnt{e}  \ottsym{\}}   :  \ottnt{A}   \Rightarrow  ^ { \varepsilon }  \ottnt{B} 
    }\quad\rname{H}{Op}
  \end{mathpar}
  \caption{Type-and-effect system of {\lang}.}
  \label{fig:typing}
\end{figure}

Before showing type-and-effect system of {\lang}, we show the auxiliary definition.
\begin{definition}[Typelike Extraction Function]\label{def:delta}
  \begin{align*}
     \Delta   \ottsym{(}   \emptyset   \ottsym{)}  =  \emptyset  \qquad  \Delta   \ottsym{(}   \Gamma  \ottsym{,}  \mathit{x}  \ottsym{:}  \ottnt{A}   \ottsym{)}  =  \Delta   \ottsym{(}   \Gamma   \ottsym{)}  \qquad  \Delta   \ottsym{(}   \Gamma  \ottsym{,}  \alpha  \ottsym{:}  \ottnt{K}   \ottsym{)}  =  \Delta   \ottsym{(}   \Gamma   \ottsym{)}   \ottsym{,}  \alpha  \ottsym{:}  \ottnt{K}
  \end{align*}
\end{definition}
The typelike extraction function accumulates only typelike variable bindings from a typing context.

We show the type-and-effect system of {\lang} in Fig~\ref{fig:typing}. Typing relations are the form of $\Gamma  \vdash  \ottnt{e}  \ottsym{:}  \ottnt{A}  \mid  \varepsilon$, meaning that an expression $\ottnt{e}$ is typed as $\ottnt{A}$ under a context $\Gamma$ and the evaluation of $\ottnt{e}$ may cause effect $\varepsilon$. The rules for variables, function abstractions, function applications, type abstractions, type applications, and let bindings are standard.

The subtyping rule means: if the typing relation $\Gamma  \vdash  \ottnt{e}  \ottsym{:}  \ottnt{A}  \mid  \varepsilon$ holds, $\ottnt{A}$ is a subtype of $\ottnt{A'}$, and $\varepsilon$ is a subeffect of $\varepsilon'$, then the typing relation $\Gamma  \vdash  \ottnt{e}  \ottsym{:}  \ottnt{A'}  \mid  \varepsilon'$ holds. We show the subtyping relation in Fig~\ref{fig:subtyping}. The subtyping rules are standard except for the subeffecting. We represent the subeffecting relation via ARE: $ \Delta   \ottsym{(}   \Gamma   \ottsym{)}   \vdash   \varepsilon  \olessthan  \varepsilon' $.

\begin{figure}[t]
  \phantom{}\\
  \textbf{Subtyping}\tquad\fbox{$\Gamma  \vdash  \ottnt{A}  <:  \ottnt{B}$}\hfill\phantom{}
  \begin{mathpar}
    \inferrule{\Gamma  \vdash  \ottnt{A}  \ottsym{:}   \mathbf{Typ} }{\Gamma  \vdash  \ottnt{A}  <:  \ottnt{A}} \quad\rname{ST}{Refl}

    \inferrule
    {\Gamma  \vdash  \ottnt{A_{{\mathrm{2}}}}  <:  \ottnt{A_{{\mathrm{1}}}} \\  \Delta   \ottsym{(}   \Gamma   \ottsym{)}   \vdash   \varepsilon_{{\mathrm{1}}}  \olessthan  \varepsilon_{{\mathrm{2}}}  \\ \Gamma  \vdash  \ottnt{B_{{\mathrm{1}}}}  <:  \ottnt{B_{{\mathrm{2}}}}}
    {\Gamma  \vdash   \ottnt{A_{{\mathrm{1}}}}    \rightarrow_{ \varepsilon_{{\mathrm{1}}} }    \ottnt{B_{{\mathrm{1}}}}   <:   \ottnt{A_{{\mathrm{2}}}}    \rightarrow_{ \varepsilon_{{\mathrm{2}}} }    \ottnt{B_{{\mathrm{2}}}} }
    \quad\rname{ST}{Fun}

    \inferrule
    {\Gamma  \ottsym{,}  \alpha  \ottsym{:}  \ottnt{K}  \vdash  \ottnt{A_{{\mathrm{1}}}}  <:  \ottnt{A_{{\mathrm{2}}}} \\  \Delta   \ottsym{(}   \Gamma  \ottsym{,}  \alpha  \ottsym{:}  \ottnt{K}   \ottsym{)}   \vdash   \varepsilon_{{\mathrm{1}}}  \olessthan  \varepsilon_{{\mathrm{2}}} }
    {\Gamma  \vdash    \forall   \alpha  \ottsym{:}  \ottnt{K}   \ottsym{.}    \ottnt{A_{{\mathrm{1}}}}    ^{ \varepsilon_{{\mathrm{1}}} }    <:    \forall   \alpha  \ottsym{:}  \ottnt{K}   \ottsym{.}    \ottnt{A_{{\mathrm{2}}}}    ^{ \varepsilon_{{\mathrm{2}}} }  }
    \quad \rname{ST}{Poly}
  \end{mathpar}
  \caption{Subtyping rules of {\lang}.}
  \label{fig:subtyping}
\end{figure}

The rules \rname{T}{Op} and \rname{T}{Handling} are for algebraic effect  handlers.

\rname{T}{Op} is the rule for operations. $ \mathit{l}  ::    \forall    {\bm{ \alpha } }^{ \mathit{I} } : {\bm{ \ottnt{K} } }^{ \mathit{I} }    \ottsym{.}    \sigma    \in   \Xi $ and $ \mathsf{op}  \ottsym{:}    \forall    {\bm{ \beta } }^{ \mathit{J} } : {\bm{ \ottnt{K'} } }^{ \mathit{J} }    \ottsym{.}    \ottnt{A}   \Rightarrow   \ottnt{B}    \in   \sigma \,  \! [ {\bm{ { S } } }^{ \mathit{I} } / {\bm{ \alpha } }^{ \mathit{I} } ]  $ in the premise of \rname{T}{Op} mean that: the 3-tuple composed of a label $\mathit{l}$, its typelike parameters $ {\bm{ \alpha } }^{ \mathit{I} } : {\bm{ \ottnt{K} } }^{ \mathit{I} } $, and its operation signature $\sigma$ belongs to the effect context, and the pair of an operation $\mathsf{op}$ and its signature $  \forall    {\bm{ \beta } }^{ \mathit{J} } : {\bm{ \ottnt{K'} } }^{ \mathit{J} }    \ottsym{.}    \ottnt{A}   \Rightarrow   \ottnt{B} $ belongs to the operation signature $\sigma \,  \! [ {\bm{ { S } } }^{ \mathit{I} } / {\bm{ \alpha } }^{ \mathit{I} } ] $. In order to call a well-typed operation, we have to give well-kinded typelikes $ \bm{ { S } } ^ {  \mathit{I}  } $ and $ \bm{ { T } } ^ {  \mathit{J}  } $ to the parameters of a label and an operation, that is $ \bm{ { \alpha } } ^ {  \mathit{I}  } $ and $ \bm{ { \beta } } ^ {  \mathit{J}  } $. It is noteworthy that the types $\ottnt{A}$ and $\ottnt{B}$ result from the substitution $ \bm{ { S } } ^ {  \mathit{I}  } $ for $ {\bm{ \alpha } }^{ \mathit{I} } : {\bm{ \ottnt{K} } }^{ \mathit{I} } $. As a result, an operation is typed as $ \ottsym{(}  \ottnt{A} \,  \! [ {\bm{ { T } } }^{ \mathit{J} } / {\bm{ \beta } }^{ \mathit{J} } ]   \ottsym{)}    \rightarrow_{  \lift{ \mathit{l} \,  \bm{ { S } } ^ {  \mathit{I}  }  }  }    \ottsym{(}  \ottnt{B} \,  \! [ {\bm{ { T } } }^{ \mathit{J} } / {\bm{ \beta } }^{ \mathit{J} } ]   \ottsym{)} $. $\vdash  \Gamma$ is required to guarantee that all the typing judgments are carried under the well-formed typing context. Note that the effect of a single label $\mathit{l} \,  \bm{ { S } } ^ {  \mathit{I}  } $ is annotated to this function type. This fact suggests that no other effects cannot arise from this operation.

\rname{T}{Handling} is the rule for handling expressions. The rule supposes that the expression $\ottnt{e}$ enclosed by a handler is typed as $\ottnt{A}$ and has effect $\varepsilon'$. The type of handling expression $\ottnt{B}$ is determined by the handler $\ottnt{h}$ such that $ \Gamma  \vdash _{ \sigma \,  \! [ {\bm{ { S } } }^{ \mathit{N} } / {\bm{ \alpha } }^{ \mathit{N} } ]  }  \ottnt{h}  :  \ottnt{A}   \Rightarrow  ^ { \varepsilon }  \ottnt{B} $. The effect of handling expression $\varepsilon$ is intuitively the result of removing a label $\mathit{l} \,  \bm{ { S } } ^ {  \mathit{I}  } $ from the effect $\varepsilon'$. This ``label-removing relation'' is represented as ARE: $ \Delta   \ottsym{(}   \Gamma   \ottsym{)}   \vdash     \lift{ \mathit{l} \,  \bm{ { S } } ^ {  \mathit{I}  }  }   \mathop{ \odot }  \varepsilon'    \sim   \varepsilon $. For example, this relation is concretized as $\{\mathit{l} \,  \bm{ { S } } ^ {  \mathit{I}  } \} \cup \varepsilon = \varepsilon'$ in a set-based effect system and $\langle  \mathit{l} \,  \bm{ { S } } ^ {  \mathit{I}  }   \mid  \varepsilon  \rangle \simeq \varepsilon'$ in a row-based effect system.

\rname{H}{Return} and \rname{H}{Handling} are the rules for handlers. These rules take the form $ \Gamma  \vdash _{ \sigma }  \ottnt{h}  :  \ottnt{A}   \Rightarrow  ^ { \varepsilon }  \ottnt{B} $. It means that under the context $\Gamma$ and the operation signature $\sigma$, the handler consists of two parts: the one return clause that takes the handled expression's result typed as $\ottnt{A}$ and returns the expression $\ottnt{e_{\ottmv{r}}}$ with the type $\ottnt{B}$ and effect $\varepsilon$; and some operation clauses any of which takes typelike parameters $ {\bm{ \beta } }^{ \mathit{J} } : {\bm{ \ottnt{K} } }^{ \mathit{J} } $, an argument $\mathit{p}$, and a delimited continuation $\mathit{k}$ of $\mathsf{op}$ in $\sigma$, and then returns the expression $\ottnt{e}$ with the type $\ottnt{B}$ and effect $\varepsilon$.
}

\OLD{
  \subsection{Programming and Type Checking Process}

  We show the programming and type checking process in {\lang}. Since {\lang} has effect signatures and ARE as parameters, we cannot write programs and check types as usual. The syntax of effects depends on effect signatures and the typing rules on ARE. Thus, the programming and type checking process in {\lang} is as follows.
  \begin{enumerate}
    \item Give the instances of effect signatures and ARE.
    \item Write a program and check types.
  \end{enumerate}

  Before writing programs, we need to provide the instance of effect signatures. We can get the whole concrete syntax of {\lang} only by determining what effect signatures are. However, the rules remain unfinished because we do not know what ARE is. Hence, we also need to give the instance of ARE. We only can get concrete language via the procedure written above.

  After we get a concrete syntax and type-and-effect system, we can write a program and check its type and effect. For example, we consider giving an effect system using Example~\ref{exa:effrow}. In this case, the following program is typed as $  \forall   \rho  \ottsym{:}   \mathbf{Eff}    \ottsym{.}      \mathsf{Int}     \rightarrow_{ \langle  \rangle }     \ottsym{(}    \mathsf{Int}     \rightarrow_{ \langle  \mathsf{Exc}  \mid  \rho  \rangle }     \mathsf{Int}    \ottsym{)}    \rightarrow_{ \rho }     \mathsf{Int}    $ under the empty typing context.
  \begin{flalign*}
     & \Lambda  \rho  \ottsym{:}   \mathbf{Eff}   \ottsym{.}  \lambda  \mathit{x}  \ottsym{:}   \mathsf{Int}   \ottsym{.}  \lambda  \mathit{g}  \ottsym{:}    \mathsf{Int}     \rightarrow_{ \langle  \mathsf{Exc}  \mid  \rho  \rangle }     \mathsf{Int}    \ottsym{.}   & \\ &    \quad    \mathbf{handle}_{ \mathsf{Exc} }  \,   \mathbf{if} \, \mathit{x}  \ottsym{=}  0 \, \mathbf{then} \,  \mathsf{raise} _{ \mathsf{Exc} }  \,  \mathsf{Int}  \,  ()  \, \mathbf{else} \, \mathit{g} \, \mathit{x}   \, \mathbf{with} \,  \ottsym{\{} \, \mathbf{return} \, \mathit{z}  \mapsto  \mathit{z}  \ottsym{\}}   \uplus   \ottsym{\{}  \mathsf{raise} \, \alpha  \ottsym{:}   \mathbf{Typ}  \, \_ \, \_  \mapsto  0  \ottsym{\}}   
  \end{flalign*}
}
\section{Safety Properties}\label{sec:safety}

This section shows the safety properties of {\lang}.
The proofs rely on safety conditions, which are
requirements on effect algebras.
Under the assumption that a given effect algebra meets the safety conditions, we prove
type-and-effect safety of {\lang}.
\OLD{
In the present work, effect safety is stated as ``if the effect $ \bbZero $ is
assigned to a program, it never causes unhandled operations.''
}

\subsection{Safety Conditions}
To prove type-and-effect safety, a given effect algebra must meet safety conditions shown in the following.
We write $ \varepsilon_{{\mathrm{1}}}  \olessthan  \varepsilon_{{\mathrm{2}}} $ to state that $  \varepsilon_{{\mathrm{1}}}  \mathop{ \odot }  \varepsilon    \sim   \varepsilon_{{\mathrm{2}}} $ for some $\varepsilon$.

\TY{Removed safety condition for substitution.}
\begin{definition}[Safety Conditions]\label{def:safe_cond}
  \phantom{}
  \begin{enumerate}
    \item\label{def:safe_cond:label_notemp}
          For any $\ottnt{L}$, $  \lift{ \ottnt{L} }   \olessthan   \bbZero  $ does not hold.

    \item\label{def:safe_cond:pres}
          If $  \lift{ \ottnt{L} }   \olessthan  \varepsilon $ and $   \lift{ \ottnt{L'} }   \mathop{ \odot }  \varepsilon'    \sim   \varepsilon $ and $\ottnt{L} \neq \ottnt{L'}$, then $  \lift{ \ottnt{L} }   \olessthan  \varepsilon' $.


          \setcounter{safecondcounter}{\value{enumi}}
  \end{enumerate}
\end{definition}

Condition~\ref{def:safe_cond:label_notemp} disallows the subeffecting to hide an invoked effect label $\ottnt{L}$ as if it were not performed.
Condition~\ref{def:safe_cond:pres} means that,
if an expression invoking a label $\ottnt{L}$ is given an effect $\varepsilon$, and an effect handler for a different label $\ottnt{L'}$ handles the expression,
then the information of $\ottnt{L}$ still remains in the effect $\varepsilon'$ assigned to the handling expression (that is, it is observable from the outer context).

\OLD{
\TY{Intuitive explanations here.}
Condition~\ref{def:safe_cond:label_notemp} prohibits overapproximating a single label effect $ \lift{ \ottnt{L} } $ as the empty effect $ \bbZero $.
Condition~\ref{def:safe_cond:pres} means that,
if we can remove $\ottnt{L}$ from $\varepsilon$ and actually remove $\ottnt{L'} (\neq \ottnt{L})$ from $\varepsilon$,
then we can also remove $\ottnt{L}$ from the remaining effect $\varepsilon'$.
%
}

To understand problems excluded by safety conditions
\ref{def:safe_cond:label_notemp} and \ref{def:safe_cond:pres},
we consider effect algebras that violate one of the conditions, and then show unsafe programs being typeable under the algebras.
\begin{example}[Unsafe Effect Algebras]\label{exa:necessity}
  \phantom{}
  \begin{description}
    \item[\quad Effect algebra violating safety condition~\ref{def:safe_cond:label_notemp}]
          Consider an effect algebra such that $\emptyset  \vdash    \lift{ \mathit{l} }   \olessthan   \bbZero  $ holds for some $\mathit{l}$.
          Clearly, this effect algebra violates safety condition~\ref{def:safe_cond:label_notemp}.
          In this case,
            $\emptyset  \vdash   \mathsf{op} _{ \mathit{l} }  \, \ottnt{v}  \ottsym{:}  \ottnt{A}  \mid   \bbZero $
          can be derived for some $\ottnt{A}$ (if $ \mathsf{op} _{ \mathit{l} }  \, \ottnt{v}$ is well typed)
          because $ \mathsf{op} _{ \mathit{l} }  \, \ottnt{v}$ is given the effect $ \lift{ \mathit{l} } $ and
          the subeffecting $\emptyset  \vdash    \lift{ \mathit{l} }   \olessthan   \bbZero  $ holds.
          However, the operation call is not handled.

    \item[\quad Effect algebra violating safety condition~\ref{def:safe_cond:pres}]
          Consider an effect algebra such that
          safety condition~\ref{def:safe_cond:label_notemp},
               $  \lift{ \mathit{l} }   \olessthan   \lift{ \mathit{l'} }  $, and
               $   \lift{ \mathit{l'} }   \mathop{ \odot }   \bbZero     \sim    \lift{ \mathit{l'} }  $
          hold for some $\mathit{l}$ and $\mathit{l'}$ such that $\mathit{l} \neq \mathit{l'}$.
          This effect algebra must violate safety condition~\ref{def:safe_cond:pres}:
          if safety condition~\ref{def:safe_cond:pres} were met, we would have $  \lift{ \mathit{l} }   \olessthan   \bbZero  $, but
          it is contradictory with safety condition~\ref{def:safe_cond:label_notemp}.

          \OLD{
          First, we show that the effect algebra allowing these assumptions violates safety condition~\ref{def:safe_cond:pres}.
          If safety condition~\ref{def:safe_cond:pres} is met, we would have $  \lift{ \mathit{l} }   \olessthan   \bbZero  $.
          However, it is contradictory with safety condition~\ref{def:safe_cond:label_notemp}.
          }

          This effect algebra allows assigning the empty effect $ \bbZero $ to the expression $ \mathbf{handle}_{ \mathit{l'} }  \,  \mathsf{op} _{ \mathit{l} }  \,  {}  \, \ottnt{v} \, \mathbf{with} \, \ottnt{h}$ as illustrated by the following typing derivation:
          %
          \[
            \inferrule* [Right=T\_Handling] {
            \cdots \\
               \lift{ \mathit{l'} }   \mathop{ \odot }   \bbZero     \sim    \lift{ \mathit{l'} }  \\
            \inferrule* [Right=T\_Sub] {
            \emptyset  \vdash   \mathsf{op} _{ \mathit{l} }  \,  {}  \, \ottnt{v}  \ottsym{:}  \ottnt{A}  \mid   \lift{ \mathit{l} }  \\
            \emptyset  \vdash  \ottnt{A}  \mid   \lift{ \mathit{l} }   <:  \ottnt{A}  \mid   \lift{ \mathit{l'} } 
            }{
            \emptyset  \vdash   \mathsf{op} _{ \mathit{l} }  \,  {}  \, \ottnt{v}  \ottsym{:}  \ottnt{A}  \mid   \lift{ \mathit{l'} } 
            }
            }{
            \emptyset  \vdash   \mathbf{handle}_{ \mathit{l'} }  \,  \mathsf{op} _{ \mathit{l} }  \,  {}  \, \ottnt{v} \, \mathbf{with} \, \ottnt{h}  \ottsym{:}  \ottnt{B}  \mid   \bbZero 
            }
          \]
          However, the operation call in it is not handled.
  \end{description}
\end{example}

\subsection{Type-and-Effect Safety}
\label{subsec:type-and-effect-safety}
This section shows type-and-effect safety.
To prove it, we assume that an effect algebra meets the safety conditions and
an effect context is proper, which means that it is consistent with a given label
signature $\Slabel$ and the types of operations in it are well
formed.
\begin{definition}[Proper Effect Contexts]\label{rem:proper_effctx}
  An effect context $\Xi$ is \emph{proper} if,
  for any $ \mathit{l}  ::    \forall    {\bm{ \alpha } }^{ \mathit{I} } : {\bm{ \ottnt{K} } }^{ \mathit{I} }    \ottsym{.}    \sigma    \in   \Xi $, the following holds:
  \begin{itemize}
    \item $ \mathit{l}   \ottsym{:}    \Pi {\bm{ { \ottnt{K} } } }^{ \mathit{I} }   \rightarrow   \mathbf{Lab}    \in   \Slabel $;
    \item the type schemes $  \forall    {\bm{ \alpha_{{\mathrm{0}}} } }^{ \mathit{I_{{\mathrm{0}}}} } : {\bm{ \ottnt{K_{{\mathrm{0}}}} } }^{ \mathit{I_{{\mathrm{0}}}} }    \ottsym{.}    \sigma_{{\mathrm{0}}} $ associated with $\mathit{l}$ by $\Xi$ are uniquely determined; and

    \item for any $ \mathsf{op}  \ottsym{:}    \forall    {\bm{ \beta } }^{ \mathit{J} } : {\bm{ \ottnt{K_{{\mathrm{0}}}} } }^{ \mathit{J} }    \ottsym{.}    \ottnt{A}   \Rightarrow   \ottnt{B}    \in   \sigma $ and $\ottnt{C} \in \{ \ottnt{A}, \ottnt{B} \}$,
          $ {\bm{ \alpha } }^{ \mathit{I} } : {\bm{ \ottnt{K} } }^{ \mathit{I} }   \ottsym{,}   {\bm{ \beta } }^{ \mathit{J} } : {\bm{ \ottnt{K_{{\mathrm{0}}}} } }^{ \mathit{J} }   \vdash  \ottnt{C}  \ottsym{:}   \mathbf{Typ} $.
  \end{itemize}
\end{definition}

\subsubsection{Type Safety}

The statement of type safety is as follows.
We write $ \longrightarrow ^*$ for the reflexive, transitive closure of $ \longrightarrow $
and $\ottnt{e}\centernot \longrightarrow $ to denote that there is no $\ottnt{e'}$ such that $\ottnt{e}  \longrightarrow  \ottnt{e'}$.
\begin{lemma}[Type Safety]\label{lem:typsafe}
  If $\emptyset  \vdash  \ottnt{e}  \ottsym{:}  \ottnt{A}  \mid  \varepsilon$ and $\ottnt{e}  \longrightarrow  ^ * \ottnt{e'} \centernot \longrightarrow $, then one of the following holds:
  \begin{itemize}
    \item $\ottnt{e'} = \ottnt{v}$ for some value $\ottnt{v}$ such that $\emptyset  \vdash  \ottnt{v}  \ottsym{:}  \ottnt{A}  \mid  \varepsilon$; or
    \item $\ottnt{e'} = \ottnt{E}  \ottsym{[}   \mathsf{op} _{ \mathit{l} \,  \bm{ { S } } ^ {  \mathit{I}  }  }  \,  \bm{ { T } } ^ {  \mathit{J}  }  \, \ottnt{v}  \ottsym{]}$
          for some $\ottnt{E}$, $\mathit{l}$, $ \bm{ { S } } ^ {  \mathit{I}  } $, $\mathsf{op}$, $ \bm{ { T } } ^ {  \mathit{J}  } $, and $\ottnt{v}$ such that
          $ 0  \mathrm{-free} ( \mathit{l} \,  \bm{ { S } } ^ {  \mathit{I}  }  ,  \ottnt{E} ) $.

  \end{itemize}
\end{lemma}
While the type safety guarantees that the result of a program, if any, has the
same type as the program, it does not ensure that all operations are handled
even if the effect $ \bbZero $, which denotes that no unhandled operation
remains, is assigned to the program: as shown shortly, the latter property is
guaranteed by effect safety.
%

Type safety is proven via progress and preservation as usual~\cite{wright_syntactic_1994}.
\begin{lemma}[Progress]\label{lem:progress}
  If $\emptyset  \vdash  \ottnt{e}  \ottsym{:}  \ottnt{A}  \mid  \varepsilon$, then one of the following holds:
 $\ottnt{e}$ is a value;
 $\ottnt{e}  \longrightarrow  \ottnt{e'}$ for some $\ottnt{e'}$; or
 $\ottnt{e} = \ottnt{E}  \ottsym{[}   \mathsf{op} _{ \mathit{l} \,  \bm{ { S } } ^ {  \mathit{I}  }  }  \,  \bm{ { T } } ^ {  \mathit{J}  }  \, \ottnt{v}  \ottsym{]}$
          for some $\ottnt{E}$, $\mathit{l}$, $ \bm{ { S } } ^ {  \mathit{I}  } $, $\mathsf{op}$, $ \bm{ { T } } ^ {  \mathit{J}  } $, and $\ottnt{v}$ such that
          $ 0  \mathrm{-free} ( \mathit{l} \,  \bm{ { S } } ^ {  \mathit{I}  }  ,  \ottnt{E} ) $.
          %
\end{lemma}
\begin{lemma}[Preservation]\label{lem:preservation}
  If $\emptyset  \vdash  \ottnt{e}  \ottsym{:}  \ottnt{A}  \mid  \varepsilon$ and $\ottnt{e}  \longrightarrow  \ottnt{e'}$, then $\emptyset  \vdash  \ottnt{e'}  \ottsym{:}  \ottnt{A}  \mid  \varepsilon$.
\end{lemma}

\subsubsection{Effect Safety}

Effect safety is stated as follows.
%
%
\begin{lemma}[Effect Safety]\label{lem:effsafe}
  If $\Gamma  \vdash  \ottnt{e}  \ottsym{:}  \ottnt{A}  \mid   \bbZero $,
  then there exist no $\ottnt{E}$, $\mathit{l}$, $ \bm{ { S } } ^ {  \mathit{I}  } $, $\mathsf{op}$, $ \bm{ { T } } ^ {  \mathit{J}  } $, and $\ottnt{v}$ such that
  both
  $\ottnt{e} = \ottnt{E}  \ottsym{[}   \mathsf{op} _{ \mathit{l} \,  \bm{ { S } } ^ {  \mathit{I}  }  }  \,  \bm{ { T } } ^ {  \mathit{J}  }  \, \ottnt{v}  \ottsym{]}$ and
  $ 0  \mathrm{-free} ( \mathit{l} \,  \bm{ { S } } ^ {  \mathit{I}  }  ,  \ottnt{E} ) $ hold.
\end{lemma}
This lemma means, if an expression is assigned to $ \bbZero $, no unhandled operation call remains there.

\subsubsection{Type-and-Effect Safety}
We obtain type-and-effect safety---terminating programs with effect
$ \bbZero $ always evaluates to values---as a corollary from type safety
and effect safety.
\begin{theorem}[Type-and-Effect Safety]
  \label{thm:type-and-eff-safe}
  If $\emptyset  \vdash  \ottnt{e}  \ottsym{:}  \ottnt{A}  \mid   \bbZero $ and $\ottnt{e}  \longrightarrow  ^ * \ottnt{e'} \centernot \longrightarrow $, then $\ottnt{e'} = \ottnt{v}$ for some $\ottnt{v}$.
\end{theorem}
%
%

\subsubsection{The Safety of Instances}

The three effect algebras {\eaSet}, {\eaSimpleRow}, and {\eaScopedRow} presented
thus far meet the safety conditions as stated below, which derives that the
effect systems with these algebras enjoy the type-and-effect safety just as
corollaries of Theorem~\ref{thm:type-and-eff-safe}.
\begin{theorem}
 The effect algebras {\eaSet}, {\eaSimpleRow}, and {\eaScopedRow} meet safety conditions~\ref{def:safe_cond:label_notemp} and \ref{def:safe_cond:pres}.
\end{theorem}
%
%

\OLD{
  This section shows safety conditions to show the meta properties of {\lang}, type safety and effect safety, and how to prove the meta properties of concrete systems using {\lang}.

  \subsection{Safety Conditions}
  To prove type and effect safety, ARE must meet safety conditions shown in the following.

  \begin{definition}[Safety Conditions]\label{def:safe_cond}
    \phantom{}
    \begin{enumerate}
      \item\label{def:safe_cond:equiv} $\forall \Delta, \varepsilon . (\Delta  \vdash  \varepsilon  \ottsym{:}   \mathbf{Eff}  \imply \Delta  \vdash   \varepsilon   \cong   \varepsilon )$.
      \item\label{def:safe_cond:assoc1} $\begin{aligned}[t]
                 & \forall \Delta, \varepsilon_{{\mathrm{1}}}, \varepsilon_{{\mathrm{2}}}, \varepsilon_{{\mathrm{3}}}, \varepsilon', \varepsilon'' . ((\Delta  \vdash    \varepsilon_{{\mathrm{1}}}  \mathop{ \odot }  \varepsilon_{{\mathrm{2}}}    \sim   \varepsilon'  \tand \Delta  \vdash    \varepsilon'  \mathop{ \odot }  \varepsilon_{{\mathrm{3}}}    \sim   \varepsilon'' ) \\
                 & \quad \imply \exists \varepsilon''' . (\Delta  \vdash    \varepsilon_{{\mathrm{2}}}  \mathop{ \odot }  \varepsilon_{{\mathrm{3}}}    \sim   \varepsilon'''  \tand \Delta  \vdash    \varepsilon_{{\mathrm{1}}}  \mathop{ \odot }  \varepsilon'''    \sim   \varepsilon''  )).
              \end{aligned}$
      \item\label{def:safe_cond:no_inv} $\forall \Delta, \varepsilon_{{\mathrm{1}}}, \varepsilon_{{\mathrm{2}}} . (\Delta  \vdash    \varepsilon_{{\mathrm{1}}}  \mathop{ \odot }  \varepsilon_{{\mathrm{2}}}    \sim    \bbZero   \imply (\Delta  \vdash   \varepsilon_{{\mathrm{1}}}   \cong    \bbZero   \tand \Delta  \vdash   \varepsilon_{{\mathrm{2}}}   \cong    \bbZero  ))$.
      \item\label{def:safe_cond:label_notemp} $\forall \Delta, \mathit{l},  \bm{ { S } } ^ {  \mathit{I}  }  . (\Delta  \vdash  \mathit{l} \,  \bm{ { S } } ^ {  \mathit{I}  }   \ottsym{:}   \mathbf{Lab}  \imply \Delta  \not\vdash    \lift{ \mathit{l} \,  \bm{ { S } } ^ {  \mathit{I}  }  }    \cong    \bbZero  )$.
      \item\label{def:safe_cond:label_uniq} $\begin{aligned}[t]
                 & \forall \Delta, \mathit{l},  \bm{ { S } } ^ {  \mathit{I}  } , \varepsilon_{{\mathrm{1}}}, \varepsilon_{{\mathrm{2}}} . (\Delta  \vdash    \varepsilon_{{\mathrm{1}}}  \mathop{ \odot }  \varepsilon_{{\mathrm{2}}}    \sim    \lift{ \mathit{l} \,  \bm{ { S } } ^ {  \mathit{I}  }  }                            \\
                 & \quad \imply ((\Delta  \vdash   \varepsilon_{{\mathrm{1}}}   \cong    \bbZero   \tand \Delta  \vdash   \varepsilon_{{\mathrm{2}}}   \cong    \lift{ \mathit{l} \,  \bm{ { S } } ^ {  \mathit{I}  }  }  ) \tor \Delta  \vdash   \varepsilon_{{\mathrm{1}}}   \cong    \lift{ \mathit{l} \,  \bm{ { S } } ^ {  \mathit{I}  }  }   ) ).
              \end{aligned}$
      \item\label{def:safe_cond:label_uniq_sim} $\begin{aligned}[t]
                 & \forall \Delta, \mathit{l}, \mathit{l'},  \bm{ { S } } ^ {  \mathit{I}  } ,  \bm{ { S' } } ^ {  \mathit{I'}  } , \varepsilon . ((\Delta  \vdash   \varepsilon   \cong    \lift{ \mathit{l} \,  \bm{ { S } } ^ {  \mathit{I}  }  }   \tand \Delta  \vdash   \varepsilon   \cong    \lift{ \mathit{l'} \,  \bm{ { S' } } ^ {  \mathit{I'}  }  }  ) \imply \mathit{l} \,  \bm{ { S } } ^ {  \mathit{I}  }  = \mathit{l'} \,  \bm{ { S' } } ^ {  \mathit{I'}  }  ).
              \end{aligned}$
      \item\label{def:safe_cond:pres} $\begin{aligned}[t]
                 & \forall \Delta, \mathit{l},  \bm{ { S } } ^ {  \mathit{I}  } , \varepsilon_{{\mathrm{1}}}, \varepsilon_{{\mathrm{2}}}, \varepsilon_{{\mathrm{3}}} . ((\Delta  \vdash    \varepsilon_{{\mathrm{1}}}  \mathop{ \odot }  \varepsilon_{{\mathrm{2}}}    \sim   \varepsilon_{{\mathrm{3}}}  \tand \Delta  \vdash    \lift{ \mathit{l} \,  \bm{ { S } } ^ {  \mathit{I}  }  }   \olessthan  \varepsilon_{{\mathrm{3}}} ) \\
                 & \quad \imply \Delta  \vdash    \lift{ \mathit{l} \,  \bm{ { S } } ^ {  \mathit{I}  }  }   \olessthan  \varepsilon_{{\mathrm{1}}}  \tor \Delta  \vdash    \lift{ \mathit{l} \,  \bm{ { S } } ^ {  \mathit{I}  }  }   \olessthan  \varepsilon_{{\mathrm{2}}} ).
              \end{aligned}$
      \item\label{def:safe_cond:weakening} $\begin{aligned}[t]
                 & \forall \Delta_{{\mathrm{1}}}, \Delta_{{\mathrm{2}}}, \Delta_{{\mathrm{3}}}, \varepsilon_{{\mathrm{1}}}, \varepsilon_{{\mathrm{2}}}, \varepsilon_{{\mathrm{3}}} . ((\Delta_{{\mathrm{1}}}  \ottsym{,}  \Delta_{{\mathrm{3}}}  \vdash    \varepsilon_{{\mathrm{1}}}  \mathop{ \odot }  \varepsilon_{{\mathrm{2}}}    \sim   \varepsilon_{{\mathrm{3}}}  \tand  \mathrm{dom}   \ottsym{(}   \Delta_{{\mathrm{1}}}  \ottsym{,}  \Delta_{{\mathrm{3}}}   \ottsym{)}  \cap  \mathrm{dom}   \ottsym{(}   \Delta_{{\mathrm{2}}}   \ottsym{)}  = \emptyset ) \\
                 & \quad \imply \Delta_{{\mathrm{1}}}  \ottsym{,}  \Delta_{{\mathrm{2}}}  \ottsym{,}  \Delta_{{\mathrm{3}}}  \vdash    \varepsilon_{{\mathrm{1}}}  \mathop{ \odot }  \varepsilon_{{\mathrm{2}}}    \sim   \varepsilon_{{\mathrm{3}}} ).
              \end{aligned}$
      \item\label{def:safe_cond:subst} $\begin{aligned}[t]
                 & \forall \Delta_{{\mathrm{1}}}, \Delta_{{\mathrm{2}}}, \alpha, \ottnt{K}, S, \varepsilon_{{\mathrm{1}}}, \varepsilon_{{\mathrm{2}}}, \varepsilon_{{\mathrm{3}}} . ((\Delta_{{\mathrm{1}}}  \ottsym{,}  \alpha  \ottsym{:}  \ottnt{K}  \ottsym{,}  \Delta_{{\mathrm{2}}}  \vdash    \varepsilon_{{\mathrm{1}}}  \mathop{ \odot }  \varepsilon_{{\mathrm{2}}}    \sim   \varepsilon_{{\mathrm{3}}}  \tand \Delta_{{\mathrm{1}}}  \vdash  S  \ottsym{:}  \ottnt{K}) \\
                 & \quad \imply \Delta_{{\mathrm{1}}}  \ottsym{,}  \Delta_{{\mathrm{2}}} \,  \! [  S  /  \alpha   ]   \vdash    \varepsilon_{{\mathrm{1}}} \,  \! [  S  /  \alpha   ]   \mathop{ \odot }  \varepsilon_{{\mathrm{2}}}  \,  \! [  S  /  \alpha   ]    \sim   \varepsilon_{{\mathrm{3}}} \,  \! [  S  /  \alpha   ]  ).
              \end{aligned}$
    \end{enumerate}
  \end{definition}
  Condition~\ref{def:safe_cond:equiv} represents reflexivity of $\Delta  \vdash   \ottsym{-}   \cong   \ottsym{-} $. This condition derives the reflexivity of $\Delta  \vdash   \ottsym{-}  \olessthan  \ottsym{-} $. Conditions~\ref{def:safe_cond:assoc1} derives transitivity of $\Delta  \vdash   \ottsym{-}  \olessthan  \ottsym{-} $. These reflexivity and transitivity are needed to show the reflexivity and transitivity of subtyping, and we use the reflexivity and transitivity of subtyping to show inversion lemma. Conditions~\ref{def:safe_cond:weakening} and \ref{def:safe_cond:subst} are needed for weakening and substitution lemmas. We provide bad examples to explain the reasons why we need conditions~\ref{def:safe_cond:no_inv}, \ref{def:safe_cond:label_notemp}, \ref{def:safe_cond:label_uniq}, \ref{def:safe_cond:label_uniq_sim}, and \ref{def:safe_cond:pres}. These bad examples are allowed by lacking conditions~\ref{def:safe_cond:no_inv}, \ref{def:safe_cond:label_notemp}, \ref{def:safe_cond:label_uniq}, \ref{def:safe_cond:label_uniq_sim}, and \ref{def:safe_cond:pres}, respectively.
  \begin{example}[Bad Programs]\label{exa:necessity}
    \phantom{}
    \begin{enumerate}[start=3]
      \item Suppose that there are $\mathit{l}$, $\varepsilon$, and $\Gamma$ such that $ \Delta   \ottsym{(}   \Gamma   \ottsym{)}   \vdash     \lift{ \mathit{l} }   \mathop{ \odot }  \varepsilon    \sim    \bbZero  $. In this case,
            \begin{align*}
              \Gamma  \vdash   \mathsf{op} _{ \mathit{l} }  \,  {}  \, \ottnt{v}  \ottsym{:}  \ottnt{B}  \mid   \bbZero 
            \end{align*}
            holds by \rname{T}{Sub}. However, this expression is neither a value nor reducible.
      \item Suppose that there are $\mathit{l}$ such that $ \Delta   \ottsym{(}   \Gamma   \ottsym{)}   \vdash    \lift{ \mathit{l} }    \cong    \bbZero  $. In this case,
            \begin{align*}
              \Gamma  \vdash   \mathsf{op} _{ \mathit{l} }  \,  {}  \, \ottnt{v}  \ottsym{:}  \ottnt{B}  \mid   \bbZero 
            \end{align*}
            holds by \rname{T}{Sub}. However, this expression is neither a value nor reducible.
      \item Suppose that there are $\mathit{l_{{\mathrm{1}}}}$, $\mathit{l_{{\mathrm{2}}}}$, $\varepsilon$ and $\Gamma$ such that $ \Delta   \ottsym{(}   \Gamma   \ottsym{)}   \vdash     \lift{ \mathit{l_{{\mathrm{1}}}} }   \mathop{ \odot }  \varepsilon    \sim    \lift{ \mathit{l_{{\mathrm{2}}}} }  $, $\mathit{l_{{\mathrm{1}}}} \neq \mathit{l_{{\mathrm{2}}}}$. In this case,
            \begin{align*}
              \Gamma  \vdash   \mathbf{handle}_{ \mathit{l_{{\mathrm{2}}}} }  \,  \mathsf{op_{{\mathrm{1}}}} _{ \mathit{l_{{\mathrm{1}}}} }  \,  {}  \, \ottnt{v} \, \mathbf{with} \, \ottnt{h_{{\mathrm{2}}}}  \ottsym{:}  \ottnt{B}  \mid   \bbZero 
            \end{align*}
            holds by \rname{T}{Sub} and \rname{T}{Handling}. However, this expression is neither a value nor reducible.
      \item Suppose that there are $\mathit{l_{{\mathrm{1}}}}$, $\mathit{l_{{\mathrm{2}}}}$, and $\Gamma$ such that $ \Delta   \ottsym{(}   \Gamma   \ottsym{)}   \vdash    \lift{ \mathit{l_{{\mathrm{1}}}} }    \cong    \lift{ \mathit{l_{{\mathrm{2}}}} }  $ and $\mathit{l_{{\mathrm{1}}}} \neq \mathit{l_{{\mathrm{2}}}}$. In this case,
            \begin{align*}
              \Gamma  \vdash   \mathbf{handle}_{ \mathit{l_{{\mathrm{2}}}} }  \,  \mathsf{op} _{ \mathit{l_{{\mathrm{1}}}} }  \,  {}  \, \ottnt{v} \, \mathbf{with} \, \ottnt{h_{{\mathrm{2}}}}  \ottsym{:}  \ottnt{C}  \mid   \bbZero 
            \end{align*}
            holds by \rname{T}{Sub} and \rname{T}{Handling}. However, this expression is neither a value nor reducible.
      \item Suppose that there are $\mathit{l_{\ottmv{i}}}$ ($i = 1, 2, 3, 4$) and $\Gamma$ such that
            \begin{itemize}
              \item $\mathit{l_{\ottmv{i}}} \neq \mathit{l_{\ottmv{j}}}$ ($\ottmv{i} \neq \ottmv{j}$),
              \item $ \Delta   \ottsym{(}   \Gamma   \ottsym{)}   \vdash     \lift{ \mathit{l_{{\mathrm{1}}}} }   \mathop{ \odot }   \lift{ \mathit{l_{{\mathrm{2}}}} }     \sim   \varepsilon $,
              \item $ \Delta   \ottsym{(}   \Gamma   \ottsym{)}   \vdash     \lift{ \mathit{l_{{\mathrm{3}}}} }   \mathop{ \odot }   \lift{ \mathit{l_{{\mathrm{4}}}} }     \sim   \varepsilon $, and
              \item $ \Delta   \ottsym{(}   \Gamma   \ottsym{)}   \vdash     \lift{ \mathit{l_{{\mathrm{4}}}} }   \mathop{ \odot }   \lift{ \mathit{l_{{\mathrm{3}}}} }     \sim   \varepsilon $.
            \end{itemize}
            In this case,
            \begin{align*}
              \Gamma  \vdash   \mathbf{handle}_{ \mathit{l_{{\mathrm{2}}}} }  \,  \mathbf{handle}_{ \mathit{l_{{\mathrm{1}}}} }  \, \mathbf{let} \, \mathit{x}  \ottsym{=}   \mathsf{op_{{\mathrm{3}}}} _{ \mathit{l_{{\mathrm{3}}}} }  \,  {}  \, \ottnt{v_{{\mathrm{3}}}} \, \mathbf{in} \,  \mathsf{op_{{\mathrm{4}}}} _{ \mathit{l_{{\mathrm{4}}}} }  \,  {}  \, \ottnt{v_{{\mathrm{4}}}} \, \mathbf{with} \, \ottnt{h_{{\mathrm{1}}}} \, \mathbf{with} \, \ottnt{h_{{\mathrm{2}}}}  \ottsym{:}  \ottnt{C}  \mid   \bbZero 
            \end{align*}
            holds by \rname{T}{Handling}. However, this expression is neither a value nor reducible.
    \end{enumerate}
  \end{example}

  \subsection{Type Safety and Effect Safety}
  If ARE meets safety conditions defined above, we can prove type safety and effect safety.

  \TS{
    \begin{definition}[Proper Effect Contexts]\label{rem:proper_effctx}
      For any effect context $\Xi$, and any $ \mathit{l}  ::    \forall    {\bm{ \alpha } }^{ \mathit{I} } : {\bm{ \ottnt{K} } }^{ \mathit{I} }    \ottsym{.}    \sigma    \in   \Xi $,
      we assume the following:
      \begin{itemize}
        \item $ \mathit{l}   \ottsym{:}    {\bm{ { \ottnt{K} } } }^{ \mathit{I} }   \rightarrow   \mathbf{Lab}    \in   \Sbase $,
        \item if $ \mathit{l}  ::    \forall    {\bm{ \alpha' } }^{ \mathit{I'} } : {\bm{ \ottnt{K'} } }^{ \mathit{I'} }    \ottsym{.}    \sigma'    \in   \Xi $, then $ {\bm{ \alpha } }^{ \mathit{I} } : {\bm{ \ottnt{K} } }^{ \mathit{I} }  =  {\bm{ \alpha' } }^{ \mathit{I'} } : {\bm{ \ottnt{K'} } }^{ \mathit{I'} } $ and $\sigma = \sigma'$; and
        \item for any $ \mathsf{op}  \ottsym{:}    \forall    {\bm{ \beta } }^{ \mathit{J} } : {\bm{ \ottnt{K'} } }^{ \mathit{J} }    \ottsym{.}    \ottnt{A}   \Rightarrow   \ottnt{B}    \in   \sigma $, the well-formedness judgments
              \[
                 {\bm{ \alpha } }^{ \mathit{I} } : {\bm{ \ottnt{K} } }^{ \mathit{I} }   \ottsym{,}   {\bm{ \beta } }^{ \mathit{J} } : {\bm{ \ottnt{K'} } }^{ \mathit{J} }   \vdash  \ottnt{A}  \ottsym{:}   \mathbf{Typ}  \quad \tand \quad  {\bm{ \alpha } }^{ \mathit{I} } : {\bm{ \ottnt{K} } }^{ \mathit{I} }   \ottsym{,}   {\bm{ \beta } }^{ \mathit{J} } : {\bm{ \ottnt{K'} } }^{ \mathit{J} }   \vdash  \ottnt{B}  \ottsym{:}   \mathbf{Typ} 
              \] are derivable.
      \end{itemize}
    \end{definition}
  }

  \subsubsection{Type Safety}
  The statement of type safety is as follows.
  \begin{lemma}[Type Safety]\label{lem:typsafe}
    If $\emptyset  \vdash  \ottnt{e}  \ottsym{:}  \ottnt{A}  \mid  \varepsilon$ and $\ottnt{e}  \longrightarrow  ^ * \ottnt{e'} \centernot \longrightarrow $, then one of the following holds:
    \begin{itemize}
      \item There is $\ottnt{v}$ such that $\ottnt{e'} = \ottnt{v}$
      \item There are $\mathit{l}$, $\mathsf{op}$, $\ottnt{A'}$, $\ottnt{B'}$, $\ottnt{v}$, and $\texttt{\textcolor{red}{<<no parses (char 8): E(l \{U\})*** >>}}$ such that satisfy the following conditions:
            \begin{itemize}
              \item $ \mathit{l}  ::    \forall    \bm{ \alpha } : \bm{ \ottnt{K} }    \ottsym{.}    \sigma    \in   \Xi $
              \item $ \mathsf{op}  \ottsym{:}    \forall    \bm{ \beta } : \bm{ \ottnt{K'} }    \ottsym{.}    \ottnt{A'}   \Rightarrow   \ottnt{B'}    \in   \sigma $
              \item $\ottnt{e} = \texttt{\textcolor{red}{<<no parses (char 27):  E(l \{U\})[op \{l \{U\}\} \{V\} v]***  >>}}$.
            \end{itemize}
    \end{itemize}
  \end{lemma}
  The second statement of this lemma means there is no handler to handle the operation of the evaluation target. This situation is called control stuck. This lemma does not imply that any pure expression evaluates to a value when the evaluation terminates. In other words, type safety does not ensure that no pure expression gets control stuck.

  Type safety is proven via progress and preservation lemmas as usual. Their statements are as follows.
  \begin{lemma}[Progress]\label{lem:progress}
    If $\emptyset  \vdash  \ottnt{e}  \ottsym{:}  \ottnt{A}  \mid  \varepsilon$, then one of the following holds:
    \begin{itemize}
      \item $\ottnt{e}$ is a value,
      \item there is $\ottnt{e'}$ such that $\ottnt{e}  \longrightarrow  \ottnt{e'}$, and
      \item there are $\mathit{l}$, $\mathsf{op}$, $\ottnt{A'}$, $\ottnt{B'}$, $\ottnt{v}$, and $\texttt{\textcolor{red}{<<no parses (char 8): E(l \{U\})*** >>}}$ such that satisfy the following conditions:
            \begin{itemize}
              \item $ \mathit{l}  ::    \forall    \bm{ \alpha } : \bm{ \ottnt{K} }    \ottsym{.}    \sigma    \in   \Xi $
              \item $ \mathsf{op}  \ottsym{:}    \forall    \bm{ \beta } : \bm{ \ottnt{K'} }    \ottsym{.}    \ottnt{A'}   \Rightarrow   \ottnt{B'}    \in   \sigma $
              \item $\ottnt{e} = \texttt{\textcolor{red}{<<no parses (char 27):  E(l \{U\})[op \{l \{U\}\} \{V\} v]***  >>}}$.
            \end{itemize}
    \end{itemize}
  \end{lemma}
  \begin{lemma}[Preservation]\label{lem:preservation}
    If $\emptyset  \vdash  \ottnt{e}  \ottsym{:}  \ottnt{A}  \mid  \varepsilon$ and $\ottnt{e}  \longrightarrow  \ottnt{e'}$, then $\emptyset  \vdash  \ottnt{e'}  \ottsym{:}  \ottnt{A}  \mid  \varepsilon$.
  \end{lemma}

  \subsubsection{Effect Safety}
  The statement of effect safety is as follows.
  \begin{lemma}[Effect Safety]\label{lem:effsafe}
    If $\texttt{\textcolor{red}{<<no parses (char 38): G \mbox{$\mid$}- E(l \{U I\}) [op \{l \{U I\}\} \{V\} v] :*** A \mbox{$\mid$} epsilon >>}}$ and $ \mathit{l}  ::    \forall    {\bm{ \alpha } }^{ \mathit{I} } : {\bm{ \ottnt{K} } }^{ \mathit{I} }    \ottsym{.}    \sigma    \in   \Xi $ and $ \mathsf{op}  \ottsym{:}    \forall    \bm{ \beta } : \bm{ \ottnt{K'} }    \ottsym{.}    \ottnt{A'}   \Rightarrow   \ottnt{B'}    \in   \sigma $, then $ \Delta   \ottsym{(}   \Gamma   \ottsym{)}   \not\vdash   \varepsilon   \cong    \bbZero  $.
  \end{lemma}
  This lemma intuitively means that the effect must have the information of unhandled operations.

  \subsubsection{Type and Effect Safety}
  We get type and effect safety as the corollary from type safety and effect safety. The statement is as follows.
  \begin{theorem}[Type and Effect Safety]
    If $\emptyset  \vdash  \ottnt{e}  \ottsym{:}  \ottnt{A}  \mid   \bbZero $ and $\ottnt{e}  \longrightarrow  ^ * \ottnt{e'} \centernot \longrightarrow $, then there is $\ottnt{v}$ such that $\ottnt{e'} = \ottnt{v}$.
  \end{theorem}
  This theorem implies that any pure expression evaluates to a value when the evaluation terminates.

  \subsection{Type and Effect Safety of Concrete Systems}
  We show how to prove type and effect safety of concrete systems using {\lang}. We take as examples the three instances in Section~\ref{sec:effdef}, i.e. Example~\ref{exa:effset}, Example~\ref{exa:eff_simple_row}, and Example~\ref{exa:effrow}. The meta properties of these effect systems are already proven, but we propose a new way of proof: checking that the instance of ARE meets safety conditions.
  \begin{theorem}
    Example~\ref{exa:effset} meets safety conditions.
  \end{theorem}
  \begin{theorem}
    Example~\ref{exa:eff_simple_row} meets safety conditions.
  \end{theorem}
  \begin{theorem}
    Example~\ref{exa:effrow} meets safety conditions.
  \end{theorem}
  Because we already have type and effect safety of {\lang} when ARE meets safety conditions,
  these theorems lead to type and effect safety of these three effect systems as corollaries
  without showing something like progress and preservation lemmas.
}
\section{Formal Relationships between {\lang} and The Existing Systems}\label{sec:comparisons}

\begin{table}[t]
 %
 \caption{Comparison of the effectful aspects in {\lang} and the existing
 works. The mark {\xmark} means ``not supported,''  and ``explicit*'' in
 the column ``polymorphism'' for {\links} indicates that {\links} supports
 not only explicit type and effect polymorphism, but also
 row polymorphism in the style of \citet{remy_type_1993} at the effect-level.
 }
 \label{tbl:comparison-lang}
 \begin{tabular}{@{}c@{\ \ }|@{\ \ }c@{\ \ }|@{\ \ }c@{\ \ }|@{\ \ }c@{\ \ }|@{\ \ }c@{}}
                                                 & effect collections               & collected effects          & effect contexts' assignment      & polymorphism \\ \hline

  {\lang}                               & effect algebras                & label              & global             & explicit      \\
  {\pretnars}                           & sets                           & operation          & global             & {\xmark}      \\
  {\links}                              & simple rows                    & operation          & local              & explicit*      \\
  {\koka}                               & scoped rows                    & label              & global             & implicit      \\





 \end{tabular}
\end{table}

This section shows that {\lang} soundly models the key aspects of the existing effect
systems.
As targets, we select the effect systems of
\citet{pretnar_introduction_2015}, \citet{hillerstrom_continuation_2017}, and
\citet{leijen_type_2017}, which employ sets, simple rows, and scoped rows,
respectively, to represent effect collections.
We call them {\pretnars}, {\links}, and {\koka} because they model
the core part of the programming languages Eff~\cite{eff_lang}, Links~\cite{links_lang}, and
Koka~\cite{koka_lang}, respectively.\footnote{The core effect system of Links was
first presented by \citet{hillerstrom_liberating_2016}, but it seems to have a
minor flaw in the typing of sequential composition. We thus refer to
\citet{hillerstrom_continuation_2017} where the flaw is fixed.}

\subsection{Differences between {\lang} and The Selected Systems}
We aim to establish the formal connection between each of the existing systems
and {\lang}, but there exist some gaps between them.
First, the existing systems adopt their own syntax not only for effects but also
for types and programs, which hinders the formal comparison.
To address this problem, we define a syntactic translation
$\trans{\mathcal{E}}$ from each $\mathcal{E}$ of the selected systems to the instance of
{\lang} with the corresponding effect algebra.
For example, operation calls in {\pretnars} take the form $\mathsf{op}(v, y.c)$, carrying continuations $y.c$.
The translator $\trans{\pretnars}$ converts it to the expression
$\mathbf{let} \, y = \mathsf{op}_{l} \, \trans{\pretnars}(v) \, \mathbf{in} \, \trans{\pretnars}(c)$ in {\lang} using some appropriate label $\mathit{l}$.
%
%
Readers interested in the complete definitions of the translations are referred
to the supplementary material.

The remaining gaps between {\lang} and the existing systems are summarized in
Table~\ref{tbl:comparison-lang}. Because addressing the gaps other than the
representation of effect collections is beyond the scope of the present work, we
impose certain assumptions on the existing systems for the comparison.  In what follows, we detail
the gaps and how we address them.

\paragraph{Collected effects.}
In {\lang}, effect collections gather effect labels, which are sets of
operations of some specific effects.  For example, the effect for state can be
expressed by a label $\mathsf{State}$ equipped with operations $\mathsf{get}$ and
$\mathsf{set}$ for getting and updating, respectively, the current state.
In this style, which we call \emph{label-based}, an operation call is given an
effect collection including the effect label to which the called operation
belongs, and a handler is required to handle all the operations of a specified
label.
{\lang} and {\koka} employ the label-based style.
By contrast, {\pretnars} and {\links} adopt the \emph{operation-based} style, where
effect collections gather operations.
In this style, an operation call is given an effect collection including the
called operation (not labels), and effect handlers can implement any operation
freely.
To address this difference, when translating {\pretnars} and {\links} in the
operation-based style to {\lang} in the label-based style, we assume that some
labels are given and any effect collection appearing in {\pretnars} and {\links}
can be decomposed into a subset of the given labels.

\paragraph{Effect contexts' assignment.}
Our language {\lang} supposes that an effect context $\Xi$ is fixed during
typechecking one program.
We call this assignment of $\Xi$ \emph{global}.
{\pretnars} and {\koka} employ the same assignment style for effect contexts.
In contrast, in {\links}, effect contexts can change during the typechecking.
For example, consider the following program.
\begin{flalign*}
   &  \mathbf{handle}  \,   (  \mathbf{if} \,  \mathsf{ask}  \,  ()  \, \mathbf{then} \, 0 \, \mathbf{else} \,  (   \mathbf{handle}  \,   \mathsf{ask}  \,  ()   +  1  \, \mathbf{with} \,  \ottsym{\{} \, \mathbf{return} \, \mathit{x}  \mapsto  \mathit{x}  \ottsym{\}}   \uplus   \ottsym{\{}  \mathsf{ask} \, \mathit{z} \, \mathit{k}  \mapsto  \mathit{k} \, 2  \ottsym{\}}   )   )    & \\ &  \, \mathbf{with} \,  \ottsym{\{} \, \mathbf{return} \, \mathit{x}  \mapsto  \mathit{x}  \ottsym{\}}   \uplus   \ottsym{\{}  \mathsf{ask} \,  {}  \, \mathit{z} \, \mathit{k}  \mapsto  \mathit{k} \,  \mathsf{true}   \ottsym{\}} 
\end{flalign*}
In this program, both $\mathsf{ask}$ operation calls take the unit value,
but the first and second ones return Booleans and integers, respectively.
This program cannot be typechecked if an effect context is globally fixed.
{\links} can typecheck it because {\links} allows enclosing handlers to modify
effect contexts; namely, effect contexts are assigned \emph{locally}.
To
address the local assignment of effect contexts, we assume that every
operation has a unique, closed type in {\links}, which enables determining the
types of operations globally.

\paragraph{Polymorphism.}
The languages {\lang}, {\links}, and {\koka} support type and effect polymorphism.
Among them, only the polymorphism in {\koka} is \emph{implicit}, that is,
no term constructor for type abstraction and application is given.
Unfortunately, it is not straightforward to translate a program (or its typing
derivation) with implicit polymorphism in {\koka} to one with \emph{explicit}
polymorphism in {\lang} while preserving the meaning of the program because
{\koka} does not adopt \emph{value
restriction}~\cite{Tofte_1990_IC,Wright_1995_LSC}.
Our approach to this difference in polymorphism is simply to forbid the use of
implicit polymorphism in {\koka} and instead introduce explicit polymorphism by
equipping {\koka} with term constructors for type abstraction and application
as in {\lang} and {\links}.
It is also noteworthy that {\links} supports more advanced polymorphism,
inspired by row polymorphism proposed by \citet{remy_type_1993}.
It introduces \emph{presence types}, which can state that a specific label is
present or absent in a row, \emph{presence polymorphism}, and effect variables
constrained by which labels are present or absent.
This form of polymorphism facilitates solving unification problems in the
composition of effect handlers~\cite{hillerstrom_liberating_2016}.
Our translation from {\links} to {\lang} addresses these unique features in
{\links} as follows: first, present labels remain in the translated row but
labels with the absent flag do not; second, the constraints on effect variables
are ignored; third, we assume that programs to be translated do not use
presence polymorphism.
%
%
We left the support for presence polymorphism as future work: it seems to be
motivated by unification and type inference, which are beyond the scope of the
present work.

\subsection{Type-and-Effect Preservation of Translations}
We show that the translations preserve well-typedness under the aforementioned
assumptions.
%
%
\begin{theorem}
 \label{thm:trans-sound}
 Let $(\mathcal{E}, \mathcal{A}) \in \{ (\pretnars, \eaSet), (\links, \eaSimpleRow), (\koka, \eaScopedRow) \}$.
 If a program $c$ in the system $\mathcal{E}$ is well typed at an effect $\epsilon$,
 then $\trans{\mathcal{E}}(c)$ is well typed at effect $\trans{\mathcal{E}}(\epsilon)$
 in $\lang$ with $\mathcal{A}$.
\end{theorem}
This result guarantees that, for each $\mathcal{E}$ of the selected systems, the
programs in $\mathcal{E}$ can be safely executed in the semantics of {\lang}.
In other words, {\lang} can work as an intermediate language that ensures
type-and-effect safety.
%
%
Note that
the equivalence relation on scoped rows in {\koka} is more restrictive than
$ \sim_{\eanameScopedRow} $ in {\eaScopedRow} because the row equivalence in {\koka}
allows swapping effect labels $\mathit{l_{{\mathrm{1}}}} \,  \bm{ { S_{{\mathrm{1}}} } } $ and $\mathit{l_{{\mathrm{2}}}} \,  \bm{ { S_{{\mathrm{2}}} } } $ only if
$\mathit{l} \neq \mathit{l'}$, whereas $ \sim_{\eanameScopedRow} $ allows their swapping if
the label names $\mathit{l_{{\mathrm{1}}}}$ and $\mathit{l_{{\mathrm{2}}}}$, \emph{or} the type arguments $ \bm{ { S_{{\mathrm{1}}} } } $ and $ \bm{ { S_{{\mathrm{2}}} } } $
are different.
This gap does not prevent proving Theorem~\ref{thm:trans-sound} because it only
means that {\lang} with {\eaScopedRow} may accept more programs than {\koka}.
We will show an effect algebra with the row equivalence in {\koka} in Section~\ref{sec:erasure}.

\OLD{

This section shows that {\lang} models some existing effect systems,
specifically the effect systems of
\citet{pretnar_introduction_2015} (we call it {\pretnars}),
\citet{hillerstrom_continuation_2017} (we call it {\links}), and
\citet{leijen_type_2017} (we call it {\koka}).
We make some assumptions that fill the gaps about formalizations
and define the syntactic translations from the existing systems to instances of {\lang}.
Target instances of {\pretnars}, {\links}, and {\koka} are
the set-based instance (Example~\ref{exa:effset-effsig} and \ref{exa:effset}),
the simple-row-based instance (Example~\ref{exa:eff_simple_row}), and
the scoped-row-based instance (Example~\ref{exa:effrow}), respectively.
\TY{The names of the existing effect systems are tentative.}
However, we focus on explaining the gaps and assumptions that fill them in this section.
Readers interested in the details of assumptions and translations
are referred to the supplementary material.

\subsection{Effect System of \citet{pretnar_introduction_2015}}

First, we explain the difference between {\pretnars} and {\lang}.
{\pretnars} has Boolean types and if-expressions, and treats handlers as first-class objects
in contrast to {\lang}.
Most importantly, the effect representation of {\pretnars} is \emph{operation-based},
while the one of {\lang} is \emph{label-based}.
We explain this difference using examples.
Consider the following program written in {\lang}.
\begin{flalign*}
   & \effdecl{ \mathsf{State}  ::  \ottsym{\{}  \mathsf{get}  \ottsym{:}    \mathsf{Unit}    \Rightarrow    \mathsf{Int}    \ottsym{,}  \mathsf{set}  \ottsym{:}    \mathsf{Int}    \Rightarrow    \mathsf{Unit}    \ottsym{\}} } & \\
   &  (   \mathbf{handle}_{ \mathsf{State} }  \,   & \\ &    \quad    \mathsf{set} _{ \mathsf{State} }  \,  {}    \, 42  \ottsym{;}   \mathsf{get} _{ \mathsf{State} }  \,  {}  \,  ()    & \\ &  \, \mathbf{with} \,   \ottsym{\{} \, \mathbf{return} \, \mathit{x}  \mapsto  \lambda  \_  \ottsym{.}  \mathit{x}  \ottsym{\}}   \uplus   \ottsym{\{}  \mathsf{get} \,  {}  \, \_ \, \mathit{k}  \mapsto  \lambda  \mathit{x}  \ottsym{.}  \mathit{k} \, \mathit{x} \, \mathit{x}  \ottsym{\}}    \uplus   \ottsym{\{}  \mathsf{set} \,  {}  \, \mathit{x} \, \mathit{k}  \mapsto  \lambda  \_  \ottsym{.}  \mathit{k} \,  ()  \, \mathit{x}  \ottsym{\}}   )  \, 0
\end{flalign*}
In this program, operations $\mathsf{get}$ and $\mathsf{set}$ belong to the label $\mathsf{State}$ and
our effect system attach the effect $ \lift{ \mathsf{State} } $
to the handled expression $ \mathsf{set} _{ \mathsf{State} }  \,  {}  \, 42  \ottsym{;}   \mathsf{get} _{ \mathsf{State} }  \,  {}  \,  () $.
We call this style of effect representation a label-based style.
However, in the similar program written in {\pretnars},
the effect system of {\pretnars} attach the set $\{\mathsf{get}, \mathsf{set}\}$ to the handled expression.
We call this style of effect representation an operation-based style.
In an operation-based style, the following program is well-typed.
\begin{flalign*}
   &  \mathbf{handle}  \,   & \\ &    \quad    \mathbf{handle}  \,    & \\ &    \quad    \quad    \mathsf{set}  \,  {}     \, 42  \ottsym{;}   \mathsf{get}  \,  {}  \,  ()    & \\ &    \quad  \, \mathbf{with} \,  \ottsym{\{} \, \mathbf{return} \, \mathit{x}  \mapsto  \mathit{x}  \ottsym{\}}   \uplus   \ottsym{\{}  \mathsf{get} \,  {}  \, \_ \, \mathit{k}  \mapsto  0  \ottsym{\}}      & \\ &  \, \mathbf{with} \,  \ottsym{\{} \, \mathbf{return} \, \mathit{x}  \mapsto  \mathit{x}  \ottsym{\}}   \uplus   \ottsym{\{}  \mathsf{set} \,  {}  \, \mathit{x} \, \mathit{k}  \mapsto  \mathit{x}  \ottsym{\}} 
\end{flalign*}
In other words, an operation-based style allows handlers for
only $\mathsf{get}$, only $\mathsf{set}$, and both $\mathsf{get}$ and $\mathsf{set}$,
while a label-based style only allows a handler for both $\mathsf{get}$ and $\mathsf{set}$.

To define a translation from {\pretnars} to the set-based instance of {\lang},
we need to fill the aforementioned gaps.
Concerning Boolean types, if-expressions, and first-class handlers,
we simply omit them from {\pretnars}.
The difference between operation-based and label-based styles is problematic
because we do not have a straightforward translation from the former to the latter.
Although the formal comparison between these two styles is left open problem,
it is out of the scope of this paper.
\TY{Should I try to define such a translation?}
Thus, we set this problem aside simply by assuming that
any effect in {\pretnars}, which is a set of operations, uniquely corresponds to a set of labels.
This assumption makes simple a translation of operation sets and handlers.

\subsection{Effect System of \citet{hillerstrom_continuation_2017}}

The differences between {\links} and {\lang} are summarized as follows.
First, {\links} has variant and record types, and presence, handler, and computation kinds,
in contrast to {\lang}.
Secondly, {\links} represents effects in an operation-based style.
Finally, {\links} allows an operation to have two or more types.
For example, consider the following program, which is well-typed under a type system like {\links}.
\begin{flalign*}
   &  \mathbf{handle}  \,      & \\ &    \quad    \mathbf{handle}  \,     & \\ &    \quad    \quad    \mathsf{ask}  \,  {}     \,  ()   +  1    & \\ &    \quad  \, \mathbf{with} \,  \ottsym{\{} \, \mathbf{return} \, \mathit{x}  \mapsto  \mathit{x}  \ottsym{\}}   \uplus   \ottsym{\{}  \mathsf{ask} \,  {}  \, \_ \, \mathit{k}  \mapsto  \mathit{k} \, 2  \ottsym{\}}      & \\ &    \quad   +  \mathbf{if} \,  \mathsf{ask}  \,  {}  \,  ()  \, \mathbf{then} \, 3 \, \mathbf{else} \, 0    & \\ &  \, \mathbf{with} \,  \ottsym{\{} \, \mathbf{return} \, \mathit{x}  \mapsto  \mathit{x}  \ottsym{\}}   \uplus   \ottsym{\{}  \mathsf{ask} \,  {}  \, \_ \, \mathit{k}  \mapsto  \mathit{k} \,  \mathsf{true}   \ottsym{\}} 
\end{flalign*}
In this program, both two $\mathsf{ask}$ operations take $ \mathsf{Unit} $ type,
but the former returns $ \mathsf{Int} $ type and the latter does $ \mathsf{Bool} $ type.
In {\links}, types of operations are determined by enclosing handlers, not given globally.

To fill these gaps, we assume the following.
Concerning the first gap, we simply omit the components such as variant and record types.
Since the second gap is the same kind of gap that occurs between {\pretnars} and {\lang},
we pose a similar assumption.
At last, we assume that every operation has a unique and closed type in {\links},
because types of operations in {\lang} are globally defined.

\subsection{Effect System of \citet{leijen_type_2017}}

The main difference between {\koka} and {\lang} is in polymorphism.
  {\koka} adopts implicit polymorphism, while {\lang} does explicit polymorphism.
Thus, we adopt an explicit polymorphic version of {\koka} as a source language.
The other differences are so minor that they are omitted in this paper.
}
\section{Extensions of {\lang}}\label{sec:extensions}

This section extends {\lang} and safety conditions to lift coercions and
type-erasure semantics.  We also introduce effect algebras safe for these
extensions (including a new one based on multisets) and discuss how
adaptable each effect representation addressed in this paper---sets, multisets,
simple rows, and scoped rows---is for the extensions.

  {
    \let\section\subsection
    \let\subsection\subsubsection
    \section{Lift Coercions}\label{sec:coercions}

\TY{New section}

This section shows an extension to \emph{lift
coercions}~\cite{biernacki_handle_2018, biernacki_abstracting_2019} (also known
as injection~\cite{leijen2018algebraic} or masking~\cite{koka_lang}).
Given an effect label, a lift coercion forbids the innermost handler for the
label to handle any operation of the label.
They can prevent \emph{accidental handling}, a situation that an effect
handler handles an operation call against the programmer's intention.
This paper focuses on how {\lang} is extended with lift coercions;
see the prior work~\cite{biernacki_handle_2018,
biernacki_abstracting_2019,leijen2018algebraic} for the detail of the accidental
handling and how lift coercions work to address it.
We also show that the effect algebras {\eaSet} and {\eaSimpleRow} are
unsafe in the extension and that {\eaScopedRow} and a new effect
algebra for \emph{multisets} are safe.
Note that \citet{biernacki_abstracting_2019} introduce coercions in other forms.
We do not support them because they can be encoded with lift
coercions (if label polymorphism is not used)~\cite{biernacki_handle_2018,biernacki_abstracting_2019}.
%

\OLD{

\subsection{Review: Lift Coercions and Accidental Handling}

\TY{Should I reduce this part?}
A Lift coercion, introduced by \citet{biernacki_handle_2018, biernacki_abstracting_2019},
explicitly adds a label at the top of an effect and
forbids the innermost handler for the label to handle any operation of the label.
They can prevent \emph{accidental handling}, which is the situation where
a handler handles an operation call against a programmer's intention.
To show how lift coercions work and what is accidental handling,
we introduce two example functions, $\mathit{bad\_count\_f}$ and $\mathit{count\_f}$,
both of which are intended to count the number of times
when function $\mathit{f}$, whose type is $  \forall   \rho  \ottsym{:}   \mathbf{Eff}    \ottsym{.}     \ottsym{(}    \mathsf{Unit}     \rightarrow_{ \rho }     \mathsf{Unit}    \ottsym{)}    \rightarrow_{ \rho }     \mathsf{Unit}   $, uses its argument.

First, we show the definition of $\mathit{bad\_count\_f}$ in the scoped row syntax.
\begin{flalign*}
   & \effdecl{ \mathsf{Tick}  ::  \ottsym{\{}  \mathsf{tick}  \ottsym{:}    \mathsf{Unit}    \Rightarrow    \mathsf{Unit}    \ottsym{\}} } & \\
   & \letdecl{\mathit{bad\_count\_f}}
  \Lambda  \rho  \ottsym{:}   \mathbf{Eff}   \ottsym{.}  \lambda  \mathit{g}  \ottsym{:}    \mathsf{Unit}     \rightarrow_{ \langle  \mathsf{Tick}  \mid  \rho  \rangle }     \mathsf{Unit}    \ottsym{.}   & \\ &    \quad    \quad    (   \mathbf{handle}_{ \mathsf{Tick} }  \,    \mathit{f} \, \langle  \mathsf{Tick}  \mid  \rho  \rangle \,  (  \lambda  \mathit{x}  \ottsym{:}   \mathsf{Unit}   \ottsym{.}   \mathsf{tick} _{ \mathsf{Tick} }  \,  ()   \ottsym{;}  \mathit{g} \, \mathit{x}  )    & \\ &    \quad    \quad  \, \mathbf{with} \,  \ottsym{\{} \, \mathbf{return} \, \_  \mapsto  \lambda  \mathit{x}  \ottsym{:}   \mathsf{Int}   \ottsym{.}  \mathit{x}  \ottsym{\}}   \uplus   \ottsym{\{}  \mathsf{tick} \,  {}  \, \_ \, \mathit{k}  \mapsto  \lambda  \mathit{x}  \ottsym{:}   \mathsf{Int}   \ottsym{.}  \mathit{k} \,  ()  \,  (   \mathit{x}  +  1   )   \ottsym{\}}   )     \, 0
\end{flalign*}
$\mathit{bad\_count\_f}$ takes an effect variable $\rho$ and a function $\mathit{g}$,
and calls $\mathit{f}$.
While evaluating the calling of $\mathit{f}$, operation $\mathsf{tick}$ arises every time
$\mathit{f}$ calls its function argument $\lambda  \mathit{x}  \ottsym{:}   \mathsf{Unit}   \ottsym{.}   \mathsf{tick} _{ \mathsf{Tick} }  \,  ()   \ottsym{;}  \mathit{g} \, \mathit{x}$.
The handler surrounding the calling of $\mathit{f}$ counts up the number of times
when $\mathsf{tick}$ is called, in other words, when $\mathit{f}$ uses its function argument.

However, $\mathit{bad\_count\_f}$ cannot count the accurate number when $\mathit{g}$ calls $\mathsf{tick}$ in its body.
For example, let $\mathit{f}$ be $\Lambda  \rho  \ottsym{:}   \mathbf{Eff}   \ottsym{.}  \lambda  \mathit{g}  \ottsym{:}    \mathsf{Unit}     \rightarrow_{ \rho }     \mathsf{Unit}    \ottsym{.}  \mathit{g} \,  () $,
which uses its function argument just once.
In this case, $\mathit{bad\_count\_f} \, \langle  \rangle \,  \mathsf{tick} _{ \mathsf{Tick} } $ evaluates to $2$, because
the handled expression is $\mathit{f} \, \langle  \mathsf{Tick}  \mid  \langle  \rangle  \rangle \,  (  \lambda  \mathit{x}  \ottsym{:}   \mathsf{Unit}   \ottsym{.}   \mathsf{tick} _{ \mathsf{Tick} }  \,  ()   \ottsym{;}   \mathsf{tick} _{ \mathsf{Tick} }  \, \mathit{x}  ) $
and evaluates to $ \mathsf{tick} _{ \mathsf{Tick} }  \,  ()   \ottsym{;}   \mathsf{tick} _{ \mathsf{Tick} }  \,  () $.
This result is against the programmer's intention.
The aforementioned situation is called accidental handling.

In contrast to $\mathit{bad\_count\_f}$, we can define $\mathit{count\_f}$ accordingly to the programmer's will
using a lift coercion.
The definition of $\mathit{count\_f}$ is as follows.
\begin{flalign*}
   & \effdecl{ \mathsf{Tick}  ::  \ottsym{\{}  \mathsf{tick}  \ottsym{:}    \mathsf{Unit}    \Rightarrow    \mathsf{Unit}    \ottsym{\}} } & \\
   & \letdecl{\mathit{count\_f}}
  \Lambda  \rho  \ottsym{:}   \mathbf{Eff}   \ottsym{.}  \lambda  \mathit{g}  \ottsym{:}   \ottnt{A}    \rightarrow_{ \rho }    \ottnt{B}   \ottsym{.}   & \\ &    \quad    \quad    (   \mathbf{handle}_{ \mathsf{Tick} }  \,    \mathit{f} \, \ottsym{(}  \langle  \mathsf{Tick}  \mid  \rho  \rangle  \ottsym{)} \,  (  \lambda  \mathit{x}  \ottsym{:}  \ottnt{A}  \ottsym{.}   \mathsf{tick} _{ \mathsf{Tick} }  \,  ()   \ottsym{;}   [  \mathit{g} \, \mathit{x}  ] _{ \mathsf{Tick} }   )    & \\ &    \quad    \quad  \, \mathbf{with} \,  \ottsym{\{} \, \mathbf{return} \, \_  \mapsto  \lambda  \mathit{x}  \ottsym{:}   \mathsf{Int}   \ottsym{.}  \mathit{x}  \ottsym{\}}   \uplus   \ottsym{\{}  \mathsf{tick} \,  {}  \, \_ \, \mathit{k}  \mapsto  \lambda  \mathit{x}  \ottsym{:}   \mathsf{Int}   \ottsym{.}  \mathit{k} \,  ()  \,  (   \mathit{x}  +  1   )   \ottsym{\}}   )     \, 0
\end{flalign*}
In this program, we use a lift coercion for $\mathsf{Tick}$ at $ [  \mathit{g} \, \mathit{x}  ] _{ \mathsf{Tick} } $.
This effect coercion adds the label $\mathsf{Tick}$ to the top of row $\rho$
Furthermore, concerning about semantics, if $\mathsf{tick}$ operation is called in $\mathit{g} \, \mathit{x}$,
then the innermost handler for $\mathsf{Tick}$ cannot handle this operation call.
Thus, the handler in $\mathit{count\_f}$ only handles $\mathsf{tick}$ before $ [  \mathit{g} \, \mathit{x}  ] _{ \mathsf{Tick} } $.
For example, if $\mathit{f}$ is $\Lambda  \rho  \ottsym{:}   \mathbf{Eff}   \ottsym{.}  \lambda  \mathit{g}  \ottsym{:}    \mathsf{Unit}     \rightarrow_{ \rho }     \mathsf{Unit}    \ottsym{.}  \mathit{g} \,  () $,
then
\begin{flalign*}
   &  \mathbf{handle}_{ \mathsf{Tick} }  \, \mathit{count\_f} \, \langle  \rangle \,  \mathsf{tick} _{ \mathsf{Tick} }  \, \mathbf{with} \,  \ottsym{\{} \, \mathbf{return} \, \mathit{x}  \mapsto  \mathit{x}  \ottsym{\}}   \uplus   \ottsym{\{}  \mathsf{tick} \,  {}  \, \_ \, \mathit{k}  \mapsto  \mathit{k} \,  ()   \ottsym{\}} 
\end{flalign*}
evaluates to $1$, which is the required result.
The outer handler in this example handles $\mathsf{tick}$ in the lift coercion and simply ignores it.

\paragraph{Remark.}
\citet{biernacki_abstracting_2019} introduce \emph{cons coercions} and \emph{swap coercions}
in addition to lift coercions.
We do not take these kinds of coercions into {\lang}
because \citet{biernacki_handle_2018} propose
encoding cons and swap coercions with lift coercions.
Although the reason why \citet{biernacki_abstracting_2019} adopt swap coercions is that
\citet{biernacki_handle_2018}'s encoding cannot work in the presence of polymorphic labels,
we conjecture that introducing a special handler construct that can be parameterized by a label variable
remedies this problem.

}

\subsection{Extending {\lang} to Lift Coercions}

\begin{figure}
  \[
    \begin{array}{rcl@{\qquad}rcl}
      \ottnt{e} & \Coloneqq & \cdots  \mid   [  \ottnt{e}  ] _{ \ottnt{L} }  \quad \text{(expressions)} &
      \ottnt{E} & \Coloneqq & \cdots  \mid   [  \ottnt{E}  ] _{ \ottnt{L} }  \quad \text{(evaluation contexts)}
    \end{array}
  \]
  \textbf{Freeness of labels}\tquad\fbox{$ \mathit{n}  \mathrm{-free} ( \ottnt{L} ,  \ottnt{E} ) $}\hfill\phantom{}
  \begin{mathpar}
    \inferrule{
       \ottsym{(}  \mathit{n}  \ottsym{+}  1  \ottsym{)}  \mathrm{-free} ( \ottnt{L} ,  \ottnt{E} ) 
    }{
       \mathit{n}  \mathrm{-free} ( \ottnt{L} ,   \mathbf{handle}_{ \ottnt{L} }  \, \ottnt{E} \, \mathbf{with} \, \ottnt{h} ) 
    }

    \inferrule{
       \mathit{n}  \mathrm{-free} ( \ottnt{L} ,  \ottnt{E} ) 
    }{
       \mathit{n}  \ottsym{+}  1  \mathrm{-free} ( \ottnt{L} ,   [  \ottnt{E}  ] _{ \ottnt{L} }  ) 
    }

    \inferrule{
     \mathit{n}  \mathrm{-free} ( \ottnt{L} ,  \ottnt{E} )  \\ \ottnt{L} \neq \ottnt{L'}
    }{
     \mathit{n}  \mathrm{-free} ( \ottnt{L} ,   [  \ottnt{E}  ] _{ \ottnt{L'} }  ) 
    }
  \end{mathpar}
  \\[1ex]
  \begin{minipage}[t]{.3\textwidth}
  \textbf{Reduction}\tquad\fbox{$\ottnt{e}  \longmapsto  \ottnt{e'}$}\hfill\phantom{}
  \begin{mathpar}
    \inferrule{
    }{
       [  \ottnt{v}  ] _{ \ottnt{L} }   \longmapsto  \ottnt{v}
    }\ \rname{R}{Lift}
  \end{mathpar}
  \end{minipage}
  \begin{minipage}[t]{.68\textwidth}
  \textbf{Typing}\tquad\fbox{$\Gamma  \vdash  \ottnt{e}  \ottsym{:}  \ottnt{A}  \mid  \varepsilon$}\hfill\phantom{}
  \begin{mathpar}
    \inferrule{
    \Gamma  \vdash  \ottnt{e}  \ottsym{:}  \ottnt{A}  \mid  \varepsilon' \\ \Gamma  \vdash  \ottnt{L}  \ottsym{:}   \mathbf{Lab}  \\    \lift{ \ottnt{L} }   \mathop{ \odot }  \varepsilon'    \sim   \varepsilon 
    }{
    \Gamma  \vdash   [  \ottnt{e}  ] _{ \ottnt{L} }   \ottsym{:}  \ottnt{A}  \mid  \varepsilon
    }\ \rname{T}{Lift}
   \end{mathpar}
  \end{minipage}
  \caption{The extension for lift coercions.}
  \label{fig:lift_extension}
\end{figure}

We show the extended part of {\lang} in Figure~\ref{fig:lift_extension}.
Expressions and evaluation contexts are extended with lift coercions $ [\,   \!\textnormal{--}\!  \, ] _{ \ottnt{L} } $.
To define the semantics of lift coercions, we generalize $0$-freeness to
\emph{$n$-freeness} for an arbitrary natural number $n$ by following
\citet{biernacki_handle_2018}.
The predicate $ \mathit{n}  \mathrm{-free} ( \ottnt{L} ,  \ottnt{E} ) $ is defined by the rules in
Figure~\ref{fig:lift_extension} in addition to the ones given previously
(Figure~\ref{fig:semantics}).
Intuitively, $ \mathit{n}  \mathrm{-free} ( \ottnt{L} ,  \ottnt{E} ) $ means that,
for an operation $\mathsf{op}$ of $\ottnt{L}$,
the operation call in $\ottnt{E}  \ottsym{[}   \mathsf{op} _{ \ottnt{L} }  \,  \bm{ { T } } ^ {  \mathit{J}  }  \, \ottnt{v}  \ottsym{]}$ will be handled by the
$(\mathit{n}  \ottsym{+}  1)$-th innermost enclosing handler for $\ottnt{L}$.
For example,
$ 1  \mathrm{-free} ( \ottnt{L} ,   [  \Box  ] _{ \ottnt{L} }  ) $ and
$ 0  \mathrm{-free} ( \ottnt{L} ,   \mathbf{handle}_{ \ottnt{L} }  \,  [  \Box  ] _{ \ottnt{L} }  \, \mathbf{with} \, \ottnt{h_{{\mathrm{1}}}} ) $ hold.
Because the semantics of the effect handling (specifically, the reduction rule
\rname{R}{Handle2} in Figure~\ref{fig:semantics}) requires the label of the
handled operation call to be $0$-free in the evaluation context enclosing the
operation call, the operation call in
$ \mathbf{handle}_{ \ottnt{L} }  \,  \mathbf{handle}_{ \ottnt{L} }  \,  [   \mathsf{op} _{ \ottnt{L} }  \, \ottnt{v}  ] _{ \ottnt{L} }  \, \mathbf{with} \, \ottnt{h_{{\mathrm{1}}}} \, \mathbf{with} \, \ottnt{h_{{\mathrm{2}}}}$ will be handled by $\ottnt{h_{{\mathrm{2}}}}$.
If a lift coercion is given a value, it returns the value as it is (\rname{R}{Lift}).
The type-and-effect system is extended with the rule \rname{T}{Lift}, which
allows the information $\varepsilon'$ of effects of an expression $\ottnt{e}$ to pass through the
innermost effect handler for a label $\ottnt{L}$ by prepending $\ottnt{L}$ to $\varepsilon'$.

\subsection{Safety Conditions and Type-and-Effect Safety}
\label{subsec:coercion-safety-cond}

To ensure the safety of programs in the presence of lift coercions, we introduce
a new safety condition in addition to the ones given in
Section~\ref{sec:safety}.

\begin{definition}[Safety Condition for Lift Coercions]\label{def:safe_cond_lift}
  The safety condition added for lift coercions is:
  \begin{enumerate*}
    \setcounter{enumi}{\value{safecondcounter}}
    \item\label{def:safe_cond_lift:removal}
          If
          $   \lift{ \ottnt{L} }   \mathop{ \odot }  \varepsilon_{{\mathrm{1}}}    \sim       \lift{ \ottnt{L_{{\mathrm{1}}}} }   \mathop{ \odot } \cdots \mathop{ \odot }   \lift{ \ottnt{L_{\ottmv{n}}} }    \mathop{ \odot }   \lift{ \ottnt{L} }    \mathop{ \odot }  \varepsilon_{{\mathrm{2}}}  $
          and $\ottnt{L} \notin \{  \ottnt{L_{{\mathrm{1}}}}  \ottsym{,}  \ldots  \ottsym{,}  \ottnt{L_{\ottmv{n}}}  \}$,
          then $ \varepsilon_{{\mathrm{1}}}   \sim      \lift{ \ottnt{L_{{\mathrm{1}}}} }   \mathop{ \odot } \cdots \mathop{ \odot }   \lift{ \ottnt{L_{\ottmv{n}}} }    \mathop{ \odot }  \varepsilon_{{\mathrm{2}}}  $.
          \setcounter{safecondcounter}{\value{enumi}}
  \end{enumerate*}
\end{definition}
This new condition can be understood as follows.
First, let $\varepsilon_{{\mathrm{2}}}$ be an effect of an expression $\ottnt{e}$.
Then, the effect of the expression $ [ \cdots [   [  \ottnt{e}  ] _{ \ottnt{L} }   ] _{ \ottnt{L_{\ottmv{n}}} } \cdots ]_{ \ottnt{L_{{\mathrm{1}}}} } $
is given as $    \lift{ \ottnt{L_{{\mathrm{1}}}} }   \mathop{ \odot } \cdots \mathop{ \odot }   \lift{ \ottnt{L_{\ottmv{n}}} }    \mathop{ \odot }   \lift{ \ottnt{L} }    \mathop{ \odot }  \varepsilon_{{\mathrm{2}}} $.
Assume that the expression is handled by an effect handler for $\ottnt{L}$ and the remaining effect is $\varepsilon_{{\mathrm{1}}}$.
Then, $\varepsilon_{{\mathrm{1}}}$ should retain the information that $\ottnt{e}$ is surrounded by
lift coercions for $\ottnt{L_{{\mathrm{1}}}}, \cdots, \ottnt{L_{\ottmv{n}}}$ because
the handling expression may be enclosed by effect handlers for $\ottnt{L_{{\mathrm{1}}}}, \cdots, \ottnt{L_{\ottmv{n}}}$.
Such information is described by $   \lift{ \ottnt{L_{{\mathrm{1}}}} }   \mathop{ \odot } \cdots \mathop{ \odot }   \lift{ \ottnt{L_{\ottmv{n}}} }    \mathop{ \odot }  \varepsilon_{{\mathrm{2}}} $.
Thus, safety condition~\ref{def:safe_cond_lift:removal} requires
$ \varepsilon_{{\mathrm{1}}}   \sim      \lift{ \ottnt{L_{{\mathrm{1}}}} }   \mathop{ \odot } \cdots \mathop{ \odot }   \lift{ \ottnt{L_{\ottmv{n}}} }    \mathop{ \odot }  \varepsilon_{{\mathrm{2}}}  $.

\OLD{
This new condition means that the remaining effects after handling must be equivalent even if lift coercions are inserted.
For example, consider the expression $ \mathbf{handle}_{ \ottnt{L} }  \,  [   [  \ottnt{e}  ] _{ \ottnt{L} }   ] _{ \ottnt{L'} }  \, \mathbf{with} \, \ottnt{h}$,
assuming that $\ottnt{e}$ has the effect $\varepsilon$.
In this expression, every operation belonging to the label $\ottnt{L}$ and called in $\ottnt{e}$
cannot be handled by $\ottnt{h}$.
Therefore, the effect of $ \mathbf{handle}_{ \ottnt{L} }  \,  [   [  \ottnt{e}  ] _{ \ottnt{L} }   ] _{ \ottnt{L'} }  \, \mathbf{with} \, \ottnt{h}$ must have the information about $\varepsilon$.
Furthermore, that effect must have the information about the lift coercion for $\ottnt{L'}$.
Thus, safety condition~\ref{def:safe_cond_lift:removal} requires that
the whole expression $ \mathbf{handle}_{ \ottnt{L} }  \,  [   [  \ottnt{e}  ] _{ \ottnt{L} }   ] _{ \ottnt{L'} }  \, \mathbf{with} \, \ottnt{h}$ must have the effect equivalent to
$  \lift{ \ottnt{L'} }   \mathop{ \odot }  \varepsilon $.
}

To see the importance of the new safety condition more concretely, we show that the
effect algebras {\eaSet} and {\eaSimpleRow} violate this new condition and then
present how they make some unsafe programs typeable.
\begin{theorem}[Unsafe Effect Algebras with Lift Coercions]
 The effect algebras {\eaSet} and {\eaSimpleRow} do not meet safety condition~\ref{def:safe_cond_lift:removal}.
 Furthermore, there exists an expression that is well typed under {\eaSet} and {\eaSimpleRow} and gets stuck.
\end{theorem}
\begin{proof}
          We consider only {\eaSet} here; a similar discussion can be applied to {\eaSimpleRow}.
          Recall that the operation $ \odot $ in {\eaSet} is implemented by the
          set union, so it meets idempotence: $  \{  \ottnt{L}  \}  \,\underline{ \cup }\,  \{  \ottnt{L}  \}    \sim   \{  \ottnt{L}  \} $.
          Furthermore, we can use the empty set as the identity element, so
          $  \{  \ottnt{L}  \}  \,\underline{ \cup }\,  \{  \ottnt{L}  \}    \sim    \{  \ottnt{L}  \}  \,\underline{ \cup }\,  \{  \}  $.
          If safety condition~\ref{def:safe_cond_lift:removal} was met,
          $ \{  \ottnt{L}  \}   \sim   \{  \} $ (where $\{  \ottnt{L}  \}$, $\{  \}$, and $0$ are
          taken as $\varepsilon_{{\mathrm{1}}}$, $\varepsilon_{{\mathrm{2}}}$, and $\ottmv{n}$, respectively, in Definition~\ref{def:safe_cond_lift}).
          However, the equivalence does not hold.

          As a program that is typeable under {\eaSet}, consider $ \mathbf{handle}_{ \mathsf{Exc} }  \,  [   \mathsf{raise} _{ \mathsf{Exc} }  \,  \mathsf{Unit}  \,  ()   ] _{ \mathsf{Exc} }  \, \mathbf{with} \, \ottnt{h}$
          where
          $ \mathsf{Exc}  ::  \ottsym{\{}  \mathsf{raise}  \ottsym{:}    \forall   \alpha  \ottsym{:}   \mathbf{Typ}    \ottsym{.}     \mathsf{Unit}    \Rightarrow   \alpha   \ottsym{\}} $.
          This program can be typechecked under an appropriate assumption as illustrated by the following typing derivation:
           \[
            \hspace{-5em}
            \inferrule* [Right=T\_Handling] {
            \cdots \\
              \{  \mathsf{Exc}  \}  \,\underline{ \cup }\,  \{  \}    \sim   \{  \mathsf{Exc}  \} \\
            \inferrule* [Right=T\_Lift] {
            \emptyset  \vdash   \mathsf{raise} _{ \mathsf{Exc} }  \,  \mathsf{Unit}  \,  ()   \ottsym{:}  \ottnt{A}  \mid  \{  \mathsf{Exc}  \} \\
              \{  \mathsf{Exc}  \}  \,\underline{ \cup }\,  \{  \mathsf{Exc}  \}    \sim   \{  \mathsf{Exc}  \} 
            }{
            \emptyset  \vdash   [   \mathsf{raise} _{ \mathsf{Exc} }  \,  \mathsf{Unit}  \,  ()   ] _{ \mathsf{Exc} }   \ottsym{:}  \ottnt{A}  \mid  \{  \mathsf{Exc}  \}
            }
            }{
            \emptyset  \vdash   \mathbf{handle}_{ \mathsf{Exc} }  \,  [   \mathsf{raise} _{ \mathsf{Exc} }  \,  \mathsf{Unit}  \,  ()   ] _{ \mathsf{Exc} }  \, \mathbf{with} \, \ottnt{h}  \ottsym{:}  \ottnt{B}  \mid  \{  \}
            }
          \]
          However, the call to $\mathsf{raise}$ is not handled as it needs to be handled by the \emph{second} closest handler.

          \OLD{
          Recall Example~\ref{exa:effset} and Example~\ref{exa:eff_simple_row}.
          Because they allow the collapsing of the same labels
          as indicated by $  \{  \ottnt{L}  \}  \,\underline{ \cup }\,  \{  \ottnt{L}  \}    \sim   \{  \ottnt{L}  \} $
          Then, we have $$  \{  \ottnt{L}  \}  \,\underline{ \cup }\,  \{  \ottnt{L}  \}    \sim   \{  \ottnt{L}  \} $$
          and $ \langle  \ottnt{L}  \mid  \langle  \ottnt{L}  \mid  \langle  \rangle  \rangle  \rangle   \sim   \langle  \ottnt{L}  \mid  \langle  \rangle  \rangle $,
          the handlings of $\ottnt{L}$ from $\{  \ottnt{L}  \}$ and $\langle  \ottnt{L}  \mid  \langle  \rangle  \rangle$ result
          $\{  \}$ or $\{  \ottnt{L}  \}$, and $\langle  \rangle$ or $\langle  \ottnt{L}  \mid  \langle  \rangle  \rangle$, respectively.
          Thus, both of them violate coercion safety condition~\ref{def:safe_cond_lift:removal}.
          In these cases, the type $ \mathsf{Int} $ and the effect $ \bbZero $ can be assigned to
          \begin{flalign*}
            \hphantom{\text{Exampl}}                                          &  \mathbf{handle}_{ \mathsf{Exc} }  \,  [   \mathsf{raise} _{ \mathsf{Exc} }  \,  \mathsf{Unit}  \,  ()   ] _{ \mathsf{Exc} }  \, \mathbf{with} \,  \ottsym{\{} \, \mathbf{return} \, \_  \mapsto   ()   \ottsym{\}}   \uplus   \ottsym{\{}  \mathsf{raise} \, \alpha  \ottsym{:}   \mathbf{Typ}  \, \_ \, \mathit{k}  \mapsto   ()   \ottsym{\}}  &
          \end{flalign*}
          where $ \mathsf{Exc}  ::  \ottsym{\{}  \mathsf{raise}  \ottsym{:}    \forall   \alpha  \ottsym{:}   \mathbf{Typ}    \ottsym{.}     \mathsf{Unit}    \Rightarrow   \alpha   \ottsym{\}} $,
          because
          $ \mathsf{raise} _{ \mathsf{Exc} }  \,  \mathsf{Unit}  \,  () $ can have effect $\{  \mathsf{Exc}  \}$ and $\langle  \mathsf{Exc}  \mid  \langle  \rangle  \rangle$ due to
          $  \{  \mathsf{Exc}  \}  \,\underline{ \cup }\,  \{  \mathsf{Exc}  \}    \sim   \{  \mathsf{Exc}  \} $ and
          $ \langle  \mathsf{Exc}  \mid  \langle  \mathsf{Exc}  \mid  \langle  \rangle  \rangle  \rangle   \sim   \langle  \mathsf{Exc}  \mid  \langle  \rangle  \rangle $, respectively.
          However, this program is the stuck term.
          }
\end{proof}

In contrast, the effect algebra {\eaScopedRow} for scoped rows satisfies safety
condition~\ref{def:safe_cond_lift:removal}. The point is that
$ \odot_\mathrm{ScpR} $ in {\eaScopedRow} is \emph{not} idempotent. Therefore, they
can represent how many lift coercions are used and how many effect handlers are
necessary to handle expressions as the information of effects.
This observation gives us a new effect algebra with \emph{multisets}.
Multisets can have multiple instances of the same element and their sum operation is also nonidempotent.
Thus, we can expect---and it is the case---that the algebra for multisets meets
safety condition~\ref{def:safe_cond_lift:removal} as well as the other
conditions.
\begin{example}[Effect Multisets]\label{exa:effmultiset}
  The effect signature $\SbaseMSet$ of effect multisets is given by
  $\{  \}  \ottsym{:}   \mathbf{Eff} $,
  $\{  \ottsym{-}  \}  \ottsym{:}   \mathbf{Lab}   \rightarrow   \mathbf{Eff} $, and
  $ \ottsym{-}  \,\underline{ \sqcup }\,  \ottsym{-}   \ottsym{:}   \mathbf{Eff}   \times   \mathbf{Eff}   \rightarrow   \mathbf{Eff} $ (which is the sum operation for multisets).
  An effect algebra {\eaMSet} for multisets is defined by
  $\langle \SbaseMSet,  \ottsym{-}  \,\underline{ \sqcup }\,  \ottsym{-} , \{  \}, \{  \ottsym{-}  \},  \sim_{\eanameMSet}  \rangle$
  where $ \sim_{\eanameMSet} $ is the least equivalence relation satisfying the same rules
  as $ \sim_{\eanameSet} $ except for the idempotence rule.



\end{example}
%
%
\begin{theorem}
 The effect algebras {\eaScopedRow} and {\eaMSet} meet safety conditions~~\ref{def:safe_cond:label_notemp}--\ref{def:safe_cond_lift:removal}.
\end{theorem}

The type-and-effect safety of {\lang} with lift coercions is proven similarly to Theorem~\ref{thm:type-and-eff-safe}
provided that an effect algebra meets safety conditions~\ref{def:safe_cond:label_notemp}--\ref{def:safe_cond_lift:removal}.

\begin{theorem}[Type-and-Effect Safety]
  \label{thm:type-and-eff-safe-lift}
  Assume that a given effect algebra meets safety conditions~\ref{def:safe_cond:label_notemp}--\ref{def:safe_cond_lift:removal}.
  If $\emptyset  \vdash  \ottnt{e}  \ottsym{:}  \ottnt{A}  \mid   \bbZero $ and $\ottnt{e}  \longrightarrow  ^ * \ottnt{e'} \centernot \longrightarrow $, then $\ottnt{e'} = \ottnt{v}$ for some $\ottnt{v}$.
\end{theorem}

\OLD{

\subsection{Type-and-Effect Safety}

\TY{Should we mention lemmas about $ \ottnt{L}  \olessthan^{ \mathit{n} }  \varepsilon $?}

We prove that type-and-effect safety similarly to Theorem~\ref{thm:type-and-eff-safe}
provided that an effect algebra meets the safety conditions in Definition~\ref{def:safe_cond}
and the condition in Definition~\ref{def:safe_cond_lift}.

\begin{theorem}[Type-and-Effect Safety]
  \label{thm:type-and-eff-safe-lift}
  If $\emptyset  \vdash  \ottnt{e}  \ottsym{:}  \ottnt{A}  \mid   \bbZero $ and $\ottnt{e}  \longrightarrow  ^ * \ottnt{e'} \centernot \longrightarrow $, then $\ottnt{e'} = \ottnt{v}$ for some $\ottnt{v}$.
\end{theorem}

\subsection{Coercible Instances}

As shown in Example~\ref{exa:unsafe_instances_coercions},
sets and simple rows violate safety condition~\ref{def:safe_cond_lift:removal}
and are indeed unsound with lift coercions.
In contrast to them, the effect algebra of scoped rows in Example~\ref{exa:effrow}
meets safety condition~\ref{def:safe_cond_lift:removal}.
\begin{theorem}
  Example~\ref{exa:effrow} meets safety conditions (for lift coercions).
\end{theorem}
This theorem demonstrates that row-style effect representations have some variants
that can be soundly combined with lift coercions.

However, not only row-style but also set-style effect representations
can be soundly adapted to lift coercions.
The observation that collapsing the same labels is problematic
as indicated by Example~\ref{exa:unsafe_instances_coercions}
leads to a sound variant of set-styles, \emph{multisets}.
\begin{example}[Effect Multisets]\label{exa:effmultiset}
  The effect signature of effect sets is given as follows, except for label names:
  \begin{itemize}
    \item $\{  \}  \ottsym{:}   \mathbf{Eff} $,
    \item $\{  \ottsym{-}  \}  \ottsym{:}   \mathbf{Lab}   \rightarrow   \mathbf{Eff} $, and
    \item $ \ottsym{-}  \,\underline{ \sqcup }\,  \ottsym{-}   \ottsym{:}   \mathbf{Eff}   \times   \mathbf{Eff}   \rightarrow   \mathbf{Eff} $.
  \end{itemize}
  The constructors $ \bbZero $ and $ \lift{ \ottsym{-} } $ are defined to
  be $\{  \}$ and $\{  \ottsym{-}  \}$, respectively.
  An effect algebra for effect sets is defined as follows:
  $ \varepsilon_{{\mathrm{1}}}  \mathop{ \odot }  \varepsilon_{{\mathrm{2}}}  =  \varepsilon_{{\mathrm{1}}}  \,\underline{ \sqcup }\,  \varepsilon_{{\mathrm{2}}} $ and
  $ \sim $ is the least equivalence relation satisfying the following rules:
  \begin{mathpar}
    \inferrule{ }{  \varepsilon_{{\mathrm{1}}}  \,\underline{ \sqcup }\,  \varepsilon_{{\mathrm{2}}}    \sim    \varepsilon_{{\mathrm{2}}}  \,\underline{ \sqcup }\,  \varepsilon_{{\mathrm{1}}}  }

    \inferrule{ }{  \varepsilon  \,\underline{ \sqcup }\,  \{  \}    \sim   \varepsilon }

    \inferrule{ }{  \ottsym{(}   \varepsilon_{{\mathrm{1}}}  \,\underline{ \sqcup }\,  \varepsilon_{{\mathrm{2}}}   \ottsym{)}  \,\underline{ \sqcup }\,  \varepsilon_{{\mathrm{3}}}    \sim    \varepsilon_{{\mathrm{1}}}  \,\underline{ \sqcup }\,  \ottsym{(}   \varepsilon_{{\mathrm{2}}}  \,\underline{ \sqcup }\,  \varepsilon_{{\mathrm{3}}}   \ottsym{)}  }

    \inferrule{
     \varepsilon_{{\mathrm{1}}}   \sim   \varepsilon_{{\mathrm{2}}}  \\  \varepsilon_{{\mathrm{3}}}   \sim   \varepsilon_{{\mathrm{4}}} 
    }{
      \varepsilon_{{\mathrm{1}}}  \,\underline{ \sqcup }\,  \varepsilon_{{\mathrm{3}}}    \sim    \varepsilon_{{\mathrm{2}}}  \,\underline{ \sqcup }\,  \varepsilon_{{\mathrm{4}}}  
    }
  \end{mathpar}
  The only difference from sets is the removal of the idempotence rule.
\end{example}
This effect algebra meets safety conditions shown in
Definition~\ref{def:safe_cond} and \ref{def:safe_cond_lift}.
\begin{theorem}
  Example~\ref{exa:effmultiset} meets type-erasure safety conditions (for lift coercions).
\end{theorem}

}
    \section{Type-Erasure Semantics}\label{sec:erasure}

This section shows an adaption of {\lang} to type-erasure semantics, which is
different from those given in Sections~\ref{sec:lang} and \ref{sec:coercions} in
that it does not rely on type arguments of label names in seeking effect handlers
matching with called operations.
Type erasure semantics is helpful to develop efficient implementations of effect
handlers with parametric
effects~\cite{biernacki_abstracting_2019}.

\subsection{Formal Definition of Type-Erasure Semantics}

\begin{figure}[t]
  \textbf{Freeness of label names}\tquad\fbox{$ \mathit{n}  \mathrm{-free} ( \mathit{l} ,  \ottnt{E} ) $} \hfill \phantom{}
  \begin{mathpar}
    \inferrule{
    }{
       0  \mathrm{-free} ( \mathit{l} ,  \Box ) 
    }

    \inferrule{
       \mathit{n}  \mathrm{-free} ( \mathit{l} ,  \ottnt{E} ) 
    }{
       \mathit{n}  \mathrm{-free} ( \mathit{l} ,  \mathbf{let} \, \mathit{x}  \ottsym{=}  \ottnt{E} \, \mathbf{in} \, \ottnt{e} ) 
    }

    \inferrule{
     \mathit{n}  \mathrm{-free} ( \mathit{l} ,  \ottnt{E} )  \\ \mathit{l} \neq \mathit{l'}
    }{
     \mathit{n}  \mathrm{-free} ( \mathit{l} ,   \mathbf{handle}_{ \mathit{l'} \,  \bm{ { S } } ^ {  \mathit{I}  }  }  \, \ottnt{E} \, \mathbf{with} \, \ottnt{h} ) 
    }
  \end{mathpar}

  \phantom{}\\
  \textbf{Reduction}\tquad\fbox{$\ottnt{e}  \longmapsto  \ottnt{e'}$} \hfill\phantom{}
  \begin{mathpar}
    \inferrule{
     \mathsf{op} \,  {\bm{ \beta } }^{ \mathit{J} } : {\bm{ \ottnt{K} } }^{ \mathit{J} }  \, \mathit{p} \, \mathit{k}  \mapsto  \ottnt{e}   \in   \ottnt{h}  \\ \ottnt{v_{\ottmv{cont}}} = \lambda  \mathit{z}  \ottsym{.}   \mathbf{handle}_{ \mathit{l} \,  \bm{ { S } } ^ {  \mathit{I}  }  }  \, \ottnt{E}  \ottsym{[}  \mathit{z}  \ottsym{]} \, \mathbf{with} \, \ottnt{h} \\  0  \mathrm{-free} ( \mathit{l} ,  \ottnt{E} ) 
    }{
     \mathbf{handle}_{ \mathit{l} \,  \bm{ { S } } ^ {  \mathit{I}  }  }  \, \ottnt{E}  \ottsym{[}   \mathsf{op} _{ \mathit{l} \,  \bm{ { S' } } ^ {  \mathit{I}  }  }  \,  \bm{ { T } } ^ {  \mathit{J}  }  \, \ottnt{v}  \ottsym{]} \, \mathbf{with} \, \ottnt{h}  \longmapsto  \ottnt{e} \,  \! [ {\bm{ { T } } }^{ \mathit{J} } / {\bm{ \beta } }^{ \mathit{J} } ]  \,  \! [  \ottnt{v}  /  \mathit{p}  ]  \,  \! [  \ottnt{v_{\ottmv{cont}}}  /  \mathit{k}  ] 
    } \quad \rname{R}{Handle2'}
  \end{mathpar}
  \caption{Type-erasure semantics.}
  \label{fig:erasure-semantics}
\end{figure}

The part modified to support the type-erasure semantics is shown in Figure~\ref{fig:erasure-semantics}.
The label freeness in the type-erasure semantics refers only to label names,
while the original definition in Figure~\ref{fig:semantics} refers to entire
labels.
The only change in the semantics is that the reduction rule \rname{R}{Handle2}
is replaced by \rname{R}{Handle2'} presented in
Figure~\ref{fig:erasure-semantics}.
For instance, consider an expression
$ \mathbf{handle}_{ \mathsf{State} \,  \mathsf{Int}  }  \,  (   \mathbf{handle}_{ \mathsf{State} \,  \mathsf{Bool}  }  \,  (   \mathsf{set} _{ \mathsf{State} \, \ottnt{A} }  \, \ottnt{v}  )  \, \mathbf{with} \, \ottnt{h_{{\mathrm{1}}}}  )  \, \mathbf{with} \, \ottnt{h_{{\mathrm{2}}}}$.
%
In the original semantics, it depends on the type argument $\ottnt{A}$ which of $\ottnt{h_{{\mathrm{1}}}}$ and $\ottnt{h_{{\mathrm{2}}}}$ handles the operation call.
By contrast, in the type-erasure semantics, the handler $\ottnt{h_{{\mathrm{1}}}}$ will be chosen regardless of $\ottnt{A}$.
The type-and-effect system is not changed.

\subsection{Safety Conditions and Type-and-Effect Safety}
\label{subsec:erasure-safety-cond}

To ensure the safety in the type-erasure semantics, we need an additional safety condition.
\begin{definition}[Safety Condition for Type-Erasure]\label{def:safe_cond_erasure}
  The safety condition added for the type-erasure semantics is:
  \begin{enumerate*}
    \setcounter{enumi}{\value{safecondcounter}}
    \item\label{def:safe_cond_erasure:uniq}
          If $  \lift{ \mathit{l} \,  \bm{ { S_{{\mathrm{1}}} } } ^ {  \mathit{I}  }  }   \olessthan  \varepsilon $ and $   \lift{ \mathit{l} \,  \bm{ { S_{{\mathrm{2}}} } } ^ {  \mathit{I}  }  }   \mathop{ \odot }  \varepsilon'    \sim   \varepsilon $,
          then $ \bm{ { S_{{\mathrm{1}}} } } ^ {  \mathit{I}  }  =  \bm{ { S_{{\mathrm{2}}} } } ^ {  \mathit{I}  } $.
  \end{enumerate*}
\end{definition}
To understand this condition, assume that an operation of label name $\mathit{l}$ is called with typelike parameters $ \bm{ { S_{{\mathrm{1}}} } } ^ {  \mathit{I}  } $ and some effect $\varepsilon_{{\mathrm{1}}}$ such that $  \lift{ \mathit{l} \,  \bm{ { S_{{\mathrm{1}}} } } ^ {  \mathit{I}  }  }   \olessthan  \varepsilon $ is assigned to the operation call via subtyping. When the operation call is handled by an effect handler for effect label $\mathit{l} \,  \bm{ { S_{{\mathrm{2}}} } } ^ {  \mathit{I}  } $,
the typelike parameters $ \bm{ { S_{{\mathrm{1}}} } } ^ {  \mathit{I}  } $ for the operation call and $ \bm{ { S_{{\mathrm{2}}} } } ^ {  \mathit{I}  } $ for the handler must be matched.
None of the effect algebras {\eaSet}, {\eaSimpleRow}, {\eaScopedRow}, and {\eaMSet}
presented thus far meets this new condition, and, even worse, they can
accept some programs unsafe in the type-erasure semantics.

%
\begin{theorem}[Unsafe Effect Algebras in Type-Erasure Semantics]
  \label{thm:unsafe_instances_erasure}
 The effect algebras {\eaSet}, {\eaMSet}, {\eaSimpleRow}, and {\eaScopedRow} do not meet safety condition~\ref{def:safe_cond_erasure:uniq}.
 Furthermore, there exists an expression that is well typed under these algebras and gets stuck.
\end{theorem}
          %
\begin{proof}
          Here we focus on the effect algebra {\eaSet}, but a similar discussion can be applied to
          the other algebras.
          Recall that $ \odot $ in {\eaSet} is implemented by the union operation for sets, and therefore it is commutative
          (i.e., it allows exchanging labels in a set no matter what label names and what type arguments are in the labels).
          Hence, for example, $  \{  \mathit{l} \,  \mathsf{Int}   \}  \,\underline{ \cup }\,  \{  \mathit{l} \,  \mathsf{Bool}   \}     \sim_{\eanameSet}     \{  \mathit{l} \,  \mathsf{Bool}   \}  \,\underline{ \cup }\,  \{  \mathit{l} \,  \mathsf{Int}   \}  $
          for a label name $\mathit{l}$ taking one type parameter.
          It means that {\eaSet} violates safety condition~\ref{def:safe_cond_erasure:uniq}.
          %
          %
          %

          To give a program that is typeable under {\eaSet} but unsafe in the type-erasure semantics,
          consider the following which uses an effect label $ \mathsf{Writer}  ::    \forall   \alpha  \ottsym{:}   \mathbf{Typ}    \ottsym{.}    \ottsym{\{}  \mathsf{tell}  \ottsym{:}   \alpha   \Rightarrow    \mathsf{Unit}    \ottsym{\}}  $:
          \begin{flalign*}
            &  \mathbf{handle}_{ \mathsf{Writer} \,  \mathsf{Int}  }  \,   \mathbf{handle}_{ \mathsf{Writer} \,  \mathsf{Bool}  }  \,  & \\ &    \quad    \quad        \mathsf{tell} _{ \mathsf{Writer} \,  \mathsf{Int}  }  \,  {}  \, 42    & \\ &    \quad      \, \mathbf{with} \,  \ottsym{\{} \, \mathbf{return} \, \mathit{x}  \mapsto  0  \ottsym{\}}   \uplus   \ottsym{\{}  \mathsf{tell} \,  {}  \, \mathit{p} \, \mathit{k}  \mapsto  \mathbf{if} \, \mathit{p} \, \mathbf{then} \, 0 \, \mathbf{else} \, 42  \ottsym{\}}    & \\ &  \, \mathbf{with} \,  \ottsym{\{} \, \mathbf{return} \, \mathit{x}  \mapsto  \mathit{x}  \ottsym{\}}   \uplus   \ottsym{\{}  \mathsf{tell} \,  {}  \, \mathit{p} \, \mathit{k}  \mapsto  \mathit{p}  \ottsym{\}} 
          \end{flalign*}
          This program is well typed because
          \begin{itemize}
           \item the operation call $ \mathsf{tell} _{ \mathsf{Writer} \,  \mathsf{Int}  }  \,  {}  \, 42$ can have effect $ \{  \mathsf{Writer} \,  \mathsf{Bool}   \}  \,\underline{ \cup }\,  \{  \mathsf{Writer} \,  \mathsf{Int}   \} $ via subeffecting $ \{  \mathsf{Writer} \,  \mathsf{Int}   \}  \olessthan   \{  \mathsf{Writer} \,  \mathsf{Bool}   \}  \,\underline{ \cup }\,  \{  \mathsf{Writer} \,  \mathsf{Int}   \}  $ (which holds because $\mathsf{Writer} \,  \mathsf{Int} $ and $\mathsf{Writer} \,  \mathsf{Bool} $ are exchangeable),
           \item the inner handling expression is well typed and its effect is $\{  \mathsf{Writer} \,  \mathsf{Int}   \}$, and
           \item the outer one is well typed and its effect is $\{  \}$.
          \end{itemize}
          Note that this typing rests on the fact that the inner handler assumes that the argument variable $\mathit{p}$ of its $\mathsf{tell}$ clause will be replaced by Boolean values as indicated by the type argument $ \mathsf{Bool} $ to $\mathsf{Writer}$.
          However, the variable $\mathit{p}$ will be replaced by integer $42$ and the program will get stuck.
\end{proof}

The proof of Theorem~\ref{thm:unsafe_instances_erasure} relies on the commutativity of $ \odot $ in each effect algebra.
This observation indicates that an effect algebra with \emph{non}commutative
$ \odot $ can be safe even in the type-erasure semantics.
In fact, the previous work~\cite{leijen_type_2017,leijen2018algebraic,biernacki_abstracting_2019} has given an instance of such an effect algebra.
By following it, we can adapt the effect algebras defined thus far to be safe in the type-erasure semantics;
we call the effect collections in such effect algebras \emph{erasable}.
\begin{example}[Erasable Effect Algebras]\label{exa:eff_simple_row_erasure}
  An effect algebra {\eaEraseSet} for erasable sets is defined similarly to {\eaSet}.
  The only difference is that the equivalence relation $ \sim_\mathrm{ESet} $ of {\eaEraseSet} is defined as
  $ \sim_{\eanameSet} $, but the commutativity rule used in the definition of $ \sim_{\eanameSet} $ is replaced with
  \begin{mathpar}
    \inferrule{
     \mathit{l_{{\mathrm{1}}}}   \neq   \mathit{l_{{\mathrm{2}}}} 
    }{
      \{  \mathit{l_{{\mathrm{1}}}} \,  \bm{ { S_{{\mathrm{1}}} } } ^ {  \mathit{I_{{\mathrm{1}}}}  }   \}  \,\underline{ \cup }\,  \{  \mathit{l_{{\mathrm{2}}}} \,  \bm{ { S_{{\mathrm{2}}} } } ^ {  \mathit{I_{{\mathrm{2}}}}  }   \}     \sim_\mathrm{ESet}     \{  \mathit{l_{{\mathrm{2}}}} \,  \bm{ { S_{{\mathrm{2}}} } } ^ {  \mathit{I_{{\mathrm{2}}}}  }   \}  \,\underline{ \cup }\,  \{  \mathit{l_{{\mathrm{1}}}} \,  \bm{ { S_{{\mathrm{1}}} } } ^ {  \mathit{I_{{\mathrm{1}}}}  }   \}  
    }
  \end{mathpar}
  which only allows exchanging labels with different names.
  %
  Effect algebras {\eaEraseSet}, {\eaEraseMSet}, and {\eaEraseScopedRow} for erasable sets, multisets, and scoped rows, respectively, are defined similarly.
\end{example}

\begin{theorem}
  The effect algebras {\eaEraseSet}, {\eaEraseMSet}, {\eaEraseSimpleRow}, and {\eaEraseScopedRow} meet safety conditions~\ref{def:safe_cond:label_notemp}, \ref{def:safe_cond:pres}, and \ref{def:safe_cond_erasure:uniq}.
\end{theorem}
Note that some equivalence properties holding on nonerasable effect collections do not hold on erasable ones.
For instance, $  \{  \mathsf{Writer} \,  \mathsf{Int}   \}  \,\underline{ \cup }\,  \{  \mathsf{Writer} \,  \mathsf{Bool}   \}    \sim    \{  \mathsf{Writer} \,  \mathsf{Bool}   \}  \,\underline{ \cup }\,  \{  \mathsf{Writer} \,  \mathsf{Int}   \}  $ and $  \rho_{{\mathrm{1}}}  \,\underline{ \cup }\,  \rho_{{\mathrm{2}}}    \sim    \rho_{{\mathrm{2}}}  \,\underline{ \cup }\,  \rho_{{\mathrm{1}}}  $ do not hold in erasable sets.
The latter equivalence is not allowed because $\rho_{{\mathrm{1}}}$ and $\rho_{{\mathrm{2}}}$ may be replaced with, e.g., $\{  \mathsf{Writer} \,  \mathsf{Int}   \}$ and $\{  \mathsf{Writer} \,  \mathsf{Bool}   \}$, respectively.
This limitation could be relaxed by supporting qualified
types~\cite{jones_theory_1992}.
%
\TS{Have we discussed?}
%

Finally, we can prove the type-and-effect safety of {\lang} with the
type-erasure semantics as Theorem~\ref{thm:type-and-eff-safe} provided that an
effect algebra meets safety conditions~\ref{def:safe_cond:label_notemp},
\ref{def:safe_cond:pres}, and \ref{def:safe_cond_erasure:uniq}.
\begin{theorem}[Type-and-Effect Safety]
  \label{thm:type-and-eff-safe-erasure}
  Assume that a given effect algebra meets safety conditions~\ref{def:safe_cond:label_notemp}, \ref{def:safe_cond:pres}, and \ref{def:safe_cond_erasure:uniq}.
  If $\emptyset  \vdash  \ottnt{e}  \ottsym{:}  \ottnt{A}  \mid   \bbZero $ and $\ottnt{e}  \longrightarrow  ^ * \ottnt{e'} \centernot \longrightarrow $, then $\ottnt{e'} = \ottnt{v}$ for some $\ottnt{v}$.
\end{theorem}

\OLD{

\subsection{Safety Condition for Type-Erasure}
\label{subsec:erasure-safety-cond}

To deal with the type-erasure semantics, we need to add a new condition to safety conditions mentioned in Section~\ref{sec:safety}.
The new safety condition is defined as follows.
\begin{definition}[Safety Condition for Type-Erasure]\label{def:safe_cond_erasure}
  The safety condition added for the type-erasure semantics is the following:
  \begin{enumerate}
    \setcounter{enumi}{\value{safecondcounter}}
    \item\label{def:safe_cond_erasure:uniq}
          If $  \lift{ \mathit{l} \,  \bm{ { S_{{\mathrm{1}}} } } ^ {  \mathit{I_{{\mathrm{1}}}}  }  }   \olessthan  \varepsilon $ and $  \lift{ \mathit{l} \,  \bm{ { S_{{\mathrm{2}}} } } ^ {  \mathit{I_{{\mathrm{2}}}}  }  }   \olessthan  \varepsilon $,
          then $ \bm{ { S_{{\mathrm{1}}} } } ^ {  \mathit{I_{{\mathrm{1}}}}  }  =  \bm{ { S_{{\mathrm{2}}} } } ^ {  \mathit{I_{{\mathrm{2}}}}  } $.
  \end{enumerate}
\end{definition}
%
\TS{Add the intuitive explanation of the above new condition and the example below.}
This new condition means that, if two handlers are installed for the same handled expression with effect $\varepsilon$, the typelike parameters $ \bm{ { S_{{\mathrm{1}}} } } ^ {  \mathit{I_{{\mathrm{1}}}}  } $ and $ \bm{ { S_{{\mathrm{2}}} } } ^ {  \mathit{I_{{\mathrm{2}}}}  } $ of the effect label $\mathit{l}$ they handle must be the same.

To understand the importance of the new safety condition,
we consider effect algebras that violate this new condition, and then show what unsafe program becomes typeable.
\TY{This example has minor changes.}
\begin{example}[An Unsafe Effect Algebra with Type-Erasure Semantics]
  \label{exa:unsafe_instances_erasure}
  \phantom{}
  \begin{description}
    \item[Effect algebra violating type-erasure safety condition~\ref{def:safe_cond_erasure:uniq}]
          %
          Recall Example~\ref{exa:effset}.
          Because they allow exchanging labels with the same label name as indicated by
          $  \{  \mathit{l} \,  \bm{ { S_{{\mathrm{1}}} } } ^ {  \mathit{I_{{\mathrm{1}}}}  }   \}  \,\underline{ \cup }\,  \{  \mathit{l} \,  \bm{ { S_{{\mathrm{2}}} } } ^ {  \mathit{I_{{\mathrm{2}}}}  }   \}    \sim    \{  \mathit{l} \,  \bm{ { S_{{\mathrm{2}}} } } ^ {  \mathit{I_{{\mathrm{2}}}}  }   \}  \,\underline{ \cup }\,  \{  \mathit{l} \,  \bm{ { S_{{\mathrm{1}}} } } ^ {  \mathit{I_{{\mathrm{1}}}}  }   \}  $,
          this effect algebra violate safety condition~\ref{def:safe_cond_erasure:uniq}.
          %
          %
          %
          In this case, 
          the type $ \mathsf{Int} $ and the effect $ \bbZero $ can be assigned to
          \begin{flalign*}
            \hphantom{\text{Exampl}} &  \mathbf{handle}_{ \mathsf{Writer} \,  \mathsf{Int}  }  \,   \mathbf{handle}_{ \mathsf{Writer} \,  \mathsf{String}  }  \,  & \\ &    \quad    \quad       \mathbf{let} \, \_  \ottsym{=}   \mathsf{tell} _{ \mathsf{Writer} \,  \mathsf{Int}  }  \,  {}  \, 42 \, \mathbf{in} \, \ottnt{e}    & \\ &    \quad      \, \mathbf{with} \,  \ottsym{\{} \, \mathbf{return} \, \_  \mapsto  0  \ottsym{\}}   \uplus   \ottsym{\{}  \mathsf{tell} \,  {}  \, \mathit{p} \, \mathit{k}  \mapsto   \textnormal{\ttfamily length}  \, \mathit{p}  \ottsym{\}}    & \\ &  \, \mathbf{with} \,  \ottsym{\{} \, \mathbf{return} \, \mathit{x}  \mapsto  \mathit{x}  \ottsym{\}}   \uplus   \ottsym{\{}  \mathsf{tell} \,  {}  \, \mathit{p} \, \mathit{k}  \mapsto  \mathit{p}  \ottsym{\}} 
          \end{flalign*}
          where $ \mathsf{Writer}  ::    \forall   \alpha  \ottsym{:}   \mathbf{Typ}    \ottsym{.}    \ottsym{\{}  \mathsf{tell}  \ottsym{:}   \alpha   \Rightarrow    \mathsf{Unit}    \ottsym{\}}  $,
          because
          $ \mathsf{tell} _{ \mathsf{Writer} \,  \mathsf{Int}  }  \,  {}  \, 42$ can have the effect
          $ \{  \mathsf{Writer} \,  \mathsf{String}   \}  \,\underline{ \cup }\,  \{  \mathsf{Writer} \,  \mathsf{Int}   \} $
          due to
          $  \{  \mathsf{Writer} \,  \mathsf{Int}   \}  \,\underline{ \cup }\,  \{  \mathsf{Writer} \,  \mathsf{String}   \}    \sim    \{  \mathsf{Writer} \,  \mathsf{String}   \}  \,\underline{ \cup }\,  \{  \mathsf{Writer} \,  \mathsf{Int}   \}  $.
          %
          However, in the type-erasure semantics, this program evaluates to the stuck term:
          \begin{flalign*}
            \hphantom{\text{Exampl}} &  \mathbf{handle}_{ \mathsf{Writer} \,  \mathsf{Int}  }  \,  \textnormal{\ttfamily length}  \, 42 \, \mathbf{with} \,  \ottsym{\{} \, \mathbf{return} \, \mathit{x}  \mapsto  \mathit{x}  \ottsym{\}}   \uplus   \ottsym{\{}  \mathsf{tell} \,  {}  \, \mathit{p} \, \mathit{k}  \mapsto  \mathit{p}  \ottsym{\}}  ~. &
          \end{flalign*}
          The similar discussions can be applied to
          Example~\ref{exa:eff_simple_row},
          Example~\ref{exa:effrow}, and Example~\ref{exa:effmultiset}.
  \end{description}
\end{example}

\subsection{Type-and-Effect Safety}
We show type and effect safety under the type-erasure semantics by assuming the type-erasure safety conditions on effect algebras.
Neither the statement nor the proof of effect safety changes.
The statement of type safety changes so that reflecting the change of definition of freeness.
%
%
\begin{lemma}[Type Safety]\label{lem:typsafe_erasure}
  If $\emptyset  \vdash  \ottnt{e}  \ottsym{:}  \ottnt{A}  \mid  \varepsilon$ and $\ottnt{e}  \longrightarrow  ^ * \ottnt{e'} \centernot \longrightarrow $, then one of the following holds:
  \begin{itemize}
    \item $\ottnt{e'} = \ottnt{v}$ for some value $\ottnt{v}$ such that $\emptyset  \vdash  \ottnt{v}  \ottsym{:}  \ottnt{A}  \mid  \varepsilon$; or
    \item $\ottnt{e'} = \ottnt{E}  \ottsym{[}   \mathsf{op} _{ \mathit{l} \,  \bm{ { S } } ^ {  \mathit{I}  }  }  \,  \bm{ { T } } ^ {  \mathit{J}  }  \, \ottnt{v}  \ottsym{]}$ and $ \mathit{n}  \mathrm{-free} ( \mathit{l} ,  \ottnt{E} ) $
          for some $\ottnt{E}$, $\mathit{l}$, $ \bm{ { S } } ^ {  \mathit{I}  } $, $\mathsf{op}$, $ \bm{ { T } } ^ {  \mathit{J}  } $, $\ottnt{v}$, and $\mathit{n}$ such that
          operation $\mathsf{op}$ belongs to effect label name $\mathit{l}$ in effect context $\Xi$.

  \end{itemize}
\end{lemma}
We can prove this lemma in a way similar to the type safety in Section~\ref{sec:safety}.
The only exception is in the case of proving that the reduction by \rname{R}{Handle2'}
preserves typing.
Consider expression $ \mathbf{handle}_{ \mathit{l} \,  \bm{ { S } } ^ {  \mathit{I}  }  }  \, \ottnt{E}  \ottsym{[}   \mathsf{op} _{ \mathit{l} \,  \bm{ { S' } } ^ {  \mathit{I}  }  }  \,  \bm{ { T } }  \, \ottnt{v}  \ottsym{]} \, \mathbf{with} \, \ottnt{h}$ to be reduced
by \rname{R}{Handle2'}.
In this case, typelike arguments $ \bm{ { S } } ^ {  \mathit{I}  } $ of label name $\mathit{l}$ have to
be the same as typelike arguments $ \bm{ { S' } } ^ {  \mathit{I}  } $ of operation $\mathsf{op}$.
Safety condition~\ref{def:safe_cond_erasure:uniq} ensures it.
Note that it seems that this equation cannot be derived only by the (non-type-erasure)
safety conditions.



Using the type and effect safety lemmas, we can prove type-and-effect safety for
the type-erasure semantics as a corollary.
\begin{theorem}[Type-and-Effect Safety]
  \label{thm:type-and-eff-safe-erasure}
  If $\emptyset  \vdash  \ottnt{e}  \ottsym{:}  \ottnt{A}  \mid   \bbZero $ and $\ottnt{e}  \longrightarrow  ^ * \ottnt{e'} \centernot \longrightarrow $, then $\ottnt{e'} = \ottnt{v}$ for some $\ottnt{v}$.
\end{theorem}

\subsection{Type-Erasable Instances}
As shown in Section~\ref{subsec:erasure-safety-cond},
the effect algebras presented thus far do not satisfy type-erasure safety condition~\ref{def:safe_cond_erasure:uniq}.
In this section, we first show a variant of each of simple and scoped rows that satisfies the type-erasure safety conditions.

\begin{example}[Erasable Simple Rows]\label{exa:eff_simple_row_erasure}
  Erasable simple rows are defined similarly to simple rows shown in Example~\ref{exa:eff_simple_row}.
  Only the difference is in the definition of equivalence $ \sim $.
  The equivalence $ \sim $ for erasable simple rows is defined as the least equivalence relation satisfying the following rules:
  \begin{mathpar}
    \inferrule{ \varepsilon_{{\mathrm{1}}}   \sim   \varepsilon_{{\mathrm{2}}} }{ \langle  \ottnt{L}  \mid  \varepsilon_{{\mathrm{1}}}  \rangle   \sim   \langle  \ottnt{L}  \mid  \varepsilon_{{\mathrm{2}}}  \rangle }

    \inferrule{ \mathit{l_{{\mathrm{1}}}}   \neq   \mathit{l_{{\mathrm{2}}}} }{ \langle  \mathit{l_{{\mathrm{1}}}} \,  \bm{ { S_{{\mathrm{1}}} } } ^ {  \mathit{I_{{\mathrm{1}}}}  }   \mid  \langle  \mathit{l_{{\mathrm{2}}}} \,  \bm{ { S_{{\mathrm{2}}} } } ^ {  \mathit{I_{{\mathrm{2}}}}  }   \mid  \varepsilon  \rangle  \rangle   \sim   \langle  \mathit{l_{{\mathrm{2}}}} \,  \bm{ { S_{{\mathrm{2}}} } } ^ {  \mathit{I_{{\mathrm{2}}}}  }   \mid  \langle  \mathit{l_{{\mathrm{1}}}} \,  \bm{ { S_{{\mathrm{1}}} } } ^ {  \mathit{I_{{\mathrm{1}}}}  }   \mid  \varepsilon  \rangle  \rangle }

    \inferrule{ }{ \langle  \mathit{l} \,  \bm{ { S_{{\mathrm{1}}} } } ^ {  \mathit{I_{{\mathrm{1}}}}  }   \mid  \langle  \mathit{l} \,  \bm{ { S_{{\mathrm{2}}} } } ^ {  \mathit{I_{{\mathrm{2}}}}  }   \mid  \varepsilon  \rangle  \rangle   \sim   \langle  \mathit{l} \,  \bm{ { S_{{\mathrm{1}}} } } ^ {  \mathit{I_{{\mathrm{1}}}}  }   \mid  \varepsilon  \rangle }

    \inferrule{ }{ \langle  \mathit{l} \,  \bm{ { S_{{\mathrm{1}}} } } ^ {  \mathit{I_{{\mathrm{1}}}}  }   \mid  \varepsilon  \rangle   \sim   \langle  \mathit{l} \,  \bm{ { S_{{\mathrm{1}}} } } ^ {  \mathit{I_{{\mathrm{1}}}}  }   \mid  \langle  \mathit{l} \,  \bm{ { S_{{\mathrm{2}}} } } ^ {  \mathit{I_{{\mathrm{2}}}}  }   \mid  \varepsilon  \rangle  \rangle }
  \end{mathpar}
  The top-right rule means that, unlike in simple rows, we can only exchange
  labels with different label names in erasable simple rows.
  By the last two rules, we can collapse the occurrences of $\mathit{l} \,  \bm{ { S } } ^ {  \mathit{I}  } $ and $\mathit{l} \,  \bm{ { S' } } ^ {  \mathit{I'}  } $
  even if $  \bm{ { S } } ^ {  \mathit{I}  }    \neq    \bm{ { S' } } ^ {  \mathit{I'}  }  $.
\end{example}
\begin{example}[Erasable Scoped Rows]\label{exa:effrow_erasure}
  Erasable scoped rows are defined similarly to scoped rows shown in Example~\ref{exa:effrow}.
  Only the difference is in the definition of equivalence $ \sim $.
  The equivalence $ \sim $ for erasable scoped rows is defined as the least equivalence relation satisfying the following rules:
  \begin{mathpar}
    \inferrule{ \varepsilon_{{\mathrm{1}}}   \sim   \varepsilon_{{\mathrm{2}}} }{ \langle  \ottnt{L}  \mid  \varepsilon_{{\mathrm{1}}}  \rangle   \sim   \langle  \ottnt{L}  \mid  \varepsilon_{{\mathrm{2}}}  \rangle }

    \inferrule{ \mathit{l_{{\mathrm{1}}}}   \neq   \mathit{l_{{\mathrm{2}}}} }{ \langle  \mathit{l_{{\mathrm{1}}}} \,  \bm{ { S_{{\mathrm{1}}} } } ^ {  \mathit{I_{{\mathrm{1}}}}  }   \mid  \langle  \mathit{l_{{\mathrm{2}}}} \,  \bm{ { S_{{\mathrm{2}}} } } ^ {  \mathit{I_{{\mathrm{2}}}}  }   \mid  \varepsilon  \rangle  \rangle   \sim   \langle  \mathit{l_{{\mathrm{2}}}} \,  \bm{ { S_{{\mathrm{2}}} } } ^ {  \mathit{I_{{\mathrm{2}}}}  }   \mid  \langle  \mathit{l_{{\mathrm{1}}}} \,  \bm{ { S_{{\mathrm{1}}} } } ^ {  \mathit{I_{{\mathrm{1}}}}  }   \mid  \varepsilon  \rangle  \rangle }
  \end{mathpar}
  Unlike in scoped rows---and like erasable simple rows---we can only exchange
  labels with different label names in erasable scoped rows.
\end{example}

As shown in Example~\ref{exa:unsafe_instances_erasure},
the commutativity prevents sound extension to type-erasure semantics.
Therefore, we can give sound instances based on set-style and multiset-style effect representations by restricting their commutativity.
\TY{The name ``erasable sets'' and ``erasable multisets'' are tentative (and should be changed.)}
\begin{example}[Erasable Sets]\label{exa:effset_erasure}
  Erasable sets are defined similarly to sets shown in Example~\ref{exa:effset}.
  Only the difference is in the definition of equivalence $ \sim $.
  The equivalence $ \sim $ for erasable sets is defined as the least equivalence relation satisfying the following rules:
  \begin{mathpar}
    \inferrule{
     \mathit{l_{{\mathrm{1}}}}   \neq   \mathit{l_{{\mathrm{2}}}} 
    }{
      \{  \mathit{l_{{\mathrm{1}}}} \,  \bm{ { S_{{\mathrm{1}}} } } ^ {  \mathit{I_{{\mathrm{1}}}}  }   \}  \,\underline{ \cup }\,  \{  \mathit{l_{{\mathrm{2}}}} \,  \bm{ { S_{{\mathrm{2}}} } } ^ {  \mathit{I_{{\mathrm{2}}}}  }   \}    \sim    \{  \mathit{l_{{\mathrm{1}}}} \,  \bm{ { S_{{\mathrm{1}}} } } ^ {  \mathit{I_{{\mathrm{1}}}}  }   \}  \,\underline{ \cup }\,  \{  \mathit{l_{{\mathrm{2}}}} \,  \bm{ { S_{{\mathrm{2}}} } } ^ {  \mathit{I_{{\mathrm{2}}}}  }   \}  
    }

    \inferrule{ }{  \varepsilon  \,\underline{ \cup }\,  \{  \}    \sim   \varepsilon }

    \inferrule{ }{  \{  \}  \,\underline{ \cup }\,  \varepsilon    \sim   \varepsilon }

    \inferrule{ }{  \ottsym{(}   \varepsilon_{{\mathrm{1}}}  \,\underline{ \cup }\,  \varepsilon_{{\mathrm{2}}}   \ottsym{)}  \,\underline{ \cup }\,  \varepsilon_{{\mathrm{3}}}    \sim    \varepsilon_{{\mathrm{1}}}  \,\underline{ \cup }\,  \ottsym{(}   \varepsilon_{{\mathrm{2}}}  \,\underline{ \cup }\,  \varepsilon_{{\mathrm{3}}}   \ottsym{)}  }

    \inferrule{
     \varepsilon_{{\mathrm{1}}}   \sim   \varepsilon_{{\mathrm{2}}}  \\  \varepsilon_{{\mathrm{3}}}   \sim   \varepsilon_{{\mathrm{4}}} 
    }{
      \varepsilon_{{\mathrm{1}}}  \,\underline{ \cup }\,  \varepsilon_{{\mathrm{3}}}    \sim    \varepsilon_{{\mathrm{2}}}  \,\underline{ \cup }\,  \varepsilon_{{\mathrm{4}}}  
    }
  \end{mathpar}
  The top-left rule means that, unlike in sets, we can only exchange
  single label effects with different label names in erasable sets.
  Because of restricting the commutativity of sets,
  we add a new rule to ensure that $\{  \}$ acts as the unit.
\end{example}
\begin{example}[Erasable Multisets]\label{exa:effmultiset_erasure}
  Erasable multisets are defined similarly to multisets shown in Example~\ref{exa:effmultiset}.
  Only the difference is in the definition of equivalence $ \sim $.
  The equivalence $ \sim $ for erasable multisets is defined as the least equivalence relation satisfying the following rules:
  \begin{mathpar}
    \inferrule{
     \mathit{l_{{\mathrm{1}}}}   \neq   \mathit{l_{{\mathrm{2}}}} 
    }{
      \{  \mathit{l_{{\mathrm{1}}}} \,  \bm{ { S_{{\mathrm{1}}} } } ^ {  \mathit{I_{{\mathrm{1}}}}  }   \}  \,\underline{ \sqcup }\,  \{  \mathit{l_{{\mathrm{2}}}} \,  \bm{ { S_{{\mathrm{2}}} } } ^ {  \mathit{I_{{\mathrm{2}}}}  }   \}    \sim    \{  \mathit{l_{{\mathrm{1}}}} \,  \bm{ { S_{{\mathrm{1}}} } } ^ {  \mathit{I_{{\mathrm{1}}}}  }   \}  \,\underline{ \sqcup }\,  \{  \mathit{l_{{\mathrm{2}}}} \,  \bm{ { S_{{\mathrm{2}}} } } ^ {  \mathit{I_{{\mathrm{2}}}}  }   \}  
    }

    \inferrule{ }{  \varepsilon  \,\underline{ \sqcup }\,  \{  \}    \sim   \varepsilon }

    \inferrule{ }{  \{  \}  \,\underline{ \sqcup }\,  \varepsilon    \sim   \varepsilon }

    \inferrule{ }{  \ottsym{(}   \varepsilon_{{\mathrm{1}}}  \,\underline{ \sqcup }\,  \varepsilon_{{\mathrm{2}}}   \ottsym{)}  \,\underline{ \sqcup }\,  \varepsilon_{{\mathrm{3}}}    \sim    \varepsilon_{{\mathrm{1}}}  \,\underline{ \sqcup }\,  \ottsym{(}   \varepsilon_{{\mathrm{2}}}  \,\underline{ \sqcup }\,  \varepsilon_{{\mathrm{3}}}   \ottsym{)}  }

    \inferrule{
     \varepsilon_{{\mathrm{1}}}   \sim   \varepsilon_{{\mathrm{2}}}  \\  \varepsilon_{{\mathrm{3}}}   \sim   \varepsilon_{{\mathrm{4}}} 
    }{
      \varepsilon_{{\mathrm{1}}}  \,\underline{ \sqcup }\,  \varepsilon_{{\mathrm{3}}}    \sim    \varepsilon_{{\mathrm{2}}}  \,\underline{ \sqcup }\,  \varepsilon_{{\mathrm{4}}}  
    }
  \end{mathpar}
  The change of rules is similar to Example~\ref{exa:effset_erasure}.
\end{example}

The effect systems with these instances enjoy the type-and-effect safety because of the following theorem.
\begin{theorem}
  Example~\ref{exa:eff_simple_row_erasure},
  \ref{exa:effrow_erasure}, \ref{exa:effset_erasure}, and \ref{exa:effmultiset_erasure}
  meet safety conditions (for type-erasure semantics).
\end{theorem}

}

\OLD{
\section{Type-Erasure Semantics}\label{sec:erasure}

This section shows how to adapt {\lang} to type-erasure semantics. Type-erasure semantics is helpful for some implementations. 

In preparation for defining type-erasure semantics, we define another evaluation context as follows:
\begin{align*}
  \texttt{\textcolor{red}{<<no parses (char 4): E(l)*** >>}} \Coloneqq  \Box   \mid  \texttt{\textcolor{red}{<<no parses (char 17): let x = E(l) in e*** >>}}  \mid  \texttt{\textcolor{red}{<<no parses (char 29): handle l' \{U' I'\} E(l) with h***  >>}} \quad (\mathit{l} \neq \mathit{l'})
\end{align*}
This evaluation context is parameterized by a label name $\mathit{l}$ and omits handlers handling operations of labels $\mathit{l} \,  \bm{ { S } } ^ {  \mathit{I}  } $ for any $ \bm{ { S } } ^ {  \mathit{I}  } $. Using this evaluation context, we define type-erasure semantics in Fig~\ref{fig:erasure}.
\begin{figure}[t]
  \begin{mathpar}
    \inferrule
    { \mathit{l}  ::    \forall    {\bm{ \alpha } }^{ \mathit{I} } : {\bm{ \ottnt{K} } }^{ \mathit{I} }    \ottsym{.}    \sigma    \in   \Xi  \\  \mathsf{op}  \ottsym{:}    \forall    {\bm{ \beta } }^{ \mathit{J} } : {\bm{ \ottnt{K} } }^{ \mathit{J} }    \ottsym{.}    \ottnt{A}   \Rightarrow   \ottnt{B}    \in   \sigma \,  \! [ {\bm{ { S } } }^{ \mathit{I} } / {\bm{ \alpha } }^{ \mathit{I} } ]   \\  \mathsf{op} \,  {\bm{ \beta } }^{ \mathit{J} } : {\bm{ \ottnt{K} } }^{ \mathit{J} }  \, \mathit{p} \, \mathit{k}  \mapsto  \ottnt{e}   \in   \ottnt{h}  \\\\ \ottnt{v_{\ottmv{cont}}} = \texttt{\textcolor{red}{<<no parses (char 51): \mbox{$\backslash{}$}z : B[\{V/beta J\}] . handle l \{U I\} E(l) [z] with h***  >>}}}
    {\texttt{\textcolor{red}{<<no parses (char 46): handle l \{U I\} E(l) [op \{l \{U' I\} \} \{V J\} v] w***ith h \mbox{$\mid$}--> e [\{V/beta J\}] [v/p] [vcont / k]  >>}}} \quad \rname{R}{Handle2'}
  \end{mathpar}
  \caption{Type-erasure semantics of {\lang}.}
  \label{fig:erasure}
\end{figure}
We omit other rules because they are the same as the ones in Fig~\ref{fig:semantics} except for \rname{R}{Handle2}. In this section, we use this operational semantics.

\subsection{Type-Erasure Safety Conditions}
To deal with type-erasure semantics, we need to change safety conditions mentioned in Section~\ref{sec:safety}. New safety conditions, which we call type-erasure safety conditions, are defined as follows.
\begin{definition}[Type-erasure Safety Conditions]\label{def:safe_cond_erasure}
  Type-erasure safety conditions consist of safety conditions and the following one.
  \begin{enumerate}[start=10]
    \item\label{def:safe_cond_erasure:uniq} $\forall  \Delta , \mathit{l},  \bm{ { S_{{\mathrm{1}}} } } ^ {  \mathit{I_{{\mathrm{1}}}}  } ,  \bm{ { S_{{\mathrm{2}}} } } ^ {  \mathit{I_{{\mathrm{2}}}}  } , \varepsilon . ((\Delta  \vdash    \lift{ \mathit{l} \,  \bm{ { S_{{\mathrm{1}}} } } ^ {  \mathit{I_{{\mathrm{1}}}}  }  }   \olessthan  \varepsilon  \tand \Delta  \vdash    \lift{ \mathit{l} \,  \bm{ { S_{{\mathrm{2}}} } } ^ {  \mathit{I_{{\mathrm{2}}}}  }  }   \olessthan  \varepsilon ) \imply  \bm{ { S_{{\mathrm{1}}} } } ^ {  \mathit{I_{{\mathrm{1}}}}  }  =  \bm{ { S_{{\mathrm{2}}} } } ^ {  \mathit{I_{{\mathrm{2}}}}  } )$.
  \end{enumerate}
\end{definition}
Type-erasure safety conditions are simply safety conditions plus the condition \ref{def:safe_cond_erasure:uniq}.
If ARE meets Definition~\ref{def:safe_cond_erasure}, we can prove type and effect safety with type-erasure semantics.

We show bad examples caused by lacking type-erasure safety conditions: effect systems defined in Example~\ref{exa:effset}, Example~\ref{exa:eff_simple_row}, and Example~\ref{exa:effrow}.
These instances meet safety conditions but violate condition \ref{def:safe_cond_erasure:uniq}.
They accept the following program as a safe one.
\begin{flalign*}
   &  \mathbf{handle}_{ \mathsf{IO} \,  \mathsf{Int}  }  \,   \mathbf{handle}_{ \mathsf{IO} \,  \mathsf{String}  }  \,  & \\ &    \quad    \quad       \mathbf{let} \, \_  \ottsym{=}   \mathsf{print} _{ \mathsf{IO} \,  \mathsf{Int}  }  \,  {}  \, 42 \, \mathbf{in} \,  \mathsf{print} _{ \mathsf{IO} \,  \mathsf{String}  }  \,  {}  \,  \textnormal{\texttt{"foo"} }     & \\ &    \quad      \, \mathbf{with} \,  \ottsym{\{} \, \mathbf{return} \, \_  \mapsto  0  \ottsym{\}}   \uplus   \ottsym{\{}  \mathsf{print} \,  {}  \, \mathit{p} \, \mathit{k}  \mapsto   \textnormal{\ttfamily length}  \, \mathit{p}  \ottsym{\}}    & \\ &  \, \mathbf{with} \,  \ottsym{\{} \, \mathbf{return} \, \mathit{x}  \mapsto  \mathit{x}  \ottsym{\}}   \uplus   \ottsym{\{}  \mathsf{print} \,  {}  \, \mathit{p} \, \mathit{k}  \mapsto  \mathit{p}  \ottsym{\}} 
\end{flalign*}
In normal operational semantics, that is, operational semantics non-erasing types, this expression evaluates to $42$. However, in type-erasure semantics, this expression evaluates to the stuck term:
\begin{flalign*}
   &  \mathbf{handle}_{ \mathsf{IO} \,  \mathsf{Int}  }  \,  \textnormal{\ttfamily length}  \, 42 \, \mathbf{with} \,  \ottsym{\{} \, \mathbf{return} \, \mathit{x}  \mapsto  \mathit{x}  \ottsym{\}}   \uplus   \ottsym{\{}  \mathsf{print} \,  {}  \, \mathit{p} \, \mathit{k}  \mapsto  \mathit{p}  \ottsym{\}} 
\end{flalign*}

\subsection{Changes of Type Safety and Effect Safety}
We show type safety and effect safety with type-erasure semantics. Effect safety has no change in both statement and proof. The statement of type safety is changed as follows.
\begin{lemma}[Type Safety]\label{lem:typsafe_erasure}
  If $\emptyset  \vdash  \ottnt{e}  \ottsym{:}  \ottnt{A}  \mid  \varepsilon$ and $\ottnt{e}  \longrightarrow  ^ * \ottnt{e'} \centernot \longrightarrow $, then one of the following holds:
  \begin{itemize}
    \item There is $\ottnt{v}$ such that $\ottnt{e'} = \ottnt{v}$
    \item There are $\mathit{l}$, $\mathsf{op}$, $\ottnt{A'}$, $\ottnt{B'}$, $\ottnt{v}$, and $\texttt{\textcolor{red}{<<no parses (char 4): E(l)*** >>}}$ such that satisfy the following conditions:
          \begin{itemize}
            \item $ \mathit{l}  ::    \forall    \bm{ \alpha } : \bm{ \ottnt{K} }    \ottsym{.}    \sigma    \in   \Xi $
            \item $ \mathsf{op}  \ottsym{:}    \forall    \bm{ \beta } : \bm{ \ottnt{K'} }    \ottsym{.}    \ottnt{A'}   \Rightarrow   \ottnt{B'}    \in   \sigma $
            \item $\ottnt{e} = \texttt{\textcolor{red}{<<no parses (char 23):  E(l)[op \{l \{U\}\} \{V\} v]***  >>}}$.
          \end{itemize}
  \end{itemize}
\end{lemma}
We can prove this lemma likely as in Section~\ref{sec:safety}.

The progress lemma is changed as follows.
\begin{lemma}[Progress]\label{lem:progress_erasure}
  If $\emptyset  \vdash  \ottnt{e}  \ottsym{:}  \ottnt{A}  \mid  \varepsilon$, then one of the following holds:
  \begin{itemize}
    \item $\ottnt{e}$ is a value,
    \item There are $\ottnt{e'}$ such that $\ottnt{e}  \longrightarrow  \ottnt{e'}$,
    \item There are $\mathit{l}$, $\mathsf{op}$, $\ottnt{A'}$, $\ottnt{B'}$, $\ottnt{v}$, and $\texttt{\textcolor{red}{<<no parses (char 4): E(l)*** >>}}$ such that satisfy the following conditions:
          \begin{itemize}
            \item $ \mathit{l}  ::    \forall    \bm{ \alpha } : \bm{ \ottnt{K} }    \ottsym{.}    \sigma    \in   \Xi $
            \item $ \mathsf{op}  \ottsym{:}    \forall    \bm{ \beta } : \bm{ \ottnt{K'} }    \ottsym{.}    \ottnt{A'}   \Rightarrow   \ottnt{B'}    \in   \sigma $
            \item $\ottnt{e} = \texttt{\textcolor{red}{<<no parses (char 23):  E(l)[op \{l \{U\}\} \{V\} v]***  >>}}$.
          \end{itemize}
  \end{itemize}
\end{lemma}
The proof of this is essentially not changed.

The preservation lemma has no change in its statement but needs change for its proof.
\begin{lemma}[Preservation]\label{lem:preservation_erasure}
  If $\emptyset  \vdash  \ottnt{e}  \ottsym{:}  \ottnt{A}  \mid  \varepsilon$ and $\ottnt{e}  \longrightarrow  \ottnt{e'}$, then $\emptyset  \vdash  \ottnt{e'}  \ottsym{:}  \ottnt{A}  \mid  \varepsilon$.
\end{lemma}
The critical case is when the reduction is \rname{R}{Handle2'}. If $\ottnt{e} = \texttt{\textcolor{red}{<<no parses (char 42): handle l \{U I\} E(l)[op \{l \{U' I\}\} \{V\} v] w***ith h >>}}$ and $\ottnt{e}  \longmapsto  \ottnt{e'}$ is carried by \rname{R}{Handle2'}, then $ \bm{ { S } } ^ {  \mathit{I}  }  =  \bm{ { S' } } ^ {  \mathit{I}  } $ must be ensured. This equation does not always hold, even when ARE meets safety conditions. Thus, we modify safety conditions to type-erasure safety conditions.

Therefore, we can prove type and effect safety with type-erasure semantics as a corollary.
\begin{theorem}
  If $\emptyset  \vdash  \ottnt{e}  \ottsym{:}  \ottnt{A}  \mid   \bbZero $ and $\ottnt{e}  \longrightarrow  ^ * \ottnt{e'}$ and $\ottnt{e} \centernot \longrightarrow $, then there is $\ottnt{v}$ such that $\ottnt{e'} = \ottnt{v}$.
\end{theorem}

\subsection{Type-Erasable Instances}
We show the instances satisfying type-erasure safety conditions.

\begin{example}[Erasable Simple Rows]\label{exa:eff_simple_row_erasure}
  Erasable simple rows are defined similarly to simple rows shown in Example~\ref{exa:eff_simple_row}.
  Only the difference is in the definition of equivalence $\simeq$.
  The equivalence $\simeq$ for erasable scoped rows is defined as the least reflexive, transitive relation satisfying the following rules:
  \begin{mathpar}
    \qquad\qquad\qquad\inferrule{\varepsilon_{{\mathrm{1}}} \simeq \varepsilon_{{\mathrm{2}}}}{\langle  \ottnt{L}  \mid  \varepsilon_{{\mathrm{1}}}  \rangle \simeq \langle  \ottnt{L}  \mid  \varepsilon_{{\mathrm{2}}}  \rangle}

    \qquad\inferrule{\mathit{l_{{\mathrm{1}}}} \neq \mathit{l_{{\mathrm{2}}}}}{\langle  \mathit{l_{{\mathrm{1}}}} \,  \bm{ { S_{{\mathrm{1}}} } } ^ {  \mathit{I_{{\mathrm{1}}}}  }   \mid  \langle  \mathit{l_{{\mathrm{2}}}} \,  \bm{ { S_{{\mathrm{2}}} } } ^ {  \mathit{I_{{\mathrm{2}}}}  }   \mid  \varepsilon  \rangle  \rangle \simeq \langle  \mathit{l_{{\mathrm{2}}}} \,  \bm{ { S_{{\mathrm{2}}} } } ^ {  \mathit{I_{{\mathrm{2}}}}  }   \mid  \langle  \mathit{l_{{\mathrm{1}}}} \,  \bm{ { S_{{\mathrm{1}}} } } ^ {  \mathit{I_{{\mathrm{1}}}}  }   \mid  \varepsilon  \rangle  \rangle}
    \\
    \inferrule{ }{\langle  \mathit{l} \,  \bm{ { S_{{\mathrm{1}}} } } ^ {  \mathit{I_{{\mathrm{1}}}}  }   \mid  \langle  \mathit{l} \,  \bm{ { S_{{\mathrm{2}}} } } ^ {  \mathit{I_{{\mathrm{2}}}}  }   \mid  \varepsilon  \rangle  \rangle \simeq \langle  \mathit{l} \,  \bm{ { S_{{\mathrm{1}}} } } ^ {  \mathit{I_{{\mathrm{1}}}}  }   \mid  \varepsilon  \rangle}

    \inferrule{ }{\langle  \mathit{l} \,  \bm{ { S_{{\mathrm{1}}} } } ^ {  \mathit{I_{{\mathrm{1}}}}  }   \mid  \varepsilon  \rangle \simeq \langle  \mathit{l} \,  \bm{ { S_{{\mathrm{1}}} } } ^ {  \mathit{I_{{\mathrm{1}}}}  }   \mid  \langle  \mathit{l} \,  \bm{ { S_{{\mathrm{2}}} } } ^ {  \mathit{I_{{\mathrm{2}}}}  }   \mid  \varepsilon  \rangle  \rangle}
  \end{mathpar}
  Unlike in simple rows, we can only exchange labels with different label names in erasable scoped rows, and we can not only collapse the same labels but also labels with the same label name.
\end{example}
\begin{example}[Erasable Scoped Rows]\label{exa:effrow_erasure}
  Erasable scoped rows are defined similarly to scoped rows shown in Example~\ref{exa:effrow}.
  Only the difference is in the definition of equivalence $\simeq$.
  The equivalence $\simeq$ for erasable scoped rows is defined as the least reflexive, transitive relation satisfying the following rules:
  \begin{mathpar}
    \inferrule{\varepsilon_{{\mathrm{1}}} \simeq \varepsilon_{{\mathrm{2}}}}{\langle  \ottnt{L}  \mid  \varepsilon_{{\mathrm{1}}}  \rangle \simeq \langle  \ottnt{L}  \mid  \varepsilon_{{\mathrm{2}}}  \rangle}

    \inferrule{\mathit{l_{{\mathrm{1}}}} \neq \mathit{l_{{\mathrm{2}}}}}{\langle  \mathit{l_{{\mathrm{1}}}} \,  \bm{ { S_{{\mathrm{1}}} } } ^ {  \mathit{I_{{\mathrm{1}}}}  }   \mid  \langle  \mathit{l_{{\mathrm{2}}}} \,  \bm{ { S_{{\mathrm{2}}} } } ^ {  \mathit{I_{{\mathrm{2}}}}  }   \mid  \varepsilon  \rangle  \rangle \simeq \langle  \mathit{l_{{\mathrm{2}}}} \,  \bm{ { S_{{\mathrm{2}}} } } ^ {  \mathit{I_{{\mathrm{2}}}}  }   \mid  \langle  \mathit{l_{{\mathrm{1}}}} \,  \bm{ { S_{{\mathrm{1}}} } } ^ {  \mathit{I_{{\mathrm{1}}}}  }   \mid  \varepsilon  \rangle  \rangle}
  \end{mathpar}
  Unlike in scoped rows, we can only exchange labels with different label names in erasable scoped rows.
\end{example}

The effect systems with these instances are type and effect safety because of the following theorems.

\begin{theorem}
  Example~\ref{exa:eff_simple_row_erasure} meets the type-erasure safety conditions.
\end{theorem}
\begin{theorem}
  Example~\ref{exa:effrow_erasure} meets the type-erasure safety conditions.
\end{theorem}
}

  }

\subsection{Mixing Lift Coercions and Type-Erasure Semantics}
It is easy to extend {\lang} with both lift coercions and type-erasure semantics
and prove its type-and-effect safety if a given effect algebra is assumed to meet safety
conditions~\ref{def:safe_cond:label_notemp}--\ref{def:safe_cond_erasure:uniq}.
Among the effect algebras presented in the paper, only {\eaEraseScopedRow}
satisfies these conditions, and so {\lang} instantiated with it is
type-and-effect safe.
See the supplementary material for the detail of the combination.
%

\section{Comparison of Effect Algebras}
\label{sec:compare-effect-algebras}
\begin{table}[t]
 \caption{Comparison of the effect algebras.}
 \label{tbl:compare-effect-algebras}
 \begin{tabular}{l|c|c|c}
                 & Lift coercions & Adaptable to type-erasure & Multiple effect variables \\ \hline
  {\eaSet}       & \xmark         & \cmark        & \cmark \\
  {\eaMSet}      & \cmark         & \cmark        & \cmark \\
  {\eaSimpleRow} & \xmark         & \cmark        & \xmark \\
  {\eaScopedRow} & \cmark         & \cmark        & \xmark \\
 \end{tabular}
\end{table}

In this section, we discuss how different the effect algebras {\eaSet},
{\eaMSet}, {\eaSimpleRow}, and {\eaScopedRow} are; it is summarized in
Table~\ref{tbl:compare-effect-algebras}.
The first column in Table~\ref{tbl:compare-effect-algebras} presents whether the
effect algebras are safe in the presence of lift coercions.
As shown in Section~\ref{sec:coercions}, {\eaSet} and {\eaSimpleRow} are unsafe and {\eaMSet} and {\eaScopedRow} are safe.
The second column indicates whether the effect algebras can be adapted to the type-erasure semantics.
As discussed in Section~\ref{sec:erasure}, none of the compared effect algebras
is safe as it is, but all of them become safe if we can admit restricting the
commutativity of the concatenation on effect collections.
The third column shows whether each effect algebra allows multiple effect
variables to appear in one effect collection.
While {\eaSimpleRow} and {\eaScopedRow} disallow it because effect variables can
appear only at the end of rows, neither {\eaSet} nor {\eaMSet} has such a
restriction.

Allowing multiple effect variables in one effect collection in {\eaSet} and
{\eaMSet} leads to more powerful abstraction of effect collections.
For example, consider a module interface \textsf{IntSet} for integer sets, which is given using {\eaSet}:
\[\begin{array}{ll}
   \exists \alpha  \ottsym{:}   \mathbf{Typ}  . \exists \rho  \ottsym{:}   \mathbf{Eff}  . \{ &
    \mathit{empty} : \alpha ,
    \quad \mathit{add} :   \mathsf{Int}     \rightarrow_{ \{  \} }     \alpha    \rightarrow_{ \{  \} }    \alpha  ,
    \quad \cdots,
    \quad
    \mathit{choose} :  \alpha    \rightarrow_{ \rho }     \mathsf{Int}  ,
    \\ &
    \namecollect :   \forall   \beta  \ottsym{:}   \mathbf{Typ}    \ottsym{.}      \forall   \rho'  \ottsym{:}   \mathbf{Eff}    \ottsym{.}     \ottsym{(}    \mathsf{Unit}     \rightarrow_{  \rho  \,\underline{ \cup }\,  \rho'  }    \beta   \ottsym{)}    \rightarrow_{ \rho' }     \beta \,\mathsf{List}    
   \quad \}
  \end{array}
\]
In this type, type variable $\alpha$ is an abstract type representing integer
sets, and the fields represent the operations on integer sets.
The interface $\mathsf{IntSet}$ requires modules to implement, in addition to
the basic operations on sets (e.g., the empty set $\mathit{empty}$ and the
addition of integers to sets $\mathit{add}$), two additional functions for
nondeterministic computation.  Function $\mathit{choose}$ takes an integer set,
performs some abstract effect $\rho$, and returns one element of it.
Intuitively, the abstract effect $\rho$ has a role of notifying that
$\mathit{choose}$ is called.
The other function $\namecollect$ is a higher-order function, taking a
function that may perform effects $\rho$ and $\rho'$.
Thus, the argument function may call $\mathit{choose}$.
Intuitively, function $\namecollect$ collects all the results of the
computation that the argument function performs with some value in a set passed to
$\mathit{choose}$ during its execution.
The argument function may invoke any effect $\rho'$, which leaks to the call
side of $\namecollect$.
The type of the argument function represents that it may invoke two
\emph{abstract} effects $\rho$ and $\rho'$ using the union operation $\underline{\cup}$ in
{\eaSet}.
The use of effect variable $\rho$ enables abstracting module implementations
over not only what effect labels are used in the implementations but also \emph{how
many labels are used there}.
We provide some implementation examples of $\mathsf{IntSet}$ with different
numbers of effect labels in the supplementary material.
Note that the type interface $\mathsf{IntSet}$ cannot be expressed in
{\eaSimpleRow} nor {\eaScopedRow} because only one effect variable may appear in
a row.

However, this is not the end of the story: some existing works have discussed
benefits of using rows as effect collections.
\citet{hillerstrom_liberating_2016} demonstrated that simple rows with row
polymorphism in the style of \citet{remy_type_1993} are useful to solve the
unification occurring in the composition of effect handlers.
\citet{leijen_type_2017} implemented a sound and complete type inference for the
effect system with polymorphism by utilizing scoped rows.
The current form of our theoretical framework, effect algebras, does not provide
a means to discuss unification and type inference for algebraic effects and
handlers, and it is left open how we can address it in an abstract manner.

\section{Related Work}\label{sec:related}

We have explained the existing effect systems for effect handlers in
Section~\ref{sec:overview}, compared some of them with the instances of our
effect system in Section~\ref{sec:comparisons}.
We will also discuss what aspect of effect handlers our framework does
not support in Section~\ref{sec:conclusion}.

Although, as far as we know, there is no prior work on abstracting effect
systems for effect handlers with nor without algebraic structures, the research
on generic effect systems that can reason about the use of a wide range of effects (such as
file resource usage, memory usage and management, and exception checking) has
been conducted.
\citet{Marino/Millstein_2009_TLDI} proposed a monomorphic type-and-effect system
that tracks a set of capabilities (or privileges) to perform effectful
operations such as memory manipulation and exception raising.
Their effect system is generic in that it is parameterized over the forms of
capabilities as well as the adjustments and checkings of capabilities per
context.
It assumes that capabilities are gathered into a set and its typing discipline
relies on the set operations (e.g., the subeffecting is implemented by
set inclusion).
\citet{Rytz/Odersky/Haller_2012_ECOOP} generalized
\citeANP{Marino/Millstein_2009_TLDI}'s effect system by allowing the use of a
join semilattice to represent collections of capabilities and introducing effect
polymorphism.
%
Join-semilattices are underlying structures of effects in effect systems for
\emph{may analysis}. In such a system, the join operation $\sqcup$ and the
ordering relation $\sqsubseteq$ in a join semilattice are used to merge multiple
effects into one and to introduce effect overapproximation as subeffecting, respectively.
As $\sqsubseteq$ can be induced by $\sqcup$
($x \sqsubseteq y \iff x \sqcup y = y$), we define the subeffecting $\olessthan$ using $ \odot $ in an effect
algebra ($ \varepsilon_{{\mathrm{1}}}  \olessthan  \varepsilon_{{\mathrm{2}}}  {\iff} \exists \varepsilon .\,   \varepsilon_{{\mathrm{1}}}  \mathop{ \odot }  \varepsilon    \sim   \varepsilon_{{\mathrm{2}}} $).
Thus, the role of $ \odot $ is similar to that of $\sqcup$, but $ \odot $ is
not required to be commutative nor idempotent, unlike $\sqcup$ (note that join
operations are characterized by associativity, commutativity, and idempotence).
In fact, $ \odot $ in the effect algebra $\eaScopedRow$ or $\eaMSet$ is
nonidempotent, which is key to support lift coercions
(Section~\ref{subsec:coercion-safety-cond}), and $ \odot $ in each of the
effect algebras being safe in the type-erasure semantics is noncommutative
(Section~\ref{subsec:erasure-safety-cond}).
%

Recent developments of generic effect systems have focused on \emph{sequential
effect
systems}~\cite{Atkey_2009_JFP,Tate_2013_POPL,Katsumata_2014_ICFP,Mycroft/Orchard/Petricek_2016_SLC,Gordon_2017_ECOOP,Gordon_2021_TOPLAS},
which aim to reason about the properties where the order of effects matters
(e.g., whether no closed file will be read nor written).
An approach common in the prior work on sequential effect systems is to introduce sequential composition
$\triangleright$, an operation to compose effects happening sequentially.  For
example, given expressions $\ottnt{e_{{\mathrm{1}}}}$ with effect $\varepsilon_{{\mathrm{1}}}$ and $\ottnt{e_{{\mathrm{2}}}}$ with
$\varepsilon_{{\mathrm{2}}}$, the effect of a let-expression $\mathbf{let} \, \mathit{x}  \ottsym{=}  \ottnt{e_{{\mathrm{1}}}} \, \mathbf{in} \, \ottnt{e_{{\mathrm{2}}}}$ is given
by $\varepsilon_{{\mathrm{1}}} \triangleright \varepsilon_{{\mathrm{2}}}$.
The sequential composition can be characterized as a (partial) monoid.
%
%
Thus, it might look similar to $ \odot $ in an effect algebra, but their roles
are significantly different: $ \odot $ is used to expand (i.e.,
overapproximate) effects and remove specific labels from effects, whereas the
sequential composition $\triangleright$ is used to compose the effects of expressions executed
sequentially.
In fact, if we were use $ \odot $ to sequence effects, the safety of {\lang} in
the type-erasure semantics would not hold even in the effect algebra
{\eaEraseScopedRow} for erasable scoped rows.
For example, assume that an expression $\mathbf{let} \, \mathit{x}  \ottsym{=}  \ottnt{e_{{\mathrm{1}}}} \, \mathbf{in} \, \ottnt{e_{{\mathrm{2}}}}$ is given effect
$ \varepsilon_{{\mathrm{1}}}  \mathop{ \odot }  \varepsilon_{{\mathrm{2}}} $ if the effects $\varepsilon_{{\mathrm{1}}}$ and $\varepsilon_{{\mathrm{2}}}$
are assigned to $\ottnt{e_{{\mathrm{1}}}}$ and $\ottnt{e_{{\mathrm{2}}}}$.
Then, the expression $\ottnt{e} \defeq \mathbf{let} \, \mathit{x}  \ottsym{=}   \mathsf{tell} _{ \mathsf{Writer} \,  \mathsf{Bool}  }  \,  \mathsf{true}  \, \mathbf{in} \,  \mathsf{tell} _{ \mathsf{Writer} \,  \mathsf{Int}  }  \, 1$ could have the effect
$ \{  \mathsf{Writer} \,  \mathsf{Bool}   \}  \,\underline{ \sqcup }\,  \{  \mathsf{Writer} \,  \mathsf{Int}   \} $ under {\eaEraseScopedRow}.
Thus, the expression $ \mathbf{handle}_{ \mathsf{Writer} \,  \mathsf{Int}  }  \,  (   \mathbf{handle}_{ \mathsf{Writer} \,  \mathsf{Bool}  }  \, \ottnt{e} \, \mathbf{with} \, \ottnt{h_{{\mathrm{1}}}}  )  \, \mathbf{with} \, \ottnt{h_{{\mathrm{2}}}}$ would be well typed (for some appropriate effect handlers $\ottnt{h_{{\mathrm{1}}}}$
and $\ottnt{h_{{\mathrm{2}}}}$), although it may get stuck in the type-erasure semantics because
the operation call $ \mathsf{tell} _{ \mathsf{Writer} \,  \mathsf{Int}  }  \, 1$ will be handled by the effect
handler $\ottnt{h_{{\mathrm{1}}}}$ for $\mathsf{Writer} \,  \mathsf{Bool} $.
Readers might wonder why $ \odot $ cannot work as a sequential
composition despite the fact that join operations, which are also used to overapproximate
effects, can.
We think that this is because the assumptions on $ \odot $ are weaker than those on join
operations as discussed above.
Making $ \odot $ in effect algebras and $\triangleright$ in
sequential effect systems coexist is a promising future direction, motivated by the
recent study on sequential effect systems for control
operators~\cite{Gordon_2020_ECOOP,Song/Foo/Chin_2022_APLAS,Sekiyama/Unno_2023_POPL}.

\OLD{

\TY{I do not write yet: Parametric effect monads and semantics of effect systems (Katsumata), Logical Relation, https://dl.acm.org/doi/10.1145/3450272}
\subsection{Type-and-Effect Systems for Algebraic Effects and Handlers}
We compare previous research with our work.
\citet{kammar_handlers_2013} introduced a set-based effect system.
We can derive this concrete language as an instance of {\lang}, except for the following differences:
first, their language employs a \emph{call-by-push-value} semantics, while {\lang} employs a call-by-value semantics; second, their language does not support effect polymorphism, while {\lang} does.
\citet{bauer_effect_2013} proposed the core effect system of programming language Eff~\cite{bauer_programming_2012, bauer_programming_2015}.
Their system differs from {\lang} regarding \emph{effect instances}.
Effect instances are (globally allocated) resources associated to algebraic effects;
every effect handler specifies both an operation and effect instance to be handled.
Our effect system does not support effect instances still.
\citet{hillerstrom_liberating_2016} introduced an effect system with simple rows.
Their effect system is based on simple rows but prohibits effects from having multiple occurrences of the same label.
Thus, we can derive their effect system by extending our language to qualified types, as discussed in Section~\ref{subsec:q_simple_rows}.
\citet{leijen_type_2017} proposed an effect system with scoped rows.
That effect system does not allow subtyping in contrast to ours, but we can derive almost the same one, except for such a difference, as an instance.
\citet{vilhena_type_2023} introduced a dynamic way to generate labels for
implementing lexically scoped handlers.  The semantics of their language is
complicated and different from ours.  Their row representation is more flexible
than simple rows, but disallows duplicate labels to appear in a row.


\subsection{Lexically Scoped Handlers}
A \emph{lexically scoped handler} \citep{biernacki_binders_2020, brachthauser_effects_2020}
creates a fresh effect that can be available in its scope, and then handles it.
For example, an exception handler is written as $\mathbf{handle}\, \mathsf{Exc} \Longrightarrow \ottnt{e} \,\mathbf{with}\, \ottnt{h}$.
This handler generates and handles a new label $\mathsf{Exc}$.  In other words, there
is a one-to-one correspondence between a lexically scoped handler and the operations it gives implementations.
Lexically scoped handlers help use the same effects differently and prevent operations from not intended handling.
  {\lang} does not support the mechanism.
However, extending {\lang} to local effects can support lexically scoped handlers
because we can encode lexically scoped handlers via local effects,
assuming that labels of local effects are fresh \citep{vilhena_type_2023}.
The encoding of the above example is as follows.
\begin{flalign*}
   & \mathbf{effect}\,  \mathsf{Exc}  ::  \ottsym{\{}  \mathsf{raise}  \ottsym{:}    \forall   \alpha  \ottsym{:}   \mathbf{Typ}    \ottsym{.}     \mathsf{Unit}    \Rightarrow   \alpha   \ottsym{\}} \,\mathbf{in}\,  \mathbf{handle}_{ \mathsf{Exc} }  \, \ottnt{e} \, \mathbf{with} \, \ottnt{h} &
\end{flalign*}
Thus, we can extend our language to support lexically scoped handlers by extending with local effects as discussed in Section~\ref{subsec:local_eff}.

}

\OLD{
  \TY{I do not write yet: Parametric effect monads and semantics of effect systems (Katsumata), Logical Relation, https://dl.acm.org/doi/10.1145/3450272}
  \subsection{Type-and-Effect Systems for Algebraic Effects and Handlers}
  We compare previous research with our work.
  \citet{kammar_handlers_2013} introduced a set-based effect system.
  We can get this concrete language as an instance of {\lang}, except for the following differences:
  on the one hand, the language of \citet{kammar_handlers_2013} takes \emph{call-by-push-value} semantics and does not allow effect polymorphism;
  on the other hand, an instance of {\lang} takes call-by-value semantics and allows effect polymorphism.
  \citet{bauer_effect_2013} proposed the core language and effect system of Eff.
  Its language differs from an instance of {\lang} regarding \emph{effect instances}.
  Effect instances are the technique to make several copies of algebraic effects.
  Our effect system does not support effect instances.
  \citet{hillerstrom_liberating_2016} introduced the language and an effect system with simple rows.
  Its effect system is based on simple rows but prohibits effects from having multiple labels.
  Thus, we can get the core of their language by extending our language to qualified types discussed in Section~\ref{subsec:q_simple_rows}.
  \citet{leijen_type_2017} proposed an effect system with scoped rows.
  Its effect system does not allow subtyping in contrast to ours, but we can get almost the same one as an instance.
  \citet{vilhena_type_2023} introduced a dynamic way to generate labels for implementing lexically scoped handlers.
  Its semantics is complicated and different from ours,
  but its effect system TES is similar to an effect system with free rows shown in Section~\ref{sec:effseq}.
  Extending an effect system with free rows to lexically scoped handlers in the way discussed below leads to an instance of the core of TES.

  \subsection{Lexically Scoped Handlers}
  A \emph{lexically scoped handler} \citep{biernacki_binders_2020, brachthauser_effects_2020}
  creates its responsible and fresh effect that can be available in its scope.
  For example, an exception handler is written as follows.
  \begin{flalign*}
     & \mathbf{handle}\, \mathsf{Exc} \Longrightarrow \ottnt{e} \,\mathbf{with}\, \ottnt{h} &
  \end{flalign*}
  This handler generates and handles a new label $\mathsf{Exc}$.
  In other words, a lexically scoped handler and its responsible operations are one-to-one correspondence.
  Lexically scoped handlers help use the same effects differently and prevent handlers from not intended handling.
    {\lang} does not support the mechanism.
  However, extending {\lang} to local effects can support lexically scoped handlers
  because we can encode lexically scoped handlers via local effects,
  assuming that labels of local effects are fresh \citep{vilhena_type_2023}.
  The encoding of the above example is as follows.
  \begin{flalign*}
     & \mathbf{effect}\,  \mathsf{Exc}  ::  \ottsym{\{}  \mathsf{raise}  \ottsym{:}    \forall   \alpha  \ottsym{:}   \mathbf{Typ}    \ottsym{.}     \mathsf{Unit}    \Rightarrow   \alpha   \ottsym{\}} \,\mathbf{in}\,  \mathbf{handle}_{ \mathsf{Exc} }  \, \ottnt{e} \, \mathbf{with} \, \ottnt{h} &
  \end{flalign*}
  Thus, we can extend our language to support lexically scoped handlers via local effects, as discussed in Section~\ref{subsec:local_eff}.
}
\section{Conclusion and Future Work}\label{sec:conclusion}

In this paper, we give {\lang} equipped with the abstract effect system that can
be instantiated to concrete effect systems, define safety conditions on effect algebras, and prove the type-and-effect safety of {\lang}
by assuming that a given effect algebra satisfies the conditions.
As far as we know, no research formalizes the differences among effect systems for effect handlers
nor the requirements for the effect systems to prove safety properties.
We reveal these essences via the abstraction of effect systems by effect algebras, and the formalization of the safety conditions.
The safety conditions added for lift coercions or type-erasure semantics clarify the differences among effect algebras.
%
%
In the rest of the paper, we discuss possible directions for future work.
%
%
%
\paragraph{Abstraction of handling mechanisms.}\label{subsec:local_eff}
Although the framework in the paper targets deep effect handlers,
adapting it to shallow effect handlers is easy.  In fact, we have provided this
adaption and proved its safety under the same safety conditions as the ones
given in the main paper; interested readers are referred to the
supplementary material.
In the literature, there are other proposals of the effect handling, especially
for resolving the problem with \emph{accidental handling} without relying on
lift coercions. For instance, local effects~\cite{biernacki_abstracting_2019},
tunneling~\cite{Zhang/Myers_2019_POP}, and lexically scoped effect
handlers~\cite{biernacki_binders_2020, brachthauser_effects_2020} have been
proposed.
These approaches can be applied to address the accidental handling, but they
employ significantly different styles. For example, lexically scoped effect
handlers can enable a new notion of effect polymorphism, called \emph{contextual
polymorphism}~\cite{brachthauser_effects_2020}.
Exploring abstraction to accommodate all of these mechanisms is a
challenging but interesting direction.

\paragraph{Abstraction for unification and type inference.}
%
As mentioned in Section~\ref{sec:comparisons}, our framework has not yet exposed
the essential roles of rows in their main application---unification and type
inference.
One of our ambitious goals for future research is to give a theoretical
framework that can discover differences among effect representations in
unification and type-inference, which have been well explored with concrete
effect representations, such as sets~\cite{Pretnar_2014_LMCS}, simple
rows~\cite{hillerstrom_liberating_2016}, and scoped rows~\cite{leijen_type_2017},
but not in an abstract manner.

\paragraph{Abstraction of constrained effect collections.}
Another interesting direction is to abstract \emph{constrained} effect
collections. For example, \citet{hillerstrom_liberating_2016} introduce
\citeANP{remy_type_1993}'s row polymorphism, which can state that some labels
are present or absent in row variables, for effective unification and
\citet{Tang/Hillerstrom/Lindley/Morris_2024_POPL} propose an effect system that
allows type abstraction over subtyping constraints on row variables.
Row constraints have been extensively studied for programming with records and
variants~\cite{Cardelli/Mitchell_1989_MFPS,Harper/Pierce_1991_POPL,jones_theory_1992,remy_type_1993}.
\citet{morris_abstracting_2019} proposed a type system which treats rows and
constraints on them abstractly.
Integrating the idea of their work with our framework is a promising approach.

\paragraph{Abstraction of implementation techniques.}
One approach to implementing effect handlers is to apply type-directed translation into an
intermediate language~\cite{leijen_type_2017,hillerstrom_continuation_2017,Schuster/Brachthauser/Muller/Ostermann_2022_PLDI,Xie/Brachthauser/Hillerstrom/Schuster/Leijen_2020_ICFP}.
Exploring the type-directed translations and
optimization techniques proposed thus far, such as a selective translation into continuation
passing style (CPS)~\cite{leijen_type_2017}, in an abstract manner may lead to a
common implementation infrastructure for languages with different effect systems or give an insight into
the influence of effect representations on efficiency.

\OLD{
\subsection{Sets vs. Simple Rows}
We summarize the difference between effect systems with sets and simple rows.

First, we compare Example~\ref{exa:effset} and Example~\ref{exa:eff_simple_row}.
The main differences are the occurrences of effect variables and the label exchange rule.
The set instance allows effect variables to occur anywhere in effects,
but the simple row instance restricts effect variables at the end of rows.
This difference is trivial because we can easily create another instance of simple rows
that allows non-restricted occurrences of effect variables.
Unlike effect sets, simple rows have the label exchange rule whose one premise
compares $\ottnt{L_{{\mathrm{1}}}}$ and $\ottnt{L_{{\mathrm{2}}}}$. However, this premise is trivial because
the prohibited exchange, that is, the exchange of the same labels, is meaningless.
Therefore, there is no essential difference between effect sets and simple rows.

Secondly, we compare Example~\ref{exa:effset} and Example~\ref{exa:eff_simple_row_erasure}.
We describe how these differences appear in some settings.
When there are no parametric effects, the two instances are equal.
Since label names do not take parameters in this setting,
any label exchange is allowed, even in simple rows.
Furthermore, the choice of semantics does not cause differences
because the type-erasure semantics is the same as the non-type-erasure semantics in this setting.
When parametric effects exist, the type-erasure semantics causes the crucial difference, as seen in Section~\ref{sec:erasure}.
Even in the non-type-erasure semantics, they are still different because of the label exchange rule in simple rows.
For example, $\mathbf{let} \, \_  \ottsym{=}   \mathsf{set} _{ \mathsf{State} \,  \mathsf{Int}  }  \,  {}  \, 0 \, \mathbf{in} \,  \mathsf{get} _{ \mathsf{State} \,  \mathsf{Bool}  }  \,  {}  \,  () $ is not well-typed under
the effect system with simple rows, but the one with sets accepts it.

\subsection{Qualified Simple Rows}\label{subsec:q_simple_rows}

The original effect system with simple rows \citep{hillerstrom_liberating_2016}
differs from the one we introduced in Example~\ref{exa:eff_simple_row} and Example~\ref{exa:eff_simple_row_erasure}
at the two points: prohibiting multiple labels and limiting instantiations of effect variables.
To distinguish them, we call what the effect system of \citet{hillerstrom_liberating_2016} is based on
\emph{qualified simple rows}.

An effect system with qualified simple rows realizes the features via its kind system shown below,
in the setting where there are no parametric effects.
\begin{mathpar}
  \inferrule{
  \Gamma \vdash P : \mathbf{Presence} \\
  \Gamma \vdash R : \mathbf{Row}_{\mathcal{L} \uplus \{l\} }
  }{
  \Gamma \vdash l : P ; R : \mathbf{Row}_{\mathcal{L}}
  }

  \inferrule{
    \Gamma \vdash R : \mathbf{Row}_{\emptyset}
  }{
    \Gamma \vdash \{ R \} : \mathbf{Eff}
  }
\end{mathpar}
$P$ is the presence with types $\mathsf{Pre}(A)$ or the absence $\mathsf{Abs}$.
$P = \mathsf{Pre}(A)$ represents that the row $l : P; R$ includes label $l$ typed as $A$.
$P = \mathsf{Abs}$ represents that the row $l : P; R$ does not include label $l$.
The set $\mathcal{L}$ attached to $\mathbf{Row}$ denotes the labels uncovered by the row.
If $\mathcal{L}$ is the empty set, then the row can be the effect.
Therefore, all effects do not have multiple labels.
Furthermore, the instantiation of row variables is limited by this kind system.
For example, consider the effect $\{l : \mathsf{Pre}(\ottnt{A}  \rightarrow  \ottnt{B}) ; \alpha\}$.
We cannot substitute rows that have the field $l$ for the row variable $\alpha$,
because the kind of $\alpha$ is $\mathbf{Row}_{\mathcal{L} \setminus \{l\}}$
where $\mathcal{L}$ is the whole set of labels.

To create an instance for qualified simple rows, we must extend {\lang}.
Specifically, we need to add \emph{qualified types} to limit substitutions for effect variables.
Adding qualified types makes an ARE of qualified simple rows more complex
because predicates in a context can be used.
For example, an ARE for qualified simple rows at least satisfies the following: if $\{l\} \not\!\!\olessthan \alpha$ is in $ \Delta $,
then $ \Delta  \vdash \{l\} \odot \alpha \sim \{l, \alpha\}$ holds.
However, attaching qualified types to {\lang} seemingly does not affect safety conditions.
Thus, the extension of {\lang} to cover qualified simple rows does not seem complex.

\subsection{Local Effects}\label{subsec:local_eff}

\emph{Local effects} are algebraic effects declared locally \citep{biernacki_abstracting_2019, vilhena_type_2023}.
For example, we can define the function $\mathit{count\_g}$ that counts
the number of times $\mathit{g}$ uses its argument using a local effect $\mathsf{Tick}$,
where $\mathit{g}$ has the type $  \forall   \rho  \ottsym{:}   \mathbf{Eff}    \ottsym{.}     \ottsym{(}   \ottnt{A}    \rightarrow_{ \rho }    \ottnt{B}   \ottsym{)}    \rightarrow_{ \rho }     \mathsf{Unit}   $.
\begin{flalign*}
  \mathbf{let}\, \mathit{count\_g}
   & = \Lambda \rho  \ottsym{:}   \mathbf{Eff}  . \lambda \mathit{g'}  \ottsym{:}   \ottnt{A}    \rightarrow_{ \rho }    \ottnt{B}  .                                                                                                               & \\
   & \qquad \mathbf{effect}\,\mathsf{Tick} = \{ \mathsf{tick} : \mathsf{Unit} \Rightarrow \mathsf{Unit} \} \,\mathbf{in}                                                   & \\
   & \qquad (\mathbf{handle}_{\mathsf{Tick}} \, \mathit{g}\,([\mathsf{Tick}];\rho)\, (\lambda \mathit{x}  \ottsym{:}  \ottnt{A} . \mathsf{tick}\, ();\, \mathit{g'} \, \mathit{x} )                                   & \\
   & \qquad \mathbf{with} \{\mathbf{return}\, \_ \mapsto \lambda  \mathit{x}  \ottsym{:}   \mathsf{Int}   \ottsym{.}  \mathit{x} \} \uplus \{\mathsf{tick} \, \_ \, k \mapsto \lambda x : \mathsf{Int} . k\, ()\, (x + 1)\})\, 0 &
\end{flalign*}
In this program, we use the syntax of free rows and declare the local effect $\mathsf{Tick}$ at
$\mathbf{effect}\,\mathsf{Tick} = \{ \mathsf{tick} : \mathsf{Unit} \Rightarrow \mathsf{Unit} \} \,\mathbf{in}$.
If we declare $\mathsf{Tick}$ globally, the handler in $\mathit{coun\_g}$ handles $\mathsf{Tick}$
caused by evaluating $\mathit{g'} \, \mathit{x}$ against the programmer's will.

We have mainly two remaining works to extend {\lang} to local effects.
First, we must define a new syntax, typing rule, and reduction rule for local effects.
This work may be trivial, thanks to \citet{biernacki_abstracting_2019}.
Secondly, it is required to check whether we need new safety conditions
to prove the safety properties of the extended language.
\citet{biernacki_abstracting_2019} use an effect system with scoped rows,
and the core equivalence rules over rows are the same as ours except for the one considering local effects.
Because an instance, not a safety condition, gives such a new equivalence rule,
we conjecture there is no need to add safety conditions.

\subsection{Effect Coercions}
\emph{Effect coercions}, introduced by \citet{biernacki_handle_2018, biernacki_abstracting_2019},
explicitly change the order of labels in rows.
For example, using effect coercions,  we can also define the function $\mathit{count\_g}$ mentioned above.
\begin{flalign*}
  \mathbf{let}\,\mathit{count\_g}
   & = \Lambda \rho  \ottsym{:}   \mathbf{Eff}  . \lambda \mathit{g'}  \ottsym{:}   \ottnt{A}    \rightarrow_{ \rho }    \ottnt{B}  .                                                                                                                   & \\
   & \qquad (\mathbf{handle}_{\mathsf{Tick}} \, \mathit{g}\,(\langle \mathsf{Tick} \mid \rho \rangle)\, (\lambda \mathit{x}  \ottsym{:}  \ottnt{A} . \mathsf{tick}\, ();\,[ \mathit{g'} \, \mathit{x} ]_{\mathsf{Tick}} ) & \\
   & \qquad \mathbf{with} \{\mathbf{return}\, \_ \mapsto \lambda  \mathit{x}  \ottsym{:}   \mathsf{Int}   \ottsym{.}  \mathit{x} \} \uplus \{\mathsf{tick} \, \_ \, k \mapsto \lambda x : \mathsf{Int} . k\, ()\, (x + 1)\})\, 0     &
\end{flalign*}
In this program, we use the syntax of scoped rows and suppose that $\mathsf{Tick}$ is declared globally.
We use a \emph{lift coercion} at $[ \mathit{g'} \, \mathit{x} ]_{\mathsf{Tick}}$.
This coercion means adding the label $\mathsf{Tick}$ to the top of row $\rho$.
Thus, the handler in $\mathit{count\_g}$ cannot handle $\mathsf{Tick}$ caused by evaluating $\mathit{g'} \, \mathit{x}$.

In order to extend our language to effect coercions, we must extend the syntax
and change the semantics to the one using \emph{effect freeness}.
Effect freeness takes the form $n\textrm{-free}(\mathit{l}, \ottnt{E})$,
which intuitively means that the $n$-th $\mathit{l}$ handler enclosing the evaluation context $\ottnt{E}$
handles a $\mathit{l}$'s operation in the hole of $\ottnt{E}$.
Effect freeness has a close relationship with operational semantics and effect coercions.
Changing the syntax and giving an instance reflecting effect coercions may be easy.
However, the change in semantics can cause difficulty in proving safety properties
because we must consider the relationship between effect freeness and coercions in the abstract setting.
We must carefully check proofs and find new conditions to add to safety conditions.
}

\OLD{
  In this paper,
  we give {\lang} equipped with the abstract effect system that can be instantiated to concrete effect systems;
  define safety conditions that the effect signatures and ARE must meet;
  prove type-and-effect safety, assuming that effect signatures and ARE satisfy all safety conditions.
  In our understanding, no research formalizes the difference between effect systems
  and the requirement for effect systems to prove safety properties.
  We reveal these essences via abstracting effect systems and finding the required conditions.
  Building safety conditions as layers, that is, safety conditions and type-erasure safety conditions,
  clarifies the difference between effect systems such as the set-based and row-based ones.
  We can adopt this notion of layered conditions to study how extending languages with effect handlers affects safety properties.

  We have several directions for future work.
  In one direction, we extend our language to other mechanisms, such as local effects discussed in Section~\ref{subsec:local_eff}.
  This direction helps design a new effect system with complicated mechanisms.
  In another, we implement the automated tool that verifies whether the given effect signatures and ARE meet safety conditions.
  To achieve this goal, we must consider the suitable structure to implement effect signatures and ARE,
  define the suitable form of safety conditions for the verifier, and devise the verification algorithm.
  After we finished to create the verification algorithm, we can put {\lang} with user-defined effect signatures and ARE into practice.
}

\bibliographystyle{ACM-Reference-Format}
\bibliography{paper}


\begin{thebibliography}{3}


\ifx \showCODEN    \undefined \def \showCODEN     #1{\unskip}     \fi
\ifx \showDOI      \undefined \def \showDOI       #1{#1}\fi
\ifx \showISBNx    \undefined \def \showISBNx     #1{\unskip}     \fi
\ifx \showISBNxiii \undefined \def \showISBNxiii  #1{\unskip}     \fi
\ifx \showISSN     \undefined \def \showISSN      #1{\unskip}     \fi
\ifx \showLCCN     \undefined \def \showLCCN      #1{\unskip}     \fi
\ifx \shownote     \undefined \def \shownote      #1{#1}          \fi
\ifx \showarticletitle \undefined \def \showarticletitle #1{#1}   \fi
\ifx \showURL      \undefined \def \showURL       {\relax}        \fi
\providecommand\bibfield[2]{#2}
\providecommand\bibinfo[2]{#2}
\providecommand\natexlab[1]{#1}
\providecommand\showeprint[2][]{arXiv:#2}

\bibitem[Hillerström et~al\mbox{.}(2017)]%
        {hillerstrom_continuation_2017}
\bibfield{author}{\bibinfo{person}{Daniel Hillerström}, \bibinfo{person}{Sam
  Lindley}, \bibinfo{person}{Robert Atkey}, {and} \bibinfo{person}{K.~C.
  Sivaramakrishnan}.} \bibinfo{year}{2017}\natexlab{}.
\newblock \showarticletitle{Continuation {Passing} {Style} for {Effect}
  {Handlers}}. In \bibinfo{booktitle}{\emph{2nd {International} {Conference} on
  {Formal} {Structures} for {Computation} and {Deduction}, {FSCD} 2017,
  {September} 3-9, 2017, {Oxford}, {UK}}} \emph{(\bibinfo{series}{{LIPIcs}},
  Vol.~\bibinfo{volume}{84})}, \bibfield{editor}{\bibinfo{person}{Dale Miller}}
  (Ed.). \bibinfo{publisher}{Schloss Dagstuhl - Leibniz-Zentrum für
  Informatik}, \bibinfo{pages}{18:1--18:19}.
\newblock
\urldef\tempurl%
\url{https://doi.org/10.4230/LIPIcs.FSCD.2017.18}
\showDOI{\tempurl}


\bibitem[Leijen(2017)]%
        {leijen_type_2017}
\bibfield{author}{\bibinfo{person}{Daan Leijen}.}
  \bibinfo{year}{2017}\natexlab{}.
\newblock \showarticletitle{Type directed compilation of row-typed algebraic
  effects}. In \bibinfo{booktitle}{\emph{Proceedings of the 44th {ACM}
  {SIGPLAN} {Symposium} on {Principles} of {Programming} {Languages}, {POPL}
  2017, {Paris}, {France}, {January} 18-20, 2017}},
  \bibfield{editor}{\bibinfo{person}{Giuseppe Castagna} {and}
  \bibinfo{person}{Andrew~D. Gordon}} (Eds.). \bibinfo{publisher}{ACM},
  \bibinfo{pages}{486--499}.
\newblock
\urldef\tempurl%
\url{https://doi.org/10.1145/3009837.3009872}
\showDOI{\tempurl}


\bibitem[Pretnar(2015)]%
        {pretnar_introduction_2015}
\bibfield{author}{\bibinfo{person}{Matija Pretnar}.}
  \bibinfo{year}{2015}\natexlab{}.
\newblock \showarticletitle{An {Introduction} to {Algebraic} {Effects} and
  {Handlers}. {Invited} tutorial paper}. In \bibinfo{booktitle}{\emph{The 31st
  {Conference} on the {Mathematical} {Foundations} of {Programming}
  {Semantics}, {MFPS} 2015, {Nijmegen}, {The} {Netherlands}, {June} 22-25,
  2015}} \emph{(\bibinfo{series}{Electronic {Notes} in {Theoretical} {Computer}
  {Science}}, Vol.~\bibinfo{volume}{319})},
  \bibfield{editor}{\bibinfo{person}{Dan~R. Ghica}} (Ed.).
  \bibinfo{publisher}{Elsevier}, \bibinfo{pages}{19--35}.
\newblock
\urldef\tempurl%
\url{https://doi.org/10.1016/j.entcs.2015.12.003}
\showDOI{\tempurl}


\end{thebibliography}



\begin{thebibliography}{51}


\ifx \showCODEN    \undefined \def \showCODEN     #1{\unskip}     \fi
\ifx \showDOI      \undefined \def \showDOI       #1{#1}\fi
\ifx \showISBNx    \undefined \def \showISBNx     #1{\unskip}     \fi
\ifx \showISBNxiii \undefined \def \showISBNxiii  #1{\unskip}     \fi
\ifx \showISSN     \undefined \def \showISSN      #1{\unskip}     \fi
\ifx \showLCCN     \undefined \def \showLCCN      #1{\unskip}     \fi
\ifx \shownote     \undefined \def \shownote      #1{#1}          \fi
\ifx \showarticletitle \undefined \def \showarticletitle #1{#1}   \fi
\ifx \showURL      \undefined \def \showURL       {\relax}        \fi
\providecommand\bibfield[2]{#2}
\providecommand\bibinfo[2]{#2}
\providecommand\natexlab[1]{#1}
\providecommand\showeprint[2][]{arXiv:#2}

\bibitem[Atkey(2009)]%
        {Atkey_2009_JFP}
\bibfield{author}{\bibinfo{person}{Robert Atkey}.}
  \bibinfo{year}{2009}\natexlab{}.
\newblock \showarticletitle{Parameterised notions of computation}.
\newblock \bibinfo{journal}{\emph{J. Funct. Program.}} \bibinfo{volume}{19},
  \bibinfo{number}{3-4} (\bibinfo{year}{2009}), \bibinfo{pages}{335--376}.
\newblock
\urldef\tempurl%
\url{https://doi.org/10.1017/S095679680900728X}
\showDOI{\tempurl}


\bibitem[Bauer and Pretnar(2013)]%
        {bauer_effect_2013}
\bibfield{author}{\bibinfo{person}{Andrej Bauer} {and} \bibinfo{person}{Matija
  Pretnar}.} \bibinfo{year}{2013}\natexlab{}.
\newblock \showarticletitle{An {Effect} {System} for {Algebraic} {Effects} and
  {Handlers}}. In \bibinfo{booktitle}{\emph{Algebra and {Coalgebra} in
  {Computer} {Science} - 5th {International} {Conference}, {CALCO} 2013,
  {Warsaw}, {Poland}, {September} 3-6, 2013. {Proceedings}}}
  \emph{(\bibinfo{series}{Lecture {Notes} in {Computer} {Science}},
  Vol.~\bibinfo{volume}{8089})}, \bibfield{editor}{\bibinfo{person}{Reiko
  Heckel} {and} \bibinfo{person}{Stefan Milius}} (Eds.).
  \bibinfo{publisher}{Springer}, \bibinfo{pages}{1--16}.
\newblock
\urldef\tempurl%
\url{https://doi.org/10.1007/978-3-642-40206-7_1}
\showDOI{\tempurl}


\bibitem[Bauer and Pretnar(2021)]%
        {eff_lang}
\bibfield{author}{\bibinfo{person}{Andrej Bauer} {and} \bibinfo{person}{Matija
  Pretnar}.} \bibinfo{year}{2021}\natexlab{}.
\newblock \bibinfo{title}{{Eff}, version 5.1}.
\newblock
\newblock
\urldef\tempurl%
\url{https://www.eff-lang.org/}
\showURL{%
\tempurl}


\bibitem[Biernacki et~al\mbox{.}(2018)]%
        {biernacki_handle_2018}
\bibfield{author}{\bibinfo{person}{Dariusz Biernacki}, \bibinfo{person}{Maciej
  Piróg}, \bibinfo{person}{Piotr Polesiuk}, {and} \bibinfo{person}{Filip
  Sieczkowski}.} \bibinfo{year}{2018}\natexlab{}.
\newblock \showarticletitle{Handle with care: relational interpretation of
  algebraic effects and handlers}.
\newblock \bibinfo{journal}{\emph{Proc. ACM Program. Lang.}}
  \bibinfo{volume}{2}, \bibinfo{number}{POPL} (\bibinfo{year}{2018}),
  \bibinfo{pages}{8:1--8:30}.
\newblock
\urldef\tempurl%
\url{https://doi.org/10.1145/3158096}
\showDOI{\tempurl}


\bibitem[Biernacki et~al\mbox{.}(2019)]%
        {biernacki_abstracting_2019}
\bibfield{author}{\bibinfo{person}{Dariusz Biernacki}, \bibinfo{person}{Maciej
  Piróg}, \bibinfo{person}{Piotr Polesiuk}, {and} \bibinfo{person}{Filip
  Sieczkowski}.} \bibinfo{year}{2019}\natexlab{}.
\newblock \showarticletitle{Abstracting algebraic effects}.
\newblock \bibinfo{journal}{\emph{Proc. ACM Program. Lang.}}
  \bibinfo{volume}{3}, \bibinfo{number}{POPL} (\bibinfo{year}{2019}),
  \bibinfo{pages}{6:1--6:28}.
\newblock
\urldef\tempurl%
\url{https://doi.org/10.1145/3290319}
\showDOI{\tempurl}


\bibitem[Biernacki et~al\mbox{.}(2020)]%
        {biernacki_binders_2020}
\bibfield{author}{\bibinfo{person}{Dariusz Biernacki}, \bibinfo{person}{Maciej
  Piróg}, \bibinfo{person}{Piotr Polesiuk}, {and} \bibinfo{person}{Filip
  Sieczkowski}.} \bibinfo{year}{2020}\natexlab{}.
\newblock \showarticletitle{Binders by day, labels by night: effect instances
  via lexically scoped handlers}.
\newblock \bibinfo{journal}{\emph{Proc. ACM Program. Lang.}}
  \bibinfo{volume}{4}, \bibinfo{number}{POPL} (\bibinfo{year}{2020}),
  \bibinfo{pages}{48:1--48:29}.
\newblock
\urldef\tempurl%
\url{https://doi.org/10.1145/3371116}
\showDOI{\tempurl}


\bibitem[Brachthäuser et~al\mbox{.}(2020)]%
        {brachthauser_effects_2020}
\bibfield{author}{\bibinfo{person}{Jonathan~Immanuel Brachthäuser},
  \bibinfo{person}{Philipp Schuster}, {and} \bibinfo{person}{Klaus Ostermann}.}
  \bibinfo{year}{2020}\natexlab{}.
\newblock \showarticletitle{Effects as capabilities: effect handlers and
  lightweight effect polymorphism}.
\newblock \bibinfo{journal}{\emph{Proc. ACM Program. Lang.}}
  \bibinfo{volume}{4}, \bibinfo{number}{OOPSLA} (\bibinfo{year}{2020}),
  \bibinfo{pages}{126:1--126:30}.
\newblock
\urldef\tempurl%
\url{https://doi.org/10.1145/3428194}
\showDOI{\tempurl}


\bibitem[Cardelli and Mitchell(1989)]%
        {Cardelli/Mitchell_1989_MFPS}
\bibfield{author}{\bibinfo{person}{Luca Cardelli} {and}
  \bibinfo{person}{John~C. Mitchell}.} \bibinfo{year}{1989}\natexlab{}.
\newblock \showarticletitle{Operations on Records}. In
  \bibinfo{booktitle}{\emph{Mathematical Foundations of Programming Semantics,
  5th International Conference, Tulane University, New Orleans, Louisiana, USA,
  March 29 - April 1, 1989, Proceedings}} \emph{(\bibinfo{series}{Lecture Notes
  in Computer Science}, Vol.~\bibinfo{volume}{442})},
  \bibfield{editor}{\bibinfo{person}{Michael~G. Main}, \bibinfo{person}{Austin
  Melton}, \bibinfo{person}{Michael~W. Mislove}, {and}
  \bibinfo{person}{David~A. Schmidt}} (Eds.). \bibinfo{publisher}{Springer},
  \bibinfo{pages}{22--52}.
\newblock
\urldef\tempurl%
\url{https://doi.org/10.1007/BFB0040253}
\showDOI{\tempurl}


\bibitem[Forster et~al\mbox{.}(2017)]%
        {forster_expressive_2017}
\bibfield{author}{\bibinfo{person}{Yannick Forster}, \bibinfo{person}{Ohad
  Kammar}, \bibinfo{person}{Sam Lindley}, {and} \bibinfo{person}{Matija
  Pretnar}.} \bibinfo{year}{2017}\natexlab{}.
\newblock \showarticletitle{On the expressive power of user-defined effects:
  effect handlers, monadic reflection, delimited control}.
\newblock \bibinfo{journal}{\emph{Proc. ACM Program. Lang.}}
  \bibinfo{volume}{1}, \bibinfo{number}{ICFP} (\bibinfo{year}{2017}),
  \bibinfo{pages}{13:1--13:29}.
\newblock
\urldef\tempurl%
\url{https://doi.org/10.1145/3110257}
\showDOI{\tempurl}


\bibitem[Foulis and Bennett(1994)]%
        {foulis_effect_1994}
\bibfield{author}{\bibinfo{person}{D.~J. Foulis} {and} \bibinfo{person}{M.~K.
  Bennett}.} \bibinfo{year}{1994}\natexlab{}.
\newblock \showarticletitle{Effect algebras and unsharp quantum logics}.
\newblock \bibinfo{journal}{\emph{Foundations of Physics}}
  \bibinfo{volume}{24}, \bibinfo{number}{10} (\bibinfo{date}{01 Oct}
  \bibinfo{year}{1994}), \bibinfo{pages}{1331--1352}.
\newblock
\showISSN{1572-9516}
\urldef\tempurl%
\url{https://doi.org/10.1007/BF02283036}
\showDOI{\tempurl}


\bibitem[Gordon(2017)]%
        {Gordon_2017_ECOOP}
\bibfield{author}{\bibinfo{person}{Colin~S. Gordon}.}
  \bibinfo{year}{2017}\natexlab{}.
\newblock \showarticletitle{A Generic Approach to Flow-Sensitive Polymorphic
  Effects}. In \bibinfo{booktitle}{\emph{31st European Conference on
  Object-Oriented Programming, {ECOOP} 2017, June 19-23, 2017, Barcelona,
  Spain}} \emph{(\bibinfo{series}{LIPIcs}, Vol.~\bibinfo{volume}{74})},
  \bibfield{editor}{\bibinfo{person}{Peter M{\"{u}}ller}} (Ed.).
  \bibinfo{publisher}{Schloss Dagstuhl - Leibniz-Zentrum f{\"{u}}r Informatik},
  \bibinfo{pages}{13:1--13:31}.
\newblock
\urldef\tempurl%
\url{https://doi.org/10.4230/LIPICS.ECOOP.2017.13}
\showDOI{\tempurl}


\bibitem[Gordon(2020)]%
        {Gordon_2020_ECOOP}
\bibfield{author}{\bibinfo{person}{Colin~S. Gordon}.}
  \bibinfo{year}{2020}\natexlab{}.
\newblock \showarticletitle{Lifting Sequential Effects to Control Operators}.
  In \bibinfo{booktitle}{\emph{34th European Conference on Object-Oriented
  Programming, {ECOOP} 2020}} \emph{(\bibinfo{series}{LIPIcs},
  Vol.~\bibinfo{volume}{166})}, \bibfield{editor}{\bibinfo{person}{Robert
  Hirschfeld} {and} \bibinfo{person}{Tobias Pape}} (Eds.).
  \bibinfo{publisher}{Schloss Dagstuhl - Leibniz-Zentrum f{\"{u}}r Informatik},
  \bibinfo{pages}{23:1--23:30}.
\newblock
\urldef\tempurl%
\url{https://doi.org/10.4230/LIPIcs.ECOOP.2020.23}
\showDOI{\tempurl}


\bibitem[Gordon(2021)]%
        {Gordon_2021_TOPLAS}
\bibfield{author}{\bibinfo{person}{Colin~S. Gordon}.}
  \bibinfo{year}{2021}\natexlab{}.
\newblock \showarticletitle{Polymorphic Iterable Sequential Effect Systems}.
\newblock \bibinfo{journal}{\emph{{ACM} Trans. Program. Lang. Syst.}}
  \bibinfo{volume}{43}, \bibinfo{number}{1} (\bibinfo{year}{2021}),
  \bibinfo{pages}{4:1--4:79}.
\newblock
\urldef\tempurl%
\url{https://doi.org/10.1145/3450272}
\showDOI{\tempurl}


\bibitem[Harper and Pierce(1991)]%
        {Harper/Pierce_1991_POPL}
\bibfield{author}{\bibinfo{person}{Robert Harper} {and}
  \bibinfo{person}{Benjamin~C. Pierce}.} \bibinfo{year}{1991}\natexlab{}.
\newblock \showarticletitle{A Record Calculus Based on Symmetric
  Concatenation}. In \bibinfo{booktitle}{\emph{Conference Record of the
  Eighteenth Annual {ACM} Symposium on Principles of Programming Languages,
  Orlando, Florida, USA, January 21-23, 1991}},
  \bibfield{editor}{\bibinfo{person}{David~S. Wise}} (Ed.).
  \bibinfo{publisher}{{ACM} Press}, \bibinfo{pages}{131--142}.
\newblock
\urldef\tempurl%
\url{https://doi.org/10.1145/99583.99603}
\showDOI{\tempurl}


\bibitem[Hillerström and Lindley(2016)]%
        {hillerstrom_liberating_2016}
\bibfield{author}{\bibinfo{person}{Daniel Hillerström} {and}
  \bibinfo{person}{Sam Lindley}.} \bibinfo{year}{2016}\natexlab{}.
\newblock \showarticletitle{Liberating effects with rows and handlers}. In
  \bibinfo{booktitle}{\emph{Proceedings of the 1st {International} {Workshop}
  on {Type}-{Driven} {Development}, {TyDe}@{ICFP} 2016, {Nara}, {Japan},
  {September} 18, 2016}}, \bibfield{editor}{\bibinfo{person}{James Chapman}
  {and} \bibinfo{person}{Wouter Swierstra}} (Eds.). \bibinfo{publisher}{ACM},
  \bibinfo{pages}{15--27}.
\newblock
\urldef\tempurl%
\url{https://doi.org/10.1145/2976022.2976033}
\showDOI{\tempurl}


\bibitem[Hillerström et~al\mbox{.}(2017)]%
        {hillerstrom_continuation_2017}
\bibfield{author}{\bibinfo{person}{Daniel Hillerström}, \bibinfo{person}{Sam
  Lindley}, \bibinfo{person}{Robert Atkey}, {and} \bibinfo{person}{K.~C.
  Sivaramakrishnan}.} \bibinfo{year}{2017}\natexlab{}.
\newblock \showarticletitle{Continuation {Passing} {Style} for {Effect}
  {Handlers}}. In \bibinfo{booktitle}{\emph{2nd {International} {Conference} on
  {Formal} {Structures} for {Computation} and {Deduction}, {FSCD} 2017,
  {September} 3-9, 2017, {Oxford}, {UK}}} \emph{(\bibinfo{series}{{LIPIcs}},
  Vol.~\bibinfo{volume}{84})}, \bibfield{editor}{\bibinfo{person}{Dale Miller}}
  (Ed.). \bibinfo{publisher}{Schloss Dagstuhl - Leibniz-Zentrum für
  Informatik}, \bibinfo{pages}{18:1--18:19}.
\newblock
\urldef\tempurl%
\url{https://doi.org/10.4230/LIPIcs.FSCD.2017.18}
\showDOI{\tempurl}


\bibitem[Jones(1992)]%
        {jones_theory_1992}
\bibfield{author}{\bibinfo{person}{Mark~P. Jones}.}
  \bibinfo{year}{1992}\natexlab{}.
\newblock \showarticletitle{A {Theory} of {Qualified} {Types}}. In
  \bibinfo{booktitle}{\emph{{ESOP} '92, 4th {European} {Symposium} on
  {Programming}, {Rennes}, {France}, {February} 26-28, 1992, {Proceedings}}}
  \emph{(\bibinfo{series}{Lecture {Notes} in {Computer} {Science}},
  Vol.~\bibinfo{volume}{582})}, \bibfield{editor}{\bibinfo{person}{Bernd
  Krieg-Brückner}} (Ed.). \bibinfo{publisher}{Springer},
  \bibinfo{pages}{287--306}.
\newblock
\urldef\tempurl%
\url{https://doi.org/10.1007/3-540-55253-7_17}
\showDOI{\tempurl}


\bibitem[Kammar et~al\mbox{.}(2013)]%
        {kammar_handlers_2013}
\bibfield{author}{\bibinfo{person}{Ohad Kammar}, \bibinfo{person}{Sam Lindley},
  {and} \bibinfo{person}{Nicolas Oury}.} \bibinfo{year}{2013}\natexlab{}.
\newblock \showarticletitle{Handlers in action}. In
  \bibinfo{booktitle}{\emph{{ACM} {SIGPLAN} {International} {Conference} on
  {Functional} {Programming}, {ICFP}'13, {Boston}, {MA}, {USA} - {September} 25
  - 27, 2013}}, \bibfield{editor}{\bibinfo{person}{Greg Morrisett} {and}
  \bibinfo{person}{Tarmo Uustalu}} (Eds.). \bibinfo{publisher}{ACM},
  \bibinfo{pages}{145--158}.
\newblock
\urldef\tempurl%
\url{https://doi.org/10.1145/2500365.2500590}
\showDOI{\tempurl}


\bibitem[Kammar and Pretnar(2017)]%
        {kammar_no_2017}
\bibfield{author}{\bibinfo{person}{Ohad Kammar} {and} \bibinfo{person}{Matija
  Pretnar}.} \bibinfo{year}{2017}\natexlab{}.
\newblock \showarticletitle{No value restriction is needed for algebraic
  effects and handlers}.
\newblock \bibinfo{journal}{\emph{J. Funct. Program.}}  \bibinfo{volume}{27}
  (\bibinfo{year}{2017}), \bibinfo{pages}{e7}.
\newblock
\urldef\tempurl%
\url{https://doi.org/10.1017/S0956796816000320}
\showDOI{\tempurl}


\bibitem[Katsumata(2014)]%
        {Katsumata_2014_ICFP}
\bibfield{author}{\bibinfo{person}{Shin{-}ya Katsumata}.}
  \bibinfo{year}{2014}\natexlab{}.
\newblock \showarticletitle{Parametric effect monads and semantics of effect
  systems}. In \bibinfo{booktitle}{\emph{The 41st Annual {ACM} {SIGPLAN-SIGACT}
  Symposium on Principles of Programming Languages, {POPL} '14, San Diego, CA,
  USA, January 20-21, 2014}}, \bibfield{editor}{\bibinfo{person}{Suresh
  Jagannathan} {and} \bibinfo{person}{Peter Sewell}} (Eds.).
  \bibinfo{publisher}{{ACM}}, \bibinfo{pages}{633--646}.
\newblock
\urldef\tempurl%
\url{https://doi.org/10.1145/2535838.2535846}
\showDOI{\tempurl}


\bibitem[Kawamata et~al\mbox{.}(2024)]%
        {kawamata_answer_2024}
\bibfield{author}{\bibinfo{person}{Fuga Kawamata}, \bibinfo{person}{Hiroshi
  Unno}, \bibinfo{person}{Taro Sekiyama}, {and} \bibinfo{person}{Tachio
  Terauchi}.} \bibinfo{year}{2024}\natexlab{}.
\newblock \showarticletitle{Answer {Refinement} {Modification}: {Refinement}
  {Type} {System} for {Algebraic} {Effects} and {Handlers}}.
\newblock \bibinfo{journal}{\emph{Proc. ACM Program. Lang.}}
  \bibinfo{volume}{8}, \bibinfo{number}{POPL} (\bibinfo{year}{2024}),
  \bibinfo{pages}{115--147}.
\newblock
\urldef\tempurl%
\url{https://doi.org/10.1145/3633280}
\showDOI{\tempurl}


\bibitem[Leijen(2017)]%
        {leijen_type_2017}
\bibfield{author}{\bibinfo{person}{Daan Leijen}.}
  \bibinfo{year}{2017}\natexlab{}.
\newblock \showarticletitle{Type directed compilation of row-typed algebraic
  effects}. In \bibinfo{booktitle}{\emph{Proceedings of the 44th {ACM}
  {SIGPLAN} {Symposium} on {Principles} of {Programming} {Languages}, {POPL}
  2017, {Paris}, {France}, {January} 18-20, 2017}},
  \bibfield{editor}{\bibinfo{person}{Giuseppe Castagna} {and}
  \bibinfo{person}{Andrew~D. Gordon}} (Eds.). \bibinfo{publisher}{ACM},
  \bibinfo{pages}{486--499}.
\newblock
\urldef\tempurl%
\url{https://doi.org/10.1145/3009837.3009872}
\showDOI{\tempurl}


\bibitem[Leijen(2018)]%
        {leijen2018algebraic}
\bibfield{author}{\bibinfo{person}{Daan Leijen}.}
  \bibinfo{year}{2018}\natexlab{}.
\newblock \bibinfo{booktitle}{\emph{Algebraic Effect Handlers with Resources
  and Deep Finalization}}.
\newblock \bibinfo{type}{{T}echnical {R}eport} MSR-TR-2018-10.
  \bibinfo{pages}{35} pages.
\newblock
\urldef\tempurl%
\url{https://www.microsoft.com/en-us/research/publication/algebraic-effect-handlers-resources-deep-finalization/}
\showURL{%
\tempurl}


\bibitem[Leijen(2024)]%
        {koka_lang}
\bibfield{author}{\bibinfo{person}{Daan Leijen}.}
  \bibinfo{year}{2024}\natexlab{}.
\newblock \bibinfo{title}{{Koka}: a Functional Language with Effects, version
  3.1.0}.
\newblock
\newblock
\urldef\tempurl%
\url{https://koka-lang.github.io/}
\showURL{%
\tempurl}


\bibitem[Lindley et~al\mbox{.}(2023)]%
        {links_lang}
\bibfield{author}{\bibinfo{person}{Sam Lindley}, \bibinfo{person}{Daniel
  Hillerström}, \bibinfo{person}{Simon Fowler}, \bibinfo{person}{James
  Cheney}, \bibinfo{person}{Jan Stolarek}, \bibinfo{person}{Frank Emrich},
  \bibinfo{person}{Rudi Horn}, \bibinfo{person}{Vashti Galpin},
  \bibinfo{person}{Wilmer Ricciotti}, {and} \bibinfo{person}{Philip~and
  Wadler}.} \bibinfo{year}{2023}\natexlab{}.
\newblock \bibinfo{title}{{Links}: Linking Theory to Practice for the Web,
  version 0.9.8}.
\newblock
\newblock
\urldef\tempurl%
\url{https://links-lang.org/}
\showURL{%
\tempurl}


\bibitem[Marino and Millstein(2009)]%
        {Marino/Millstein_2009_TLDI}
\bibfield{author}{\bibinfo{person}{Daniel Marino} {and}
  \bibinfo{person}{Todd~D. Millstein}.} \bibinfo{year}{2009}\natexlab{}.
\newblock \showarticletitle{A generic type-and-effect system}. In
  \bibinfo{booktitle}{\emph{Proceedings of TLDI'09: 2009 {ACM} {SIGPLAN}
  International Workshop on Types in Languages Design and Implementation,
  Savannah, GA, USA, January 24, 2009}},
  \bibfield{editor}{\bibinfo{person}{Andrew Kennedy} {and}
  \bibinfo{person}{Amal Ahmed}} (Eds.). \bibinfo{publisher}{{ACM}},
  \bibinfo{pages}{39--50}.
\newblock
\urldef\tempurl%
\url{https://doi.org/10.1145/1481861.1481868}
\showDOI{\tempurl}


\bibitem[Moggi(1991)]%
        {moggi_notions_1991}
\bibfield{author}{\bibinfo{person}{Eugenio Moggi}.}
  \bibinfo{year}{1991}\natexlab{}.
\newblock \showarticletitle{Notions of {{Computation}} and {{Monads}}}.
\newblock \bibinfo{journal}{\emph{Inf. Comput.}} \bibinfo{volume}{93},
  \bibinfo{number}{1} (\bibinfo{year}{1991}), \bibinfo{pages}{55--92}.
\newblock
\urldef\tempurl%
\url{https://doi.org/10.1016/0890-5401(91)90052-4}
\showDOI{\tempurl}


\bibitem[Morris and McKinna(2019)]%
        {morris_abstracting_2019}
\bibfield{author}{\bibinfo{person}{J.~Garrett Morris} {and}
  \bibinfo{person}{James McKinna}.} \bibinfo{year}{2019}\natexlab{}.
\newblock \showarticletitle{Abstracting extensible data types: or, rows by any
  other name}.
\newblock \bibinfo{journal}{\emph{Proc. ACM Program. Lang.}}
  \bibinfo{volume}{3}, \bibinfo{number}{POPL} (\bibinfo{year}{2019}),
  \bibinfo{pages}{12:1--12:28}.
\newblock
\urldef\tempurl%
\url{https://doi.org/10.1145/3290325}
\showDOI{\tempurl}


\bibitem[Mycroft et~al\mbox{.}(2016)]%
        {Mycroft/Orchard/Petricek_2016_SLC}
\bibfield{author}{\bibinfo{person}{Alan Mycroft}, \bibinfo{person}{Dominic~A.
  Orchard}, {and} \bibinfo{person}{Tomas Petricek}.}
  \bibinfo{year}{2016}\natexlab{}.
\newblock \showarticletitle{Effect Systems Revisited - Control-Flow Algebra and
  Semantics}. In \bibinfo{booktitle}{\emph{Semantics, Logics, and Calculi -
  Essays Dedicated to Hanne Riis Nielson and Flemming Nielson on the Occasion
  of Their 60th Birthdays}} \emph{(\bibinfo{series}{Lecture Notes in Computer
  Science}, Vol.~\bibinfo{volume}{9560})},
  \bibfield{editor}{\bibinfo{person}{Christian~W. Probst},
  \bibinfo{person}{Chris Hankin}, {and} \bibinfo{person}{Ren{\'{e}}~Rydhof
  Hansen}} (Eds.). \bibinfo{publisher}{Springer}, \bibinfo{pages}{1--32}.
\newblock
\urldef\tempurl%
\url{https://doi.org/10.1007/978-3-319-27810-0\_1}
\showDOI{\tempurl}


\bibitem[Plotkin and Power(2003)]%
        {plotkin_algebraic_2003}
\bibfield{author}{\bibinfo{person}{Gordon~D. Plotkin} {and}
  \bibinfo{person}{John Power}.} \bibinfo{year}{2003}\natexlab{}.
\newblock \showarticletitle{Algebraic {{Operations}} and {{Generic Effects}}}.
\newblock \bibinfo{journal}{\emph{Appl. Categorical Struct.}}
  \bibinfo{volume}{11}, \bibinfo{number}{1} (\bibinfo{year}{2003}),
  \bibinfo{pages}{69--94}.
\newblock
\urldef\tempurl%
\url{https://doi.org/10.1023/A:1023064908962}
\showDOI{\tempurl}


\bibitem[Plotkin and Pretnar(2009)]%
        {plotkin_handlers_2009}
\bibfield{author}{\bibinfo{person}{Gordon~D. Plotkin} {and}
  \bibinfo{person}{Matija Pretnar}.} \bibinfo{year}{2009}\natexlab{}.
\newblock \showarticletitle{Handlers of {Algebraic} {Effects}}. In
  \bibinfo{booktitle}{\emph{Programming {Languages} and {Systems}, 18th
  {European} {Symposium} on {Programming}, {ESOP} 2009, {Held} as {Part} of the
  {Joint} {European} {Conferences} on {Theory} and {Practice} of {Software},
  {ETAPS} 2009, {York}, {UK}, {March} 22-29, 2009. {Proceedings}}}
  \emph{(\bibinfo{series}{Lecture {Notes} in {Computer} {Science}},
  Vol.~\bibinfo{volume}{5502})}, \bibfield{editor}{\bibinfo{person}{Giuseppe
  Castagna}} (Ed.). \bibinfo{publisher}{Springer}, \bibinfo{pages}{80--94}.
\newblock
\urldef\tempurl%
\url{https://doi.org/10.1007/978-3-642-00590-9_7}
\showDOI{\tempurl}


\bibitem[Plotkin and Pretnar(2013)]%
        {plotkin_handling_2013}
\bibfield{author}{\bibinfo{person}{Gordon~D. Plotkin} {and}
  \bibinfo{person}{Matija Pretnar}.} \bibinfo{year}{2013}\natexlab{}.
\newblock \showarticletitle{Handling {Algebraic} {Effects}}.
\newblock \bibinfo{journal}{\emph{Log. Methods Comput. Sci.}}
  \bibinfo{volume}{9}, \bibinfo{number}{4} (\bibinfo{year}{2013}).
\newblock
\urldef\tempurl%
\url{https://doi.org/10.2168/LMCS-9(4:23)2013}
\showDOI{\tempurl}


\bibitem[Pretnar(2014)]%
        {Pretnar_2014_LMCS}
\bibfield{author}{\bibinfo{person}{Matija Pretnar}.}
  \bibinfo{year}{2014}\natexlab{}.
\newblock \showarticletitle{Inferring Algebraic Effects}.
\newblock \bibinfo{journal}{\emph{Log. Methods Comput. Sci.}}
  \bibinfo{volume}{10}, \bibinfo{number}{3} (\bibinfo{year}{2014}).
\newblock
\urldef\tempurl%
\url{https://doi.org/10.2168/LMCS-10(3:21)2014}
\showDOI{\tempurl}


\bibitem[Pretnar(2015)]%
        {pretnar_introduction_2015}
\bibfield{author}{\bibinfo{person}{Matija Pretnar}.}
  \bibinfo{year}{2015}\natexlab{}.
\newblock \showarticletitle{An {Introduction} to {Algebraic} {Effects} and
  {Handlers}. {Invited} tutorial paper}. In \bibinfo{booktitle}{\emph{The 31st
  {Conference} on the {Mathematical} {Foundations} of {Programming}
  {Semantics}, {MFPS} 2015, {Nijmegen}, {The} {Netherlands}, {June} 22-25,
  2015}} \emph{(\bibinfo{series}{Electronic {Notes} in {Theoretical} {Computer}
  {Science}}, Vol.~\bibinfo{volume}{319})},
  \bibfield{editor}{\bibinfo{person}{Dan~R. Ghica}} (Ed.).
  \bibinfo{publisher}{Elsevier}, \bibinfo{pages}{19--35}.
\newblock
\urldef\tempurl%
\url{https://doi.org/10.1016/j.entcs.2015.12.003}
\showDOI{\tempurl}


\bibitem[Rytz et~al\mbox{.}(2012)]%
        {Rytz/Odersky/Haller_2012_ECOOP}
\bibfield{author}{\bibinfo{person}{Lukas Rytz}, \bibinfo{person}{Martin
  Odersky}, {and} \bibinfo{person}{Philipp Haller}.}
  \bibinfo{year}{2012}\natexlab{}.
\newblock \showarticletitle{Lightweight Polymorphic Effects}. In
  \bibinfo{booktitle}{\emph{{ECOOP} 2012 - Object-Oriented Programming - 26th
  European Conference, Beijing, China, June 11-16, 2012. Proceedings}}
  \emph{(\bibinfo{series}{Lecture Notes in Computer Science},
  Vol.~\bibinfo{volume}{7313})}, \bibfield{editor}{\bibinfo{person}{James
  Noble}} (Ed.). \bibinfo{publisher}{Springer}, \bibinfo{pages}{258--282}.
\newblock
\urldef\tempurl%
\url{https://doi.org/10.1007/978-3-642-31057-7\_13}
\showDOI{\tempurl}


\bibitem[Rémy(1994)]%
        {remy_type_1993}
\bibfield{author}{\bibinfo{person}{Didier Rémy}.}
  \bibinfo{year}{1994}\natexlab{}.
\newblock \showarticletitle{Type {Inference} for {Records} in a {Natural}
  {Extension} of {ML}}.
\newblock In \bibinfo{booktitle}{\emph{Theoretical {Aspects} {Of}
  {Object}-{Oriented} {Programming}. {Types}, {Semantics} and {Language}
  {Design}}}, \bibfield{editor}{\bibinfo{person}{Carl~A. Gunter} {and}
  \bibinfo{person}{John~C. Mitchell}} (Eds.). \bibinfo{publisher}{MIT Press}.
\newblock


\bibitem[Saleh et~al\mbox{.}(2018)]%
        {saleh_explicit_2018}
\bibfield{author}{\bibinfo{person}{Amr~Hany Saleh}, \bibinfo{person}{Georgios
  Karachalias}, \bibinfo{person}{Matija Pretnar}, {and} \bibinfo{person}{Tom
  Schrijvers}.} \bibinfo{year}{2018}\natexlab{}.
\newblock \showarticletitle{Explicit {Effect} {Subtyping}}. In
  \bibinfo{booktitle}{\emph{Programming {Languages} and {Systems} - 27th
  {European} {Symposium} on {Programming}, {ESOP} 2018, {Held} as {Part} of the
  {European} {Joint} {Conferences} on {Theory} and {Practice} of {Software},
  {ETAPS} 2018, {Thessaloniki}, {Greece}, {April} 14-20, 2018, {Proceedings}}}
  \emph{(\bibinfo{series}{Lecture {Notes} in {Computer} {Science}},
  Vol.~\bibinfo{volume}{10801})}, \bibfield{editor}{\bibinfo{person}{Amal
  Ahmed}} (Ed.). \bibinfo{publisher}{Springer}, \bibinfo{pages}{327--354}.
\newblock
\urldef\tempurl%
\url{https://doi.org/10.1007/978-3-319-89884-1_12}
\showDOI{\tempurl}


\bibitem[Schuster et~al\mbox{.}(2022)]%
        {Schuster/Brachthauser/Muller/Ostermann_2022_PLDI}
\bibfield{author}{\bibinfo{person}{Philipp Schuster},
  \bibinfo{person}{Jonathan~Immanuel Brachth{\"{a}}user},
  \bibinfo{person}{Marius M{\"{u}}ller}, {and} \bibinfo{person}{Klaus
  Ostermann}.} \bibinfo{year}{2022}\natexlab{}.
\newblock \showarticletitle{A typed continuation-passing translation for
  lexical effect handlers}. In \bibinfo{booktitle}{\emph{{PLDI} '22: 43rd {ACM}
  {SIGPLAN} International Conference on Programming Language Design and
  Implementation, San Diego, CA, USA, June 13 - 17, 2022}},
  \bibfield{editor}{\bibinfo{person}{Ranjit Jhala} {and} \bibinfo{person}{Isil
  Dillig}} (Eds.). \bibinfo{publisher}{{ACM}}, \bibinfo{pages}{566--579}.
\newblock
\urldef\tempurl%
\url{https://doi.org/10.1145/3519939.3523710}
\showDOI{\tempurl}


\bibitem[Sekiyama and Igarashi(2019)]%
        {Sekiyama/Igarashi_2019_ESOP}
\bibfield{author}{\bibinfo{person}{Taro Sekiyama} {and}
  \bibinfo{person}{Atsushi Igarashi}.} \bibinfo{year}{2019}\natexlab{}.
\newblock \showarticletitle{Handling Polymorphic Algebraic Effects}. In
  \bibinfo{booktitle}{\emph{Programming Languages and Systems - 28th European
  Symposium on Programming, {ESOP} 2019, Held as Part of the European Joint
  Conferences on Theory and Practice of Software, {ETAPS} 2019, Prague, Czech
  Republic, April 6-11, 2019, Proceedings}} \emph{(\bibinfo{series}{Lecture
  Notes in Computer Science}, Vol.~\bibinfo{volume}{11423})},
  \bibfield{editor}{\bibinfo{person}{Lu{\'{\i}}s Caires}} (Ed.).
  \bibinfo{publisher}{Springer}, \bibinfo{pages}{353--380}.
\newblock
\urldef\tempurl%
\url{https://doi.org/10.1007/978-3-030-17184-1\_13}
\showDOI{\tempurl}


\bibitem[Sekiyama et~al\mbox{.}(2020)]%
        {sekiyama_signature_2020}
\bibfield{author}{\bibinfo{person}{Taro Sekiyama}, \bibinfo{person}{Takeshi
  Tsukada}, {and} \bibinfo{person}{Atsushi Igarashi}.}
  \bibinfo{year}{2020}\natexlab{}.
\newblock \showarticletitle{Signature restriction for polymorphic algebraic
  effects}.
\newblock \bibinfo{journal}{\emph{Proc. ACM Program. Lang.}}
  \bibinfo{volume}{4}, \bibinfo{number}{ICFP} (\bibinfo{year}{2020}),
  \bibinfo{pages}{117:1--117:30}.
\newblock
\urldef\tempurl%
\url{https://doi.org/10.1145/3408999}
\showDOI{\tempurl}


\bibitem[Sekiyama and Unno(2023)]%
        {Sekiyama/Unno_2023_POPL}
\bibfield{author}{\bibinfo{person}{Taro Sekiyama} {and}
  \bibinfo{person}{Hiroshi Unno}.} \bibinfo{year}{2023}\natexlab{}.
\newblock \showarticletitle{Temporal Verification with Answer-Effect
  Modification: Dependent Temporal Type-and-Effect System with Delimited
  Continuations}.
\newblock \bibinfo{journal}{\emph{Proc. {ACM} Program. Lang.}}
  \bibinfo{volume}{7}, \bibinfo{number}{{POPL}}, Article
  \bibinfo{articleno}{71} (\bibinfo{year}{2023}), \bibinfo{numpages}{32}~pages.
\newblock
\urldef\tempurl%
\url{https://doi.org/10.1145/3571264}
\showDOI{\tempurl}


\bibitem[Song et~al\mbox{.}(2022)]%
        {Song/Foo/Chin_2022_APLAS}
\bibfield{author}{\bibinfo{person}{Yahui Song}, \bibinfo{person}{Darius Foo},
  {and} \bibinfo{person}{Wei{-}Ngan Chin}.} \bibinfo{year}{2022}\natexlab{}.
\newblock \showarticletitle{Automated Temporal Verification for Algebraic
  Effects}. In \bibinfo{booktitle}{\emph{Programming Languages and Systems -
  20th Asian Symposium, {APLAS} 2022, Auckland, New Zealand, December 5, 2022,
  Proceedings}} \emph{(\bibinfo{series}{Lecture Notes in Computer Science},
  Vol.~\bibinfo{volume}{13658})}, \bibfield{editor}{\bibinfo{person}{Ilya
  Sergey}} (Ed.). \bibinfo{publisher}{Springer}, \bibinfo{pages}{88--109}.
\newblock
\urldef\tempurl%
\url{https://doi.org/10.1007/978-3-031-21037-2\_5}
\showDOI{\tempurl}


\bibitem[Tang et~al\mbox{.}(2024)]%
        {Tang/Hillerstrom/Lindley/Morris_2024_POPL}
\bibfield{author}{\bibinfo{person}{Wenhao Tang}, \bibinfo{person}{Daniel
  Hillerstr{\"{o}}m}, \bibinfo{person}{Sam Lindley}, {and}
  \bibinfo{person}{J.~Garrett Morris}.} \bibinfo{year}{2024}\natexlab{}.
\newblock \showarticletitle{Soundly Handling Linearity}.
\newblock \bibinfo{journal}{\emph{Proc. {ACM} Program. Lang.}}
  \bibinfo{volume}{8}, \bibinfo{number}{{POPL}} (\bibinfo{year}{2024}),
  \bibinfo{pages}{1600--1628}.
\newblock
\urldef\tempurl%
\url{https://doi.org/10.1145/3632896}
\showDOI{\tempurl}


\bibitem[Tate(2013)]%
        {Tate_2013_POPL}
\bibfield{author}{\bibinfo{person}{Ross Tate}.}
  \bibinfo{year}{2013}\natexlab{}.
\newblock \showarticletitle{The sequential semantics of producer effect
  systems}. In \bibinfo{booktitle}{\emph{The 40th Annual {ACM} {SIGPLAN-SIGACT}
  Symposium on Principles of Programming Languages, {POPL} '13, Rome, Italy -
  January 23 - 25, 2013}}, \bibfield{editor}{\bibinfo{person}{Roberto
  Giacobazzi} {and} \bibinfo{person}{Radhia Cousot}} (Eds.).
  \bibinfo{publisher}{{ACM}}, \bibinfo{pages}{15--26}.
\newblock
\urldef\tempurl%
\url{https://doi.org/10.1145/2429069.2429074}
\showDOI{\tempurl}


\bibitem[Tofte(1990)]%
        {Tofte_1990_IC}
\bibfield{author}{\bibinfo{person}{Mads Tofte}.}
  \bibinfo{year}{1990}\natexlab{}.
\newblock \showarticletitle{Type Inference for Polymorphic References}.
\newblock \bibinfo{journal}{\emph{Inf. Comput.}} \bibinfo{volume}{89},
  \bibinfo{number}{1} (\bibinfo{year}{1990}), \bibinfo{pages}{1--34}.
\newblock
\urldef\tempurl%
\url{https://doi.org/10.1016/0890-5401(90)90018-D}
\showDOI{\tempurl}


\bibitem[Wadler(1998)]%
        {wadler_marriage_1998}
\bibfield{author}{\bibinfo{person}{Philip Wadler}.}
  \bibinfo{year}{1998}\natexlab{}.
\newblock \showarticletitle{The {{Marriage}} of {{Effects}} and {{Monads}}}. In
  \bibinfo{booktitle}{\emph{Proceedings of the Third {{ACM SIGPLAN
  International Conference}} on {{Functional Programming}} ({{ICFP}} '98),
  {{Baltimore}}, {{Maryland}}, {{USA}}, {{September}} 27-29, 1998}},
  \bibfield{editor}{\bibinfo{person}{Matthias Felleisen}, \bibinfo{person}{Paul
  Hudak}, {and} \bibinfo{person}{Christian Queinnec}} (Eds.).
  \bibinfo{publisher}{{ACM}}, \bibinfo{pages}{63--74}.
\newblock
\urldef\tempurl%
\url{https://doi.org/10.1145/289423.289429}
\showDOI{\tempurl}


\bibitem[Wright(1995)]%
        {Wright_1995_LSC}
\bibfield{author}{\bibinfo{person}{Andrew~K. Wright}.}
  \bibinfo{year}{1995}\natexlab{}.
\newblock \showarticletitle{Simple Imperative Polymorphism}.
\newblock \bibinfo{journal}{\emph{Lisp and Symbolic Computation}}
  \bibinfo{volume}{8}, \bibinfo{number}{4} (\bibinfo{year}{1995}),
  \bibinfo{pages}{343--355}.
\newblock


\bibitem[Wright and Felleisen(1994)]%
        {wright_syntactic_1994}
\bibfield{author}{\bibinfo{person}{Andrew~K. Wright} {and}
  \bibinfo{person}{Matthias Felleisen}.} \bibinfo{year}{1994}\natexlab{}.
\newblock \showarticletitle{A {Syntactic} {Approach} to {Type} {Soundness}}.
\newblock \bibinfo{journal}{\emph{Inf. Comput.}} \bibinfo{volume}{115},
  \bibinfo{number}{1} (\bibinfo{year}{1994}), \bibinfo{pages}{38--94}.
\newblock
\urldef\tempurl%
\url{https://doi.org/10.1006/inco.1994.1093}
\showDOI{\tempurl}


\bibitem[Xie et~al\mbox{.}(2020)]%
        {Xie/Brachthauser/Hillerstrom/Schuster/Leijen_2020_ICFP}
\bibfield{author}{\bibinfo{person}{Ningning Xie},
  \bibinfo{person}{Jonathan~Immanuel Brachth{\"{a}}user},
  \bibinfo{person}{Daniel Hillerstr{\"{o}}m}, \bibinfo{person}{Philipp
  Schuster}, {and} \bibinfo{person}{Daan Leijen}.}
  \bibinfo{year}{2020}\natexlab{}.
\newblock \showarticletitle{Effect handlers, evidently}.
\newblock \bibinfo{journal}{\emph{Proc. {ACM} Program. Lang.}}
  \bibinfo{volume}{4}, \bibinfo{number}{{ICFP}} (\bibinfo{year}{2020}),
  \bibinfo{pages}{99:1--99:29}.
\newblock
\urldef\tempurl%
\url{https://doi.org/10.1145/3408981}
\showDOI{\tempurl}


\bibitem[Xie et~al\mbox{.}(2022)]%
        {xie_first-class_2022}
\bibfield{author}{\bibinfo{person}{Ningning Xie}, \bibinfo{person}{Youyou
  Cong}, \bibinfo{person}{Kazuki Ikemori}, {and} \bibinfo{person}{Daan
  Leijen}.} \bibinfo{year}{2022}\natexlab{}.
\newblock \showarticletitle{First-class names for effect handlers}.
\newblock \bibinfo{journal}{\emph{Proc. ACM Program. Lang.}}
  \bibinfo{volume}{6}, \bibinfo{number}{OOPSLA2} (\bibinfo{year}{2022}),
  \bibinfo{pages}{30--59}.
\newblock
\urldef\tempurl%
\url{https://doi.org/10.1145/3563289}
\showDOI{\tempurl}


\bibitem[Zhang and Myers(2019)]%
        {Zhang/Myers_2019_POP}
\bibfield{author}{\bibinfo{person}{Yizhou Zhang} {and}
  \bibinfo{person}{Andrew~C. Myers}.} \bibinfo{year}{2019}\natexlab{}.
\newblock \showarticletitle{Abstraction-safe effect handlers via tunneling}.
\newblock \bibinfo{journal}{\emph{Proc. {ACM} Program. Lang.}}
  \bibinfo{volume}{3}, \bibinfo{number}{{POPL}} (\bibinfo{year}{2019}),
  \bibinfo{pages}{5:1--5:29}.
\newblock
\urldef\tempurl%
\url{https://doi.org/10.1145/3290318}
\showDOI{\tempurl}


\end{thebibliography}

\end{document}


\maketitle

\documentclass[]{paper}

\begin{document}

\section{Definitions}

\TS{
  \begin{itemize}
    \item The typing rule for operation clauses is too large and complex.
  \end{itemize}
}

\begin{remark}[Notation]
  We write $ \bm{ { \alpha } } ^ {  \mathit{I}  } $ for a finite sequence $\alpha_{{\mathrm{0}}}  \ottsym{,}  \ldots  \ottsym{,}  \alpha_{\ottmv{n}}$ with an index set $\mathit{I} = \{0, \ldots, n\}$, where $\alpha$ is any metavariable.
  We also write $\{   \bm{ { \alpha } } ^ {  \mathit{I}  }   \}$ for the set consisting of the elements of $ \bm{ { \alpha } } ^ {  \mathit{I}  } $.
\end{remark}

\begin{definition}[Kinds]
  Kinds are defined as $\ottnt{K} \Coloneqq  \mathbf{Typ}   \mid   \mathbf{Lab}   \mid   \mathbf{Eff} $.
\end{definition}


\begin{definition}[Signatures]\label{def:effsig}
  Given a set $S$ of label names, a label signature $ \Slabel $ is a functional relation whose domain $\labels$ is $S$.
  The codomain of $ \Slabel $ is the set of functional kinds of the form $ \Pi   _{ \ottmv{i}  \in  \mathit{I} }    \ottnt{K_{\ottmv{i}}}   \rightarrow   \mathbf{Lab} $ for some $\mathit{I}$ and $\ottnt{K}_i^{i \in \mathit{I}}$
  (if $\mathit{I} = \emptyset$, it means $ \mathbf{Lab} $ simply).
  Similarly, given a set $S$ of effect constructors, an effect signature $ \Sbase $ is a function relation whose domain $\EC$ is $S$ and
  its codomain is the set of functional kinds of the form $ \Pi   _{ \ottmv{i}  \in  \mathit{I} }    \ottnt{K_{\ottmv{i}}}   \rightarrow   \mathbf{Eff} $ for some $\mathit{I}$ and $\ottnt{K}_i^{i \in \mathit{I}}$.
  A signature $\Sigma$ is the disjoint union of a label signature and an effect signature.
  We write $ \Pi {\bm{ { \ottnt{K} } } }^{ \mathit{I} }   \rightarrow  \ottnt{K}$, and more simply, $ \Pi {\bm{ { \ottnt{K} } } }   \rightarrow  \ottnt{K}$
  as an abbreviation for $ \Pi   _{ \ottmv{i}  \in  \mathit{I} }    \ottnt{K_{\ottmv{i}}}   \rightarrow  \ottnt{K}$.
  %
\end{definition}

\begin{remark}
  We write $\mathcal{C} :  \Pi {\bm{ { \ottnt{K} } } }   \rightarrow  \ottnt{K}$ to denote the pair $\langle \mathcal{C},  \Pi {\bm{ { \ottnt{K} } } }   \rightarrow  \ottnt{K} \rangle$
  for label name or effect constructor $\ottnt{C}$.
\end{remark}

\begin{definition}[The Syntax of {\lang}]\label{def:syntax}
  Given a signature $\Sigma =  \Slabel   \uplus   \Sbase $,
  the syntax of {\lang} is defined as follows.
  \upshape
  \[
    \begin{array}{l@{\ \ }l@{\qquad}l@{\ \ }l}
      \mathit{I}, \mathit{J}, \mathit{N}                                        & \text{(index sets)}          &
      \ottmv{i}, \ottmv{j}, \ottmv{n}, \ottmv{r}                                 & \text{(indices)}               \\
      \mathit{f}, \mathit{g}, \mathit{x}, \mathit{y}, \mathit{z}, \mathit{p}, \mathit{k}            & \text{(variables)}           &
      \alpha, \beta, \gamma, \tau, \iota, \rho & \text{(typelike variables)}    \\
      \mathsf{op}                                                     & \text{(operation names)}     &
      \mathit{l} \in \labels                                          & \text{(label names)}           \\
      \mathcal{F} \in \EC                                              & \text{(effect constructors)} &
      \mathcal{C} \in \labels \cup \EC
    \end{array}
  \]
  \[
    \begin{array}{rcll}
      \ottnt{K}               & \Coloneqq &  \mathbf{Typ}   \mid   \mathbf{Lab}   \mid   \mathbf{Eff}                                                               & \text{(kinds)}                \\
      S, T        & \Coloneqq & \ottnt{A}  \mid  \ottnt{L}  \mid  \varepsilon                                                              & \text{(typelikes)}            \\
      \ottnt{A}, \ottnt{B}, \ottnt{C} & \Coloneqq & \tau  \mid   \ottnt{A}    \rightarrow_{ \varepsilon }    \ottnt{B}   \mid    \forall   \alpha  \ottsym{:}  \ottnt{K}   \ottsym{.}    \ottnt{A}    ^{ \varepsilon }                          & \text{(types)}                \\
      \ottnt{L}               & \Coloneqq & \iota  \mid  \mathit{l} \,  \bm{ { S } } ^ {  \mathit{I}  }                                                                        & \text{(labels)}               \\
      \varepsilon         & \Coloneqq & \rho  \mid  \mathcal{F} \,  \bm{ { S } } ^ {  \mathit{I}  }                                                                         & \text{(effects)}              \\
      \sigma             & \Coloneqq &  \{\}   \mid   \sigma   \uplus   \ottsym{\{}  \mathsf{op}  \ottsym{:}    \forall    {\bm{ \beta } }^{ \mathit{J} } : {\bm{ \ottnt{K} } }^{ \mathit{J} }    \ottsym{.}    \ottnt{A}   \Rightarrow   \ottnt{B}   \ottsym{\}}                             & \text{(operation signatures)} \\
      \Xi               & \Coloneqq &  \emptyset   \mid  \Xi  \ottsym{,}   \mathit{l}  ::    \forall    {\bm{ \alpha } }^{ \mathit{I} } : {\bm{ \ottnt{K} } }^{ \mathit{I} }    \ottsym{.}    \sigma                                              & \text{(effect contexts)}      \\
      \Gamma               & \Coloneqq &  \emptyset   \mid  \Gamma  \ottsym{,}  \mathit{x}  \ottsym{:}  \ottnt{A}  \mid  \Gamma  \ottsym{,}  \alpha  \ottsym{:}  \ottnt{K}                                              & \text{(typing contexts)}      \\
      \ottnt{e}               & \Coloneqq & \ottnt{v}  \mid  \ottnt{v_{{\mathrm{1}}}} \, \ottnt{v_{{\mathrm{2}}}}  \mid  \ottnt{v} \, S  \mid  \mathbf{let} \, \mathit{x}  \ottsym{=}  \ottnt{e_{{\mathrm{1}}}} \, \mathbf{in} \, \ottnt{e_{{\mathrm{2}}}}  \mid   \mathbf{handle}_{ \mathit{l} \,  \bm{ { S } } ^ {  \mathit{I}  }  }  \, \ottnt{e} \, \mathbf{with} \, \ottnt{h} & \text{(expressions)}          \\
      \ottnt{v}               & \Coloneqq & \mathit{x}  \mid  \ottkw{fun} \, \ottsym{(}  \mathit{f}  \ottsym{,}  \mathit{x}  \ottsym{,}  \ottnt{e}  \ottsym{)}  \mid  \Lambda  \alpha  \ottsym{:}  \ottnt{K}  \ottsym{.}  \ottnt{e}  \mid   \mathsf{op} _{ \mathit{l} \,  \bm{ { S } } ^ {  \mathit{I}  }  }  \,  \bm{ { T } } ^ {  \mathit{J}  }                                        & \text{(values)}                               \\
      \ottnt{h}               & \Coloneqq & \ottsym{\{} \, \mathbf{return} \, \mathit{x}  \mapsto  \ottnt{e}  \ottsym{\}}  \mid   \ottnt{h}   \uplus   \ottsym{\{}  \mathsf{op} \,  {\bm{ \beta } }^{ \mathit{J} } : {\bm{ \ottnt{K} } }^{ \mathit{J} }  \, \mathit{p} \, \mathit{k}  \mapsto  \ottnt{e}  \ottsym{\}}                                & \text{(handlers)}             \\
      \ottnt{E}               & \Coloneqq &  \Box   \mid  \mathbf{let} \, \mathit{x}  \ottsym{=}  \ottnt{E} \, \mathbf{in} \, \ottnt{e}  \mid   \mathbf{handle}_{ \mathit{l} \,  \bm{ { S } } ^ {  \mathit{I}  }  }  \, \ottnt{E} \, \mathbf{with} \, \ottnt{h}                              & \text{(evaluation contexts)}
    \end{array}
  \]
\end{definition}

\TY{General coercions are removed.

  \begin{definition}[The Syntax of {\lang} with General Coercions]
    \TY{I must change the metavariable $c$ used to denote general coercions.}
    The syntax of {\lang} extended by general coercions is the same as Definition~\ref{def:syntax} except for the following.
    \[
      \begin{array}{rclr}
        \ottnt{e} & \Coloneqq & \cdots  \mid  \langle  c  \rangle  \ottnt{e} & \text{(expressions)}         \\
        \ottnt{E} & \Coloneqq & \cdots  \mid  \langle  c  \rangle  \ottnt{E} & \text{(evaluation contexts)}
      \end{array}
    \]
  \end{definition}
}

\TY{Variable handlers are removed.
  \begin{definition}[The Syntax of {\lang} with Variable Handlers]
    The syntax of {\lang} extended by variable handlers is the same as Definition~\ref{def:syntax_lift} except for the following.
    \[
      \begin{array}{rclr}
        \ottnt{e} & \Coloneqq & \cdots  \mid   {\mathbf{handle}\mathrm{-}\mathbf{var} }_{ \ottnt{L} }  \, \ottnt{e} \, \mathbf{with} \,  \{ \mathbf{return}\, x \mapsto  \ottnt{e_{\ottmv{r}}}  \}  \uplus  \{ ( \mathit{o}  :  \alpha   \Rightarrow   \beta )\, p\, k \mapsto  \ottnt{e_{{\mathrm{0}}}}  \}  & \text{(expressions)}         \\
        \ottnt{E} & \Coloneqq & \cdots  \mid   {\mathbf{handle}\mathrm{-}\mathbf{var} }_{ \ottnt{L} }  \, \ottnt{E} \, \mathbf{with} \,  \{ \mathbf{return}\, x \mapsto  \ottnt{e_{\ottmv{r}}}  \}  \uplus  \{ ( \mathit{o}  :  \alpha   \Rightarrow   \beta )\, p\, k \mapsto  \ottnt{e_{{\mathrm{0}}}}  \}  & \text{(evaluation contexts)}
      \end{array}
    \]
  \end{definition}
}

\begin{remark}
  We write $\lambda  \mathit{x}  \ottsym{.}  \ottnt{e}$ for $\ottkw{fun} \, \ottsym{(}  \mathit{f}  \ottsym{,}  \mathit{x}  \ottsym{,}  \ottnt{e}  \ottsym{)}$ if variable $\mathit{f}$ does not occur in expression $\ottnt{e}$.
\end{remark}

\begin{definition}[Free Variables]\label{def:free_var}
  The notion of free variables is defined as usual.
  We write
  $ \mathrm{FV}   \ottsym{(}   \ottnt{e}   \ottsym{)} $ for the set of free variables in expression $\ottnt{e}$.
\end{definition}

\begin{definition}[Free Typelike Variables]\label{def:free_typelike_var}
  The notion of free typelike variables is defined as usual.
  We write $ \mathrm{FTV}   \ottsym{(}   \ottnt{e}   \ottsym{)} $ and $ \mathrm{FTV}   \ottsym{(}   S   \ottsym{)} $ for the sets of free typelike variables in expression $\ottnt{e}$ and typelike $S$, respectively.
\end{definition}

\begin{definition}[Value Substitution]\label{def:subst_value}
  Substitution $\ottnt{e} \,  \! [  \ottnt{v}  /  \mathit{x}  ] $ and $\ottnt{h} \,  \! [  \ottnt{v}  /  \mathit{x}  ] $ of value $\ottnt{v}$ for variable
  $\mathit{x}$ in expression $\ottnt{e}$ and handler $\ottnt{h}$, respectively, are defined as follows:
  \upshape
  \begin{align*}
    \mathit{x} \,  \! [  \ottnt{v}  /  \mathit{x}  ]                                         & = \ottnt{v}                                                                                                     \\
    \mathit{y} \,  \! [  \ottnt{v}  /  \mathit{x}  ]                                         & = \mathit{y} \quad (\tif  \mathit{x}   \neq   \mathit{y} )                                                                            \\
    \ottkw{fun} \, \ottsym{(}  \mathit{f}  \ottsym{,}  \mathit{y}  \ottsym{,}  \ottnt{e}  \ottsym{)} \,  \! [  \ottnt{v}  /  \mathit{x}  ]                             & = \ottkw{fun} \, \ottsym{(}  \mathit{f}  \ottsym{,}  \mathit{y}  \ottsym{,}  \ottnt{e} \,  \! [  \ottnt{v}  /  \mathit{x}  ]   \ottsym{)} \quad (\tif \mathit{f},  \mathit{y}   \notin     \mathrm{FV}   \ottsym{(}   \ottnt{v}   \ottsym{)}    \cup   \{  \mathit{x}  \}   )                                \\
    \ottsym{(}  \Lambda  \alpha  \ottsym{:}  \ottnt{K}  \ottsym{.}  \ottnt{e}  \ottsym{)} \,  \! [  \ottnt{v}  /  \mathit{x}  ]                                                            & = \Lambda  \alpha  \ottsym{:}  \ottnt{K}  \ottsym{.}  \ottnt{e} \,  \! [  \ottnt{v}  /  \mathit{x}  ]  \quad (\tif  \alpha   \notin    \mathrm{FTV}   \ottsym{(}   \ottnt{v}   \ottsym{)}  )                                                                                                              \\
     \mathsf{op} _{ \mathit{l} \,  \bm{ { S } } ^ {  \mathit{I}  }  }  \,  \bm{ { T } } ^ {  \mathit{J}  }  \,  \! [  \ottnt{v}  /  \mathit{x}  ]                       & =  \mathsf{op} _{ \mathit{l} \,  \bm{ { S } } ^ {  \mathit{I}  }  }  \,  \bm{ { T } } ^ {  \mathit{J}  }                                                                                   \\
     (  \ottnt{v_{{\mathrm{1}}}} \, \ottnt{v_{{\mathrm{2}}}}  )  \,  \! [  \ottnt{v}  /  \mathit{x}  ]                                   & =  (  \ottnt{v_{{\mathrm{1}}}} \,  \! [  \ottnt{v}  /  \mathit{x}  ]   )  \,  (  \ottnt{v_{{\mathrm{2}}}} \,  \! [  \ottnt{v}  /  \mathit{x}  ]   )                                                                                  \\
     (  \ottnt{v'} \, S  )  \,  \! [  \ottnt{v}  /  \mathit{x}  ]                                   & = \ottsym{(}  \ottnt{v'} \,  \! [  \ottnt{v}  /  \mathit{x}  ]   \ottsym{)} \, S                                                                                          \\
     (   \mathbf{handle}_{ \mathit{l} \,  \bm{ { S } } ^ {  \mathit{N}  }  }  \, \ottnt{e} \, \mathbf{with} \, \ottnt{h}  )  \,  \! [  \ottnt{v}  /  \mathit{x}  ]                & =  \mathbf{handle}_{ \mathit{l} \,  \bm{ { S } } ^ {  \mathit{N}  }  }  \, \ottnt{e} \,  \! [  \ottnt{v}  /  \mathit{x}  ]  \, \mathbf{with} \, \ottsym{(}  \ottnt{h} \,  \! [  \ottnt{v}  /  \mathit{x}  ]   \ottsym{)}                                                                  \\
     (  \mathbf{let} \, \mathit{y}  \ottsym{=}  \ottnt{e_{{\mathrm{1}}}} \, \mathbf{in} \, \ottnt{e_{{\mathrm{2}}}}  )  \,  \! [  \ottnt{v}  /  \mathit{x}  ]                       & = \mathbf{let} \, \mathit{y}  \ottsym{=}  \ottnt{e_{{\mathrm{1}}}} \,  \! [  \ottnt{v}  /  \mathit{x}  ]  \, \mathbf{in} \, \ottnt{e_{{\mathrm{2}}}} \,  \! [  \ottnt{v}  /  \mathit{x}  ]                                                                            \\
                                                        & \qquad (\tif  \mathit{y}   \neq   \mathit{x}  \tand  \mathit{y}   \notin    \mathrm{FV}   \ottsym{(}   \ottnt{v}   \ottsym{)}  )                                                           \\
     (   [  \ottnt{e}  ] _{ \ottnt{L} }   )  \,  \! [  \ottnt{v}  /  \mathit{x}  ]                                 & =  [  \ottnt{e} \,  \! [  \ottnt{v}  /  \mathit{x}  ]   ] _{ \ottnt{L} }                                                                                        \\
    \ottsym{\{} \, \mathbf{return} \, \mathit{y}  \mapsto  \ottnt{e_{\ottmv{r}}}  \ottsym{\}} \,  \! [  \ottnt{v}  /  \mathit{x}  ]                        & = \ottsym{\{} \, \mathbf{return} \, \mathit{y}  \mapsto  \ottnt{e_{\ottmv{r}}} \,  \! [  \ottnt{v}  /  \mathit{x}  ]   \ottsym{\}}                                                                            \\
                                                        & \qquad (\tif  \mathit{y}   \neq   \mathit{x}  \tand  \mathit{y}   \notin    \mathrm{FV}   \ottsym{(}   \ottnt{v}   \ottsym{)}  )                                                           \\
    \ottsym{(}   \ottnt{h}   \uplus   \ottsym{\{}  \mathsf{op} \,  {\bm{ \beta } }^{ \mathit{J} } : {\bm{ \ottnt{K} } }^{ \mathit{J} }  \, \mathit{p} \, \mathit{k}  \mapsto  \ottnt{e}  \ottsym{\}}   \ottsym{)} \,  \! [  \ottnt{v}  /  \mathit{x}  ]  & =  \ottnt{h} \,  \! [  \ottnt{v}  /  \mathit{x}  ]    \uplus   \ottsym{\{}  \mathsf{op} \,  {\bm{ \beta } }^{ \mathit{J} } : {\bm{ \ottnt{K} } }^{ \mathit{J} }  \, \mathit{p} \, \mathit{k}  \mapsto  \ottnt{e} \,  \! [  \ottnt{v}  /  \mathit{x}  ]   \ottsym{\}}                                                      \\
                                                        & \qquad (\tif  \mathit{x}   \neq   \mathit{p} , \mathit{k} \tand \mathit{p},  \mathit{k}   \notin    \mathrm{FV}   \ottsym{(}   \ottnt{v}   \ottsym{)}   \tand  \{   \bm{ { \beta } } ^ {  \mathit{J}  }   \}   \cap    \mathrm{FTV}   \ottsym{(}   \ottnt{v}   \ottsym{)}    \ottsym{=}  \emptyset)
  \end{align*}
\end{definition}

\begin{definition}[Typelike Substitution]\label{def:subst_typelike}
  Substitution $\ottnt{e} \,  \! [  S  /  \alpha   ] $, $\ottnt{h} \,  \! [  S  /  \alpha   ] $, $T \,  \! [  S  /  \alpha   ] $,
  and $\Gamma \,  \! [  S  /  \alpha   ] $ of typelike $S$ for typelike variable $\alpha$ in
  expression $\ottnt{e}$, handler $\ottnt{h}$, typelike $T$, and typing context
  $\Gamma$, respectively, are defined as follows:
  \upshape
  \begin{align*}
    \mathit{x} \,  \! [  S  /  \alpha   ]                                               & = \mathit{x}                                                                         \\
    \ottsym{(}  \ottkw{fun} \, \ottsym{(}  \mathit{f}  \ottsym{,}  \mathit{x}  \ottsym{,}  \ottnt{e}  \ottsym{)}  \ottsym{)} \,  \! [  S  /  \alpha   ]                                 & = \ottkw{fun} \, \ottsym{(}  \mathit{f}  \ottsym{,}  \mathit{x}  \ottsym{,}  \ottnt{e} \,  \! [  S  /  \alpha   ]   \ottsym{)}                                                   \\
    \ottsym{(}  \Lambda  \beta  \ottsym{:}  \ottnt{K}  \ottsym{.}  \ottnt{e}  \ottsym{)} \,  \! [  S  /  \alpha   ]  & = \Lambda  \beta  \ottsym{:}  \ottnt{K}  \ottsym{.}   (  \ottnt{e} \,  \! [  S  /  \alpha   ]   )  \quad (\tif  \alpha   \neq   \beta  \tand  \beta   \notin    \mathrm{FTV}   \ottsym{(}   S   \ottsym{)}   ) \\
    \ottsym{(}   \mathsf{op} _{ \mathit{l} \,  \bm{ { S' } } ^ {  \mathit{I}  }  }  \,  \bm{ { T } } ^ {  \mathit{J}  }   \ottsym{)} \,  \! [  S  /  \alpha   ]                           & =  \mathsf{op} _{ \mathit{l} \,  \bm{ { S' \,  \! [  S  /  \alpha   ]  } } ^ {  \mathit{I}  }  }  \,  \bm{ { T \,  \! [  S  /  \alpha   ]  } } ^ {  \mathit{J}  }                                     \\
     (  \ottnt{v_{{\mathrm{1}}}} \, \ottnt{v_{{\mathrm{2}}}}  )  \,  \! [  S  /  \alpha   ]                                         & =  (  \ottnt{v_{{\mathrm{1}}}} \,  \! [  S  /  \alpha   ]   )  \,  (  \ottnt{v_{{\mathrm{2}}}} \,  \! [  S  /  \alpha   ]   )                                              \\
     (  \ottnt{v} \, T  )  \,  \! [  S  /  \alpha   ]                                           & = \ottsym{(}  \ottnt{v} \,  \! [  S  /  \alpha   ]   \ottsym{)} \, \ottsym{(}  T \,  \! [  S  /  \alpha   ]   \ottsym{)}                                               \\
     (   \mathbf{handle}_{ \mathit{l} \,  \bm{ { T } } ^ {  \mathit{N}  }  }  \, \ottnt{e} \, \mathbf{with} \, \ottnt{h}  )  \,  \! [  S  /  \alpha   ]                       & =  \mathbf{handle}_{ \mathit{l} \,  \bm{ { T \,  \! [  S  /  \alpha   ]  } } ^ {  \mathit{N}  }  }  \, \ottnt{e} \,  \! [  S  /  \alpha   ]  \, \mathbf{with} \, \ottsym{(}  \ottnt{h} \,  \! [  S  /  \alpha   ]   \ottsym{)}                    \\
     (  \mathbf{let} \, \mathit{x}  \ottsym{=}  \ottnt{e_{{\mathrm{1}}}} \, \mathbf{in} \, \ottnt{e_{{\mathrm{2}}}}  )  \,  \! [  S  /  \alpha   ]                              & = \mathbf{let} \, \mathit{x}  \ottsym{=}  \ottnt{e_{{\mathrm{1}}}} \,  \! [  S  /  \alpha   ]  \, \mathbf{in} \, \ottnt{e_{{\mathrm{2}}}} \,  \! [  S  /  \alpha   ]                                       \\
     (   [  \ottnt{e}  ] _{ \ottnt{L} }   )  \,  \! [  S  /  \alpha   ]                                        & =  [  \ottnt{e} \,  \! [  S  /  \alpha   ]   ] _{ \ottnt{L} \,  \! [  S  /  \alpha   ]  }                                                \\
    \ottsym{\{} \, \mathbf{return} \, \mathit{x}  \mapsto  \ottnt{e_{\ottmv{r}}}  \ottsym{\}} \,  \! [  S  /  \alpha   ]                               & = \ottsym{\{} \, \mathbf{return} \, \mathit{x}  \mapsto  \ottnt{e_{\ottmv{r}}} \,  \! [  S  /  \alpha   ]   \ottsym{\}}                                              \\
    \ottsym{(}   \ottnt{h}   \uplus   \ottsym{\{}  \mathsf{op} \,  {\bm{ \beta } }^{ \mathit{J} } : {\bm{ \ottnt{K} } }^{ \mathit{J} }  \, \mathit{p} \, \mathit{k}  \mapsto  \ottnt{e}  \ottsym{\}}   \ottsym{)} \,  \! [  S  /  \alpha   ]         & =  \ottnt{h} \,  \! [  S  /  \alpha   ]    \uplus   \ottsym{\{}  \mathsf{op} \,  {\bm{ \beta } }^{ \mathit{J} } : {\bm{ \ottnt{K} } }^{ \mathit{J} }  \, \mathit{p} \, \mathit{k}  \mapsto  \ottnt{e} \,  \! [  S  /  \alpha   ]   \ottsym{\}}                  \\
                                                                  & \qquad (\tif  \{   \bm{ { \beta } } ^ {  \mathit{J}  }   \}   \cap   \ottsym{(}   \{  \alpha  \}   \cup    \mathrm{FTV}   \ottsym{(}   S   \ottsym{)}    \ottsym{)}   \ottsym{=}  \emptyset )          \\[2ex]
    \alpha \,  \! [  S  /  \alpha   ]                                           & = S                                                                         \\
    \beta \,  \! [  S  /  \alpha   ]                                            & = \beta \quad (\tif  \alpha   \neq   \beta )                                      \\
    \ottsym{(}   \ottnt{A}    \rightarrow_{ \varepsilon }    \ottnt{B}   \ottsym{)} \,  \! [  S  /  \alpha   ]                                & =  \ottsym{(}  \ottnt{A} \,  \! [  S  /  \alpha   ]   \ottsym{)}    \rightarrow_{ \varepsilon \,  \! [  S  /  \alpha   ]  }    \ottsym{(}  \ottnt{B} \,  \! [  S  /  \alpha   ]   \ottsym{)}                            \\
    \ottsym{(}    \forall   \beta  \ottsym{:}  \ottnt{K}   \ottsym{.}    \ottnt{A}    ^{ \varepsilon }    \ottsym{)} \,  \! [  S  /  \alpha   ]                 & =   \forall   \beta  \ottsym{:}  \ottnt{K}   \ottsym{.}    \ottnt{A} \,  \! [  S  /  \alpha   ]     ^{ \ottsym{(}  \varepsilon \,  \! [  S  /  \alpha   ]   \ottsym{)} }                         \\
                                                                  & \qquad (\tif  \alpha   \neq   \beta  \tand  \beta   \notin    \mathrm{FTV}   \ottsym{(}   S   \ottsym{)}   )                   \\
    \ottsym{(}  \mathcal{C} \,  \bm{ { T } } ^ {  \mathit{I}  }   \ottsym{)} \,  \! [  S  /  \alpha   ]                                        & = \mathcal{C} \,  \bm{ { T \,  \! [  S  /  \alpha   ]  } } ^ {  \mathit{I}  }                                                           \\[2ex]
     \{\}  \,  \! [  S  /  \alpha   ]                                        & =  \{\}                                                                  \\
    \ottsym{(}   \sigma   \uplus   \ottsym{\{}  \mathsf{op}  \ottsym{:}    \forall    {\bm{ \beta } }^{ \mathit{J} } : {\bm{ \ottnt{K} } }^{ \mathit{J} }    \ottsym{.}    \ottnt{A}   \Rightarrow   \ottnt{B}   \ottsym{\}}   \ottsym{)} \,  \! [  S  /  \alpha   ]  & =  \sigma \,  \! [  S  /  \alpha   ]    \uplus   \ottsym{\{}  \mathsf{op}  \ottsym{:}    \forall    {\bm{ \beta } }^{ \mathit{J} } : {\bm{ \ottnt{K} } }^{ \mathit{J} }    \ottsym{.}    \ottnt{A} \,  \! [  S  /  \alpha   ]    \Rightarrow   \ottnt{B} \,  \! [  S  /  \alpha   ]    \ottsym{\}}  \\
                                                                  & \qquad  (\tif  \{   \bm{ { \beta } } ^ {  \mathit{J}  }   \}   \cap    \mathrm{FTV}   \ottsym{(}   S   \ottsym{)}    \ottsym{=}  \emptyset)                            \\[2ex]
    \emptyset \,  \! [  S  /  \alpha   ]                                          & =  \emptyset                                                                    \\
    \ottsym{(}  \Gamma  \ottsym{,}  \mathit{x}  \ottsym{:}  \ottnt{A}  \ottsym{)} \,  \! [  S  /  \alpha   ]                                      & = \Gamma \,  \! [  S  /  \alpha   ]   \ottsym{,}  \mathit{x}  \ottsym{:}  \ottnt{A} \,  \! [  S  /  \alpha   ]                                               \\
    \ottsym{(}  \Gamma  \ottsym{,}  \beta  \ottsym{:}  \ottnt{K}  \ottsym{)} \,  \! [  S  /  \alpha   ]                                   & = \Gamma \,  \! [  S  /  \alpha   ]   \ottsym{,}  \beta  \ottsym{:}  \ottnt{K} \quad (\tif  \alpha   \neq   \beta )                    \\
  \end{align*}
\end{definition}

\begin{definition}[Typelike Extraction Function]\label{def:delta}
  A typelike context $ \Delta   \ottsym{(}   \Gamma   \ottsym{)} $ extracted from a typing context $\Gamma$ is defined as follows:
  \begin{align*}
     \Delta   \ottsym{(}   \emptyset   \ottsym{)}  =  \emptyset  \qquad  \Delta   \ottsym{(}   \Gamma  \ottsym{,}  \mathit{x}  \ottsym{:}  \ottnt{A}   \ottsym{)}  =  \Delta   \ottsym{(}   \Gamma   \ottsym{)}  \qquad  \Delta   \ottsym{(}   \Gamma  \ottsym{,}  \alpha  \ottsym{:}  \ottnt{K}   \ottsym{)}  =  \Delta   \ottsym{(}   \Gamma   \ottsym{)}   \ottsym{,}  \alpha  \ottsym{:}  \ottnt{K} ~.
  \end{align*}
\end{definition}

\begin{definition}[Domains of Typing Contexts]\label{def:domain}
  The set $ \mathrm{dom}   \ottsym{(}   \Gamma   \ottsym{)} $ of variables and typelike variables bound by a typing context $\Gamma$ is defined as follows:
  \begin{align*}
     \mathrm{dom}   \ottsym{(}   \emptyset   \ottsym{)}   \ottsym{=}  \emptyset \qquad  \mathrm{dom}   \ottsym{(}   \Gamma  \ottsym{,}  \mathit{x}  \ottsym{:}  \ottnt{A}   \ottsym{)}   \ottsym{=}    \mathrm{dom}   \ottsym{(}   \Gamma   \ottsym{)}    \cup   \{  \mathit{x}  \}  \qquad
     \mathrm{dom}   \ottsym{(}   \Gamma  \ottsym{,}  \alpha  \ottsym{:}  \ottnt{K}   \ottsym{)}   \ottsym{=}    \mathrm{dom}   \ottsym{(}   \Gamma   \ottsym{)}    \cup   \{  \alpha  \}  ~ . 
  \end{align*}
\end{definition}

\begin{definition}[Context Well-formedness and Kinding Rules]
  \addtolength{\jot}{0.5ex}
  \phantom{}\\
  \textnormal{\bfseries Contexts Well-formedness}\tquad\fbox{$\vdash  \Gamma$}
  \begin{mathpar}
    \inferrule{ }{\vdash  \emptyset}\ \rname{C}{Empty}

    \inferrule{ \mathit{x}   \notin    \mathrm{dom}   \ottsym{(}   \Gamma   \ottsym{)}   \\ \Gamma  \vdash  \ottnt{A}  \ottsym{:}   \mathbf{Typ} }{\vdash  \Gamma  \ottsym{,}  \mathit{x}  \ottsym{:}  \ottnt{A}}\ \rname{C}{Var}

    \inferrule{\vdash  \Gamma \\  \alpha   \notin    \mathrm{dom}   \ottsym{(}   \Gamma   \ottsym{)}  }{\vdash  \Gamma  \ottsym{,}  \alpha  \ottsym{:}  \ottnt{K}}\ \rname{C}{TVar}
  \end{mathpar}
  \textnormal{\bfseries Kinding}\tquad\fbox{$\Gamma  \vdash  S  \ottsym{:}  \ottnt{K}$}\tqquad\fbox{$\Gamma  \vdash   \bm{ { S } }^{ \mathit{I} } : \bm{ \ottnt{K} }^{ \mathit{I} } $} $\phantom{} \iff \forall \ottmv{i} \in \mathit{I} . (\Gamma  \vdash  S_{\ottmv{i}}  \ottsym{:}  \ottnt{K_{\ottmv{i}}})$
  \begin{mathpar}
    %
    %
    %
    \inferrule{\vdash  \Gamma \\  \alpha   \ottsym{:}   \ottnt{K}   \in   \Gamma }{\Gamma  \vdash  \alpha  \ottsym{:}  \ottnt{K}}\ \rname{K}{Var}

    \inferrule{\vdash  \Gamma \\  \mathcal{C}   \ottsym{:}    \Pi {\bm{ { \ottnt{K} } } }^{ \mathit{I} }   \rightarrow  \ottnt{K}   \in   \Sigma  \\ \Gamma  \vdash   \bm{ { S } }^{ \mathit{I} } : \bm{ \ottnt{K} }^{ \mathit{I} } }{\Gamma  \vdash  \mathcal{C} \,  \bm{ { S } } ^ {  \mathit{I}  }   \ottsym{:}  \ottnt{K}}\ \rname{K}{Cons}

    \inferrule{\Gamma  \vdash  \ottnt{A}  \ottsym{:}   \mathbf{Typ}  \\ \Gamma  \vdash  \varepsilon  \ottsym{:}   \mathbf{Eff}  \\ \Gamma  \vdash  \ottnt{B}  \ottsym{:}   \mathbf{Typ} }{\Gamma  \vdash   \ottnt{A}    \rightarrow_{ \varepsilon }    \ottnt{B}   \ottsym{:}   \mathbf{Typ} }\ \rname{K}{Fun}

    \inferrule{\Gamma  \ottsym{,}  \alpha  \ottsym{:}  \ottnt{K}  \vdash  \ottnt{A}  \ottsym{:}   \mathbf{Typ}  \\ \Gamma  \ottsym{,}  \alpha  \ottsym{:}  \ottnt{K}  \vdash  \varepsilon  \ottsym{:}   \mathbf{Eff} }{\Gamma  \vdash    \forall   \alpha  \ottsym{:}  \ottnt{K}   \ottsym{.}    \ottnt{A}    ^{ \varepsilon }    \ottsym{:}   \mathbf{Typ} }\ \rname{K}{Poly}
    %
  \end{mathpar}
\end{definition}

\begin{definition}[Proper Effect Contexts]\label{def:proper_effctx}
  An effect context $\Xi$ is proper if,
  for any $ \mathit{l}  ::    \forall    {\bm{ \alpha } }^{ \mathit{I} } : {\bm{ \ottnt{K} } }^{ \mathit{I} }    \ottsym{.}    \sigma    \in   \Xi $, the following holds:
  \begin{itemize}
    \item $ \mathit{l}   \ottsym{:}    \Pi {\bm{ { \ottnt{K} } } }^{ \mathit{I} }   \rightarrow   \mathbf{Lab}    \in   \Slabel $;
    \item for any $ \bm{ { \alpha_{{\mathrm{0}}} } } ^ {  \mathit{I_{{\mathrm{0}}}}  } $, $ {\bm{ { \ottnt{K_{{\mathrm{0}}}} } } }^{ \mathit{I_{{\mathrm{0}}}} } $, and $\sigma_{{\mathrm{0}}}$,
          if $ \mathit{l}  ::    \forall    {\bm{ \alpha_{{\mathrm{0}}} } }^{ \mathit{I_{{\mathrm{0}}}} } : {\bm{ \ottnt{K_{{\mathrm{0}}}} } }^{ \mathit{I_{{\mathrm{0}}}} }    \ottsym{.}    \sigma_{{\mathrm{0}}}    \in   \Xi $, then
          $ {\bm{ \alpha } }^{ \mathit{I} } : {\bm{ \ottnt{K} } }^{ \mathit{I} }  =  {\bm{ \alpha_{{\mathrm{0}}} } }^{ \mathit{I_{{\mathrm{0}}}} } : {\bm{ \ottnt{K_{{\mathrm{0}}}} } }^{ \mathit{I_{{\mathrm{0}}}} } $ and $\sigma = \sigma_{{\mathrm{0}}}$; and
    \item for any $ \mathsf{op}  \ottsym{:}    \forall    {\bm{ \beta } }^{ \mathit{J} } : {\bm{ \ottnt{K_{{\mathrm{0}}}} } }^{ \mathit{J} }    \ottsym{.}    \ottnt{A}   \Rightarrow   \ottnt{B}    \in   \sigma $,
          \[
             {\bm{ \alpha } }^{ \mathit{I} } : {\bm{ \ottnt{K} } }^{ \mathit{I} }   \ottsym{,}   {\bm{ \beta } }^{ \mathit{J} } : {\bm{ \ottnt{K_{{\mathrm{0}}}} } }^{ \mathit{J} }   \vdash  \ottnt{A}  \ottsym{:}   \mathbf{Typ}  \quad \tand \quad  {\bm{ \alpha } }^{ \mathit{I} } : {\bm{ \ottnt{K} } }^{ \mathit{I} }   \ottsym{,}   {\bm{ \beta } }^{ \mathit{J} } : {\bm{ \ottnt{K_{{\mathrm{0}}}} } }^{ \mathit{J} }   \vdash  \ottnt{B}  \ottsym{:}   \mathbf{Typ}  ~.
          \]
  \end{itemize}
\end{definition}

\TY{New definition of ARE}
\TS{TODO: Appending Relation on Effects (ARE) $\rightarrow$ Effect Algebra}

\begin{definition}[Well-Formedness-Preserving Functions]
  Given a signature $\Sigma$, a (possibly partial) function $f \in \KSet{\ottnt{K_{\ottmv{i}}}}{\Sigma}^{i \in \{ 1, \ldots, n\}} \rightharpoonup \KSet{\ottnt{K}}{\Sigma}$
  preserves well-formedness if
  \[
    \forall \Gamma, S_{{\mathrm{1}}}, \ldots, S_{\ottmv{n}} . \, \Gamma  \vdash  S_{{\mathrm{1}}}  \ottsym{:}  \ottnt{K_{{\mathrm{1}}}} \land \cdots \land \Gamma  \vdash  S_{\ottmv{n}}  \ottsym{:}  \ottnt{K_{\ottmv{n}}} \land f(S_{{\mathrm{1}}}, \ldots, S_{\ottmv{n}}) \in \KSet{\ottnt{K}}{\Sigma} \implies \Gamma \vdash f(S_{{\mathrm{1}}}, \ldots, S_{\ottmv{n}}) : \ottnt{K} ~.
  \]
  Similarly, $f \in \KSet{\ottnt{K}}{\Sigma}$ preserves well-formedness if $\Gamma \vdash f : \ottnt{K}$ for any $\Gamma$.
\end{definition}

\begin{definition}
  We write $ \alpha   \mapsto   T   \vdash    \bm{ { S } }   :  \ottnt{K_{{\mathrm{0}}}} $ for a quadruple
  $\langle \alpha, T,  \bm{ { S } } , \ottnt{K_{{\mathrm{0}}}} \rangle$ such that
  $\exists \Gamma_{{\mathrm{1}}}, \ottnt{K}, \Gamma_{{\mathrm{2}}} . \,
    (\forall S_{{\mathrm{0}}} \in  \bm{ { S } }  . \, \Gamma_{{\mathrm{1}}}  \ottsym{,}  \alpha  \ottsym{:}  \ottnt{K}  \ottsym{,}  \Gamma_{{\mathrm{2}}}  \vdash  S_{{\mathrm{0}}}  \ottsym{:}  \ottnt{K_{{\mathrm{0}}}}) \land \Gamma_{{\mathrm{1}}}  \vdash  T  \ottsym{:}  \ottnt{K}$.
\end{definition}

\begin{definition}[Effect algebras]\label{def:relation}
  {\sloppy{
  Given a label signature $ \Slabel $, an effect algebra is a quintuple $\langle  \Sbase ,  \odot ,  \bbZero ,  \lift{ \ottsym{-} } ,  \sim  \rangle$
  satisfying the following, where we let $\Sigma =  \Slabel   \uplus   \Sbase $.
  }}
  \begin{itemize}
    \item $ \odot  \in \EffSet{\Sigma} \times \EffSet{\Sigma} \rightharpoonup \EffSet{\Sigma}$,
          $ \bbZero  \in \EffSet{\Sigma}$, and
          $ \lift{ \ottsym{-} }  \in \LabelSet{\Sigma} \rightarrow \EffSet{\Sigma}$ preserve well-formedness.
          Furthermore,
          $ \sim $ is an equivalence relation on $\EffSet{\Sigma}$ and preserves well-formedness, that is,
          $\forall \varepsilon_{{\mathrm{1}}}, \varepsilon_{{\mathrm{2}}} . \,  \varepsilon_{{\mathrm{1}}}   \sim   \varepsilon_{{\mathrm{2}}}  \implies (\forall \Gamma . \, \Gamma  \vdash  \varepsilon_{{\mathrm{1}}}  \ottsym{:}   \mathbf{Eff}  \iff \Gamma  \vdash  \varepsilon_{{\mathrm{2}}}  \ottsym{:}   \mathbf{Eff} )$.

    \item  $\langle \EffSet{\Sigma},  \odot ,  \bbZero  \rangle$ is a partial
          monoid under $ \sim $, that is, the following holds:
          \begin{itemize}
            \item $\forall \varepsilon \in \EffSet{\Sigma} . \,
                      \varepsilon  \mathop{ \odot }   \bbZero     \sim   \varepsilon  \land    \bbZero   \mathop{ \odot }  \varepsilon    \sim   \varepsilon $; and

            \item $\forall \varepsilon_{{\mathrm{1}}}, \varepsilon_{{\mathrm{2}}}, \varepsilon_{{\mathrm{3}}} \in \EffSet{\Sigma} . \,$ \\
                  $ \ottsym{(}   \varepsilon_{{\mathrm{1}}}  \mathop{ \odot }  \varepsilon_{{\mathrm{2}}}   \ottsym{)}  \mathop{ \odot }  \varepsilon_{{\mathrm{3}}}  \in \EffSet{\Sigma} \lor
                     \varepsilon_{{\mathrm{1}}}  \mathop{ \odot }  \ottsym{(}   \varepsilon_{{\mathrm{2}}}  \mathop{ \odot }  \varepsilon_{{\mathrm{3}}}   \ottsym{)}  \in \EffSet{\Sigma}
                    \implies
                      \ottsym{(}   \varepsilon_{{\mathrm{1}}}  \mathop{ \odot }  \varepsilon_{{\mathrm{2}}}   \ottsym{)}  \mathop{ \odot }  \varepsilon_{{\mathrm{3}}}    \sim    \varepsilon_{{\mathrm{1}}}  \mathop{ \odot }  \ottsym{(}   \varepsilon_{{\mathrm{2}}}  \mathop{ \odot }  \varepsilon_{{\mathrm{3}}}   \ottsym{)}  $.
          \end{itemize}

    \item Typelike substitution respecting well-formedness is a homomorphism
          for $ \odot $, $ \lift{ \ottsym{-} } $, and $ \sim $, that is,
          the following holds:
          \begin{itemize}
            \item $\forall \alpha, S, \varepsilon_{{\mathrm{1}}}, \varepsilon_{{\mathrm{2}}} . \,
                     \alpha   \mapsto   S   \vdash   \varepsilon_{{\mathrm{1}}}  \ottsym{,}  \varepsilon_{{\mathrm{2}}}  :   \mathbf{Eff}   \land
                     \varepsilon_{{\mathrm{1}}}  \mathop{ \odot }  \varepsilon_{{\mathrm{2}}}  \in \EffSet{\Sigma}
                    \implies \ottsym{(}   \varepsilon_{{\mathrm{1}}}  \mathop{ \odot }  \varepsilon_{{\mathrm{2}}}   \ottsym{)} \,  \! [  S  /  \alpha   ]  =  \varepsilon_{{\mathrm{1}}} \,  \! [  S  /  \alpha   ]   \mathop{ \odot }  \varepsilon_{{\mathrm{2}}}  \,  \! [  S  /  \alpha   ] $;
            \item $\forall \alpha, S, \ottnt{L} . \,
                     \alpha   \mapsto   S   \vdash   \ottnt{L}  :   \mathbf{Lab}   \implies   \lift{ \ottnt{L} }   \,  \! [  S  /  \alpha   ]  =  \lift{  \ottnt{L} \,  \! [  S  /  \alpha   ]   } $; and
            \item $\forall \alpha, S, \varepsilon_{{\mathrm{1}}}, \varepsilon_{{\mathrm{2}}} . \,
                     \alpha   \mapsto   S   \vdash   \varepsilon_{{\mathrm{1}}}  \ottsym{,}  \varepsilon_{{\mathrm{2}}}  :   \mathbf{Eff}   \land  \varepsilon_{{\mathrm{1}}}   \sim   \varepsilon_{{\mathrm{2}}}  \implies  \varepsilon_{{\mathrm{1}}} \,  \! [  S  /  \alpha   ]    \sim   \varepsilon \,  \! [  S  /  \alpha   ]  $.
          \end{itemize}
  \end{itemize}
\end{definition}

\begin{remark}
  For readability, we introduce the following abbreviations.
  \begin{itemize}
    \item $ \varepsilon_{{\mathrm{1}}}  \olessthan  \varepsilon_{{\mathrm{2}}}  \defeq \exists \varepsilon . \,   \varepsilon_{{\mathrm{1}}}  \mathop{ \odot }  \varepsilon    \sim   \varepsilon_{{\mathrm{2}}} $ and
    \item $\Gamma  \vdash   \varepsilon_{{\mathrm{1}}}  \olessthan  \varepsilon_{{\mathrm{2}}}  \defeq \exists \varepsilon . \,   \varepsilon_{{\mathrm{1}}}  \mathop{ \odot }  \varepsilon    \sim   \varepsilon_{{\mathrm{2}}}  \land (\forall \varepsilon' \in \{ \varepsilon_{{\mathrm{1}}}, \varepsilon_{{\mathrm{2}}}, \varepsilon \} . \, \Gamma  \vdash  \varepsilon'  \ottsym{:}   \mathbf{Eff} )$.
  \end{itemize}



\end{remark}

\begin{remark}[Parameters of {\lang}]
  {\lang} takes
  a label signature in Definition~\ref{def:effsig},
  an effect algebra over that label signature in Definition~\ref{def:relation}, and
  an effect context as parameters.
\end{remark}

\begin{example}[Effect Signature for Effect Sets]
  The effect signature $\SbaseSet$ for effect sets consists of the pairs
  $\{  \}  \ottsym{:}   \mathbf{Eff} $, $\{  \ottsym{-}  \}  \ottsym{:}   \mathbf{Lab}   \rightarrow   \mathbf{Eff} $, and $ \ottsym{-}  \,\underline{ \cup }\,  \ottsym{-}   \ottsym{:}   \mathbf{Eff}   \times   \mathbf{Eff}   \rightarrow   \mathbf{Eff} $.
\end{example}

\begin{example}[Effect Signature for Effect Multisets]
  The effect signature $\SbaseMSet$ for effect multisets consists of the pairs
  $\{  \}  \ottsym{:}   \mathbf{Eff} $, $\{  \ottsym{-}  \}  \ottsym{:}   \mathbf{Lab}   \rightarrow   \mathbf{Eff} $, and $ \ottsym{-}  \,\underline{ \sqcup }\,  \ottsym{-}   \ottsym{:}   \mathbf{Eff}   \times   \mathbf{Eff}   \rightarrow   \mathbf{Eff} $.
\end{example}

\begin{example}[Effect Signature for Rows]
  The effect signature $\SbaseRow$ for both simple rows and scoped rows consists of the pairs
  $\langle  \rangle  \ottsym{:}   \mathbf{Eff} $ and $\langle  \ottsym{-}  \mid  \ottsym{-}  \rangle  \ottsym{:}   \mathbf{Lab}   \times   \mathbf{Eff}   \rightarrow   \mathbf{Eff} $.
\end{example}

\begin{example}[Effect Sets]\label{exa:effset}
  %
  An effect algebra {\eaSet} for effect sets is defined by
  $\langle \SbaseSet,  \ottsym{-}  \,\underline{ \cup }\,  \ottsym{-} , \{  \}, \{  \ottsym{-}  \},  \sim_{\eanameSet}  \rangle$
  where $ \sim_{\eanameSet} $ is the least equivalence relation satisfying the following rules:
  \begin{mathpar}
    \inferrule{ }{  \varepsilon  \,\underline{ \cup }\,  \{  \}     \sim_{\eanameSet}    \varepsilon }

    \inferrule{ }{  \varepsilon_{{\mathrm{1}}}  \,\underline{ \cup }\,  \varepsilon_{{\mathrm{2}}}     \sim_{\eanameSet}     \varepsilon_{{\mathrm{2}}}  \,\underline{ \cup }\,  \varepsilon_{{\mathrm{1}}}  }

    \inferrule{ }{  \varepsilon  \,\underline{ \cup }\,  \varepsilon     \sim_{\eanameSet}    \varepsilon }

    \inferrule{ }{  \ottsym{(}   \varepsilon_{{\mathrm{1}}}  \,\underline{ \cup }\,  \varepsilon_{{\mathrm{2}}}   \ottsym{)}  \,\underline{ \cup }\,  \varepsilon_{{\mathrm{3}}}     \sim_{\eanameSet}     \varepsilon_{{\mathrm{1}}}  \,\underline{ \cup }\,  \ottsym{(}   \varepsilon_{{\mathrm{2}}}  \,\underline{ \cup }\,  \varepsilon_{{\mathrm{3}}}   \ottsym{)}  }

    \inferrule{
     \varepsilon_{{\mathrm{1}}}    \sim_{\eanameSet}    \varepsilon_{{\mathrm{2}}}  \\  \varepsilon_{{\mathrm{3}}}    \sim_{\eanameSet}    \varepsilon_{{\mathrm{4}}} 
    }{
      \varepsilon_{{\mathrm{1}}}  \,\underline{ \cup }\,  \varepsilon_{{\mathrm{3}}}     \sim_{\eanameSet}     \varepsilon_{{\mathrm{2}}}  \,\underline{ \cup }\,  \varepsilon_{{\mathrm{4}}}  
    }
  \end{mathpar}
\end{example}

\begin{example}[Effect Multisets]\label{exa:effmultiset}
  %
  An effect algebra {\eaMSet} for effect multisets is defined by
  $\langle \SbaseMSet,  \ottsym{-}  \,\underline{ \sqcup }\,  \ottsym{-} , \{  \},\allowbreak \{  \ottsym{-}  \},  \sim_{\eanameMSet}  \rangle$
  where $ \sim_{\eanameMSet} $ is the least equivalence relation satisfying the following rules:
  \begin{mathpar}
    \inferrule{ }{  \varepsilon  \,\underline{ \sqcup }\,  \{  \}     \sim_{\eanameMSet}    \varepsilon }

    \inferrule{ }{  \varepsilon_{{\mathrm{1}}}  \,\underline{ \sqcup }\,  \varepsilon_{{\mathrm{2}}}     \sim_{\eanameMSet}     \varepsilon_{{\mathrm{2}}}  \,\underline{ \sqcup }\,  \varepsilon_{{\mathrm{1}}}  }

    \inferrule{ }{  \ottsym{(}   \varepsilon_{{\mathrm{1}}}  \,\underline{ \sqcup }\,  \varepsilon_{{\mathrm{2}}}   \ottsym{)}  \,\underline{ \sqcup }\,  \varepsilon_{{\mathrm{3}}}     \sim_{\eanameMSet}     \varepsilon_{{\mathrm{1}}}  \,\underline{ \sqcup }\,  \ottsym{(}   \varepsilon_{{\mathrm{2}}}  \,\underline{ \sqcup }\,  \varepsilon_{{\mathrm{3}}}   \ottsym{)}  }

    \inferrule{
     \varepsilon_{{\mathrm{1}}}    \sim_{\eanameMSet}    \varepsilon_{{\mathrm{2}}}  \\  \varepsilon_{{\mathrm{3}}}    \sim_{\eanameMSet}    \varepsilon_{{\mathrm{4}}} 
    }{
      \varepsilon_{{\mathrm{1}}}  \,\underline{ \sqcup }\,  \varepsilon_{{\mathrm{3}}}     \sim_{\eanameMSet}     \varepsilon_{{\mathrm{2}}}  \,\underline{ \sqcup }\,  \varepsilon_{{\mathrm{4}}}  
    }
  \end{mathpar}
\end{example}

\begin{example}[Simple Rows]\label{exa:eff_simple_row}
  %
  An effect algebra {\eaSimpleRow} for simple rows is defined by
  $\langle \SbaseRow,  \odot_\mathrm{SimpR} , \langle  \rangle, \langle  \ottsym{-}  \mid  \langle  \rangle  \rangle,  \sim_{\eanameSimpleRow}  \rangle$
  where
  \begin{align*}
     \varepsilon_{{\mathrm{1}}}  \mathop{  \odot_\mathrm{SimpR}  }  \varepsilon_{{\mathrm{2}}}  \defeq
    \begin{dcases}
      \, \langle \ottnt{L_{{\mathrm{1}}}} \mid \langle \cdots \langle  \ottnt{L_{\ottmv{n}}}  \mid  \varepsilon_{{\mathrm{2}}}  \rangle \rangle \rangle
       & (\tif \varepsilon_{{\mathrm{1}}} = \langle \ottnt{L_{{\mathrm{1}}}} \mid \langle \cdots \langle  \ottnt{L_{\ottmv{n}}}  \mid  \langle  \rangle  \rangle \rangle \rangle )                           \\
      \, \varepsilon_{{\mathrm{1}}}
       & (\tif \varepsilon_{{\mathrm{1}}} = \langle \ottnt{L_{{\mathrm{1}}}} \mid \langle \cdots \langle  \ottnt{L_{\ottmv{n}}}  \mid  \rho  \rangle \rangle \rangle \tand \varepsilon_{{\mathrm{2}}} = \langle  \rangle ) \\
      \mathit{undefined}
       & (\mathrm{otherwise})
    \end{dcases}
  \end{align*}
  and $ \sim_{\eanameSimpleRow} $ is the least equivalence relation satisfying the following.
  \begin{mathpar}
    \inferrule{ \varepsilon_{{\mathrm{1}}}    \sim_{\eanameSimpleRow}    \varepsilon_{{\mathrm{2}}} }{ \langle  \ottnt{L}  \mid  \varepsilon_{{\mathrm{1}}}  \rangle    \sim_{\eanameSimpleRow}    \langle  \ottnt{L}  \mid  \varepsilon_{{\mathrm{2}}}  \rangle }

    \inferrule{ \ottnt{L_{{\mathrm{1}}}}   \neq   \ottnt{L_{{\mathrm{2}}}} }{ \langle  \ottnt{L_{{\mathrm{1}}}}  \mid  \langle  \ottnt{L_{{\mathrm{2}}}}  \mid  \varepsilon  \rangle  \rangle    \sim_{\eanameSimpleRow}    \langle  \ottnt{L_{{\mathrm{2}}}}  \mid  \langle  \ottnt{L_{{\mathrm{1}}}}  \mid  \varepsilon  \rangle  \rangle }

    \inferrule{ }{ \langle  \ottnt{L}  \mid  \varepsilon  \rangle    \sim_{\eanameSimpleRow}    \langle  \ottnt{L}  \mid  \langle  \ottnt{L}  \mid  \varepsilon  \rangle  \rangle }
  \end{mathpar}
\end{example}

\begin{example}[Scoped Rows]\label{exa:effrow}
  %
  An effect algebra {\eaScopedRow} for scoped rows is defined by
  $\langle \SbaseRow,  \odot_\mathrm{ScpR} , \langle  \rangle, \langle  \ottsym{-}  \mid  \langle  \rangle  \rangle,  \sim_{\eanameScopedRow}  \rangle$
  where
  \begin{align*}
     \varepsilon_{{\mathrm{1}}}  \mathop{  \odot_\mathrm{ScpR}  }  \varepsilon_{{\mathrm{2}}}  \defeq
    \begin{dcases}
      \, \langle \ottnt{L_{{\mathrm{1}}}} \mid \langle \cdots \langle  \ottnt{L_{\ottmv{n}}}  \mid  \varepsilon_{{\mathrm{2}}}  \rangle \rangle \rangle
       & (\tif \varepsilon_{{\mathrm{1}}} = \langle \ottnt{L_{{\mathrm{1}}}} \mid \langle \cdots \langle  \ottnt{L_{\ottmv{n}}}  \mid  \langle  \rangle  \rangle \rangle \rangle )                           \\
      \, \varepsilon_{{\mathrm{1}}}
       & (\tif \varepsilon_{{\mathrm{1}}} = \langle \ottnt{L_{{\mathrm{1}}}} \mid \langle \cdots \langle  \ottnt{L_{\ottmv{n}}}  \mid  \rho  \rangle \rangle \rangle \tand \varepsilon_{{\mathrm{2}}} = \langle  \rangle ) \\
      \mathit{undefined}
       & (\mathrm{otherwise})
    \end{dcases}
  \end{align*}
  and $ \sim_{\eanameScopedRow} $ is the least equivalence relation satisfying the following.
  \begin{mathpar}
    \inferrule{ \varepsilon_{{\mathrm{1}}}    \sim_{\eanameScopedRow}    \varepsilon_{{\mathrm{2}}} }{ \langle  \ottnt{L}  \mid  \varepsilon_{{\mathrm{1}}}  \rangle    \sim_{\eanameScopedRow}    \langle  \ottnt{L}  \mid  \varepsilon_{{\mathrm{2}}}  \rangle }

    \inferrule{ \ottnt{L_{{\mathrm{1}}}}   \neq   \ottnt{L_{{\mathrm{2}}}} }{ \langle  \ottnt{L_{{\mathrm{1}}}}  \mid  \langle  \ottnt{L_{{\mathrm{2}}}}  \mid  \varepsilon  \rangle  \rangle    \sim_{\eanameScopedRow}    \langle  \ottnt{L_{{\mathrm{2}}}}  \mid  \langle  \ottnt{L_{{\mathrm{1}}}}  \mid  \varepsilon  \rangle  \rangle }
  \end{mathpar}
\end{example}

\begin{example}[Erasable Sets]\label{exa:effset_erasure}
  %
  An effect algebra {\eaEraseSet} for effect sets is defined by
  $\langle \SbaseSet,  \ottsym{-}  \,\underline{ \cup }\,  \ottsym{-} , \{  \}, \{  \ottsym{-}  \},  \sim_\mathrm{ESet}  \rangle$
  where $ \sim_\mathrm{ESet} $ is the least equivalence relation satisfying the following rules:
  \begin{mathpar}
    \inferrule{
     \mathit{l_{{\mathrm{1}}}}   \neq   \mathit{l_{{\mathrm{2}}}} 
    }{
      \{  \mathit{l_{{\mathrm{1}}}} \,  \bm{ { S_{{\mathrm{1}}} } } ^ {  \mathit{I_{{\mathrm{1}}}}  }   \}  \,\underline{ \cup }\,  \{  \mathit{l_{{\mathrm{2}}}} \,  \bm{ { S_{{\mathrm{2}}} } } ^ {  \mathit{I_{{\mathrm{2}}}}  }   \}     \sim_\mathrm{ESet}     \{  \mathit{l_{{\mathrm{2}}}} \,  \bm{ { S_{{\mathrm{2}}} } } ^ {  \mathit{I_{{\mathrm{2}}}}  }   \}  \,\underline{ \cup }\,  \{  \mathit{l_{{\mathrm{1}}}} \,  \bm{ { S_{{\mathrm{1}}} } } ^ {  \mathit{I_{{\mathrm{1}}}}  }   \}  
    }

    \inferrule{ }{  \{  \mathit{l} \,  \bm{ { S_{{\mathrm{1}}} } } ^ {  \mathit{I_{{\mathrm{1}}}}  }   \}  \,\underline{ \cup }\,  \{  \mathit{l} \,  \bm{ { S_{{\mathrm{2}}} } } ^ {  \mathit{I_{{\mathrm{2}}}}  }   \}     \sim_\mathrm{ESet}    \{  \mathit{l} \,  \bm{ { S_{{\mathrm{1}}} } } ^ {  \mathit{I_{{\mathrm{1}}}}  }   \} }

    \inferrule{ }{  \varepsilon  \,\underline{ \cup }\,  \{  \}     \sim_\mathrm{ESet}    \varepsilon }

    \inferrule{ }{  \{  \}  \,\underline{ \cup }\,  \varepsilon     \sim_\mathrm{ESet}    \varepsilon }

    \inferrule{ }{  \ottsym{(}   \varepsilon_{{\mathrm{1}}}  \,\underline{ \cup }\,  \varepsilon_{{\mathrm{2}}}   \ottsym{)}  \,\underline{ \cup }\,  \varepsilon_{{\mathrm{3}}}     \sim_\mathrm{ESet}     \varepsilon_{{\mathrm{1}}}  \,\underline{ \cup }\,  \ottsym{(}   \varepsilon_{{\mathrm{2}}}  \,\underline{ \cup }\,  \varepsilon_{{\mathrm{3}}}   \ottsym{)}  }

    \inferrule{
     \varepsilon_{{\mathrm{1}}}    \sim_\mathrm{ESet}    \varepsilon_{{\mathrm{2}}}  \\  \varepsilon_{{\mathrm{3}}}    \sim_\mathrm{ESet}    \varepsilon_{{\mathrm{4}}} 
    }{
      \varepsilon_{{\mathrm{1}}}  \,\underline{ \cup }\,  \varepsilon_{{\mathrm{3}}}     \sim_\mathrm{ESet}     \varepsilon_{{\mathrm{2}}}  \,\underline{ \cup }\,  \varepsilon_{{\mathrm{4}}}  
    }
  \end{mathpar}
\end{example}

\begin{example}[Erasable Multisets]\label{exa:effmultiset_erasure}
  %
  An effect algebra {\eaEraseMSet} for effect multisets is defined by
  $\langle \SbaseMSet,  \ottsym{-}  \,\underline{ \sqcup }\,  \ottsym{-} , \{  \},\allowbreak \{  \ottsym{-}  \},  \sim_\mathrm{EMSet}  \rangle$
  where $ \sim_\mathrm{EMSet} $ is the least equivalence relation satisfying the following rules:
  \begin{mathpar}
    \inferrule{
     \mathit{l_{{\mathrm{1}}}}   \neq   \mathit{l_{{\mathrm{2}}}} 
    }{
      \{  \mathit{l_{{\mathrm{1}}}} \,  \bm{ { S_{{\mathrm{1}}} } } ^ {  \mathit{I_{{\mathrm{1}}}}  }   \}  \,\underline{ \sqcup }\,  \{  \mathit{l_{{\mathrm{2}}}} \,  \bm{ { S_{{\mathrm{2}}} } } ^ {  \mathit{I_{{\mathrm{2}}}}  }   \}     \sim_\mathrm{EMSet}     \{  \mathit{l_{{\mathrm{2}}}} \,  \bm{ { S_{{\mathrm{2}}} } } ^ {  \mathit{I_{{\mathrm{2}}}}  }   \}  \,\underline{ \sqcup }\,  \{  \mathit{l_{{\mathrm{1}}}} \,  \bm{ { S_{{\mathrm{1}}} } } ^ {  \mathit{I_{{\mathrm{1}}}}  }   \}  
    }

    \inferrule{ }{  \varepsilon  \,\underline{ \sqcup }\,  \{  \}     \sim_\mathrm{EMSet}    \varepsilon }

    \inferrule{ }{  \{  \}  \,\underline{ \sqcup }\,  \varepsilon     \sim_\mathrm{EMSet}    \varepsilon }

    \inferrule{ }{  \ottsym{(}   \varepsilon_{{\mathrm{1}}}  \,\underline{ \sqcup }\,  \varepsilon_{{\mathrm{2}}}   \ottsym{)}  \,\underline{ \sqcup }\,  \varepsilon_{{\mathrm{3}}}     \sim_\mathrm{EMSet}     \varepsilon_{{\mathrm{1}}}  \,\underline{ \sqcup }\,  \ottsym{(}   \varepsilon_{{\mathrm{2}}}  \,\underline{ \sqcup }\,  \varepsilon_{{\mathrm{3}}}   \ottsym{)}  }

    \inferrule{
     \varepsilon_{{\mathrm{1}}}    \sim_\mathrm{EMSet}    \varepsilon_{{\mathrm{2}}}  \\  \varepsilon_{{\mathrm{3}}}    \sim_\mathrm{EMSet}    \varepsilon_{{\mathrm{4}}} 
    }{
      \varepsilon_{{\mathrm{1}}}  \,\underline{ \sqcup }\,  \varepsilon_{{\mathrm{3}}}     \sim_\mathrm{EMSet}     \varepsilon_{{\mathrm{2}}}  \,\underline{ \sqcup }\,  \varepsilon_{{\mathrm{4}}}  
    }
  \end{mathpar}
\end{example}

\begin{example}[Erasable Simple Rows]\label{exa:eff_simple_row_erasure}
  %
  An effect algebra {\eaEraseSimpleRow} for erasable simple rows is defined by
  $\langle \SbaseRow,  \odot_\mathrm{ESimpR} , \langle  \rangle, \langle  \ottsym{-}  \mid  \langle  \rangle  \rangle,  \sim_\mathrm{ESimpR}  \rangle$
  where
  \begin{align*}
     \varepsilon_{{\mathrm{1}}}  \mathop{  \odot_\mathrm{ESimpR}  }  \varepsilon_{{\mathrm{2}}}  \defeq
    \begin{dcases}
      \, \langle \ottnt{L_{{\mathrm{1}}}} \mid \langle \cdots \langle  \ottnt{L_{\ottmv{n}}}  \mid  \varepsilon_{{\mathrm{2}}}  \rangle \rangle \rangle
       & (\tif \varepsilon_{{\mathrm{1}}} = \langle \ottnt{L_{{\mathrm{1}}}} \mid \langle \cdots \langle  \ottnt{L_{\ottmv{n}}}  \mid  \langle  \rangle  \rangle \rangle \rangle )                           \\
      \, \varepsilon_{{\mathrm{1}}}
       & (\tif \varepsilon_{{\mathrm{1}}} = \langle \ottnt{L_{{\mathrm{1}}}} \mid \langle \cdots \langle  \ottnt{L_{\ottmv{n}}}  \mid  \rho  \rangle \rangle \rangle \tand \varepsilon_{{\mathrm{2}}} = \langle  \rangle ) \\
      \mathit{undefined}
       & (\mathrm{otherwise})
    \end{dcases}
  \end{align*}
  and $ \sim_\mathrm{ESimpR} $ is the least equivalence relation satisfying the following.
  \begin{mathpar}
    \inferrule{ \varepsilon_{{\mathrm{1}}}    \sim_{\eanameSimpleRow}    \varepsilon_{{\mathrm{2}}} }{ \langle  \ottnt{L}  \mid  \varepsilon_{{\mathrm{1}}}  \rangle    \sim_{\eanameSimpleRow}    \langle  \ottnt{L}  \mid  \varepsilon_{{\mathrm{2}}}  \rangle }

    \inferrule{ \mathit{l_{{\mathrm{1}}}}   \neq   \mathit{l_{{\mathrm{2}}}} }{ \langle  \mathit{l_{{\mathrm{1}}}} \,  \bm{ { S_{{\mathrm{1}}} } } ^ {  \mathit{I_{{\mathrm{1}}}}  }   \mid  \langle  \mathit{l_{{\mathrm{2}}}} \,  \bm{ { S_{{\mathrm{2}}} } } ^ {  \mathit{I_{{\mathrm{2}}}}  }   \mid  \varepsilon  \rangle  \rangle    \sim_{\eanameSimpleRow}    \langle  \mathit{l_{{\mathrm{2}}}} \,  \bm{ { S_{{\mathrm{2}}} } } ^ {  \mathit{I_{{\mathrm{2}}}}  }   \mid  \langle  \mathit{l_{{\mathrm{1}}}} \,  \bm{ { S_{{\mathrm{1}}} } } ^ {  \mathit{I_{{\mathrm{1}}}}  }   \mid  \varepsilon  \rangle  \rangle }

    \inferrule{ }{ \langle  \mathit{l} \,  \bm{ { S_{{\mathrm{1}}} } } ^ {  \mathit{I_{{\mathrm{1}}}}  }   \mid  \varepsilon  \rangle    \sim_{\eanameSimpleRow}    \langle  \mathit{l} \,  \bm{ { S_{{\mathrm{1}}} } } ^ {  \mathit{I_{{\mathrm{1}}}}  }   \mid  \langle  \mathit{l} \,  \bm{ { S_{{\mathrm{2}}} } } ^ {  \mathit{I_{{\mathrm{2}}}}  }   \mid  \varepsilon  \rangle  \rangle }
  \end{mathpar}
\end{example}

\begin{example}[Erasable Scoped Rows]\label{exa:effrow_erasure}
  %
  An effect algebra {\eaEraseScopedRow} for scoped rows is defined by
  $\langle \SbaseRow,  \odot_\mathrm{EScpR} ,\allowbreak \langle  \rangle,\allowbreak \langle  \ottsym{-}  \mid  \langle  \rangle  \rangle,  \sim_\mathrm{EScpR}  \rangle$
  where
  \begin{align*}
     \varepsilon_{{\mathrm{1}}}  \mathop{  \odot_\mathrm{EScpR}  }  \varepsilon_{{\mathrm{2}}}  \defeq
    \begin{dcases}
      \, \langle \ottnt{L_{{\mathrm{1}}}} \mid \langle \cdots \langle  \ottnt{L_{\ottmv{n}}}  \mid  \varepsilon_{{\mathrm{2}}}  \rangle \rangle \rangle
       & (\tif \varepsilon_{{\mathrm{1}}} = \langle \ottnt{L_{{\mathrm{1}}}} \mid \langle \cdots \langle  \ottnt{L_{\ottmv{n}}}  \mid  \langle  \rangle  \rangle \rangle \rangle )                           \\
      \, \varepsilon_{{\mathrm{1}}}
       & (\tif \varepsilon_{{\mathrm{1}}} = \langle \ottnt{L_{{\mathrm{1}}}} \mid \langle \cdots \langle  \ottnt{L_{\ottmv{n}}}  \mid  \rho  \rangle \rangle \rangle \tand \varepsilon_{{\mathrm{2}}} = \langle  \rangle ) \\
      \mathit{undefined}
       & (\mathrm{otherwise})
    \end{dcases}
  \end{align*}
  and $ \sim_\mathrm{EScpR} $ is the least equivalence relation satisfying the following.
  \begin{mathpar}
    \inferrule{ \varepsilon_{{\mathrm{1}}}    \sim_\mathrm{EScpR}    \varepsilon_{{\mathrm{2}}} }{ \langle  \ottnt{L}  \mid  \varepsilon_{{\mathrm{1}}}  \rangle    \sim_\mathrm{EScpR}    \langle  \ottnt{L}  \mid  \varepsilon_{{\mathrm{2}}}  \rangle }

    \inferrule{ \mathit{l_{{\mathrm{1}}}}   \neq   \mathit{l_{{\mathrm{2}}}} }{ \langle  \mathit{l_{{\mathrm{1}}}} \,  \bm{ { S_{{\mathrm{1}}} } } ^ {  \mathit{I_{{\mathrm{1}}}}  }   \mid  \langle  \mathit{l_{{\mathrm{2}}}} \,  \bm{ { S_{{\mathrm{2}}} } } ^ {  \mathit{I_{{\mathrm{2}}}}  }   \mid  \varepsilon  \rangle  \rangle    \sim_\mathrm{EScpR}    \langle  \mathit{l_{{\mathrm{2}}}} \,  \bm{ { S_{{\mathrm{2}}} } } ^ {  \mathit{I_{{\mathrm{2}}}}  }   \mid  \langle  \mathit{l_{{\mathrm{1}}}} \,  \bm{ { S_{{\mathrm{1}}} } } ^ {  \mathit{I_{{\mathrm{1}}}}  }   \mid  \varepsilon  \rangle  \rangle }
  \end{mathpar}
\end{example}

\begin{definition}[Freeness of Labels]\label{def:freeness}
  \phantom{}\\
  \textnormal{\bfseries Freeness of labels}\tquad\fbox{$ \mathit{n}  \mathrm{-free} ( \ottnt{L} ,  \ottnt{E} ) $}
  \begin{mathpar}
    \inferrule{
    }{
       0  \mathrm{-free} ( \ottnt{L} ,  \Box ) 
    }

    \inferrule{
       \mathit{n}  \mathrm{-free} ( \ottnt{L} ,  \ottnt{E} ) 
    }{
       \mathit{n}  \mathrm{-free} ( \ottnt{L} ,  \mathbf{let} \, \mathit{x}  \ottsym{=}  \ottnt{E} \, \mathbf{in} \, \ottnt{e} ) 
    }

    \inferrule{
       \mathit{n}  \ottsym{+}  1  \mathrm{-free} ( \ottnt{L} ,  \ottnt{E} ) 
    }{
       \mathit{n}  \mathrm{-free} ( \ottnt{L} ,   \mathbf{handle}_{ \ottnt{L} }  \, \ottnt{E} \, \mathbf{with} \, \ottnt{h} ) 
    }

    \inferrule{
     \mathit{n}  \mathrm{-free} ( \ottnt{L} ,  \ottnt{E} )  \\ \ottnt{L} \neq \ottnt{L'}
    }{
     \mathit{n}  \mathrm{-free} ( \ottnt{L} ,   \mathbf{handle}_{ \ottnt{L'} }  \, \ottnt{E} \, \mathbf{with} \, \ottnt{h} ) 
    }
  \end{mathpar}
\end{definition}

\begin{definition}[Operational Semantics]\label{def:semantics}
  \phantom{}\\
  \textnormal{\bfseries Reduction}\tquad\fbox{$\ottnt{e}  \longmapsto  \ottnt{e'}$}
  \begin{mathpar}
    \inferrule{
    }{
      \ottkw{fun} \, \ottsym{(}  \mathit{f}  \ottsym{,}  \mathit{x}  \ottsym{,}  \ottnt{e}  \ottsym{)} \, \ottnt{v}  \longmapsto  \ottnt{e} \,  \! [  \ottkw{fun} \, \ottsym{(}  \mathit{f}  \ottsym{,}  \mathit{x}  \ottsym{,}  \ottnt{e}  \ottsym{)}  /  \mathit{f}  ]  \,  \! [  \ottnt{v}  /  \mathit{x}  ] 
    } \ \rname{R}{App}

    \inferrule{
    }{
      \ottsym{(}  \Lambda  \alpha  \ottsym{:}  \ottnt{K}  \ottsym{.}  \ottnt{e}  \ottsym{)} \, S  \longmapsto  \ottnt{e} \,  \! [  S  /  \alpha   ] 
    } \ \rname{R}{TApp}
    %

    \inferrule{
    }{
      \mathbf{let} \, \mathit{x}  \ottsym{=}  \ottnt{v} \, \mathbf{in} \, \ottnt{e}  \longmapsto  \ottnt{e} \,  \! [  \ottnt{v}  /  \mathit{x}  ] 
    } \ \rname{R}{Let}

    \inferrule{
       \mathbf{return} \, \mathit{x}  \mapsto  \ottnt{e_{\ottmv{r}}}   \in   \ottnt{h} 
    }{
       \mathbf{handle}_{ \mathit{l} \,  \bm{ { S } } ^ {  \mathit{I}  }  }  \, \ottnt{v} \, \mathbf{with} \, \ottnt{h}  \longmapsto  \ottnt{e_{\ottmv{r}}} \,  \! [  \ottnt{v}  /  \mathit{x}  ] 
    } \ \rname{R}{Handle1}

    \inferrule{
     \mathsf{op} \,  {\bm{ \beta } }^{ \mathit{J} } : {\bm{ \ottnt{K} } }^{ \mathit{J} }  \, \mathit{p} \, \mathit{k}  \mapsto  \ottnt{e}   \in   \ottnt{h}  \\
    \ottnt{v_{\ottmv{cont}}} = \lambda  \mathit{z}  \ottsym{.}   \mathbf{handle}_{ \mathit{l} \,  \bm{ { S } } ^ {  \mathit{I}  }  }  \, \ottnt{E}  \ottsym{[}  \mathit{z}  \ottsym{]} \, \mathbf{with} \, \ottnt{h} \\  0  \mathrm{-free} ( \mathit{l} \,  \bm{ { S } } ^ {  \mathit{I}  }  ,  \ottnt{E} ) 
    }{
     \mathbf{handle}_{ \mathit{l} \,  \bm{ { S } } ^ {  \mathit{I}  }  }  \, \ottnt{E}  \ottsym{[}   \mathsf{op} _{ \mathit{l} \,  \bm{ { S } } ^ {  \mathit{I}  }  }  \,  \bm{ { T } } ^ {  \mathit{J}  }  \, \ottnt{v}  \ottsym{]} \, \mathbf{with} \, \ottnt{h}  \longmapsto  \ottnt{e} \,  \! [ {\bm{ { T } } }^{ \mathit{J} } / {\bm{ \beta } }^{ \mathit{J} } ]  \,  \! [  \ottnt{v}  /  \mathit{p}  ]  \,  \! [  \ottnt{v_{\ottmv{cont}}}  /  \mathit{k}  ] 
    }\ \rname{R}{Handle2}
  \end{mathpar}
  \textnormal{\bfseries Evaluation}\tquad\fbox{$\ottnt{e}  \longrightarrow  \ottnt{e'}$}
  \begin{mathpar}
    \inferrule{
      \ottnt{e_{{\mathrm{1}}}}  \longmapsto  \ottnt{e_{{\mathrm{2}}}}
    }{
      \ottnt{E}  \ottsym{[}  \ottnt{e_{{\mathrm{1}}}}  \ottsym{]}  \longrightarrow  \ottnt{E}  \ottsym{[}  \ottnt{e_{{\mathrm{2}}}}  \ottsym{]}
    }\ \rname{E}{Eval}
  \end{mathpar}
\end{definition}

\begin{definition}\label{def:semantics_notation}
  We write $ \longrightarrow  ^ *$ for the reflexive, transitive closure of $ \longrightarrow $. We also write $\ottnt{e}\centernot \longrightarrow $ to denote that there is no $\ottnt{e'}$ such that $\ottnt{e}  \longrightarrow  \ottnt{e'}$.
\end{definition}

\begin{definition}[Typing and Subtyping Rules]\label{def:typing}
  \phantom{}\\
  \textnormal{\bfseries Typing}\tquad\fbox{$\Gamma  \vdash  \ottnt{e}  \ottsym{:}  \ottnt{A}  \mid  \varepsilon$}
  \begin{mathpar}
    \inferrule{
    \vdash  \Gamma \\
     \mathit{x}   \ottsym{:}   \ottnt{A}   \in   \Gamma 
    }{
    \Gamma  \vdash  \mathit{x}  \ottsym{:}  \ottnt{A}  \mid   \bbZero 
    }\ \rname{T}{Var}

    \inferrule{
      \Gamma  \ottsym{,}  \mathit{f}  \ottsym{:}   \ottnt{A}    \rightarrow_{ \varepsilon }    \ottnt{B}   \ottsym{,}  \mathit{x}  \ottsym{:}  \ottnt{A}  \vdash  \ottnt{e}  \ottsym{:}  \ottnt{B}  \mid  \varepsilon
    }{
      \Gamma  \vdash  \ottkw{fun} \, \ottsym{(}  \mathit{f}  \ottsym{,}  \mathit{x}  \ottsym{,}  \ottnt{e}  \ottsym{)}  \ottsym{:}   \ottnt{A}    \rightarrow_{ \varepsilon }    \ottnt{B}   \mid   \bbZero 
    }\ \rname{T}{Abs}

    \inferrule{
    \Gamma  \vdash  \ottnt{v_{{\mathrm{1}}}}  \ottsym{:}   \ottnt{A}    \rightarrow_{ \varepsilon }    \ottnt{B}   \mid   \bbZero  \\ \Gamma  \vdash  \ottnt{v_{{\mathrm{2}}}}  \ottsym{:}  \ottnt{A}  \mid   \bbZero 
    }{
    \Gamma  \vdash  \ottnt{v_{{\mathrm{1}}}} \, \ottnt{v_{{\mathrm{2}}}}  \ottsym{:}  \ottnt{B}  \mid  \varepsilon
    }\ \rname{T}{App}

    \inferrule{
      \Gamma  \ottsym{,}  \alpha  \ottsym{:}  \ottnt{K}  \vdash  \ottnt{e}  \ottsym{:}  \ottnt{A}  \mid  \varepsilon
    }{
      \Gamma  \vdash  \Lambda  \alpha  \ottsym{:}  \ottnt{K}  \ottsym{.}  \ottnt{e}  \ottsym{:}    \forall   \alpha  \ottsym{:}  \ottnt{K}   \ottsym{.}    \ottnt{A}    ^{ \varepsilon }    \mid   \bbZero 
    }\ \rname{T}{TAbs}

    \inferrule{
    \Gamma  \vdash  \ottnt{v}  \ottsym{:}    \forall   \alpha  \ottsym{:}  \ottnt{K}   \ottsym{.}    \ottnt{A}    ^{ \varepsilon }    \mid   \bbZero  \\ \Gamma  \vdash  S  \ottsym{:}  \ottnt{K}
    }{
    \Gamma  \vdash  \ottnt{v} \, S  \ottsym{:}  \ottnt{A} \,  \! [  S  /  \alpha   ]   \mid  \varepsilon \,  \! [  S  /  \alpha   ] 
    }\ \rname{T}{TApp}
    %
    %

    \inferrule{
    \Gamma  \vdash  \ottnt{e_{{\mathrm{1}}}}  \ottsym{:}  \ottnt{A}  \mid  \varepsilon \\ \Gamma  \ottsym{,}  \mathit{x}  \ottsym{:}  \ottnt{A}  \vdash  \ottnt{e_{{\mathrm{2}}}}  \ottsym{:}  \ottnt{B}  \mid  \varepsilon
    }{
    \Gamma  \vdash  \mathbf{let} \, \mathit{x}  \ottsym{=}  \ottnt{e_{{\mathrm{1}}}} \, \mathbf{in} \, \ottnt{e_{{\mathrm{2}}}}  \ottsym{:}  \ottnt{B}  \mid  \varepsilon
    }\ \rname{T}{Let}

    \inferrule{
    \Gamma  \vdash  \ottnt{e}  \ottsym{:}  \ottnt{A}  \mid  \varepsilon \\ \Gamma  \vdash  \ottnt{A}  \mid  \varepsilon  <:  \ottnt{A'}  \mid  \varepsilon'
    }{
    \Gamma  \vdash  \ottnt{e}  \ottsym{:}  \ottnt{A'}  \mid  \varepsilon'
    }\ \rname{T}{Sub}

    \inferrule{
     \mathit{l}  ::    \forall    {\bm{ \alpha } }^{ \mathit{I} } : {\bm{ \ottnt{K} } }^{ \mathit{I} }    \ottsym{.}    \sigma    \in   \Xi  \\  \mathsf{op}  \ottsym{:}    \forall    {\bm{ \beta } }^{ \mathit{J} } : {\bm{ \ottnt{K'} } }^{ \mathit{J} }    \ottsym{.}    \ottnt{A}   \Rightarrow   \ottnt{B}    \in   \sigma \,  \! [ {\bm{ { S } } }^{ \mathit{I} } / {\bm{ \alpha } }^{ \mathit{I} } ]   \\\\
    \vdash  \Gamma \\
    \Gamma  \vdash   \bm{ { S } }^{ \mathit{I} } : \bm{ \ottnt{K} }^{ \mathit{I} }  \\ \Gamma  \vdash   \bm{ { T } }^{ \mathit{J} } : \bm{ \ottnt{K'} }^{ \mathit{J} }  \\
    }{
    \Gamma  \vdash   \mathsf{op} _{ \mathit{l} \,  \bm{ { S } } ^ {  \mathit{I}  }  }  \,  \bm{ { T } } ^ {  \mathit{J}  }   \ottsym{:}   \ottsym{(}  \ottnt{A} \,  \! [ {\bm{ { T } } }^{ \mathit{J} } / {\bm{ \beta } }^{ \mathit{J} } ]   \ottsym{)}    \rightarrow_{  \lift{ \mathit{l} \,  \bm{ { S } } ^ {  \mathit{I}  }  }  }    \ottsym{(}  \ottnt{B} \,  \! [ {\bm{ { T } } }^{ \mathit{I} } / {\bm{ \beta } }^{ \mathit{I} } ]   \ottsym{)}   \mid   \bbZero 
    }\ \rname{T}{Op}
    %

    \inferrule{
    \Gamma  \vdash  \ottnt{e}  \ottsym{:}  \ottnt{A}  \mid  \varepsilon' \\ \\  \mathit{l}  ::    \forall    {\bm{ \alpha } }^{ \mathit{I} } : {\bm{ \ottnt{K} } }^{ \mathit{I} }    \ottsym{.}    \sigma    \in   \Xi  \\ \\ \Gamma  \vdash   \bm{ { S } }^{ \mathit{I} } : \bm{ \ottnt{K} }^{ \mathit{I} }  \\\\
     \Gamma  \vdash _{ \sigma \,  \! [ {\bm{ { S } } }^{ \mathit{I} } / {\bm{ \alpha } }^{ \mathit{I} } ]  }  \ottnt{h}  :  \ottnt{A}   \Rightarrow  ^ { \varepsilon }  \ottnt{B}  \\    \lift{ \mathit{l} \,  \bm{ { S } } ^ {  \mathit{I}  }  }   \mathop{ \odot }  \varepsilon    \sim   \varepsilon' 
    }{
    \Gamma  \vdash   \mathbf{handle}_{ \mathit{l} \,  \bm{ { S } } ^ {  \mathit{I}  }  }  \, \ottnt{e} \, \mathbf{with} \, \ottnt{h}  \ottsym{:}  \ottnt{B}  \mid  \varepsilon
    }\ \rname{T}{Handling}
  \end{mathpar}
  \textnormal{\bfseries Handler Typing}\tquad\fbox{$ \Gamma  \vdash _{ \sigma }  \ottnt{h}  :  \ottnt{A}   \Rightarrow  ^ { \varepsilon }  \ottnt{B} $}
  \begin{mathpar}
    \inferrule{
    \Gamma  \ottsym{,}  \mathit{x}  \ottsym{:}  \ottnt{A}  \vdash  \ottnt{e_{\ottmv{r}}}  \ottsym{:}  \ottnt{B}  \mid  \varepsilon
    }{
     \Gamma  \vdash _{  \{\}  }  \ottsym{\{} \, \mathbf{return} \, \mathit{x}  \mapsto  \ottnt{e_{\ottmv{r}}}  \ottsym{\}}  :  \ottnt{A}   \Rightarrow  ^ { \varepsilon }  \ottnt{B} 
    }\ \rname{H}{Return}

    \inferrule{
    \sigma  \ottsym{=}   \sigma'   \uplus   \ottsym{\{}  \mathsf{op}  \ottsym{:}    \forall    {\bm{ \beta } }^{ \mathit{J} } : {\bm{ \ottnt{K} } }^{ \mathit{J} }    \ottsym{.}    \ottnt{A'}   \Rightarrow   \ottnt{B'}   \ottsym{\}}  \\\\
     \Gamma  \vdash _{ \sigma' }  \ottnt{h}  :  \ottnt{A}   \Rightarrow  ^ { \varepsilon }  \ottnt{B}  \\
    \Gamma  \ottsym{,}   {\bm{ \beta } }^{ \mathit{J} } : {\bm{ \ottnt{K} } }^{ \mathit{J} }   \ottsym{,}  \mathit{p}  \ottsym{:}  \ottnt{A'}  \ottsym{,}  \mathit{k}  \ottsym{:}   \ottnt{B'}    \rightarrow_{ \varepsilon }    \ottnt{B}   \vdash  \ottnt{e}  \ottsym{:}  \ottnt{B}  \mid  \varepsilon
    }{
     \Gamma  \vdash _{ \sigma }   \ottnt{h}   \uplus   \ottsym{\{}  \mathsf{op} \,  {\bm{ \beta } }^{ \mathit{J} } : {\bm{ \ottnt{K} } }^{ \mathit{J} }  \, \mathit{p} \, \mathit{k}  \mapsto  \ottnt{e}  \ottsym{\}}   :  \ottnt{A}   \Rightarrow  ^ { \varepsilon }  \ottnt{B} 
    }\ \rname{H}{Op}
  \end{mathpar}
  \textnormal{\bfseries Subtyping}\tquad\fbox{$\Gamma  \vdash  \ottnt{A}  <:  \ottnt{B}$}
  \begin{mathpar}
    \inferrule{\Gamma  \vdash  \ottnt{A}  \ottsym{:}   \mathbf{Typ} }{\Gamma  \vdash  \ottnt{A}  <:  \ottnt{A}} \ \rname{ST}{Refl}

    \inferrule{
    \Gamma  \vdash  \ottnt{A_{{\mathrm{2}}}}  <:  \ottnt{A_{{\mathrm{1}}}} \\ \Gamma  \vdash  \ottnt{B_{{\mathrm{1}}}}  \mid  \varepsilon_{{\mathrm{1}}}  <:  \ottnt{B_{{\mathrm{2}}}}  \mid  \varepsilon_{{\mathrm{2}}}
    }{
    \Gamma  \vdash   \ottnt{A_{{\mathrm{1}}}}    \rightarrow_{ \varepsilon_{{\mathrm{1}}} }    \ottnt{B_{{\mathrm{1}}}}   <:   \ottnt{A_{{\mathrm{2}}}}    \rightarrow_{ \varepsilon_{{\mathrm{2}}} }    \ottnt{B_{{\mathrm{2}}}} 
    } \ \rname{ST}{Fun}

    \inferrule{\Gamma  \ottsym{,}  \alpha  \ottsym{:}  \ottnt{K}  \vdash  \ottnt{A_{{\mathrm{1}}}}  \mid  \varepsilon_{{\mathrm{1}}}  <:  \ottnt{A_{{\mathrm{2}}}}  \mid  \varepsilon_{{\mathrm{2}}}
    }{
      \Gamma  \vdash    \forall   \alpha  \ottsym{:}  \ottnt{K}   \ottsym{.}    \ottnt{A_{{\mathrm{1}}}}    ^{ \varepsilon_{{\mathrm{1}}} }    <:    \forall   \alpha  \ottsym{:}  \ottnt{K}   \ottsym{.}    \ottnt{A_{{\mathrm{2}}}}    ^{ \varepsilon_{{\mathrm{2}}} }  
    } \  \rname{ST}{Poly}

    \inferrule{
    \Gamma  \vdash  \ottnt{A_{{\mathrm{1}}}}  <:  \ottnt{B} \\ \Gamma  \vdash   \varepsilon_{{\mathrm{1}}}  \olessthan  \varepsilon_{{\mathrm{2}}} 
    }{
    \Gamma  \vdash  \ottnt{A}  \mid  \varepsilon_{{\mathrm{1}}}  <:  \ottnt{B}  \mid  \varepsilon_{{\mathrm{2}}}
    } \  \rname{ST}{Comp}
  \end{mathpar}
\end{definition}

\begin{definition}[Semantics of Shallow Handlers]\label{def:shallow_semantics}
  The semantics for shallow handlers consists of the reduction and evaluation relations defined by the following rule \rname{R}{SHandle} and those in Definition~\ref{def:semantics} except for \rname{R}{Handle2}.
  \begin{mathpar}
    \inferrule{
     \mathsf{op} \,  {\bm{ \beta } }^{ \mathit{J} } : {\bm{ \ottnt{K} } }^{ \mathit{J} }  \, \mathit{p} \, \mathit{k}  \mapsto  \ottnt{e}   \in   \ottnt{h}  \\
    \ottnt{v_{\ottmv{cont}}} = \lambda  \mathit{z}  \ottsym{.}  \ottnt{E}  \ottsym{[}  \mathit{z}  \ottsym{]} \\  0  \mathrm{-free} ( \mathit{l} \,  \bm{ { S } } ^ {  \mathit{I}  }  ,  \ottnt{E} ) 
    }{
     \mathbf{handle}_{ \mathit{l} \,  \bm{ { S } } ^ {  \mathit{I}  }  }  \, \ottnt{E}  \ottsym{[}   \mathsf{op} _{ \mathit{l} \,  \bm{ { S } } ^ {  \mathit{I}  }  }  \,  \bm{ { T } } ^ {  \mathit{J}  }  \, \ottnt{v}  \ottsym{]} \, \mathbf{with} \, \ottnt{h}  \longmapsto  \ottnt{e} \,  \! [ {\bm{ { T } } }^{ \mathit{J} } / {\bm{ \beta } }^{ \mathit{J} } ]  \,  \! [  \ottnt{v}  /  \mathit{p}  ]  \,  \! [  \ottnt{v_{\ottmv{cont}}}  /  \mathit{k}  ] 
    }\ \rname{R}{SHandle}
  \end{mathpar}
\end{definition}

\begin{definition}[Typing of Shallow Handlers]\label{def:shallow_typing}
  The typing rules of shallow handlers consist of the rules defined by the following rules \rname{T}{SHandling}, \rname{SH}{Return}, and \rname{SH}{Op}, and those in Definition~\ref{def:typing} except for \rname{T}{Handling}, \rname{H}{Return}, and \rname{H}{Op}.
  \phantom{}\\
  \textnormal{\bfseries Typing}\tquad\fbox{$\Gamma  \vdash  \ottnt{e}  \ottsym{:}  \ottnt{A}  \mid  \varepsilon$}
  \begin{mathpar}
    \inferrule{
    \\ \Gamma  \vdash  \ottnt{e}  \ottsym{:}  \ottnt{A}  \mid  \varepsilon' \\ \\  \mathit{l}  ::    \forall    {\bm{ \alpha } }^{ \mathit{N} } : {\bm{ \ottnt{K} } }^{ \mathit{N} }    \ottsym{.}    \sigma    \in   \Xi  \\ \\ \Gamma  \vdash   \bm{ { S } }^{ \mathit{N} } : \bm{ \ottnt{K} }^{ \mathit{N} }  \\\\
     \Gamma  \vdash _{ \sigma \,  \! [ {\bm{ { S } } }^{ \mathit{N} } / {\bm{ \alpha } }^{ \mathit{N} } ]  }  \ottnt{h}  :  \ottnt{A}  ^ { \varepsilon' }  \Rightarrow  ^ { \varepsilon }  \ottnt{B}  \\    \lift{ \mathit{l} \,  \bm{ { S } } ^ {  \mathit{N}  }  }   \mathop{ \odot }  \varepsilon    \sim   \varepsilon' 
    }{
    \Gamma  \vdash   \mathbf{handle}_{ \mathit{l} \,  \bm{ { S } } ^ {  \mathit{N}  }  }  \, \ottnt{e} \, \mathbf{with} \, \ottnt{h}  \ottsym{:}  \ottnt{B}  \mid  \varepsilon
    }\ \rname{T}{SHandling}
  \end{mathpar}
  \textnormal{\bfseries Shallow Handler Typing}\tquad\fbox{$ \Gamma  \vdash _{ \sigma }  \ottnt{h}  :  \ottnt{A}  ^ { \varepsilon' }  \Rightarrow  ^ { \varepsilon }  \ottnt{B} $}
  \begin{mathpar}
    \inferrule{
    \Gamma  \ottsym{,}  \mathit{x}  \ottsym{:}  \ottnt{A}  \vdash  \ottnt{e_{\ottmv{r}}}  \ottsym{:}  \ottnt{B}  \mid  \varepsilon \\ \Gamma  \vdash  \varepsilon'  \ottsym{:}   \mathbf{Eff} 
    }{
     \Gamma  \vdash _{  \{\}  }  \ottsym{\{} \, \mathbf{return} \, \mathit{x}  \mapsto  \ottnt{e_{\ottmv{r}}}  \ottsym{\}}  :  \ottnt{A}  ^ { \varepsilon' }  \Rightarrow  ^ { \varepsilon }  \ottnt{B} 
    }\ \rname{SH}{Return}

    \inferrule{
    \sigma  \ottsym{=}   \sigma'   \uplus   \ottsym{\{}  \mathsf{op}  \ottsym{:}    \forall    {\bm{ \beta } }^{ \mathit{J} } : {\bm{ \ottnt{K} } }^{ \mathit{J} }    \ottsym{.}    \ottnt{A'}   \Rightarrow   \ottnt{B'}   \ottsym{\}}  \\\\
     \Gamma  \vdash _{ \sigma' }  \ottnt{h}  :  \ottnt{A}  ^ { \varepsilon' }  \Rightarrow  ^ { \varepsilon }  \ottnt{B}  \\
    \Gamma  \ottsym{,}   {\bm{ \beta } }^{ \mathit{J} } : {\bm{ \ottnt{K} } }^{ \mathit{J} }   \ottsym{,}  \mathit{p}  \ottsym{:}  \ottnt{A'}  \ottsym{,}  \mathit{k}  \ottsym{:}   \ottnt{B'}    \rightarrow_{ \varepsilon' }    \ottnt{A}   \vdash  \ottnt{e}  \ottsym{:}  \ottnt{B}  \mid  \varepsilon
    }{
     \Gamma  \vdash _{ \sigma }   \ottnt{h}   \uplus   \ottsym{\{}  \mathsf{op} \,  {\bm{ \beta } }^{ \mathit{J} } : {\bm{ \ottnt{K} } }^{ \mathit{J} }  \, \mathit{p} \, \mathit{k}  \mapsto  \ottnt{e}  \ottsym{\}}   :  \ottnt{A}  ^ { \varepsilon' }  \Rightarrow  ^ { \varepsilon }  \ottnt{B} 
    }\ \rname{SH}{Op}
  \end{mathpar}
\end{definition}

\begin{definition}[The Syntax of {\lang} with Lift Coercions]\label{def:syntax_lift}
  The syntax of {\lang} extended by lift coercions is the same as Definition~\ref{def:syntax} except for the following.
  \upshape
  \[
    \begin{array}{rcll@{\qquad}rcll}
      \ottnt{e} & \Coloneqq & \cdots  \mid   [  \ottnt{e}  ] _{ \ottnt{L} }  & \text{(expressions)}         &
      \ottnt{E} & \Coloneqq & \cdots  \mid   [  \ottnt{E}  ] _{ \ottnt{L} }  & \text{(evaluation contexts)}
    \end{array}
  \]
\end{definition}

\begin{definition}[Freeness of Labels with Lift Coercions]\label{def:freeness_lift}
  The rules of freeness of labels for {\lang} extended by lift coercions consist of the rules in Definition~\ref{def:freeness} and the following rules.
  \\
  \textnormal{\bfseries Freeness of labels}\tquad\fbox{$ \mathit{n}  \mathrm{-free} ( \ottnt{L} ,  \ottnt{E} ) $}
  \begin{mathpar}
    \inferrule{
       \mathit{n}  \mathrm{-free} ( \ottnt{L} ,  \ottnt{E} ) 
    }{
       \mathit{n}  \ottsym{+}  1  \mathrm{-free} ( \ottnt{L} ,   [  \ottnt{E}  ] _{ \ottnt{L} }  ) 
    }

    \inferrule{
     \mathit{n}  \mathrm{-free} ( \ottnt{L} ,  \ottnt{E} )  \\ \ottnt{L} \neq \ottnt{L'}
    }{
     \mathit{n}  \mathrm{-free} ( \ottnt{L} ,   [  \ottnt{E}  ] _{ \ottnt{L'} }  ) 
    }
  \end{mathpar}
\end{definition}

\begin{definition}[Semantics with Lift Coercions]\label{def:semantics_lift}
  The semantics for {\lang} extended by lift coercions consists of the reduction and evaluation relations defined by the following rule \rname{R}{Lift} and those in Definition~\ref{def:semantics} except for \rname{R}{Handle2}.
  \\
  \textnormal{\bfseries Reduction}\tquad\fbox{$\ottnt{e}  \longmapsto  \ottnt{e'}$}
  \begin{mathpar}
    \inferrule{
    }{
       [  \ottnt{v}  ] _{ \ottnt{L} }   \longmapsto  \ottnt{v}
    }\ \rname{R}{Lift}
  \end{mathpar}
\end{definition}

\begin{definition}[Typing of Lift Coercions]\label{def:typing_lift}
  The typing rules of {\lang} extended by lift coercions consist of the rules in Definition~\ref{def:typing} and the following rule.
  \begin{mathpar}
    \inferrule{
    \Gamma  \vdash  \ottnt{e}  \ottsym{:}  \ottnt{A}  \mid  \varepsilon' \\ \Gamma  \vdash  \ottnt{L}  \ottsym{:}   \mathbf{Lab}  \\    \lift{ \ottnt{L} }   \mathop{ \odot }  \varepsilon'    \sim   \varepsilon 
    }{
    \Gamma  \vdash   [  \ottnt{e}  ] _{ \ottnt{L} }   \ottsym{:}  \ottnt{A}  \mid  \varepsilon
    }\ \rname{T}{Lift}
  \end{mathpar}
\end{definition}

\begin{definition}[Freeness of Label Names]\label{def:freeness_erasure}
  \phantom{}\\
  \textnormal{\bfseries Freeness of label names}\tquad\fbox{$ \mathit{n}  \mathrm{-free} ( \ottnt{L} ,  \ottnt{E} ) $} \hfill \phantom{}
  \begin{mathpar}
    \inferrule{
    }{
       0  \mathrm{-free} ( \mathit{l} ,  \Box ) 
    }

    \inferrule{
       \mathit{n}  \mathrm{-free} ( \mathit{l} ,  \ottnt{E} ) 
    }{
       \mathit{n}  \mathrm{-free} ( \mathit{l} ,  \mathbf{let} \, \mathit{x}  \ottsym{=}  \ottnt{E} \, \mathbf{in} \, \ottnt{e} ) 
    }

    \inferrule{
       \mathit{n}  \ottsym{+}  1  \mathrm{-free} ( \mathit{l} ,  \ottnt{E} ) 
    }{
       \mathit{n}  \mathrm{-free} ( \mathit{l} ,   \mathbf{handle}_{ \mathit{l} \,  \bm{ { S } } ^ {  \mathit{I}  }  }  \, \ottnt{E} \, \mathbf{with} \, \ottnt{h} ) 
    }

    \inferrule{
     \mathit{n}  \mathrm{-free} ( \mathit{l} ,  \ottnt{E} )  \\ \mathit{l} \neq \mathit{l'}
    }{
     \mathit{n}  \mathrm{-free} ( \mathit{l} ,   \mathbf{handle}_{ \mathit{l'} \,  \bm{ { S } } ^ {  \mathit{I}  }  }  \, \ottnt{E} \, \mathbf{with} \, \ottnt{h} ) 
    }
  \end{mathpar}
\end{definition}

\begin{definition}[Operational Semantics with Type-Erasure]\label{def:semantics_erasure}
  The type-erasure semantics consists of the reduction and evaluation relations defined by the following rule \rname{R}{Handle2'} and those in Definition~\ref{def:semantics} except for \rname{R}{Handle2}.
  \begin{mathpar}
    \inferrule{
     \mathsf{op} \,  {\bm{ \beta } }^{ \mathit{J} } : {\bm{ \ottnt{K} } }^{ \mathit{J} }  \, \mathit{p} \, \mathit{k}  \mapsto  \ottnt{e}   \in   \ottnt{h}  \\
    \ottnt{v_{\ottmv{cont}}} = \lambda  \mathit{z}  \ottsym{.}   \mathbf{handle}_{ \mathit{l} \,  \bm{ { S } } ^ {  \mathit{I}  }  }  \, \ottnt{E}  \ottsym{[}  \mathit{z}  \ottsym{]} \, \mathbf{with} \, \ottnt{h} \\  0  \mathrm{-free} ( \mathit{l} ,  \ottnt{E} ) 
    }{
     \mathbf{handle}_{ \mathit{l} \,  \bm{ { S } } ^ {  \mathit{I}  }  }  \, \ottnt{E}  \ottsym{[}   \mathsf{op} _{ \mathit{l} \,  \bm{ { S' } } ^ {  \mathit{I}  }  }  \,  \bm{ { T } } ^ {  \mathit{J}  }  \, \ottnt{v}  \ottsym{]} \, \mathbf{with} \, \ottnt{h}  \longmapsto  \ottnt{e} \,  \! [ {\bm{ { T } } }^{ \mathit{J} } / {\bm{ \beta } }^{ \mathit{J} } ]  \,  \! [  \ottnt{v}  /  \mathit{p}  ]  \,  \! [  \ottnt{v_{\ottmv{cont}}}  /  \mathit{k}  ] 
    }\ \rname{R}{Handle2'}
  \end{mathpar}
\end{definition}

\begin{definition}[Freeness of Label Names with Lift Coercions]\label{def:freeness_lift_erasure}
  \phantom{}\\
  The rules of freeness of label names for {\lang} extended by lift coercions consist of the rules in Definition~\ref{def:freeness_erasure} and the following rules.
  \\
  \textnormal{\bfseries Freeness of label names}\tquad\fbox{$ \mathit{n}  \mathrm{-free} ( \mathit{l} ,  \ottnt{E} ) $}
  \begin{mathpar}
    \inferrule{
       \mathit{n}  \mathrm{-free} ( \mathit{l} ,  \ottnt{E} ) 
    }{
       \mathit{n}  \ottsym{+}  1  \mathrm{-free} ( \mathit{l} ,   [  \ottnt{E}  ] _{ \mathit{l} \,  \bm{ { S } } ^ {  \mathit{I}  }  }  ) 
    }

    \inferrule{
     \mathit{n}  \mathrm{-free} ( \mathit{l} ,  \ottnt{E} )  \\ \ottnt{L} \neq \mathit{l} \,  \bm{ { S } } ^ {  \mathit{I}  } 
    }{
     \mathit{n}  \mathrm{-free} ( \mathit{l} ,   [  \ottnt{E}  ] _{ \ottnt{L} }  ) 
    }
  \end{mathpar}
\end{definition}

\begin{definition}[Semantics with Lift Coercions and Type-Erasure]\label{def:semantics_lift_erasure}
  The semantics for lift coercions consists of the reduction and evaluation relations defined by
  the rule \rname{R}{Handle2'} defined in Definition~\ref{def:semantics_erasure} and
  those in Definition~\ref{def:semantics_lift} except for \rname{R}{Handle2}.
\end{definition}

\TY{Variable handlers are removed.
  \begin{definition}[Typing of Variable Handlers]\label{def:handlevar_typing}
    The typing rules of {\lang} with variable handlers consist of the rules defined by
    the following rule \rname{T}{VarHandling} and those in Definition~\ref{def:typing_lift}.
    \phantom{}\\
    \textnormal{\bfseries Typing}\tquad\fbox{$\Gamma  \vdash  \ottnt{e}  \ottsym{:}  \ottnt{A}  \mid  \varepsilon$}
    \begin{mathpar}
      \inferrule{
      \Gamma  \vdash  \ottnt{e}  \ottsym{:}  \ottnt{A}  \mid  \varepsilon' \\ \Gamma  \ottsym{,}  \mathit{x}  \ottsym{:}  \ottnt{A}  \vdash  \ottnt{e_{\ottmv{r}}}  \ottsym{:}  \ottnt{B}  \mid  \varepsilon \\ 
         \lift{ \ottnt{L} }   \mathop{ \odot }  \varepsilon    \sim   \varepsilon'  \\\\
      \Gamma  \ottsym{,}  \alpha  \ottsym{:}   \mathbf{Typ}   \ottsym{,}  \beta  \ottsym{:}   \mathbf{Typ}   \ottsym{,}  \mathit{o}  \ottsym{:}   \alpha    \rightarrow_{  \lift{ \ottnt{L} }  }    \beta   \ottsym{,}  \mathit{p}  \ottsym{:}  \alpha  \ottsym{,}  \mathit{k}  \ottsym{:}   \beta    \rightarrow_{ \varepsilon }    \ottnt{B}   \vdash  \ottnt{e_{{\mathrm{0}}}}  \ottsym{:}  \ottnt{B}  \mid  \varepsilon
      }{
      \Gamma  \vdash   {\mathbf{handle}\mathrm{-}\mathbf{var} }_{ \ottnt{L} }  \, \ottnt{e} \, \mathbf{with} \,  \{ \mathbf{return}\, x \mapsto  \ottnt{e_{\ottmv{r}}}  \}  \uplus  \{ ( \mathit{o}  :  \alpha   \Rightarrow   \beta )\, p\, k \mapsto  \ottnt{e_{{\mathrm{0}}}}  \}   \ottsym{:}  \ottnt{B}  \mid  \varepsilon
      }\ \rname{T}{VarHandling}
    \end{mathpar}
  \end{definition}

  \begin{definition}[Freeness with Variable Handlers]\label{def:freeness_handlevar}
    The rules of freeness with variable handlers consist of the following rules and those in Definition~\ref{def:freeness}.
    \begin{mathpar}
      \inferrule{
         \mathit{n}  \ottsym{+}  1  \mathrm{-free} ( \ottnt{L} ,  \ottnt{E} ) 
      }{
         \mathit{n}  \mathrm{-free} ( \ottnt{L} ,   {\mathbf{handle}\mathrm{-}\mathbf{var} }_{ \ottnt{L} }  \, \ottnt{E} \, \mathbf{with} \,  \{ \mathbf{return}\, x \mapsto  \ottnt{e_{\ottmv{r}}}  \}  \uplus  \{ ( \mathit{o}  :  \alpha   \Rightarrow   \beta )\, p\, k \mapsto  \ottnt{e_{{\mathrm{0}}}}  \}  ) 
      }

      \inferrule{
       \mathit{n}  \mathrm{-free} ( \ottnt{L} ,  \ottnt{E} )  \\ \ottnt{L} \neq \ottnt{L'}
      }{
       \mathit{n}  \mathrm{-free} ( \ottnt{L} ,   {\mathbf{handle}\mathrm{-}\mathbf{var} }_{ \ottnt{L'} }  \, \ottnt{E} \, \mathbf{with} \,  \{ \mathbf{return}\, x \mapsto  \ottnt{e_{\ottmv{r}}}  \}  \uplus  \{ ( \mathit{o}  :  \alpha   \Rightarrow   \beta )\, p\, k \mapsto  \ottnt{e_{{\mathrm{0}}}}  \}  ) 
      }
    \end{mathpar}
  \end{definition}

  \begin{definition}[Freeness with Variable Handlers and Type-Erasure]\label{def:freeness_erasure_handlevar}
    The rules of freeness with variable handlers and type-erasure consist of the following rules and those in Definition~\ref{def:freeness_erasure}.
    \begin{mathpar}
      \inferrule{
         \mathit{n}  \ottsym{+}  1  \mathrm{-free} ( \mathit{l} ,  \ottnt{E} ) 
      }{
         \mathit{n}  \mathrm{-free} ( \mathit{l} ,   {\mathbf{handle}\mathrm{-}\mathbf{var} }_{ \mathit{l} \,  \bm{ { S } } ^ {  \mathit{I}  }  }  \, \ottnt{E} \, \mathbf{with} \,  \{ \mathbf{return}\, x \mapsto  \ottnt{e_{\ottmv{r}}}  \}  \uplus  \{ ( \mathit{o}  :  \alpha   \Rightarrow   \beta )\, p\, k \mapsto  \ottnt{e_{{\mathrm{0}}}}  \}  ) 
      }

      \inferrule{
         \mathit{n}  \mathrm{-free} ( \mathit{l} ,  \ottnt{E} )  \\ \forall  \bm{ { S } } ^ {  \mathit{I}  }  . (\ottnt{L} \neq \mathit{l} \,  \bm{ { S } } ^ {  \mathit{I}  } )
      }{
         \mathit{n}  \mathrm{-free} ( \mathit{l} ,   {\mathbf{handle}\mathrm{-}\mathbf{var} }_{ \ottnt{L} }  \, \ottnt{E} \, \mathbf{with} \,  \{ \mathbf{return}\, x \mapsto  \ottnt{e_{\ottmv{r}}}  \}  \uplus  \{ ( \mathit{o}  :  \alpha   \Rightarrow   \beta )\, p\, k \mapsto  \ottnt{e_{{\mathrm{0}}}}  \}  ) 
      }
    \end{mathpar}
  \end{definition}

  \begin{definition}[Semantics of Variable Handlers]\label{def:handlevar_semantics}
    The semantics for variable handlers consists of the reduction and evaluation relations defined by the following rules \rname{R}{VarHandle1} and \rname{R}{VarHandle2} and those in Definition~\ref{def:semantics_lift}.
    \begin{mathpar}
      \inferrule{ }{
         {\mathbf{handle}\mathrm{-}\mathbf{var} }_{ \mathit{l} \,  \bm{ { S } } ^ {  \mathit{I}  }  }  \, \ottnt{v} \, \mathbf{with} \,  \{ \mathbf{return}\, x \mapsto  \ottnt{e_{\ottmv{r}}}  \}  \uplus  \{ ( \mathit{o}  :  \alpha'   \Rightarrow   \beta' )\, p\, k \mapsto  \ottnt{e_{{\mathrm{0}}}}  \}   \longmapsto  \ottnt{e_{\ottmv{r}}} \,  \! [  \ottnt{v}  /  \mathit{x}  ] 
      } \ \rname{R}{VarHandle1}

      \inferrule{
       \mathit{l}  ::    \forall    {\bm{ \alpha } }^{ \mathit{I} } : {\bm{ \ottnt{K} } }^{ \mathit{I} }    \ottsym{.}    \sigma    \in   \Xi  \\  \mathsf{op}  \ottsym{:}    \forall    {\bm{ \beta } }^{ \mathit{J} } : {\bm{ \ottnt{K} } }^{ \mathit{J} }    \ottsym{.}    \ottnt{A}   \Rightarrow   \ottnt{B}    \in   \sigma  \\  0  \mathrm{-free} ( \mathit{l} \,  \bm{ { S } } ^ {  \mathit{I}  }  ,  \ottnt{E} )  \\\\
      \ottnt{v_{\ottmv{cont}}} = \lambda  \mathit{z}  \ottsym{.}   {\mathbf{handle}\mathrm{-}\mathbf{var} }_{ \mathit{l} \,  \bm{ { S } } ^ {  \mathit{I}  }  }  \, \ottnt{E}  \ottsym{[}  \mathit{z}  \ottsym{]} \, \mathbf{with} \,  \{ \mathbf{return}\, x \mapsto  \ottnt{e_{\ottmv{r}}}  \}  \uplus  \{ ( \mathit{o}  :  \alpha'   \Rightarrow   \beta' )\, p\, k \mapsto  \ottnt{e_{{\mathrm{0}}}}  \}  
      }{
      {\begin{aligned}
             &   {\mathbf{handle}\mathrm{-}\mathbf{var} }_{ \mathit{l} \,  \bm{ { S } } ^ {  \mathit{I}  }  }  \, \ottnt{E}  \ottsym{[}   \mathsf{op} _{ \mathit{l} \,  \bm{ { S } } ^ {  \mathit{I}  }  }  \,  \bm{ { T } } ^ {  \mathit{J}  }  \, \ottnt{v}  \ottsym{]} \, \mathbf{with} \,  \{ \mathbf{return}\, x \mapsto  \ottnt{e_{\ottmv{r}}}  \}  \uplus  \{ ( \mathit{o}  :  \alpha'   \Rightarrow   \beta' )\, p\, k \mapsto  \ottnt{e_{{\mathrm{0}}}}  \}    & \\ &   \longmapsto  \ottnt{e_{{\mathrm{0}}}} \,  \! [  \ottnt{A} \,  \! [ {\bm{ { S } } }^{ \mathit{I} } / {\bm{ \alpha } }^{ \mathit{I} } ]  \,  \! [ {\bm{ { T } } }^{ \mathit{J} } / {\bm{ \beta } }^{ \mathit{J} } ]   /  \alpha'   ]  \,  \! [  \ottnt{B} \,  \! [ {\bm{ { S } } }^{ \mathit{I} } / {\bm{ \alpha } }^{ \mathit{I} } ]  \,  \! [ {\bm{ { T } } }^{ \mathit{J} } / {\bm{ \beta } }^{ \mathit{J} } ]   /  \beta'   ]  \,  \! [   \mathsf{op} _{ \mathit{l} \,  \bm{ { S } } ^ {  \mathit{I}  }  }  \,  \bm{ { T } } ^ {  \mathit{J}  }   /  \mathit{o}  ]  \,  \! [  \ottnt{v}  /  \mathit{p}  ]  \,  \! [  \ottnt{v_{\ottmv{cont}}}  /  \mathit{k}  ] 
          \end{aligned}}
      } \ \rname{R}{VarHandle2}
    \end{mathpar}
  \end{definition}

  \begin{definition}[Semantics of Variable Handlers and Type-Erasure]\label{def:handlevar_semantics_erasure}
    The semantics for variable handlers consists of the reduction and evaluation relations defined by the following rule \rname{R}{VarHandle2'} and those in Definition~\ref{def:semantics_lift_erasure} except for \rname{R}{VarHandle2}.
    \begin{mathpar}
      \inferrule{
       \mathit{l}  ::    \forall    {\bm{ \alpha } }^{ \mathit{I} } : {\bm{ \ottnt{K} } }^{ \mathit{I} }    \ottsym{.}    \sigma    \in   \Xi  \\  \mathsf{op}  \ottsym{:}    \forall    {\bm{ \beta } }^{ \mathit{J} } : {\bm{ \ottnt{K} } }^{ \mathit{J} }    \ottsym{.}    \ottnt{A}   \Rightarrow   \ottnt{B}    \in   \sigma  \\  0  \mathrm{-free} ( \mathit{l} ,  \ottnt{E} )  \\\\
      \ottnt{v_{\ottmv{cont}}} = \lambda  \mathit{z}  \ottsym{.}   {\mathbf{handle}\mathrm{-}\mathbf{var} }_{ \mathit{l} \,  \bm{ { S } } ^ {  \mathit{I}  }  }  \, \ottnt{E}  \ottsym{[}  \mathit{z}  \ottsym{]} \, \mathbf{with} \,  \{ \mathbf{return}\, x \mapsto  \ottnt{e_{\ottmv{r}}}  \}  \uplus  \{ ( \mathit{o}  :  \alpha'   \Rightarrow   \beta' )\, p\, k \mapsto  \ottnt{e_{{\mathrm{0}}}}  \}  
      }{
      {\begin{aligned}
             &   {\mathbf{handle}\mathrm{-}\mathbf{var} }_{ \mathit{l} \,  \bm{ { S } } ^ {  \mathit{I}  }  }  \, \ottnt{E}  \ottsym{[}   \mathsf{op} _{ \mathit{l} \,  \bm{ { S' } } ^ {  \mathit{I'}  }  }  \,  \bm{ { T } } ^ {  \mathit{J}  }  \, \ottnt{v}  \ottsym{]} \, \mathbf{with} \,  \{ \mathbf{return}\, x \mapsto  \ottnt{e_{\ottmv{r}}}  \}  \uplus  \{ ( \mathit{o}  :  \alpha'   \Rightarrow   \beta' )\, p\, k \mapsto  \ottnt{e_{{\mathrm{0}}}}  \}    & \\ &   \longmapsto  \ottnt{e_{{\mathrm{0}}}} \,  \! [  \ottnt{A} \,  \! [ {\bm{ { S } } }^{ \mathit{I} } / {\bm{ \alpha } }^{ \mathit{I} } ]  \,  \! [ {\bm{ { T } } }^{ \mathit{J} } / {\bm{ \beta } }^{ \mathit{J} } ]   /  \alpha'   ]  \,  \! [  \ottnt{B} \,  \! [ {\bm{ { S } } }^{ \mathit{I} } / {\bm{ \alpha } }^{ \mathit{I} } ]  \,  \! [ {\bm{ { T } } }^{ \mathit{J} } / {\bm{ \beta } }^{ \mathit{J} } ]   /  \beta'   ]  \,  \! [   \mathsf{op} _{ \mathit{l} \,  \bm{ { S } } ^ {  \mathit{I}  }  }  \,  \bm{ { T } } ^ {  \mathit{J}  }   /  \mathit{o}  ]  \,  \! [  \ottnt{v}  /  \mathit{p}  ]  \,  \! [  \ottnt{v_{\ottmv{cont}}}  /  \mathit{k}  ] 
          \end{aligned}}
      } \ \rname{R}{VarHandle2'}
    \end{mathpar}
  \end{definition}
}

\TY{New safety conditions}
\begin{definition}[Safety Conditions]\label{def:safe_cond}
  \phantom{}
  \begin{enumerate}
    \item\label{def:safe_cond:label_notemp}
          For any $\ottnt{L}$, $  \lift{ \ottnt{L} }   \olessthan   \bbZero  $ does not hold.

    \item\label{def:safe_cond:pres} If $  \lift{ \ottnt{L} }   \olessthan  \varepsilon $ and $   \lift{ \ottnt{L'} }   \mathop{ \odot }  \varepsilon'    \sim   \varepsilon $ and $\ottnt{L} \neq \ottnt{L'}$, then $  \lift{ \ottnt{L} }   \olessthan  \varepsilon' $.

          \setcounter{safecondcounter}{\value{enumi}}
  \end{enumerate}
\end{definition}


\TS{TODO: Change the statement of type/effect safety according to the change of the following definition.}
\begin{definition}[Safety Condition for Lift Coercions]\label{def:safe_cond_lift}
  The safety condition added for lift coercions is the following:
  \TS{TODO: Replace the following two conditions by a single, merged condition.}
  \TY{Old two conditions are commented out.}
  \begin{enumerate}
    \setcounter{enumi}{\value{safecondcounter}}


    \item\label{def:safe_cond_lift:removal}
          If
          $   \lift{ \ottnt{L} }   \mathop{ \odot }  \varepsilon_{{\mathrm{1}}}    \sim       \lift{ \ottnt{L_{{\mathrm{1}}}} }   \mathop{ \odot } \cdots \mathop{ \odot }   \lift{ \ottnt{L_{\ottmv{n}}} }    \mathop{ \odot }   \lift{ \ottnt{L} }    \mathop{ \odot }  \varepsilon_{{\mathrm{2}}}  $
          and $\ottnt{L} \notin \{  \ottnt{L_{{\mathrm{1}}}}  \ottsym{,}  \ldots  \ottsym{,}  \ottnt{L_{\ottmv{n}}}  \}$,
          then $ \varepsilon_{{\mathrm{1}}}   \sim      \lift{ \ottnt{L_{{\mathrm{1}}}} }   \mathop{ \odot } \cdots \mathop{ \odot }   \lift{ \ottnt{L_{\ottmv{n}}} }    \mathop{ \odot }  \varepsilon_{{\mathrm{2}}}  $.
          \setcounter{safecondcounter}{\value{enumi}}
  \end{enumerate}
\end{definition}

\TS{TODO: Change the statement of type/effect safety according to the change of the following definition.}
\begin{definition}[Safety Condition for Type-Erasure]\label{def:safe_cond_erasure}
  The safety condition added for the type-erasure semantics is the following:
  \begin{enumerate}
    \setcounter{enumi}{\value{safecondcounter}}
    \item\label{def:safe_cond_erasure:uniq}
          If $  \lift{ \mathit{l} \,  \bm{ { S_{{\mathrm{1}}} } } ^ {  \mathit{I_{{\mathrm{1}}}}  }  }   \olessthan  \varepsilon $ and $  \lift{ \mathit{l} \,  \bm{ { S_{{\mathrm{2}}} } } ^ {  \mathit{I_{{\mathrm{2}}}}  }  }   \olessthan  \varepsilon $, then $ \bm{ { S_{{\mathrm{1}}} } } ^ {  \mathit{I_{{\mathrm{1}}}}  }  =  \bm{ { S_{{\mathrm{2}}} } } ^ {  \mathit{I_{{\mathrm{2}}}}  } $.
          \setcounter{safecondcounter}{\value{enumi}}
  \end{enumerate}
\end{definition}

\begin{example}[Unsafe Effect Algebras]\label{exa:necessity}
  \phantom{}
  \begin{description}
    \item[\quad Effect algebra violating safety condition~\ref{def:safe_cond:label_notemp}]
          Consider an effect algebra such that $\emptyset  \vdash    \lift{ \mathit{l} }   \olessthan   \bbZero  $ holds for some $\mathit{l}$.
          Clearly, this effect algebra violates safety condition~\ref{def:safe_cond:label_notemp}.
          In this case,
          $\emptyset  \vdash   \mathsf{op} _{ \mathit{l} }  \, \ottnt{v}  \ottsym{:}  \ottnt{A}  \mid   \bbZero $
          can be derived for some $\ottnt{A}$ (if $ \mathsf{op} _{ \mathit{l} }  \, \ottnt{v}$ is well typed)
          because $ \mathsf{op} _{ \mathit{l} }  \, \ottnt{v}$ is given the effect $ \lift{ \mathit{l} } $ and
          the subeffecting $\emptyset  \vdash    \lift{ \mathit{l} }   \olessthan   \bbZero  $ holds.
          However, the operation call is not handled.

    \item[\quad Effect algebra violating safety condition~\ref{def:safe_cond:pres}]
          Consider an effect algebra such that
          safety condition~\ref{def:safe_cond:label_notemp},
          $  \lift{ \mathit{l} }   \olessthan   \lift{ \mathit{l'} }  $, and
          $   \lift{ \mathit{l'} }   \mathop{ \odot }   \bbZero     \sim    \lift{ \mathit{l'} }  $
          hold for some $\mathit{l}$ and $\mathit{l'}$ such that $\mathit{l} \neq \mathit{l'}$.
          This effect algebra violates safety condition~\ref{def:safe_cond:pres}:
          if safety condition~\ref{def:safe_cond:pres} is met, we would have $  \lift{ \mathit{l} }   \olessthan   \bbZero  $, but
          it is contradictory with safety condition~\ref{def:safe_cond:label_notemp}.

          \OLD{
            First, we show that the effect algebra allowing these assumptions violates safety condition~\ref{def:safe_cond:pres}.
            If safety condition~\ref{def:safe_cond:pres} is met, we would have $  \lift{ \mathit{l} }   \olessthan   \bbZero  $.
            However, it is contradictory with safety condition~\ref{def:safe_cond:label_notemp}.
          }

          This effect algebra allows assigning the empty effect $ \bbZero $ to the expression $ \mathbf{handle}_{ \mathit{l'} }  \,  \mathsf{op} _{ \mathit{l} }  \,  {}  \, \ottnt{v} \, \mathbf{with} \, \ottnt{h}$ as illustrated by the following typing derivation:
          %
          \[
            \inferrule* [Right=T\_Handling] {
            \cdots \\
               \lift{ \mathit{l'} }   \mathop{ \odot }   \bbZero     \sim    \lift{ \mathit{l'} }  \\
            \inferrule* [Right=T\_Sub] {
            \emptyset  \vdash   \mathsf{op} _{ \mathit{l} }  \,  {}  \, \ottnt{v}  \ottsym{:}  \ottnt{A}  \mid   \lift{ \mathit{l} }  \\
            \emptyset  \vdash  \ottnt{A}  \mid   \lift{ \mathit{l} }   <:  \ottnt{A}  \mid   \lift{ \mathit{l'} } 
            }{
            \emptyset  \vdash   \mathsf{op} _{ \mathit{l} }  \,  {}  \, \ottnt{v}  \ottsym{:}  \ottnt{A}  \mid   \lift{ \mathit{l'} } 
            }
            }{
            \emptyset  \vdash   \mathbf{handle}_{ \mathit{l'} }  \,  \mathsf{op} _{ \mathit{l} }  \,  {}  \, \ottnt{v} \, \mathbf{with} \, \ottnt{h}  \ottsym{:}  \ottnt{B}  \mid   \bbZero 
            }
          \]
          However, the operation call in it is not handled.
  \end{description}
\end{example}

\end{document}

\clearpage
\section{Example}

We present a motivating example of allowing multiple effect variables in one effect collection.
In this example, we use {\eaSet} and offer two modules of type \textsf{IntSet},
which is an interface of implementations for integer sets defined in Figure~\ref{fig:intset_type}.

We show two implementations of \textsf{IntSet} in Figure~\ref{fig:impl}.
The former implementation assumes the effect context $ \mathsf{Selection}  ::    \forall   \alpha  \ottsym{:}   \mathbf{Typ}    \ottsym{.}    \ottsym{\{}  \mathsf{select}  \ottsym{:}    \alpha \,\mathsf{List}    \Rightarrow   \alpha   \ottsym{\}}  $, and
concretizes $\rho$ by $\mathsf{Selection} \,  \mathsf{Int} $.
The latter implementation assumes the effect context
$ \mathsf{Choice}  ::  \ottsym{\{}  \mathsf{decide}  \ottsym{:}    \mathsf{Unit}    \Rightarrow    \mathsf{Bool}    \ottsym{\}}   \ottsym{,}   \mathsf{Fail}  ::  \ottsym{\{}  \mathsf{fail}  \ottsym{:}    \forall   \alpha  \ottsym{:}   \mathbf{Typ}    \ottsym{.}     \mathsf{Unit}    \Rightarrow   \alpha   \ottsym{\}} $, and
concretizes $\rho$ by $ \{  \mathsf{Fail}  \}  \,\underline{ \cup }\,  \{  \mathsf{Choice}  \} $.
Because the concrete effect of the latter implementation consists of two labels,
it needs to be abstracted by a row variable, not by a label variable.

\begin{figure}[t]
  \begin{align*}
     & \exists \alpha  \ottsym{:}   \mathbf{Typ}  . \exists \rho  \ottsym{:}   \mathbf{Eff}  . \{                                                              \\
     & \quad \begin{aligned}
                & \mathit{empty} : \alpha ,
               \quad \mathit{add} :   \mathsf{Int}     \rightarrow_{ \{  \} }     \alpha    \rightarrow_{ \{  \} }    \alpha  ,
               \quad \mathit{size} :  \alpha    \rightarrow_{ \{  \} }     \mathsf{Int}  ,
               \quad \mathit{find} :   \mathsf{Int}     \rightarrow_{ \{  \} }     \alpha    \rightarrow_{ \{  \} }     \mathsf{Bool}   ,                                                  \\
                & \mathit{filter} :  \ottsym{(}    \mathsf{Int}     \rightarrow_{ \{  \} }     \mathsf{Bool}    \ottsym{)}    \rightarrow_{ \{  \} }     \alpha    \rightarrow_{ \{  \} }    \alpha  ,
               \quad \mathit{choose} :  \alpha    \rightarrow_{ \rho }     \mathsf{Int}  ,                                                               \\
                & \namecollect :   \forall   \beta  \ottsym{:}   \mathbf{Typ}    \ottsym{.}      \forall   \rho'  \ottsym{:}   \mathbf{Eff}    \ottsym{.}     \ottsym{(}    \mathsf{Unit}     \rightarrow_{  \rho  \,\underline{ \cup }\,  \rho'  }    \beta   \ottsym{)}    \rightarrow_{ \rho' }     \beta \,\mathsf{List}    
             \end{aligned} \\
     & \}
  \end{align*}
  \caption{Module Interface \textsf{IntSet}}
  \label{fig:intset_type}
\end{figure}

\begin{figure}[t]
  \begin{flalign*}
     & \mathbf{pack} (  \mathsf{Int}  \,\mathsf{List} , \{  \mathsf{Selection} \,  \mathsf{Int}   \}, \{ \cdots                                                                                                                                &   \\
     & \qquad \begin{aligned}
                 & \mathit{choose} =  \mathsf{select} _{ \mathsf{Selection} \,  \mathsf{Int}  }  \,  {}                                                                                       \\
                 & \namecollect = \Lambda \beta  \ottsym{:}   \mathbf{Typ}  . \Lambda \rho'  \ottsym{:}   \mathbf{Eff}  . \lambda \mathit{f}  \ottsym{:}    \mathsf{Unit}     \rightarrow_{  \{  \mathsf{Selection} \,  \mathsf{Int}   \}  \,\underline{ \cup }\,  \rho'  }    \beta  .        \\
                 & \qquad \qquad \qquad \qquad  \mathbf{handle}_{ \mathsf{Selection} \,  \mathsf{Int}  }  \, \mathit{f} \,  ()  \, \mathbf{with} \,  \ottsym{\{} \, \mathbf{return} \, \mathit{x}  \mapsto  \ottsym{[}  \mathit{x}  \ottsym{]}  \ottsym{\}}   \uplus   \ottsym{\{}  \mathsf{select} \,  {}  \, \mathit{xs} \, \mathit{k}  \mapsto   \textnormal{\ttfamily concat}  \,  (   \textnormal{\ttfamily map}  \, \mathit{k} \, \mathit{xs}  )   \ottsym{\}} 
              \end{aligned} \\
     & \})
  \end{flalign*}
  \begin{flalign*}
     & \mathbf{pack} (  \mathsf{Int}  \,\mathsf{List} ,  \{  \mathsf{Fail}  \}  \,\underline{ \cup }\,  \{  \mathsf{Choice}  \} , \{ \cdots                                                                                                                                                                   &   \\
     & \qquad \begin{aligned}
                 & \mathit{choose} = \mathbf{fun}(\mathit{aux}, \mathit{xs}, \mathbf{match}\; \mathit{xs} \;\mathbf{with}                                                                                  \\
                 & \qquad \qquad \qquad \qquad \quad \mid [] \rightarrow  \mathsf{fail} _{ \mathsf{Fail} }  \,  \mathsf{Int}  \,  ()                                                                                                 \\
                 & \qquad \qquad \qquad \qquad \quad \mid \mathit{y} :: \mathit{ys} \rightarrow \mathbf{if}\;  \mathsf{decide} _{ \mathsf{Choice} }  \,  {}  \,  ()  \;\mathbf{then}\; \mathit{y} \;\mathbf{else}\; \mathit{aux}\;\mathit{ys}), \\
                 & \namecollect = \Lambda \beta  \ottsym{:}   \mathbf{Typ}  . \Lambda \rho'  \ottsym{:}   \mathbf{Eff}  . \lambda \mathit{f}  \ottsym{:}    \mathsf{Unit}     \rightarrow_{   \{  \mathsf{Fail}  \}  \,\underline{ \cup }\,  \{  \mathsf{Choice}  \}   \,\underline{ \cup }\,  \rho'  }    \beta  .                                         \\
                 & \qquad \qquad \qquad \qquad  \mathbf{handle}_{ \mathsf{Choice} }                                                                                                                                  \\
                 & \qquad \qquad \qquad \qquad \qquad  \mathbf{handle}_{ \mathsf{Fail} }                                                                                                                             \\
                 & \qquad \qquad \qquad \qquad \qquad \qquad \mathit{f} \,  ()                                                                                                                           \\
                 & \qquad \qquad \qquad \qquad \qquad  \mathbf{with}  \,  \ottsym{\{} \, \mathbf{return} \, \mathit{x}  \mapsto  \ottsym{[}  \mathit{x}  \ottsym{]}  \ottsym{\}}   \uplus   \ottsym{\{}  \mathsf{fail} \, \alpha  \ottsym{:}   \mathbf{Typ}  \, \_ \, \_  \mapsto   []   \ottsym{\}}                                                                    \\
                 & \qquad \qquad \qquad \qquad  \mathbf{with}  \,  \ottsym{\{} \, \mathbf{return} \, \mathit{x}  \mapsto  \mathit{x}  \ottsym{\}}   \uplus   \ottsym{\{}  \mathsf{decide} \,  {}  \, \_ \, \mathit{k}  \mapsto   \mathit{k} \,  \mathsf{true}  \,@\, \mathit{k}  \,  \mathsf{false}   \ottsym{\}}                                                                        \\
              \end{aligned} \\
     & \})
  \end{flalign*}
  \caption{Two implementation of \textsf{IntSet}}
  \label{fig:impl}
\end{figure}

We define the function $\mathit{search\_path}$ using this package as follows.
\begin{flalign*}
  \mathit{search\_path}
   & = \lambda \mathit{sets} : \mathsf{IntSet}\,\mathsf{List} . \lambda s :  \mathsf{Int}  . \lambda t :  \mathsf{Int}  .                                                                                                                  & \\
   & \qquad \mathbf{fun}(\mathit{aux}, p, \lambda \mathit{path} : \mathsf{Int}\,\mathsf{List} .                                                                                                                              & \\
   & \qquad \qquad \mathbf{if}\, p = t \,\mathbf{then}\, \mathit{path}                                                                                                                                                       & \\
   & \qquad \qquad \mathbf{else}\,\mathbf{let}\, x = \mathit{choose} \, (\mathit{filter}\, (\lambda y : \mathsf{Int} . \mathit{not}\,(\mathtt{exists}\,(\lambda  \mathit{z}  \ottsym{.}  \mathit{z}  \ottsym{=}  \mathit{y})\,\mathit{path}))\, (\mathit{nth}\,p\,\mathit{sets})) & \\
   & \qquad \qquad \qquad \mathbf{in}\, \mathit{aux}\,x\,(x :: \mathit{path}) )                                                                                                                                              & \\
   & \qquad s\,[s]                                                                                                                                                                                                           &
\end{flalign*}

We show the example program using $\mathit{search\_path}$ as follows.
\begin{flalign*}
                                                                                 & \mathit{graph} = [ \mathit{add}\, 1 \, (\mathit{add}\, 2 \, \mathit{empty});\,
  \mathit{add}\, 0 \, (\mathit{add}\, 2 \, (\mathit{add}\, 3\, \mathit{empty}));\,
  \mathit{add}\, 0 \, (\mathit{add}\, 1 \, (\mathit{add}\, 4\, \mathit{empty})); &                                                                                                  \\
                                                                                 & \qquad \qquad \mathit{add}\, 1 \, (\mathit{add}\, 4 \, (\mathit{add}\, 5\, \mathit{empty}));\,
  \mathit{add}\, 2 \, (\mathit{add}\, 3 \, (\mathit{add}\, 6 \, \mathit{empty}));\,
  \mathit{add}\, 3\, \mathit{empty};\, \mathit{add}\, 4 \,\mathit{empty} ]
                                                                                 &                                                                                                  \\
                                                                                 & \mathit{clean} \, (\lambda \_ :  \mathsf{Unit}  . \mathit{search\_path}\, \mathit{graph}\, 0\, 5)     & \\
                                                                                 & \mathit{clean} \, (\lambda \_ :  \mathsf{Unit}  . \mathit{search\_path}\, \mathit{graph}\, 0\, 5)     &
\end{flalign*}
The evaluation results are as follows.
\begin{flalign*}
   & [ [5;\, 3;\, 4;\, 2;\, 1;\, 0];\, [5;\, 3;\, 1;\, 0];\, [5;\, 3;\, 4;\, 2;\, 0];\, [5;\, 3;\, 1;\, 2;\, 0] ] & \\
   & [ [6;\, 4;\, 2;\, 0;\, 1];\, [6;\, 4;\, 2;\, 1];\, [6;\, 4;\, 3;\, 1] ]                                      &
\end{flalign*}

\clearpage
\section{Properties}

\subsection{Properties with Deep Handlers}

This section assumes that the safety conditions in Definition~\ref{def:safe_cond} hold.

\begin{lemma}[Well-formedness of context in judgement]\label{lem:wf}
  If $\Gamma  \vdash  S  \ottsym{:}  \ottnt{K}$, then $\vdash  \Gamma$.
\end{lemma}

\begin{proof}
  By induction on a derivation of $\Gamma  \vdash  S  \ottsym{:}  \ottnt{K}$. We proceed by case analysis on the kinding rule applied lastly to this derivation.
  \begin{divcases}
    \item[\rname{K}{Var}] Clearly.

    \item[\rname{K}{Fun}] $S =  \ottnt{A}    \rightarrow_{ \varepsilon }    \ottnt{B} $, $\Gamma  \vdash  \ottnt{A}  \ottsym{:}   \mathbf{Typ} $, $\Gamma  \vdash  \varepsilon  \ottsym{:}   \mathbf{Eff} $, and $\Gamma  \vdash  \ottnt{B}  \ottsym{:}   \mathbf{Typ} $ are given. By the induction hypothesis, we have $\vdash  \Gamma$.

    \item[\rname{K}{Poly}] $S =   \forall   \alpha  \ottsym{:}  \ottnt{K}   \ottsym{.}    \ottnt{A}    ^{ \varepsilon }  $, $\Gamma  \ottsym{,}  \alpha  \ottsym{:}  \ottnt{K}  \vdash  \ottnt{A}  \ottsym{:}   \mathbf{Typ} $, and $\Gamma  \ottsym{,}  \alpha  \ottsym{:}  \ottnt{K}  \vdash  \varepsilon  \ottsym{:}   \mathbf{Eff} $ are given. By the induction hypothesis, we have $\vdash  \Gamma  \ottsym{,}  \alpha  \ottsym{:}  \ottnt{K}$. Since only \rname{C}{TVar} can derive $\vdash  \Gamma  \ottsym{,}  \alpha  \ottsym{:}  \ottnt{K}$, the required result $\vdash  \Gamma$ is achieved.

    \item[\rname{K}{Cons}] Clearly.
  \end{divcases}
\end{proof}

\begin{lemma}\label{lem:delta_context}
  \phantom{}
  \begin{enumerate}
    \item\label{lem:delta_context:wf} If $\vdash  \Gamma$, then $\vdash   \Delta   \ottsym{(}   \Gamma   \ottsym{)} $.
    \item\label{lem:delta_context:kinding} If $\Gamma  \vdash  S  \ottsym{:}  \ottnt{K}$, then $ \Delta   \ottsym{(}   \Gamma   \ottsym{)}   \vdash  S  \ottsym{:}  \ottnt{K}$.
  \end{enumerate}
\end{lemma}

\begin{proof}
  \phantom{}
  By mutual induction on the derivations. We proceed by case analysis on the rule applied lastly to the derivation.
  \begin{divcases}
    \item[\rname{C}{Empty}] Clearly because of $ \Delta   \ottsym{(}   \emptyset   \ottsym{)}  =  \emptyset $.
    \item[\rname{C}{Var}] For some $\Gamma'$, $\mathit{x}$, and $\ottnt{A}$, the following are given:
    \begin{itemize}
      \item $\Gamma = \Gamma'  \ottsym{,}  \mathit{x}  \ottsym{:}  \ottnt{A}$ and
      \item $\Gamma'  \vdash  \ottnt{A}  \ottsym{:}   \mathbf{Typ} $.
    \end{itemize}
    By the induction hypothesis, we have $ \Delta   \ottsym{(}   \Gamma'   \ottsym{)}   \vdash  \ottnt{A}  \ottsym{:}   \mathbf{Typ} $. By Lemma~\ref{lem:wf}, we have $\vdash   \Delta   \ottsym{(}   \Gamma'   \ottsym{)} $. Thus, we get $\vdash   \Delta   \ottsym{(}   \Gamma   \ottsym{)} $ as required because of $ \Delta   \ottsym{(}   \Gamma'  \ottsym{,}  \mathit{x}  \ottsym{:}  \ottnt{A}   \ottsym{)}  =  \Delta   \ottsym{(}   \Gamma'   \ottsym{)} $.
    \item[\rname{C}{TVar}] For some $\Gamma'$, $\alpha$, and $\ottnt{K}$, the following are given:
    \begin{itemize}
      \item $\Gamma = \Gamma'  \ottsym{,}  \alpha  \ottsym{:}  \ottnt{K}$,
      \item $\vdash  \Gamma'$, and
      \item $ \alpha   \notin    \mathrm{dom}   \ottsym{(}   \Gamma'   \ottsym{)}  $.
    \end{itemize}
    By the induction hypothesis, we have $\vdash   \Delta   \ottsym{(}   \Gamma'   \ottsym{)} $. By $ \alpha   \notin    \mathrm{dom}   \ottsym{(}   \Gamma'   \ottsym{)}  $, we have $ \alpha   \notin    \mathrm{dom}   \ottsym{(}    \Delta   \ottsym{(}   \Gamma'   \ottsym{)}    \ottsym{)}  $ because $  \mathrm{dom}   \ottsym{(}    \Delta   \ottsym{(}   \Gamma'   \ottsym{)}    \ottsym{)}    \subseteq    \mathrm{dom}   \ottsym{(}   \Gamma'   \ottsym{)}  $. Thus, \rname{C}{TVar} derives $\vdash   \Delta   \ottsym{(}   \Gamma'   \ottsym{)}   \ottsym{,}  \alpha  \ottsym{:}  \ottnt{K}$ as required.
    \item[\rname{K}{Var}] $\vdash  \Gamma$, $ \alpha   \ottsym{:}   \ottnt{K}   \in   \Gamma $, and $S = \alpha$ are given for some $\alpha$. By the induction hypothesis, we have $\vdash   \Delta   \ottsym{(}   \Gamma   \ottsym{)} $. By Definition~\ref{def:delta}, we have $ \alpha   \ottsym{:}   \ottnt{K}   \in    \Delta   \ottsym{(}   \Gamma   \ottsym{)}  $. Thus, \rname{K}{Var} derives $ \Delta   \ottsym{(}   \Gamma   \ottsym{)}   \vdash  \alpha  \ottsym{:}  \ottnt{K}$.
    \item[\rname{K}{Cons}] For some $\mathcal{C}$, $ \bm{ { S } } ^ {  \mathit{I}  } $, and $ {\bm{ { \ottnt{K} } } }^{ \mathit{I} } $, the following are given:
    \begin{itemize}
      \item $S = \mathcal{C} \,  \bm{ { S } } ^ {  \mathit{I}  } $,
      \item $\vdash  \Gamma$,
      \item $ \mathcal{C}   \ottsym{:}    \Pi {\bm{ { \ottnt{K} } } }^{ \mathit{I} }   \rightarrow  \ottnt{K}   \in   \Sigma $, and
      \item $\Gamma  \vdash   \bm{ { S } }^{ \mathit{I} } : \bm{ \ottnt{K} }^{ \mathit{I} } $.
    \end{itemize}
    By the induction hypothesis, we have $\vdash   \Delta   \ottsym{(}   \Gamma   \ottsym{)} $ and $ \Delta   \ottsym{(}   \Gamma   \ottsym{)}   \vdash   \bm{ { S } }^{ \mathit{I} } : \bm{ \ottnt{K} }^{ \mathit{I} } $. Thus, \rname{K}{Cons} derives $ \Delta   \ottsym{(}   \Gamma   \ottsym{)}   \vdash  \mathcal{C} \,  \bm{ { S } } ^ {  \mathit{I}  }   \ottsym{:}  \ottnt{K}$ as required.
    \item[\rname{K}{Fun}] For some $\ottnt{A}$, $\varepsilon$, and $\ottnt{B}$, the following are given:
    \begin{itemize}
      \item $S =  \ottnt{A}    \rightarrow_{ \varepsilon }    \ottnt{B} $,
      \item $\ottnt{K} =  \mathbf{Typ} $,
      \item $\Gamma  \vdash  \ottnt{A}  \ottsym{:}   \mathbf{Typ} $,
      \item $\Gamma  \vdash  \varepsilon  \ottsym{:}   \mathbf{Eff} $, and
      \item $\Gamma  \vdash  \ottnt{B}  \ottsym{:}   \mathbf{Typ} $.
    \end{itemize}
    By the induction hypothesis, we have
    \begin{itemize}
      \item $ \Delta   \ottsym{(}   \Gamma   \ottsym{)}   \vdash  \ottnt{A}  \ottsym{:}   \mathbf{Typ} $,
      \item $ \Delta   \ottsym{(}   \Gamma   \ottsym{)}   \vdash  \varepsilon  \ottsym{:}   \mathbf{Eff} $, and
      \item $ \Delta   \ottsym{(}   \Gamma   \ottsym{)}   \vdash  \ottnt{B}  \ottsym{:}   \mathbf{Typ} $.
    \end{itemize}
    Thus, \rname{K}{Fun} derives $ \Delta   \ottsym{(}   \Gamma   \ottsym{)}   \vdash   \ottnt{A}    \rightarrow_{ \varepsilon }    \ottnt{B}   \ottsym{:}   \mathbf{Typ} $.
    \item[\rname{K}{Poly}] For some $\alpha$, $\ottnt{K'}$, $\ottnt{A}$, and $\varepsilon$, the following are given:
    \begin{itemize}
      \item $S =   \forall   \alpha  \ottsym{:}  \ottnt{K'}   \ottsym{.}    \ottnt{A}    ^{ \varepsilon }  $,
      \item $\ottnt{K} =  \mathbf{Typ} $ ,
      \item $\Gamma  \ottsym{,}  \alpha  \ottsym{:}  \ottnt{K'}  \vdash  \ottnt{A}  \ottsym{:}   \mathbf{Typ} $, and
      \item $\Gamma  \ottsym{,}  \alpha  \ottsym{:}  \ottnt{K'}  \vdash  \varepsilon  \ottsym{:}   \mathbf{Eff} $.
    \end{itemize}
    By the induction hypothesis, we have
    \begin{itemize}
      \item $ \Delta   \ottsym{(}   \Gamma  \ottsym{,}  \alpha  \ottsym{:}  \ottnt{K'}   \ottsym{)}   \vdash  \ottnt{A}  \ottsym{:}   \mathbf{Typ} $ and
      \item $ \Delta   \ottsym{(}   \Gamma  \ottsym{,}  \alpha  \ottsym{:}  \ottnt{K'}   \ottsym{)}   \vdash  \varepsilon  \ottsym{:}   \mathbf{Eff} $.
    \end{itemize}
    By Definition~\ref{def:delta}, we have $ \Delta   \ottsym{(}   \Gamma  \ottsym{,}  \alpha  \ottsym{:}  \ottnt{K'}   \ottsym{)}  =  \Delta   \ottsym{(}   \Gamma   \ottsym{)}   \ottsym{,}  \alpha  \ottsym{:}  \ottnt{K'}$. Thus, \rname{K}{Poly} derives $ \Delta   \ottsym{(}   \Gamma   \ottsym{)}   \vdash    \forall   \alpha  \ottsym{:}  \ottnt{K'}   \ottsym{.}    \ottnt{A}    ^{ \varepsilon }    \ottsym{:}   \mathbf{Typ} $ as required.
  \end{divcases}
\end{proof}

\begin{lemma}\label{lem:entailment}
  \phantom{}
  \begin{enumerate}
    \item\label{lem:entailment:refl}
          For any $\Gamma$ and $\varepsilon$,
          if $\Gamma  \vdash  \varepsilon  \ottsym{:}   \mathbf{Eff} $,
          then $\Gamma  \vdash   \varepsilon  \olessthan  \varepsilon $ holds.

    \item\label{lem:entailment:trans}
          For any $\Gamma$, $\varepsilon_{{\mathrm{1}}}$, $\varepsilon_{{\mathrm{2}}}$, and $\varepsilon_{{\mathrm{3}}}$,
          if $\Gamma  \vdash   \varepsilon_{{\mathrm{1}}}  \olessthan  \varepsilon_{{\mathrm{2}}} $ and $\Gamma  \vdash   \varepsilon_{{\mathrm{2}}}  \olessthan  \varepsilon_{{\mathrm{3}}} $,
          then $\Gamma  \vdash   \varepsilon_{{\mathrm{1}}}  \olessthan  \varepsilon_{{\mathrm{3}}} $.
  \end{enumerate}
\end{lemma}

\begin{proof}
  \phantom{}
  \begin{enumerate}
    \item Clearly because of Lemma~\ref{lem:delta_context}\ref{lem:delta_context:kinding} and because $ \bbZero $ is a unit element.

    \item Clearly because $ \odot $ is associative and preserves well-formendness.
  \end{enumerate}
\end{proof}

\begin{lemma}[Transitivity of Subtyping]\label{lem:trans_subtyping}
  \phantom{}\\
  \begin{enumerate}
    \item If $\Gamma  \vdash  \ottnt{A_{{\mathrm{1}}}}  <:  \ottnt{A_{{\mathrm{2}}}}$ and $\Gamma  \vdash  \ottnt{A_{{\mathrm{2}}}}  <:  \ottnt{A_{{\mathrm{3}}}}$, then $\Gamma  \vdash  \ottnt{A_{{\mathrm{1}}}}  <:  \ottnt{A_{{\mathrm{3}}}}$.
    \item If $\Gamma  \vdash  \ottnt{A_{{\mathrm{1}}}}  \mid  \varepsilon_{{\mathrm{1}}}  <:  \ottnt{A_{{\mathrm{2}}}}  \mid  \varepsilon_{{\mathrm{2}}}$ and $\Gamma  \vdash  \ottnt{A_{{\mathrm{2}}}}  \mid  \varepsilon_{{\mathrm{2}}}  <:  \ottnt{A_{{\mathrm{3}}}}  \mid  \varepsilon_{{\mathrm{3}}}$, then $\Gamma  \vdash  \ottnt{A_{{\mathrm{1}}}}  \mid  \varepsilon_{{\mathrm{1}}}  <:  \ottnt{A_{{\mathrm{3}}}}  \mid  \varepsilon_{{\mathrm{3}}}$.
  \end{enumerate}
\end{lemma}

\begin{proof}
  By the structural induction on the summation of the sizes of $\ottnt{A_{{\mathrm{1}}}}$, $\ottnt{A_{{\mathrm{2}}}}$, and $\ottnt{A_{{\mathrm{3}}}}$.
  If either $\Gamma  \vdash  \ottnt{A_{{\mathrm{1}}}}  <:  \ottnt{A_{{\mathrm{2}}}}$ or $\Gamma  \vdash  \ottnt{A_{{\mathrm{2}}}}  <:  \ottnt{A_{{\mathrm{3}}}}$ is derived by \rname{ST}{Refl},
  then we have $\Gamma  \vdash  \ottnt{A_{{\mathrm{1}}}}  <:  \ottnt{A_{{\mathrm{3}}}}$ immediately.
  Thus, we suppose that neither $\Gamma  \vdash  \ottnt{A_{{\mathrm{1}}}}  <:  \ottnt{A_{{\mathrm{2}}}}$ nor $\Gamma  \vdash  \ottnt{A_{{\mathrm{2}}}}  <:  \ottnt{A_{{\mathrm{3}}}}$ is derived by \rname{ST}{Refl} in the following.
  We proceed by case analysis on what form $\ottnt{A_{{\mathrm{1}}}}$ has.
  \begin{divcases}
    \item[$\ottnt{A_{{\mathrm{1}}}} = \tau$] No rules other than \rname{ST}{Refl} can derive $\Gamma  \vdash  \ottnt{A_{{\mathrm{1}}}}  <:  \ottnt{A_{{\mathrm{2}}}}$.

    \item[$\ottnt{A_{{\mathrm{1}}}} =  \ottnt{B_{{\mathrm{1}}}}    \rightarrow_{ \varepsilon_{{\mathrm{1}}} }    \ottnt{C_{{\mathrm{1}}}} $]
    Since only \rname{ST}{Fun} can derive $\Gamma  \vdash   \ottnt{B_{{\mathrm{1}}}}    \rightarrow_{ \varepsilon_{{\mathrm{1}}} }    \ottnt{C_{{\mathrm{1}}}}   <:  \ottnt{A_{{\mathrm{2}}}}$,
    we have $\ottnt{A_{{\mathrm{2}}}} =  \ottnt{B_{{\mathrm{2}}}}    \rightarrow_{ \varepsilon_{{\mathrm{2}}} }    \ottnt{C_{{\mathrm{2}}}} $ for some $\ottnt{B_{{\mathrm{2}}}}$, $\varepsilon_{{\mathrm{2}}}$, and $\ottnt{C_{{\mathrm{2}}}}$
    such that
    \begin{itemize}
      \item $\Gamma  \vdash  \ottnt{B_{{\mathrm{2}}}}  <:  \ottnt{B_{{\mathrm{1}}}}$ and
      \item $\Gamma  \vdash  \ottnt{C_{{\mathrm{1}}}}  \mid  \varepsilon_{{\mathrm{1}}}  <:  \ottnt{C_{{\mathrm{2}}}}  \mid  \varepsilon_{{\mathrm{2}}}$.
    \end{itemize}
    Since only \rname{ST}{Fun} can derive $\Gamma  \vdash   \ottnt{B_{{\mathrm{2}}}}    \rightarrow_{ \varepsilon_{{\mathrm{2}}} }    \ottnt{C_{{\mathrm{2}}}}   <:  \ottnt{A_{{\mathrm{3}}}}$,
    we have $\ottnt{A_{{\mathrm{3}}}} =  \ottnt{B_{{\mathrm{3}}}}    \rightarrow_{ \varepsilon_{{\mathrm{3}}} }    \ottnt{C_{{\mathrm{3}}}} $ for some $\ottnt{B_{{\mathrm{3}}}}$, $\varepsilon_{{\mathrm{3}}}$, and $\ottnt{C_{{\mathrm{3}}}}$
    such that
    \begin{itemize}
      \item $\Gamma  \vdash  \ottnt{B_{{\mathrm{3}}}}  <:  \ottnt{B_{{\mathrm{2}}}}$ and
      \item $\Gamma  \vdash  \ottnt{C_{{\mathrm{2}}}}  \mid  \varepsilon_{{\mathrm{2}}}  <:  \ottnt{C_{{\mathrm{3}}}}  \mid  \varepsilon_{{\mathrm{3}}}$.
    \end{itemize}
    Since only \rname{ST}{Comp} can derive
    $\Gamma  \vdash  \ottnt{C_{{\mathrm{2}}}}  \mid  \varepsilon_{{\mathrm{2}}}  <:  \ottnt{C_{{\mathrm{3}}}}  \mid  \varepsilon_{{\mathrm{3}}}$ and $\Gamma  \vdash  \ottnt{C_{{\mathrm{1}}}}  \mid  \varepsilon_{{\mathrm{1}}}  <:  \ottnt{C_{{\mathrm{2}}}}  \mid  \varepsilon_{{\mathrm{2}}}$,
    we have
    \begin{itemize}
      \item $\Gamma  \vdash  \ottnt{C_{{\mathrm{1}}}}  <:  \ottnt{C_{{\mathrm{2}}}}$,
      \item $\Gamma  \vdash  \ottnt{C_{{\mathrm{2}}}}  <:  \ottnt{C_{{\mathrm{3}}}}$,
      \item $\Gamma  \vdash   \varepsilon_{{\mathrm{1}}}  \olessthan  \varepsilon_{{\mathrm{2}}} $, and
      \item $\Gamma  \vdash   \varepsilon_{{\mathrm{2}}}  \olessthan  \varepsilon_{{\mathrm{3}}} $.
    \end{itemize}
    By the induction hypothesis and Lemma~\ref{lem:entailment}\ref{lem:entailment:trans},
    we have
    \begin{itemize}
      \item $\Gamma  \vdash  \ottnt{B_{{\mathrm{3}}}}  <:  \ottnt{B_{{\mathrm{1}}}}$,
      \item $\Gamma  \vdash   \varepsilon_{{\mathrm{1}}}  \olessthan  \varepsilon_{{\mathrm{3}}} $, and
      \item $\Gamma  \vdash  \ottnt{C_{{\mathrm{1}}}}  <:  \ottnt{C_{{\mathrm{3}}}}$.
    \end{itemize}
    Thus, we have $\Gamma  \vdash  \ottnt{A_{{\mathrm{1}}}}  <:  \ottnt{A_{{\mathrm{3}}}}$ by \rname{ST}{Fun} as required.

    \item[$\ottnt{A_{{\mathrm{1}}}} =   \forall   \alpha  \ottsym{:}  \ottnt{K}   \ottsym{.}    \ottnt{B_{{\mathrm{1}}}}    ^{ \varepsilon_{{\mathrm{1}}} }  $]
    Since only \rname{ST}{Poly} can derive $\Gamma  \vdash    \forall   \alpha  \ottsym{:}  \ottnt{K}   \ottsym{.}    \ottnt{B_{{\mathrm{1}}}}    ^{ \varepsilon_{{\mathrm{1}}} }    <:  \ottnt{A_{{\mathrm{2}}}}$,
    we have $\ottnt{A_{{\mathrm{2}}}} =   \forall   \alpha  \ottsym{:}  \ottnt{K}   \ottsym{.}    \ottnt{B_{{\mathrm{2}}}}    ^{ \varepsilon_{{\mathrm{2}}} }  $ for some $\ottnt{B_{{\mathrm{2}}}}$ and $\varepsilon_{{\mathrm{2}}}$ such that $\Gamma  \ottsym{,}  \alpha  \ottsym{:}  \ottnt{K}  \vdash  \ottnt{B_{{\mathrm{1}}}}  \mid  \varepsilon_{{\mathrm{1}}}  <:  \ottnt{B_{{\mathrm{2}}}}  \mid  \varepsilon_{{\mathrm{2}}}$.
    Since only \rname{ST}{Poly} can derive $\Gamma  \vdash    \forall   \alpha  \ottsym{:}  \ottnt{K}   \ottsym{.}    \ottnt{B_{{\mathrm{2}}}}    ^{ \varepsilon_{{\mathrm{2}}} }    <:  \ottnt{A_{{\mathrm{3}}}}$,
    we have $\ottnt{A_{{\mathrm{3}}}} =   \forall   \alpha  \ottsym{:}  \ottnt{K}   \ottsym{.}    \ottnt{B_{{\mathrm{3}}}}    ^{ \varepsilon_{{\mathrm{3}}} }  $ for some $\ottnt{B_{{\mathrm{3}}}}$ and $\varepsilon_{{\mathrm{3}}}$ such that $\Gamma  \ottsym{,}  \alpha  \ottsym{:}  \ottnt{K}  \vdash  \ottnt{B_{{\mathrm{2}}}}  \mid  \varepsilon_{{\mathrm{2}}}  <:  \ottnt{B_{{\mathrm{3}}}}  \mid  \varepsilon_{{\mathrm{3}}}$.
    Since only \rname{ST}{Comp} can derive
    $\Gamma  \ottsym{,}  \alpha  \ottsym{:}  \ottnt{K}  \vdash  \ottnt{B_{{\mathrm{1}}}}  \mid  \varepsilon_{{\mathrm{1}}}  <:  \ottnt{B_{{\mathrm{2}}}}  \mid  \varepsilon_{{\mathrm{2}}}$ and
    $\Gamma  \ottsym{,}  \alpha  \ottsym{:}  \ottnt{K}  \vdash  \ottnt{B_{{\mathrm{2}}}}  \mid  \varepsilon_{{\mathrm{2}}}  <:  \ottnt{B_{{\mathrm{3}}}}  \mid  \varepsilon_{{\mathrm{3}}}$,
    we have
    \begin{itemize}
      \item $\Gamma  \ottsym{,}  \alpha  \ottsym{:}  \ottnt{K}  \vdash  \ottnt{B_{{\mathrm{1}}}}  <:  \ottnt{B_{{\mathrm{2}}}}$,
      \item $\Gamma  \ottsym{,}  \alpha  \ottsym{:}  \ottnt{K}  \vdash   \varepsilon_{{\mathrm{1}}}  \olessthan  \varepsilon_{{\mathrm{2}}} $,
      \item $\Gamma  \ottsym{,}  \alpha  \ottsym{:}  \ottnt{K}  \vdash  \ottnt{B_{{\mathrm{2}}}}  <:  \ottnt{B_{{\mathrm{3}}}}$, and
      \item $\Gamma  \ottsym{,}  \alpha  \ottsym{:}  \ottnt{K}  \vdash   \varepsilon_{{\mathrm{2}}}  \olessthan  \varepsilon_{{\mathrm{3}}} $.
    \end{itemize}
    By the induction hypothesis and Lemma~\ref{lem:entailment}\ref{lem:entailment:trans},
    we have
    \begin{itemize}
      \item $\Gamma  \ottsym{,}  \alpha  \ottsym{:}  \ottnt{K}  \vdash  \ottnt{B_{{\mathrm{1}}}}  <:  \ottnt{B_{{\mathrm{3}}}}$ and
      \item $\Gamma  \ottsym{,}  \alpha  \ottsym{:}  \ottnt{K}  \vdash   \varepsilon_{{\mathrm{1}}}  \olessthan  \varepsilon_{{\mathrm{3}}} $.
    \end{itemize}
    Thus, we have $\Gamma  \vdash  \ottnt{A_{{\mathrm{1}}}}  <:  \ottnt{A_{{\mathrm{3}}}}$ by \rname{ST}{Poly} as required.
  \end{divcases}
\end{proof}

\begin{lemma}[Weakening]\label{lem:weakening}
  Suppose that $\vdash  \Gamma_{{\mathrm{1}}}  \ottsym{,}  \Gamma_{{\mathrm{2}}}$ and $  \mathrm{dom}   \ottsym{(}   \Gamma_{{\mathrm{2}}}   \ottsym{)}    \cap    \mathrm{dom}   \ottsym{(}   \Gamma_{{\mathrm{3}}}   \ottsym{)}    \ottsym{=}  \emptyset$.
  \begin{enumerate}
    \item\label{lem:weakening:wf} If $\vdash  \Gamma_{{\mathrm{1}}}  \ottsym{,}  \Gamma_{{\mathrm{3}}}$, then $\vdash  \Gamma_{{\mathrm{1}}}  \ottsym{,}  \Gamma_{{\mathrm{2}}}  \ottsym{,}  \Gamma_{{\mathrm{3}}}$.

    \item\label{lem:weakening:kinding} If $\Gamma_{{\mathrm{1}}}  \ottsym{,}  \Gamma_{{\mathrm{3}}}  \vdash  S  \ottsym{:}  \ottnt{K}$, then $\Gamma_{{\mathrm{1}}}  \ottsym{,}  \Gamma_{{\mathrm{2}}}  \ottsym{,}  \Gamma_{{\mathrm{3}}}  \vdash  S  \ottsym{:}  \ottnt{K}$.

    \item\label{lem:weakening:subtyping} If $\Gamma_{{\mathrm{1}}}  \ottsym{,}  \Gamma_{{\mathrm{3}}}  \vdash  \ottnt{A}  <:  \ottnt{B}$, then $\Gamma_{{\mathrm{1}}}  \ottsym{,}  \Gamma_{{\mathrm{2}}}  \ottsym{,}  \Gamma_{{\mathrm{3}}}  \vdash  \ottnt{A}  <:  \ottnt{B}$.

    \item\label{lem:weakening:subtyping_comp} If $\Gamma_{{\mathrm{1}}}  \ottsym{,}  \Gamma_{{\mathrm{3}}}  \vdash  \ottnt{A_{{\mathrm{1}}}}  \mid  \varepsilon_{{\mathrm{1}}}  <:  \ottnt{A_{{\mathrm{2}}}}  \mid  \varepsilon_{{\mathrm{2}}}$, then $\Gamma_{{\mathrm{1}}}  \ottsym{,}  \Gamma_{{\mathrm{2}}}  \ottsym{,}  \Gamma_{{\mathrm{3}}}  \vdash  \ottnt{A_{{\mathrm{1}}}}  \mid  \varepsilon_{{\mathrm{1}}}  <:  \ottnt{A_{{\mathrm{2}}}}  \mid  \varepsilon_{{\mathrm{2}}}$.

    \item\label{lem:weakening:typing} If $\Gamma_{{\mathrm{1}}}  \ottsym{,}  \Gamma_{{\mathrm{3}}}  \vdash  \ottnt{e}  \ottsym{:}  \ottnt{A}  \mid  \varepsilon$, then $\Gamma_{{\mathrm{1}}}  \ottsym{,}  \Gamma_{{\mathrm{2}}}  \ottsym{,}  \Gamma_{{\mathrm{3}}}  \vdash  \ottnt{e}  \ottsym{:}  \ottnt{A}  \mid  \varepsilon$.

    \item\label{lem:weakening:handling} If $ \Gamma_{{\mathrm{1}}}  \ottsym{,}  \Gamma_{{\mathrm{3}}}  \vdash _{ \sigma }  \ottnt{h}  :  \ottnt{A}   \Rightarrow  ^ { \varepsilon }  \ottnt{B} $, then $ \Gamma_{{\mathrm{1}}}  \ottsym{,}  \Gamma_{{\mathrm{2}}}  \ottsym{,}  \Gamma_{{\mathrm{3}}}  \vdash _{ \sigma }  \ottnt{h}  :  \ottnt{A}   \Rightarrow  ^ { \varepsilon }  \ottnt{B} $.
  \end{enumerate}
\end{lemma}

\begin{proof}
  \phantom{}
  \begin{itemize}
    \item[(1)(2)] By mutual induction on derivations of the judgments. We proceed by case analysis on the rule applied lastly to the derivation.
          \begin{divcases}
            \item[\rname{C}{Empty}] Clearly because of $\vdash  \Gamma_{{\mathrm{1}}}  \ottsym{,}  \Gamma_{{\mathrm{2}}}$ and $\Gamma_{{\mathrm{1}}} = \Gamma_{{\mathrm{3}}} =  \emptyset $.
            \item[\rname{C}{Var}] If $\Gamma_{{\mathrm{3}}} =  \emptyset $, then $\vdash  \Gamma_{{\mathrm{1}}}  \ottsym{,}  \Gamma_{{\mathrm{2}}}  \ottsym{,}  \Gamma_{{\mathrm{3}}}$ holds immediately. If $\Gamma_{{\mathrm{3}}} \neq  \emptyset $, then for some $\Gamma'_{{\mathrm{3}}}$, $\mathit{x}$, and $\ottnt{A}$, the following are given:
            \begin{itemize}
              \item $\Gamma_{{\mathrm{3}}} = \Gamma'_{{\mathrm{3}}}  \ottsym{,}  \mathit{x}  \ottsym{:}  \ottnt{A}$,
              \item $ \mathit{x}   \notin    \mathrm{dom}   \ottsym{(}   \Gamma_{{\mathrm{1}}}  \ottsym{,}  \Gamma'_{{\mathrm{3}}}   \ottsym{)}  $, and
              \item $\Gamma_{{\mathrm{1}}}  \ottsym{,}  \Gamma'_{{\mathrm{3}}}  \vdash  \ottnt{A}  \ottsym{:}   \mathbf{Typ} $.
            \end{itemize}
            Since $  \mathrm{dom}   \ottsym{(}   \Gamma_{{\mathrm{2}}}   \ottsym{)}    \cap    \mathrm{dom}   \ottsym{(}   \Gamma'_{{\mathrm{3}}}   \ottsym{)}    \ottsym{=}  \emptyset$ holds, we have $\Gamma_{{\mathrm{1}}}  \ottsym{,}  \Gamma_{{\mathrm{2}}}  \ottsym{,}  \Gamma'_{{\mathrm{3}}}  \vdash  \ottnt{A}  \ottsym{:}   \mathbf{Typ} $ by the induction hypothesis. By $ \mathit{x}   \notin    \mathrm{dom}   \ottsym{(}   \Gamma_{{\mathrm{1}}}  \ottsym{,}  \Gamma'_{{\mathrm{3}}}   \ottsym{)}  $ and $  \mathrm{dom}   \ottsym{(}   \Gamma_{{\mathrm{2}}}   \ottsym{)}    \cap    \mathrm{dom}   \ottsym{(}   \Gamma'_{{\mathrm{3}}}  \ottsym{,}  \mathit{x}  \ottsym{:}  \ottnt{A}   \ottsym{)}    \ottsym{=}  \emptyset$, we have $ \mathit{x}   \notin    \mathrm{dom}   \ottsym{(}   \Gamma_{{\mathrm{1}}}  \ottsym{,}  \Gamma_{{\mathrm{2}}}  \ottsym{,}  \Gamma'_{{\mathrm{3}}}   \ottsym{)}  $. Thus, \rname{C}{Var} derives $\vdash  \Gamma_{{\mathrm{1}}}  \ottsym{,}  \Gamma_{{\mathrm{2}}}  \ottsym{,}  \Gamma'_{{\mathrm{3}}}  \ottsym{,}  \mathit{x}  \ottsym{:}  \ottnt{A}$.
            \item[\rname{C}{TVar}] If $\Gamma_{{\mathrm{3}}} =  \emptyset $, then $\vdash  \Gamma_{{\mathrm{1}}}  \ottsym{,}  \Gamma_{{\mathrm{2}}}$ holds immediately. If $\Gamma_{{\mathrm{3}}} \neq  \emptyset $, then for some $\Gamma'_{{\mathrm{3}}}$, $\alpha$, and $\ottnt{K}$, the following are given:
            \begin{itemize}
              \item $\Gamma_{{\mathrm{3}}} = \Gamma'_{{\mathrm{3}}}  \ottsym{,}  \alpha  \ottsym{:}  \ottnt{K}$,
              \item $ \alpha   \notin    \mathrm{dom}   \ottsym{(}   \Gamma_{{\mathrm{1}}}  \ottsym{,}  \Gamma'_{{\mathrm{3}}}   \ottsym{)}  $, and
              \item $\vdash  \Gamma_{{\mathrm{1}}}  \ottsym{,}  \Gamma'_{{\mathrm{3}}}$.
            \end{itemize}
            Since $  \mathrm{dom}   \ottsym{(}   \Gamma_{{\mathrm{2}}}   \ottsym{)}    \cap    \mathrm{dom}   \ottsym{(}   \Gamma'_{{\mathrm{3}}}   \ottsym{)}    \ottsym{=}  \emptyset$, we have $\vdash  \Gamma_{{\mathrm{1}}}  \ottsym{,}  \Gamma_{{\mathrm{2}}}  \ottsym{,}  \Gamma'_{{\mathrm{3}}}$ by the induction hypothesis. By $ \alpha   \notin    \mathrm{dom}   \ottsym{(}   \Gamma_{{\mathrm{1}}}  \ottsym{,}  \Gamma'_{{\mathrm{3}}}   \ottsym{)}  $ and $  \mathrm{dom}   \ottsym{(}   \Gamma_{{\mathrm{2}}}   \ottsym{)}    \cap    \mathrm{dom}   \ottsym{(}   \Gamma'_{{\mathrm{3}}}  \ottsym{,}  \alpha  \ottsym{:}  \ottnt{K}   \ottsym{)}    \ottsym{=}  \emptyset$, we have $ \alpha   \notin    \mathrm{dom}   \ottsym{(}   \Gamma_{{\mathrm{1}}}  \ottsym{,}  \Gamma_{{\mathrm{2}}}  \ottsym{,}  \Gamma'_{{\mathrm{3}}}   \ottsym{)}  $ Thus, \rname{C}{TVar} derives $\vdash  \Gamma_{{\mathrm{1}}}  \ottsym{,}  \Gamma_{{\mathrm{2}}}  \ottsym{,}  \Gamma'_{{\mathrm{3}}}  \ottsym{,}  \alpha  \ottsym{:}  \ottnt{K}$.
            \item[\rname{K}{Var}] For some $\alpha$, the following are given:
            \begin{itemize}
              \item $S = \alpha$,
              \item $\vdash  \Gamma_{{\mathrm{1}}}  \ottsym{,}  \Gamma_{{\mathrm{3}}}$, and
              \item $ \alpha   \ottsym{:}   \ottnt{K}   \in   \Gamma_{{\mathrm{1}}}  \ottsym{,}  \Gamma_{{\mathrm{3}}} $.
            \end{itemize}
            By the induction hypothesis, we have $\vdash  \Gamma_{{\mathrm{1}}}  \ottsym{,}  \Gamma_{{\mathrm{2}}}  \ottsym{,}  \Gamma_{{\mathrm{3}}}$. Thus, $\Gamma_{{\mathrm{1}}}  \ottsym{,}  \Gamma_{{\mathrm{2}}}  \ottsym{,}  \Gamma_{{\mathrm{3}}}  \vdash  \alpha  \ottsym{:}  \ottnt{K}$ holds because of $ \alpha   \ottsym{:}   \ottnt{K}   \in   \Gamma_{{\mathrm{1}}}  \ottsym{,}  \Gamma_{{\mathrm{2}}}  \ottsym{,}  \Gamma_{{\mathrm{3}}} $.
            \item[\rname{K}{Fun}] For some $\ottnt{A}$, $\ottnt{B}$, and $\varepsilon$, the following are given:
            \begin{itemize}
              \item $S =  \ottnt{A}    \rightarrow_{ \varepsilon }    \ottnt{B} $,
              \item $\ottnt{K} =  \mathbf{Typ} $,
              \item $\Gamma_{{\mathrm{1}}}  \ottsym{,}  \Gamma_{{\mathrm{3}}}  \vdash  \ottnt{A}  \ottsym{:}   \mathbf{Typ} $,
              \item $\Gamma_{{\mathrm{1}}}  \ottsym{,}  \Gamma_{{\mathrm{3}}}  \vdash  \varepsilon  \ottsym{:}   \mathbf{Eff} $, and
              \item $\Gamma_{{\mathrm{1}}}  \ottsym{,}  \Gamma_{{\mathrm{3}}}  \vdash  \ottnt{B}  \ottsym{:}   \mathbf{Typ} $.
            \end{itemize}
            By the induction hypothesis, we have
            \begin{itemize}
              \item $\Gamma_{{\mathrm{1}}}  \ottsym{,}  \Gamma_{{\mathrm{2}}}  \ottsym{,}  \Gamma_{{\mathrm{3}}}  \vdash  \ottnt{A}  \ottsym{:}   \mathbf{Typ} $,
              \item $\Gamma_{{\mathrm{1}}}  \ottsym{,}  \Gamma_{{\mathrm{2}}}  \ottsym{,}  \Gamma_{{\mathrm{3}}}  \vdash  \varepsilon  \ottsym{:}   \mathbf{Eff} $, and
              \item $\Gamma_{{\mathrm{1}}}  \ottsym{,}  \Gamma_{{\mathrm{2}}}  \ottsym{,}  \Gamma_{{\mathrm{3}}}  \vdash  \ottnt{B}  \ottsym{:}   \mathbf{Typ} $.
            \end{itemize}
            Thus, \rname{K}{Fun} derives $\Gamma_{{\mathrm{1}}}  \ottsym{,}  \Gamma_{{\mathrm{2}}}  \ottsym{,}  \Gamma_{{\mathrm{3}}}  \vdash   \ottnt{A}    \rightarrow_{ \varepsilon }    \ottnt{B}   \ottsym{:}   \mathbf{Typ} $.
            \item[\rname{K}{Poly}] Without loss of generality, we can choose $\alpha$ such that $ \alpha   \notin    \mathrm{dom}   \ottsym{(}   \Gamma_{{\mathrm{2}}}   \ottsym{)}  $. For some $\ottnt{K'}$, $\ottnt{A}$, and $\varepsilon$, the following are given:
            \begin{itemize}
              \item $S =   \forall   \alpha  \ottsym{:}  \ottnt{K'}   \ottsym{.}    \ottnt{A}    ^{ \varepsilon }  $,
              \item $\ottnt{K} =  \mathbf{Typ} $,
              \item $\Gamma_{{\mathrm{1}}}  \ottsym{,}  \Gamma_{{\mathrm{3}}}  \ottsym{,}  \alpha  \ottsym{:}  \ottnt{K'}  \vdash  \ottnt{A}  \ottsym{:}   \mathbf{Typ} $, and
              \item $\Gamma_{{\mathrm{1}}}  \ottsym{,}  \Gamma_{{\mathrm{3}}}  \ottsym{,}  \alpha  \ottsym{:}  \ottnt{K'}  \vdash  \varepsilon  \ottsym{:}   \mathbf{Eff} $.
            \end{itemize}
            Since $  \mathrm{dom}   \ottsym{(}   \Gamma_{{\mathrm{2}}}   \ottsym{)}    \cap    \mathrm{dom}   \ottsym{(}   \Gamma_{{\mathrm{3}}}  \ottsym{,}  \alpha  \ottsym{:}  \ottnt{K'}   \ottsym{)}    \ottsym{=}  \emptyset$, we have
            \begin{itemize}
              \item $\Gamma_{{\mathrm{1}}}  \ottsym{,}  \Gamma_{{\mathrm{2}}}  \ottsym{,}  \Gamma_{{\mathrm{3}}}  \ottsym{,}  \alpha  \ottsym{:}  \ottnt{K'}  \vdash  \ottnt{A}  \ottsym{:}   \mathbf{Typ} $ and
              \item $\Gamma_{{\mathrm{1}}}  \ottsym{,}  \Gamma_{{\mathrm{2}}}  \ottsym{,}  \Gamma_{{\mathrm{3}}}  \ottsym{,}  \alpha  \ottsym{:}  \ottnt{K'}  \vdash  \varepsilon  \ottsym{:}   \mathbf{Eff} $
            \end{itemize}
            by the induction hypothesis. Thus, \rname{K}{Poly} derives $\Gamma_{{\mathrm{1}}}  \ottsym{,}  \Gamma_{{\mathrm{2}}}  \ottsym{,}  \Gamma_{{\mathrm{3}}}  \vdash    \forall   \alpha  \ottsym{:}  \ottnt{K'}   \ottsym{.}    \ottnt{A}    ^{ \varepsilon }    \ottsym{:}   \mathbf{Typ} $.
            \item[\rname{K}{Cons}] For some $\mathcal{C}$, $ \bm{ { S } } ^ {  \mathit{I}  } $, and $ {\bm{ { \ottnt{K} } } }^{ \mathit{I} } $, the following are given:
            \begin{itemize}
              \item $S = \mathcal{C} \,  \bm{ { S } } ^ {  \mathit{I}  } $,
              \item $ \mathcal{C}   \ottsym{:}    \Pi {\bm{ { \ottnt{K} } } }^{ \mathit{I} }   \rightarrow  \ottnt{K}   \in   \Sigma $,
              \item $\vdash  \Gamma_{{\mathrm{1}}}  \ottsym{,}  \Gamma_{{\mathrm{3}}}$, and
              \item $\Gamma_{{\mathrm{1}}}  \ottsym{,}  \Gamma_{{\mathrm{3}}}  \vdash   \bm{ { S } }^{ \mathit{I} } : \bm{ \ottnt{K} }^{ \mathit{I} } $.
            \end{itemize}
            By the induction hypothesis, we have $\vdash  \Gamma_{{\mathrm{1}}}  \ottsym{,}  \Gamma_{{\mathrm{2}}}  \ottsym{,}  \Gamma_{{\mathrm{3}}}$ and $\Gamma_{{\mathrm{1}}}  \ottsym{,}  \Gamma_{{\mathrm{2}}}  \ottsym{,}  \Gamma_{{\mathrm{3}}}  \vdash   \bm{ { S } }^{ \mathit{I} } : \bm{ \ottnt{K} }^{ \mathit{I} } $. Thus, \rname{K}{Cons} derives $\Gamma_{{\mathrm{1}}}  \ottsym{,}  \Gamma_{{\mathrm{2}}}  \ottsym{,}  \Gamma_{{\mathrm{3}}}  \vdash  \mathcal{C} \,  \bm{ { S } } ^ {  \mathit{I}  }   \ottsym{:}  \ottnt{K}$.
          \end{divcases}

    \item[(3)(4)] By mutual induction on derivations of the judgments.
          We proceed by case analysis on the rule applied lastly to the derivation.
          \begin{divcases}
            \item[\rname{ST}{Refl}]
            $\ottnt{A} = \ottnt{B}$ and $\Gamma_{{\mathrm{1}}}  \ottsym{,}  \Gamma_{{\mathrm{3}}}  \vdash  \ottnt{A}  \ottsym{:}   \mathbf{Typ} $ are given.
            By case~\ref{lem:weakening:kinding}, we have $\Gamma_{{\mathrm{1}}}  \ottsym{,}  \Gamma_{{\mathrm{2}}}  \ottsym{,}  \Gamma_{{\mathrm{3}}}  \vdash  \ottnt{A}  \ottsym{:}   \mathbf{Typ} $. Thus, \rname{ST}{Refl} derives $\Gamma_{{\mathrm{1}}}  \ottsym{,}  \Gamma_{{\mathrm{2}}}  \ottsym{,}  \Gamma_{{\mathrm{3}}}  \vdash  \ottnt{A}  <:  \ottnt{A}$.

            \item[\rname{ST}{Fun}]
            For some $\ottnt{A_{{\mathrm{1}}}}$, $\varepsilon_{{\mathrm{1}}}$, $\ottnt{B_{{\mathrm{1}}}}$, $\ottnt{A_{{\mathrm{2}}}}$, $\varepsilon_{{\mathrm{2}}}$, and $\ottnt{B_{{\mathrm{2}}}}$,
            the following are given:
            \begin{itemize}
              \item $\ottnt{A} =  \ottnt{A_{{\mathrm{1}}}}    \rightarrow_{ \varepsilon_{{\mathrm{1}}} }    \ottnt{B_{{\mathrm{1}}}} $,
              \item $\ottnt{B} =  \ottnt{A_{{\mathrm{2}}}}    \rightarrow_{ \varepsilon_{{\mathrm{2}}} }    \ottnt{B_{{\mathrm{2}}}} $,
              \item $\Gamma_{{\mathrm{1}}}  \ottsym{,}  \Gamma_{{\mathrm{3}}}  \vdash  \ottnt{A_{{\mathrm{2}}}}  <:  \ottnt{A_{{\mathrm{1}}}}$, and
              \item $\Gamma_{{\mathrm{1}}}  \ottsym{,}  \Gamma_{{\mathrm{3}}}  \vdash  \ottnt{B_{{\mathrm{1}}}}  \mid  \varepsilon_{{\mathrm{1}}}  <:  \ottnt{B_{{\mathrm{2}}}}  \mid  \varepsilon_{{\mathrm{2}}}$.
            \end{itemize}
            By the induction hypothesis, we have $\Gamma_{{\mathrm{1}}}  \ottsym{,}  \Gamma_{{\mathrm{2}}}  \ottsym{,}  \Gamma_{{\mathrm{3}}}  \vdash  \ottnt{B_{{\mathrm{1}}}}  \mid  \varepsilon_{{\mathrm{1}}}  <:  \ottnt{B_{{\mathrm{2}}}}  \mid  \varepsilon_{{\mathrm{2}}}$.
            Thus, \rname{ST}{Fun} derives
            \begin{align*}
              \Gamma_{{\mathrm{1}}}  \ottsym{,}  \Gamma_{{\mathrm{2}}}  \ottsym{,}  \Gamma_{{\mathrm{3}}}  \vdash   \ottnt{A_{{\mathrm{1}}}}    \rightarrow_{ \varepsilon_{{\mathrm{1}}} }    \ottnt{B_{{\mathrm{1}}}}   <:   \ottnt{A_{{\mathrm{2}}}}    \rightarrow_{ \varepsilon_{{\mathrm{2}}} }    \ottnt{B_{{\mathrm{2}}}} 
            \end{align*}
            as required.

            \item[\rname{ST}{Poly}]
            Without loss of generality, we can choose $\alpha$ such that $ \alpha   \notin    \mathrm{dom}   \ottsym{(}   \Gamma_{{\mathrm{2}}}   \ottsym{)}  $.
            For some $\ottnt{K}$, $\ottnt{A_{{\mathrm{1}}}}$, $\varepsilon_{{\mathrm{1}}}$, $\ottnt{A_{{\mathrm{2}}}}$, and $\varepsilon_{{\mathrm{2}}}$, the following are given:
            \begin{itemize}
              \item $\ottnt{A} =   \forall   \alpha  \ottsym{:}  \ottnt{K}   \ottsym{.}    \ottnt{A_{{\mathrm{1}}}}    ^{ \varepsilon_{{\mathrm{1}}} }  $,
              \item $\ottnt{B} =   \forall   \alpha  \ottsym{:}  \ottnt{K}   \ottsym{.}    \ottnt{A_{{\mathrm{2}}}}    ^{ \varepsilon_{{\mathrm{2}}} }  $, and
              \item $\Gamma_{{\mathrm{1}}}  \ottsym{,}  \Gamma_{{\mathrm{3}}}  \ottsym{,}  \alpha  \ottsym{:}  \ottnt{K}  \vdash  \ottnt{A_{{\mathrm{1}}}}  \mid  \varepsilon_{{\mathrm{1}}}  <:  \ottnt{A_{{\mathrm{2}}}}  \mid  \varepsilon_{{\mathrm{2}}}$.
            \end{itemize}
            By the induction hypothesis, we have
            $\Gamma_{{\mathrm{1}}}  \ottsym{,}  \Gamma_{{\mathrm{2}}}  \ottsym{,}  \Gamma_{{\mathrm{3}}}  \ottsym{,}  \alpha  \ottsym{:}  \ottnt{K}  \vdash  \ottnt{A_{{\mathrm{1}}}}  \mid  \varepsilon_{{\mathrm{1}}}  <:  \ottnt{A_{{\mathrm{2}}}}  \mid  \varepsilon_{{\mathrm{2}}}$.
            Thus, \rname{ST}{Poly} derives
            \begin{align*}
              \Gamma_{{\mathrm{1}}}  \ottsym{,}  \Gamma_{{\mathrm{2}}}  \ottsym{,}  \Gamma_{{\mathrm{3}}}  \vdash    \forall   \alpha  \ottsym{:}  \ottnt{K}   \ottsym{.}    \ottnt{A_{{\mathrm{1}}}}    ^{ \varepsilon_{{\mathrm{1}}} }    <:    \forall   \alpha  \ottsym{:}  \ottnt{K}   \ottsym{.}    \ottnt{A_{{\mathrm{2}}}}    ^{ \varepsilon_{{\mathrm{2}}} }  
            \end{align*}
            as required.

            \item[\rname{ST}{Comp}]
            We have $\Gamma_{{\mathrm{1}}}  \ottsym{,}  \Gamma_{{\mathrm{3}}}  \vdash  \ottnt{A_{{\mathrm{1}}}}  <:  \ottnt{A_{{\mathrm{2}}}}$ and $\Gamma_{{\mathrm{1}}}  \ottsym{,}  \Gamma_{{\mathrm{3}}}  \vdash   \varepsilon_{{\mathrm{1}}}  \olessthan  \varepsilon_{{\mathrm{2}}} $.
            By the induction hypothesis, we have $\Gamma_{{\mathrm{1}}}  \ottsym{,}  \Gamma_{{\mathrm{2}}}  \ottsym{,}  \Gamma_{{\mathrm{3}}}  \vdash  \ottnt{A_{{\mathrm{1}}}}  <:  \ottnt{A_{{\mathrm{2}}}}$.
            By case~\ref{lem:weakening:kinding},
            we have $\Gamma_{{\mathrm{1}}}  \ottsym{,}  \Gamma_{{\mathrm{2}}}  \ottsym{,}  \Gamma_{{\mathrm{3}}}  \vdash   \varepsilon_{{\mathrm{1}}}  \olessthan  \varepsilon_{{\mathrm{2}}} $.
            Thus, \rname{ST}{Comp} derives $\Gamma_{{\mathrm{1}}}  \ottsym{,}  \Gamma_{{\mathrm{2}}}  \ottsym{,}  \Gamma_{{\mathrm{3}}}  \vdash  \ottnt{A_{{\mathrm{1}}}}  \mid  \varepsilon_{{\mathrm{1}}}  <:  \ottnt{A_{{\mathrm{2}}}}  \mid  \varepsilon_{{\mathrm{2}}}$
            as required.
          \end{divcases}

    \item[(5)(6)]
          By mutual induction on derivations of the judgments.
          We proceed by case analysis on the rule applied lastly to the derivation.
          \begin{divcases}
            \item[\rname{T}{Var}]
            For some $\mathit{x}$, the following are given:
            \begin{itemize}
              \item $\ottnt{e} = \mathit{x}$,
              \item $\varepsilon =  \bbZero $,
              \item $\vdash  \Gamma_{{\mathrm{1}}}  \ottsym{,}  \Gamma_{{\mathrm{3}}}$, and
              \item $ \mathit{x}   \ottsym{:}   \ottnt{A}   \in   \Gamma_{{\mathrm{1}}}  \ottsym{,}  \Gamma_{{\mathrm{3}}} $.
            \end{itemize}
            By case \ref{lem:weakening:wf}, we have $\vdash  \Gamma_{{\mathrm{1}}}  \ottsym{,}  \Gamma_{{\mathrm{2}}}  \ottsym{,}  \Gamma_{{\mathrm{3}}}$.
            Thus, \rname{T}{Var} derives $\Gamma_{{\mathrm{1}}}  \ottsym{,}  \Gamma_{{\mathrm{2}}}  \ottsym{,}  \Gamma_{{\mathrm{3}}}  \vdash  \mathit{x}  \ottsym{:}  \ottnt{A}  \mid   \bbZero $
            because of $ \mathit{x}   \ottsym{:}   \ottnt{A}   \in   \Gamma_{{\mathrm{1}}}  \ottsym{,}  \Gamma_{{\mathrm{2}}}  \ottsym{,}  \Gamma_{{\mathrm{3}}} $.

            \item[\rname{T}{Abs}]
            Without loss of generality, we can choose $\mathit{f}$ and $\mathit{x}$
            such that $ \mathit{f}   \notin    \mathrm{dom}   \ottsym{(}   \Gamma_{{\mathrm{2}}}   \ottsym{)}  $ and $ \mathit{x}   \notin    \mathrm{dom}   \ottsym{(}   \Gamma_{{\mathrm{2}}}   \ottsym{)}  $.
            For some $\ottnt{e'}$, $\ottnt{A'}$, $\ottnt{B'}$, and $\varepsilon'$, the following are given:
            \begin{itemize}
              \item $\ottnt{e} = \ottkw{fun} \, \ottsym{(}  \mathit{f}  \ottsym{,}  \mathit{x}  \ottsym{,}  \ottnt{e'}  \ottsym{)}$,
              \item $\ottnt{A} =  \ottnt{A'}    \rightarrow_{ \varepsilon' }    \ottnt{B'} $,
              \item $\varepsilon =  \bbZero $, and
              \item $\Gamma_{{\mathrm{1}}}  \ottsym{,}  \Gamma_{{\mathrm{3}}}  \ottsym{,}  \mathit{f}  \ottsym{:}   \ottnt{A'}    \rightarrow_{ \varepsilon' }    \ottnt{B'}   \ottsym{,}  \mathit{x}  \ottsym{:}  \ottnt{A'}  \vdash  \ottnt{e'}  \ottsym{:}  \ottnt{B'}  \mid  \varepsilon'$.
            \end{itemize}
            By the induction hypothesis,
            we have $\Gamma_{{\mathrm{1}}}  \ottsym{,}  \Gamma_{{\mathrm{2}}}  \ottsym{,}  \Gamma_{{\mathrm{3}}}  \ottsym{,}  \mathit{f}  \ottsym{:}   \ottnt{A'}    \rightarrow_{ \varepsilon' }    \ottnt{B'}   \ottsym{,}  \mathit{x}  \ottsym{:}  \ottnt{A'}  \vdash  \ottnt{e'}  \ottsym{:}  \ottnt{B'}  \mid  \varepsilon'$
            because of $  \mathrm{dom}   \ottsym{(}   \Gamma_{{\mathrm{2}}}   \ottsym{)}    \cap    \mathrm{dom}   \ottsym{(}   \Gamma_{{\mathrm{3}}}  \ottsym{,}  \mathit{f}  \ottsym{:}   \ottnt{A'}    \rightarrow_{ \varepsilon' }    \ottnt{B'}   \ottsym{,}  \mathit{x}  \ottsym{:}  \ottnt{A'}   \ottsym{)}    \ottsym{=}  \emptyset$.
            Thus, \rname{T}{Abs} derives
            $\Gamma_{{\mathrm{1}}}  \ottsym{,}  \Gamma_{{\mathrm{2}}}  \ottsym{,}  \Gamma_{{\mathrm{3}}}  \vdash  \ottkw{fun} \, \ottsym{(}  \mathit{f}  \ottsym{,}  \mathit{x}  \ottsym{,}  \ottnt{e'}  \ottsym{)}  \ottsym{:}   \ottnt{A'}    \rightarrow_{ \varepsilon' }    \ottnt{B'}   \mid   \bbZero $.

            \item[\rname{T}{App}]
            For some $\ottnt{v_{{\mathrm{1}}}}$, $\ottnt{v_{{\mathrm{2}}}}$, and $\ottnt{C}$, the following are given:
            \begin{itemize}
              \item $\ottnt{e} = \ottnt{v_{{\mathrm{1}}}} \, \ottnt{v_{{\mathrm{2}}}}$,
              \item $\Gamma_{{\mathrm{1}}}  \ottsym{,}  \Gamma_{{\mathrm{3}}}  \vdash  \ottnt{v_{{\mathrm{1}}}}  \ottsym{:}   \ottnt{B}    \rightarrow_{ \varepsilon }    \ottnt{A}   \mid   \bbZero $, and
              \item $\Gamma_{{\mathrm{1}}}  \ottsym{,}  \Gamma_{{\mathrm{3}}}  \vdash  \ottnt{v_{{\mathrm{2}}}}  \ottsym{:}  \ottnt{B}  \mid   \bbZero $.
            \end{itemize}
            By the induction hypothesis, we have
            \begin{itemize}
              \item $\Gamma_{{\mathrm{1}}}  \ottsym{,}  \Gamma_{{\mathrm{2}}}  \ottsym{,}  \Gamma_{{\mathrm{3}}}  \vdash  \ottnt{v_{{\mathrm{1}}}}  \ottsym{:}   \ottnt{B}    \rightarrow_{ \varepsilon }    \ottnt{A}   \mid   \bbZero $ and
              \item $\Gamma_{{\mathrm{1}}}  \ottsym{,}  \Gamma_{{\mathrm{2}}}  \ottsym{,}  \Gamma_{{\mathrm{3}}}  \vdash  \ottnt{v_{{\mathrm{2}}}}  \ottsym{:}  \ottnt{B}  \mid   \bbZero $.
            \end{itemize}
            Thus, \rname{T}{App} derives $\Gamma_{{\mathrm{1}}}  \ottsym{,}  \Gamma_{{\mathrm{2}}}  \ottsym{,}  \Gamma_{{\mathrm{3}}}  \vdash  \ottnt{v_{{\mathrm{1}}}} \, \ottnt{v_{{\mathrm{2}}}}  \ottsym{:}  \ottnt{A}  \mid  \varepsilon$.

            \item[\rname{T}{TAbs}]
            Without loss of generality, we can choose $\alpha$ such that $ \alpha   \notin    \mathrm{dom}   \ottsym{(}   \Gamma_{{\mathrm{2}}}   \ottsym{)}  $.
            For some $\ottnt{K}$, $\ottnt{e'}$, $\ottnt{B'}$, and $\varepsilon'$, the following are given:
            \begin{itemize}
              \item $\ottnt{e} = \Lambda  \alpha  \ottsym{:}  \ottnt{K}  \ottsym{.}  \ottnt{e'}$,
              \item $\ottnt{A} =   \forall   \alpha  \ottsym{:}  \ottnt{K}   \ottsym{.}    \ottnt{A'}    ^{ \varepsilon' }  $,
              \item $\varepsilon =  \bbZero $, and
              \item $\Gamma_{{\mathrm{1}}}  \ottsym{,}  \Gamma_{{\mathrm{3}}}  \ottsym{,}  \alpha  \ottsym{:}  \ottnt{K}  \vdash  \ottnt{e'}  \ottsym{:}  \ottnt{A'}  \mid  \varepsilon'$.
            \end{itemize}
            By the induction hypothesis,
            we have $\Gamma_{{\mathrm{1}}}  \ottsym{,}  \Gamma_{{\mathrm{2}}}  \ottsym{,}  \Gamma_{{\mathrm{3}}}  \ottsym{,}  \alpha  \ottsym{:}  \ottnt{K}  \vdash  \ottnt{e'}  \ottsym{:}  \ottnt{A'}  \mid  \varepsilon'$
            because of $  \mathrm{dom}   \ottsym{(}   \Gamma_{{\mathrm{2}}}   \ottsym{)}    \cap    \mathrm{dom}   \ottsym{(}   \Gamma_{{\mathrm{3}}}  \ottsym{,}  \alpha  \ottsym{:}  \ottnt{K}   \ottsym{)}    \ottsym{=}  \emptyset$.
            Thus, \rname{T}{TAbs} derives
            $\Gamma_{{\mathrm{1}}}  \ottsym{,}  \Gamma_{{\mathrm{2}}}  \ottsym{,}  \Gamma_{{\mathrm{3}}}  \vdash  \Lambda  \alpha  \ottsym{:}  \ottnt{K}  \ottsym{.}  \ottnt{e'}  \ottsym{:}    \forall   \alpha  \ottsym{:}  \ottnt{K}   \ottsym{.}    \ottnt{A'}    ^{ \varepsilon' }    \mid   \bbZero $.

            \item[\rname{T}{TApp}]
            For some $\ottnt{v}$, $S$, $\alpha$, $\ottnt{A'}$, $\varepsilon'$, and $\ottnt{K}$,
            the following are given:
            \begin{itemize}
              \item $\ottnt{e} = \ottnt{v} \, S$,
              \item $\ottnt{A} = \ottnt{A'} \,  \! [  S  /  \alpha   ] $,
              \item $\varepsilon = \varepsilon' \,  \! [  S  /  \alpha   ] $,
              \item $\Gamma_{{\mathrm{1}}}  \ottsym{,}  \Gamma_{{\mathrm{3}}}  \vdash  \ottnt{v}  \ottsym{:}    \forall   \alpha  \ottsym{:}  \ottnt{K}   \ottsym{.}    \ottnt{A'}    ^{ \varepsilon' }    \mid   \bbZero $, and
              \item $\Gamma_{{\mathrm{1}}}  \ottsym{,}  \Gamma_{{\mathrm{3}}}  \vdash  S  \ottsym{:}  \ottnt{K}$.
            \end{itemize}
            By the induction hypothesis and case \ref{lem:weakening:kinding},
            we have
            \begin{itemize}
              \item $\Gamma_{{\mathrm{1}}}  \ottsym{,}  \Gamma_{{\mathrm{2}}}  \ottsym{,}  \Gamma_{{\mathrm{3}}}  \vdash  \ottnt{v}  \ottsym{:}    \forall   \alpha  \ottsym{:}  \ottnt{K}   \ottsym{.}    \ottnt{A'}    ^{ \varepsilon' }    \mid   \bbZero $ and
              \item $\Gamma_{{\mathrm{1}}}  \ottsym{,}  \Gamma_{{\mathrm{2}}}  \ottsym{,}  \Gamma_{{\mathrm{3}}}  \vdash  S  \ottsym{:}  \ottnt{K}$.
            \end{itemize}
            Thus, \rname{T}{TApp} derives $\Gamma_{{\mathrm{1}}}  \ottsym{,}  \Gamma_{{\mathrm{2}}}  \ottsym{,}  \Gamma_{{\mathrm{3}}}  \vdash  \ottnt{v} \, S  \ottsym{:}  \ottnt{A'} \,  \! [  S  /  \alpha   ]   \mid  \varepsilon' \,  \! [  S  /  \alpha   ] $.

            \item[\rname{T}{Let}]
            Without loss of generality, we can choose $\mathit{x}$ such that $ \mathit{x}   \notin    \mathrm{dom}   \ottsym{(}   \Gamma_{{\mathrm{2}}}   \ottsym{)}  $.
            For some $\ottnt{e_{{\mathrm{1}}}}$, $\ottnt{e_{{\mathrm{2}}}}$, and $\ottnt{B}$, the following are given:
            \begin{itemize}
              \item $\ottnt{e} = (\mathbf{let} \, \mathit{x}  \ottsym{=}  \ottnt{e_{{\mathrm{1}}}} \, \mathbf{in} \, \ottnt{e_{{\mathrm{2}}}})$,
              \item $\Gamma_{{\mathrm{1}}}  \ottsym{,}  \Gamma_{{\mathrm{3}}}  \vdash  \ottnt{e_{{\mathrm{1}}}}  \ottsym{:}  \ottnt{B}  \mid  \varepsilon$, and
              \item $\Gamma_{{\mathrm{1}}}  \ottsym{,}  \Gamma_{{\mathrm{3}}}  \ottsym{,}  \mathit{x}  \ottsym{:}  \ottnt{B}  \vdash  \ottnt{e_{{\mathrm{2}}}}  \ottsym{:}  \ottnt{A}  \mid  \varepsilon$.
            \end{itemize}
            By the induction hypothesis, we have
            \begin{itemize}
              \item $\Gamma_{{\mathrm{1}}}  \ottsym{,}  \Gamma_{{\mathrm{2}}}  \ottsym{,}  \Gamma_{{\mathrm{3}}}  \vdash  \ottnt{e_{{\mathrm{1}}}}  \ottsym{:}  \ottnt{B}  \mid  \varepsilon$ and
              \item $\Gamma_{{\mathrm{1}}}  \ottsym{,}  \Gamma_{{\mathrm{2}}}  \ottsym{,}  \Gamma_{{\mathrm{3}}}  \ottsym{,}  \mathit{x}  \ottsym{:}  \ottnt{B}  \vdash  \ottnt{e_{{\mathrm{2}}}}  \ottsym{:}  \ottnt{A}  \mid  \varepsilon$
            \end{itemize}
            because of $  \mathrm{dom}   \ottsym{(}   \Gamma_{{\mathrm{2}}}   \ottsym{)}    \cap    \mathrm{dom}   \ottsym{(}   \Gamma_{{\mathrm{3}}}  \ottsym{,}  \mathit{x}  \ottsym{:}  \ottnt{B}   \ottsym{)}    \ottsym{=}  \emptyset$.
            Thus, \rname{T}{Let} derives $\Gamma_{{\mathrm{1}}}  \ottsym{,}  \Gamma_{{\mathrm{2}}}  \ottsym{,}  \Gamma_{{\mathrm{3}}}  \vdash  \mathbf{let} \, \mathit{x}  \ottsym{=}  \ottnt{e_{{\mathrm{1}}}} \, \mathbf{in} \, \ottnt{e_{{\mathrm{2}}}}  \ottsym{:}  \ottnt{A}  \mid  \varepsilon$.

            \item[\rname{T}{Sub}]
            For some $\ottnt{A'}$ and $\varepsilon'$, the following are given:
            \begin{itemize}
              \item $\Gamma_{{\mathrm{1}}}  \ottsym{,}  \Gamma_{{\mathrm{3}}}  \vdash  \ottnt{e}  \ottsym{:}  \ottnt{A'}  \mid  \varepsilon'$ and
              \item $\Gamma_{{\mathrm{1}}}  \ottsym{,}  \Gamma_{{\mathrm{3}}}  \vdash  \ottnt{A'}  \mid  \varepsilon'  <:  \ottnt{A}  \mid  \varepsilon$.
            \end{itemize}
            By the induction hypothesis and case \ref{lem:weakening:subtyping_comp}, we have
            \begin{itemize}
              \item $\Gamma_{{\mathrm{1}}}  \ottsym{,}  \Gamma_{{\mathrm{2}}}  \ottsym{,}  \Gamma_{{\mathrm{3}}}  \vdash  \ottnt{e}  \ottsym{:}  \ottnt{A'}  \mid  \varepsilon'$ and
              \item $\Gamma_{{\mathrm{1}}}  \ottsym{,}  \Gamma_{{\mathrm{2}}}  \ottsym{,}  \Gamma_{{\mathrm{3}}}  \vdash  \ottnt{A'}  \mid  \varepsilon'  <:  \ottnt{A}  \mid  \varepsilon$.
            \end{itemize}
            Thus, \rname{T}{Sub} derives $\Gamma_{{\mathrm{1}}}  \ottsym{,}  \Gamma_{{\mathrm{2}}}  \ottsym{,}  \Gamma_{{\mathrm{3}}}  \vdash  \ottnt{e}  \ottsym{:}  \ottnt{A}  \mid  \varepsilon$.

            \item[\rname{T}{Op}]
            For some $\mathsf{op}$, $\mathit{l}$, $\ottnt{A'}$, $\ottnt{B'}$, $\mathit{I}$, and $\mathit{J}$,
            the following are given:
            \begin{itemize}
              \item $\ottnt{e} =  \mathsf{op} _{ \mathit{l} \,  \bm{ { S } } ^ {  \mathit{I}  }  }  \,  \bm{ { T } } ^ {  \mathit{J}  } $,
              \item $\ottnt{A} =  \ottsym{(}  \ottnt{A'} \,  \! [ {\bm{ { T } } }^{ \mathit{J} } / {\bm{ \beta } }^{ \mathit{J} } ]   \ottsym{)}    \rightarrow_{  \lift{ \mathit{l} \,  \bm{ { S } } ^ {  \mathit{I}  }  }  }    \ottsym{(}  \ottnt{B'} \,  \! [ {\bm{ { T } } }^{ \mathit{J} } / {\bm{ \beta } }^{ \mathit{J} } ]   \ottsym{)} $,
              \item $\varepsilon =  \bbZero $,
              \item $ \mathit{l}  ::    \forall    {\bm{ \alpha } }^{ \mathit{I} } : {\bm{ \ottnt{K} } }^{ \mathit{I} }    \ottsym{.}    \sigma    \in   \Xi $,
              \item $ \mathsf{op}  \ottsym{:}    \forall    {\bm{ \beta } }^{ \mathit{J} } : {\bm{ \ottnt{K'} } }^{ \mathit{J} }    \ottsym{.}    \ottnt{A'}   \Rightarrow   \ottnt{B'}    \in   \sigma \,  \! [ {\bm{ { S } } }^{ \mathit{I} } / {\bm{ \alpha } }^{ \mathit{I} } ]  $,
              \item $\vdash  \Gamma_{{\mathrm{1}}}  \ottsym{,}  \Gamma_{{\mathrm{3}}}$,
              \item $\Gamma_{{\mathrm{1}}}  \ottsym{,}  \Gamma_{{\mathrm{3}}}  \vdash   \bm{ { S } }^{ \mathit{I} } : \bm{ \ottnt{K} }^{ \mathit{I} } $, and
              \item $\Gamma_{{\mathrm{1}}}  \ottsym{,}  \Gamma_{{\mathrm{3}}}  \vdash   \bm{ { T } }^{ \mathit{J} } : \bm{ \ottnt{K'} }^{ \mathit{J} } $.
            \end{itemize}
            By cases \ref{lem:weakening:wf} and \ref{lem:weakening:kinding}, we have
            \begin{itemize}
              \item $\vdash  \Gamma_{{\mathrm{1}}}  \ottsym{,}  \Gamma_{{\mathrm{2}}}  \ottsym{,}  \Gamma_{{\mathrm{3}}}$,
              \item $\Gamma_{{\mathrm{1}}}  \ottsym{,}  \Gamma_{{\mathrm{2}}}  \ottsym{,}  \Gamma_{{\mathrm{3}}}  \vdash   \bm{ { S } }^{ \mathit{I} } : \bm{ \ottnt{K} }^{ \mathit{I} } $, and
              \item $\Gamma_{{\mathrm{1}}}  \ottsym{,}  \Gamma_{{\mathrm{2}}}  \ottsym{,}  \Gamma_{{\mathrm{3}}}  \vdash   \bm{ { T } }^{ \mathit{J} } : \bm{ \ottnt{K'} }^{ \mathit{J} } $.
            \end{itemize}
            Thus, \rname{T}{Op} derives
            \begin{align*}
              \Gamma_{{\mathrm{1}}}  \ottsym{,}  \Gamma_{{\mathrm{2}}}  \ottsym{,}  \Gamma_{{\mathrm{3}}}  \vdash   \mathsf{op} _{ \mathit{l} \,  \bm{ { S } } ^ {  \mathit{I}  }  }  \,  \bm{ { T } } ^ {  \mathit{I}  }   \ottsym{:}   \ottsym{(}  \ottnt{A'} \,  \! [ {\bm{ { T } } }^{ \mathit{J} } / {\bm{ \beta } }^{ \mathit{J} } ]   \ottsym{)}    \rightarrow_{  \lift{ \mathit{l} \,  \bm{ { S } } ^ {  \mathit{I}  }  }  }    \ottsym{(}  \ottnt{B'} \,  \! [ {\bm{ { T } } }^{ \mathit{J} } / {\bm{ \beta } }^{ \mathit{J} } ]   \ottsym{)}   \mid   \bbZero .
            \end{align*}

            \item[\rname{T}{Handling}]
            For some $\mathit{N}$, $\ottnt{e'}$, $\ottnt{A'}$, $\varepsilon'$, $\mathit{l}$, $ \bm{ { S } } ^ {  \mathit{N}  } $, $ {\bm{ { \ottnt{K} } } }^{ \mathit{N} } $, $\ottnt{h}$, and $\sigma$, the following are given:
            \begin{itemize}
              \item $\ottnt{e} =  \mathbf{handle}_{ \mathit{l} \,  \bm{ { S } } ^ {  \mathit{N}  }  }  \, \ottnt{e'} \, \mathbf{with} \, \ottnt{h}$,
              \item $\Gamma_{{\mathrm{1}}}  \ottsym{,}  \Gamma_{{\mathrm{3}}}  \vdash  \ottnt{e'}  \ottsym{:}  \ottnt{A'}  \mid  \varepsilon'$,
              \item $ \mathit{l}  ::    \forall    {\bm{ \alpha } }^{ \mathit{N} } : {\bm{ \ottnt{K} } }^{ \mathit{N} }    \ottsym{.}    \sigma    \in   \Xi $,
              \item $ \Gamma_{{\mathrm{1}}}  \ottsym{,}  \Gamma_{{\mathrm{3}}}  \vdash _{ \sigma \,  \! [ {\bm{ { S } } }^{ \mathit{N} } / {\bm{ \alpha } }^{ \mathit{N} } ]  }  \ottnt{h}  :  \ottnt{A'}   \Rightarrow  ^ { \varepsilon }  \ottnt{A} $,
              \item $\Gamma_{{\mathrm{1}}}  \ottsym{,}  \Gamma_{{\mathrm{3}}}  \vdash   \bm{ { S } }^{ \mathit{N} } : \bm{ \ottnt{K} }^{ \mathit{N} } $, and
              \item $   \lift{ \mathit{l} \,  \bm{ { S } } ^ {  \mathit{N}  }  }   \mathop{ \odot }  \varepsilon    \sim   \varepsilon' $.
            \end{itemize}
            By the induction hypothesis and case~\ref{lem:weakening:kinding},
            we have
            \begin{itemize}
              \item $\Gamma_{{\mathrm{1}}}  \ottsym{,}  \Gamma_{{\mathrm{2}}}  \ottsym{,}  \Gamma_{{\mathrm{3}}}  \vdash  \ottnt{e'}  \ottsym{:}  \ottnt{A'}  \mid  \varepsilon'$,
              \item $ \Gamma_{{\mathrm{1}}}  \ottsym{,}  \Gamma_{{\mathrm{2}}}  \ottsym{,}  \Gamma_{{\mathrm{3}}}  \vdash _{ \sigma \,  \! [ {\bm{ { S } } }^{ \mathit{N} } / {\bm{ \alpha } }^{ \mathit{N} } ]  }  \ottnt{h}  :  \ottnt{A'}   \Rightarrow  ^ { \varepsilon }  \ottnt{A} $, and
              \item $\Gamma_{{\mathrm{1}}}  \ottsym{,}  \Gamma_{{\mathrm{2}}}  \ottsym{,}  \Gamma_{{\mathrm{3}}}  \vdash   \bm{ { S } }^{ \mathit{N} } : \bm{ \ottnt{K} }^{ \mathit{N} } $.
            \end{itemize}
            Thus, \rname{T}{Handling} derives
            \begin{align*}
              \Gamma_{{\mathrm{1}}}  \ottsym{,}  \Gamma_{{\mathrm{2}}}  \ottsym{,}  \Gamma_{{\mathrm{3}}}  \vdash   \mathbf{handle}_{ \mathit{l} \,  \bm{ { S } } ^ {  \mathit{N}  }  }  \, \ottnt{e} \, \mathbf{with} \, \ottnt{h}  \ottsym{:}  \ottnt{A}  \mid  \varepsilon.
            \end{align*}

            \item[\rname{H}{Return}]
            Without loss of generality, we can choose $\mathit{x}$ such that $ \mathit{x}   \notin    \mathrm{dom}   \ottsym{(}   \Gamma_{{\mathrm{2}}}   \ottsym{)}  $.
            For some $\ottnt{e_{\ottmv{r}}}$, the following are given:
            \begin{itemize}
              \item $\ottnt{h} = \ottsym{\{} \, \mathbf{return} \, \mathit{x}  \mapsto  \ottnt{e_{\ottmv{r}}}  \ottsym{\}}$,
              \item $\sigma =  \{\} $, and
              \item $\Gamma_{{\mathrm{1}}}  \ottsym{,}  \Gamma_{{\mathrm{3}}}  \ottsym{,}  \mathit{x}  \ottsym{:}  \ottnt{A}  \vdash  \ottnt{e_{\ottmv{r}}}  \ottsym{:}  \ottnt{B}  \mid  \varepsilon$.
            \end{itemize}
            By the induction hypothesis, we have $\Gamma_{{\mathrm{1}}}  \ottsym{,}  \Gamma_{{\mathrm{2}}}  \ottsym{,}  \Gamma_{{\mathrm{3}}}  \ottsym{,}  \mathit{x}  \ottsym{:}  \ottnt{A}  \vdash  \ottnt{e_{\ottmv{r}}}  \ottsym{:}  \ottnt{B}  \mid  \varepsilon$.
            Thus, \rname{H}{Return} derives $ \Gamma_{{\mathrm{1}}}  \ottsym{,}  \Gamma_{{\mathrm{2}}}  \ottsym{,}  \Gamma_{{\mathrm{3}}}  \vdash _{  \{\}  }  \ottsym{\{} \, \mathbf{return} \, \mathit{x}  \mapsto  \ottnt{e_{\ottmv{r}}}  \ottsym{\}}  :  \ottnt{A}   \Rightarrow  ^ { \varepsilon }  \ottnt{B} $.

            \item[\rname{H}{Op}]
            Without loss of generality, we can choose $ \bm{ { \beta } } ^ {  \mathit{J}  } $ and $\mathit{p}$ and $\mathit{k}$ such that:
            \begin{itemize}
              \item $ \{   \bm{ { \beta } } ^ {  \mathit{J}  }   \}   \cap    \mathrm{dom}   \ottsym{(}   \Gamma_{{\mathrm{2}}}   \ottsym{)}    \ottsym{=}  \emptyset$,
              \item $ \mathit{p}   \notin    \mathrm{dom}   \ottsym{(}   \Gamma_{{\mathrm{2}}}   \ottsym{)}  $, and
              \item $ \mathit{k}   \notin    \mathrm{dom}   \ottsym{(}   \Gamma_{{\mathrm{2}}}   \ottsym{)}  $.
            \end{itemize}
            For some $\ottnt{h'}$, $\sigma'$, $\mathsf{op}$, $\ottnt{A'}$, $\ottnt{B'}$, and $\ottnt{e}$, the following are given:
            \begin{itemize}
              \item $\ottnt{h} =  \ottnt{h'}   \uplus   \ottsym{\{}  \mathsf{op} \,  {\bm{ \beta } }^{ \mathit{J} } : {\bm{ \ottnt{K} } }^{ \mathit{J} }  \, \mathit{p} \, \mathit{k}  \mapsto  \ottnt{e}  \ottsym{\}} $,
              \item $\sigma =  \sigma'   \uplus   \ottsym{\{}  \mathsf{op}  \ottsym{:}    \forall    {\bm{ \beta } }^{ \mathit{J} } : {\bm{ \ottnt{K} } }^{ \mathit{J} }    \ottsym{.}    \ottnt{A'}   \Rightarrow   \ottnt{B'}   \ottsym{\}} $,
              \item $ \Gamma_{{\mathrm{1}}}  \ottsym{,}  \Gamma_{{\mathrm{3}}}  \vdash _{ \sigma' }  \ottnt{h'}  :  \ottnt{A}   \Rightarrow  ^ { \varepsilon }  \ottnt{B} $, and
              \item $\Gamma_{{\mathrm{1}}}  \ottsym{,}  \Gamma_{{\mathrm{3}}}  \ottsym{,}   {\bm{ \beta } }^{ \mathit{J} } : {\bm{ \ottnt{K} } }^{ \mathit{J} }   \ottsym{,}  \mathit{p}  \ottsym{:}  \ottnt{A'}  \ottsym{,}  \mathit{k}  \ottsym{:}   \ottnt{B'}    \rightarrow_{ \varepsilon }    \ottnt{B}   \vdash  \ottnt{e}  \ottsym{:}  \ottnt{B}  \mid  \varepsilon$.
            \end{itemize}
            By the induction hypothesis, we have
            \begin{itemize}
              \item $ \Gamma_{{\mathrm{1}}}  \ottsym{,}  \Gamma_{{\mathrm{2}}}  \ottsym{,}  \Gamma_{{\mathrm{3}}}  \vdash _{ \sigma' }  \ottnt{h'}  :  \ottnt{A}   \Rightarrow  ^ { \varepsilon }  \ottnt{B} $ and
              \item $\Gamma_{{\mathrm{1}}}  \ottsym{,}  \Gamma_{{\mathrm{2}}}  \ottsym{,}  \Gamma_{{\mathrm{3}}}  \ottsym{,}   {\bm{ \beta } }^{ \mathit{J} } : {\bm{ \ottnt{K} } }^{ \mathit{J} }   \ottsym{,}  \mathit{p}  \ottsym{:}  \ottnt{A'}  \ottsym{,}  \mathit{k}  \ottsym{:}   \ottnt{B'}    \rightarrow_{ \varepsilon }    \ottnt{B}   \vdash  \ottnt{e}  \ottsym{:}  \ottnt{B}  \mid  \varepsilon$.
            \end{itemize}
            Thus, \rname{H}{Op} derives $ \Gamma_{{\mathrm{1}}}  \ottsym{,}  \Gamma_{{\mathrm{2}}}  \ottsym{,}  \Gamma_{{\mathrm{3}}}  \vdash _{ \sigma }   \ottnt{h'}   \uplus   \ottsym{\{}  \mathsf{op} \,  {\bm{ \beta } }^{ \mathit{J} } : {\bm{ \ottnt{K} } }^{ \mathit{J} }  \, \mathit{p} \, \mathit{k}  \mapsto  \ottnt{e}  \ottsym{\}}   :  \ottnt{A}   \Rightarrow  ^ { \varepsilon }  \ottnt{B} $.
          \end{divcases}
  \end{itemize}
\end{proof}

\begin{lemma}\label{lem:delta_weakening}
  For any $\Gamma_{{\mathrm{1}}}$, $\Gamma_{{\mathrm{2}}}$, $S$, and $\ottnt{K}$, if $ \Delta   \ottsym{(}   \Gamma_{{\mathrm{1}}}   \ottsym{)}   \ottsym{,}  \Gamma_{{\mathrm{2}}}  \vdash  S  \ottsym{:}  \ottnt{K}$ and $\vdash  \Gamma_{{\mathrm{1}}}$ and $  \mathrm{dom}   \ottsym{(}   \Gamma_{{\mathrm{1}}}   \ottsym{)}    \cap    \mathrm{dom}   \ottsym{(}   \Gamma_{{\mathrm{2}}}   \ottsym{)}    \ottsym{=}  \emptyset$, then $\Gamma_{{\mathrm{1}}}  \ottsym{,}  \Gamma_{{\mathrm{2}}}  \vdash  S  \ottsym{:}  \ottnt{K}$.
\end{lemma}

\begin{proof}
  By induction on the size of $\Gamma_{{\mathrm{1}}}$. We proceed by case analysis on the rule lastly applied to this derivation.
  \begin{divcases}
    \item[\rname{C}{Empty}] Clearly.
    \item[\rname{C}{Var}] For some $\Gamma'_{{\mathrm{1}}}$, $\mathit{x}$, and $\ottnt{A}$, we have
    \begin{itemize}
      \item $\Gamma_{{\mathrm{1}}} = \Gamma'_{{\mathrm{1}}}  \ottsym{,}  \mathit{x}  \ottsym{:}  \ottnt{A}$,
      \item $ \mathit{x}   \notin    \mathrm{dom}   \ottsym{(}   \Gamma'_{{\mathrm{1}}}   \ottsym{)}  $, and
      \item $\Gamma'_{{\mathrm{1}}}  \vdash  \ottnt{A}  \ottsym{:}   \mathbf{Typ} $.
    \end{itemize}
    By Lemma~\ref{lem:wf}, we have $\vdash  \Gamma'_{{\mathrm{1}}}$.
    By Definition~\ref{def:delta}, we have $ \Delta   \ottsym{(}   \Gamma'_{{\mathrm{1}}}   \ottsym{)}   \ottsym{,}  \Gamma_{{\mathrm{2}}}  \vdash  S  \ottsym{:}  \ottnt{K}$.
    By $  \mathrm{dom}   \ottsym{(}   \Gamma'_{{\mathrm{1}}}   \ottsym{)}    \subseteq    \mathrm{dom}   \ottsym{(}   \Gamma_{{\mathrm{1}}}   \ottsym{)}  $, we have $  \mathrm{dom}   \ottsym{(}   \Gamma'_{{\mathrm{1}}}   \ottsym{)}    \cap    \mathrm{dom}   \ottsym{(}   \Gamma_{{\mathrm{2}}}   \ottsym{)}    \ottsym{=}  \emptyset$.
    By the induction hypothesis, we have $\Gamma'_{{\mathrm{1}}}  \ottsym{,}  \Gamma_{{\mathrm{2}}}  \vdash  S  \ottsym{:}  \ottnt{K}$.
    By Lemma~\ref{lem:weakening}\ref{lem:weakening:kinding}, we have $\Gamma'_{{\mathrm{1}}}  \ottsym{,}  \mathit{x}  \ottsym{:}  \ottnt{A}  \ottsym{,}  \Gamma_{{\mathrm{2}}}  \vdash  S  \ottsym{:}  \ottnt{K}$ as required.
    \item[\rname{C}{TVar}] For some $\Gamma'_{{\mathrm{1}}}$, $\alpha$, and $\ottnt{K'}$, we have
    \begin{itemize}
      \item $\Gamma_{{\mathrm{1}}} = \Gamma'_{{\mathrm{1}}}  \ottsym{,}  \alpha  \ottsym{:}  \ottnt{K'}$,
      \item $\vdash  \Gamma'_{{\mathrm{1}}}$, and
      \item $ \alpha   \notin    \mathrm{dom}   \ottsym{(}   \Gamma'_{{\mathrm{1}}}   \ottsym{)}  $.
    \end{itemize}
    By Definition~\ref{def:delta}, we have $ \Delta   \ottsym{(}   \Gamma'_{{\mathrm{1}}}   \ottsym{)}   \ottsym{,}  \alpha  \ottsym{:}  \ottnt{K'}  \ottsym{,}  \Gamma_{{\mathrm{2}}}  \vdash  S  \ottsym{:}  \ottnt{K}$.
    By $ \alpha   \notin    \mathrm{dom}   \ottsym{(}   \Gamma'_{{\mathrm{1}}}   \ottsym{)}  $ and $  \mathrm{dom}   \ottsym{(}   \Gamma'_{{\mathrm{1}}}   \ottsym{)}    \subseteq    \mathrm{dom}   \ottsym{(}   \Gamma_{{\mathrm{1}}}   \ottsym{)}  $, we have $  \mathrm{dom}   \ottsym{(}   \Gamma'_{{\mathrm{1}}}   \ottsym{)}    \cap    \mathrm{dom}   \ottsym{(}   \alpha  \ottsym{:}  \ottnt{K'}  \ottsym{,}  \Gamma'_{{\mathrm{2}}}   \ottsym{)}    \ottsym{=}  \emptyset$.
    By the induction hypothesis, we have $\Gamma'_{{\mathrm{1}}}  \ottsym{,}  \alpha  \ottsym{:}  \ottnt{K'}  \ottsym{,}  \Gamma_{{\mathrm{2}}}  \vdash  S  \ottsym{:}  \ottnt{K}$ as required.
  \end{divcases}
\end{proof}
\begin{lemma}[Substitution of values]\label{lem:subst_value}
  Suppose that $\Gamma_{{\mathrm{1}}}  \vdash  \ottnt{v}  \ottsym{:}  \ottnt{A}  \mid   \bbZero $.
  \begin{enumerate}
    \item\label{lem:subst_value:wf} If $\vdash  \Gamma_{{\mathrm{1}}}  \ottsym{,}  \mathit{x}  \ottsym{:}  \ottnt{A}  \ottsym{,}  \Gamma_{{\mathrm{2}}}$, then $\vdash  \Gamma_{{\mathrm{1}}}  \ottsym{,}  \Gamma_{{\mathrm{2}}}$.
    \item\label{lem:subst_value:kinding} If $\Gamma_{{\mathrm{1}}}  \ottsym{,}  \mathit{x}  \ottsym{:}  \ottnt{A}  \ottsym{,}  \Gamma_{{\mathrm{2}}}  \vdash  S  \ottsym{:}  \ottnt{K}$, then $\Gamma_{{\mathrm{1}}}  \ottsym{,}  \Gamma_{{\mathrm{2}}}  \vdash  S  \ottsym{:}  \ottnt{K}$.
    \item\label{lem:subst_value:subtyping} If $\Gamma_{{\mathrm{1}}}  \ottsym{,}  \mathit{x}  \ottsym{:}  \ottnt{A}  \ottsym{,}  \Gamma_{{\mathrm{2}}}  \vdash  \ottnt{B}  <:  \ottnt{C}$, then $\Gamma_{{\mathrm{1}}}  \ottsym{,}  \Gamma_{{\mathrm{2}}}  \vdash  \ottnt{B}  <:  \ottnt{C}$.
    \item\label{lem:subst_value:subtyping_comp} If $\Gamma_{{\mathrm{1}}}  \ottsym{,}  \mathit{x}  \ottsym{:}  \ottnt{A}  \ottsym{,}  \Gamma_{{\mathrm{2}}}  \vdash  \ottnt{B_{{\mathrm{1}}}}  \mid  \varepsilon_{{\mathrm{1}}}  <:  \ottnt{B_{{\mathrm{2}}}}  \mid  \varepsilon_{{\mathrm{2}}}$, then $\Gamma_{{\mathrm{1}}}  \ottsym{,}  \Gamma_{{\mathrm{2}}}  \vdash  \ottnt{B_{{\mathrm{1}}}}  \mid  \varepsilon_{{\mathrm{1}}}  <:  \ottnt{B_{{\mathrm{2}}}}  \mid  \varepsilon_{{\mathrm{2}}}$.
    \item\label{lem:subst_value:typing} If $\Gamma_{{\mathrm{1}}}  \ottsym{,}  \mathit{x}  \ottsym{:}  \ottnt{A}  \ottsym{,}  \Gamma_{{\mathrm{2}}}  \vdash  \ottnt{e}  \ottsym{:}  \ottnt{B}  \mid  \varepsilon$, then $\Gamma_{{\mathrm{1}}}  \ottsym{,}  \Gamma_{{\mathrm{2}}}  \vdash  \ottnt{e} \,  \! [  \ottnt{v}  /  \mathit{x}  ]   \ottsym{:}  \ottnt{B}  \mid  \varepsilon$.
    \item\label{lem:subst_value:handling} If $ \Gamma_{{\mathrm{1}}}  \ottsym{,}  \mathit{x}  \ottsym{:}  \ottnt{A}  \ottsym{,}  \Gamma_{{\mathrm{2}}}  \vdash _{ \sigma }  \ottnt{h}  :  \ottnt{B}   \Rightarrow  ^ { \varepsilon }  \ottnt{C} $, then $ \Gamma_{{\mathrm{1}}}  \ottsym{,}  \Gamma_{{\mathrm{2}}}  \vdash _{ \sigma }  \ottnt{h} \,  \! [  \ottnt{v}  /  \mathit{x}  ]   :  \ottnt{B}   \Rightarrow  ^ { \varepsilon }  \ottnt{C} $.
  \end{enumerate}
\end{lemma}

\begin{proof}
  \phantom{}
  \begin{itemize}
    \item[(1)(2)] By mutual induction on derivations of the judgments. We proceed by case analysis on the rule applied lastly to the derivation.
          \begin{divcases}
            \item[\rname{C}{Empty}] Cannot happen.
            \item[\rname{C}{Var}] If $\Gamma_{{\mathrm{2}}} =  \emptyset $, then we have $\Gamma_{{\mathrm{1}}}  \vdash  \ottnt{A}  \ottsym{:}   \mathbf{Typ} $. By Lemma~\ref{lem:wf}, $\vdash  \Gamma_{{\mathrm{1}}}$ holds. If $\Gamma_{{\mathrm{2}}} \neq  \emptyset $, then we have
            \begin{itemize}
              \item $\Gamma_{{\mathrm{2}}} = \Gamma'_{{\mathrm{2}}}  \ottsym{,}  \mathit{y}  \ottsym{:}  \ottnt{B}$,
              \item $\Gamma_{{\mathrm{1}}}  \ottsym{,}  \mathit{x}  \ottsym{:}  \ottnt{A}  \ottsym{,}  \Gamma'_{{\mathrm{2}}}  \vdash  \ottnt{B}  \ottsym{:}   \mathbf{Typ} $, and
              \item $ \mathit{y}   \notin    \mathrm{dom}   \ottsym{(}   \Gamma_{{\mathrm{1}}}  \ottsym{,}  \mathit{x}  \ottsym{:}  \ottnt{A}  \ottsym{,}  \Gamma'_{{\mathrm{2}}}   \ottsym{)}  $,
            \end{itemize}
            for some $\Gamma'_{{\mathrm{2}}}$, $\mathit{y}$, and $\ottnt{B}$. By the induction hypothesis, we have $\Gamma_{{\mathrm{1}}}  \ottsym{,}  \Gamma'_{{\mathrm{2}}}  \vdash  \ottnt{B}  \ottsym{:}   \mathbf{Typ} $. Thus, \rname{C}{Var} derives $\vdash  \Gamma_{{\mathrm{1}}}  \ottsym{,}  \Gamma_{{\mathrm{2}}}$ because $ \mathit{y}   \notin    \mathrm{dom}   \ottsym{(}   \Gamma_{{\mathrm{1}}}  \ottsym{,}  \Gamma'_{{\mathrm{2}}}   \ottsym{)}  $.
            \item[\rname{C}{TVar}] Since $\Gamma_{{\mathrm{2}}}$ cannot be $ \emptyset $, we have
            \begin{itemize}
              \item $\Gamma_{{\mathrm{2}}} = \Gamma'_{{\mathrm{2}}}  \ottsym{,}  \alpha  \ottsym{:}  \ottnt{K}$,
              \item $\vdash  \Gamma_{{\mathrm{1}}}  \ottsym{,}  \mathit{x}  \ottsym{:}  \ottnt{A}  \ottsym{,}  \Gamma'_{{\mathrm{2}}}$, and
              \item $ \alpha   \notin    \mathrm{dom}   \ottsym{(}   \Gamma_{{\mathrm{1}}}  \ottsym{,}  \mathit{x}  \ottsym{:}  \ottnt{A}  \ottsym{,}  \Gamma'_{{\mathrm{2}}}   \ottsym{)}  $,
            \end{itemize}
            for some $\Gamma_{{\mathrm{2}}}$, $\alpha$, and $\ottnt{K}$. By the induction hypothesis, we have $\vdash  \Gamma_{{\mathrm{1}}}  \ottsym{,}  \Gamma'_{{\mathrm{2}}}$. Thus, \rname{C}{TVar} derives $\vdash  \Gamma_{{\mathrm{1}}}  \ottsym{,}  \Gamma_{{\mathrm{2}}}$ because $ \alpha   \notin    \mathrm{dom}   \ottsym{(}   \Gamma_{{\mathrm{1}}}  \ottsym{,}  \Gamma'_{{\mathrm{2}}}   \ottsym{)}  $.
            \item[\rname{K}{Var}] For some $\alpha$, the following are given:
            \begin{itemize}
              \item $S = \alpha$,
              \item $\vdash  \Gamma_{{\mathrm{1}}}  \ottsym{,}  \mathit{x}  \ottsym{:}  \ottnt{A}  \ottsym{,}  \Gamma_{{\mathrm{2}}}$, and
              \item $ \alpha   \ottsym{:}   \ottnt{K}   \in   \Gamma_{{\mathrm{1}}}  \ottsym{,}  \mathit{x}  \ottsym{:}  \ottnt{A}  \ottsym{,}  \Gamma_{{\mathrm{2}}} $.
            \end{itemize}
            By the induction hypothesis, we have $\vdash  \Gamma_{{\mathrm{1}}}  \ottsym{,}  \Gamma_{{\mathrm{2}}}$. Thus, \rname{K}{Var} derives $\Gamma_{{\mathrm{1}}}  \ottsym{,}  \Gamma_{{\mathrm{2}}}  \vdash  \alpha  \ottsym{:}  \ottnt{K}$ because of $ \alpha   \ottsym{:}   \ottnt{K}   \in   \Gamma_{{\mathrm{1}}}  \ottsym{,}  \Gamma_{{\mathrm{2}}} $.
            \item[\rname{K}{Fun}] For some $\ottnt{B}$, $\ottnt{C}$, and $\varepsilon$, the following are given:
            \begin{itemize}
              \item $S =  \ottnt{B}    \rightarrow_{ \varepsilon }    \ottnt{C} $,
              \item $\ottnt{K} =  \mathbf{Typ} $,
              \item $\Gamma_{{\mathrm{1}}}  \ottsym{,}  \mathit{x}  \ottsym{:}  \ottnt{A}  \ottsym{,}  \Gamma_{{\mathrm{2}}}  \vdash  \ottnt{B}  \ottsym{:}   \mathbf{Typ} $,
              \item $\Gamma_{{\mathrm{1}}}  \ottsym{,}  \mathit{x}  \ottsym{:}  \ottnt{A}  \ottsym{,}  \Gamma_{{\mathrm{2}}}  \vdash  \varepsilon  \ottsym{:}   \mathbf{Eff} $, and
              \item $\Gamma_{{\mathrm{1}}}  \ottsym{,}  \mathit{x}  \ottsym{:}  \ottnt{A}  \ottsym{,}  \Gamma_{{\mathrm{2}}}  \vdash  \ottnt{C}  \ottsym{:}   \mathbf{Typ} $.
            \end{itemize}
            By the induction hypothesis, we have
            \begin{itemize}
              \item $\Gamma_{{\mathrm{1}}}  \ottsym{,}  \Gamma_{{\mathrm{2}}}  \vdash  \ottnt{B}  \ottsym{:}   \mathbf{Typ} $,
              \item $\Gamma_{{\mathrm{1}}}  \ottsym{,}  \Gamma_{{\mathrm{2}}}  \vdash  \varepsilon  \ottsym{:}   \mathbf{Eff} $, and
              \item $\Gamma_{{\mathrm{1}}}  \ottsym{,}  \Gamma_{{\mathrm{2}}}  \vdash  \ottnt{C}  \ottsym{:}   \mathbf{Typ} $.
            \end{itemize}
            Thus, \rname{K}{Fun} derives $\Gamma_{{\mathrm{1}}}  \ottsym{,}  \Gamma_{{\mathrm{2}}}  \vdash   \ottnt{B}    \rightarrow_{ \varepsilon }    \ottnt{C}   \ottsym{:}   \mathbf{Typ} $.
            \item[\rname{K}{Poly}] For some $\alpha$, $\ottnt{K'}$, $\ottnt{A'}$, and $\varepsilon$, the following are given:
            \begin{itemize}
              \item $S =   \forall   \alpha  \ottsym{:}  \ottnt{K'}   \ottsym{.}    \ottnt{A'}    ^{ \varepsilon }  $,
              \item $\ottnt{K} =  \mathbf{Typ} $,
              \item $\Gamma_{{\mathrm{1}}}  \ottsym{,}  \mathit{x}  \ottsym{:}  \ottnt{A}  \ottsym{,}  \Gamma_{{\mathrm{2}}}  \ottsym{,}  \alpha  \ottsym{:}  \ottnt{K'}  \vdash  \ottnt{A'}  \ottsym{:}   \mathbf{Typ} $, and
              \item $\Gamma_{{\mathrm{1}}}  \ottsym{,}  \mathit{x}  \ottsym{:}  \ottnt{A}  \ottsym{,}  \Gamma_{{\mathrm{2}}}  \ottsym{,}  \alpha  \ottsym{:}  \ottnt{K'}  \vdash  \varepsilon  \ottsym{:}   \mathbf{Eff} $.
            \end{itemize}
            By the induction hypothesis, we have
            \begin{itemize}
              \item $\Gamma_{{\mathrm{1}}}  \ottsym{,}  \Gamma_{{\mathrm{2}}}  \ottsym{,}  \alpha  \ottsym{:}  \ottnt{K'}  \vdash  \ottnt{A'}  \ottsym{:}   \mathbf{Typ} $, and
              \item $\Gamma_{{\mathrm{1}}}  \ottsym{,}  \Gamma_{{\mathrm{2}}}  \ottsym{,}  \alpha  \ottsym{:}  \ottnt{K'}  \vdash  \varepsilon  \ottsym{:}   \mathbf{Eff} $.
            \end{itemize}
            Thus, \rname{K}{Poly} derives $\Gamma_{{\mathrm{1}}}  \ottsym{,}  \Gamma_{{\mathrm{2}}}  \vdash    \forall   \alpha  \ottsym{:}  \ottnt{K'}   \ottsym{.}    \ottnt{A'}    ^{ \varepsilon }    \ottsym{:}   \mathbf{Typ} $.
            \item[\rname{K}{Cons}] For some $\mathcal{C}$, $ \bm{ { S } } ^ {  \mathit{I}  } $ and $ {\bm{ { \ottnt{K} } } }^{ \mathit{I} } $, the following are given:
            \begin{itemize}
              \item $S = \mathcal{C} \,  \bm{ { S } } ^ {  \mathit{I}  } $,
              \item $ \mathcal{C}   \ottsym{:}    \Pi {\bm{ { \ottnt{K} } } }^{ \mathit{I} }   \rightarrow  \ottnt{K}   \in   \Sigma $,
              \item $\vdash  \Gamma_{{\mathrm{1}}}  \ottsym{,}  \mathit{x}  \ottsym{:}  \ottnt{A}  \ottsym{,}  \Gamma_{{\mathrm{2}}}$, and
              \item $\Gamma_{{\mathrm{1}}}  \ottsym{,}  \mathit{x}  \ottsym{:}  \ottnt{A}  \ottsym{,}  \Gamma_{{\mathrm{2}}}  \vdash   \bm{ { S } }^{ \mathit{I} } : \bm{ \ottnt{K} }^{ \mathit{I} } $.
            \end{itemize}
            By the induction hypothesis, we have $\vdash  \Gamma_{{\mathrm{1}}}  \ottsym{,}  \Gamma_{{\mathrm{2}}}$ and $\Gamma_{{\mathrm{1}}}  \ottsym{,}  \Gamma_{{\mathrm{2}}}  \vdash   \bm{ { S } }^{ \mathit{I} } : \bm{ \ottnt{K} }^{ \mathit{I} } $. Thus, \rname{K}{Cons} derives $\Gamma_{{\mathrm{1}}}  \ottsym{,}  \Gamma_{{\mathrm{2}}}  \vdash  \mathcal{C} \,  \bm{ { S } } ^ {  \mathit{I}  }   \ottsym{:}  \ottnt{K}$.
          \end{divcases}

    \item[(3)(4)] By mutual induction on derivations of the judgments.
          We proceed by case analysis on the rule applied lastly to the derivation.
          \begin{divcases}
            \item[\rname{ST}{Refl}]
            $\ottnt{B} = \ottnt{C}$ and $\Gamma_{{\mathrm{1}}}  \ottsym{,}  \mathit{x}  \ottsym{:}  \ottnt{A}  \ottsym{,}  \Gamma_{{\mathrm{2}}}  \vdash  \ottnt{B}  \ottsym{:}   \mathbf{Typ} $ are given.
            By case \ref{lem:subst_value:kinding}, we have $\Gamma_{{\mathrm{1}}}  \ottsym{,}  \Gamma_{{\mathrm{2}}}  \vdash  \ottnt{B}  \ottsym{:}   \mathbf{Typ} $.
            Thus, \rname{ST}{Refl} derives $\Gamma_{{\mathrm{1}}}  \ottsym{,}  \Gamma_{{\mathrm{2}}}  \vdash  \ottnt{B}  <:  \ottnt{B}$.

            \item[\rname{ST}{Fun}]
            For some $\ottnt{A_{{\mathrm{11}}}}$, $\varepsilon_{{\mathrm{1}}}$, $\ottnt{A_{{\mathrm{12}}}}$, $\ottnt{A_{{\mathrm{21}}}}$, $\varepsilon_{{\mathrm{2}}}$, and $\ottnt{A_{{\mathrm{22}}}}$, the following are given:
            \begin{itemize}
              \item $\ottnt{B} =  \ottnt{A_{{\mathrm{11}}}}    \rightarrow_{ \varepsilon_{{\mathrm{1}}} }    \ottnt{A_{{\mathrm{12}}}} $,
              \item $\ottnt{C} =  \ottnt{A_{{\mathrm{21}}}}    \rightarrow_{ \varepsilon_{{\mathrm{2}}} }    \ottnt{A_{{\mathrm{22}}}} $,
              \item $\Gamma_{{\mathrm{1}}}  \ottsym{,}  \mathit{x}  \ottsym{:}  \ottnt{A}  \ottsym{,}  \Gamma_{{\mathrm{2}}}  \vdash  \ottnt{A_{{\mathrm{21}}}}  <:  \ottnt{A_{{\mathrm{11}}}}$, and
              \item $\Gamma_{{\mathrm{1}}}  \ottsym{,}  \mathit{x}  \ottsym{:}  \ottnt{A}  \ottsym{,}  \Gamma_{{\mathrm{2}}}  \vdash  \ottnt{A_{{\mathrm{12}}}}  \mid  \varepsilon_{{\mathrm{1}}}  <:  \ottnt{A_{{\mathrm{22}}}}  \mid  \varepsilon_{{\mathrm{2}}}$.
            \end{itemize}
            By the induction hypothesis, we have $\Gamma_{{\mathrm{1}}}  \ottsym{,}  \Gamma_{{\mathrm{2}}}  \vdash  \ottnt{A_{{\mathrm{21}}}}  <:  \ottnt{A_{{\mathrm{11}}}}$ and
            $\Gamma_{{\mathrm{1}}}  \ottsym{,}  \Gamma_{{\mathrm{2}}}  \vdash  \ottnt{A_{{\mathrm{12}}}}  \mid  \varepsilon_{{\mathrm{1}}}  <:  \ottnt{A_{{\mathrm{22}}}}  \mid  \varepsilon_{{\mathrm{2}}}$.
            Thus, \rname{ST}{Fun} derives $\Gamma_{{\mathrm{1}}}  \ottsym{,}  \Gamma_{{\mathrm{2}}}  \vdash   \ottnt{A_{{\mathrm{11}}}}    \rightarrow_{ \varepsilon_{{\mathrm{1}}} }    \ottnt{A_{{\mathrm{12}}}}   <:   \ottnt{A_{{\mathrm{21}}}}    \rightarrow_{ \varepsilon_{{\mathrm{2}}} }    \ottnt{A_{{\mathrm{22}}}} $.

            \item[\rname{ST}{Poly}]
            For some $\alpha$, $\ottnt{K}$, $\ottnt{A_{{\mathrm{1}}}}$, $\varepsilon_{{\mathrm{1}}}$, $\ottnt{A_{{\mathrm{2}}}}$, and $\varepsilon_{{\mathrm{2}}}$, the following are given:
            \begin{itemize}
              \item $\ottnt{B} =   \forall   \alpha  \ottsym{:}  \ottnt{K}   \ottsym{.}    \ottnt{A_{{\mathrm{1}}}}    ^{ \varepsilon_{{\mathrm{1}}} }  $,
              \item $\ottnt{C} =   \forall   \alpha  \ottsym{:}  \ottnt{K}   \ottsym{.}    \ottnt{A_{{\mathrm{2}}}}    ^{ \varepsilon_{{\mathrm{2}}} }  $, and
              \item $\Gamma_{{\mathrm{1}}}  \ottsym{,}  \mathit{x}  \ottsym{:}  \ottnt{A}  \ottsym{,}  \Gamma_{{\mathrm{2}}}  \ottsym{,}  \alpha  \ottsym{:}  \ottnt{K}  \vdash  \ottnt{A_{{\mathrm{1}}}}  \mid  \varepsilon_{{\mathrm{1}}}  <:  \ottnt{A_{{\mathrm{2}}}}  \mid  \varepsilon_{{\mathrm{2}}}$.
            \end{itemize}
            By the induction hypothesis, we have
            $\Gamma_{{\mathrm{1}}}  \ottsym{,}  \Gamma_{{\mathrm{2}}}  \ottsym{,}  \alpha  \ottsym{:}  \ottnt{K}  \vdash  \ottnt{A_{{\mathrm{1}}}}  \mid  \varepsilon_{{\mathrm{1}}}  <:  \ottnt{A_{{\mathrm{2}}}}  \mid  \varepsilon_{{\mathrm{2}}}$.
            Thus, \rname{ST}{Poly} derives $\Gamma_{{\mathrm{1}}}  \ottsym{,}  \Gamma_{{\mathrm{2}}}  \vdash    \forall   \alpha  \ottsym{:}  \ottnt{K}   \ottsym{.}    \ottnt{A_{{\mathrm{1}}}}    ^{ \varepsilon_{{\mathrm{1}}} }    <:    \forall   \alpha  \ottsym{:}  \ottnt{K}   \ottsym{.}    \ottnt{A_{{\mathrm{2}}}}    ^{ \varepsilon_{{\mathrm{2}}} }  $.

            \item[\rname{ST}{Comp}]
            We have $\Gamma_{{\mathrm{1}}}  \ottsym{,}  \mathit{x}  \ottsym{:}  \ottnt{A}  \ottsym{,}  \Gamma_{{\mathrm{2}}}  \vdash  \ottnt{B_{{\mathrm{1}}}}  <:  \ottnt{B_{{\mathrm{2}}}}$ and $\Gamma_{{\mathrm{1}}}  \ottsym{,}  \mathit{x}  \ottsym{:}  \ottnt{A}  \ottsym{,}  \Gamma_{{\mathrm{2}}}  \vdash   \varepsilon_{{\mathrm{1}}}  \olessthan  \varepsilon_{{\mathrm{2}}} $.
            By the induction hypothesis, we have $\Gamma_{{\mathrm{1}}}  \ottsym{,}  \Gamma_{{\mathrm{2}}}  \vdash  \ottnt{B_{{\mathrm{1}}}}  <:  \ottnt{B_{{\mathrm{2}}}}$.
            By case~\ref{lem:subst_value:kinding},
            we have $\Gamma_{{\mathrm{1}}}  \ottsym{,}  \Gamma_{{\mathrm{2}}}  \vdash   \varepsilon_{{\mathrm{1}}}  \olessthan  \varepsilon_{{\mathrm{2}}} $.
            Thus, \rname{ST}{Comp} derives $\Gamma_{{\mathrm{1}}}  \ottsym{,}  \Gamma_{{\mathrm{2}}}  \vdash  \ottnt{B_{{\mathrm{1}}}}  \mid  \varepsilon_{{\mathrm{1}}}  <:  \ottnt{B_{{\mathrm{2}}}}  \mid  \varepsilon_{{\mathrm{2}}}$
            as required.
          \end{divcases}

    \item[(5)(6)]
          By mutual induction on derivations of the judgments.
          We proceed by case analysis on the rule applied lastly to the derivation.
          \begin{divcases}
            \item[\rname{T}{Var}]
            For some $\mathit{y}$, the following are given:
            \begin{itemize}
              \item $\ottnt{e} = \mathit{y}$,
              \item $\varepsilon =  \bbZero $,
              \item $\vdash  \Gamma_{{\mathrm{1}}}  \ottsym{,}  \mathit{x}  \ottsym{:}  \ottnt{A}  \ottsym{,}  \Gamma_{{\mathrm{2}}}$, and
              \item $ \mathit{y}   \ottsym{:}   \ottnt{B}   \in   \Gamma_{{\mathrm{1}}}  \ottsym{,}  \mathit{x}  \ottsym{:}  \ottnt{A}  \ottsym{,}  \Gamma_{{\mathrm{2}}} $.
            \end{itemize}
            By case \ref{lem:subst_value:wf}, we have $\vdash  \Gamma_{{\mathrm{1}}}  \ottsym{,}  \Gamma_{{\mathrm{2}}}$.

            If $\mathit{y} = \mathit{x}$, then $\Gamma_{{\mathrm{1}}}  \ottsym{,}  \Gamma_{{\mathrm{2}}}  \vdash  \ottnt{v}  \ottsym{:}  \ottnt{A}  \mid   \bbZero $ holds because of $\Gamma_{{\mathrm{1}}}  \vdash  \ottnt{v}  \ottsym{:}  \ottnt{A}  \mid   \bbZero $ and Lemma~\ref{lem:weakening}\ref{lem:weakening:typing}.

            If $\mathit{y} \neq \mathit{x}$, then we have $ \mathit{y}   \ottsym{:}   \ottnt{B}   \in   \Gamma_{{\mathrm{1}}}  \ottsym{,}  \Gamma_{{\mathrm{2}}} $.
            Thus, \rname{T}{Var} derives $\Gamma_{{\mathrm{1}}}  \ottsym{,}  \Gamma_{{\mathrm{2}}}  \vdash  \mathit{y}  \ottsym{:}  \ottnt{B}  \mid   \bbZero $.

            \item[\rname{T}{Abs}]
            Without loss of generality, we can choose $\mathit{f}$ and $\mathit{y}$ such that
            $\mathit{f}, \mathit{y} \neq \mathit{x}$ and $\mathit{f},  \mathit{y}   \notin    \mathrm{FV}   \ottsym{(}   \ottnt{v}   \ottsym{)}  $.
            For some $\ottnt{e'}$, $\ottnt{A'}$, $\ottnt{B'}$, and $\varepsilon'$, the following are given:
            \begin{itemize}
              \item $\ottnt{e} = \ottkw{fun} \, \ottsym{(}  \mathit{f}  \ottsym{,}  \mathit{y}  \ottsym{,}  \ottnt{e'}  \ottsym{)}$,
              \item $\ottnt{B} =  \ottnt{A'}    \rightarrow_{ \varepsilon' }    \ottnt{B'} $,
              \item $\varepsilon =  \bbZero $, and
              \item $\Gamma_{{\mathrm{1}}}  \ottsym{,}  \mathit{x}  \ottsym{:}  \ottnt{A}  \ottsym{,}  \Gamma_{{\mathrm{2}}}  \ottsym{,}  \mathit{f}  \ottsym{:}   \ottnt{A'}    \rightarrow_{ \varepsilon' }    \ottnt{B'}   \ottsym{,}  \mathit{y}  \ottsym{:}  \ottnt{A'}  \vdash  \ottnt{e'}  \ottsym{:}  \ottnt{B'}  \mid  \varepsilon'$.
            \end{itemize}
            By the induction hypothesis, we have
            $\Gamma_{{\mathrm{1}}}  \ottsym{,}  \Gamma_{{\mathrm{2}}}  \ottsym{,}  \mathit{g}  \ottsym{:}   \ottnt{A'}    \rightarrow_{ \varepsilon' }    \ottnt{B'}   \ottsym{,}  \mathit{y}  \ottsym{:}  \ottnt{A'}  \vdash  \ottnt{e'} \,  \! [  \ottnt{v}  /  \mathit{x}  ]   \ottsym{:}  \ottnt{B'}  \mid  \varepsilon'$.
            Thus, \rname{T}{Abs} derives $\Gamma_{{\mathrm{1}}}  \ottsym{,}  \Gamma_{{\mathrm{2}}}  \vdash  \ottkw{fun} \, \ottsym{(}  \mathit{f}  \ottsym{,}  \mathit{y}  \ottsym{,}  \ottnt{e'} \,  \! [  \ottnt{v}  /  \mathit{x}  ]   \ottsym{)}  \ottsym{:}   \ottnt{A'}    \rightarrow_{ \varepsilon' }    \ottnt{B'}   \mid   \bbZero $, and
            since $\ottsym{(}  \ottkw{fun} \, \ottsym{(}  \mathit{f}  \ottsym{,}  \mathit{y}  \ottsym{,}  \ottnt{e'}  \ottsym{)}  \ottsym{)} \,  \! [  \ottnt{v}  /  \mathit{x}  ]  = \ottkw{fun} \, \ottsym{(}  \mathit{f}  \ottsym{,}  \mathit{y}  \ottsym{,}  \ottnt{e'} \,  \! [  \ottnt{v}  /  \mathit{x}  ]   \ottsym{)}$,
            the required result is achieved.

            \item[\rname{T}{App}]
            For some $\ottnt{v_{{\mathrm{1}}}}$, $\ottnt{v_{{\mathrm{2}}}}$, and $\ottnt{C}$, the following are given:
            \begin{itemize}
              \item $\ottnt{e} = \ottnt{v_{{\mathrm{1}}}} \, \ottnt{v_{{\mathrm{2}}}}$,
              \item $\Gamma_{{\mathrm{1}}}  \ottsym{,}  \mathit{x}  \ottsym{:}  \ottnt{A}  \ottsym{,}  \Gamma_{{\mathrm{2}}}  \vdash  \ottnt{v_{{\mathrm{1}}}}  \ottsym{:}   \ottnt{C}    \rightarrow_{ \varepsilon }    \ottnt{B}   \mid   \bbZero $, and
              \item $\Gamma_{{\mathrm{1}}}  \ottsym{,}  \mathit{x}  \ottsym{:}  \ottnt{A}  \ottsym{,}  \Gamma_{{\mathrm{2}}}  \vdash  \ottnt{v_{{\mathrm{2}}}}  \ottsym{:}  \ottnt{C}  \mid   \bbZero $.
            \end{itemize}
            By the induction hypothesis, we have
            \begin{itemize}
              \item $\Gamma_{{\mathrm{1}}}  \ottsym{,}  \Gamma_{{\mathrm{2}}}  \vdash  \ottnt{v_{{\mathrm{1}}}} \,  \! [  \ottnt{v}  /  \mathit{x}  ]   \ottsym{:}   \ottnt{C}    \rightarrow_{ \varepsilon }    \ottnt{B}   \mid   \bbZero $
              \item and $\Gamma_{{\mathrm{1}}}  \ottsym{,}  \Gamma_{{\mathrm{2}}}  \vdash  \ottnt{v_{{\mathrm{2}}}} \,  \! [  \ottnt{v}  /  \mathit{x}  ]   \ottsym{:}  \ottnt{C}  \mid   \bbZero $.
            \end{itemize}
            Thus, \rname{T}{App} derives $\Gamma_{{\mathrm{1}}}  \ottsym{,}  \Gamma_{{\mathrm{2}}}  \vdash   (  \ottnt{v_{{\mathrm{1}}}} \,  \! [  \ottnt{v}  /  \mathit{x}  ]   )  \,  (  \ottnt{v_{{\mathrm{2}}}} \,  \! [  \ottnt{v}  /  \mathit{x}  ]   )   \ottsym{:}  \ottnt{B}  \mid  \varepsilon$, and
            since $ (  \ottnt{v_{{\mathrm{1}}}} \, \ottnt{v_{{\mathrm{2}}}}  )  \,  \! [  \ottnt{v}  /  \mathit{x}  ]  =  (  \ottnt{v_{{\mathrm{1}}}} \,  \! [  \ottnt{v}  /  \mathit{x}  ]   )  \,  (  \ottnt{v_{{\mathrm{2}}}} \,  \! [  \ottnt{v}  /  \mathit{x}  ]   ) $,
            the required result is achieved.

            \item[\rname{T}{TAbs}]
            Without loss of generality, we can choose $\alpha$ such that $ \alpha   \notin    \mathrm{FTV}   \ottsym{(}   \ottnt{v}   \ottsym{)}  $.
            For some $\ottnt{K}$, $\ottnt{e'}$, $\ottnt{B'}$, and $\varepsilon'$, the following are given:
            \begin{itemize}
              \item $\ottnt{e} = \Lambda  \alpha  \ottsym{:}  \ottnt{K}  \ottsym{.}  \ottnt{e'}$,
              \item $\ottnt{B} =   \forall   \alpha  \ottsym{:}  \ottnt{K}   \ottsym{.}    \ottnt{B'}    ^{ \varepsilon' }  $,
              \item $\varepsilon =  \bbZero $, and
              \item $\Gamma_{{\mathrm{1}}}  \ottsym{,}  \mathit{x}  \ottsym{:}  \ottnt{A}  \ottsym{,}  \Gamma_{{\mathrm{2}}}  \ottsym{,}  \alpha  \ottsym{:}  \ottnt{K}  \vdash  \ottnt{e'}  \ottsym{:}  \ottnt{B'}  \mid  \varepsilon'$.
            \end{itemize}
            By the induction hypothesis, we have
            $\Gamma_{{\mathrm{1}}}  \ottsym{,}  \Gamma_{{\mathrm{2}}}  \ottsym{,}  \alpha  \ottsym{:}  \ottnt{K}  \vdash  \ottnt{e'} \,  \! [  \ottnt{v}  /  \mathit{x}  ]   \ottsym{:}  \ottnt{B'}  \mid  \varepsilon'$.
            Thus, \rname{T}{TAbs} derives $\Gamma_{{\mathrm{1}}}  \ottsym{,}  \Gamma_{{\mathrm{2}}}  \vdash  \Lambda  \alpha  \ottsym{:}  \ottnt{K}  \ottsym{.}  \ottnt{e'} \,  \! [  \ottnt{v}  /  \mathit{x}  ]   \ottsym{:}    \forall   \alpha  \ottsym{:}  \ottnt{K}   \ottsym{.}    \ottnt{B'}    ^{ \varepsilon' }    \mid   \bbZero $, and
            since $\ottsym{(}  \Lambda  \alpha  \ottsym{:}  \ottnt{K}  \ottsym{.}  \ottnt{e'}  \ottsym{)} \,  \! [  \ottnt{v}  /  \mathit{x}  ]  = \Lambda  \alpha  \ottsym{:}  \ottnt{K}  \ottsym{.}  \ottnt{e'} \,  \! [  \ottnt{v}  /  \mathit{x}  ] $,
            the required result is achieved.

            \item[\rname{T}{TApp}]
            For some $\ottnt{v'}$, $S$, $\alpha$, $\ottnt{B'}$, $\varepsilon'$, and $\ottnt{K}$,
            the following are given:
            \begin{itemize}
              \item $\ottnt{e} = \ottnt{v'} \, S$,
              \item $\ottnt{B} = \ottnt{B'} \,  \! [  S  /  \alpha   ] $,
              \item $\varepsilon = \varepsilon' \,  \! [  S  /  \alpha   ] $,
              \item $\Gamma_{{\mathrm{1}}}  \ottsym{,}  \mathit{x}  \ottsym{:}  \ottnt{A}  \ottsym{,}  \Gamma_{{\mathrm{2}}}  \vdash  \ottnt{v'}  \ottsym{:}    \forall   \alpha  \ottsym{:}  \ottnt{K}   \ottsym{.}    \ottnt{B'}    ^{ \varepsilon' }    \mid   \bbZero $, and
              \item $\Gamma_{{\mathrm{1}}}  \ottsym{,}  \mathit{x}  \ottsym{:}  \ottnt{A}  \ottsym{,}  \Gamma_{{\mathrm{2}}}  \vdash  S  \ottsym{:}  \ottnt{K}$.
            \end{itemize}
            By the induction hypothesis and case \ref{lem:subst_value:kinding}, we have
            \begin{itemize}
              \item $\Gamma_{{\mathrm{1}}}  \ottsym{,}  \Gamma_{{\mathrm{2}}}  \vdash  \ottnt{v'} \,  \! [  \ottnt{v}  /  \mathit{x}  ]   \ottsym{:}    \forall   \alpha  \ottsym{:}  \ottnt{K}   \ottsym{.}    \ottnt{B'}    ^{ \varepsilon }    \mid   \bbZero $ and
              \item $\Gamma_{{\mathrm{1}}}  \ottsym{,}  \Gamma_{{\mathrm{2}}}  \vdash  S  \ottsym{:}  \ottnt{K}$.
            \end{itemize}
            Thus, \rname{T}{TApp} derives $\Gamma_{{\mathrm{1}}}  \ottsym{,}  \Gamma_{{\mathrm{2}}}  \vdash  \ottnt{v'} \,  \! [  \ottnt{v}  /  \mathit{x}  ]  \, S  \ottsym{:}  \ottnt{B'} \,  \! [  S  /  \alpha   ]   \mid  \varepsilon' \,  \! [  S  /  \alpha   ] $, and
            since $ (  \ottnt{v'} \, S  )  \,  \! [  \ottnt{v}  /  \mathit{x}  ]  = \ottnt{v'} \,  \! [  \ottnt{v}  /  \mathit{x}  ]  \, S$, the required result is achieved.

            \item[\rname{T}{Let}]
            Without loss of generality, we can choose $\mathit{y}$ such that $\mathit{y} \neq \mathit{x}$ and $ \mathit{y}   \notin    \mathrm{FV}   \ottsym{(}   \ottnt{v}   \ottsym{)}  $.
            For some $\ottnt{e_{{\mathrm{1}}}}$, $\ottnt{e_{{\mathrm{2}}}}$, and $\ottnt{C}$, the following are given:
            \begin{itemize}
              \item $\ottnt{e} = (\mathbf{let} \, \mathit{y}  \ottsym{=}  \ottnt{e_{{\mathrm{1}}}} \, \mathbf{in} \, \ottnt{e_{{\mathrm{2}}}})$,
              \item $\Gamma_{{\mathrm{1}}}  \ottsym{,}  \mathit{x}  \ottsym{:}  \ottnt{A}  \ottsym{,}  \Gamma_{{\mathrm{2}}}  \vdash  \ottnt{e_{{\mathrm{1}}}}  \ottsym{:}  \ottnt{C}  \mid  \varepsilon$, and
              \item $\Gamma_{{\mathrm{1}}}  \ottsym{,}  \mathit{x}  \ottsym{:}  \ottnt{A}  \ottsym{,}  \Gamma_{{\mathrm{2}}}  \ottsym{,}  \mathit{y}  \ottsym{:}  \ottnt{C}  \vdash  \ottnt{e_{{\mathrm{2}}}}  \ottsym{:}  \ottnt{B}  \mid  \varepsilon$.
            \end{itemize}
            By the induction hypothesis, we have
            \begin{itemize}
              \item $\Gamma_{{\mathrm{1}}}  \ottsym{,}  \Gamma_{{\mathrm{2}}}  \vdash  \ottnt{e_{{\mathrm{1}}}} \,  \! [  \ottnt{v}  /  \mathit{x}  ]   \ottsym{:}  \ottnt{C}  \mid  \varepsilon$ and
              \item $\Gamma_{{\mathrm{1}}}  \ottsym{,}  \Gamma_{{\mathrm{2}}}  \ottsym{,}  \mathit{y}  \ottsym{:}  \ottnt{C}  \vdash  \ottnt{e_{{\mathrm{2}}}} \,  \! [  \ottnt{v}  /  \mathit{x}  ]   \ottsym{:}  \ottnt{B}  \mid  \varepsilon$.
            \end{itemize}
            Thus, \rname{T}{Let} derives $\Gamma_{{\mathrm{1}}}  \ottsym{,}  \Gamma_{{\mathrm{2}}}  \vdash  \mathbf{let} \, \mathit{y}  \ottsym{=}  \ottnt{e_{{\mathrm{1}}}} \,  \! [  \ottnt{v}  /  \mathit{x}  ]  \, \mathbf{in} \, \ottnt{e_{{\mathrm{2}}}} \,  \! [  \ottnt{v}  /  \mathit{x}  ]   \ottsym{:}  \ottnt{B}  \mid  \varepsilon$, and
            since $ (  \mathbf{let} \, \mathit{y}  \ottsym{=}  \ottnt{e_{{\mathrm{1}}}} \, \mathbf{in} \, \ottnt{e_{{\mathrm{2}}}}  )  \,  \! [  \ottnt{v}  /  \mathit{x}  ]  = \mathbf{let} \, \mathit{y}  \ottsym{=}  \ottnt{e_{{\mathrm{1}}}} \,  \! [  \ottnt{v}  /  \mathit{x}  ]  \, \mathbf{in} \, \ottnt{e_{{\mathrm{2}}}} \,  \! [  \ottnt{v}  /  \mathit{x}  ] $,
            the required result is achieved.

            \item[\rname{T}{Sub}]
            For some $\ottnt{B'}$ and $\varepsilon'$, the following are given:
            \begin{itemize}
              \item $\Gamma_{{\mathrm{1}}}  \ottsym{,}  \mathit{x}  \ottsym{:}  \ottnt{A}  \ottsym{,}  \Gamma_{{\mathrm{2}}}  \vdash  \ottnt{e}  \ottsym{:}  \ottnt{B'}  \mid  \varepsilon'$ and
              \item $\Gamma_{{\mathrm{1}}}  \ottsym{,}  \mathit{x}  \ottsym{:}  \ottnt{A}  \ottsym{,}  \Gamma_{{\mathrm{2}}}  \vdash  \ottnt{B'}  \mid  \varepsilon'  <:  \ottnt{B}  \mid  \varepsilon$.
            \end{itemize}
            By the induction hypothesis and case \ref{lem:subst_value:subtyping_comp},
            we have $\Gamma_{{\mathrm{1}}}  \ottsym{,}  \Gamma_{{\mathrm{2}}}  \vdash  \ottnt{e} \,  \! [  \ottnt{v}  /  \mathit{x}  ]   \ottsym{:}  \ottnt{B'}  \mid  \varepsilon'$ and
            $\Gamma_{{\mathrm{1}}}  \ottsym{,}  \Gamma_{{\mathrm{2}}}  \vdash  \ottnt{B'}  \mid  \varepsilon'  <:  \ottnt{B}  \mid  \varepsilon$.
            Thus, \rname{T}{Sub} derives $\Gamma_{{\mathrm{1}}}  \ottsym{,}  \Gamma_{{\mathrm{2}}}  \vdash  \ottnt{e} \,  \! [  \ottnt{v}  /  \mathit{x}  ]   \ottsym{:}  \ottnt{B}  \mid  \varepsilon$.

            \item[\rname{T}{Op}]
            For some $ \mathsf{op} _{ \mathit{l} \,  \bm{ { S } } ^ {  \mathit{I}  }  }  \,  \bm{ { T } } ^ {  \mathit{J}  } $, $\ottnt{A'}$, and $\ottnt{B'}$, the following are given:
            \begin{itemize}
              \item $\ottnt{e} =  \mathsf{op} _{ \mathit{l} \,  \bm{ { S } } ^ {  \mathit{I}  }  }  \,  \bm{ { T } } ^ {  \mathit{J}  } $,
              \item $\ottnt{B} =  \ottsym{(}  \ottnt{A'} \,  \! [ {\bm{ { T } } }^{ \mathit{J} } / {\bm{ \beta } }^{ \mathit{J} } ]   \ottsym{)}    \rightarrow_{  \lift{ \mathit{l} \,  \bm{ { S } } ^ {  \mathit{I}  }  }  }    \ottsym{(}  \ottnt{B'} \,  \! [ {\bm{ { T } } }^{ \mathit{J} } / {\bm{ \beta } }^{ \mathit{J} } ]   \ottsym{)} $,
              \item $\varepsilon =  \bbZero $,
              \item $ \mathit{l}  ::    \forall    {\bm{ \alpha } }^{ \mathit{I} } : {\bm{ \ottnt{K} } }^{ \mathit{I} }    \ottsym{.}    \sigma    \in   \Xi $
              \item $ \mathsf{op}  \ottsym{:}    \forall    {\bm{ \beta } }^{ \mathit{J} } : {\bm{ \ottnt{K'} } }^{ \mathit{J} }    \ottsym{.}    \ottnt{A'}   \Rightarrow   \ottnt{B'}    \in   \sigma \,  \! [ {\bm{ { S } } }^{ \mathit{I} } / {\bm{ \alpha } }^{ \mathit{I} } ]  $,
              \item $\vdash  \Gamma_{{\mathrm{1}}}  \ottsym{,}  \mathit{x}  \ottsym{:}  \ottnt{A}  \ottsym{,}  \Gamma_{{\mathrm{2}}}$,
              \item $\Gamma_{{\mathrm{1}}}  \ottsym{,}  \mathit{x}  \ottsym{:}  \ottnt{A}  \ottsym{,}  \Gamma_{{\mathrm{2}}}  \vdash   \bm{ { S } }^{ \mathit{I} } : \bm{ \ottnt{K} }^{ \mathit{I} } $, and
              \item $\Gamma_{{\mathrm{1}}}  \ottsym{,}  \mathit{x}  \ottsym{:}  \ottnt{A}  \ottsym{,}  \Gamma_{{\mathrm{2}}}  \vdash   \bm{ { T } }^{ \mathit{J} } : \bm{ \ottnt{K'} }^{ \mathit{J} } $.
            \end{itemize}
            By cases~\ref{lem:subst_value:wf} and \ref{lem:subst_value:kinding}, we have
            \begin{itemize}
              \item $\vdash  \Gamma_{{\mathrm{1}}}  \ottsym{,}  \Gamma_{{\mathrm{2}}}$,
              \item $\Gamma_{{\mathrm{1}}}  \ottsym{,}  \Gamma_{{\mathrm{2}}}  \vdash   \bm{ { S } }^{ \mathit{I} } : \bm{ \ottnt{K} }^{ \mathit{I} } $, and
              \item $\Gamma_{{\mathrm{1}}}  \ottsym{,}  \Gamma_{{\mathrm{2}}}  \vdash   \bm{ { T } }^{ \mathit{J} } : \bm{ \ottnt{K'} }^{ \mathit{J} } $.
            \end{itemize}
            Thus, \rname{T}{Op} derives
            \begin{align*}
              \Gamma_{{\mathrm{1}}}  \ottsym{,}  \Gamma_{{\mathrm{2}}}  \vdash   \mathsf{op} _{ \mathit{l} \,  \bm{ { S } } ^ {  \mathit{I}  }  }  \,  \bm{ { T } } ^ {  \mathit{J}  }   \ottsym{:}   \ottsym{(}  \ottnt{A'} \,  \! [ {\bm{ { T } } }^{ \mathit{J} } / {\bm{ \beta } }^{ \mathit{J} } ]   \ottsym{)}    \rightarrow_{  \lift{ \mathit{l} \,  \bm{ { S } } ^ {  \mathit{I}  }  }  }    \ottsym{(}  \ottnt{B'} \,  \! [ {\bm{ { T } } }^{ \mathit{J} } / {\bm{ \beta } }^{ \mathit{J} } ]   \ottsym{)}   \mid   \bbZero .
            \end{align*}

            \item[\rname{T}{Handling}]
            For some $\mathit{N}$, $\ottnt{e'}$, $\ottnt{A'}$, $\varepsilon'$, $\mathit{l}$, $ \bm{ { S } } ^ {  \mathit{N}  } $, $ \bm{ { \alpha } } ^ {  \mathit{N}  } $, $ {\bm{ { \ottnt{K} } } }^{ \mathit{N} } $, $\ottnt{h}$, and $\sigma$,
            the following are given:
            \begin{itemize}
              \item $\ottnt{e} =  \mathbf{handle}_{ \mathit{l} \,  \bm{ { S } } ^ {  \mathit{N}  }  }  \, \ottnt{e'} \, \mathbf{with} \, \ottnt{h}$,
              \item $\Gamma_{{\mathrm{1}}}  \ottsym{,}  \mathit{x}  \ottsym{:}  \ottnt{A}  \ottsym{,}  \Gamma_{{\mathrm{2}}}  \vdash  \ottnt{e'}  \ottsym{:}  \ottnt{A'}  \mid  \varepsilon'$,
              \item $ \mathit{l}  ::    \forall    {\bm{ \alpha } }^{ \mathit{N} } : {\bm{ \ottnt{K} } }^{ \mathit{N} }    \ottsym{.}    \sigma    \in   \Xi $,
              \item $\Gamma_{{\mathrm{1}}}  \ottsym{,}  \mathit{x}  \ottsym{:}  \ottnt{A}  \ottsym{,}  \Gamma_{{\mathrm{2}}}  \vdash   \bm{ { S } }^{ \mathit{N} } : \bm{ \ottnt{K} }^{ \mathit{N} } $,
              \item $ \Gamma_{{\mathrm{1}}}  \ottsym{,}  \mathit{x}  \ottsym{:}  \ottnt{A}  \ottsym{,}  \Gamma_{{\mathrm{2}}}  \vdash _{ \sigma \,  \! [ {\bm{ { S } } }^{ \mathit{N} } / {\bm{ \alpha } }^{ \mathit{N} } ]  }  \ottnt{h}  :  \ottnt{A'}   \Rightarrow  ^ { \varepsilon }  \ottnt{B} $, and
              \item $   \lift{ \mathit{l} \,  \bm{ { S } } ^ {  \mathit{N}  }  }   \mathop{ \odot }  \varepsilon    \sim   \varepsilon' $.
            \end{itemize}
            By the induction hypothesis and case~\ref{lem:subst_value:kinding}, we have
            \begin{itemize}
              \item $\Gamma_{{\mathrm{1}}}  \ottsym{,}  \Gamma_{{\mathrm{2}}}  \vdash   \bm{ { S } }^{ \mathit{N} } : \bm{ \ottnt{K} }^{ \mathit{N} } $,
              \item $\Gamma_{{\mathrm{1}}}  \ottsym{,}  \Gamma_{{\mathrm{2}}}  \vdash  \ottnt{e'} \,  \! [  \ottnt{v}  /  \mathit{x}  ]   \ottsym{:}  \ottnt{A'}  \mid  \varepsilon'$, and
              \item $ \Gamma_{{\mathrm{1}}}  \ottsym{,}  \Gamma_{{\mathrm{2}}}  \vdash _{ \sigma \,  \! [ {\bm{ { S } } }^{ \mathit{N} } / {\bm{ \alpha } }^{ \mathit{N} } ]  }  \ottnt{h} \,  \! [  \ottnt{v}  /  \mathit{x}  ]   :  \ottnt{A'}   \Rightarrow  ^ { \varepsilon }  \ottnt{A} $.
            \end{itemize}
            Thus, \rname{T}{Handling} derives
            \begin{align*}
              \Gamma_{{\mathrm{1}}}  \ottsym{,}  \Gamma_{{\mathrm{2}}}  \vdash   \mathbf{handle}_{ \mathit{l} \,  \bm{ { S } } ^ {  \mathit{N}  }  }  \, \ottnt{e'} \,  \! [  \ottnt{v}  /  \mathit{x}  ]  \, \mathbf{with} \, \ottnt{h} \,  \! [  \ottnt{v}  /  \mathit{x}  ]   \ottsym{:}  \ottnt{B}  \mid  \varepsilon.
            \end{align*}

            \item[\rname{H}{Return}]
            Without loss of generality, we can choose $\mathit{y}$ such that
            $\mathit{y} \neq \mathit{x}$ and $ \mathit{y}   \notin    \mathrm{FV}   \ottsym{(}   \ottnt{v}   \ottsym{)}  $.
            For some $\ottnt{e_{\ottmv{r}}}$, the following are given:
            \begin{itemize}
              \item $\ottnt{h} = \ottsym{\{} \, \mathbf{return} \, \mathit{y}  \mapsto  \ottnt{e_{\ottmv{r}}}  \ottsym{\}}$,
              \item $\sigma =  \{\} $, and
              \item $\Gamma_{{\mathrm{1}}}  \ottsym{,}  \mathit{x}  \ottsym{:}  \ottnt{A}  \ottsym{,}  \Gamma_{{\mathrm{2}}}  \ottsym{,}  \mathit{y}  \ottsym{:}  \ottnt{B}  \vdash  \ottnt{e_{\ottmv{r}}}  \ottsym{:}  \ottnt{C}  \mid  \varepsilon$.
            \end{itemize}
            By the induction hypothesis, we have
            \begin{itemize}
              \item $\Gamma_{{\mathrm{1}}}  \ottsym{,}  \Gamma_{{\mathrm{2}}}  \ottsym{,}  \mathit{y}  \ottsym{:}  \ottnt{B}  \vdash  \ottnt{e_{\ottmv{r}}} \,  \! [  \ottnt{v}  /  \mathit{x}  ]   \ottsym{:}  \ottnt{C}  \mid  \varepsilon$.
            \end{itemize}
            Thus, \rname{H}{Return} derives
            \begin{align*}
               \Gamma_{{\mathrm{1}}}  \ottsym{,}  \Gamma_{{\mathrm{2}}}  \vdash _{  \{\}  }  \ottsym{\{} \, \mathbf{return} \, \mathit{y}  \mapsto  \ottnt{e_{\ottmv{r}}} \,  \! [  \ottnt{v}  /  \mathit{x}  ]   \ottsym{\}}  :  \ottnt{B}   \Rightarrow  ^ { \varepsilon }  \ottnt{C} .
            \end{align*}

            \item[\rname{H}{Op}]
            Without loss of generality, we can choose $ \bm{ { \beta } } ^ {  \mathit{J}  } $ and $\mathit{p}$ and $\mathit{k}$ such that:
            \begin{itemize}
              \item $\mathit{p} \neq \mathit{x}$,
              \item $\mathit{k} \neq \mathit{x}$,
              \item $ \mathit{p}   \notin    \mathrm{FV}   \ottsym{(}   \ottnt{v}   \ottsym{)}  $,
              \item $ \mathit{k}   \notin    \mathrm{FV}   \ottsym{(}   \ottnt{v}   \ottsym{)}  $, and
              \item $ \{   \bm{ { \beta } } ^ {  \mathit{J}  }   \}   \cap    \mathrm{FTV}   \ottsym{(}   \ottnt{v}   \ottsym{)}    \ottsym{=}  \emptyset$.
            \end{itemize}
            For some $\ottnt{h'}$, $\sigma'$, $\mathsf{op}$, $\ottnt{A'}$, $\ottnt{B'}$, and $\ottnt{e}$, the following are given:
            \begin{itemize}
              \item $\ottnt{h} =  \ottnt{h'}   \uplus   \ottsym{\{}  \mathsf{op} \,  {\bm{ \beta } }^{ \mathit{J} } : {\bm{ \ottnt{K} } }^{ \mathit{J} }  \, \mathit{p} \, \mathit{k}  \mapsto  \ottnt{e}  \ottsym{\}} $,
              \item $\sigma =  \sigma'   \uplus   \ottsym{\{}  \mathsf{op}  \ottsym{:}    \forall    {\bm{ \beta } }^{ \mathit{J} } : {\bm{ \ottnt{K} } }^{ \mathit{J} }    \ottsym{.}    \ottnt{A'}   \Rightarrow   \ottnt{B'}   \ottsym{\}} $,
              \item $ \Gamma_{{\mathrm{1}}}  \ottsym{,}  \mathit{x}  \ottsym{:}  \ottnt{A}  \ottsym{,}  \Gamma_{{\mathrm{2}}}  \vdash _{ \sigma' }  \ottnt{h'}  :  \ottnt{B}   \Rightarrow  ^ { \varepsilon }  \ottnt{C} $, and
              \item $\Gamma_{{\mathrm{1}}}  \ottsym{,}  \mathit{x}  \ottsym{:}  \ottnt{A}  \ottsym{,}  \Gamma_{{\mathrm{2}}}  \ottsym{,}   {\bm{ \beta } }^{ \mathit{J} } : {\bm{ \ottnt{K} } }^{ \mathit{J} }   \ottsym{,}  \mathit{p}  \ottsym{:}  \ottnt{A'}  \ottsym{,}  \mathit{k}  \ottsym{:}   \ottnt{B'}    \rightarrow_{ \varepsilon }    \ottnt{C}   \vdash  \ottnt{e}  \ottsym{:}  \ottnt{C}  \mid  \varepsilon$.
            \end{itemize}
            By the induction hypothesis, we have
            \begin{itemize}
              \item $ \Gamma_{{\mathrm{1}}}  \ottsym{,}  \Gamma_{{\mathrm{2}}}  \vdash _{ \sigma' }  \ottnt{h'} \,  \! [  \ottnt{v}  /  \mathit{x}  ]   :  \ottnt{A}   \Rightarrow  ^ { \varepsilon }  \ottnt{B} $ and
              \item $\Gamma_{{\mathrm{1}}}  \ottsym{,}  \Gamma_{{\mathrm{2}}}  \ottsym{,}   {\bm{ \beta } }^{ \mathit{J} } : {\bm{ \ottnt{K} } }^{ \mathit{J} }   \ottsym{,}  \mathit{p}  \ottsym{:}  \ottnt{A'}  \ottsym{,}  \mathit{k}  \ottsym{:}   \ottnt{B'}    \rightarrow_{ \varepsilon }    \ottnt{B}   \vdash  \ottnt{e} \,  \! [  \ottnt{v}  /  \mathit{x}  ]   \ottsym{:}  \ottnt{B}  \mid  \varepsilon$.
            \end{itemize}
            Thus, \rname{H}{Op} derives
            \begin{align*}
               \Gamma_{{\mathrm{1}}}  \ottsym{,}  \Gamma_{{\mathrm{2}}}  \vdash _{ \sigma }   \ottnt{h'} \,  \! [  \ottnt{v}  /  \mathit{x}  ]    \uplus   \ottsym{\{}  \mathsf{op} \,  {\bm{ \beta } }^{ \mathit{J} } : {\bm{ \ottnt{K} } }^{ \mathit{J} }  \, \mathit{p} \, \mathit{k}  \mapsto  \ottnt{e} \,  \! [  \ottnt{v}  /  \mathit{x}  ]   \ottsym{\}}   :  \ottnt{B}   \Rightarrow  ^ { \varepsilon }  \ottnt{C} 
            \end{align*}.
          \end{divcases}
  \end{itemize}
\end{proof}

\begin{lemma}[Well-formedness of contexts in subtyping judgments]
  \label{lem:ctx-wf-subtyping}
  \phantom{}
  \begin{itemize}
    \item If $\Gamma  \vdash  \ottnt{A_{{\mathrm{1}}}}  <:  \ottnt{A_{{\mathrm{2}}}}$, then $\vdash  \Gamma$.
    \item If $\Gamma  \vdash  \ottnt{A_{{\mathrm{1}}}}  \mid  \varepsilon_{{\mathrm{1}}}  <:  \ottnt{A_{{\mathrm{2}}}}  \mid  \varepsilon_{{\mathrm{2}}}$, then $\vdash  \Gamma$.
  \end{itemize}
\end{lemma}

\begin{proof}
  Straightforward by mutual induction on the subtyping derivations
  with Lemma~\ref{lem:wf}.
\end{proof}

\begin{lemma}[Well-formedness of contexts in typing judgments]
  \label{lem:ctx-wf-typing}
  \phantom{}
  \begin{itemize}
    \item If $\Gamma  \vdash  \ottnt{e}  \ottsym{:}  \ottnt{A}  \mid  \varepsilon$, then $\vdash  \Gamma$.
    \item If $ \Gamma  \vdash _{ \sigma }  \ottnt{h}  :  \ottnt{A}   \Rightarrow  ^ { \varepsilon }  \ottnt{B} $, then $\vdash  \Gamma$.
  \end{itemize}
\end{lemma}

\begin{proof}
  Straightforward by mutual induction on the derivations
  with Lemma~\ref{lem:wf}.
\end{proof}

\begin{lemma}[Substitution of Typelikes]\label{lem:subst_type}
  Suppose that $\Gamma_{{\mathrm{1}}}  \vdash   \bm{ { S } }^{ \mathit{I} } : \bm{ \ottnt{K} }^{ \mathit{I} } $.
  \begin{enumerate}
    \item\label{lem:subst_type:wf} If $\vdash  \Gamma_{{\mathrm{1}}}  \ottsym{,}   {\bm{ \alpha } }^{ \mathit{I} } : {\bm{ \ottnt{K} } }^{ \mathit{I} }   \ottsym{,}  \Gamma_{{\mathrm{2}}}$, then $\vdash  \Gamma_{{\mathrm{1}}}  \ottsym{,}  \Gamma_{{\mathrm{2}}} \,  \! [ {\bm{ { S } } }^{ \mathit{I} } / {\bm{ \alpha } }^{ \mathit{I} } ] $.
    \item\label{lem:subst_type:kinding} If $\Gamma_{{\mathrm{1}}}  \ottsym{,}   {\bm{ \alpha } }^{ \mathit{I} } : {\bm{ \ottnt{K} } }^{ \mathit{I} }   \ottsym{,}  \Gamma_{{\mathrm{2}}}  \vdash  T  \ottsym{:}  \ottnt{K}$, then $\Gamma_{{\mathrm{1}}}  \ottsym{,}  \Gamma_{{\mathrm{2}}} \,  \! [ {\bm{ { S } } }^{ \mathit{I} } / {\bm{ \alpha } }^{ \mathit{I} } ]   \vdash  T \,  \! [ {\bm{ { S } } }^{ \mathit{I} } / {\bm{ \alpha } }^{ \mathit{I} } ]   \ottsym{:}  \ottnt{K}$.
    \item\label{lem:subst_type:subtyping} If $\Gamma_{{\mathrm{1}}}  \ottsym{,}   {\bm{ \alpha } }^{ \mathit{I} } : {\bm{ \ottnt{K} } }^{ \mathit{I} }   \ottsym{,}  \Gamma_{{\mathrm{2}}}  \vdash  \ottnt{A}  <:  \ottnt{B}$, then $\Gamma_{{\mathrm{1}}}  \ottsym{,}  \Gamma_{{\mathrm{2}}} \,  \! [ {\bm{ { S } } }^{ \mathit{I} } / {\bm{ \alpha } }^{ \mathit{I} } ]   \vdash  \ottnt{A} \,  \! [ {\bm{ { S } } }^{ \mathit{I} } / {\bm{ \alpha } }^{ \mathit{I} } ]   <:  \ottnt{B} \,  \! [ {\bm{ { S } } }^{ \mathit{I} } / {\bm{ \alpha } }^{ \mathit{I} } ] $.
    \item\label{lem:subst_type:subtyping_comp} If $\Gamma_{{\mathrm{1}}}  \ottsym{,}   {\bm{ \alpha } }^{ \mathit{I} } : {\bm{ \ottnt{K} } }^{ \mathit{I} }   \ottsym{,}  \Gamma_{{\mathrm{2}}}  \vdash  \ottnt{A_{{\mathrm{1}}}}  \mid  \varepsilon_{{\mathrm{1}}}  <:  \ottnt{A_{{\mathrm{2}}}}  \mid  \varepsilon_{{\mathrm{2}}}$, then $\Gamma_{{\mathrm{1}}}  \ottsym{,}  \Gamma_{{\mathrm{2}}} \,  \! [ {\bm{ { S } } }^{ \mathit{I} } / {\bm{ \alpha } }^{ \mathit{I} } ]   \vdash  \ottnt{A_{{\mathrm{1}}}} \,  \! [ {\bm{ { S } } }^{ \mathit{I} } / {\bm{ \alpha } }^{ \mathit{I} } ]   \mid  \varepsilon_{{\mathrm{1}}} \,  \! [ {\bm{ { S } } }^{ \mathit{I} } / {\bm{ \alpha } }^{ \mathit{I} } ]   <:  \ottnt{A_{{\mathrm{2}}}} \,  \! [ {\bm{ { S } } }^{ \mathit{I} } / {\bm{ \alpha } }^{ \mathit{I} } ]   \mid  \varepsilon_{{\mathrm{2}}} \,  \! [ {\bm{ { S } } }^{ \mathit{I} } / {\bm{ \alpha } }^{ \mathit{I} } ] $.
    \item\label{lem:subst_type:typing} If $\Gamma_{{\mathrm{1}}}  \ottsym{,}   {\bm{ \alpha } }^{ \mathit{I} } : {\bm{ \ottnt{K} } }^{ \mathit{I} }   \ottsym{,}  \Gamma_{{\mathrm{2}}}  \vdash  \ottnt{e}  \ottsym{:}  \ottnt{A}  \mid  \varepsilon$, then $\Gamma_{{\mathrm{1}}}  \ottsym{,}  \Gamma_{{\mathrm{2}}} \,  \! [ {\bm{ { S } } }^{ \mathit{I} } / {\bm{ \alpha } }^{ \mathit{I} } ]   \vdash  \ottnt{e} \,  \! [ {\bm{ { S } } }^{ \mathit{I} } / {\bm{ \alpha } }^{ \mathit{I} } ]   \ottsym{:}  \ottnt{A} \,  \! [ {\bm{ { S } } }^{ \mathit{I} } / {\bm{ \alpha } }^{ \mathit{I} } ]   \mid  \varepsilon \,  \! [ {\bm{ { S } } }^{ \mathit{I} } / {\bm{ \alpha } }^{ \mathit{I} } ] $.
    \item\label{lem:subst_type:handling} If $ \Gamma_{{\mathrm{1}}}  \ottsym{,}   {\bm{ \alpha } }^{ \mathit{I} } : {\bm{ \ottnt{K} } }^{ \mathit{I} }   \ottsym{,}  \Gamma_{{\mathrm{2}}}  \vdash _{ \sigma }  \ottnt{h}  :  \ottnt{A}   \Rightarrow  ^ { \varepsilon }  \ottnt{B} $, then $ \Gamma_{{\mathrm{1}}}  \ottsym{,}  \Gamma_{{\mathrm{2}}} \,  \! [ {\bm{ { S } } }^{ \mathit{I} } / {\bm{ \alpha } }^{ \mathit{I} } ]   \vdash _{ \sigma \,  \! [ \bm{ { S } } / \bm{ \alpha } ]  }  \ottnt{h} \,  \! [ \bm{ { S } } / \bm{ \alpha } ]   :  \ottnt{A} \,  \! [ {\bm{ { S } } }^{ \mathit{I} } / {\bm{ \alpha } }^{ \mathit{I} } ]    \Rightarrow  ^ { \varepsilon \,  \! [ {\bm{ { S } } }^{ \mathit{I} } / {\bm{ \alpha } }^{ \mathit{I} } ]  }  \ottnt{B} \,  \! [ {\bm{ { S } } }^{ \mathit{I} } / {\bm{ \alpha } }^{ \mathit{I} } ]  $.
  \end{enumerate}
\end{lemma}

\begin{proof}
  \phantom{}
  \begin{itemize}
    \item[(1)(2)] By mutual induction on derivations of the judgments.  We proceed by case analysis on the rule applied lastly to the derivations.
          \begin{divcases}
            \item[\rname{C}{Empty}] Cannot happen.
            \item[\rname{C}{Var}] Since $\Gamma_{{\mathrm{2}}}$ cannot be $ \emptyset $, for some $\Gamma'_{{\mathrm{2}}}$, $\mathit{x}$, and $\ottnt{A}$, the following are given:
            \begin{itemize}
              \item $\Gamma_{{\mathrm{2}}} = \Gamma'_{{\mathrm{2}}}  \ottsym{,}  \mathit{x}  \ottsym{:}  \ottnt{A}$,
              \item $ \mathit{x}   \notin    \mathrm{dom}   \ottsym{(}   \Gamma_{{\mathrm{1}}}  \ottsym{,}   {\bm{ \alpha } }^{ \mathit{I} } : {\bm{ \ottnt{K} } }^{ \mathit{I} }   \ottsym{,}  \Gamma'_{{\mathrm{2}}}   \ottsym{)}  $, and
              \item $\Gamma_{{\mathrm{1}}}  \ottsym{,}   {\bm{ \alpha } }^{ \mathit{I} } : {\bm{ \ottnt{K} } }^{ \mathit{I} }   \ottsym{,}  \Gamma'_{{\mathrm{2}}}  \vdash  \ottnt{A}  \ottsym{:}   \mathbf{Typ} $.
            \end{itemize}
            By the induction hypothesis, we have $\Gamma_{{\mathrm{1}}}  \ottsym{,}  \ottsym{(}  \Gamma'_{{\mathrm{2}}} \,  \! [ {\bm{ { S } } }^{ \mathit{I} } / {\bm{ \alpha } }^{ \mathit{I} } ]   \ottsym{)}  \vdash  \ottnt{A} \,  \! [ {\bm{ { S } } }^{ \mathit{I} } / {\bm{ \alpha } }^{ \mathit{I} } ]   \ottsym{:}   \mathbf{Typ} $.
            By $ \mathit{x}   \notin    \mathrm{dom}   \ottsym{(}   \Gamma_{{\mathrm{1}}}  \ottsym{,}  \ottsym{(}  \Gamma'_{{\mathrm{2}}} \,  \! [ {\bm{ { S } } }^{ \mathit{I} } / {\bm{ \alpha } }^{ \mathit{I} } ]   \ottsym{)}   \ottsym{)}  $, \rname{C}{Var} derives $\vdash  \Gamma_{{\mathrm{1}}}  \ottsym{,}  \ottsym{(}  \Gamma'_{{\mathrm{2}}} \,  \! [ {\bm{ { S } } }^{ \mathit{I} } / {\bm{ \alpha } }^{ \mathit{I} } ]   \ottsym{)}  \ottsym{,}  \mathit{x}  \ottsym{:}  \ottnt{A} \,  \! [ {\bm{ { S } } }^{ \mathit{I} } / {\bm{ \alpha } }^{ \mathit{I} } ] $, and since $\Gamma_{{\mathrm{2}}} \,  \! [ {\bm{ { S } } }^{ \mathit{I} } / {\bm{ \alpha } }^{ \mathit{I} } ]  = \Gamma'_{{\mathrm{2}}} \,  \! [ {\bm{ { S } } }^{ \mathit{I} } / {\bm{ \alpha } }^{ \mathit{I} } ]   \ottsym{,}  \mathit{x}  \ottsym{:}  \ottnt{A} \,  \! [ {\bm{ { S } } }^{ \mathit{I} } / {\bm{ \alpha } }^{ \mathit{I} } ] $ holds, the required result is achieved.
            \item[\rname{C}{TVar}] If $\Gamma_{{\mathrm{2}}} =  \emptyset $, we have
            \begin{itemize}
              \item $ {\bm{ \alpha } }^{ \mathit{I} } : {\bm{ \ottnt{K} } }^{ \mathit{I} }  =  {\bm{ \alpha } }^{ \mathit{J} } : {\bm{ \ottnt{K} } }^{ \mathit{J} }   \ottsym{,}  \alpha_{\ottmv{i}}  \ottsym{:}  \ottnt{K_{\ottmv{i}}}$,
              \item $\vdash  \Gamma_{{\mathrm{1}}}  \ottsym{,}   {\bm{ \alpha } }^{ \mathit{J} } : {\bm{ \ottnt{K} } }^{ \mathit{J} } $, and
              \item $ \alpha_{\ottmv{i}}   \notin    \mathrm{dom}   \ottsym{(}   \Gamma_{{\mathrm{1}}}  \ottsym{,}   {\bm{ \alpha } }^{ \mathit{J} } : {\bm{ \ottnt{K} } }^{ \mathit{J} }    \ottsym{)}  $,
            \end{itemize}
            for some $\mathit{J}$, $ \bm{ { \alpha } } ^ {  \mathit{J}  } $, $ {\bm{ { \ottnt{K} } } }^{ \mathit{J} } $, $\ottmv{i}$, $\alpha_{\ottmv{i}}$, and $\ottnt{K_{\ottmv{i}}}$.
            By the induction hypothesis, we have $\vdash  \Gamma_{{\mathrm{1}}}$.

            If $\Gamma_{{\mathrm{2}}} \neq  \emptyset $, for some $\Gamma_{{\mathrm{2}}}$, $\beta$, and $\ottnt{K'}$, the following are given:
            \begin{itemize}
              \item $\Gamma_{{\mathrm{2}}} = \Gamma'_{{\mathrm{2}}}  \ottsym{,}  \beta  \ottsym{:}  \ottnt{K'}$,
              \item $\vdash  \Gamma_{{\mathrm{1}}}  \ottsym{,}   {\bm{ \alpha } }^{ \mathit{I} } : {\bm{ \ottnt{K} } }^{ \mathit{I} }   \ottsym{,}  \Gamma'_{{\mathrm{2}}}$, and
              \item $ \beta   \notin    \mathrm{dom}   \ottsym{(}   \Gamma_{{\mathrm{1}}}  \ottsym{,}   {\bm{ \alpha } }^{ \mathit{I} } : {\bm{ \ottnt{K} } }^{ \mathit{I} }   \ottsym{,}  \Gamma'_{{\mathrm{2}}}   \ottsym{)}  $.
            \end{itemize}
            By the induction hypothesis, we have $\vdash  \Gamma_{{\mathrm{1}}}  \ottsym{,}  \ottsym{(}  \Gamma'_{{\mathrm{2}}} \,  \! [ {\bm{ { S } } }^{ \mathit{I} } / {\bm{ \alpha } }^{ \mathit{I} } ]   \ottsym{)}$. Thus, \rname{C}{TVar} derives $\vdash  \Gamma_{{\mathrm{1}}}  \ottsym{,}  \ottsym{(}  \Gamma'_{{\mathrm{2}}} \,  \! [ {\bm{ { S } } }^{ \mathit{I} } / {\bm{ \alpha } }^{ \mathit{I} } ]   \ottsym{)}  \ottsym{,}  \beta  \ottsym{:}  \ottnt{K'}$, and since
            \begin{align*}
              \Gamma_{{\mathrm{1}}}  \ottsym{,}  \Gamma_{{\mathrm{2}}} \,  \! [ {\bm{ { S } } }^{ \mathit{I} } / {\bm{ \alpha } }^{ \mathit{I} } ]  = \Gamma_{{\mathrm{1}}}  \ottsym{,}  \ottsym{(}  \Gamma'_{{\mathrm{2}}} \,  \! [ {\bm{ { S } } }^{ \mathit{I} } / {\bm{ \alpha } }^{ \mathit{I} } ]   \ottsym{)}  \ottsym{,}  \beta  \ottsym{:}  \ottnt{K'}
            \end{align*}
            holds, the required result is achieved.

            \item[\rname{K}{Var}] For some $\beta$, the following are given:
            \begin{itemize}
              \item $T = \beta$,
              \item $\vdash  \Gamma_{{\mathrm{1}}}  \ottsym{,}   {\bm{ \alpha } }^{ \mathit{I} } : {\bm{ \ottnt{K} } }^{ \mathit{I} }   \ottsym{,}  \Gamma_{{\mathrm{2}}}$, and
              \item $ \beta   \ottsym{:}   \ottnt{K}   \in   \Gamma_{{\mathrm{1}}}  \ottsym{,}   {\bm{ \alpha } }^{ \mathit{I} } : {\bm{ \ottnt{K} } }^{ \mathit{I} }   \ottsym{,}  \Gamma_{{\mathrm{2}}} $.
            \end{itemize}
            By the induction hypothesis, we have $\vdash  \Gamma_{{\mathrm{1}}}  \ottsym{,}  \Gamma_{{\mathrm{2}}} \,  \! [ {\bm{ { S } } }^{ \mathit{I} } / {\bm{ \alpha } }^{ \mathit{I} } ] $.

            If $\beta = \alpha_{\ottmv{i}}$ for some $\ottmv{i} \in \mathit{I}$, then $\Gamma_{{\mathrm{1}}}  \ottsym{,}  \Gamma_{{\mathrm{2}}} \,  \! [ {\bm{ { S } } }^{ \mathit{I} } / {\bm{ \alpha } }^{ \mathit{I} } ]   \vdash  \beta \,  \! [ {\bm{ { S } } }^{ \mathit{I} } / {\bm{ \alpha } }^{ \mathit{I} } ]   \ottsym{:}  \ottnt{K}$ holds because of the following:
            \begin{itemize}
              \item $\Gamma_{{\mathrm{1}}}  \vdash   \bm{ { S } }^{ \mathit{I} } : \bm{ \ottnt{K} }^{ \mathit{I} } $,
              \item Lemma~\ref{lem:weakening}\ref{lem:weakening:kinding},
              \item $S_{\ottmv{i}} = \beta \,  \! [ {\bm{ { S } } }^{ \mathit{I} } / {\bm{ \alpha } }^{ \mathit{I} } ] $, and
              \item $\ottnt{K_{\ottmv{i}}} = \ottnt{K}$.
            \end{itemize}
            If $\beta \neq \alpha_{\ottmv{i}}$ for any $\ottmv{i} \in \mathit{I}$, then \rname{K}{Var} derives $\Gamma_{{\mathrm{1}}}  \ottsym{,}  \Gamma_{{\mathrm{2}}} \,  \! [ {\bm{ { S } } }^{ \mathit{I} } / {\bm{ \alpha } }^{ \mathit{I} } ]   \vdash  \beta  \ottsym{:}  \ottnt{K}$ because of $ \beta   \ottsym{:}   \ottnt{K}   \in   \Gamma_{{\mathrm{1}}}  \ottsym{,}  \Gamma_{{\mathrm{2}}} \,  \! [ {\bm{ { S } } }^{ \mathit{I} } / {\bm{ \alpha } }^{ \mathit{I} } ]  $.
            Since $\beta = \beta \,  \! [ {\bm{ { S } } }^{ \mathit{I} } / {\bm{ \alpha } }^{ \mathit{I} } ] $, the required result is achieved.
            \item[\rname{K}{Fun}] For some $\ottnt{A}$, $\ottnt{B}$, and $\varepsilon$, the following are given:
            \begin{itemize}
              \item $S =  \ottnt{A}    \rightarrow_{ \varepsilon }    \ottnt{B} $,
              \item $\ottnt{K} =  \mathbf{Typ} $,
              \item $\Gamma_{{\mathrm{1}}}  \ottsym{,}   {\bm{ \alpha } }^{ \mathit{I} } : {\bm{ \ottnt{K} } }^{ \mathit{I} }   \ottsym{,}  \Gamma_{{\mathrm{2}}}  \vdash  \ottnt{A}  \ottsym{:}   \mathbf{Typ} $,
              \item $\Gamma_{{\mathrm{1}}}  \ottsym{,}   {\bm{ \alpha } }^{ \mathit{I} } : {\bm{ \ottnt{K} } }^{ \mathit{I} }   \ottsym{,}  \Gamma_{{\mathrm{2}}}  \vdash  \varepsilon  \ottsym{:}   \mathbf{Eff} $, and
              \item $\Gamma_{{\mathrm{1}}}  \ottsym{,}   {\bm{ \alpha } }^{ \mathit{I} } : {\bm{ \ottnt{K} } }^{ \mathit{I} }   \ottsym{,}  \Gamma_{{\mathrm{2}}}  \vdash  \ottnt{B}  \ottsym{:}   \mathbf{Eff} $.
            \end{itemize}
            By the induction hypothesis, we have
            \begin{itemize}
              \item $\Gamma_{{\mathrm{1}}}  \ottsym{,}  \Gamma_{{\mathrm{2}}} \,  \! [ {\bm{ { S } } }^{ \mathit{I} } / {\bm{ \alpha } }^{ \mathit{I} } ]   \vdash  \ottnt{A} \,  \! [ {\bm{ { S } } }^{ \mathit{I} } / {\bm{ \alpha } }^{ \mathit{I} } ]   \ottsym{:}   \mathbf{Typ} $,
              \item $\Gamma_{{\mathrm{1}}}  \ottsym{,}  \Gamma_{{\mathrm{2}}} \,  \! [ {\bm{ { S } } }^{ \mathit{I} } / {\bm{ \alpha } }^{ \mathit{I} } ]   \vdash  \varepsilon \,  \! [ {\bm{ { S } } }^{ \mathit{I} } / {\bm{ \alpha } }^{ \mathit{I} } ]   \ottsym{:}   \mathbf{Eff} $, and
              \item $\Gamma_{{\mathrm{1}}}  \ottsym{,}  \Gamma_{{\mathrm{2}}} \,  \! [ {\bm{ { S } } }^{ \mathit{I} } / {\bm{ \alpha } }^{ \mathit{I} } ]   \vdash  \ottnt{B} \,  \! [ {\bm{ { S } } }^{ \mathit{I} } / {\bm{ \alpha } }^{ \mathit{I} } ]   \ottsym{:}   \mathbf{Eff} $.
            \end{itemize}
            Thus, \rname{K}{Fun} derives
            \begin{align*}
              \Gamma_{{\mathrm{1}}}  \ottsym{,}  \Gamma_{{\mathrm{2}}} \,  \! [ {\bm{ { S } } }^{ \mathit{I} } / {\bm{ \alpha } }^{ \mathit{I} } ]   \vdash   \ottsym{(}  \ottnt{A} \,  \! [ {\bm{ { S } } }^{ \mathit{I} } / {\bm{ \alpha } }^{ \mathit{I} } ]   \ottsym{)}    \rightarrow_{ \varepsilon \,  \! [ {\bm{ { S } } }^{ \mathit{I} } / {\bm{ \alpha } }^{ \mathit{I} } ]  }    \ottsym{(}  \ottnt{B} \,  \! [ {\bm{ { S } } }^{ \mathit{I} } / {\bm{ \alpha } }^{ \mathit{I} } ]   \ottsym{)}   \ottsym{:}   \mathbf{Typ} ,
            \end{align*}
            and since
            \begin{align*}
              \ottsym{(}   \ottnt{A}    \rightarrow_{ \varepsilon }    \ottnt{B}   \ottsym{)} \,  \! [ {\bm{ { S } } }^{ \mathit{I} } / {\bm{ \alpha } }^{ \mathit{I} } ]  =  \ottsym{(}  \ottnt{A} \,  \! [ {\bm{ { S } } }^{ \mathit{I} } / {\bm{ \alpha } }^{ \mathit{I} } ]   \ottsym{)}    \rightarrow_{ \varepsilon \,  \! [ {\bm{ { S } } }^{ \mathit{I} } / {\bm{ \alpha } }^{ \mathit{I} } ]  }    \ottsym{(}  \ottnt{B} \,  \! [ {\bm{ { S } } }^{ \mathit{I} } / {\bm{ \alpha } }^{ \mathit{I} } ]   \ottsym{)} 
            \end{align*}
            holds, the required result is achieved.
            \item[\rname{K}{Poly}] For some $\beta$, $\ottnt{K'}$, $\ottnt{A}$, and $\varepsilon$, the following are given:
            \begin{itemize}
              \item $S =   \forall   \beta  \ottsym{:}  \ottnt{K'}   \ottsym{.}    \ottnt{A}    ^{ \varepsilon }  $,
              \item $\ottnt{K} =  \mathbf{Typ} $,
              \item $\Gamma_{{\mathrm{1}}}  \ottsym{,}   {\bm{ \alpha } }^{ \mathit{I} } : {\bm{ \ottnt{K} } }^{ \mathit{I} }   \ottsym{,}  \Gamma_{{\mathrm{2}}}  \ottsym{,}  \beta  \ottsym{:}  \ottnt{K'}  \vdash  \ottnt{A}  \ottsym{:}   \mathbf{Typ} $, and
              \item $\Gamma_{{\mathrm{1}}}  \ottsym{,}   {\bm{ \alpha } }^{ \mathit{I} } : {\bm{ \ottnt{K} } }^{ \mathit{I} }   \ottsym{,}  \Gamma_{{\mathrm{2}}}  \ottsym{,}  \beta  \ottsym{:}  \ottnt{K'}  \vdash  \varepsilon  \ottsym{:}   \mathbf{Eff} $.
            \end{itemize}
            By the induction hypothesis, we have
            \begin{itemize}
              \item $\Gamma_{{\mathrm{1}}}  \ottsym{,}  \Gamma_{{\mathrm{2}}} \,  \! [ {\bm{ { S } } }^{ \mathit{I} } / {\bm{ \alpha } }^{ \mathit{I} } ]   \ottsym{,}  \beta  \ottsym{:}  \ottnt{K'}  \vdash  \ottnt{A} \,  \! [ {\bm{ { S } } }^{ \mathit{I} } / {\bm{ \alpha } }^{ \mathit{I} } ]   \ottsym{:}   \mathbf{Typ} $ and
              \item $\Gamma_{{\mathrm{1}}}  \ottsym{,}  \Gamma_{{\mathrm{2}}} \,  \! [ {\bm{ { S } } }^{ \mathit{I} } / {\bm{ \alpha } }^{ \mathit{I} } ]   \ottsym{,}  \beta  \ottsym{:}  \ottnt{K'}  \vdash  \varepsilon \,  \! [ {\bm{ { S } } }^{ \mathit{I} } / {\bm{ \alpha } }^{ \mathit{I} } ]   \ottsym{:}   \mathbf{Eff} $.
            \end{itemize}
            Thus, \rname{K}{Poly} derives
            \begin{align*}
              \Gamma_{{\mathrm{1}}}  \ottsym{,}  \Gamma_{{\mathrm{2}}} \,  \! [ {\bm{ { S } } }^{ \mathit{I} } / {\bm{ \alpha } }^{ \mathit{I} } ]   \vdash    \forall   \beta  \ottsym{:}  \ottnt{K'}   \ottsym{.}    \ottnt{A} \,  \! [ {\bm{ { S } } }^{ \mathit{I} } / {\bm{ \alpha } }^{ \mathit{I} } ]     ^{ \ottsym{(}  \varepsilon \,  \! [ {\bm{ { S } } }^{ \mathit{I} } / {\bm{ \alpha } }^{ \mathit{I} } ]   \ottsym{)} }    \ottsym{:}   \mathbf{Typ}  ~.
            \end{align*}
            Since
            we can assume that $\beta$ does not occur in $ \bm{ { S } } ^ {  \mathit{I}  } $ and $ \bm{ { \alpha } } ^ {  \mathit{I}  } $ without loss of generality,
            we have
            \begin{align*}
              \ottsym{(}    \forall   \beta  \ottsym{:}  \ottnt{K'}   \ottsym{.}    \ottnt{A}    ^{ \varepsilon }    \ottsym{)} \,  \! [ {\bm{ { S } } }^{ \mathit{I} } / {\bm{ \alpha } }^{ \mathit{I} } ]  =   \forall   \beta  \ottsym{:}  \ottnt{K'}   \ottsym{.}    \ottnt{A} \,  \! [ {\bm{ { S } } }^{ \mathit{I} } / {\bm{ \alpha } }^{ \mathit{I} } ]     ^{ \ottsym{(}  \varepsilon \,  \! [ {\bm{ { S } } }^{ \mathit{I} } / {\bm{ \alpha } }^{ \mathit{I} } ]   \ottsym{)} }   ~.
            \end{align*}
            Therefore, the required result is achieved.

            \item[\rname{K}{Cons}] For some $\mathcal{C}$, $ \bm{ { S' } } ^ {  \mathit{J}  } $, and $ {\bm{ { \ottnt{K'} } } }^{ \mathit{J} } $, the following are given:
            \begin{itemize}
              \item $S = \mathcal{C} \,  \bm{ { S' } } ^ {  \mathit{J}  } $,
              \item $ \mathcal{C}   \ottsym{:}    \Pi {\bm{ { \ottnt{K'} } } }^{ \mathit{J} }   \rightarrow  \ottnt{K}   \in   \Sigma $,
              \item $\vdash  \Gamma_{{\mathrm{1}}}  \ottsym{,}   {\bm{ \alpha } }^{ \mathit{I} } : {\bm{ \ottnt{K} } }^{ \mathit{I} }   \ottsym{,}  \Gamma_{{\mathrm{2}}}$, and
              \item $\Gamma_{{\mathrm{1}}}  \ottsym{,}   {\bm{ \alpha } }^{ \mathit{I} } : {\bm{ \ottnt{K} } }^{ \mathit{I} }   \ottsym{,}  \Gamma_{{\mathrm{2}}}  \vdash   \bm{ { S' } }^{ \mathit{J} } : \bm{ \ottnt{K'} }^{ \mathit{J} } $
            \end{itemize}
            By the induction hypothesis, we have $\vdash  \Gamma_{{\mathrm{1}}}  \ottsym{,}  \Gamma_{{\mathrm{2}}} \,  \! [ {\bm{ { S } } }^{ \mathit{I} } / {\bm{ \alpha } }^{ \mathit{I} } ] $ and $\Gamma_{{\mathrm{1}}}  \ottsym{,}  \Gamma_{{\mathrm{2}}} \,  \! [ {\bm{ { S } } }^{ \mathit{I} } / {\bm{ \alpha } }^{ \mathit{I} } ]   \vdash   \bm{ { S' \,  \! [ {\bm{ { S } } }^{ \mathit{I} } / {\bm{ \alpha } }^{ \mathit{I} } ]  } }^{ \mathit{J} } : \bm{ \ottnt{K'} }^{ \mathit{J} } $. Thus, \rname{K}{Cons} derives $\Gamma_{{\mathrm{1}}}  \ottsym{,}  \Gamma_{{\mathrm{2}}} \,  \! [ {\bm{ { S } } }^{ \mathit{I} } / {\bm{ \alpha } }^{ \mathit{I} } ]   \vdash  \mathcal{C} \,  \bm{ { S' \,  \! [ {\bm{ { S } } }^{ \mathit{I} } / {\bm{ \alpha } }^{ \mathit{I} } ]  } } ^ {  \mathit{J}  }   \ottsym{:}  \ottnt{K}$.
          \end{divcases}

    \item[(3)(4)] By mutual induction on derivations of the judgments.
          We proceed by case analysis on the rule applied lastly to the derivation.
          \begin{divcases}
            \item[\rname{ST}{Refl}]
            $\ottnt{A} = \ottnt{B}$ and $\Gamma_{{\mathrm{1}}}  \ottsym{,}   {\bm{ \alpha } }^{ \mathit{I} } : {\bm{ \ottnt{K} } }^{ \mathit{I} }   \ottsym{,}  \Gamma_{{\mathrm{2}}}  \vdash  \ottnt{A}  \ottsym{:}   \mathbf{Typ} $ are given.
            By case \ref{lem:subst_type:kinding}, we have
            $\Gamma_{{\mathrm{1}}}  \ottsym{,}  \Gamma_{{\mathrm{2}}} \,  \! [ {\bm{ { S } } }^{ \mathit{I} } / {\bm{ \alpha } }^{ \mathit{I} } ]   \vdash  \ottnt{A} \,  \! [ {\bm{ { S } } }^{ \mathit{I} } / {\bm{ \alpha } }^{ \mathit{I} } ]   \ottsym{:}   \mathbf{Typ} $.
            Thus, \rname{ST}{Refl} derives
            \begin{align*}
              \Gamma_{{\mathrm{1}}}  \ottsym{,}  \Gamma_{{\mathrm{2}}} \,  \! [ {\bm{ { S } } }^{ \mathit{I} } / {\bm{ \alpha } }^{ \mathit{I} } ]   \vdash  \ottnt{A} \,  \! [ {\bm{ { S } } }^{ \mathit{I} } / {\bm{ \alpha } }^{ \mathit{I} } ]   <:  \ottnt{A} \,  \! [ {\bm{ { S } } }^{ \mathit{I} } / {\bm{ \alpha } }^{ \mathit{I} } ] .
            \end{align*}

            \item[\rname{ST}{Fun}]
            For some $\ottnt{A_{{\mathrm{11}}}}$, $\varepsilon_{{\mathrm{1}}}$, $\ottnt{A_{{\mathrm{12}}}}$, $\ottnt{A_{{\mathrm{21}}}}$, $\varepsilon_{{\mathrm{2}}}$, $\ottnt{B_{{\mathrm{22}}}}$,
            the following are given:
            \begin{itemize}
              \item $\ottnt{A} =  \ottnt{A_{{\mathrm{11}}}}    \rightarrow_{ \varepsilon_{{\mathrm{1}}} }    \ottnt{A_{{\mathrm{12}}}} $,
              \item $\ottnt{B} =  \ottnt{A_{{\mathrm{21}}}}    \rightarrow_{ \varepsilon_{{\mathrm{2}}} }    \ottnt{A_{{\mathrm{22}}}} $,
              \item $\Gamma_{{\mathrm{1}}}  \ottsym{,}   {\bm{ \alpha } }^{ \mathit{I} } : {\bm{ \ottnt{K} } }^{ \mathit{I} }   \ottsym{,}  \Gamma_{{\mathrm{2}}}  \vdash  \ottnt{A_{{\mathrm{21}}}}  <:  \ottnt{A_{{\mathrm{11}}}}$, and
              \item $\Gamma_{{\mathrm{1}}}  \ottsym{,}   {\bm{ \alpha } }^{ \mathit{I} } : {\bm{ \ottnt{K} } }^{ \mathit{I} }   \ottsym{,}  \Gamma_{{\mathrm{2}}}  \vdash  \ottnt{A_{{\mathrm{12}}}}  \mid  \varepsilon_{{\mathrm{1}}}  <:  \ottnt{A_{{\mathrm{22}}}}  \mid  \varepsilon_{{\mathrm{2}}}$.
            \end{itemize}
            By the induction hypothesis, we have
            \begin{itemize}
              \item $\Gamma_{{\mathrm{1}}}  \ottsym{,}  \Gamma_{{\mathrm{2}}} \,  \! [ {\bm{ { S } } }^{ \mathit{I} } / {\bm{ \alpha } }^{ \mathit{I} } ]   \vdash  \ottnt{A_{{\mathrm{21}}}} \,  \! [ {\bm{ { S } } }^{ \mathit{I} } / {\bm{ \alpha } }^{ \mathit{I} } ]   <:  \ottnt{A_{{\mathrm{11}}}} \,  \! [ {\bm{ { S } } }^{ \mathit{I} } / {\bm{ \alpha } }^{ \mathit{I} } ] $ and
              \item $\Gamma_{{\mathrm{1}}}  \ottsym{,}  \Gamma_{{\mathrm{2}}} \,  \! [ {\bm{ { S } } }^{ \mathit{I} } / {\bm{ \alpha } }^{ \mathit{I} } ]   \vdash  \ottnt{A_{{\mathrm{12}}}} \,  \! [ {\bm{ { S } } }^{ \mathit{I} } / {\bm{ \alpha } }^{ \mathit{I} } ]   \mid  \varepsilon_{{\mathrm{1}}} \,  \! [ {\bm{ { S } } }^{ \mathit{I} } / {\bm{ \alpha } }^{ \mathit{I} } ]   <:  \ottnt{A_{{\mathrm{22}}}} \,  \! [ {\bm{ { S } } }^{ \mathit{I} } / {\bm{ \alpha } }^{ \mathit{I} } ]   \mid  \varepsilon_{{\mathrm{2}}} \,  \! [ {\bm{ { S } } }^{ \mathit{I} } / {\bm{ \alpha } }^{ \mathit{I} } ] $.
            \end{itemize}
            Thus, \rname{ST}{Fun} drives
            \begin{align*}
              \Gamma_{{\mathrm{1}}}  \ottsym{,}  \Gamma_{{\mathrm{2}}} \,  \! [ {\bm{ { S } } }^{ \mathit{I} } / {\bm{ \alpha } }^{ \mathit{I} } ]   \vdash   \ottsym{(}  \ottnt{A_{{\mathrm{11}}}} \,  \! [ {\bm{ { S } } }^{ \mathit{I} } / {\bm{ \alpha } }^{ \mathit{I} } ]   \ottsym{)}    \rightarrow_{ \varepsilon_{{\mathrm{1}}} \,  \! [ {\bm{ { S } } }^{ \mathit{I} } / {\bm{ \alpha } }^{ \mathit{I} } ]  }    \ottsym{(}  \ottnt{A_{{\mathrm{12}}}} \,  \! [ {\bm{ { S } } }^{ \mathit{I} } / {\bm{ \alpha } }^{ \mathit{I} } ]   \ottsym{)}   <:   \ottsym{(}  \ottnt{A_{{\mathrm{21}}}} \,  \! [ {\bm{ { S } } }^{ \mathit{I} } / {\bm{ \alpha } }^{ \mathit{I} } ]   \ottsym{)}    \rightarrow_{ \varepsilon_{{\mathrm{2}}} \,  \! [ {\bm{ { S } } }^{ \mathit{I} } / {\bm{ \alpha } }^{ \mathit{I} } ]  }    \ottsym{(}  \ottnt{A_{{\mathrm{22}}}} \,  \! [ {\bm{ { S } } }^{ \mathit{I} } / {\bm{ \alpha } }^{ \mathit{I} } ]   \ottsym{)} 
            \end{align*}
            and since, for any $\ottmv{i} \in \{1,2\}$,
            \begin{align*}
              \ottsym{(}   \ottnt{A_{\ottmv{i}\,{\mathrm{1}}}}    \rightarrow_{ \varepsilon_{\ottmv{i}} }    \ottnt{A_{\ottmv{i}\,{\mathrm{2}}}}   \ottsym{)} \,  \! [ {\bm{ { S } } }^{ \mathit{I} } / {\bm{ \alpha } }^{ \mathit{I} } ]  =  \ottsym{(}  \ottnt{A_{\ottmv{i}\,{\mathrm{1}}}} \,  \! [ {\bm{ { S } } }^{ \mathit{I} } / {\bm{ \alpha } }^{ \mathit{I} } ]   \ottsym{)}    \rightarrow_{ \varepsilon_{\ottmv{i}} \,  \! [ {\bm{ { S } } }^{ \mathit{I} } / {\bm{ \alpha } }^{ \mathit{I} } ]  }    \ottsym{(}  \ottnt{A_{\ottmv{i}\,{\mathrm{2}}}} \,  \! [ {\bm{ { S } } }^{ \mathit{I} } / {\bm{ \alpha } }^{ \mathit{I} } ]   \ottsym{)} 
            \end{align*}
            holds, the required result is achieved.

            \item[\rname{ST}{Poly}]
            For some $\beta$, $\ottnt{K}$, $\ottnt{A_{{\mathrm{1}}}}$, $\varepsilon_{{\mathrm{1}}}$, $\ottnt{A_{{\mathrm{2}}}}$, and $\varepsilon_{{\mathrm{2}}}$,
            the following are given:
            \begin{itemize}
              \item $\ottnt{A} =   \forall   \beta  \ottsym{:}  \ottnt{K}   \ottsym{.}    \ottnt{A_{{\mathrm{1}}}}    ^{ \varepsilon_{{\mathrm{1}}} }  $,
              \item $\ottnt{B} =   \forall   \beta  \ottsym{:}  \ottnt{K}   \ottsym{.}    \ottnt{A_{{\mathrm{2}}}}    ^{ \varepsilon_{{\mathrm{2}}} }  $, and
              \item $\Gamma_{{\mathrm{1}}}  \ottsym{,}   {\bm{ \alpha } }^{ \mathit{I} } : {\bm{ \ottnt{K} } }^{ \mathit{I} }   \ottsym{,}  \Gamma_{{\mathrm{2}}}  \ottsym{,}  \beta  \ottsym{:}  \ottnt{K}  \vdash  \ottnt{A_{{\mathrm{1}}}}  \mid  \varepsilon_{{\mathrm{1}}}  <:  \ottnt{A_{{\mathrm{2}}}}  \mid  \varepsilon$.
            \end{itemize}
            By the induction hypothesis, we have
            $\Gamma_{{\mathrm{1}}}  \ottsym{,}  \Gamma_{{\mathrm{2}}} \,  \! [ {\bm{ { S } } }^{ \mathit{I} } / {\bm{ \alpha } }^{ \mathit{I} } ]   \ottsym{,}  \beta  \ottsym{:}  \ottnt{K}  \vdash  \ottnt{A_{{\mathrm{1}}}} \,  \! [ {\bm{ { S } } }^{ \mathit{I} } / {\bm{ \alpha } }^{ \mathit{I} } ]   \mid  \varepsilon_{{\mathrm{1}}} \,  \! [ {\bm{ { S } } }^{ \mathit{I} } / {\bm{ \alpha } }^{ \mathit{I} } ]   <:  \ottnt{A_{{\mathrm{2}}}} \,  \! [ {\bm{ { S } } }^{ \mathit{I} } / {\bm{ \alpha } }^{ \mathit{I} } ]   \mid  \varepsilon_{{\mathrm{2}}} \,  \! [ {\bm{ { S } } }^{ \mathit{I} } / {\bm{ \alpha } }^{ \mathit{I} } ] $.
            Thus, \rname{ST}{Poly} derives
            \begin{align*}
              \Gamma_{{\mathrm{1}}}  \ottsym{,}  \Gamma_{{\mathrm{2}}} \,  \! [ {\bm{ { S } } }^{ \mathit{I} } / {\bm{ \alpha } }^{ \mathit{I} } ]   \vdash    \forall   \beta  \ottsym{:}  \ottnt{K}   \ottsym{.}    \ottnt{A_{{\mathrm{1}}}} \,  \! [ {\bm{ { S } } }^{ \mathit{I} } / {\bm{ \alpha } }^{ \mathit{I} } ]     ^{ \ottsym{(}  \varepsilon_{{\mathrm{1}}} \,  \! [ {\bm{ { S } } }^{ \mathit{I} } / {\bm{ \alpha } }^{ \mathit{I} } ]   \ottsym{)} }    <:    \forall   \beta  \ottsym{:}  \ottnt{K}   \ottsym{.}    \ottnt{A_{{\mathrm{2}}}} \,  \! [ {\bm{ { S } } }^{ \mathit{I} } / {\bm{ \alpha } }^{ \mathit{I} } ]     ^{ \ottsym{(}  \varepsilon_{{\mathrm{2}}} \,  \! [ {\bm{ { S } } }^{ \mathit{I} } / {\bm{ \alpha } }^{ \mathit{I} } ]   \ottsym{)} }  
            \end{align*}
            and since
            \begin{itemize}
              \item $\ottsym{(}    \forall   \beta  \ottsym{:}  \ottnt{K}   \ottsym{.}    \ottnt{A_{{\mathrm{1}}}}    ^{ \varepsilon_{{\mathrm{1}}} }    \ottsym{)} \,  \! [ {\bm{ { S } } }^{ \mathit{I} } / {\bm{ \alpha } }^{ \mathit{I} } ]  =   \forall   \beta  \ottsym{:}  \ottnt{K}   \ottsym{.}    \ottnt{A_{{\mathrm{1}}}} \,  \! [ {\bm{ { S } } }^{ \mathit{I} } / {\bm{ \alpha } }^{ \mathit{I} } ]     ^{ \ottsym{(}  \varepsilon_{{\mathrm{1}}} \,  \! [ {\bm{ { S } } }^{ \mathit{I} } / {\bm{ \alpha } }^{ \mathit{I} } ]   \ottsym{)} }  $ and
              \item $\ottsym{(}    \forall   \beta  \ottsym{:}  \ottnt{K}   \ottsym{.}    \ottnt{A_{{\mathrm{2}}}}    ^{ \varepsilon_{{\mathrm{2}}} }    \ottsym{)} \,  \! [ {\bm{ { S } } }^{ \mathit{I} } / {\bm{ \alpha } }^{ \mathit{I} } ]  =   \forall   \beta  \ottsym{:}  \ottnt{K}   \ottsym{.}    \ottnt{A_{{\mathrm{2}}}} \,  \! [ {\bm{ { S } } }^{ \mathit{I} } / {\bm{ \alpha } }^{ \mathit{I} } ]     ^{ \ottsym{(}  \varepsilon_{{\mathrm{2}}} \,  \! [ {\bm{ { S } } }^{ \mathit{I} } / {\bm{ \alpha } }^{ \mathit{I} } ]   \ottsym{)} }  $
            \end{itemize}
            hold, the required result is achieved.

            \item[\rname{ST}{Comp}]
            We have $\Gamma_{{\mathrm{1}}}  \ottsym{,}   {\bm{ \alpha } }^{ \mathit{I} } : {\bm{ \ottnt{K} } }^{ \mathit{I} }   \ottsym{,}  \Gamma_{{\mathrm{2}}}  \vdash  \ottnt{A_{{\mathrm{1}}}}  <:  \ottnt{A_{{\mathrm{2}}}}$ and
            $\Gamma_{{\mathrm{1}}}  \ottsym{,}   {\bm{ \alpha } }^{ \mathit{I} } : {\bm{ \ottnt{K} } }^{ \mathit{I} }   \ottsym{,}  \Gamma_{{\mathrm{2}}}  \vdash   \varepsilon_{{\mathrm{1}}}  \olessthan  \varepsilon_{{\mathrm{2}}} $.
            By the induction hypothesis, we have
            $\Gamma_{{\mathrm{1}}}  \ottsym{,}  \Gamma_{{\mathrm{2}}} \,  \! [ {\bm{ { S } } }^{ \mathit{I} } / {\bm{ \alpha } }^{ \mathit{I} } ]   \vdash  \ottnt{A_{{\mathrm{1}}}} \,  \! [ {\bm{ { S } } }^{ \mathit{I} } / {\bm{ \alpha } }^{ \mathit{I} } ]   <:  \ottnt{A_{{\mathrm{2}}}} \,  \! [ {\bm{ { S } } }^{ \mathit{I} } / {\bm{ \alpha } }^{ \mathit{I} } ] $.
            By Lemma~\ref{lem:ctx-wf-subtyping}, $\vdash  \Gamma_{{\mathrm{1}}}  \ottsym{,}   {\bm{ \alpha } }^{ \mathit{I} } : {\bm{ \ottnt{K} } }^{ \mathit{I} }   \ottsym{,}  \Gamma_{{\mathrm{2}}}$.
            Then, by case~\ref{lem:subst_type:kinding} and
            the fact that a typelike substitution is homomorphism for $ \odot $ and $ \sim $,
            we have
            $\Gamma_{{\mathrm{1}}}  \ottsym{,}  \Gamma_{{\mathrm{2}}} \,  \! [ {\bm{ { S } } }^{ \mathit{I} } / {\bm{ \alpha } }^{ \mathit{I} } ]   \vdash   \ottsym{(}  \varepsilon_{{\mathrm{1}}} \,  \! [ {\bm{ { S } } }^{ \mathit{I} } / {\bm{ \alpha } }^{ \mathit{I} } ]   \ottsym{)}  \olessthan  \ottsym{(}  \varepsilon_{{\mathrm{2}}} \,  \! [ {\bm{ { S } } }^{ \mathit{I} } / {\bm{ \alpha } }^{ \mathit{I} } ]   \ottsym{)} $.
            Thus, \rname{ST}{Comp} derives
            $\Gamma_{{\mathrm{1}}}  \ottsym{,}  \Gamma_{{\mathrm{2}}} \,  \! [ {\bm{ { S } } }^{ \mathit{I} } / {\bm{ \alpha } }^{ \mathit{I} } ]   \vdash  \ottnt{A_{{\mathrm{1}}}} \,  \! [ {\bm{ { S } } }^{ \mathit{I} } / {\bm{ \alpha } }^{ \mathit{I} } ]   \mid  \varepsilon_{{\mathrm{1}}} \,  \! [ {\bm{ { S } } }^{ \mathit{I} } / {\bm{ \alpha } }^{ \mathit{I} } ]   <:  \ottnt{A_{{\mathrm{2}}}} \,  \! [ {\bm{ { S } } }^{ \mathit{I} } / {\bm{ \alpha } }^{ \mathit{I} } ]   \mid  \varepsilon_{{\mathrm{2}}} \,  \! [ {\bm{ { S } } }^{ \mathit{I} } / {\bm{ \alpha } }^{ \mathit{I} } ] $.
          \end{divcases}

    \item[(5)(6)]
          By mutual induction on derivations of the judgments.
          We proceed by case analysis on the rule applied lastly to the derivation.
          \begin{divcases}
            \item[\rname{T}{Var}]
            For some $\mathit{x}$, the following are given:
            \begin{itemize}
              \item $\ottnt{e} = \mathit{x}$,
              \item $\varepsilon =  \bbZero $,
              \item $\vdash  \Gamma_{{\mathrm{1}}}  \ottsym{,}   {\bm{ \alpha } }^{ \mathit{I} } : {\bm{ \ottnt{K} } }^{ \mathit{I} }   \ottsym{,}  \Gamma_{{\mathrm{2}}}$, and
              \item $ \mathit{x}   \ottsym{:}   \ottnt{A}   \in   \Gamma_{{\mathrm{1}}}  \ottsym{,}   {\bm{ \alpha } }^{ \mathit{I} } : {\bm{ \ottnt{K} } }^{ \mathit{I} }   \ottsym{,}  \Gamma_{{\mathrm{2}}} $.
            \end{itemize}
            By case~\ref{lem:subst_type:wf}, we have
            $\vdash  \Gamma_{{\mathrm{1}}}  \ottsym{,}  \Gamma_{{\mathrm{2}}} \,  \! [ {\bm{ { S } } }^{ \mathit{I} } / {\bm{ \alpha } }^{ \mathit{I} } ] $.
            \begin{divcases}
              \item[$ \mathit{x}   \ottsym{:}   \ottnt{A}   \in   \Gamma_{{\mathrm{1}}} $]
              Since $ \mathit{x}   \ottsym{:}   \ottnt{A}   \in   \Gamma_{{\mathrm{1}}}  \ottsym{,}  \Gamma_{{\mathrm{2}}} \,  \! [ {\bm{ { S } } }^{ \mathit{I} } / {\bm{ \alpha } }^{ \mathit{I} } ]  $ and
              $\ottnt{A} \,  \! [ {\bm{ { S } } }^{ \mathit{I} } / {\bm{ \alpha } }^{ \mathit{I} } ]  = \ottnt{A}$ hold,
              \rname{T}{Var} derives
              \begin{align*}
                \Gamma_{{\mathrm{1}}}  \ottsym{,}  \Gamma_{{\mathrm{2}}} \,  \! [ {\bm{ { S } } }^{ \mathit{I} } / {\bm{ \alpha } }^{ \mathit{I} } ]   \vdash  \mathit{x}  \ottsym{:}  \ottnt{A} \,  \! [ {\bm{ { S } } }^{ \mathit{I} } / {\bm{ \alpha } }^{ \mathit{I} } ]   \mid   \bbZero .
              \end{align*}

              \item[$ \mathit{x}   \ottsym{:}   \ottnt{A}   \in    {\bm{ \alpha } }^{ \mathit{I} } : {\bm{ \ottnt{K} } }^{ \mathit{I} }  $] Cannot happen.

              \item[$ \mathit{x}   \ottsym{:}   \ottnt{A}   \in   \Gamma_{{\mathrm{2}}} $]
              Since $ \mathit{x}   \ottsym{:}   \ottnt{A} \,  \! [ {\bm{ { S } } }^{ \mathit{I} } / {\bm{ \alpha } }^{ \mathit{I} } ]    \in   \Gamma_{{\mathrm{1}}}  \ottsym{,}  \Gamma_{{\mathrm{2}}} \,  \! [ {\bm{ { S } } }^{ \mathit{I} } / {\bm{ \alpha } }^{ \mathit{I} } ]  $ holds,
              \rname{T}{Var} derives
              \begin{align*}
                \Gamma_{{\mathrm{1}}}  \ottsym{,}  \Gamma_{{\mathrm{2}}} \,  \! [ {\bm{ { S } } }^{ \mathit{I} } / {\bm{ \alpha } }^{ \mathit{I} } ]   \vdash  \mathit{x}  \ottsym{:}  \ottnt{A} \,  \! [ {\bm{ { S } } }^{ \mathit{I} } / {\bm{ \alpha } }^{ \mathit{I} } ]   \mid   \bbZero .
              \end{align*}
            \end{divcases}
            Thus, the required result is achieved because of $\mathit{x} \,  \! [ {\bm{ { S } } }^{ \mathit{I} } / {\bm{ \alpha } }^{ \mathit{I} } ] = \mathit{x}$.

            \item[\rname{T}{Abs}]
            For some $\mathit{f}$, $\mathit{x}$, $\ottnt{e'}$, $\ottnt{A'}$, $\ottnt{B'}$, and $\varepsilon'$,
            the following are given:
            \begin{itemize}
              \item $\ottnt{e} = \ottkw{fun} \, \ottsym{(}  \mathit{f}  \ottsym{,}  \mathit{x}  \ottsym{,}  \ottnt{e'}  \ottsym{)}$,
              \item $\ottnt{A} =  \ottnt{A'}    \rightarrow_{ \varepsilon' }    \ottnt{B'} $,
              \item $\varepsilon =  \bbZero $, and
              \item $\Gamma_{{\mathrm{1}}}  \ottsym{,}   {\bm{ \alpha } }^{ \mathit{I} } : {\bm{ \ottnt{K} } }^{ \mathit{I} }   \ottsym{,}  \Gamma_{{\mathrm{2}}}  \ottsym{,}  \mathit{f}  \ottsym{:}   \ottnt{A'}    \rightarrow_{ \varepsilon' }    \ottnt{B'}   \ottsym{,}  \mathit{x}  \ottsym{:}  \ottnt{A'}  \vdash  \ottnt{e'}  \ottsym{:}  \ottnt{B'}  \mid  \varepsilon'$.
            \end{itemize}
            By the induction hypothesis, we have
            \begin{align*}
              \Gamma_{{\mathrm{1}}}  \ottsym{,}  \Gamma_{{\mathrm{2}}} \,  \! [ {\bm{ { S } } }^{ \mathit{I} } / {\bm{ \alpha } }^{ \mathit{I} } ]   \ottsym{,}  \mathit{f}  \ottsym{:}  \ottsym{(}   \ottnt{A'}    \rightarrow_{ \varepsilon' }    \ottnt{B'}   \ottsym{)} \,  \! [ {\bm{ { S } } }^{ \mathit{I} } / {\bm{ \alpha } }^{ \mathit{I} } ]   \ottsym{,}  \mathit{x}  \ottsym{:}  \ottnt{A'} \,  \! [ {\bm{ { S } } }^{ \mathit{I} } / {\bm{ \alpha } }^{ \mathit{I} } ]   \vdash  \ottnt{e'} \,  \! [ {\bm{ { S } } }^{ \mathit{I} } / {\bm{ \alpha } }^{ \mathit{I} } ]   \ottsym{:}  \ottnt{B'} \,  \! [ {\bm{ { S } } }^{ \mathit{I} } / {\bm{ \alpha } }^{ \mathit{I} } ]   \mid  \varepsilon' \,  \! [ {\bm{ { S } } }^{ \mathit{I} } / {\bm{ \alpha } }^{ \mathit{I} } ] .
            \end{align*}
            Since
            \begin{align*}
              \ottsym{(}   \ottnt{A'}    \rightarrow_{ \varepsilon' }    \ottnt{B'}   \ottsym{)} \,  \! [ {\bm{ { S } } }^{ \mathit{I} } / {\bm{ \alpha } }^{ \mathit{I} } ]  =  \ottsym{(}  \ottnt{A'} \,  \! [ {\bm{ { S } } }^{ \mathit{I} } / {\bm{ \alpha } }^{ \mathit{I} } ]   \ottsym{)}    \rightarrow_{ \varepsilon' \,  \! [ {\bm{ { S } } }^{ \mathit{I} } / {\bm{ \alpha } }^{ \mathit{I} } ]  }    \ottsym{(}  \ottnt{B'} \,  \! [ {\bm{ { S } } }^{ \mathit{I} } / {\bm{ \alpha } }^{ \mathit{I} } ]   \ottsym{)} 
            \end{align*}
            holds, \rname{T}{Abs} derives
            \begin{align*}
              \Gamma_{{\mathrm{1}}}  \ottsym{,}  \Gamma_{{\mathrm{2}}} \,  \! [ {\bm{ { S } } }^{ \mathit{I} } / {\bm{ \alpha } }^{ \mathit{I} } ]   \vdash  \ottkw{fun} \, \ottsym{(}  \mathit{f}  \ottsym{,}  \mathit{x}  \ottsym{,}  \ottnt{e'} \,  \! [ {\bm{ { S } } }^{ \mathit{I} } / {\bm{ \alpha } }^{ \mathit{I} } ]   \ottsym{)}  \ottsym{:}   \ottsym{(}  \ottnt{A'} \,  \! [ {\bm{ { S } } }^{ \mathit{I} } / {\bm{ \alpha } }^{ \mathit{I} } ]   \ottsym{)}    \rightarrow_{ \varepsilon' \,  \! [ {\bm{ { S } } }^{ \mathit{I} } / {\bm{ \alpha } }^{ \mathit{I} } ]  }    \ottsym{(}  \ottnt{B'} \,  \! [ {\bm{ { S } } }^{ \mathit{I} } / {\bm{ \alpha } }^{ \mathit{I} } ]   \ottsym{)}   \mid   \bbZero .
            \end{align*}
            Thus, the required result is achieved because
            \begin{align*}
              \ottsym{(}  \ottkw{fun} \, \ottsym{(}  \mathit{f}  \ottsym{,}  \mathit{x}  \ottsym{,}  \ottnt{e'}  \ottsym{)}  \ottsym{)} \,  \! [ {\bm{ { S } } }^{ \mathit{I} } / {\bm{ \alpha } }^{ \mathit{I} } ]  = \ottkw{fun} \, \ottsym{(}  \mathit{f}  \ottsym{,}  \mathit{x}  \ottsym{,}  \ottnt{e'} \,  \! [ {\bm{ { S } } }^{ \mathit{I} } / {\bm{ \alpha } }^{ \mathit{I} } ]   \ottsym{)}
            \end{align*}
            holds.

            \item[\rname{T}{App}]
            For some $\ottnt{v_{{\mathrm{1}}}}$, $\ottnt{v_{{\mathrm{2}}}}$, and $\ottnt{B}$, the following are given:
            \begin{itemize}
              \item $\ottnt{e} = \ottnt{v_{{\mathrm{1}}}} \, \ottnt{v_{{\mathrm{2}}}}$,
              \item $\Gamma_{{\mathrm{1}}}  \ottsym{,}   {\bm{ \alpha } }^{ \mathit{I} } : {\bm{ \ottnt{K} } }^{ \mathit{I} }   \ottsym{,}  \Gamma_{{\mathrm{2}}}  \vdash  \ottnt{v_{{\mathrm{1}}}}  \ottsym{:}   \ottnt{B}    \rightarrow_{ \varepsilon }    \ottnt{A}   \mid   \bbZero $, and
              \item $\Gamma_{{\mathrm{1}}}  \ottsym{,}   {\bm{ \alpha } }^{ \mathit{I} } : {\bm{ \ottnt{K} } }^{ \mathit{I} }   \ottsym{,}  \Gamma_{{\mathrm{2}}}  \vdash  \ottnt{v_{{\mathrm{2}}}}  \ottsym{:}  \ottnt{B}  \mid   \bbZero $.
            \end{itemize}
            By the induction hypothesis, we have
            \begin{itemize}
              \item $\Gamma_{{\mathrm{1}}}  \ottsym{,}  \Gamma_{{\mathrm{2}}} \,  \! [ {\bm{ { S } } }^{ \mathit{I} } / {\bm{ \alpha } }^{ \mathit{I} } ]   \vdash  \ottnt{v_{{\mathrm{1}}}} \,  \! [ {\bm{ { S } } }^{ \mathit{I} } / {\bm{ \alpha } }^{ \mathit{I} } ]   \ottsym{:}  \ottsym{(}   \ottnt{B}    \rightarrow_{ \varepsilon }    \ottnt{A}   \ottsym{)} \,  \! [ {\bm{ { S } } }^{ \mathit{I} } / {\bm{ \alpha } }^{ \mathit{I} } ]   \mid   \bbZero  \,  \! [ {\bm{ { S } } }^{ \mathit{I} } / {\bm{ \alpha } }^{ \mathit{I} } ] $ and
              \item $\Gamma_{{\mathrm{1}}}  \ottsym{,}  \Gamma_{{\mathrm{2}}} \,  \! [ {\bm{ { S } } }^{ \mathit{I} } / {\bm{ \alpha } }^{ \mathit{I} } ]   \vdash  \ottnt{v_{{\mathrm{2}}}} \,  \! [ {\bm{ { S } } }^{ \mathit{I} } / {\bm{ \alpha } }^{ \mathit{I} } ]   \ottsym{:}  \ottnt{B} \,  \! [ {\bm{ { S } } }^{ \mathit{I} } / {\bm{ \alpha } }^{ \mathit{I} } ]   \mid   \bbZero  \,  \! [ {\bm{ { S } } }^{ \mathit{I} } / {\bm{ \alpha } }^{ \mathit{I} } ] $.
            \end{itemize}
            Since
            \begin{itemize}
              \item $\ottsym{(}   \ottnt{B}    \rightarrow_{ \varepsilon }    \ottnt{A}   \ottsym{)} \,  \! [ {\bm{ { S } } }^{ \mathit{I} } / {\bm{ \alpha } }^{ \mathit{I} } ] =  \ottsym{(}  \ottnt{B} \,  \! [ {\bm{ { S } } }^{ \mathit{I} } / {\bm{ \alpha } }^{ \mathit{I} } ]   \ottsym{)}    \rightarrow_{ \varepsilon \,  \! [ {\bm{ { S } } }^{ \mathit{I} } / {\bm{ \alpha } }^{ \mathit{I} } ]  }    \ottsym{(}  \ottnt{A} \,  \! [ {\bm{ { S } } }^{ \mathit{I} } / {\bm{ \alpha } }^{ \mathit{I} } ]   \ottsym{)} $ and
              \item $ \bbZero  \,  \! [ {\bm{ { S } } }^{ \mathit{I} } / {\bm{ \alpha } }^{ \mathit{I} } ] =  \bbZero $
            \end{itemize}
            hold, \rname{T}{App} derives
            \begin{align*}
              \Gamma_{{\mathrm{1}}}  \ottsym{,}  \Gamma_{{\mathrm{2}}} \,  \! [ {\bm{ { S } } }^{ \mathit{I} } / {\bm{ \alpha } }^{ \mathit{I} } ]   \vdash   (  \ottnt{v_{{\mathrm{1}}}} \,  \! [ {\bm{ { S } } }^{ \mathit{I} } / {\bm{ \alpha } }^{ \mathit{I} } ]   )  \,  (  \ottnt{v_{{\mathrm{2}}}} \,  \! [ {\bm{ { S } } }^{ \mathit{I} } / {\bm{ \alpha } }^{ \mathit{I} } ]   )   \ottsym{:}  \ottnt{A} \,  \! [ {\bm{ { S } } }^{ \mathit{I} } / {\bm{ \alpha } }^{ \mathit{I} } ]   \mid  \varepsilon \,  \! [ {\bm{ { S } } }^{ \mathit{I} } / {\bm{ \alpha } }^{ \mathit{I} } ] 
            \end{align*}
            as required.

            \item[\rname{T}{TAbs}]
            Without loss of generality, we can choose $\beta$ such that
            $\beta \neq \alpha_{\ottmv{i}}$ and $ \beta   \notin    \mathrm{FTV}   \ottsym{(}   S_{\ottmv{i}}   \ottsym{)}  $ for any $\ottmv{i} \in \mathit{I}$.
            For some $\ottnt{K}$, $\ottnt{e'}$, $\ottnt{A'}$, and $\varepsilon'$, the following are given:
            \begin{itemize}
              \item $\ottnt{e} = \Lambda  \beta  \ottsym{:}  \ottnt{K}  \ottsym{.}  \ottnt{e'}$,
              \item $\ottnt{A} =   \forall   \beta  \ottsym{:}  \ottnt{K}   \ottsym{.}    \ottnt{A'}    ^{ \varepsilon' }  $,
              \item $\varepsilon =  \bbZero $, and
              \item $\Gamma_{{\mathrm{1}}}  \ottsym{,}   {\bm{ \alpha } }^{ \mathit{I} } : {\bm{ \ottnt{K} } }^{ \mathit{I} }   \ottsym{,}  \Gamma_{{\mathrm{2}}}  \ottsym{,}  \beta  \ottsym{:}  \ottnt{K}  \vdash  \ottnt{e'}  \ottsym{:}  \ottnt{A'}  \mid  \varepsilon'$.
            \end{itemize}
            By the induction hypothesis, we have
            \begin{align*}
              \Gamma_{{\mathrm{1}}}  \ottsym{,}  \Gamma_{{\mathrm{2}}} \,  \! [ {\bm{ { S } } }^{ \mathit{I} } / {\bm{ \alpha } }^{ \mathit{I} } ]   \ottsym{,}  \beta  \ottsym{:}  \ottnt{K}  \vdash  \ottnt{e'} \,  \! [ {\bm{ { S } } }^{ \mathit{I} } / {\bm{ \alpha } }^{ \mathit{I} } ]   \ottsym{:}  \ottnt{A'} \,  \! [ {\bm{ { S } } }^{ \mathit{I} } / {\bm{ \alpha } }^{ \mathit{I} } ]   \mid  \varepsilon' \,  \! [ {\bm{ { S } } }^{ \mathit{I} } / {\bm{ \alpha } }^{ \mathit{I} } ] 
            \end{align*}
            Thus, \rname{T}{TAbs} derives
            \begin{align*}
              \Gamma_{{\mathrm{1}}}  \ottsym{,}  \Gamma_{{\mathrm{2}}} \,  \! [ {\bm{ { S } } }^{ \mathit{I} } / {\bm{ \alpha } }^{ \mathit{I} } ]   \vdash  \Lambda  \beta  \ottsym{:}  \ottnt{K}  \ottsym{.}   (  \ottnt{e'} \,  \! [ {\bm{ { S } } }^{ \mathit{I} } / {\bm{ \alpha } }^{ \mathit{I} } ]   )   \ottsym{:}    \forall   \beta  \ottsym{:}  \ottnt{K}   \ottsym{.}    \ottnt{A'} \,  \! [ {\bm{ { S } } }^{ \mathit{I} } / {\bm{ \alpha } }^{ \mathit{I} } ]     ^{ \ottsym{(}  \varepsilon' \,  \! [ {\bm{ { S } } }^{ \mathit{I} } / {\bm{ \alpha } }^{ \mathit{I} } ]   \ottsym{)} }    \mid   \bbZero 
            \end{align*}
            and since
            \begin{align*}
              \ottsym{(}  \Lambda  \beta  \ottsym{:}  \ottnt{K}  \ottsym{.}  \ottnt{e'}  \ottsym{)} \,  \! [ {\bm{ { S } } }^{ \mathit{I} } / {\bm{ \alpha } }^{ \mathit{I} } ]  = \Lambda  \beta  \ottsym{:}  \ottnt{K}  \ottsym{.}   (  \ottnt{e'} \,  \! [ {\bm{ { S } } }^{ \mathit{I} } / {\bm{ \alpha } }^{ \mathit{I} } ]   ) 
            \end{align*}
            holds, the required result is achieved.

            \item[\rname{T}{TApp}]
            Without loss of generality, we can choose $\beta$ such that
            $\beta \neq \alpha_{\ottmv{i}}$ and $ \beta   \notin    \mathrm{FTV}   \ottsym{(}   S_{\ottmv{i}}   \ottsym{)}  $ for any $\ottmv{i} \in \mathit{I}$.
            For some $\ottnt{v}$, $T$, $\beta$, $\ottnt{A'}$, $\varepsilon'$, and $\ottnt{K}$, the following are given:
            \begin{itemize}
              \item $\ottnt{e} = \ottnt{v} \, T$,
              \item $\ottnt{A} = \ottnt{A'} \,  \! [  T  /  \beta   ] $,
              \item $\varepsilon = \varepsilon' \,  \! [  T  /  \beta   ] $,
              \item $\Gamma_{{\mathrm{1}}}  \ottsym{,}   {\bm{ \alpha } }^{ \mathit{I} } : {\bm{ \ottnt{K} } }^{ \mathit{I} }   \ottsym{,}  \Gamma_{{\mathrm{2}}}  \vdash  \ottnt{v}  \ottsym{:}    \forall   \beta  \ottsym{:}  \ottnt{K}   \ottsym{.}    \ottnt{A'}    ^{ \varepsilon' }    \mid   \bbZero $, and
              \item $\Gamma_{{\mathrm{1}}}  \ottsym{,}   {\bm{ \alpha } }^{ \mathit{I} } : {\bm{ \ottnt{K} } }^{ \mathit{I} }   \ottsym{,}  \Gamma_{{\mathrm{2}}}  \vdash  T  \ottsym{:}  \ottnt{K}$.
            \end{itemize}
            By the induction hypothesis, we have
            \begin{itemize}
              \item $\Gamma_{{\mathrm{1}}}  \ottsym{,}  \Gamma_{{\mathrm{2}}} \,  \! [ {\bm{ { S } } }^{ \mathit{I} } / {\bm{ \alpha } }^{ \mathit{I} } ]   \vdash  \ottnt{v} \,  \! [ {\bm{ { S } } }^{ \mathit{I} } / {\bm{ \alpha } }^{ \mathit{I} } ]   \ottsym{:}  \ottsym{(}    \forall   \beta  \ottsym{:}  \ottnt{K}   \ottsym{.}    \ottnt{A'}    ^{ \varepsilon' }    \ottsym{)} \,  \! [ {\bm{ { S } } }^{ \mathit{I} } / {\bm{ \alpha } }^{ \mathit{I} } ]   \mid   \bbZero  \,  \! [ {\bm{ { S } } }^{ \mathit{I} } / {\bm{ \alpha } }^{ \mathit{I} } ] $ and
              \item $\Gamma_{{\mathrm{1}}}  \ottsym{,}  \Gamma_{{\mathrm{2}}} \,  \! [ {\bm{ { S } } }^{ \mathit{I} } / {\bm{ \alpha } }^{ \mathit{I} } ]   \vdash  T \,  \! [ {\bm{ { S } } }^{ \mathit{I} } / {\bm{ \alpha } }^{ \mathit{I} } ]   \ottsym{:}  \ottnt{K}$.
            \end{itemize}
            Since
            \begin{itemize}
              \item $\ottsym{(}    \forall   \beta  \ottsym{:}  \ottnt{K}   \ottsym{.}    \ottnt{A'}    ^{ \varepsilon' }    \ottsym{)} \,  \! [ {\bm{ { S } } }^{ \mathit{I} } / {\bm{ \alpha } }^{ \mathit{I} } ]  =   \forall   \beta  \ottsym{:}  \ottnt{K}   \ottsym{.}    \ottnt{A'} \,  \! [ {\bm{ { S } } }^{ \mathit{I} } / {\bm{ \alpha } }^{ \mathit{I} } ]     ^{ \ottsym{(}  \varepsilon' \,  \! [ {\bm{ { S } } }^{ \mathit{I} } / {\bm{ \alpha } }^{ \mathit{I} } ]   \ottsym{)} }  $ and
              \item $ \bbZero  \,  \! [ {\bm{ { S } } }^{ \mathit{I} } / {\bm{ \alpha } }^{ \mathit{I} } ]  =  \bbZero $
            \end{itemize}
            hold, \rname{T}{TApp} derives
            \begin{align*}
              \Gamma_{{\mathrm{1}}}  \ottsym{,}  \Gamma_{{\mathrm{2}}} \,  \! [ {\bm{ { S } } }^{ \mathit{I} } / {\bm{ \alpha } }^{ \mathit{I} } ]   \vdash  \ottsym{(}  \ottnt{v} \,  \! [ {\bm{ { S } } }^{ \mathit{I} } / {\bm{ \alpha } }^{ \mathit{I} } ]   \ottsym{)} \, \ottsym{(}  T \,  \! [ {\bm{ { S } } }^{ \mathit{I} } / {\bm{ \alpha } }^{ \mathit{I} } ]   \ottsym{)}  \ottsym{:}  \ottsym{(}  \ottnt{A'} \,  \! [ {\bm{ { S } } }^{ \mathit{I} } / {\bm{ \alpha } }^{ \mathit{I} } ]   \ottsym{)} \,  \! [  T \,  \! [ {\bm{ { S } } }^{ \mathit{I} } / {\bm{ \alpha } }^{ \mathit{I} } ]   /  \beta   ]   \mid  \ottsym{(}  \varepsilon' \,  \! [ {\bm{ { S } } }^{ \mathit{I} } / {\bm{ \alpha } }^{ \mathit{I} } ]   \ottsym{)} \,  \! [  T \,  \! [ {\bm{ { S } } }^{ \mathit{I} } / {\bm{ \alpha } }^{ \mathit{I} } ]   /  \beta   ] .
            \end{align*}
            Finally, we have
            \begin{itemize}
              \item $ (  \ottnt{v} \, T  )  \,  \! [ {\bm{ { S } } }^{ \mathit{I} } / {\bm{ \alpha } }^{ \mathit{I} } ]  = \ottsym{(}  \ottnt{v} \,  \! [ {\bm{ { S } } }^{ \mathit{I} } / {\bm{ \alpha } }^{ \mathit{I} } ]   \ottsym{)} \, \ottsym{(}  T \,  \! [ {\bm{ { S } } }^{ \mathit{I} } / {\bm{ \alpha } }^{ \mathit{I} } ]   \ottsym{)}$,
              \item $\ottsym{(}  \ottnt{A'} \,  \! [ {\bm{ { S } } }^{ \mathit{I} } / {\bm{ \alpha } }^{ \mathit{I} } ]   \ottsym{)} \,  \! [  T \,  \! [ {\bm{ { S } } }^{ \mathit{I} } / {\bm{ \alpha } }^{ \mathit{I} } ]   /  \beta   ]  = \ottsym{(}  \ottnt{A'} \,  \! [  T  /  \beta   ]   \ottsym{)} \,  \! [ {\bm{ { S } } }^{ \mathit{I} } / {\bm{ \alpha } }^{ \mathit{I} } ] $, and
              \item $\ottsym{(}  \varepsilon' \,  \! [ {\bm{ { S } } }^{ \mathit{I} } / {\bm{ \alpha } }^{ \mathit{I} } ]   \ottsym{)} \,  \! [  T \,  \! [ {\bm{ { S } } }^{ \mathit{I} } / {\bm{ \alpha } }^{ \mathit{I} } ]   /  \beta   ]  = \ottsym{(}  \varepsilon' \,  \! [  T  /  \beta   ]   \ottsym{)} \,  \! [ {\bm{ { S } } }^{ \mathit{I} } / {\bm{ \alpha } }^{ \mathit{I} } ] $
            \end{itemize}
            because $\forall \ottmv{i} \in \mathit{I} . ( \beta   \notin    \mathrm{FTV}   \ottsym{(}   S_{\ottmv{i}}   \ottsym{)}  )$.
            Thus, the required result is achieved.

            \item[\rname{T}{Let}]
            For some $\mathit{x}$, $\ottnt{e_{{\mathrm{1}}}}$, $\ottnt{e_{{\mathrm{2}}}}$, and $\ottnt{B}$, the following are given:
            \begin{itemize}
              \item $\ottnt{e} = (\mathbf{let} \, \mathit{x}  \ottsym{=}  \ottnt{e_{{\mathrm{1}}}} \, \mathbf{in} \, \ottnt{e_{{\mathrm{2}}}})$
              \item $\Gamma_{{\mathrm{1}}}  \ottsym{,}   {\bm{ \alpha } }^{ \mathit{I} } : {\bm{ \ottnt{K} } }^{ \mathit{I} }   \ottsym{,}  \Gamma_{{\mathrm{2}}}  \vdash  \ottnt{e_{{\mathrm{1}}}}  \ottsym{:}  \ottnt{B}  \mid  \varepsilon$, and
              \item $\Gamma_{{\mathrm{1}}}  \ottsym{,}   {\bm{ \alpha } }^{ \mathit{I} } : {\bm{ \ottnt{K} } }^{ \mathit{I} }   \ottsym{,}  \Gamma_{{\mathrm{2}}}  \ottsym{,}  \mathit{x}  \ottsym{:}  \ottnt{B}  \vdash  \ottnt{e_{{\mathrm{2}}}}  \ottsym{:}  \ottnt{A}  \mid  \varepsilon$.
            \end{itemize}
            By the induction hypothesis, we have
            \begin{itemize}
              \item $\Gamma_{{\mathrm{1}}}  \ottsym{,}  \Gamma_{{\mathrm{2}}} \,  \! [ {\bm{ { S } } }^{ \mathit{I} } / {\bm{ \alpha } }^{ \mathit{I} } ]   \vdash  \ottnt{e_{{\mathrm{1}}}} \,  \! [ {\bm{ { S } } }^{ \mathit{I} } / {\bm{ \alpha } }^{ \mathit{I} } ]   \ottsym{:}  \ottnt{B} \,  \! [ {\bm{ { S } } }^{ \mathit{I} } / {\bm{ \alpha } }^{ \mathit{I} } ]   \mid  \varepsilon \,  \! [ {\bm{ { S } } }^{ \mathit{I} } / {\bm{ \alpha } }^{ \mathit{I} } ] $ and
              \item $\Gamma_{{\mathrm{1}}}  \ottsym{,}  \Gamma_{{\mathrm{2}}} \,  \! [ {\bm{ { S } } }^{ \mathit{I} } / {\bm{ \alpha } }^{ \mathit{I} } ]   \ottsym{,}  \mathit{x}  \ottsym{:}  \ottnt{B} \,  \! [ {\bm{ { S } } }^{ \mathit{I} } / {\bm{ \alpha } }^{ \mathit{I} } ]   \vdash  \ottnt{e_{{\mathrm{2}}}} \,  \! [ {\bm{ { S } } }^{ \mathit{I} } / {\bm{ \alpha } }^{ \mathit{I} } ]   \ottsym{:}  \ottnt{A} \,  \! [ {\bm{ { S } } }^{ \mathit{I} } / {\bm{ \alpha } }^{ \mathit{I} } ]   \mid  \varepsilon \,  \! [ {\bm{ { S } } }^{ \mathit{I} } / {\bm{ \alpha } }^{ \mathit{I} } ] $.
            \end{itemize}
            Thus, \rname{T}{Let} derives
            \begin{align*}
              \Gamma_{{\mathrm{1}}}  \ottsym{,}  \Gamma_{{\mathrm{2}}} \,  \! [ {\bm{ { S } } }^{ \mathit{I} } / {\bm{ \alpha } }^{ \mathit{I} } ]   \vdash  \mathbf{let} \, \mathit{x}  \ottsym{=}  \ottnt{e_{{\mathrm{1}}}} \,  \! [ {\bm{ { S } } }^{ \mathit{I} } / {\bm{ \alpha } }^{ \mathit{I} } ]  \, \mathbf{in} \,  (  \ottnt{e_{{\mathrm{2}}}} \,  \! [ {\bm{ { S } } }^{ \mathit{I} } / {\bm{ \alpha } }^{ \mathit{I} } ]   )   \ottsym{:}  \ottnt{A} \,  \! [ {\bm{ { S } } }^{ \mathit{I} } / {\bm{ \alpha } }^{ \mathit{I} } ]   \mid  \varepsilon \,  \! [ {\bm{ { S } } }^{ \mathit{I} } / {\bm{ \alpha } }^{ \mathit{I} } ] 
            \end{align*}
            and since
            \begin{align*}
               (  \mathbf{let} \, \mathit{x}  \ottsym{=}  \ottnt{e_{{\mathrm{1}}}} \, \mathbf{in} \, \ottnt{e_{{\mathrm{2}}}}  )  \,  \! [ {\bm{ { S } } }^{ \mathit{I} } / {\bm{ \alpha } }^{ \mathit{I} } ]  = \mathbf{let} \, \mathit{x}  \ottsym{=}  \ottnt{e_{{\mathrm{1}}}} \,  \! [ {\bm{ { S } } }^{ \mathit{I} } / {\bm{ \alpha } }^{ \mathit{I} } ]  \, \mathbf{in} \,  (  \ottnt{e_{{\mathrm{2}}}} \,  \! [ {\bm{ { S } } }^{ \mathit{I} } / {\bm{ \alpha } }^{ \mathit{I} } ]   ) 
            \end{align*}
            holds, the required result is achieved.

            \item[\rname{T}{Sub}]
            For some $\ottnt{A'}$ and $\varepsilon'$, the following are given:
            \begin{itemize}
              \item $\Gamma_{{\mathrm{1}}}  \ottsym{,}   {\bm{ \alpha } }^{ \mathit{I} } : {\bm{ \ottnt{K} } }^{ \mathit{I} }   \ottsym{,}  \Gamma_{{\mathrm{2}}}  \vdash  \ottnt{e}  \ottsym{:}  \ottnt{A'}  \mid  \varepsilon'$ and
              \item $\Gamma_{{\mathrm{1}}}  \ottsym{,}   {\bm{ \alpha } }^{ \mathit{I} } : {\bm{ \ottnt{K} } }^{ \mathit{I} }   \ottsym{,}  \Gamma_{{\mathrm{2}}}  \vdash  \ottnt{A'}  \mid  \varepsilon'  <:  \ottnt{A}  \mid  \varepsilon$.
            \end{itemize}
            By the induction hypothesis and case~\ref{lem:subst_type:subtyping_comp}, we have
            \begin{itemize}
              \item $\Gamma_{{\mathrm{1}}}  \ottsym{,}  \Gamma_{{\mathrm{2}}} \,  \! [ {\bm{ { S } } }^{ \mathit{I} } / {\bm{ \alpha } }^{ \mathit{I} } ]   \vdash  \ottnt{e} \,  \! [ {\bm{ { S } } }^{ \mathit{I} } / {\bm{ \alpha } }^{ \mathit{I} } ]   \ottsym{:}  \ottnt{A'} \,  \! [ {\bm{ { S } } }^{ \mathit{I} } / {\bm{ \alpha } }^{ \mathit{I} } ]   \mid  \varepsilon' \,  \! [ {\bm{ { S } } }^{ \mathit{I} } / {\bm{ \alpha } }^{ \mathit{I} } ] $ and
              \item $\Gamma_{{\mathrm{1}}}  \ottsym{,}  \Gamma_{{\mathrm{2}}} \,  \! [ {\bm{ { S } } }^{ \mathit{I} } / {\bm{ \alpha } }^{ \mathit{I} } ]   \vdash  \ottnt{A'} \,  \! [ {\bm{ { S } } }^{ \mathit{I} } / {\bm{ \alpha } }^{ \mathit{I} } ]   \mid  \varepsilon' \,  \! [ {\bm{ { S } } }^{ \mathit{I} } / {\bm{ \alpha } }^{ \mathit{I} } ]   <:  \ottnt{A} \,  \! [ {\bm{ { S } } }^{ \mathit{I} } / {\bm{ \alpha } }^{ \mathit{I} } ]   \mid  \varepsilon \,  \! [ {\bm{ { S } } }^{ \mathit{I} } / {\bm{ \alpha } }^{ \mathit{I} } ] $.
            \end{itemize}
            Thus, \rname{T}{Sub} derives
            \begin{align*}
              \Gamma_{{\mathrm{1}}}  \ottsym{,}  \Gamma_{{\mathrm{2}}} \,  \! [ {\bm{ { S } } }^{ \mathit{I} } / {\bm{ \alpha } }^{ \mathit{I} } ]   \vdash  \ottnt{e} \,  \! [ {\bm{ { S } } }^{ \mathit{I} } / {\bm{ \alpha } }^{ \mathit{I} } ]   \ottsym{:}  \ottnt{A} \,  \! [ {\bm{ { S } } }^{ \mathit{I} } / {\bm{ \alpha } }^{ \mathit{I} } ]   \mid  \varepsilon \,  \! [ {\bm{ { S } } }^{ \mathit{I} } / {\bm{ \alpha } }^{ \mathit{I} } ] 
            \end{align*}
            as required.

            \item[\rname{T}{Op}]
            For some $\mathsf{op}$, $\mathit{l}$, $ \bm{ { S_{{\mathrm{0}}} } } ^ {  \mathit{I_{{\mathrm{0}}}}  } $, $ \bm{ { T } } ^ {  \mathit{J}  } $, $\sigma$, $ \bm{ { \alpha_{{\mathrm{0}}} } } ^ {  \mathit{I_{{\mathrm{0}}}}  } $, $ {\bm{ { \ottnt{K_{{\mathrm{0}}}} } } }^{ \mathit{I_{{\mathrm{0}}}} } $, $ \bm{ { \beta } } ^ {  \mathit{J}  } $, $ {\bm{ { \ottnt{K''} } } }^{ \mathit{J} } $, $\ottnt{A'}$, and $\ottnt{B'}$,
            the following are given:
            \begin{itemize}
              \item $\ottnt{e} =  \mathsf{op} _{ \mathit{l} \,  \bm{ { S_{{\mathrm{0}}} } } ^ {  \mathit{I_{{\mathrm{0}}}}  }  }  \,  \bm{ { T } } ^ {  \mathit{J}  } $,
              \item $\ottnt{A} =  \ottsym{(}  \ottnt{A'} \,  \! [ {\bm{ { T } } }^{ \mathit{J} } / {\bm{ \beta } }^{ \mathit{J} } ]   \ottsym{)}    \rightarrow_{  \lift{ \mathit{l} \,  \bm{ { S_{{\mathrm{0}}} } } ^ {  \mathit{I_{{\mathrm{0}}}}  }  }  }    \ottsym{(}  \ottnt{B'} \,  \! [ {\bm{ { T } } }^{ \mathit{J} } / {\bm{ \beta } }^{ \mathit{J} } ]   \ottsym{)} $,
              \item $\varepsilon =  \bbZero $,
              \item $ \mathit{l}  ::    \forall    {\bm{ \alpha_{{\mathrm{0}}} } }^{ \mathit{I_{{\mathrm{0}}}} } : {\bm{ \ottnt{K_{{\mathrm{0}}}} } }^{ \mathit{I_{{\mathrm{0}}}} }    \ottsym{.}    \sigma    \in   \Xi $,
              \item $ \mathsf{op}  \ottsym{:}    \forall    {\bm{ \beta } }^{ \mathit{J} } : {\bm{ \ottnt{K''} } }^{ \mathit{J} }    \ottsym{.}    \ottnt{A'}   \Rightarrow   \ottnt{B'}    \in   \sigma \,  \! [ {\bm{ { S_{{\mathrm{0}}} } } }^{ \mathit{I_{{\mathrm{0}}}} } / {\bm{ \alpha_{{\mathrm{0}}} } }^{ \mathit{I_{{\mathrm{0}}}} } ]  $,
              \item $\vdash  \Gamma_{{\mathrm{1}}}  \ottsym{,}   {\bm{ \alpha } }^{ \mathit{I} } : {\bm{ \ottnt{K} } }^{ \mathit{I} }   \ottsym{,}  \Gamma_{{\mathrm{2}}}$,
              \item $\Gamma_{{\mathrm{1}}}  \ottsym{,}   {\bm{ \alpha } }^{ \mathit{I} } : {\bm{ \ottnt{K} } }^{ \mathit{I} }   \ottsym{,}  \Gamma_{{\mathrm{2}}}  \vdash   \bm{ { S_{{\mathrm{0}}} } }^{ \mathit{I_{{\mathrm{0}}}} } : \bm{ \ottnt{K_{{\mathrm{0}}}} }^{ \mathit{I_{{\mathrm{0}}}} } $, and
              \item $\Gamma_{{\mathrm{1}}}  \ottsym{,}   {\bm{ \alpha } }^{ \mathit{I} } : {\bm{ \ottnt{K} } }^{ \mathit{I} }   \ottsym{,}  \Gamma_{{\mathrm{2}}}  \vdash   \bm{ { T } }^{ \mathit{J} } : \bm{ \ottnt{K''} }^{ \mathit{J} } $.
            \end{itemize}
            By cases~\ref{lem:subst_type:wf} and \ref{lem:subst_type:kinding}, we have
            \begin{itemize}
              \item $\vdash  \Gamma_{{\mathrm{1}}}  \ottsym{,}  \Gamma_{{\mathrm{2}}} \,  \! [ {\bm{ { S } } }^{ \mathit{I} } / {\bm{ \alpha } }^{ \mathit{I} } ] $,
              \item $\Gamma_{{\mathrm{1}}}  \ottsym{,}  \Gamma_{{\mathrm{2}}} \,  \! [ {\bm{ { S } } }^{ \mathit{I} } / {\bm{ \alpha } }^{ \mathit{I} } ]   \vdash   \bm{ { S_{{\mathrm{0}}} \,  \! [ {\bm{ { S } } }^{ \mathit{I} } / {\bm{ \alpha } }^{ \mathit{I} } ]  } }^{ \mathit{I_{{\mathrm{0}}}} } : \bm{ \ottnt{K_{{\mathrm{0}}}} }^{ \mathit{I_{{\mathrm{0}}}} } $, and
              \item $\Gamma_{{\mathrm{1}}}  \ottsym{,}  \Gamma_{{\mathrm{2}}} \,  \! [ {\bm{ { S } } }^{ \mathit{I} } / {\bm{ \alpha } }^{ \mathit{I} } ]   \vdash   \bm{ { T \,  \! [ {\bm{ { S } } }^{ \mathit{I} } / {\bm{ \alpha } }^{ \mathit{I} } ]  } }^{ \mathit{J} } : \bm{ \ottnt{K''} }^{ \mathit{J} } $.
            \end{itemize}
            Since
            \begin{itemize}
              \item $\ottsym{(}   \lift{ \mathit{l} \,  \bm{ { S_{{\mathrm{0}}} } } ^ {  \mathit{I_{{\mathrm{0}}}}  }  }   \ottsym{)} \,  \! [ {\bm{ { S } } }^{ \mathit{I} } / {\bm{ \alpha } }^{ \mathit{I} } ]  =  \lift{ \mathit{l} \,  \bm{ { S_{{\mathrm{0}}} \,  \! [ {\bm{ { S } } }^{ \mathit{I} } / {\bm{ \alpha } }^{ \mathit{I} } ]  } } ^ {  \mathit{I_{{\mathrm{0}}}}  }  } $ and
              \item $ \bbZero  \,  \! [ {\bm{ { S } } }^{ \mathit{I} } / {\bm{ \alpha } }^{ \mathit{I} } ]  =  \bbZero $,
            \end{itemize}
            \rname{T}{Op} derives
            \begin{align*}
               & \Gamma_{{\mathrm{1}}}  \ottsym{,}  \Gamma_{{\mathrm{2}}} \,  \! [ {\bm{ { S } } }^{ \mathit{I} } / {\bm{ \alpha } }^{ \mathit{I} } ]   \vdash   \mathsf{op} _{ \mathit{l} \,  \bm{ { S_{{\mathrm{0}}} \,  \! [ {\bm{ { S } } }^{ \mathit{I} } / {\bm{ \alpha } }^{ \mathit{I} } ]  } } ^ {  \mathit{I_{{\mathrm{0}}}}  }  }  \,  \bm{ { T \,  \! [ {\bm{ { S } } }^{ \mathit{I} } / {\bm{ \alpha } }^{ \mathit{I} } ]  } } ^ {  \mathit{J}  }   \ottsym{:}   \ottnt{A'_{{\mathrm{0}}}} \,  \! [ {\bm{ { T \,  \! [ {\bm{ { S } } }^{ \mathit{I} } / {\bm{ \alpha } }^{ \mathit{I} } ]  } } }^{ \mathit{J} } / {\bm{ \beta } }^{ \mathit{J} } ]     \rightarrow_{  \lift{ \mathit{l} \,  \bm{ { S_{{\mathrm{0}}} \,  \! [ {\bm{ { S } } }^{ \mathit{I} } / {\bm{ \alpha } }^{ \mathit{I} } ]  } } ^ {  \mathit{I_{{\mathrm{0}}}}  }  }  }    \ottnt{B'_{{\mathrm{0}}}} \,  \! [ {\bm{ { T \,  \! [ {\bm{ { S } } }^{ \mathit{I} } / {\bm{ \alpha } }^{ \mathit{I} } ]  } } }^{ \mathit{J} } / {\bm{ \beta } }^{ \mathit{J} } ]    \mid   \bbZero 
            \end{align*}
            where
            \[
               \mathsf{op}  \ottsym{:}    \forall    {\bm{ \beta } }^{ \mathit{J} } : {\bm{ \ottnt{K''} } }^{ \mathit{J} }    \ottsym{.}    \ottnt{A'_{{\mathrm{0}}}}   \Rightarrow   \ottnt{B'_{{\mathrm{0}}}}    \in   \sigma \,  \! [ {\bm{ { S_{{\mathrm{0}}} \,  \! [ {\bm{ { S } } }^{ \mathit{I} } / {\bm{ \alpha } }^{ \mathit{I} } ]  } } }^{ \mathit{I_{{\mathrm{0}}}} } / {\bm{ \alpha_{{\mathrm{0}}} } }^{ \mathit{I_{{\mathrm{0}}}} } ]  .
            \]
            Without loss of generality, we can assume that, for any $\ottmv{i} \in \mathit{I}$,
            $ \alpha_{\ottmv{i}}   \notin     \mathrm{FTV}   \ottsym{(}   \ottnt{A'}   \ottsym{)}    \cup    \mathrm{FTV}   \ottsym{(}   \ottnt{B'}   \ottsym{)}   $, and
            $ \ottsym{(}   \{  \alpha_{\ottmv{i}}  \}   \cup    \mathrm{FTV}   \ottsym{(}   S_{\ottmv{i}}   \ottsym{)}    \ottsym{)}   \cap   \ottsym{(}   \{   \bm{ { \alpha_{{\mathrm{0}}} } } ^ {  \mathit{I_{{\mathrm{0}}}}  }   \}   \cup   \{   \bm{ { \beta } } ^ {  \mathit{J}  }   \}   \ottsym{)}   \ottsym{=}  \emptyset$.
            Then,
            \begin{itemize}
              \item $\ottnt{A'} \,  \! [ {\bm{ { T } } }^{ \mathit{J} } / {\bm{ \beta } }^{ \mathit{J} } ]  \,  \! [ {\bm{ { S } } }^{ \mathit{I} } / {\bm{ \alpha } }^{ \mathit{I} } ]  = \ottnt{A'_{{\mathrm{0}}}} \,  \! [ {\bm{ { T \,  \! [ {\bm{ { S } } }^{ \mathit{I} } / {\bm{ \alpha } }^{ \mathit{I} } ]  } } }^{ \mathit{J} } / {\bm{ \beta } }^{ \mathit{J} } ] $ and
              \item $\ottnt{B'} \,  \! [ {\bm{ { T } } }^{ \mathit{J} } / {\bm{ \beta } }^{ \mathit{J} } ]  \,  \! [ {\bm{ { S } } }^{ \mathit{I} } / {\bm{ \alpha } }^{ \mathit{I} } ]  = \ottnt{B'_{{\mathrm{0}}}} \,  \! [ {\bm{ { T \,  \! [ {\bm{ { S } } }^{ \mathit{I} } / {\bm{ \alpha } }^{ \mathit{I} } ]  } } }^{ \mathit{J} } / {\bm{ \beta } }^{ \mathit{J} } ] $
            \end{itemize}
            hold.
            Therefore, the required result is achieved.

            \item[\rname{T}{Handling}]
            For some $\mathit{N}$, $\ottnt{e'}$, $\ottnt{A'}$, $\varepsilon'$, $\mathit{l}$, $ \bm{ { S_{{\mathrm{0}}} } } ^ {  \mathit{N}  } $, $ \bm{ { \alpha_{{\mathrm{0}}} } } ^ {  \mathit{N}  } $, $ {\bm{ { \ottnt{K_{{\mathrm{0}}}} } } }^{ \mathit{N} } $, $\ottnt{h}$, and $\sigma$,
            the following are given:
            \begin{itemize}
              \item $\ottnt{e} =  \mathbf{handle}_{ \mathit{l} \,  \bm{ { S_{{\mathrm{0}}} } } ^ {  \mathit{N}  }  }  \, \ottnt{e'} \, \mathbf{with} \, \ottnt{h}$,
              \item $\Gamma_{{\mathrm{1}}}  \ottsym{,}   {\bm{ \alpha } }^{ \mathit{I} } : {\bm{ \ottnt{K} } }^{ \mathit{I} }   \ottsym{,}  \Gamma_{{\mathrm{2}}}  \vdash  \ottnt{e'}  \ottsym{:}  \ottnt{A'}  \mid  \varepsilon'$,
              \item $ \mathit{l}  ::    \forall    {\bm{ \alpha_{{\mathrm{0}}} } }^{ \mathit{N} } : {\bm{ \ottnt{K_{{\mathrm{0}}}} } }^{ \mathit{N} }    \ottsym{.}    \sigma    \in   \Xi $,
              \item $\Gamma_{{\mathrm{1}}}  \ottsym{,}   {\bm{ \alpha } }^{ \mathit{I} } : {\bm{ \ottnt{K} } }^{ \mathit{I} }   \ottsym{,}  \Gamma_{{\mathrm{2}}}  \vdash   \bm{ { S_{{\mathrm{0}}} } }^{ \mathit{N} } : \bm{ \ottnt{K_{{\mathrm{0}}}} }^{ \mathit{N} } $,
              \item $ \Gamma_{{\mathrm{1}}}  \ottsym{,}   {\bm{ \alpha } }^{ \mathit{I} } : {\bm{ \ottnt{K} } }^{ \mathit{I} }   \ottsym{,}  \Gamma_{{\mathrm{2}}}  \vdash _{ \sigma \,  \! [ {\bm{ { S_{{\mathrm{0}}} } } }^{ \mathit{N} } / {\bm{ \alpha_{{\mathrm{0}}} } }^{ \mathit{N} } ]  }  \ottnt{h}  :  \ottnt{A'}   \Rightarrow  ^ { \varepsilon }  \ottnt{A} $, and
              \item $   \lift{ \mathit{l} \,  \bm{ { S_{{\mathrm{0}}} } } ^ {  \mathit{N}  }  }   \mathop{ \odot }  \varepsilon    \sim   \varepsilon' $.
            \end{itemize}
            By the induction hypothesis,
            case~\ref{lem:subst_type:kinding}, and
            the fact that a typelike substitution is homomorphism for $ \odot $ and $ \sim $,we have
            \begin{itemize}
              \item $\Gamma_{{\mathrm{1}}}  \ottsym{,}  \Gamma_{{\mathrm{2}}} \,  \! [ {\bm{ { S } } }^{ \mathit{I} } / {\bm{ \alpha } }^{ \mathit{I} } ]   \vdash   \bm{ { S_{{\mathrm{0}}} \,  \! [ {\bm{ { S } } }^{ \mathit{I} } / {\bm{ \alpha } }^{ \mathit{I} } ]  } }^{ \mathit{N} } : \bm{ \ottnt{K_{{\mathrm{0}}}} }^{ \mathit{N} } $,
              \item $\Gamma_{{\mathrm{1}}}  \ottsym{,}  \Gamma_{{\mathrm{2}}} \,  \! [ {\bm{ { S } } }^{ \mathit{I} } / {\bm{ \alpha } }^{ \mathit{I} } ]   \vdash  \ottnt{e'} \,  \! [ {\bm{ { S } } }^{ \mathit{I} } / {\bm{ \alpha } }^{ \mathit{I} } ]   \ottsym{:}  \ottnt{A'} \,  \! [ {\bm{ { S } } }^{ \mathit{I} } / {\bm{ \alpha } }^{ \mathit{I} } ]   \mid  \varepsilon' \,  \! [ {\bm{ { S } } }^{ \mathit{I} } / {\bm{ \alpha } }^{ \mathit{I} } ] $,
              \item $ \Gamma_{{\mathrm{1}}}  \ottsym{,}  \Gamma_{{\mathrm{2}}} \,  \! [ {\bm{ { S } } }^{ \mathit{I} } / {\bm{ \alpha } }^{ \mathit{I} } ]   \vdash _{ \sigma \,  \! [ {\bm{ { S_{{\mathrm{0}}} } } }^{ \mathit{N} } / {\bm{ \alpha_{{\mathrm{0}}} } }^{ \mathit{N} } ]  \,  \! [ {\bm{ { S } } }^{ \mathit{I} } / {\bm{ \alpha } }^{ \mathit{I} } ]  }  \ottnt{h} \,  \! [ {\bm{ { S } } }^{ \mathit{I} } / {\bm{ \alpha } }^{ \mathit{I} } ]   :  \ottnt{A'} \,  \! [ {\bm{ { S } } }^{ \mathit{I} } / {\bm{ \alpha } }^{ \mathit{I} } ]    \Rightarrow  ^ { \varepsilon \,  \! [ {\bm{ { S } } }^{ \mathit{I} } / {\bm{ \alpha } }^{ \mathit{I} } ]  }  \ottnt{A} \,  \! [ {\bm{ { S } } }^{ \mathit{I} } / {\bm{ \alpha } }^{ \mathit{I} } ]  $, and
              \item $   \lift{ \mathit{l} \,  \bm{ { S_{{\mathrm{0}}} \,  \! [ {\bm{ { S } } }^{ \mathit{I} } / {\bm{ \alpha } }^{ \mathit{I} } ]  } } ^ {  \mathit{N}  }  }   \mathop{ \odot }  \varepsilon  \,  \! [ {\bm{ { S } } }^{ \mathit{I} } / {\bm{ \alpha } }^{ \mathit{I} } ]    \sim   \varepsilon' \,  \! [ {\bm{ { S } } }^{ \mathit{I} } / {\bm{ \alpha } }^{ \mathit{I} } ]  $.
            \end{itemize}
            Now, because we can assume that
            \begin{itemize}
              \item $ \{   \bm{ { \alpha } } ^ {  \mathit{I}  }   \}   \cap   \{   \bm{ { \alpha_{{\mathrm{0}}} } } ^ {  \mathit{N}  }   \}   \ottsym{=}  \emptyset$ and
              \item $ \{   \bm{ { \alpha_{{\mathrm{0}}} } } ^ {  \mathit{N}  }   \}   \cap    \mathrm{FTV}   \ottsym{(}    \bm{ { S } } ^ {  \mathit{I}  }    \ottsym{)}    \ottsym{=}  \emptyset$
            \end{itemize}
            without loss of generality, we have
            \begin{align*}
               \Gamma_{{\mathrm{1}}}  \ottsym{,}  \Gamma_{{\mathrm{2}}} \,  \! [ {\bm{ { S } } }^{ \mathit{I} } / {\bm{ \alpha } }^{ \mathit{I} } ]   \vdash _{ \sigma \,  \! [ {\bm{ { S_{{\mathrm{0}}} \,  \! [ {\bm{ { S } } }^{ \mathit{I} } / {\bm{ \alpha } }^{ \mathit{I} } ]  } } }^{ \mathit{N} } / {\bm{ \alpha_{{\mathrm{0}}} } }^{ \mathit{N} } ]  }  \ottnt{h} \,  \! [ {\bm{ { S } } }^{ \mathit{I} } / {\bm{ \alpha } }^{ \mathit{I} } ]   :  \ottnt{A'} \,  \! [ {\bm{ { S } } }^{ \mathit{I} } / {\bm{ \alpha } }^{ \mathit{I} } ]    \Rightarrow  ^ { \varepsilon \,  \! [ {\bm{ { S } } }^{ \mathit{I} } / {\bm{ \alpha } }^{ \mathit{I} } ]  }  \ottnt{A} \,  \! [ {\bm{ { S } } }^{ \mathit{I} } / {\bm{ \alpha } }^{ \mathit{I} } ]  .
            \end{align*}
            Thus, \rname{T}{Handling} derives
            \begin{align*}
              \Gamma_{{\mathrm{1}}}  \ottsym{,}  \Gamma_{{\mathrm{2}}} \,  \! [ {\bm{ { S } } }^{ \mathit{I} } / {\bm{ \alpha } }^{ \mathit{I} } ]   \vdash   \mathbf{handle}_{ \mathit{l} \,  \bm{ { S_{{\mathrm{0}}} \,  \! [ {\bm{ { S } } }^{ \mathit{I} } / {\bm{ \alpha } }^{ \mathit{I} } ]  } } ^ {  \mathit{N}  }  }  \, \ottnt{e} \,  \! [ {\bm{ { S } } }^{ \mathit{I} } / {\bm{ \alpha } }^{ \mathit{I} } ]  \, \mathbf{with} \, \ottnt{h} \,  \! [ {\bm{ { S } } }^{ \mathit{I} } / {\bm{ \alpha } }^{ \mathit{I} } ]   \ottsym{:}  \ottnt{B} \,  \! [ {\bm{ { S } } }^{ \mathit{I} } / {\bm{ \alpha } }^{ \mathit{I} } ]   \mid  \varepsilon \,  \! [ {\bm{ { S } } }^{ \mathit{I} } / {\bm{ \alpha } }^{ \mathit{I} } ] .
            \end{align*}

            \item[\rname{H}{Return}]
            For some $\mathit{x}$ and $\ottnt{e_{\ottmv{r}}}$, the following are given:
            \begin{itemize}
              \item $\ottnt{h} = \ottsym{\{} \, \mathbf{return} \, \mathit{y}  \mapsto  \ottnt{e_{\ottmv{r}}}  \ottsym{\}}$,
              \item $\sigma =  \{\} $, and
              \item $\Gamma_{{\mathrm{1}}}  \ottsym{,}   {\bm{ \alpha } }^{ \mathit{I} } : {\bm{ \ottnt{K} } }^{ \mathit{I} }   \ottsym{,}  \Gamma_{{\mathrm{2}}}  \ottsym{,}  \mathit{x}  \ottsym{:}  \ottnt{A}  \vdash  \ottnt{e_{\ottmv{r}}}  \ottsym{:}  \ottnt{B}  \mid  \varepsilon$.
            \end{itemize}
            By the induction hypothesis, we have
            \begin{itemize}
              \item $\Gamma_{{\mathrm{1}}}  \ottsym{,}  \Gamma_{{\mathrm{2}}} \,  \! [ {\bm{ { S } } }^{ \mathit{I} } / {\bm{ \alpha } }^{ \mathit{I} } ]   \ottsym{,}  \mathit{x}  \ottsym{:}  \ottnt{A} \,  \! [ {\bm{ { S } } }^{ \mathit{I} } / {\bm{ \alpha } }^{ \mathit{I} } ]   \vdash  \ottnt{e_{\ottmv{r}}} \,  \! [ {\bm{ { S } } }^{ \mathit{I} } / {\bm{ \alpha } }^{ \mathit{I} } ]   \ottsym{:}  \ottnt{B} \,  \! [ {\bm{ { S } } }^{ \mathit{I} } / {\bm{ \alpha } }^{ \mathit{I} } ]   \mid  \varepsilon \,  \! [ {\bm{ { S } } }^{ \mathit{I} } / {\bm{ \alpha } }^{ \mathit{I} } ] $.
            \end{itemize}
            Thus, \rname{H}{Return} derives
            \begin{align*}
               \Gamma_{{\mathrm{1}}}  \ottsym{,}  \Gamma_{{\mathrm{2}}}  \vdash _{  \{\}  }  \ottsym{\{} \, \mathbf{return} \, \mathit{x}  \mapsto  \ottnt{e_{\ottmv{r}}} \,  \! [ {\bm{ { S } } }^{ \mathit{I} } / {\bm{ \alpha } }^{ \mathit{I} } ]   \ottsym{\}}  :  \ottnt{A} \,  \! [ {\bm{ { S } } }^{ \mathit{I} } / {\bm{ \alpha } }^{ \mathit{I} } ]    \Rightarrow  ^ { \varepsilon \,  \! [ {\bm{ { S } } }^{ \mathit{I} } / {\bm{ \alpha } }^{ \mathit{I} } ]  }  \ottnt{B} \,  \! [ {\bm{ { S } } }^{ \mathit{I} } / {\bm{ \alpha } }^{ \mathit{I} } ]  .
            \end{align*}

            \item[\rname{H}{Op}]
            Without loss of generality, we can choose $ \bm{ { \beta } } ^ {  \mathit{J}  } $ such that:
            \begin{itemize}
              \item $ \{   \bm{ { \beta } } ^ {  \mathit{J}  }   \}   \cap   \{   \bm{ { \alpha } } ^ {  \mathit{I}  }   \}   \ottsym{=}  \emptyset$ and
              \item $ \{   \bm{ { \beta } } ^ {  \mathit{J}  }   \}   \cap    \mathrm{FTV}   \ottsym{(}    \bm{ { S } } ^ {  \mathit{I}  }    \ottsym{)}    \ottsym{=}  \emptyset$.
            \end{itemize}
            For some $\ottnt{h'}$, $\sigma'$, $\mathsf{op}$, $\ottnt{A'}$, $\ottnt{B'}$, and $\ottnt{e}$, the following are given:
            \begin{itemize}
              \item $\ottnt{h} =  \ottnt{h'}   \uplus   \ottsym{\{}  \mathsf{op} \,  {\bm{ \beta } }^{ \mathit{J} } : {\bm{ \ottnt{K} } }^{ \mathit{J} }  \, \mathit{p} \, \mathit{k}  \mapsto  \ottnt{e}  \ottsym{\}} $,
              \item $\sigma =  \sigma'   \uplus   \ottsym{\{}  \mathsf{op}  \ottsym{:}    \forall    {\bm{ \beta } }^{ \mathit{J} } : {\bm{ \ottnt{K} } }^{ \mathit{J} }    \ottsym{.}    \ottnt{A'}   \Rightarrow   \ottnt{B'}   \ottsym{\}} $,
              \item $ \Gamma_{{\mathrm{1}}}  \ottsym{,}   {\bm{ \alpha } }^{ \mathit{I} } : {\bm{ \ottnt{K} } }^{ \mathit{I} }   \ottsym{,}  \Gamma_{{\mathrm{2}}}  \vdash _{ \sigma' }  \ottnt{h'}  :  \ottnt{A}   \Rightarrow  ^ { \varepsilon }  \ottnt{B} $, and
              \item $\Gamma_{{\mathrm{1}}}  \ottsym{,}   {\bm{ \alpha } }^{ \mathit{I} } : {\bm{ \ottnt{K} } }^{ \mathit{I} }   \ottsym{,}  \Gamma_{{\mathrm{2}}}  \ottsym{,}   {\bm{ \beta } }^{ \mathit{J} } : {\bm{ \ottnt{K} } }^{ \mathit{J} }   \ottsym{,}  \mathit{p}  \ottsym{:}  \ottnt{A'}  \ottsym{,}  \mathit{k}  \ottsym{:}   \ottnt{B'}    \rightarrow_{ \varepsilon }    \ottnt{B}   \vdash  \ottnt{e}  \ottsym{:}  \ottnt{B}  \mid  \varepsilon$.
            \end{itemize}
            By the induction hypothesis and Definition~\ref{def:subst_typelike}, we have
            \begin{itemize}
              \item $\sigma \,  \! [ {\bm{ { S } } }^{ \mathit{I} } / {\bm{ \alpha } }^{ \mathit{I} } ]  =  \sigma' \,  \! [ {\bm{ { S } } }^{ \mathit{I} } / {\bm{ \alpha } }^{ \mathit{I} } ]    \uplus   \ottsym{\{}  \mathsf{op}  \ottsym{:}    \forall    {\bm{ \beta } }^{ \mathit{J} } : {\bm{ \ottnt{K} } }^{ \mathit{J} }    \ottsym{.}    \ottnt{A'} \,  \! [ {\bm{ { S } } }^{ \mathit{I} } / {\bm{ \alpha } }^{ \mathit{I} } ]    \Rightarrow   \ottnt{B'} \,  \! [ {\bm{ { S } } }^{ \mathit{I} } / {\bm{ \alpha } }^{ \mathit{I} } ]    \ottsym{\}} $,
              \item $ \Gamma_{{\mathrm{1}}}  \ottsym{,}  \Gamma_{{\mathrm{2}}} \,  \! [ {\bm{ { S } } }^{ \mathit{I} } / {\bm{ \alpha } }^{ \mathit{I} } ]   \vdash _{ \sigma' \,  \! [ {\bm{ { S } } }^{ \mathit{I} } / {\bm{ \alpha } }^{ \mathit{I} } ]  }  \ottnt{h'} \,  \! [ {\bm{ { S } } }^{ \mathit{I} } / {\bm{ \alpha } }^{ \mathit{I} } ]   :  \ottnt{A} \,  \! [ {\bm{ { S } } }^{ \mathit{I} } / {\bm{ \alpha } }^{ \mathit{I} } ]    \Rightarrow  ^ { \varepsilon \,  \! [ {\bm{ { S } } }^{ \mathit{I} } / {\bm{ \alpha } }^{ \mathit{I} } ]  }  \ottnt{B} \,  \! [ {\bm{ { S } } }^{ \mathit{I} } / {\bm{ \alpha } }^{ \mathit{I} } ]  $, and
              \item $\Gamma_{{\mathrm{1}}}  \ottsym{,}  \Gamma_{{\mathrm{2}}} \,  \! [ {\bm{ { S } } }^{ \mathit{I} } / {\bm{ \alpha } }^{ \mathit{I} } ]   \ottsym{,}   {\bm{ \beta } }^{ \mathit{J} } : {\bm{ \ottnt{K} } }^{ \mathit{J} }   \ottsym{,}  \mathit{p}  \ottsym{:}  \ottnt{A'} \,  \! [ {\bm{ { S } } }^{ \mathit{I} } / {\bm{ \alpha } }^{ \mathit{I} } ]   \ottsym{,}  \mathit{k}  \ottsym{:}   \ottnt{B'} \,  \! [ {\bm{ { S } } }^{ \mathit{I} } / {\bm{ \alpha } }^{ \mathit{I} } ]     \rightarrow_{ \varepsilon \,  \! [ {\bm{ { S } } }^{ \mathit{I} } / {\bm{ \alpha } }^{ \mathit{I} } ]  }    \ottnt{B} \,  \! [ {\bm{ { S } } }^{ \mathit{I} } / {\bm{ \alpha } }^{ \mathit{I} } ]    \vdash  \ottnt{e} \,  \! [ {\bm{ { S } } }^{ \mathit{I} } / {\bm{ \alpha } }^{ \mathit{I} } ]   \ottsym{:}  \ottnt{B} \,  \! [ {\bm{ { S } } }^{ \mathit{I} } / {\bm{ \alpha } }^{ \mathit{I} } ]   \mid  \varepsilon \,  \! [ {\bm{ { S } } }^{ \mathit{I} } / {\bm{ \alpha } }^{ \mathit{I} } ] $.
            \end{itemize}
            Thus, \rname{H}{Op} derives
            \begin{align*}
               \Gamma_{{\mathrm{1}}}  \ottsym{,}  \Gamma_{{\mathrm{2}}} \,  \! [ {\bm{ { S } } }^{ \mathit{I} } / {\bm{ \alpha } }^{ \mathit{I} } ]   \vdash _{ \sigma \,  \! [ {\bm{ { S } } }^{ \mathit{I} } / {\bm{ \alpha } }^{ \mathit{I} } ]  }   \ottnt{h'} \,  \! [ {\bm{ { S } } }^{ \mathit{I} } / {\bm{ \alpha } }^{ \mathit{I} } ]    \uplus   \ottsym{\{}  \mathsf{op} \,  {\bm{ \beta } }^{ \mathit{J} } : {\bm{ \ottnt{K} } }^{ \mathit{J} }  \, \mathit{p} \, \mathit{k}  \mapsto  \ottnt{e} \,  \! [ {\bm{ { S } } }^{ \mathit{I} } / {\bm{ \alpha } }^{ \mathit{I} } ]   \ottsym{\}}   :  \ottnt{A} \,  \! [ {\bm{ { S } } }^{ \mathit{I} } / {\bm{ \alpha } }^{ \mathit{I} } ]    \Rightarrow  ^ { \varepsilon \,  \! [ {\bm{ { S } } }^{ \mathit{I} } / {\bm{ \alpha } }^{ \mathit{I} } ]  }  \ottnt{B} \,  \! [ {\bm{ { S } } }^{ \mathit{I} } / {\bm{ \alpha } }^{ \mathit{I} } ]  .
            \end{align*}
          \end{divcases}
  \end{itemize}
\end{proof}

\begin{lemma}[Well-kinded of Subtyping]\label{lem:wk_subtyping}
  \phantom{}
  \begin{itemize}
    \item If $\Gamma  \vdash  \ottnt{A}  <:  \ottnt{B}$, then $\Gamma  \vdash  \ottnt{A}  \ottsym{:}   \mathbf{Typ} $ and $\Gamma  \vdash  \ottnt{B}  \ottsym{:}   \mathbf{Typ} $.
    \item If $\Gamma  \vdash  \ottnt{A_{{\mathrm{1}}}}  \mid  \varepsilon_{{\mathrm{1}}}  <:  \ottnt{A_{{\mathrm{2}}}}  \mid  \varepsilon$,
          then $\Gamma  \vdash  \ottnt{A_{\ottmv{i}}}  \ottsym{:}   \mathbf{Typ} $ and $\Gamma  \vdash  \varepsilon_{\ottmv{i}}  \ottsym{:}   \mathbf{Eff} $ for $\ottmv{i} \in \{1, 2\}$.
  \end{itemize}
\end{lemma}

\begin{proof}
  Straightforward by mutual induction on the subtyping derivations
  with Lemma~\ref{lem:delta_weakening}.
\end{proof}

\begin{lemma}[Well-kinded of Typing]\label{lem:wk}
  \phantom{}
  \begin{enumerate}
    \item\label{lem:wk:typing} If $\Gamma  \vdash  \ottnt{e}  \ottsym{:}  \ottnt{A}  \mid  \varepsilon$, then $\Gamma  \vdash  \ottnt{A}  \ottsym{:}   \mathbf{Typ} $ and $\Gamma  \vdash  \varepsilon  \ottsym{:}   \mathbf{Eff} $.
    \item\label{lem:wk:handling} If $ \Gamma  \vdash _{ \sigma }  \ottnt{h}  :  \ottnt{A}   \Rightarrow  ^ { \varepsilon }  \ottnt{B} $, then $\Gamma  \vdash  \ottnt{A}  \ottsym{:}   \mathbf{Typ} $ and $\Gamma  \vdash  \ottnt{B}  \ottsym{:}   \mathbf{Typ} $ and $\Gamma  \vdash  \varepsilon  \ottsym{:}   \mathbf{Eff} $.
  \end{enumerate}
\end{lemma}

\begin{proof}
  By mutual induction on derivations of the judgments.
  We proceed by cases on the typing rule applied lastly to the derivation.
  \begin{divcases}
    \item[\rname{T}{Var}]
    We are given
    $\varepsilon =  \bbZero $ and $\vdash  \Gamma$ and $\Gamma = \Gamma_{{\mathrm{1}}}  \ottsym{,}  \mathit{x}  \ottsym{:}  \ottnt{A}  \ottsym{,}  \Gamma_{{\mathrm{2}}}$
    for some $\mathit{x}$, $\Gamma_{{\mathrm{1}}}$, and $\Gamma_{{\mathrm{2}}}$.
    Because $\vdash  \Gamma$, it is easy to prove that $\Gamma_{{\mathrm{1}}}  \vdash  \ottnt{A}  \ottsym{:}   \mathbf{Typ} $ using Lemma~\ref{lem:wf}.
    Then, by Lemma~\ref{lem:weakening}, $\Gamma_{{\mathrm{1}}}  \ottsym{,}  \mathit{x}  \ottsym{:}  \ottnt{A}  \ottsym{,}  \Gamma_{{\mathrm{2}}}  \vdash  \ottnt{A}  \ottsym{:}   \mathbf{Typ} $.
    We also have $\Gamma  \vdash   \bbZero   \ottsym{:}   \mathbf{Eff} $ because $ \bbZero $ is well-formedness-preserving.

    \item[\rname{T}{Abs}]
    For some $\mathit{f}$, $\mathit{x}$, $\ottnt{e'}$, $\ottnt{B}$, $\ottnt{C}$, and $\varepsilon'$,
    the following are given:
    \begin{itemize}
      \item $\ottnt{e} = \ottkw{fun} \, \ottsym{(}  \mathit{f}  \ottsym{,}  \mathit{x}  \ottsym{,}  \ottnt{e'}  \ottsym{)}$,
      \item $\ottnt{A} =  \ottnt{B}    \rightarrow_{ \varepsilon' }    \ottnt{C} $,
      \item $\varepsilon =  \bbZero $, and
      \item $\Gamma  \ottsym{,}  \mathit{f}  \ottsym{:}   \ottnt{B}    \rightarrow_{ \varepsilon' }    \ottnt{C}   \ottsym{,}  \mathit{x}  \ottsym{:}  \ottnt{B}  \vdash  \ottnt{e'}  \ottsym{:}  \ottnt{C}  \mid  \varepsilon'$.
    \end{itemize}
    Since $ \bbZero $ is well-formedness-preserving, we have $\Gamma  \vdash   \bbZero   \ottsym{:}   \mathbf{Eff} $.
    By the induction hypothesis, we have $\Gamma  \ottsym{,}  \mathit{f}  \ottsym{:}   \ottnt{B}    \rightarrow_{ \varepsilon' }    \ottnt{C}   \ottsym{,}  \mathit{x}  \ottsym{:}  \ottnt{B}  \vdash  \ottnt{C}  \ottsym{:}   \mathbf{Typ} $.
    By Lemma~\ref{lem:wf},
    we have $\vdash  \Gamma  \ottsym{,}  \mathit{f}  \ottsym{:}   \ottnt{B}    \rightarrow_{ \varepsilon' }    \ottnt{C}   \ottsym{,}  \mathit{x}  \ottsym{:}  \ottnt{B}$.
    Since only \rname{C}{Var} can derive $\vdash  \Gamma  \ottsym{,}  \mathit{f}  \ottsym{:}   \ottnt{B}    \rightarrow_{ \varepsilon' }    \ottnt{C}   \ottsym{,}  \mathit{x}  \ottsym{:}  \ottnt{B}$,
    we have $\Gamma  \ottsym{,}  \mathit{f}  \ottsym{:}   \ottnt{B}    \rightarrow_{ \varepsilon' }    \ottnt{C}   \vdash  \ottnt{B}  \ottsym{:}   \mathbf{Typ} $.
    By Lemma~\ref{lem:wf}, we have $\vdash  \Gamma  \ottsym{,}  \mathit{f}  \ottsym{:}   \ottnt{B}    \rightarrow_{ \varepsilon' }    \ottnt{C} $.
    Since only \rname{C}{Var} can derive $\vdash  \Gamma  \ottsym{,}  \mathit{f}  \ottsym{:}   \ottnt{B}    \rightarrow_{ \varepsilon' }    \ottnt{C} $,
    we have $\Gamma  \vdash   \ottnt{B}    \rightarrow_{ \varepsilon' }    \ottnt{C}   \ottsym{:}   \mathbf{Typ} $.

    \item[\rname{T}{App}]
    For some $\ottnt{v_{{\mathrm{1}}}}$, $\ottnt{v_{{\mathrm{2}}}}$, and $\ottnt{B}$, the following are given:
    \begin{itemize}
      \item $\ottnt{e} = \ottnt{v_{{\mathrm{1}}}} \, \ottnt{v_{{\mathrm{2}}}}$,
      \item $\Gamma  \vdash  \ottnt{v_{{\mathrm{1}}}}  \ottsym{:}   \ottnt{B}    \rightarrow_{ \varepsilon }    \ottnt{A}   \mid   \bbZero $, and
      \item $\Gamma  \vdash  \ottnt{v_{{\mathrm{2}}}}  \ottsym{:}  \ottnt{B}  \mid   \bbZero $.
    \end{itemize}
    By the induction hypothesis, we have
    $\Gamma  \vdash   \ottnt{B}    \rightarrow_{ \varepsilon }    \ottnt{A}   \ottsym{:}   \mathbf{Typ} $ and $\Gamma  \vdash   \bbZero   \ottsym{:}   \mathbf{Eff} $.
    Since only \rname{K}{Fun} can derive $\Gamma  \vdash   \ottnt{B}    \rightarrow_{ \varepsilon }    \ottnt{A}   \ottsym{:}   \mathbf{Typ} $,
    we have $\Gamma  \vdash  \ottnt{A}  \ottsym{:}   \mathbf{Typ} $ and $\Gamma  \vdash  \varepsilon  \ottsym{:}   \mathbf{Eff} $ as required.

    \item[\rname{T}{TAbs}]
    For some $\alpha$, $\ottnt{K}$, $\ottnt{e'}$, $\ottnt{B}$, and $\varepsilon'$,
    the following are given:
    \begin{itemize}
      \item $\ottnt{e} = \Lambda  \alpha  \ottsym{:}  \ottnt{K}  \ottsym{.}  \ottnt{e'}$,
      \item $\ottnt{A} =   \forall   \alpha  \ottsym{:}  \ottnt{K}   \ottsym{.}    \ottnt{B}    ^{ \varepsilon' }  $,
      \item $\varepsilon =  \bbZero $, and
      \item $\Gamma  \ottsym{,}  \alpha  \ottsym{:}  \ottnt{K}  \vdash  \ottnt{e'}  \ottsym{:}  \ottnt{B}  \mid  \varepsilon'$.
    \end{itemize}
    Since $ \bbZero $ is well-formedness-preserving, we have $\Gamma  \vdash   \bbZero   \ottsym{:}   \mathbf{Eff} $.
    By the induction hypothesis, we have
    $\Gamma  \ottsym{,}  \alpha  \ottsym{:}  \ottnt{K}  \vdash  \ottnt{B}  \ottsym{:}   \mathbf{Typ} $ and $\Gamma  \ottsym{,}  \alpha  \ottsym{:}  \ottnt{K}  \vdash  \varepsilon'  \ottsym{:}   \mathbf{Eff} $.
    Thus, \rname{K}{Poly} derives $\Gamma  \vdash    \forall   \alpha  \ottsym{:}  \ottnt{K}   \ottsym{.}    \ottnt{B}    ^{ \varepsilon' }    \ottsym{:}   \mathbf{Typ} $.

    \item[\rname{T}{TApp}]
    For some $\ottnt{v}$, $S$, $\ottnt{A'}$, $\varepsilon'$, $\alpha$, and $\ottnt{K}$,
    the following are given:
    \begin{itemize}
      \item $\ottnt{e} = \ottnt{v} \, S$,
      \item $\ottnt{A} = \ottnt{A'} \,  \! [  S  /  \alpha   ] $,
      \item $\varepsilon = \varepsilon' \,  \! [  S  /  \alpha   ] $,
      \item $\Gamma  \vdash  \ottnt{v}  \ottsym{:}    \forall   \alpha  \ottsym{:}  \ottnt{K}   \ottsym{.}    \ottnt{A'}    ^{ \varepsilon' }    \mid   \bbZero $, and
      \item $\Gamma  \vdash  S  \ottsym{:}  \ottnt{K}$.
    \end{itemize}
    By the induction hypothesis, we have $\Gamma  \vdash    \forall   \alpha  \ottsym{:}  \ottnt{K}   \ottsym{.}    \ottnt{A'}    ^{ \varepsilon' }    \ottsym{:}   \mathbf{Typ} $.
    Since only \rname{K}{Poly} can derive $\Gamma  \vdash    \forall   \alpha  \ottsym{:}  \ottnt{K}   \ottsym{.}    \ottnt{A'}    ^{ \varepsilon' }    \ottsym{:}   \mathbf{Typ} $,
    we have $\Gamma  \ottsym{,}  \alpha  \ottsym{:}  \ottnt{K}  \vdash  \ottnt{A'}  \ottsym{:}   \mathbf{Typ} $ and $\Gamma  \ottsym{,}  \alpha  \ottsym{:}  \ottnt{K}  \vdash  \varepsilon'  \ottsym{:}   \mathbf{Eff} $.
    By Lemma~\ref{lem:subst_type}\ref{lem:subst_type:kinding}, we have
    $\Gamma  \vdash  \ottnt{A'} \,  \! [  S  /  \alpha   ]   \ottsym{:}   \mathbf{Typ} $ and $\Gamma  \vdash  \varepsilon' \,  \! [  S  /  \alpha   ]   \ottsym{:}   \mathbf{Eff} $ as required.

    \item[\rname{T}{Let}]
    For some $\mathit{x}$, $\ottnt{e_{{\mathrm{1}}}}$, $\ottnt{e_{{\mathrm{2}}}}$, and $\ottnt{B}$, the following are given:
    \begin{itemize}
      \item $\ottnt{e} = (\mathbf{let} \, \mathit{x}  \ottsym{=}  \ottnt{e_{{\mathrm{1}}}} \, \mathbf{in} \, \ottnt{e_{{\mathrm{2}}}})$,
      \item $\Gamma  \vdash  \ottnt{e_{{\mathrm{1}}}}  \ottsym{:}  \ottnt{B}  \mid  \varepsilon$, and
      \item $\Gamma  \ottsym{,}  \mathit{x}  \ottsym{:}  \ottnt{B}  \vdash  \ottnt{e_{{\mathrm{2}}}}  \ottsym{:}  \ottnt{A}  \mid  \varepsilon$.
    \end{itemize}
    By the induction hypothesis, we have $\Gamma  \ottsym{,}  \mathit{x}  \ottsym{:}  \ottnt{B}  \vdash  \ottnt{A}  \ottsym{:}   \mathbf{Typ} $ and $\Gamma  \vdash  \varepsilon  \ottsym{:}   \mathbf{Eff} $.
    By $ \Delta   \ottsym{(}   \Gamma  \ottsym{,}  \mathit{x}  \ottsym{:}  \ottnt{B}   \ottsym{)}  =  \Delta   \ottsym{(}   \Gamma   \ottsym{)} $ and Lemma~\ref{lem:delta_context}\ref{lem:delta_context:kinding} and Lemma~\ref{lem:delta_weakening},
    we have $\Gamma  \vdash  \ottnt{A}  \ottsym{:}   \mathbf{Typ} $ as required.

    \item[\rname{T}{Sub}]
    For some $\ottnt{A'}$ and $\varepsilon'$, the following are given:
    \begin{itemize}
      \item $\Gamma  \vdash  \ottnt{e}  \ottsym{:}  \ottnt{A'}  \mid  \varepsilon'$ and
      \item $\Gamma  \vdash  \ottnt{A'}  \mid  \varepsilon'  <:  \ottnt{A}  \mid  \varepsilon$.
    \end{itemize}
    By Lemma~\ref{lem:wk_subtyping}, we have $\Gamma  \vdash  \ottnt{A}  \ottsym{:}   \mathbf{Typ} $ and $\Gamma  \vdash  \varepsilon  \ottsym{:}   \mathbf{Typ} $.

    \item[\rname{T}{Op}]
    For some $\mathsf{op}$, $\mathit{l}$, $ \bm{ { S } } ^ {  \mathit{I}  } $, $ \bm{ { T } } ^ {  \mathit{J}  } $, $\sigma$, $ \bm{ { \alpha } } ^ {  \mathit{I}  } $, $ {\bm{ { \ottnt{K} } } }^{ \mathit{I} } $, $ \bm{ { \beta } } ^ {  \mathit{J}  } $, $ {\bm{ { \ottnt{K'} } } }^{ \mathit{J} } $, $\ottnt{A'}$, $\ottnt{B'}$,
    the following are given:
    \begin{itemize}
      \item $\ottnt{e} =  \mathsf{op} _{ \mathit{l} \,  \bm{ { S } } ^ {  \mathit{I}  }  }  \,  \bm{ { T } } ^ {  \mathit{J}  } $,
      \item $\ottnt{A} =  \ottsym{(}  \ottnt{A'} \,  \! [ {\bm{ { T } } }^{ \mathit{J} } / {\bm{ \beta } }^{ \mathit{J} } ]   \ottsym{)}    \rightarrow_{  \lift{ \mathit{l} \,  \bm{ { S } } ^ {  \mathit{I}  }  }  }    \ottsym{(}  \ottnt{B'} \,  \! [ {\bm{ { T } } }^{ \mathit{J} } / {\bm{ \beta } }^{ \mathit{J} } ]   \ottsym{)} $,
      \item $ \mathit{l}  ::    \forall    {\bm{ \alpha } }^{ \mathit{I} } : {\bm{ \ottnt{K} } }^{ \mathit{I} }    \ottsym{.}    \sigma    \in   \Xi $,
      \item $ \mathsf{op}  \ottsym{:}    \forall    {\bm{ \beta } }^{ \mathit{J} } : {\bm{ \ottnt{K'} } }^{ \mathit{J} }    \ottsym{.}    \ottnt{A'}   \Rightarrow   \ottnt{B'}    \in   \sigma \,  \! [ {\bm{ { S } } }^{ \mathit{I} } / {\bm{ \alpha } }^{ \mathit{I} } ]  $,
      \item $\vdash  \Gamma$,
      \item $\Gamma  \vdash   \bm{ { S } }^{ \mathit{I} } : \bm{ \ottnt{K} }^{ \mathit{I} } $, and
      \item $\Gamma  \vdash   \bm{ { T } }^{ \mathit{J} } : \bm{ \ottnt{K'} }^{ \mathit{J} } $.
    \end{itemize}
    Since $ \bbZero $ is well-formedness-preserving, we have $\Gamma  \vdash   \bbZero   \ottsym{:}   \mathbf{Eff} $.
    Without loss of generality, we can assume that $ \bm{ { \alpha } } ^ {  \mathit{I}  } $ and $ \bm{ { \beta } } ^ {  \mathit{J}  } $
    do not occur in $\Gamma$.
    Then, because there exist some $\ottnt{A''}$ and $\ottnt{B''}$ such that
    \begin{itemize}
      \item $ {\bm{ \alpha } }^{ \mathit{I} } : {\bm{ \ottnt{K} } }^{ \mathit{I} }   \ottsym{,}   {\bm{ \beta } }^{ \mathit{J} } : {\bm{ \ottnt{K'} } }^{ \mathit{J} }   \vdash  \ottnt{A''}  \ottsym{:}   \mathbf{Typ} $,
      \item $ {\bm{ \alpha } }^{ \mathit{I} } : {\bm{ \ottnt{K} } }^{ \mathit{I} }   \ottsym{,}   {\bm{ \beta } }^{ \mathit{J} } : {\bm{ \ottnt{K'} } }^{ \mathit{J} }   \vdash  \ottnt{B''}  \ottsym{:}   \mathbf{Typ} $,
      \item $\ottnt{A''} \,  \! [ {\bm{ { S } } }^{ \mathit{I} } / {\bm{ \alpha } }^{ \mathit{I} } ]  = \ottnt{A'}$, and
      \item $\ottnt{B''} \,  \! [ {\bm{ { S } } }^{ \mathit{I} } / {\bm{ \alpha } }^{ \mathit{I} } ]  = \ottnt{B'}$,
    \end{itemize}
    Lemma~\ref{lem:weakening} and \ref{lem:subst_type}\ref{lem:subst_type:kinding} imply
    $\Gamma  \vdash  \ottnt{A'} \,  \! [ {\bm{ { T } } }^{ \mathit{J} } / {\bm{ \beta } }^{ \mathit{J} } ]   \ottsym{:}   \mathbf{Typ} $ and $\Gamma  \vdash  \ottnt{B'} \,  \! [ {\bm{ { T } } }^{ \mathit{J} } / {\bm{ \beta } }^{ \mathit{J} } ]   \ottsym{:}   \mathbf{Typ} $.
    Thus, \rname{K}{Fun} derives $\Gamma  \vdash   \ottsym{(}  \ottnt{A'} \,  \! [ {\bm{ { T } } }^{ \mathit{J} } / {\bm{ \beta } }^{ \mathit{J} } ]   \ottsym{)}    \rightarrow_{  \lift{ \mathit{l} \,  \bm{ { S } } ^ {  \mathit{I}  }  }  }    \ottsym{(}  \ottnt{B'} \,  \! [ {\bm{ { T } } }^{ \mathit{J} } / {\bm{ \beta } }^{ \mathit{J} } ]   \ottsym{)}   \ottsym{:}   \mathbf{Typ} $.

    \item[\rname{T}{Handling}]
    For some $\ottnt{A'}$, $\sigma$, $\mathit{N}$, $ \bm{ { \alpha } } ^ {  \mathit{N}  } $, and $ \bm{ { S } } ^ {  \mathit{N}  } $, we have
    \begin{align*}
       \Gamma  \vdash _{ \sigma \,  \! [ {\bm{ { S } } }^{ \mathit{N} } / {\bm{ \alpha } }^{ \mathit{N} } ]  }  \ottnt{h}  :  \ottnt{A'}   \Rightarrow  ^ { \varepsilon }  \ottnt{A} .
    \end{align*}
    By the induction hypothesis, we have $\Gamma  \vdash  \ottnt{A}  \ottsym{:}   \mathbf{Typ} $ and $\Gamma  \vdash  \varepsilon  \ottsym{:}   \mathbf{Eff} $.

    \item[\rname{H}{Return}]
    For some $\mathit{x}$ and $\ottnt{e_{\ottmv{r}}}$, we have
    \begin{align*}
      \Gamma  \ottsym{,}  \mathit{x}  \ottsym{:}  \ottnt{A}  \vdash  \ottnt{e_{\ottmv{r}}}  \ottsym{:}  \ottnt{B}  \mid  \varepsilon.
    \end{align*}
    By the induction hypothesis, we have
    \begin{itemize}
      \item $\Gamma  \ottsym{,}  \mathit{x}  \ottsym{:}  \ottnt{A}  \vdash  \ottnt{B}  \ottsym{:}   \mathbf{Typ} $ and
      \item $\Gamma  \ottsym{,}  \mathit{x}  \ottsym{:}  \ottnt{A}  \vdash  \varepsilon  \ottsym{:}   \mathbf{Eff} $.
    \end{itemize}
    By Lemma~\ref{lem:delta_context}\ref{lem:delta_context:kinding}, we have
    \begin{itemize}
      \item $ \Delta   \ottsym{(}   \Gamma   \ottsym{)}   \vdash  \ottnt{B}  \ottsym{:}   \mathbf{Typ} $ and
      \item $ \Delta   \ottsym{(}   \Gamma   \ottsym{)}   \vdash  \varepsilon  \ottsym{:}   \mathbf{Eff} $.
    \end{itemize}
    By Lemma~\ref{lem:delta_weakening}, we have
    \begin{itemize}
      \item $\Gamma  \vdash  \ottnt{B}  \ottsym{:}   \mathbf{Typ} $ and
      \item $\Gamma  \vdash  \varepsilon  \ottsym{:}   \mathbf{Eff} $.
    \end{itemize}

    Now, we have $\vdash  \Gamma  \ottsym{,}  \mathit{x}  \ottsym{:}  \ottnt{A}$ by Lemma~\ref{lem:ctx-wf-typing}.
    Since only \rname{C}{Var} can derive $\vdash  \Gamma  \ottsym{,}  \mathit{x}  \ottsym{:}  \ottnt{A}$, we have $\Gamma  \vdash  \ottnt{A}  \ottsym{:}   \mathbf{Typ} $.

    \item[\rname{H}{Op}]
    For some $\ottnt{h'}$ and $\sigma'$, we have $ \Gamma  \vdash _{ \sigma' }  \ottnt{h'}  :  \ottnt{A}   \Rightarrow  ^ { \varepsilon }  \ottnt{B} $.
    By the induction hypothesis, we have $\Gamma  \vdash  \ottnt{A}  \ottsym{:}   \mathbf{Typ} $ and $\Gamma  \vdash  \ottnt{B}  \ottsym{:}   \mathbf{Typ} $ and $\Gamma  \vdash  \varepsilon  \ottsym{:}   \mathbf{Eff} $.
  \end{divcases}
\end{proof}

\begin{lemma}[Inversion of Subtyping]\label{lem:inv_st}
  \phantom{}
  \begin{enumerate}
    \item\label{lem:inv_st:abs}
          If $\Gamma  \vdash  \ottnt{C}  <:   \ottnt{A_{{\mathrm{1}}}}    \rightarrow_{ \varepsilon_{{\mathrm{1}}} }    \ottnt{B_{{\mathrm{1}}}} $ and $\Gamma  \vdash   \bbZero   \ottsym{:}   \mathbf{Eff} $,
          then $\ottnt{C} =  \ottnt{A_{{\mathrm{2}}}}    \rightarrow_{ \varepsilon_{{\mathrm{2}}} }    \ottnt{B_{{\mathrm{2}}}} $ such that $\Gamma  \vdash  \ottnt{A_{{\mathrm{1}}}}  <:  \ottnt{A_{{\mathrm{2}}}}$, $\Gamma  \vdash  \ottnt{B_{{\mathrm{2}}}}  <:  \ottnt{B_{{\mathrm{1}}}}$, and $\Gamma  \vdash   \varepsilon_{{\mathrm{2}}}  \olessthan  \varepsilon_{{\mathrm{1}}} $.

    \item\label{lem:inv_st:tabs}
          If $\Gamma  \vdash  \ottnt{C}  <:    \forall   \alpha  \ottsym{:}  \ottnt{K}   \ottsym{.}    \ottnt{A_{{\mathrm{1}}}}    ^{ \varepsilon_{{\mathrm{1}}} }  $ and $\Gamma  \vdash   \bbZero   \ottsym{:}   \mathbf{Eff} $,
          then $\ottnt{C} =   \forall   \alpha  \ottsym{:}  \ottnt{K}   \ottsym{.}    \ottnt{A_{{\mathrm{2}}}}    ^{ \varepsilon_{{\mathrm{2}}} }  $ such that $\Gamma  \ottsym{,}  \alpha  \ottsym{:}  \ottnt{K}  \vdash  \ottnt{A_{{\mathrm{2}}}}  <:  \ottnt{A_{{\mathrm{1}}}}$ and $\Gamma  \ottsym{,}  \alpha  \ottsym{:}  \ottnt{K}  \vdash   \varepsilon_{{\mathrm{2}}}  \olessthan  \varepsilon_{{\mathrm{1}}} $.
  \end{enumerate}
\end{lemma}

\begin{proof}
  \phantom{}
  \begin{enumerate}
    \item By induction on a derivation of $\Gamma  \vdash  \ottnt{C}  <:   \ottnt{A_{{\mathrm{1}}}}    \rightarrow_{ \varepsilon_{{\mathrm{1}}} }    \ottnt{B_{{\mathrm{1}}}} $.
          We proceed by case analysis on the subtyping rule applied lastly to this derivation.
          \begin{divcases}
            \item[\rname{ST}{Refl}]
            $\Gamma  \vdash   \ottnt{A_{{\mathrm{1}}}}    \rightarrow_{ \varepsilon_{{\mathrm{1}}} }    \ottnt{B_{{\mathrm{1}}}}   \ottsym{:}   \mathbf{Typ} $ and $\ottnt{C} =  \ottnt{A_{{\mathrm{1}}}}    \rightarrow_{ \varepsilon_{{\mathrm{1}}} }    \ottnt{B_{{\mathrm{1}}}} $ are given.
            Because only \rname{K}{Fun} can derive $\Gamma  \vdash   \ottnt{A_{{\mathrm{1}}}}    \rightarrow_{ \varepsilon_{{\mathrm{1}}} }    \ottnt{B_{{\mathrm{1}}}}   \ottsym{:}   \mathbf{Typ} $,
            we have $\Gamma  \vdash  \ottnt{A_{{\mathrm{1}}}}  \ottsym{:}   \mathbf{Typ} $, $\Gamma  \vdash  \varepsilon_{{\mathrm{1}}}  \ottsym{:}   \mathbf{Eff} $, and $\Gamma  \vdash  \ottnt{B_{{\mathrm{1}}}}  \ottsym{:}   \mathbf{Typ} $.
            By \rname{ST}{Refl}, $\Gamma  \vdash  \ottnt{A_{{\mathrm{1}}}}  <:  \ottnt{A_{{\mathrm{1}}}}$ and $\Gamma  \vdash  \ottnt{B_{{\mathrm{1}}}}  <:  \ottnt{B_{{\mathrm{1}}}}$ hold.
            By Lemma~\ref{lem:entailment}\ref{lem:entailment:refl},
            $\Gamma  \vdash   \varepsilon_{{\mathrm{1}}}  \olessthan  \varepsilon_{{\mathrm{1}}} $ holds.

            \item[\rname{ST}{Fun}] Clearly.

            \item[others] Cannot happen.
          \end{divcases}

    \item By induction on a derivation of $\Gamma  \vdash  \ottnt{C}  <:    \forall   \alpha  \ottsym{:}  \ottnt{K}   \ottsym{.}    \ottnt{A_{{\mathrm{1}}}}    ^{ \varepsilon_{{\mathrm{1}}} }  $.
          We proceed by case analysis on the subtyping rule applied lastly to this derivation.
          \begin{divcases}
            \item[\rname{ST}{Refl}]
            $\Gamma  \vdash    \forall   \alpha  \ottsym{:}  \ottnt{K}   \ottsym{.}    \ottnt{A_{{\mathrm{1}}}}    ^{ \varepsilon_{{\mathrm{1}}} }    \ottsym{:}   \mathbf{Typ} $ and
            $\ottnt{C} =   \forall   \alpha  \ottsym{:}  \ottnt{K}   \ottsym{.}    \ottnt{A_{{\mathrm{1}}}}    ^{ \varepsilon_{{\mathrm{1}}} }  $ are given.
            Because only \rname{K}{Poly} can derive
            $\Gamma  \vdash    \forall   \alpha  \ottsym{:}  \ottnt{K}   \ottsym{.}    \ottnt{A_{{\mathrm{1}}}}    ^{ \varepsilon_{{\mathrm{1}}} }    \ottsym{:}   \mathbf{Typ} $,
            we have $\Gamma  \ottsym{,}  \alpha  \ottsym{:}  \ottnt{K}  \vdash  \ottnt{A_{{\mathrm{1}}}}  \ottsym{:}   \mathbf{Typ} $ and $\Gamma  \ottsym{,}  \alpha  \ottsym{:}  \ottnt{K}  \vdash  \varepsilon_{{\mathrm{1}}}  \ottsym{:}   \mathbf{Eff} $.
            By \rname{ST}{Refl}, $\Gamma  \ottsym{,}  \alpha  \ottsym{:}  \ottnt{K}  \vdash  \ottnt{A_{{\mathrm{1}}}}  <:  \ottnt{A_{{\mathrm{1}}}}$ holds.
            By Lemma~\ref{lem:entailment}\ref{lem:entailment:refl},
            $\Gamma  \ottsym{,}  \alpha  \ottsym{:}  \ottnt{K}  \vdash   \varepsilon_{{\mathrm{1}}}  \olessthan  \varepsilon_{{\mathrm{1}}} $.

            \item[\rname{ST}{Poly}] Clearly.

            \item[others] Cannot happen.
          \end{divcases}
  \end{enumerate}
\end{proof}

\begin{lemma}[Inversion]\label{lem:inversion}
  \mbox{}
  \begin{enumerate}
    \item\label{lem:inversion:var}
          If $\Gamma  \vdash  \ottnt{v}  \ottsym{:}  \ottnt{A}  \mid  \varepsilon$, then $\Gamma  \vdash  \ottnt{v}  \ottsym{:}  \ottnt{A}  \mid   \bbZero $.

    \item\label{lem:inversion:abs}
          If $\Gamma  \vdash  \ottkw{fun} \, \ottsym{(}  \mathit{f}  \ottsym{,}  \mathit{x}  \ottsym{,}  \ottnt{e}  \ottsym{)}  \ottsym{:}   \ottnt{A_{{\mathrm{1}}}}    \rightarrow_{ \varepsilon_{{\mathrm{1}}} }    \ottnt{B_{{\mathrm{1}}}}   \mid  \varepsilon$,
          then $\Gamma  \ottsym{,}  \mathit{f}  \ottsym{:}   \ottnt{A_{{\mathrm{2}}}}    \rightarrow_{ \varepsilon_{{\mathrm{2}}} }    \ottnt{B_{{\mathrm{2}}}}   \ottsym{,}  \mathit{x}  \ottsym{:}  \ottnt{A_{{\mathrm{2}}}}  \vdash  \ottnt{e}  \ottsym{:}  \ottnt{B_{{\mathrm{2}}}}  \mid  \varepsilon_{{\mathrm{2}}}$
          for some $\ottnt{A_{{\mathrm{2}}}}$, $\varepsilon_{{\mathrm{2}}}$, and $\ottnt{B_{{\mathrm{2}}}}$
          such that $\Gamma  \vdash   \ottnt{A_{{\mathrm{2}}}}    \rightarrow_{ \varepsilon_{{\mathrm{2}}} }    \ottnt{B_{{\mathrm{2}}}}   <:   \ottnt{A_{{\mathrm{1}}}}    \rightarrow_{ \varepsilon_{{\mathrm{1}}} }    \ottnt{B_{{\mathrm{1}}}} $.

    \item\label{lem:inversion:tabs}
          If $\Gamma  \vdash  \Lambda  \alpha  \ottsym{:}  \ottnt{K}  \ottsym{.}  \ottnt{e}  \ottsym{:}    \forall   \alpha  \ottsym{:}  \ottnt{K}   \ottsym{.}    \ottnt{A_{{\mathrm{1}}}}    ^{ \varepsilon_{{\mathrm{1}}} }    \mid  \varepsilon$,
          then $\Gamma  \ottsym{,}  \alpha  \ottsym{:}  \ottnt{K}  \vdash  \ottnt{e}  \ottsym{:}  \ottnt{A_{{\mathrm{1}}}}  \mid  \varepsilon_{{\mathrm{1}}}$.

    \item\label{lem:inversion:op}
          If $\Gamma  \vdash   \mathsf{op} _{ \mathit{l} \,  \bm{ { S } } ^ {  \mathit{I}  }  }  \,  \bm{ { T } } ^ {  \mathit{J}  }   \ottsym{:}   \ottnt{A_{{\mathrm{1}}}}    \rightarrow_{ \varepsilon_{{\mathrm{1}}} }    \ottnt{B_{{\mathrm{1}}}}   \mid  \varepsilon$,
          then the following hold:
          \begin{itemize}
            \item $ \mathit{l}  ::    \forall    {\bm{ \alpha } }^{ \mathit{I} } : {\bm{ \ottnt{K} } }^{ \mathit{I} }    \ottsym{.}    \sigma    \in   \Xi $,
            \item $ \mathsf{op}  \ottsym{:}    \forall    {\bm{ \beta } }^{ \mathit{J} } : {\bm{ \ottnt{K'} } }^{ \mathit{J} }    \ottsym{.}    \ottnt{A}   \Rightarrow   \ottnt{B}    \in   \sigma \,  \! [ {\bm{ { S } } }^{ \mathit{I} } / {\bm{ \alpha } }^{ \mathit{I} } ]  $,
            \item $\vdash  \Gamma$,
            \item $\Gamma  \vdash   \bm{ { S } }^{ \mathit{I} } : \bm{ \ottnt{K} }^{ \mathit{I} } $,
            \item $\Gamma  \vdash   \bm{ { T } }^{ \mathit{J} } : \bm{ \ottnt{K'} }^{ \mathit{J} } $,
            \item $\Gamma  \vdash  \ottnt{A_{{\mathrm{1}}}}  <:  \ottnt{A} \,  \! [ {\bm{ { T } } }^{ \mathit{J} } / {\bm{ \beta } }^{ \mathit{J} } ] $,
            \item $\Gamma  \vdash  \ottnt{B} \,  \! [ {\bm{ { T } } }^{ \mathit{J} } / {\bm{ \beta } }^{ \mathit{J} } ]   <:  \ottnt{B_{{\mathrm{1}}}}$, and
            \item $\Gamma  \vdash    \lift{ \mathit{l} \,  \bm{ { S } } ^ {  \mathit{I}  }  }   \olessthan  \varepsilon_{{\mathrm{1}}} $
          \end{itemize}
          for some $ \bm{ { \alpha } } ^ {  \mathit{I}  } $, $ {\bm{ { \ottnt{K} } } }^{ \mathit{I} } $, $\sigma$, $ \bm{ { \beta } } ^ {  \mathit{J}  } $, $ {\bm{ { \ottnt{K'} } } }^{ \mathit{J} } $, $\ottnt{A}$, and $\ottnt{B}$.

    \item\label{lem:inversion:app}
          If $\Gamma  \vdash  \ottnt{v_{{\mathrm{1}}}} \, \ottnt{v_{{\mathrm{2}}}}  \ottsym{:}  \ottnt{B}  \mid  \varepsilon$,
          then there exists some type $\ottnt{A}$
          such that $\Gamma  \vdash  \ottnt{v_{{\mathrm{1}}}}  \ottsym{:}   \ottnt{A}    \rightarrow_{ \varepsilon }    \ottnt{B}   \mid   \bbZero $ and $\Gamma  \vdash  \ottnt{v_{{\mathrm{2}}}}  \ottsym{:}  \ottnt{A}  \mid   \bbZero $.
  \end{enumerate}
\end{lemma}

\begin{proof}
  \phantom{}
  \begin{enumerate}
    \item By induction on a derivation of $\Gamma  \vdash  \ottnt{v}  \ottsym{:}  \ottnt{A}  \mid  \varepsilon$.
          We proceed by cases on the typing rule applied lastly to this derivation.
          \begin{divcases}
            \item[\rname{T}{Var}] Clearly because of $\varepsilon =  \bbZero $.

            \item[\rname{T}{Abs}] Clearly because of $\varepsilon =  \bbZero $.

            \item[\rname{T}{TAbs}] Clearly because of $\varepsilon =  \bbZero $.

            \item[\rname{T}{Op}] Clearly because of $\varepsilon =  \bbZero $.

            \item[\rname{T}{Sub}]
            For some $\ottnt{A'}$ and $\varepsilon'$, the following are given:
            \begin{itemize}
              \item $\Gamma  \vdash  \ottnt{v}  \ottsym{:}  \ottnt{A'}  \mid  \varepsilon'$ and
              \item $\Gamma  \vdash  \ottnt{A'}  \mid  \varepsilon'  <:  \ottnt{A}  \mid  \varepsilon$.
            \end{itemize}
            By the induction hypothesis, $\Gamma  \vdash  \ottnt{v}  \ottsym{:}  \ottnt{A'}  \mid   \bbZero $.
            Since only \rname{ST}{Comp} derives $\Gamma  \vdash  \ottnt{A'}  \mid  \varepsilon'  <:  \ottnt{A}  \mid  \varepsilon$,
            we have $\Gamma  \vdash  \ottnt{A'}  <:  \ottnt{A}$ and $\Gamma  \vdash   \varepsilon'  \olessthan  \varepsilon $.
            Because of Lemma~\ref{lem:entailment}\ref{lem:entailment:refl},
            $\Gamma  \vdash    \bbZero   \olessthan   \bbZero  $ holds.
            By \rname{T}{Sub}, we have $\Gamma  \vdash  \ottnt{v}  \ottsym{:}  \ottnt{A}  \mid   \bbZero $ as required.

            \item[others] Cannot happen.
          \end{divcases}

    \item By induction on a derivation of $\Gamma  \vdash  \ottkw{fun} \, \ottsym{(}  \mathit{f}  \ottsym{,}  \mathit{x}  \ottsym{,}  \ottnt{e}  \ottsym{)}  \ottsym{:}   \ottnt{A_{{\mathrm{1}}}}    \rightarrow_{ \varepsilon_{{\mathrm{1}}} }    \ottnt{B_{{\mathrm{1}}}}   \mid  \varepsilon$.
          We proceed by cases on the typing rule applied lastly to this derivation.
          \begin{divcases}
            \item[\rname{T}{Abs}]
            $\Gamma  \ottsym{,}  \mathit{f}  \ottsym{:}   \ottnt{A_{{\mathrm{1}}}}    \rightarrow_{ \varepsilon_{{\mathrm{1}}} }    \ottnt{B_{{\mathrm{1}}}}   \ottsym{,}  \mathit{x}  \ottsym{:}  \ottnt{A_{{\mathrm{1}}}}  \vdash  \ottnt{e}  \ottsym{:}  \ottnt{B_{{\mathrm{1}}}}  \mid  \varepsilon_{{\mathrm{1}}}$ is given.
            By Lemma~\ref{lem:wk}, we have $\Gamma  \vdash   \ottnt{A_{{\mathrm{1}}}}    \rightarrow_{ \varepsilon_{{\mathrm{1}}} }    \ottnt{B_{{\mathrm{1}}}}   \ottsym{:}   \mathbf{Typ} $.
            Thus, \rname{ST}{Refl} derives $\Gamma  \vdash   \ottnt{A_{{\mathrm{1}}}}    \rightarrow_{ \varepsilon_{{\mathrm{1}}} }    \ottnt{B_{{\mathrm{1}}}}   <:   \ottnt{A_{{\mathrm{1}}}}    \rightarrow_{ \varepsilon_{{\mathrm{1}}} }    \ottnt{B_{{\mathrm{1}}}} $.

            \item[\rname{T}{Sub}]
            For some $\ottnt{C}$ and $\varepsilon'$, the following are given:
            \begin{itemize}
              \item $\Gamma  \vdash  \ottkw{fun} \, \ottsym{(}  \mathit{f}  \ottsym{,}  \mathit{x}  \ottsym{,}  \ottnt{e}  \ottsym{)}  \ottsym{:}  \ottnt{C}  \mid  \varepsilon'$ and
              \item $\Gamma  \vdash  \ottnt{C}  \mid  \varepsilon'  <:   \ottnt{A_{{\mathrm{1}}}}    \rightarrow_{ \varepsilon_{{\mathrm{1}}} }    \ottnt{B_{{\mathrm{1}}}}   \mid  \varepsilon$.
            \end{itemize}
            Since only \rname{ST}{Comp} derives $\Gamma  \vdash  \ottnt{C}  \mid  \varepsilon'  <:   \ottnt{A_{{\mathrm{1}}}}    \rightarrow_{ \varepsilon_{{\mathrm{1}}} }    \ottnt{B_{{\mathrm{1}}}}   \mid  \varepsilon$,
            we have $\Gamma  \vdash  \ottnt{C}  <:   \ottnt{A_{{\mathrm{1}}}}    \rightarrow_{ \varepsilon_{{\mathrm{1}}} }    \ottnt{B_{{\mathrm{1}}}} $.
            By Lemma~\ref{lem:inv_st}\ref{lem:inv_st:abs},
            $\ottnt{C} =  \ottnt{A_{{\mathrm{2}}}}    \rightarrow_{ \varepsilon_{{\mathrm{2}}} }    \ottnt{B_{{\mathrm{2}}}} $ for some $\ottnt{A_{{\mathrm{2}}}}$, $\varepsilon_{{\mathrm{2}}}$, and $\ottnt{B_{{\mathrm{2}}}}$.
            By the induction hypothesis and Lemma~\ref{lem:trans_subtyping},
            the required results are achieved.

            \item[others] Cannot happen.
          \end{divcases}

    \item By induction on a derivation of $\Gamma  \vdash  \Lambda  \alpha  \ottsym{:}  \ottnt{K}  \ottsym{.}  \ottnt{e}  \ottsym{:}    \forall   \alpha  \ottsym{:}  \ottnt{K}   \ottsym{.}    \ottnt{A_{{\mathrm{1}}}}    ^{ \varepsilon_{{\mathrm{1}}} }    \mid  \varepsilon$.
          We proceed by cases on the typing rule applied lastly to this derivation.
          \begin{divcases}
            \item[\rname{T}{TAbs}] Clearly.

            \item[\rname{T}{Sub}]
            For some $\ottnt{B}$ and $\varepsilon'$, the following are given:
            \begin{itemize}
              \item $\Gamma  \vdash  \Lambda  \alpha  \ottsym{:}  \ottnt{K}  \ottsym{.}  \ottnt{e}  \ottsym{:}  \ottnt{B}  \mid  \varepsilon'$ and
              \item $\Gamma  \vdash  \ottnt{B}  \mid  \varepsilon'  <:    \forall   \alpha  \ottsym{:}  \ottnt{K}   \ottsym{.}    \ottnt{A_{{\mathrm{1}}}}    ^{ \varepsilon_{{\mathrm{1}}} }    \mid  \varepsilon$.
            \end{itemize}
            Since only \rname{ST}{Comp} derives
            $\Gamma  \vdash  \ottnt{B}  \mid  \varepsilon'  <:    \forall   \alpha  \ottsym{:}  \ottnt{K}   \ottsym{.}    \ottnt{A_{{\mathrm{1}}}}    ^{ \varepsilon_{{\mathrm{1}}} }    \mid  \varepsilon$,
            we have $\Gamma  \vdash  \ottnt{B}  <:    \forall   \alpha  \ottsym{:}  \ottnt{K}   \ottsym{.}    \ottnt{A_{{\mathrm{1}}}}    ^{ \varepsilon_{{\mathrm{1}}} }  $.
            By Lemma~\ref{lem:inv_st}\ref{lem:inv_st:tabs},
            we have $\ottnt{B} =   \forall   \alpha  \ottsym{:}  \ottnt{K}   \ottsym{.}    \ottnt{A_{{\mathrm{2}}}}    ^{ \varepsilon_{{\mathrm{2}}} }  $
            for some $\ottnt{A_{{\mathrm{2}}}}$ and $\varepsilon_{{\mathrm{2}}}$ such that
            \begin{itemize}
              \item $\Gamma  \ottsym{,}  \alpha  \ottsym{:}  \ottnt{K}  \vdash  \ottnt{A_{{\mathrm{2}}}}  <:  \ottnt{A_{{\mathrm{1}}}}$ and
              \item $\Gamma  \ottsym{,}  \alpha  \ottsym{:}  \ottnt{K}  \vdash   \varepsilon_{{\mathrm{2}}}  \olessthan  \varepsilon_{{\mathrm{1}}} $.
            \end{itemize}
            By the induction hypothesis, we have $\Gamma  \ottsym{,}  \alpha  \ottsym{:}  \ottnt{K}  \vdash  \ottnt{e}  \ottsym{:}  \ottnt{A_{{\mathrm{2}}}}  \mid  \varepsilon_{{\mathrm{2}}}$.
            Thus, \rname{T}{Sub} derives $\Gamma  \ottsym{,}  \alpha  \ottsym{:}  \ottnt{K}  \vdash  \ottnt{e}  \ottsym{:}  \ottnt{A_{{\mathrm{1}}}}  \mid  \varepsilon_{{\mathrm{1}}}$,
            because \rname{ST}{Comp} derives $\Gamma  \ottsym{,}  \alpha  \ottsym{:}  \ottnt{K}  \vdash  \ottnt{A_{{\mathrm{2}}}}  \mid  \varepsilon_{{\mathrm{2}}}  <:  \ottnt{A_{{\mathrm{1}}}}  \mid  \varepsilon_{{\mathrm{1}}}$.

            \item[others] Cannot happen.
          \end{divcases}

    \item By induction on a derivation of $\Gamma  \vdash   \mathsf{op} _{ \mathit{l} \,  \bm{ { S } } ^ {  \mathit{I}  }  }  \,  \bm{ { T } } ^ {  \mathit{J}  }   \ottsym{:}   \ottnt{A_{{\mathrm{1}}}}    \rightarrow_{ \varepsilon_{{\mathrm{1}}} }    \ottnt{B_{{\mathrm{1}}}}   \mid  \varepsilon$.
          We proceed by cases on the typing rule applied lastly to this derivation.
          \begin{divcases}
            \item[\rname{T}{Op}]
            For some $ \bm{ { \alpha } } ^ {  \mathit{I}  } $, $ {\bm{ { \ottnt{K} } } }^{ \mathit{I} } $, $\sigma$, $ \bm{ { \beta } } ^ {  \mathit{J}  } $, $ {\bm{ { \ottnt{K'} } } }^{ \mathit{J} } $, $\ottnt{A}$, and $\ottnt{B}$,
            the following are given:
            \begin{itemize}
              \item $ \mathit{l}  ::    \forall    {\bm{ \alpha } }^{ \mathit{I} } : {\bm{ \ottnt{K} } }^{ \mathit{I} }    \ottsym{.}    \sigma    \in   \Xi $,
              \item $ \mathsf{op}  \ottsym{:}    \forall    {\bm{ \beta } }^{ \mathit{J} } : {\bm{ \ottnt{K'} } }^{ \mathit{J} }    \ottsym{.}    \ottnt{A}   \Rightarrow   \ottnt{B}    \in   \sigma \,  \! [ {\bm{ { S } } }^{ \mathit{I} } / {\bm{ \alpha } }^{ \mathit{I} } ]  $,
              \item $\vdash  \Gamma$,
              \item $\Gamma  \vdash   \bm{ { S } }^{ \mathit{I} } : \bm{ \ottnt{K} }^{ \mathit{I} } $,
              \item $\Gamma  \vdash   \bm{ { T } }^{ \mathit{J} } : \bm{ \ottnt{K'} }^{ \mathit{J} } $,
              \item $\ottnt{A_{{\mathrm{1}}}} = \ottnt{A} \,  \! [ {\bm{ { T } } }^{ \mathit{J} } / {\bm{ \beta } }^{ \mathit{J} } ] $,
              \item $\ottnt{B_{{\mathrm{1}}}} = \ottnt{B} \,  \! [ {\bm{ { T } } }^{ \mathit{J} } / {\bm{ \beta } }^{ \mathit{J} } ] $, and
              \item $\varepsilon_{{\mathrm{1}}} =  \lift{ \mathit{l} \,  \bm{ { S } } ^ {  \mathit{I}  }  } $.
            \end{itemize}
            By Lemma~\ref{lem:wk}, we have $\Gamma  \vdash   \ottnt{A_{{\mathrm{1}}}}    \rightarrow_{ \varepsilon_{{\mathrm{1}}} }    \ottnt{B_{{\mathrm{1}}}}   \ottsym{:}   \mathbf{Typ} $.
            Since only \rname{K}{Fun} can derive $\Gamma  \vdash   \ottnt{A_{{\mathrm{1}}}}    \rightarrow_{ \varepsilon_{{\mathrm{1}}} }    \ottnt{B_{{\mathrm{1}}}}   \ottsym{:}   \mathbf{Typ} $, we have
            \begin{itemize}
              \item $\Gamma  \vdash  \ottnt{A} \,  \! [ {\bm{ { T } } }^{ \mathit{J} } / {\bm{ \beta } }^{ \mathit{J} } ]   \ottsym{:}   \mathbf{Typ} $,
              \item $\Gamma  \vdash   \lift{ \mathit{l} \,  \bm{ { S } } ^ {  \mathit{I}  }  }   \ottsym{:}   \mathbf{Eff} $, and
              \item $\Gamma  \vdash  \ottnt{B} \,  \! [ {\bm{ { T } } }^{ \mathit{J} } / {\bm{ \beta } }^{ \mathit{J} } ]   \ottsym{:}   \mathbf{Typ} $.
            \end{itemize}
            Thus, the required results are achieved
            by \rname{ST}{Refl} and Lemma~\ref{lem:entailment}\ref{lem:entailment:refl}.

            \item[\rname{T}{Sub}]
            For some $\ottnt{C}$ and $\varepsilon'$, the following are given:
            \begin{itemize}
              \item $\Gamma  \vdash   \mathsf{op} _{ \mathit{l} \,  \bm{ { S } } ^ {  \mathit{I}  }  }  \,  \bm{ { T } } ^ {  \mathit{J}  }   \ottsym{:}  \ottnt{C}  \mid  \varepsilon'$ and
              \item $\Gamma  \vdash  \ottnt{C}  \mid  \varepsilon'  <:   \ottnt{A_{{\mathrm{1}}}}    \rightarrow_{ \varepsilon_{{\mathrm{1}}} }    \ottnt{B_{{\mathrm{1}}}}   \mid  \varepsilon$.
            \end{itemize}
            Since only \rname{ST}{Comp} can derive
            $\Gamma  \vdash  \ottnt{C}  \mid  \varepsilon'  <:   \ottnt{A_{{\mathrm{1}}}}    \rightarrow_{ \varepsilon_{{\mathrm{1}}} }    \ottnt{B_{{\mathrm{1}}}}   \mid  \varepsilon$,
            we have $\Gamma  \vdash  \ottnt{C}  <:   \ottnt{A_{{\mathrm{1}}}}    \rightarrow_{ \varepsilon_{{\mathrm{1}}} }    \ottnt{B_{{\mathrm{1}}}} $.
            By Lemma~\ref{lem:inv_st}\ref{lem:inv_st:abs},
            we have $\ottnt{C} =  \ottnt{A_{{\mathrm{2}}}}    \rightarrow_{ \varepsilon_{{\mathrm{2}}} }    \ottnt{B_{{\mathrm{2}}}} $ such that
            \begin{itemize}
              \item $\Gamma  \vdash  \ottnt{A_{{\mathrm{1}}}}  <:  \ottnt{A_{{\mathrm{2}}}}$,
              \item $\Gamma  \vdash  \ottnt{B_{{\mathrm{2}}}}  <:  \ottnt{B_{{\mathrm{1}}}}$, and
              \item $\Gamma  \vdash   \varepsilon_{{\mathrm{2}}}  \olessthan  \varepsilon_{{\mathrm{1}}} $.
            \end{itemize}
            By the induction hypothesis,
            \begin{itemize}
              \item $ \mathit{l}  ::    \forall    {\bm{ \alpha } }^{ \mathit{I} } : {\bm{ \ottnt{K} } }^{ \mathit{I} }    \ottsym{.}    \sigma    \in   \Xi $,
              \item $ \mathsf{op}  \ottsym{:}    \forall    {\bm{ \beta } }^{ \mathit{J} } : {\bm{ \ottnt{K'} } }^{ \mathit{J} }    \ottsym{.}    \ottnt{A}   \Rightarrow   \ottnt{B}    \in   \sigma \,  \! [ {\bm{ { S } } }^{ \mathit{I} } / {\bm{ \alpha } }^{ \mathit{I} } ]  $,
              \item $\vdash  \Gamma$,
              \item $\Gamma  \vdash   \bm{ { S } }^{ \mathit{I} } : \bm{ \ottnt{K} }^{ \mathit{I} } $,
              \item $\Gamma  \vdash   \bm{ { T } }^{ \mathit{J} } : \bm{ \ottnt{K'} }^{ \mathit{J} } $,
              \item $\Gamma  \vdash  \ottnt{A_{{\mathrm{2}}}}  <:  \ottnt{A} \,  \! [ {\bm{ { T } } }^{ \mathit{J} } / {\bm{ \beta } }^{ \mathit{J} } ] $,
              \item $\Gamma  \vdash  \ottnt{B} \,  \! [ {\bm{ { T } } }^{ \mathit{J} } / {\bm{ \beta } }^{ \mathit{J} } ]   <:  \ottnt{B_{{\mathrm{2}}}}$, and
              \item $\Gamma  \vdash    \lift{ \mathit{l} \,  \bm{ { S } } ^ {  \mathit{I}  }  }   \olessthan  \varepsilon_{{\mathrm{2}}} $.
            \end{itemize}
            By Lemma~\ref{lem:trans_subtyping} and Lemma~\ref{lem:entailment}\ref{lem:entailment:trans}, the required result is achieved.

            \item[others] Cannot happen.
          \end{divcases}

    \item By induction on a derivation of $\Gamma  \vdash  \ottnt{v_{{\mathrm{1}}}} \, \ottnt{v_{{\mathrm{2}}}}  \ottsym{:}  \ottnt{B}  \mid  \varepsilon$.
          We proceed by cases on the typing rule applied lastly to this derivation.
          \begin{divcases}
            \item[\rname{T}{App}] Clearly.

            \item[\rname{T}{Sub}]
            For some $\ottnt{B'}$ and $\varepsilon'$, the following are given:
            \begin{itemize}
              \item $\Gamma  \vdash  \ottnt{v_{{\mathrm{1}}}} \, \ottnt{v_{{\mathrm{2}}}}  \ottsym{:}  \ottnt{B'}  \mid  \varepsilon'$ and
              \item $\Gamma  \vdash  \ottnt{B'}  \mid  \varepsilon'  <:  \ottnt{B}  \mid  \varepsilon$.
            \end{itemize}
            By the induction hypothesis, we have
            \begin{itemize}
              \item $\Gamma  \vdash  \ottnt{v_{{\mathrm{1}}}}  \ottsym{:}   \ottnt{A}    \rightarrow_{ \varepsilon' }    \ottnt{B'}   \mid   \bbZero $ and
              \item $\Gamma  \vdash  \ottnt{v_{{\mathrm{2}}}}  \ottsym{:}  \ottnt{A}  \mid   \bbZero $
            \end{itemize}
            for some $\ottnt{A}$.
            By Lemma~\ref{lem:wk}, we have $\Gamma  \vdash  \ottnt{A}  \ottsym{:}   \mathbf{Typ} $ and $\Gamma  \vdash   \bbZero   \ottsym{:}   \mathbf{Eff} $.
            Thus, \rname{ST}{Refl} derives $\Gamma  \vdash  \ottnt{A}  <:  \ottnt{A}$ and
            Lemma~\ref{lem:entailment}\ref{lem:entailment:refl} derives $\Gamma  \vdash    \bbZero   \olessthan   \bbZero  $.
            Therefore, by \rname{ST}{Fun} and \rname{ST}{Comp}, $\Gamma  \vdash   \ottnt{A}    \rightarrow_{ \varepsilon' }    \ottnt{B'}   \mid   \bbZero   <:   \ottnt{A}    \rightarrow_{ \varepsilon }    \ottnt{B}   \mid   \bbZero $.
            Then, by \rname{T}{Sub}, $\Gamma  \vdash  \ottnt{v_{{\mathrm{1}}}}  \ottsym{:}   \ottnt{A}    \rightarrow_{ \varepsilon }    \ottnt{B}   \mid   \bbZero $.

            \item[others] Cannot happen.
          \end{divcases}
  \end{enumerate}
\end{proof}

\begin{lemma}[Canonical Form]\label{lem:canonical}
  \phantom{}
  \begin{enumerate}
    \item\label{lem:canonical:abs} If $\emptyset  \vdash  \ottnt{v}  \ottsym{:}   \ottnt{A}    \rightarrow_{ \varepsilon }    \ottnt{B}   \mid  \varepsilon'$, then either of the following holds:
          \begin{itemize}
            \item $\ottnt{v} = \ottkw{fun} \, \ottsym{(}  \mathit{f}  \ottsym{,}  \mathit{x}  \ottsym{,}  \ottnt{e}  \ottsym{)}$ for some $\mathit{f}$, $\mathit{x}$, and $\ottnt{e}$, or
            \item $\ottnt{v} =  \mathsf{op} _{ \mathit{l} \,  \bm{ { S } } ^ {  \mathit{I}  }  }  \,  \bm{ { T } } ^ {  \mathit{J}  } $ for some $\mathsf{op}$, $\mathit{l}$, $ \bm{ { S } } ^ {  \mathit{I}  } $, and $ \bm{ { T } } ^ {  \mathit{J}  } $.
          \end{itemize}
    \item\label{lem:canonical:tabs} If $\emptyset  \vdash  \ottnt{v}  \ottsym{:}    \forall   \alpha  \ottsym{:}  \ottnt{K}   \ottsym{.}    \ottnt{A}    ^{ \varepsilon }    \mid  \varepsilon'$, then $\ottnt{v} = \Lambda  \alpha  \ottsym{:}  \ottnt{K}  \ottsym{.}  \ottnt{e}$ for some $\ottnt{e}$.
  \end{enumerate}
\end{lemma}

\begin{proof}
  \phantom{}
  \begin{enumerate}
    \item By induction on a derivation of $\Gamma  \vdash  \ottnt{v}  \ottsym{:}   \ottnt{A}    \rightarrow_{ \varepsilon }    \ottnt{B}   \mid  \varepsilon'$. We proceed by cases on the typing rule applied lastly to this derivation.
          \begin{divcases}
            \item[\rname{T}{Var}] Cannot happen.

            \item[\rname{T}{Abs}] Clearly.

            \item[\rname{T}{Sub}]
            For some $\ottnt{C}$, the following are given:
            \begin{itemize}
              \item $\Gamma  \vdash  \ottnt{v}  \ottsym{:}  \ottnt{C}  \mid  \varepsilon''$ and
              \item $\Gamma  \vdash  \ottnt{C}  \mid  \varepsilon''  <:   \ottnt{A}    \rightarrow_{ \varepsilon }    \ottnt{B}   \mid  \varepsilon'$.
            \end{itemize}
            By Lemma~\ref{lem:inv_st}\ref{lem:inv_st:abs},
            we have $\ottnt{C} =  \ottnt{A_{{\mathrm{1}}}}    \rightarrow_{ \varepsilon_{{\mathrm{1}}} }    \ottnt{B_{{\mathrm{1}}}} $ for some $\ottnt{A_{{\mathrm{1}}}}$, $\varepsilon_{{\mathrm{1}}}$, and $\ottnt{B_{{\mathrm{1}}}}$.
            By the induction hypothesis, the required result is achieved.

            \item[\rname{T}{Op}] Clearly.

            \item[others] Cannot happen.
          \end{divcases}

    \item By induction on a derivation of $\Gamma  \vdash  \ottnt{v}  \ottsym{:}    \forall   \alpha  \ottsym{:}  \ottnt{K}   \ottsym{.}    \ottnt{A}    ^{ \varepsilon }    \mid  \varepsilon'$.
          We proceed by cases on the typing rule applied lastly to this derivation.
          \begin{divcases}
            \item[\rname{T}{Var}] Cannot happen.

            \item[\rname{T}{TAbs}] Clearly.

            \item[\rname{T}{Sub}]
            For some $\ottnt{B}$, the following are given:
            \begin{itemize}
              \item $\Gamma  \vdash  \ottnt{v}  \ottsym{:}  \ottnt{B}  \mid  \varepsilon''$ and
              \item $\Gamma  \vdash  \ottnt{B}  \mid  \varepsilon''  <:    \forall   \alpha  \ottsym{:}  \ottnt{K}   \ottsym{.}    \ottnt{A}    ^{ \varepsilon }    \mid  \varepsilon'$.
            \end{itemize}
            By Lemma~\ref{lem:inv_st}\ref{lem:inv_st:tabs},
            we have $\ottnt{B} =   \forall   \alpha  \ottsym{:}  \ottnt{K}   \ottsym{.}    \ottnt{A_{{\mathrm{1}}}}    ^{ \varepsilon_{{\mathrm{1}}} }  $ for some $\ottnt{A_{{\mathrm{1}}}}$ and $\varepsilon_{{\mathrm{1}}}$.
            By the induction hypothesis, the required result is achieved.

            \item[others] Cannot happen.
          \end{divcases}
  \end{enumerate}
\end{proof}

\TY{Change list (1)}
\begin{lemma}[Inversion of Handler Typing]\label{lem:inversion_handler}
  \phantom{}
  \begin{enumerate}
    \item\label{lem:inversion_handler:return}
          If $ \Gamma  \vdash _{ \sigma }  \ottnt{h}  :  \ottnt{A}   \Rightarrow  ^ { \varepsilon }  \ottnt{B} $, then there exist some $\mathit{x}$ and $\ottnt{e_{\ottmv{r}}}$
          such that $ \mathbf{return} \, \mathit{x}  \mapsto  \ottnt{e_{\ottmv{r}}}   \in   \ottnt{h} $ and $\Gamma  \ottsym{,}  \mathit{x}  \ottsym{:}  \ottnt{A}  \vdash  \ottnt{e_{\ottmv{r}}}  \ottsym{:}  \ottnt{B}  \mid  \varepsilon$.

    \item\label{lem:inversion_handler:operation}
          If $ \Gamma  \vdash _{ \sigma }  \ottnt{h}  :  \ottnt{A}   \Rightarrow  ^ { \varepsilon }  \ottnt{B} $ and $ \mathsf{op}  \ottsym{:}    \forall    {\bm{ \beta } }^{ \mathit{J} } : {\bm{ \ottnt{K} } }^{ \mathit{J} }    \ottsym{.}    \ottnt{A'}   \Rightarrow   \ottnt{B'}    \in   \sigma $, then
          \begin{itemize}
            \item $ \mathsf{op} \,  {\bm{ \beta } }^{ \mathit{J} } : {\bm{ \ottnt{K} } }^{ \mathit{J} }  \, \mathit{p} \, \mathit{k}  \mapsto  \ottnt{e}   \in   \ottnt{h} $ and
            \item $\Gamma  \ottsym{,}   {\bm{ \beta } }^{ \mathit{J} } : {\bm{ \ottnt{K} } }^{ \mathit{J} }   \ottsym{,}  \mathit{p}  \ottsym{:}  \ottnt{A'}  \ottsym{,}  \mathit{k}  \ottsym{:}   \ottnt{B'}    \rightarrow_{ \varepsilon }    \ottnt{B}   \vdash  \ottnt{e}  \ottsym{:}  \ottnt{B}  \mid  \varepsilon$
          \end{itemize}
          for some $\mathit{p}$, $\mathit{k}$, and $\ottnt{e}$.
  \end{enumerate}
\end{lemma}

\begin{proof}
  \begin{enumerate}
    \item By induction on a derivation of $ \Gamma  \vdash _{ \sigma }  \ottnt{h}  :  \ottnt{A}   \Rightarrow  ^ { \varepsilon }  \ottnt{B} $.
          We proceed by cases on the typing rule applied lastly to this derivation.
          \begin{divcases}
            \item[\rname{H}{Return}] Clearly.

            \item[\rname{H}{Op}] Clearly by the induction hypothesis.
          \end{divcases}

    \item By induction on a derivation of $ \Gamma  \vdash _{ \sigma }  \ottnt{h}  :  \ottnt{A}   \Rightarrow  ^ { \varepsilon }  \ottnt{B} $.
          We proceed by cases on the typing rule applied lastly to this derivation.
          \begin{divcases}
            \item[\rname{H}{Return}]
            Clearly because there is no operation belonging to $ \{\} $.

            \item[\rname{H}{Op}]
            For some $\ottnt{h'}$, $\sigma'$, $\mathsf{op'}$, $ \bm{ { \beta' } } ^ {  \mathit{J'}  } $, $ {\bm{ { \ottnt{K'} } } }^{ \mathit{J'} } $, $\ottnt{A''}$, $\ottnt{B''}$, $\mathit{p'}$, $\mathit{k'}$, $\ottnt{e''}$, the following are given:
            \begin{itemize}
              \item $\ottnt{h} =  \ottnt{h'}   \uplus   \ottsym{\{}  \mathsf{op'} \,  {\bm{ \beta' } }^{ \mathit{J'} } : {\bm{ \ottnt{K'} } }^{ \mathit{J'} }  \, \mathit{p'} \, \mathit{k'}  \mapsto  \ottnt{e'}  \ottsym{\}} $,
              \item $\sigma =  \sigma'   \uplus   \ottsym{\{}  \mathsf{op'}  \ottsym{:}    \forall    {\bm{ \beta' } }^{ \mathit{J'} } : {\bm{ \ottnt{K'} } }^{ \mathit{J'} }    \ottsym{.}    \ottnt{A''}   \Rightarrow   \ottnt{B''}   \ottsym{\}} $,
              \item $ \Gamma  \vdash _{ \sigma' }  \ottnt{h'}  :  \ottnt{A}   \Rightarrow  ^ { \varepsilon }  \ottnt{B} $, and
              \item $\Gamma  \ottsym{,}   {\bm{ \beta' } }^{ \mathit{J'} } : {\bm{ \ottnt{K'} } }^{ \mathit{J'} }   \ottsym{,}  \mathit{p'}  \ottsym{:}  \ottnt{A''}  \ottsym{,}  \mathit{k'}  \ottsym{:}   \ottnt{B''}    \rightarrow_{ \varepsilon }    \ottnt{B}   \vdash  \ottnt{e}  \ottsym{:}  \ottnt{B}  \mid  \varepsilon$.
            \end{itemize}

            If $\mathsf{op} = \mathsf{op'}$, then clearly.

            If $\mathsf{op} \neq \mathsf{op'}$, then clearly by the induction hypothesis.
          \end{divcases}
  \end{enumerate}
\end{proof}

\begin{lemma}[Independence of Evaluation Contexts]\label{lem:ind_ev}
  If $\Gamma  \vdash  \ottnt{E}  \ottsym{[}  \ottnt{e}  \ottsym{]}  \ottsym{:}  \ottnt{A}  \mid  \varepsilon$, then
  there exist some $\ottnt{A'}$ and $\varepsilon'$ such that
  \begin{itemize}
    \item $\Gamma  \vdash  \ottnt{e}  \ottsym{:}  \ottnt{A'}  \mid  \varepsilon'$, and
    \item $\Gamma  \ottsym{,}  \Gamma'  \vdash  \ottnt{E}  \ottsym{[}  \ottnt{e'}  \ottsym{]}  \ottsym{:}  \ottnt{A}  \mid  \varepsilon$ holds for any $\ottnt{e'}$ and $\Gamma'$ such that $\Gamma  \ottsym{,}  \Gamma'  \vdash  \ottnt{e'}  \ottsym{:}  \ottnt{A'}  \mid  \varepsilon'$.
  \end{itemize}
\end{lemma}

\begin{proof}
  By induction on a derivation of $\Gamma  \vdash  \ottnt{E}  \ottsym{[}  \ottnt{e}  \ottsym{]}  \ottsym{:}  \ottnt{A}  \mid  \varepsilon$. We proceed by cases on the typing rule applied lastly to this derivation.
  \begin{divcases}
    \item[\rname{T}{Let}]
    If $\ottnt{E} =  \Box $, then the required result is achieved immediately.

    If $\ottnt{E} \neq  \Box $, then we have
    \begin{itemize}
      \item $\ottnt{E} = (\mathbf{let} \, \mathit{x}  \ottsym{=}  \ottnt{E'} \, \mathbf{in} \, \ottnt{e_{{\mathrm{2}}}})$,
      \item $\Gamma  \vdash  \ottnt{E'}  \ottsym{[}  \ottnt{e}  \ottsym{]}  \ottsym{:}  \ottnt{B}  \mid  \varepsilon$, and
      \item $\Gamma  \ottsym{,}  \mathit{x}  \ottsym{:}  \ottnt{B}  \vdash  \ottnt{e_{{\mathrm{2}}}}  \ottsym{:}  \ottnt{A}  \mid  \varepsilon$,
    \end{itemize}
    for some $\mathit{x}$, $\ottnt{E'}$, $\ottnt{e_{{\mathrm{2}}}}$, and $\ottnt{B}$.
    By the induction hypothesis,
    there exist some $\ottnt{A'}$ and $\varepsilon'$ such that
    \begin{itemize}
      \item $\Gamma  \vdash  \ottnt{e}  \ottsym{:}  \ottnt{A'}  \mid  \varepsilon'$, and
      \item for any $\ottnt{e'}$ and $\Gamma'$ such that $\Gamma  \ottsym{,}  \Gamma'  \vdash  \ottnt{e'}  \ottsym{:}  \ottnt{A'}  \mid  \varepsilon'$, typing judgment $\Gamma  \ottsym{,}  \Gamma'  \vdash  \ottnt{E'}  \ottsym{[}  \ottnt{e'}  \ottsym{]}  \ottsym{:}  \ottnt{B}  \mid  \varepsilon$ is derivable.
    \end{itemize}
    Let $\ottnt{e'}$ be an expression and $\Gamma'$ be a typing context such that
    $\Gamma  \ottsym{,}  \Gamma'  \vdash  \ottnt{e'}  \ottsym{:}  \ottnt{A'}  \mid  \varepsilon'$.
    Without loss of generality, we can assume $ \mathit{x}   \notin    \mathrm{dom}   \ottsym{(}   \Gamma'   \ottsym{)}  $.
    The induction hypothesis result implies $\Gamma  \ottsym{,}  \Gamma'  \vdash  \ottnt{E'}  \ottsym{[}  \ottnt{e'}  \ottsym{]}  \ottsym{:}  \ottnt{B}  \mid  \varepsilon$.
    By Lemma~\ref{lem:weakening} and \rname{T}{Let}, it suffices to show that
    $\vdash  \Gamma  \ottsym{,}  \Gamma'$, which is implied by Lemma~\ref{lem:ctx-wf-typing}.

    \item[\rname{T}{Sub}]
    For some $\ottnt{A'}$ and $\varepsilon'$, given are the following:
    \begin{itemize}
      \item $\Gamma  \vdash  \ottnt{E}  \ottsym{[}  \ottnt{e}  \ottsym{]}  \ottsym{:}  \ottnt{A'}  \mid  \varepsilon'$ and
      \item $\Gamma  \vdash  \ottnt{A'}  \mid  \varepsilon'  <:  \ottnt{A}  \mid  \varepsilon$.
    \end{itemize}
    By the induction hypothesis, there exist some $\ottnt{A''}$ and $\varepsilon''$ such that
    \begin{itemize}
      \item $\Gamma  \vdash  \ottnt{e}  \ottsym{:}  \ottnt{A''}  \mid  \varepsilon''$, and
      \item for any $\ottnt{e'}$ and $\Gamma'$ such that $\Gamma  \ottsym{,}  \Gamma'  \vdash  \ottnt{e'}  \ottsym{:}  \ottnt{A''}  \mid  \varepsilon''$, typing judgment $\Gamma  \ottsym{,}  \Gamma'  \vdash  \ottnt{E}  \ottsym{[}  \ottnt{e'}  \ottsym{]}  \ottsym{:}  \ottnt{A'}  \mid  \varepsilon'$ is derivable.
    \end{itemize}
    Let $\ottnt{e'}$ be an expression and $\Gamma'$ be a typing context
    such that $\Gamma  \ottsym{,}  \Gamma'  \vdash  \ottnt{e'}  \ottsym{:}  \ottnt{A''}  \mid  \varepsilon''$.
    Since only \rname{ST}{Comp} can derive $\Gamma  \vdash  \ottnt{A'}  \mid  \varepsilon'  <:  \ottnt{A}  \mid  \varepsilon$,
    we have $\Gamma  \vdash  \ottnt{A'}  <:  \ottnt{A}$ and $\Gamma  \vdash   \varepsilon'  \olessthan  \varepsilon $.
    We have $\Gamma  \ottsym{,}  \Gamma'  \vdash  \ottnt{A'}  <:  \ottnt{A}$ and $\Gamma  \ottsym{,}  \Gamma'  \vdash   \varepsilon'  \olessthan  \varepsilon $
    by Lemma~\ref{lem:ctx-wf-typing},
    Lemma~\ref{lem:weakening}\ref{lem:weakening:kinding}, and
    Lemma~\ref{lem:weakening}\ref{lem:weakening:subtyping}.
    Thus, because $\Gamma  \ottsym{,}  \Gamma'  \vdash  \ottnt{E}  \ottsym{[}  \ottnt{e'}  \ottsym{]}  \ottsym{:}  \ottnt{A'}  \mid  \varepsilon'$ by the induction hypothesis result,
    \rname{ST}{Comp} and \rname{T}{Sub} derive $\Gamma  \ottsym{,}  \Gamma'  \vdash  \ottnt{E}  \ottsym{[}  \ottnt{e'}  \ottsym{]}  \ottsym{:}  \ottnt{A}  \mid  \varepsilon$.

    \item[\rname{T}{Handling}]
    If $\ottnt{E} =  \Box $, then the required result is achieved immediately.

    If $\ottnt{E} \neq  \Box $, then we have
    \begin{itemize}
      \item $\ottnt{E} =  \mathbf{handle}_{ \mathit{l} \,  \bm{ { S } } ^ {  \mathit{N}  }  }  \, \ottnt{E'} \, \mathbf{with} \, \ottnt{h}$,
      \item $\Gamma  \vdash  \ottnt{E'}  \ottsym{[}  \ottnt{e}  \ottsym{]}  \ottsym{:}  \ottnt{A'}  \mid  \varepsilon'$, and
      \item $   \lift{ \mathit{l} \,  \bm{ { S } } ^ {  \mathit{N}  }  }   \mathop{ \odot }  \varepsilon    \sim   \varepsilon' $,
    \end{itemize}
    for some $\mathit{l}$, $ \bm{ { S } } ^ {  \mathit{N}  } $, $\ottnt{E'}$, $\ottnt{h}$, $\ottnt{A'}$, and $\varepsilon'$.
    By the induction hypothesis, there exist some $\ottnt{A''}$ and $\varepsilon''$ such that
    \begin{itemize}
      \item $\Gamma  \vdash  \ottnt{e}  \ottsym{:}  \ottnt{A''}  \mid  \varepsilon''$, and
      \item for any $\ottnt{e'}$ and $\Gamma'$ such that $\Gamma  \ottsym{,}  \Gamma'  \vdash  \ottnt{e'}  \ottsym{:}  \ottnt{A''}  \mid  \varepsilon''$, typing judgment $\Gamma  \ottsym{,}  \Gamma'  \vdash  \ottnt{E'}  \ottsym{[}  \ottnt{e'}  \ottsym{]}  \ottsym{:}  \ottnt{A'}  \mid  \varepsilon'$ is derivable.
    \end{itemize}
    Because the premises of \rname{T}{Handling} other than the typing of handled expressions
    are independent of the handled expressions,
    the required result is achieved by Lemma~\ref{lem:ctx-wf-typing}, Lemma~\ref{lem:weakening}, and Lemma~\ref{lem:delta_context}\ref{lem:delta_context:kinding}.

    \item[others] Clearly because $\ottnt{E} =  \Box $.
  \end{divcases}
\end{proof}

\begin{lemma}[Progress]\label{lem:progress}
  If $\emptyset  \vdash  \ottnt{e}  \ottsym{:}  \ottnt{A}  \mid  \varepsilon$, then one of the following holds:
  \begin{itemize}
    \item $\ottnt{e}$ is a value;

    \item There exists some expression $\ottnt{e'}$ such that $\ottnt{e}  \longrightarrow  \ottnt{e'}$; or

    \item There exist some $\mathsf{op}$, $\mathit{l}$, $ \bm{ { S } } ^ {  \mathit{I}  } $, $ \bm{ { T } } ^ {  \mathit{J}  } $, $\ottnt{v}$, $\ottnt{E}$, and $\mathit{n}$
          such that $\ottnt{e} = \ottnt{E}  \ottsym{[}   \mathsf{op} _{ \mathit{l} \,  \bm{ { S } } ^ {  \mathit{I}  }  }  \,  \bm{ { T } } ^ {  \mathit{J}  }  \, \ottnt{v}  \ottsym{]}$ and $ \mathit{n}  \mathrm{-free} ( \mathit{l} \,  \bm{ { S } } ^ {  \mathit{I}  }  ,  \ottnt{E} ) $.
  \end{itemize}
\end{lemma}

\begin{proof}
  By induction on a derivation of $\emptyset  \vdash  \ottnt{e}  \ottsym{:}  \ottnt{A}  \mid  \varepsilon$. We proceed by cases on the typing rule applied lastly to this derivation.
  \begin{divcases}
    \item[\rname{T}{Var}] Cannot happen.

    \item[\rname{T}{Abs}] $\ottnt{e}$ is a value because of $\ottnt{e} = \ottkw{fun} \, \ottsym{(}  \mathit{f_{{\mathrm{1}}}}  \ottsym{,}  \mathit{x_{{\mathrm{1}}}}  \ottsym{,}  \ottnt{e_{{\mathrm{1}}}}  \ottsym{)}$ for some $\mathit{f_{{\mathrm{1}}}}$, $\mathit{x_{{\mathrm{1}}}}$, and $\ottnt{e_{{\mathrm{1}}}}$.

    \item[\rname{T}{App}]
    For some $\ottnt{v_{{\mathrm{1}}}}$, $\ottnt{v_{{\mathrm{2}}}}$, and $\ottnt{B}$, the following are given:
    \begin{itemize}
      \item $\ottnt{e} = \ottnt{v_{{\mathrm{1}}}} \, \ottnt{v_{{\mathrm{2}}}}$,
      \item $\emptyset  \vdash  \ottnt{v_{{\mathrm{1}}}}  \ottsym{:}   \ottnt{B}    \rightarrow_{ \varepsilon }    \ottnt{A}   \mid   \bbZero $, and
      \item $\emptyset  \vdash  \ottnt{v_{{\mathrm{2}}}}  \ottsym{:}  \ottnt{B}  \mid   \bbZero $.
    \end{itemize}
    By case analysis on the result of Lemma~\ref{lem:canonical}\ref{lem:canonical:abs} on $\emptyset  \vdash  \ottnt{v_{{\mathrm{1}}}}  \ottsym{:}   \ottnt{B}    \rightarrow_{ \varepsilon }    \ottnt{A}   \mid   \bbZero $.

    If $\ottnt{v_{{\mathrm{1}}}} = \ottkw{fun} \, \ottsym{(}  \mathit{f_{{\mathrm{1}}}}  \ottsym{,}  \mathit{x_{{\mathrm{1}}}}  \ottsym{,}  \ottnt{e_{{\mathrm{1}}}}  \ottsym{)}$ for some $\mathit{f_{{\mathrm{1}}}}$, $\mathit{x_{{\mathrm{1}}}}$, and $\ottnt{e_{{\mathrm{1}}}}$, then \rname{R}{App} derives $\ottnt{e}  \longmapsto  \ottnt{e_{{\mathrm{1}}}} \,  \! [  \ottkw{fun} \, \ottsym{(}  \mathit{f_{{\mathrm{1}}}}  \ottsym{,}  \mathit{x_{{\mathrm{1}}}}  \ottsym{,}  \ottnt{e_{{\mathrm{1}}}}  \ottsym{)}  /  \mathit{f_{{\mathrm{1}}}}  ]  \,  \! [  \ottnt{v_{{\mathrm{2}}}}  /  \mathit{x}  ] $.

    If $\ottnt{v_{{\mathrm{1}}}} =  \mathsf{op} _{ \mathit{l} \,  \bm{ { S } } ^ {  \mathit{I}  }  }  \,  \bm{ { T } } ^ {  \mathit{J}  } $ for some $\mathsf{op}$, $\mathit{l}$, $ \bm{ { S } } ^ {  \mathit{I}  } $, $ \bm{ { T } } ^ {  \mathit{J}  } $, then
    the required result is implied by Lemma~\ref{lem:inversion}\ref{lem:inversion:op} and the fact that
    $\ottnt{e} = \Box  \ottsym{[}   \mathsf{op} _{ \mathit{l} \,  \bm{ { S } } ^ {  \mathit{I}  }  }  \,  \bm{ { T } } ^ {  \mathit{J}  }  \, \ottnt{v_{{\mathrm{2}}}}  \ottsym{]}$.

    \item[\rname{T}{TAbs}] $\ottnt{e}$ is a value because of $\ottnt{e} = \Lambda  \alpha  \ottsym{:}  \ottnt{K}  \ottsym{.}  \ottnt{e_{{\mathrm{1}}}}$ for some $\alpha$, $\ottnt{K}$, and $\ottnt{e_{{\mathrm{1}}}}$.

    \item[\rname{T}{TApp}]
    For some $\ottnt{v}$, $\alpha$, $S$, $\ottnt{K}$, $\ottnt{A_{{\mathrm{1}}}}$, and $\varepsilon_{{\mathrm{1}}}$, the following are given:
    \begin{itemize}
      \item $\ottnt{e} = \ottnt{v} \, S$,
      \item $\ottnt{A} = \ottnt{A_{{\mathrm{1}}}} \,  \! [  S  /  \alpha   ] $,
      \item $\varepsilon = \varepsilon_{{\mathrm{1}}} \,  \! [  S  /  \alpha   ] $,
      \item $\emptyset  \vdash  \ottnt{v}  \ottsym{:}    \forall   \alpha  \ottsym{:}  \ottnt{K}   \ottsym{.}    \ottnt{A_{{\mathrm{1}}}}    ^{ \varepsilon_{{\mathrm{1}}} }    \mid   \bbZero $, and
      \item $\emptyset  \vdash  S  \ottsym{:}  \ottnt{K}$.
    \end{itemize}
    By Lemma~\ref{lem:canonical}\ref{lem:canonical:tabs}, we have $\ottnt{v} = \Lambda  \alpha  \ottsym{:}  \ottnt{K}  \ottsym{.}  \ottnt{e_{{\mathrm{1}}}}$ for some $\ottnt{e_{{\mathrm{1}}}}$.
    Thus, \rname{R}{TApp} derives $\ottnt{e}  \longmapsto  \ottnt{e_{{\mathrm{1}}}} \,  \! [  S  /  \alpha   ] $.

    \item[\rname{T}{Let}]
    For some $\mathit{x}$, $\ottnt{e_{{\mathrm{1}}}}$, $\ottnt{e_{{\mathrm{2}}}}$, and $\ottnt{B}$, given are the following:
    \begin{itemize}
      \item $\ottnt{e} = (\mathbf{let} \, \mathit{x}  \ottsym{=}  \ottnt{e_{{\mathrm{1}}}} \, \mathbf{in} \, \ottnt{e_{{\mathrm{2}}}})$,
      \item $\emptyset  \vdash  \ottnt{e_{{\mathrm{1}}}}  \ottsym{:}  \ottnt{B}  \mid  \varepsilon$, and
      \item $\mathit{x}  \ottsym{:}  \ottnt{B}  \vdash  \ottnt{e_{{\mathrm{2}}}}  \ottsym{:}  \ottnt{A}  \mid  \varepsilon$.
    \end{itemize}
    By the induction hypothesis, we proceed by cases on the following conditions:
    \begin{enumerate}
      \item $\ottnt{e_{{\mathrm{1}}}}$ is a value,
      \item There exists some $\ottnt{e'_{{\mathrm{1}}}}$ such that $\ottnt{e_{{\mathrm{1}}}}  \longrightarrow  \ottnt{e'_{{\mathrm{1}}}}$,
      \item There exist some $\mathsf{op}$, $\mathit{l}$, $ \bm{ { S } } ^ {  \mathit{I}  } $, $ \bm{ { T } } ^ {  \mathit{J}  } $, $\ottnt{v}$, $\ottnt{E}$, and $\mathit{n}$ such that $\ottnt{e_{{\mathrm{1}}}} = \ottnt{E}  \ottsym{[}   \mathsf{op} _{ \mathit{l} \,  \bm{ { S } } ^ {  \mathit{I}  }  }  \,  \bm{ { T } } ^ {  \mathit{J}  }  \, \ottnt{v}  \ottsym{]}$ and $ \mathit{n}  \mathrm{-free} ( \mathit{l} \,  \bm{ { S } } ^ {  \mathit{I}  }  ,  \ottnt{E} ) $.
    \end{enumerate}
    \begin{divcases}
      \item[(1)] \rname{R}{Let} derives $\ottnt{e}  \longmapsto  \ottnt{e_{{\mathrm{2}}}} \,  \! [  \ottnt{v_{{\mathrm{1}}}}  /  \mathit{x}  ] $ because $\ottnt{e_{{\mathrm{1}}}}$ is a value $\ottnt{v_{{\mathrm{1}}}}$.

      \item[(2)] Since only \rname{E}{Eval} can derive $\ottnt{e_{{\mathrm{1}}}}  \longrightarrow  \ottnt{e'_{{\mathrm{1}}}}$, we have
      \begin{itemize}
        \item $\ottnt{e_{{\mathrm{1}}}} = \ottnt{E_{{\mathrm{1}}}}  \ottsym{[}  \ottnt{e_{{\mathrm{11}}}}  \ottsym{]}$,
        \item $\ottnt{e'_{{\mathrm{1}}}} = \ottnt{E_{{\mathrm{1}}}}  \ottsym{[}  \ottnt{e_{{\mathrm{12}}}}  \ottsym{]}$, and
        \item $\ottnt{e_{{\mathrm{11}}}}  \longmapsto  \ottnt{e_{{\mathrm{12}}}}$,
      \end{itemize}
      for some $\ottnt{E_{{\mathrm{1}}}}$, $\ottnt{e_{{\mathrm{11}}}}$, and $\ottnt{e_{{\mathrm{12}}}}$.
      Let $\ottnt{E} = (\mathbf{let} \, \mathit{x}  \ottsym{=}  \ottnt{E_{{\mathrm{1}}}} \, \mathbf{in} \, \ottnt{e_{{\mathrm{2}}}})$.
      \rname{E}{Eval} derives $\ottnt{e}  \longrightarrow  \ottnt{E}  \ottsym{[}  \ottnt{e_{{\mathrm{12}}}}  \ottsym{]}$ because of $\ottnt{e} = \ottnt{E}  \ottsym{[}  \ottnt{e_{{\mathrm{11}}}}  \ottsym{]}$.

      \item[(3)] Clearly because $\ottnt{e} = (\mathbf{let} \, \mathit{x}  \ottsym{=}  \ottnt{E}  \ottsym{[}   \mathsf{op} _{ \mathit{l} \,  \bm{ { S } } ^ {  \mathit{I}  }  }  \,  \bm{ { T } } ^ {  \mathit{J}  }  \, \ottnt{v}  \ottsym{]} \, \mathbf{in} \, \ottnt{e_{{\mathrm{2}}}})$ and $ \mathit{n}  \mathrm{-free} ( \mathit{l} \,  \bm{ { S } } ^ {  \mathit{I}  }  ,  \mathbf{let} \, \mathit{x}  \ottsym{=}  \ottnt{E} \, \mathbf{in} \, \ottnt{e_{{\mathrm{2}}}} ) $.
    \end{divcases}

    \item[\rname{T}{Sub}] Clearly by the induction hypothesis.

    \item[\rname{T}{Op}] $\ottnt{e}$ is a value because of $\ottnt{e} =  \mathsf{op} _{ \mathit{l} \,  \bm{ { S } } ^ {  \mathit{I}  }  }  \,  \bm{ { T } } ^ {  \mathit{J}  } $ for some $\mathsf{op}$, $\mathit{l}$, $ \bm{ { S } } ^ {  \mathit{I}  } $, and $ \bm{ { T } } ^ {  \mathit{J}  } $.

    \item[\rname{T}{Handling}] \TY{Change list (1)}
    For some $\ottnt{e_{{\mathrm{1}}}}$, $\ottnt{h}$, $\mathit{l}$, $ \bm{ { S } } ^ {  \mathit{N}  } $, $ \bm{ { \alpha } } ^ {  \mathit{N}  } $, $ {\bm{ { \ottnt{K} } } }^{ \mathit{N} } $, $\ottnt{A_{{\mathrm{1}}}}$, and $\varepsilon_{{\mathrm{1}}}$, given are the following:
    \begin{itemize}
      \item $\ottnt{e} =  \mathbf{handle}_{ \mathit{l} \,  \bm{ { S } } ^ {  \mathit{N}  }  }  \, \ottnt{e_{{\mathrm{1}}}} \, \mathbf{with} \, \ottnt{h}$,
      \item $\emptyset  \vdash  \ottnt{e_{{\mathrm{1}}}}  \ottsym{:}  \ottnt{A_{{\mathrm{1}}}}  \mid  \varepsilon_{{\mathrm{1}}}$,
      \item $ \mathit{l}  ::    \forall    {\bm{ \alpha } }^{ \mathit{N} } : {\bm{ \ottnt{K} } }^{ \mathit{N} }    \ottsym{.}    \sigma    \in   \Xi $,
      \item $\emptyset  \vdash   \bm{ { S } }^{ \mathit{N} } : \bm{ \ottnt{K} }^{ \mathit{N} } $,
      \item $ \emptyset  \vdash _{ \sigma \,  \! [ {\bm{ { S } } }^{ \mathit{N} } / {\bm{ \alpha } }^{ \mathit{N} } ]  }  \ottnt{h}  :  \ottnt{A_{{\mathrm{1}}}}   \Rightarrow  ^ { \varepsilon }  \ottnt{A} $, and
      \item $   \lift{ \mathit{l} \,  \bm{ { S } } ^ {  \mathit{N}  }  }   \mathop{ \odot }  \varepsilon    \sim   \varepsilon_{{\mathrm{1}}} $.
    \end{itemize}
    By the induction hypothesis, we proceed by cases on the following conditions:
    \begin{enumerate}
      \item $\ottnt{e_{{\mathrm{1}}}}$ is a value,
      \item There exists some $\ottnt{e'_{{\mathrm{1}}}}$ such that $\ottnt{e_{{\mathrm{1}}}}  \longrightarrow  \ottnt{e'_{{\mathrm{1}}}}$,
      \item There exist some $\mathsf{op'}$, $\mathit{l'}$, $ \bm{ { S' } } ^ {  \mathit{N'}  } $, $ \bm{ { T } } ^ {  \mathit{J}  } $, $\ottnt{v}$, $\ottnt{E}$, and $\mathit{n}$ such that $\ottnt{e_{{\mathrm{1}}}} = \ottnt{E}  \ottsym{[}   \mathsf{op'} _{ \mathit{l'} \,  \bm{ { S' } } ^ {  \mathit{N'}  }  }  \,  \bm{ { T } } ^ {  \mathit{J}  }  \, \ottnt{v}  \ottsym{]}$ and $ \mathit{n}  \mathrm{-free} ( \mathit{l'} \,  \bm{ { S' } } ^ {  \mathit{N'}  }  ,  \ottnt{E} ) $.
    \end{enumerate}
    \begin{divcases}
      \item[(1)]
      By Lemma~\ref{lem:inversion_handler}\ref{lem:inversion_handler:return},
      there exists some $\mathit{x}$ and $\ottnt{e_{\ottmv{r}}}$ such that $ \mathbf{return} \, \mathit{x}  \mapsto  \ottnt{e_{\ottmv{r}}}   \in   \ottnt{h} $.
      Thus, \rname{R}{Handle1} derives $\ottnt{e}  \longmapsto  \ottnt{e_{\ottmv{r}}} \,  \! [  \ottnt{v_{{\mathrm{1}}}}  /  \mathit{x}  ] $
      because $\ottnt{e_{{\mathrm{1}}}}$ is a value $\ottnt{v_{{\mathrm{1}}}}$.

      \item[(2)]
      Since only \rname{E}{Eval} can derive $\ottnt{e_{{\mathrm{1}}}}  \longrightarrow  \ottnt{e'_{{\mathrm{1}}}}$, we have
      \begin{itemize}
        \item $\ottnt{e_{{\mathrm{1}}}} = \ottnt{E_{{\mathrm{1}}}}  \ottsym{[}  \ottnt{e_{{\mathrm{11}}}}  \ottsym{]}$,
        \item $\ottnt{e'_{{\mathrm{1}}}} = \ottnt{E_{{\mathrm{1}}}}  \ottsym{[}  \ottnt{e_{{\mathrm{12}}}}  \ottsym{]}$, and
        \item $\ottnt{e_{{\mathrm{11}}}}  \longmapsto  \ottnt{e_{{\mathrm{12}}}}$,
      \end{itemize}
      for some $\ottnt{E}$, $\ottnt{e_{{\mathrm{11}}}}$, and $\ottnt{e_{{\mathrm{12}}}}$.
      Let $\ottnt{E} =  \mathbf{handle}_{ \mathit{l} \,  \bm{ { S } } ^ {  \mathit{N}  }  }  \, \ottnt{E_{{\mathrm{1}}}} \, \mathbf{with} \, \ottnt{h}$.
      \rname{E}{Eval} derives $\ottnt{e}  \longrightarrow  \ottnt{E}  \ottsym{[}  \ottnt{e_{{\mathrm{12}}}}  \ottsym{]}$ because of $\ottnt{e} = \ottnt{E}  \ottsym{[}  \ottnt{e_{{\mathrm{11}}}}  \ottsym{]}$.

      \item[(3)]
      If $ \mathit{l} \,  \bm{ { S } } ^ {  \mathit{N}  }    \neq   \mathit{l'} \,  \bm{ { S' } } ^ {  \mathit{N'}  }  $,
      then $\ottnt{e} = \ottsym{(}   \mathbf{handle}_{ \mathit{l} \,  \bm{ { S } } ^ {  \mathit{N}  }  }  \, \ottnt{E} \, \mathbf{with} \, \ottnt{h}  \ottsym{)}  \ottsym{[}   \mathsf{op} _{ \mathit{l'} \,  \bm{ { S' } } ^ {  \mathit{N'}  }  }  \,  \bm{ { T } } ^ {  \mathit{J}  }  \, \ottnt{v}  \ottsym{]}$ and
      $ \mathit{n}  \mathrm{-free} ( \mathit{l'} \,  \bm{ { S' } } ^ {  \mathit{N'}  }  ,   \mathbf{handle}_{ \mathit{l} \,  \bm{ { S } } ^ {  \mathit{N}  }  }  \, \ottnt{E} \, \mathbf{with} \, \ottnt{h} ) $.

      If $\mathit{l} \,  \bm{ { S } } ^ {  \mathit{N}  }  = \mathit{l'} \,  \bm{ { S' } } ^ {  \mathit{N'}  } $,
      then by Lemma~\ref{lem:ind_ev} and \ref{lem:inversion}\ref{lem:inversion:op}, we have
      \begin{itemize}
        \item $ \mathit{l'}  ::    \forall    {\bm{ \alpha' } }^{ \mathit{N'} } : {\bm{ \ottnt{K'} } }^{ \mathit{N'} }    \ottsym{.}    \sigma'    \in   \Xi $ and
        \item $ \mathsf{op'}  \ottsym{:}    \forall    {\bm{ \beta' } }^{ \mathit{J} } : {\bm{ \ottnt{K'_{{\mathrm{0}}}} } }^{ \mathit{J} }    \ottsym{.}    \ottnt{A'}   \Rightarrow   \ottnt{B'}    \in   \sigma' \,  \! [ {\bm{ { S' } } }^{ \mathit{N'} } / {\bm{ \alpha' } }^{ \mathit{N'} } ]  $,
      \end{itemize}
      for some $ \bm{ { \alpha' } } ^ {  \mathit{N'}  } $, $ {\bm{ { \ottnt{K'} } } }^{ \mathit{N'} } $, $\sigma'$, $ \bm{ { \beta' } } ^ {  \mathit{J}  } $, $\ottnt{A'}$, and $\ottnt{B'}$.
      Therefore, since $\mathit{l} \,  \bm{ { S } } ^ {  \mathit{N}  }  = \mathit{l'} \,  \bm{ { S' } } ^ {  \mathit{N'}  } $, we have
      \begin{itemize}
        \item $\sigma = \sigma'$,
        \item $ \bm{ { \alpha } } ^ {  \mathit{N}  }  =  \bm{ { \alpha' } } ^ {  \mathit{N'}  } $, and
        \item $ {\bm{ { \ottnt{K} } } }^{ \mathit{N} }  =  {\bm{ { \ottnt{K_{{\mathrm{0}}}} } } }^{ \mathit{N'} } $.
      \end{itemize}
      By $ \emptyset  \vdash _{ \sigma \,  \! [ {\bm{ { S } } }^{ \mathit{N} } / {\bm{ \alpha } }^{ \mathit{N} } ]  }  \ottnt{h}  :  \ottnt{A_{{\mathrm{1}}}}   \Rightarrow  ^ { \varepsilon }  \ottnt{A} $ and
      $ \mathsf{op'}  \ottsym{:}    \forall    {\bm{ \beta' } }^{ \mathit{J} } : {\bm{ \ottnt{K'_{{\mathrm{0}}}} } }^{ \mathit{J} }    \ottsym{.}    \ottnt{A'}   \Rightarrow   \ottnt{B'}    \in   \sigma \,  \! [ {\bm{ { S } } }^{ \mathit{N} } / {\bm{ \alpha } }^{ \mathit{N} } ]  $ and
      Lemma~\ref{lem:inversion_handler}\ref{lem:inversion_handler:operation},
      we have
      \begin{align*}
         \mathsf{op'} \,  {\bm{ \beta' } }^{ \mathit{J} } : {\bm{ \ottnt{K'_{{\mathrm{0}}}} } }^{ \mathit{J} }  \, \mathit{p} \, \mathit{k}  \mapsto  \ottnt{e'}   \in   \ottnt{h} 
      \end{align*}
      for some $\mathit{p}$, $\mathit{k}$, and $\ottnt{e'}$.
      If $\mathit{n} = 0$, the evaluation of $\ottnt{e}$ proceeds by \rname{R}{Handle2}.
      Otherwise, there exists some $\mathit{m}$ such that
      $\mathit{n} = \mathit{m}  \ottsym{+}  1$ and $ \mathit{m}  \mathrm{-free} ( \mathit{l} \,  \bm{ { S } } ^ {  \mathit{N}  }  ,   \mathbf{handle}_{ \mathit{l} \,  \bm{ { S } } ^ {  \mathit{N}  }  }  \, \ottnt{E} \, \mathbf{with} \, \ottnt{h} ) $.
    \end{divcases}
  \end{divcases}
\end{proof}

\begin{lemma}[Preservation in Reduction]\label{lem:pres_red}
  If $\emptyset  \vdash  \ottnt{e}  \ottsym{:}  \ottnt{A}  \mid  \varepsilon$ and $\ottnt{e}  \longmapsto  \ottnt{e'}$, then $\emptyset  \vdash  \ottnt{e'}  \ottsym{:}  \ottnt{A}  \mid  \varepsilon$.
\end{lemma}

\begin{proof}
  By induction on a derivation of $\Gamma  \vdash  \ottnt{e}  \ottsym{:}  \ottnt{A}  \mid  \varepsilon$.
  We proceed by cases on the typing rule applied lastly to this derivation.
  \begin{divcases}
    \item[\rname{T}{Var}] There is no $\ottnt{e'}$ such that $\ottnt{e}  \longmapsto  \ottnt{e'}$.

    \item[\rname{T}{Abs}] There is no $\ottnt{e'}$ such that $\ottnt{e}  \longmapsto  \ottnt{e'}$.

    \item[\rname{T}{App}]
    Since only \rname{R}{App} can derive $\ottnt{e}  \longmapsto  \ottnt{e'}$, we have
    \begin{itemize}
      \item $\ottnt{e} =  (  \ottkw{fun} \, \ottsym{(}  \mathit{f_{{\mathrm{1}}}}  \ottsym{,}  \mathit{x_{{\mathrm{1}}}}  \ottsym{,}  \ottnt{e_{{\mathrm{1}}}}  \ottsym{)}  )  \, \ottnt{v_{{\mathrm{2}}}}$,
      \item $\emptyset  \vdash  \ottkw{fun} \, \ottsym{(}  \mathit{f_{{\mathrm{1}}}}  \ottsym{,}  \mathit{x_{{\mathrm{1}}}}  \ottsym{,}  \ottnt{e_{{\mathrm{1}}}}  \ottsym{)}  \ottsym{:}   \ottnt{A_{{\mathrm{1}}}}    \rightarrow_{ \varepsilon }    \ottnt{A}   \mid   \bbZero $,
      \item $\emptyset  \vdash  \ottnt{v_{{\mathrm{2}}}}  \ottsym{:}  \ottnt{A_{{\mathrm{1}}}}  \mid   \bbZero $, and
      \item $\ottnt{e'} = \ottnt{e_{{\mathrm{1}}}} \,  \! [  \ottkw{fun} \, \ottsym{(}  \mathit{f_{{\mathrm{1}}}}  \ottsym{,}  \mathit{x_{{\mathrm{1}}}}  \ottsym{,}  \ottnt{e_{{\mathrm{1}}}}  \ottsym{)}  /  \mathit{f_{{\mathrm{1}}}}  ]  \,  \! [  \ottnt{v_{{\mathrm{2}}}}  /  \mathit{x_{{\mathrm{1}}}}  ] $
    \end{itemize}
    for some $\mathit{f_{{\mathrm{1}}}}$, $\mathit{x_{{\mathrm{1}}}}$, $\ottnt{e_{{\mathrm{1}}}}$, $\ottnt{v_{{\mathrm{2}}}}$, and $\ottnt{A_{{\mathrm{1}}}}$.
    By Lemma~\ref{lem:inversion}\ref{lem:inversion:abs}, we have
    \begin{itemize}
      \item $\mathit{f_{{\mathrm{1}}}}  \ottsym{:}   \ottnt{A_{{\mathrm{2}}}}    \rightarrow_{ \varepsilon_{{\mathrm{2}}} }    \ottnt{B_{{\mathrm{2}}}}   \ottsym{,}  \mathit{x_{{\mathrm{1}}}}  \ottsym{:}  \ottnt{A_{{\mathrm{2}}}}  \vdash  \ottnt{e_{{\mathrm{1}}}}  \ottsym{:}  \ottnt{B_{{\mathrm{2}}}}  \mid  \varepsilon_{{\mathrm{2}}}$ and
      \item $\emptyset  \vdash   \ottnt{A_{{\mathrm{2}}}}    \rightarrow_{ \varepsilon_{{\mathrm{2}}} }    \ottnt{B_{{\mathrm{2}}}}   <:   \ottnt{A_{{\mathrm{1}}}}    \rightarrow_{ \varepsilon }    \ottnt{A} $.
    \end{itemize}
    for some $\ottnt{A_{{\mathrm{2}}}}$, $\varepsilon_{{\mathrm{2}}}$, and $\ottnt{B_{{\mathrm{2}}}}$.
    Thus, \rname{T}{Abs} derives
    $\emptyset  \vdash  \ottkw{fun} \, \ottsym{(}  \mathit{f_{{\mathrm{1}}}}  \ottsym{,}  \mathit{x_{{\mathrm{1}}}}  \ottsym{,}  \ottnt{e_{{\mathrm{1}}}}  \ottsym{)}  \ottsym{:}   \ottnt{A_{{\mathrm{2}}}}    \rightarrow_{ \varepsilon_{{\mathrm{2}}} }    \ottnt{B_{{\mathrm{2}}}}   \mid   \bbZero $.
    By Lemma~\ref{lem:inv_st}\ref{lem:inv_st:abs}, we have
    \begin{itemize}
      \item $\emptyset  \vdash  \ottnt{A_{{\mathrm{1}}}}  <:  \ottnt{A_{{\mathrm{2}}}}$,
      \item $\emptyset  \vdash  \ottnt{B_{{\mathrm{2}}}}  <:  \ottnt{A}$, and
      \item $\emptyset  \vdash   \varepsilon_{{\mathrm{2}}}  \olessthan  \varepsilon $.
    \end{itemize}
    By Lemma~\ref{lem:ctx-wf-typing} and Lemma~\ref{lem:weakening}\ref{lem:weakening:subtyping},
    we have $\mathit{f_{{\mathrm{1}}}}  \ottsym{:}   \ottnt{A_{{\mathrm{2}}}}    \rightarrow_{ \varepsilon_{{\mathrm{2}}} }    \ottnt{B_{{\mathrm{2}}}}   \ottsym{,}  \mathit{x_{{\mathrm{1}}}}  \ottsym{:}  \ottnt{A_{{\mathrm{2}}}}  \vdash  \ottnt{B_{{\mathrm{2}}}}  <:  \ottnt{A}$.
    Because Lemma~\ref{lem:weakening}\ref{lem:weakening:kinding},
    \rname{ST}{Comp} derives $\mathit{f_{{\mathrm{1}}}}  \ottsym{:}   \ottnt{A_{{\mathrm{2}}}}    \rightarrow_{ \varepsilon_{{\mathrm{2}}} }    \ottnt{B_{{\mathrm{2}}}}   \ottsym{,}  \mathit{x_{{\mathrm{1}}}}  \ottsym{:}  \ottnt{A_{{\mathrm{2}}}}  \vdash  \ottnt{B_{{\mathrm{2}}}}  \mid  \varepsilon_{{\mathrm{2}}}  <:  \ottnt{A}  \mid  \varepsilon$.
    Therefore, \rname{T}{Sub} derives $\mathit{f_{{\mathrm{1}}}}  \ottsym{:}   \ottnt{A_{{\mathrm{2}}}}    \rightarrow_{ \varepsilon_{{\mathrm{2}}} }    \ottnt{B_{{\mathrm{2}}}}   \ottsym{,}  \mathit{x_{{\mathrm{1}}}}  \ottsym{:}  \ottnt{A_{{\mathrm{2}}}}  \vdash  \ottnt{e_{{\mathrm{1}}}}  \ottsym{:}  \ottnt{A}  \mid  \varepsilon$.
    Since \rname{T}{Sub} derives $\emptyset  \vdash  \ottnt{v_{{\mathrm{2}}}}  \ottsym{:}  \ottnt{A_{{\mathrm{2}}}}  \mid   \bbZero $,
    Lemma~\ref{lem:subst_value}\ref{lem:subst_value:typing} makes
    $\emptyset  \vdash  \ottnt{e_{{\mathrm{1}}}} \,  \! [  \ottkw{fun} \, \ottsym{(}  \mathit{f_{{\mathrm{1}}}}  \ottsym{,}  \mathit{x_{{\mathrm{1}}}}  \ottsym{,}  \ottnt{e_{{\mathrm{1}}}}  \ottsym{)}  /  \mathit{f_{{\mathrm{1}}}}  ]  \,  \! [  \ottnt{v_{{\mathrm{2}}}}  /  \mathit{x_{{\mathrm{1}}}}  ]   \ottsym{:}  \ottnt{A}  \mid  \varepsilon$ hold as required.

    \item[\rname{T}{TAbs}] There is no $\ottnt{e'}$ such that $\ottnt{e}  \longmapsto  \ottnt{e'}$.

    \item[\rname{T}{TApp}]
    Since only \rname{R}{TApp} derives $\ottnt{e}  \longmapsto  \ottnt{e'}$, we have
    \begin{itemize}
      \item $\ottnt{e} = \ottsym{(}  \Lambda  \alpha  \ottsym{:}  \ottnt{K}  \ottsym{.}  \ottnt{e_{{\mathrm{1}}}}  \ottsym{)} \, S$,
      \item $\ottnt{A} = \ottnt{A_{{\mathrm{1}}}} \,  \! [  S  /  \alpha   ] $,
      \item $\varepsilon = \varepsilon_{{\mathrm{1}}} \,  \! [  S  /  \alpha   ] $,
      \item $\emptyset  \vdash  \Lambda  \alpha  \ottsym{:}  \ottnt{K}  \ottsym{.}  \ottnt{e_{{\mathrm{1}}}}  \ottsym{:}    \forall   \alpha  \ottsym{:}  \ottnt{K}   \ottsym{.}    \ottnt{A_{{\mathrm{1}}}}    ^{ \varepsilon_{{\mathrm{1}}} }    \mid   \bbZero $,
      \item $\emptyset  \vdash  S  \ottsym{:}  \ottnt{K}$, and
      \item $\ottnt{e'} = \ottnt{e_{{\mathrm{1}}}} \,  \! [  S  /  \alpha   ] $
    \end{itemize}
    for some $\alpha$, $\ottnt{K}$, $\ottnt{e_{{\mathrm{1}}}}$, $S$, $\ottnt{A_{{\mathrm{1}}}}$, and $\varepsilon_{{\mathrm{1}}}$.
    By Lemma~\ref{lem:inversion}\ref{lem:inversion:tabs},
    we have $\alpha  \ottsym{:}  \ottnt{K}  \vdash  \ottnt{e_{{\mathrm{1}}}}  \ottsym{:}  \ottnt{A_{{\mathrm{1}}}}  \mid  \varepsilon_{{\mathrm{1}}}$.
    Thus, Lemma~\ref{lem:subst_type}\ref{lem:subst_type:typing} makes
    $\emptyset  \vdash  \ottnt{e_{{\mathrm{1}}}} \,  \! [  S  /  \alpha   ]   \ottsym{:}  \ottnt{A_{{\mathrm{1}}}} \,  \! [  S  /  \alpha   ]   \mid  \varepsilon_{{\mathrm{1}}} \,  \! [  S  /  \alpha   ] $ hold as required.

    \item[\rname{T}{Let}]
    Since only \rname{R}{Let} derives $\ottnt{e}  \longmapsto  \ottnt{e'}$, we have
    \begin{itemize}
      \item $\ottnt{e} = (\mathbf{let} \, \mathit{x}  \ottsym{=}  \ottnt{v} \, \mathbf{in} \, \ottnt{e_{{\mathrm{1}}}})$,
      \item $\emptyset  \vdash  \ottnt{v}  \ottsym{:}  \ottnt{B}  \mid  \varepsilon$,
      \item $\mathit{x}  \ottsym{:}  \ottnt{B}  \vdash  \ottnt{e_{{\mathrm{1}}}}  \ottsym{:}  \ottnt{A}  \mid  \varepsilon$, and
      \item $\ottnt{e'} = \ottnt{e_{{\mathrm{1}}}} \,  \! [  \ottnt{v}  /  \mathit{x}  ] $
    \end{itemize}
    for some $\mathit{x}$, $\ottnt{v}$, $\ottnt{e_{{\mathrm{1}}}}$, and $\ottnt{B}$.
    By Lemma~\ref{lem:inversion}\ref{lem:inversion:var} and Lemma~\ref{lem:subst_value}\ref{lem:subst_value:typing},
    we have $\emptyset  \vdash  \ottnt{e_{{\mathrm{1}}}} \,  \! [  \ottnt{v}  /  \mathit{x}  ]   \ottsym{:}  \ottnt{A}  \mid  \varepsilon$ as required.

    \item[\rname{T}{Sub}] For some $\ottnt{A'}$ and $\varepsilon'$, we have
    \begin{itemize}
      \item $\emptyset  \vdash  \ottnt{e}  \ottsym{:}  \ottnt{A'}  \mid  \varepsilon'$ and
      \item $\emptyset  \vdash  \ottnt{A'}  \mid  \varepsilon'  <:  \ottnt{A}  \mid  \varepsilon$.
    \end{itemize}
    By the induction hypothesis, we have $\emptyset  \vdash  \ottnt{e'}  \ottsym{:}  \ottnt{A'}  \mid  \varepsilon'$.
    Thus, \rname{T}{Sub} derives $\emptyset  \vdash  \ottnt{e'}  \ottsym{:}  \ottnt{A}  \mid  \varepsilon$ as required.

    \item[\rname{T}{Op}] There is no $\ottnt{e'}$ such that $\ottnt{e}  \longmapsto  \ottnt{e'}$.

    \item[\rname{T}{Handling}] \TY{Change list (1)}
    We proceed by cases on the derivation rule which derives $\ottnt{e}  \longmapsto  \ottnt{e'}$.
    \begin{divcases}
      \item[\rname{R}{Handle1}] We have
      \begin{itemize}
        \item $\ottnt{e} =  \mathbf{handle}_{ \mathit{l} \,  \bm{ { S } } ^ {  \mathit{I}  }  }  \, \ottnt{v} \, \mathbf{with} \, \ottnt{h}$,
        \item $ \mathbf{return} \, \mathit{x}  \mapsto  \ottnt{e_{\ottmv{r}}}   \in   \ottnt{h} $,
        \item $\emptyset  \vdash  \ottnt{v}  \ottsym{:}  \ottnt{B}  \mid  \varepsilon'$,
        \item $ \mathit{l}  ::    \forall    {\bm{ \alpha } }^{ \mathit{I} } : {\bm{ \ottnt{K} } }^{ \mathit{I} }    \ottsym{.}    \sigma    \in   \Xi $,
        \item $\Gamma  \vdash   \bm{ { S } }^{ \mathit{I} } : \bm{ \ottnt{K} }^{ \mathit{I} } $,
        \item $ \emptyset  \vdash _{ \sigma \,  \! [ {\bm{ { S } } }^{ \mathit{I} } / {\bm{ \alpha } }^{ \mathit{I} } ]  }  \ottnt{h}  :  \ottnt{B}   \Rightarrow  ^ { \varepsilon }  \ottnt{A} $,
        \item $   \lift{ \mathit{l} \,  \bm{ { S } } ^ {  \mathit{I}  }  }   \mathop{ \odot }  \varepsilon    \sim   \varepsilon' $, and
        \item $\ottnt{e'} = \ottnt{e_{\ottmv{r}}} \,  \! [  \ottnt{v}  /  \mathit{x}  ] $
      \end{itemize}
      for some $\mathit{l}$, $ \bm{ { S } } ^ {  \mathit{I}  } $, $ \bm{ { \alpha } } ^ {  \mathit{I}  } $, $ {\bm{ { \ottnt{K} } } }^{ \mathit{I} } $, $\sigma$, $\ottnt{v}$, $\ottnt{h}$, $\ottnt{B}$, and $\varepsilon'$.
      By $ \emptyset  \vdash _{ \sigma \,  \! [ {\bm{ { S } } }^{ \mathit{I} } / {\bm{ \alpha } }^{ \mathit{I} } ]  }  \ottnt{h}  :  \ottnt{B}   \Rightarrow  ^ { \varepsilon }  \ottnt{A} $ and
      $ \mathbf{return} \, \mathit{x}  \mapsto  \ottnt{e_{\ottmv{r}}}   \in   \ottnt{h} $ and
      Lemma~\ref{lem:inversion_handler}\ref{lem:inversion_handler:return},
      we have
      \begin{align*}
        \mathit{x}  \ottsym{:}  \ottnt{B}  \vdash  \ottnt{e_{\ottmv{r}}}  \ottsym{:}  \ottnt{A}  \mid  \varepsilon.
      \end{align*}
      By Lemma~\ref{lem:inversion}\ref{lem:inversion:var},
      we have $\emptyset  \vdash  \ottnt{v}  \ottsym{:}  \ottnt{B}  \mid   \bbZero $.
      Thus, Lemma~\ref{lem:subst_value}\ref{lem:subst_value:typing} makes
      $\emptyset  \vdash  \ottnt{e_{\ottmv{r}}} \,  \! [  \ottnt{v}  /  \mathit{x}  ]   \ottsym{:}  \ottnt{A}  \mid  \varepsilon$ hold as required.

      \item[\rname{R}{Handle2}] We have
      \begin{itemize}
        \item $\ottnt{e} =  \mathbf{handle}_{ \mathit{l} \,  \bm{ { S } } ^ {  \mathit{N}  }  }  \, \ottnt{E}  \ottsym{[}   \mathsf{op_{{\mathrm{0}}}} _{ \mathit{l} \,  \bm{ { S } } ^ {  \mathit{N}  }  }  \,  \bm{ { T } } ^ {  \mathit{J}  }  \, \ottnt{v}  \ottsym{]} \, \mathbf{with} \, \ottnt{h}$,
        \item $ \mathit{l}  ::    \forall    {\bm{ \alpha } }^{ \mathit{N} } : {\bm{ \ottnt{K} } }^{ \mathit{N} }    \ottsym{.}    \sigma    \in   \Xi $,
        \item $\emptyset  \vdash   \bm{ { S } }^{ \mathit{N} } : \bm{ \ottnt{K} }^{ \mathit{N} } $,
        \item $ \mathsf{op_{{\mathrm{0}}}} \,  {\bm{ \beta_{{\mathrm{0}}} } }^{ \mathit{J} } : {\bm{ \ottnt{K_{{\mathrm{0}}}} } }^{ \mathit{J} }  \, \mathit{p_{{\mathrm{0}}}} \, \mathit{k_{{\mathrm{0}}}}  \mapsto  \ottnt{e_{{\mathrm{0}}}}   \in   \ottnt{h} $,
        \item $ 0  \mathrm{-free} ( \mathit{l} \,  \bm{ { S } } ^ {  \mathit{N}  }  ,  \ottnt{E} ) $,
        \item $\emptyset  \vdash  \ottnt{E}  \ottsym{[}   \mathsf{op_{{\mathrm{0}}}} _{ \mathit{l} \,  \bm{ { S } } ^ {  \mathit{N}  }  }  \,  \bm{ { T } } ^ {  \mathit{J}  }  \, \ottnt{v}  \ottsym{]}  \ottsym{:}  \ottnt{B}  \mid  \varepsilon'$,
        \item $ \emptyset  \vdash _{ \sigma \,  \! [ {\bm{ { S } } }^{ \mathit{N} } / {\bm{ \alpha } }^{ \mathit{N} } ]  }  \ottnt{h}  :  \ottnt{B}   \Rightarrow  ^ { \varepsilon }  \ottnt{A} $,
        \item $   \lift{ \mathit{l} \,  \bm{ { S } } ^ {  \mathit{N}  }  }   \mathop{ \odot }  \varepsilon    \sim   \varepsilon' $, and
        \item $\ottnt{e'} = \ottnt{e_{{\mathrm{0}}}} \,  \! [ {\bm{ { T } } }^{ \mathit{J} } / {\bm{ \beta_{{\mathrm{0}}} } }^{ \mathit{J} } ]  \,  \! [  \ottnt{v}  /  \mathit{p_{{\mathrm{0}}}}  ]  \,  \! [  \lambda  \mathit{z}  \ottsym{.}   \mathbf{handle}_{ \mathit{l} \,  \bm{ { S } } ^ {  \mathit{N}  }  }  \, \ottnt{E}  \ottsym{[}  \mathit{z}  \ottsym{]} \, \mathbf{with} \, \ottnt{h}  /  \mathit{k_{{\mathrm{0}}}}  ] $
      \end{itemize}
      for some $\mathit{l}$, $ \bm{ { S } } ^ {  \mathit{N}  } $, $\ottnt{E}$, $\mathsf{op_{{\mathrm{0}}}}$, $ \bm{ { T } } ^ {  \mathit{J}  } $, $\ottnt{v}$, $\ottnt{h}$, $ \bm{ { \alpha } } ^ {  \mathit{N}  } $, $ {\bm{ { \ottnt{K} } } }^{ \mathit{N} } $, $\sigma$, $ \bm{ { \beta_{{\mathrm{0}}} } } ^ {  \mathit{J}  } $, $ {\bm{ { \ottnt{K_{{\mathrm{0}}}} } } }^{ \mathit{J} } $, $\mathit{p_{{\mathrm{0}}}}$, $\mathit{k_{{\mathrm{0}}}}$, $\ottnt{e_{{\mathrm{0}}}}$, $\ottnt{B}$, and $\varepsilon'$.
      By Lemma~\ref{lem:ind_ev}, there exist some $\ottnt{B_{{\mathrm{1}}}}$ and $\varepsilon_{{\mathrm{1}}}$ such that
      \begin{itemize}
        \item $\emptyset  \vdash   \mathsf{op_{{\mathrm{0}}}} _{ \mathit{l} \,  \bm{ { S } } ^ {  \mathit{N}  }  }  \,  \bm{ { T } } ^ {  \mathit{J}  }  \, \ottnt{v}  \ottsym{:}  \ottnt{B_{{\mathrm{1}}}}  \mid  \varepsilon_{{\mathrm{1}}}$, and
        \item for any $\ottnt{e''}$ and $\Gamma''$,
              if $\Gamma''  \vdash  \ottnt{e''}  \ottsym{:}  \ottnt{B_{{\mathrm{1}}}}  \mid  \varepsilon_{{\mathrm{1}}}$,
              then $\Gamma''  \vdash  \ottnt{E}  \ottsym{[}  \ottnt{e''}  \ottsym{]}  \ottsym{:}  \ottnt{B}  \mid  \varepsilon'$.
      \end{itemize}
      By Lemma~\ref{lem:inversion}\ref{lem:inversion:app},
      we have $\emptyset  \vdash   \mathsf{op_{{\mathrm{0}}}} _{ \mathit{l} \,  \bm{ { S } } ^ {  \mathit{N}  }  }  \,  \bm{ { T } } ^ {  \mathit{J}  }   \ottsym{:}   \ottnt{A_{{\mathrm{1}}}}    \rightarrow_{ \varepsilon_{{\mathrm{1}}} }    \ottnt{B_{{\mathrm{1}}}}   \mid   \bbZero $ and
      $\emptyset  \vdash  \ottnt{v}  \ottsym{:}  \ottnt{A_{{\mathrm{1}}}}  \mid   \bbZero $ for some $\ottnt{A_{{\mathrm{1}}}}$.
      By Lemma~\ref{lem:inversion}\ref{lem:inversion:op} and
      \ref{lem:inversion_handler}\ref{lem:inversion_handler:operation},
      we have
      \begin{itemize}
        \item $ \mathsf{op_{{\mathrm{0}}}}  \ottsym{:}    \forall    {\bm{ \beta_{{\mathrm{0}}} } }^{ \mathit{J} } : {\bm{ \ottnt{K_{{\mathrm{0}}}} } }^{ \mathit{J} }    \ottsym{.}    \ottnt{A_{{\mathrm{0}}}}   \Rightarrow   \ottnt{B_{{\mathrm{0}}}}    \in   \sigma \,  \! [ {\bm{ { S } } }^{ \mathit{N} } / {\bm{ \alpha } }^{ \mathit{N} } ]  $,
        \item $\emptyset  \vdash   \bm{ { S } }^{ \mathit{N} } : \bm{ \ottnt{K} }^{ \mathit{N} } $,
        \item $\emptyset  \vdash   \bm{ { T } }^{ \mathit{J} } : \bm{ \ottnt{K_{{\mathrm{0}}}} }^{ \mathit{J} } $,
        \item $\emptyset  \vdash  \ottnt{A_{{\mathrm{1}}}}  <:  \ottnt{A_{{\mathrm{0}}}} \,  \! [ {\bm{ { T } } }^{ \mathit{J} } / {\bm{ \beta_{{\mathrm{0}}} } }^{ \mathit{J} } ] $,
        \item $\emptyset  \vdash  \ottnt{B_{{\mathrm{0}}}} \,  \! [ {\bm{ { T } } }^{ \mathit{J} } / {\bm{ \beta_{{\mathrm{0}}} } }^{ \mathit{J} } ]   <:  \ottnt{B_{{\mathrm{1}}}}$, and
        \item $\emptyset  \vdash    \lift{ \mathit{l} \,  \bm{ { S } } ^ {  \mathit{N}  }  }   \olessthan  \varepsilon_{{\mathrm{1}}} $.
      \end{itemize}
      for some $\ottnt{A_{{\mathrm{0}}}}$ and $\ottnt{B_{{\mathrm{0}}}}$.
      Thus, \rname{T}{Sub} with $\emptyset  \vdash    \bbZero   \olessthan   \bbZero  $ implied by Lemma~\ref{lem:entailment} derives
      \begin{align*}
        \emptyset  \vdash  \ottnt{v}  \ottsym{:}  \ottnt{A_{{\mathrm{0}}}} \,  \! [ {\bm{ { T } } }^{ \mathit{J} } / {\bm{ \beta_{{\mathrm{0}}} } }^{ \mathit{J} } ]   \mid   \bbZero .
      \end{align*}
      By Lemma~\ref{lem:wk_subtyping}, we have $\emptyset  \vdash  \ottnt{B_{{\mathrm{0}}}} \,  \! [ {\bm{ { T } } }^{ \mathit{J} } / {\bm{ \beta_{{\mathrm{0}}} } }^{ \mathit{J} } ]   \ottsym{:}   \mathbf{Typ} $.
      Thus, \rname{C}{Var} derives $\vdash  \mathit{z}  \ottsym{:}  \ottnt{B_{{\mathrm{0}}}} \,  \! [ {\bm{ { T } } }^{ \mathit{J} } / {\bm{ \beta_{{\mathrm{0}}} } }^{ \mathit{J} } ] $.
      By $\emptyset  \vdash   \bbZero   \ottsym{:}   \mathbf{Eff} $, $\emptyset  \vdash  \varepsilon_{{\mathrm{1}}}  \ottsym{:}   \mathbf{Eff} $ implied by Lemma~\ref{lem:wk}, and $   \bbZero   \mathop{ \odot }  \varepsilon_{{\mathrm{1}}}    \sim   \varepsilon_{{\mathrm{1}}} $,
      we have $\emptyset  \vdash    \bbZero   \olessthan  \varepsilon_{{\mathrm{1}}} $.
      Since \rname{T}{Var} and \rname{T}{Sub} derives $\mathit{z}  \ottsym{:}  \ottnt{B_{{\mathrm{0}}}} \,  \! [ {\bm{ { T } } }^{ \mathit{J} } / {\bm{ \beta_{{\mathrm{0}}} } }^{ \mathit{J} } ]   \vdash  \mathit{z}  \ottsym{:}  \ottnt{B_{{\mathrm{1}}}}  \mid  \varepsilon_{{\mathrm{1}}}$,
      we have
      \begin{align*}
        \mathit{z}  \ottsym{:}  \ottnt{B_{{\mathrm{0}}}} \,  \! [ {\bm{ { T } } }^{ \mathit{J} } / {\bm{ \beta_{{\mathrm{0}}} } }^{ \mathit{J} } ]   \vdash   \mathbf{handle}_{ \mathit{l} \,  \bm{ { S } } ^ {  \mathit{N}  }  }  \, \ottnt{E}  \ottsym{[}  \mathit{z}  \ottsym{]} \, \mathbf{with} \, \ottnt{h}  \ottsym{:}  \ottnt{A}  \mid  \varepsilon
      \end{align*}
      by the result of Lemma~\ref{lem:ind_ev}, Lemma~\ref{lem:weakening}, and \rname{T}{Handling}.
      Thus, \rname{T}{Abs} derives
      \begin{align*}
        \emptyset  \vdash  \lambda  \mathit{z}  \ottsym{.}   \mathbf{handle}_{ \mathit{l} \,  \bm{ { S } } ^ {  \mathit{N}  }  }  \, \ottnt{E}  \ottsym{[}  \mathit{z}  \ottsym{]} \, \mathbf{with} \, \ottnt{h}  \ottsym{:}   \ottnt{B_{{\mathrm{0}}}} \,  \! [ {\bm{ { T } } }^{ \mathit{J} } / {\bm{ \beta_{{\mathrm{0}}} } }^{ \mathit{J} } ]     \rightarrow_{ \varepsilon }    \ottnt{A}   \mid   \bbZero .
      \end{align*}
      Since
      \[
         {\bm{ \beta_{{\mathrm{0}}} } }^{ \mathit{J} } : {\bm{ \ottnt{K_{{\mathrm{0}}}} } }^{ \mathit{J} }   \ottsym{,}  \mathit{p_{{\mathrm{0}}}}  \ottsym{:}  \ottnt{A_{{\mathrm{0}}}}  \ottsym{,}  \mathit{k_{{\mathrm{0}}}}  \ottsym{:}   \ottnt{B_{{\mathrm{0}}}}    \rightarrow_{ \varepsilon }    \ottnt{A}   \vdash  \ottnt{e_{{\mathrm{0}}}}  \ottsym{:}  \ottnt{A}  \mid  \varepsilon
      \]
      by $ \emptyset  \vdash _{ \sigma \,  \! [ {\bm{ { S } } }^{ \mathit{N} } / {\bm{ \alpha } }^{ \mathit{N} } ]  }  \ottnt{h}  :  \ottnt{B}   \Rightarrow  ^ { \varepsilon }  \ottnt{A} $ and
      $ \mathsf{op_{{\mathrm{0}}}}  \ottsym{:}    \forall    {\bm{ \beta_{{\mathrm{0}}} } }^{ \mathit{J} } : {\bm{ \ottnt{K_{{\mathrm{0}}}} } }^{ \mathit{J} }    \ottsym{.}    \ottnt{A_{{\mathrm{0}}}}   \Rightarrow   \ottnt{B_{{\mathrm{0}}}}    \in   \sigma \,  \! [ {\bm{ { S } } }^{ \mathit{N} } / {\bm{ \alpha } }^{ \mathit{N} } ]  $ and
      Lemma~\ref{lem:inversion_handler}\ref{lem:inversion_handler:operation},
      Lemma~\ref{lem:subst_type}\ref{lem:subst_type:typing} and Lemma~\ref{lem:subst_value}\ref{lem:subst_value:typing} imply
      \begin{align*}
        \emptyset  \vdash  \ottnt{e_{{\mathrm{0}}}} \,  \! [ {\bm{ { T } } }^{ \mathit{J} } / {\bm{ \beta_{{\mathrm{0}}} } }^{ \mathit{J} } ]  \,  \! [  \ottnt{v}  /  \mathit{p_{{\mathrm{0}}}}  ]  \,  \! [  \lambda  \mathit{z}  \ottsym{.}   \mathbf{handle}_{ \mathit{l} \,  \bm{ { S } } ^ {  \mathit{N}  }  }  \, \ottnt{E}  \ottsym{[}  \mathit{z}  \ottsym{]} \, \mathbf{with} \, \ottnt{h}  /  \mathit{k_{{\mathrm{0}}}}  ]   \ottsym{:}  \ottnt{A}  \mid  \varepsilon
      \end{align*}
      as required.
    \end{divcases}
  \end{divcases}
\end{proof}

\begin{lemma}[Preservation]\label{lem:preservation}
  If $\emptyset  \vdash  \ottnt{e}  \ottsym{:}  \ottnt{A}  \mid  \varepsilon$ and $\ottnt{e}  \longrightarrow  \ottnt{e'}$, then $\emptyset  \vdash  \ottnt{e'}  \ottsym{:}  \ottnt{A}  \mid  \varepsilon$.
\end{lemma}

\begin{proof}
  Since only \rname{E}{Eval} derives $\ottnt{e}  \longrightarrow  \ottnt{e'}$, we have
  \begin{itemize}
    \item $\ottnt{e} = \ottnt{E}  \ottsym{[}  \ottnt{e_{{\mathrm{1}}}}  \ottsym{]}$,
    \item $\ottnt{e'} = \ottnt{E}  \ottsym{[}  \ottnt{e_{{\mathrm{2}}}}  \ottsym{]}$, and
    \item $\ottnt{e_{{\mathrm{1}}}}  \longmapsto  \ottnt{e_{{\mathrm{2}}}}$.
  \end{itemize}
  By Lemma~\ref{lem:ind_ev}, there exist some $\ottnt{A'}$ and $\varepsilon'$ such that
  \begin{itemize}
    \item $\emptyset  \vdash  \ottnt{e_{{\mathrm{1}}}}  \ottsym{:}  \ottnt{A'}  \mid  \varepsilon'$, and
    \item for any $\ottnt{e'_{{\mathrm{1}}}}$ and $\Gamma'$, if $\Gamma'  \vdash  \ottnt{e'_{{\mathrm{1}}}}  \ottsym{:}  \ottnt{A'}  \mid  \varepsilon'$,
          then $\Gamma'  \vdash  \ottnt{E}  \ottsym{[}  \ottnt{e'_{{\mathrm{1}}}}  \ottsym{]}  \ottsym{:}  \ottnt{A}  \mid  \varepsilon$.
  \end{itemize}
  By Lemma~\ref{lem:pres_red}, we have $\emptyset  \vdash  \ottnt{e_{{\mathrm{2}}}}  \ottsym{:}  \ottnt{A'}  \mid  \varepsilon'$.
  Thus, $\emptyset  \vdash  \ottnt{E}  \ottsym{[}  \ottnt{e_{{\mathrm{2}}}}  \ottsym{]}  \ottsym{:}  \ottnt{A}  \mid  \varepsilon$ holds as required.
\end{proof}

\begin{lemma}\label{lem:zero_freeness}
  If $ \mathit{n}  \mathrm{-free} ( \ottnt{L} ,  \ottnt{E} ) $, then $\mathit{n} = \mathtt{0}$.
\end{lemma}

\begin{proof}
  Straightforward by the induction on the derivation of $ \mathit{n}  \mathrm{-free} ( \ottnt{L} ,  \ottnt{E} ) $.
\end{proof}

\begin{lemma}\label{lem:effsafe_aux}
  If $\Gamma  \vdash  \ottnt{E}  \ottsym{[}   \mathsf{op} _{ \mathit{l} \,  \bm{ { S } } ^ {  \mathit{I}  }  }  \,  \bm{ { T } } ^ {  \mathit{J}  }  \, \ottnt{v}  \ottsym{]}  \ottsym{:}  \ottnt{A}  \mid  \varepsilon$ and
  $ \mathit{n}  \mathrm{-free} ( \mathit{l} \,  \bm{ { S } } ^ {  \mathit{I}  }  ,  \ottnt{E} ) $,
  then $  \lift{ \mathit{l} \,  \bm{ { S } } ^ {  \mathit{I}  }  }   \olessthan  \varepsilon $.
\end{lemma}

\begin{proof}
  By induction on a derivation of $\Gamma  \vdash  \ottnt{E}  \ottsym{[}   \mathsf{op} _{ \mathit{l} \,  \bm{ { S } } ^ {  \mathit{I}  }  }  \,  \bm{ { T } } ^ {  \mathit{J}  }  \, \ottnt{v}  \ottsym{]}  \ottsym{:}  \ottnt{A}  \mid  \varepsilon$.
  We proceed by case analysis on the typing rule applied lastly to this derivation.
  \begin{divcases}
    \item[\rname{T}{App}]
    For some $\ottnt{B}$, we have
    \begin{itemize}
      \item $\ottnt{E} =  \Box $,
      \item $\Gamma  \vdash   \mathsf{op} _{ \mathit{l} \,  \bm{ { S } } ^ {  \mathit{I}  }  }  \,  \bm{ { T } } ^ {  \mathit{J}  }   \ottsym{:}   \ottnt{B}    \rightarrow_{ \varepsilon }    \ottnt{A}   \mid   \bbZero $, and
      \item $\Gamma  \vdash  \ottnt{v}  \ottsym{:}  \ottnt{B}  \mid   \bbZero $.
    \end{itemize}
    By Lemma~\ref{lem:inversion}\ref{lem:inversion:op},
    we have $\Gamma  \vdash    \lift{ \mathit{l} \,  \bm{ { S } } ^ {  \mathit{I}  }  }   \olessthan  \varepsilon $.
    Thus, the required result is achieved.

    \item[\rname{T}{Let}]
    For some $\mathit{x}$, $\ottnt{E_{{\mathrm{1}}}}$, $\ottnt{e}$, and $\ottnt{B}$, we have
    \begin{itemize}
      \item $\ottnt{E} = (\mathbf{let} \, \mathit{x}  \ottsym{=}  \ottnt{E_{{\mathrm{1}}}} \, \mathbf{in} \, \ottnt{e})$,
      \item $\Gamma  \vdash  \ottnt{E_{{\mathrm{1}}}}  \ottsym{[}   \mathsf{op} _{ \mathit{l} \,  \bm{ { S } } ^ {  \mathit{I}  }  }  \,  \bm{ { T } } ^ {  \mathit{J}  }  \, \ottnt{v}  \ottsym{]}  \ottsym{:}  \ottnt{B}  \mid  \varepsilon$, and
      \item $\Gamma  \ottsym{,}  \mathit{x}  \ottsym{:}  \ottnt{B}  \vdash  \ottnt{e}  \ottsym{:}  \ottnt{A}  \mid  \varepsilon$.
    \end{itemize}
    By $ \mathit{n}  \mathrm{-free} ( \mathit{l} \,  \bm{ { S } } ^ {  \mathit{I}  }  ,  \ottnt{E_{{\mathrm{1}}}} ) $ and the induction hypothesis,
    we have $  \lift{ \mathit{l} \,  \bm{ { S } } ^ {  \mathit{I}  }  }   \olessthan  \varepsilon $ as required.

    \item[\rname{T}{Sub}]
    For some $\ottnt{A'}$ and $\varepsilon'$, we have
    \begin{itemize}
      \item $\Gamma  \vdash  \ottnt{E}  \ottsym{[}   \mathsf{op} _{ \mathit{l} \,  \bm{ { S } } ^ {  \mathit{I}  }  }  \,  \bm{ { T } } ^ {  \mathit{J}  }  \, \ottnt{v}  \ottsym{]}  \ottsym{:}  \ottnt{A'}  \mid  \varepsilon'$ and
      \item $\Gamma  \vdash  \ottnt{A'}  \mid  \varepsilon'  <:  \ottnt{A}  \mid  \varepsilon$.
    \end{itemize}
    Since only \rname{ST}{Comp} can derive $\Gamma  \vdash  \ottnt{A'}  \mid  \varepsilon'  <:  \ottnt{A}  \mid  \varepsilon$,
    we have $\Gamma  \vdash   \varepsilon'  \olessthan  \varepsilon $.
    By the induction hypothesis, we have $  \lift{ \mathit{l} \,  \bm{ { S } } ^ {  \mathit{I}  }  }   \olessthan  \varepsilon' $.
    By the associativity of $ \odot $, we have $  \lift{ \mathit{l} \,  \bm{ { S } } ^ {  \mathit{I}  }  }   \olessthan  \varepsilon $ as required.

    \item[\rname{T}{Handling}]
    For some $\mathit{l'}$, $ \bm{ { S' } } ^ {  \mathit{I'}  } $, $\ottnt{E_{{\mathrm{1}}}}$, $\ottnt{h}$, $\ottnt{B}$, and $\varepsilon'$, we have
    \begin{itemize}
      \item $\ottnt{E} =  \mathbf{handle}_{ \mathit{l'} \,  \bm{ { S' } } ^ {  \mathit{I'}  }  }  \, \ottnt{E_{{\mathrm{1}}}} \, \mathbf{with} \, \ottnt{h}$,
      \item $\Gamma  \vdash  \ottnt{E_{{\mathrm{1}}}}  \ottsym{[}   \mathsf{op} _{ \mathit{l} \,  \bm{ { S } } ^ {  \mathit{I}  }  }  \,  \bm{ { T } } ^ {  \mathit{J}  }  \, \ottnt{v}  \ottsym{]}  \ottsym{:}  \ottnt{B}  \mid  \varepsilon'$, and
      \item $   \lift{ \mathit{l'} \,  \bm{ { S' } } ^ {  \mathit{I'}  }  }   \mathop{ \odot }  \varepsilon    \sim   \varepsilon' $.
    \end{itemize}
    By Lemma~\ref{lem:zero_freeness},
    we have $ \mathit{l} \,  \bm{ { S } } ^ {  \mathit{I}  }    \neq   \mathit{l'} \,  \bm{ { S' } } ^ {  \mathit{I'}  }  $ and $ 0  \mathrm{-free} ( \mathit{l} \,  \bm{ { S } } ^ {  \mathit{I}  }  ,  \ottnt{E_{{\mathrm{1}}}} ) $.
    By the induction hypothesis, we have $  \lift{ \mathit{l} \,  \bm{ { S } } ^ {  \mathit{I}  }  }   \olessthan  \varepsilon' $.
    Thus, safety condition \ref{def:safe_cond:pres} makes $  \lift{ \mathit{l} \,  \bm{ { S } } ^ {  \mathit{I}  }  }   \olessthan  \varepsilon $ hold as required.

    \item[others] Cannot happen.
  \end{divcases}
\end{proof}

\begin{lemma}[Effect Safety]\label{lem:effsafe}
  If $\Gamma  \vdash  \ottnt{E}  \ottsym{[}   \mathsf{op} _{ \mathit{l} \,  \bm{ { S } } ^ {  \mathit{I}  }  }  \,  \bm{ { T } } ^ {  \mathit{J}  }  \, \ottnt{v}  \ottsym{]}  \ottsym{:}  \ottnt{A}  \mid  \varepsilon$ and
  $ \mathit{n}  \mathrm{-free} ( \mathit{l} \,  \bm{ { S } } ^ {  \mathit{I}  }  ,  \ottnt{E} ) $,
  then $ \varepsilon   \nsim    \bbZero  $.
\end{lemma}

\begin{proof}
  Assume that $ \varepsilon   \sim    \bbZero  $.
  By Lemma~\ref{lem:effsafe_aux}, we have $  \lift{ \mathit{l} \,  \bm{ { S } } ^ {  \mathit{I}  }  }   \olessthan  \varepsilon $.
  Therefore, we have $   \lift{ \mathit{l} \,  \bm{ { S } } ^ {  \mathit{I}  }  }   \mathop{ \odot }  \varepsilon'    \sim    \bbZero  $ for some $\varepsilon'$.
  However, this is contradictory with safety condition \ref{def:safe_cond:label_notemp}.
\end{proof}

\begin{theorem}[Type and Effect Safety]\label{thm:safety}
  If $\emptyset  \vdash  \ottnt{e}  \ottsym{:}  \ottnt{A}  \mid   \bbZero $ and $\ottnt{e}  \longrightarrow  ^ * \ottnt{e'}$ and $\ottnt{e'} \centernot \longrightarrow $, then $\ottnt{e'}$ is a value.
\end{theorem}

\begin{proof}
  By Lemma~\ref{lem:preservation},
  $\emptyset  \vdash  \ottnt{e'}  \ottsym{:}  \ottnt{A}  \mid   \bbZero $
  (it is easy to extend Lemma~\ref{lem:preservation} to multi-step evaluation).
  By Lemma~\ref{lem:effsafe}, $\ottnt{e'} \not= \ottnt{E}  \ottsym{[}   \mathsf{op} _{ \mathit{l} \,  \bm{ { S } } ^ {  \mathit{N}  }  }  \,  \bm{ { T } } ^ {  \mathit{J}  }  \, \ottnt{v}  \ottsym{]}$
  for any $\ottnt{E}$, $\mathit{l}$, $ \bm{ { S } } ^ {  \mathit{N}  } $, $\mathsf{op}$, $ \bm{ { T } } ^ {  \mathit{J}  } $, and $\ottnt{v}$ such that
  $ \mathit{n}  \mathrm{-free} ( \mathit{l} \,  \bm{ { S } } ^ {  \mathit{I}  }  ,  \ottnt{E} ) $ for some $\mathit{n}$.
  Thus, by Lemma~\ref{lem:progress}, we have the fact that $\ottnt{e'}$ is a value.
\end{proof}

\subsection{Properties with Shallow Handlers}

This section assumes that the safety conditions in Definition~\ref{def:safe_cond} hold.

\begin{lemma}[Weakening]\label{lem:weakening_shallow}
  Suppose that $\vdash  \Gamma_{{\mathrm{1}}}  \ottsym{,}  \Gamma_{{\mathrm{2}}}$ and $  \mathrm{dom}   \ottsym{(}   \Gamma_{{\mathrm{2}}}   \ottsym{)}    \cap    \mathrm{dom}   \ottsym{(}   \Gamma_{{\mathrm{3}}}   \ottsym{)}    \ottsym{=}  \emptyset$.
  \begin{enumerate}
    \item\label{lem:weakening_shallow:wf} If $\vdash  \Gamma_{{\mathrm{1}}}  \ottsym{,}  \Gamma_{{\mathrm{3}}}$, then $\vdash  \Gamma_{{\mathrm{1}}}  \ottsym{,}  \Gamma_{{\mathrm{2}}}  \ottsym{,}  \Gamma_{{\mathrm{3}}}$.
    \item\label{lem:weakening_shallow:kinding} If $\Gamma_{{\mathrm{1}}}  \ottsym{,}  \Gamma_{{\mathrm{3}}}  \vdash  S  \ottsym{:}  \ottnt{K}$, then $\Gamma_{{\mathrm{1}}}  \ottsym{,}  \Gamma_{{\mathrm{2}}}  \ottsym{,}  \Gamma_{{\mathrm{3}}}  \vdash  S  \ottsym{:}  \ottnt{K}$.
    \item\label{lem:weakening_shallow:subtyping} If $\Gamma_{{\mathrm{1}}}  \ottsym{,}  \Gamma_{{\mathrm{3}}}  \vdash  \ottnt{A}  <:  \ottnt{B}$, then $\Gamma_{{\mathrm{1}}}  \ottsym{,}  \Gamma_{{\mathrm{2}}}  \ottsym{,}  \Gamma_{{\mathrm{3}}}  \vdash  \ottnt{A}  <:  \ottnt{B}$.
    \item\label{lem:weakening_shallow:subtyping_comp} If $\Gamma_{{\mathrm{1}}}  \ottsym{,}  \Gamma_{{\mathrm{3}}}  \vdash  \ottnt{A_{{\mathrm{1}}}}  \mid  \varepsilon_{{\mathrm{1}}}  <:  \ottnt{A_{{\mathrm{2}}}}  \mid  \varepsilon_{{\mathrm{2}}}$, then $\Gamma_{{\mathrm{1}}}  \ottsym{,}  \Gamma_{{\mathrm{2}}}  \ottsym{,}  \Gamma_{{\mathrm{3}}}  \vdash  \ottnt{A_{{\mathrm{1}}}}  \mid  \varepsilon_{{\mathrm{1}}}  <:  \ottnt{A_{{\mathrm{2}}}}  \mid  \varepsilon_{{\mathrm{2}}}$.
    \item\label{lem:weakening_shallow:typing} If $\Gamma_{{\mathrm{1}}}  \ottsym{,}  \Gamma_{{\mathrm{3}}}  \vdash  \ottnt{e}  \ottsym{:}  \ottnt{A}  \mid  \varepsilon$, then $\Gamma_{{\mathrm{1}}}  \ottsym{,}  \Gamma_{{\mathrm{2}}}  \ottsym{,}  \Gamma_{{\mathrm{3}}}  \vdash  \ottnt{e}  \ottsym{:}  \ottnt{A}  \mid  \varepsilon$.
    \item\label{lem:weakening_shallow:handling} If $ \Gamma_{{\mathrm{1}}}  \ottsym{,}  \Gamma_{{\mathrm{3}}}  \vdash _{ \sigma }  \ottnt{h}  :  \ottnt{A}  ^ { \varepsilon' }  \Rightarrow  ^ { \varepsilon }  \ottnt{B} $, then $ \Gamma_{{\mathrm{1}}}  \ottsym{,}  \Gamma_{{\mathrm{2}}}  \ottsym{,}  \Gamma_{{\mathrm{3}}}  \vdash _{ \sigma }  \ottnt{h}  :  \ottnt{A}  ^ { \varepsilon' }  \Rightarrow  ^ { \varepsilon }  \ottnt{B} $.
  \end{enumerate}
\end{lemma}

\begin{proof}
  \phantom{}
  \begin{itemize}
    \item[(1)(2)] Similarly to Lemma~\ref{lem:weakening}\ref{lem:weakening:wf} and \ref{lem:weakening:kinding}.

    \item[(3)(4)] Similarly to Lemma~\ref{lem:weakening}\ref{lem:weakening:subtyping} and \ref{lem:weakening:subtyping_comp}.

    \item[(5)(6)]
          By mutual induction on derivations of the judgments.
          We proceed by case analysis on the rule applied lastly to the derivation.
          \begin{divcases}
            \item[\rname{T}{SHandling}]
            For some $\mathit{N}$, $\ottnt{e'}$, $\ottnt{A'}$, $\varepsilon'$, $\mathit{l}$, $ \bm{ { S } } ^ {  \mathit{N}  } $, $ {\bm{ { \ottnt{K} } } }^{ \mathit{N} } $, $\ottnt{h}$, and $\sigma$,
            the following are given:
            \begin{itemize}
              \item $\ottnt{e} =  \mathbf{handle}_{ \mathit{l} \,  \bm{ { S } } ^ {  \mathit{N}  }  }  \, \ottnt{e'} \, \mathbf{with} \, \ottnt{h}$,
              \item $\Gamma_{{\mathrm{1}}}  \ottsym{,}  \Gamma_{{\mathrm{3}}}  \vdash  \ottnt{e'}  \ottsym{:}  \ottnt{A'}  \mid  \varepsilon'$,
              \item $ \mathit{l}  ::    \forall    {\bm{ \alpha } }^{ \mathit{N} } : {\bm{ \ottnt{K} } }^{ \mathit{N} }    \ottsym{.}    \sigma    \in   \Xi $,
              \item $\Gamma_{{\mathrm{1}}}  \ottsym{,}  \Gamma_{{\mathrm{3}}}  \vdash   \bm{ { S } }^{ \mathit{N} } : \bm{ \ottnt{K} }^{ \mathit{N} } $,
              \item $ \Gamma_{{\mathrm{1}}}  \ottsym{,}  \Gamma_{{\mathrm{3}}}  \vdash _{ \sigma \,  \! [ {\bm{ { S } } }^{ \mathit{N} } / {\bm{ \alpha } }^{ \mathit{N} } ]  }  \ottnt{h}  :  \ottnt{A'}  ^ { \varepsilon' }  \Rightarrow  ^ { \varepsilon }  \ottnt{A} $, and
              \item $   \lift{ \mathit{l} \,  \bm{ { S } } ^ {  \mathit{N}  }  }   \mathop{ \odot }  \varepsilon    \sim   \varepsilon' $.
            \end{itemize}
            By the induction hypothesis and case~\ref{lem:weakening_shallow:kinding}, we have
            \begin{itemize}
              \item $\Gamma_{{\mathrm{1}}}  \ottsym{,}  \Gamma_{{\mathrm{2}}}  \ottsym{,}  \Gamma_{{\mathrm{3}}}  \vdash  \ottnt{e'}  \ottsym{:}  \ottnt{A'}  \mid  \varepsilon'$,
              \item $\Gamma_{{\mathrm{1}}}  \ottsym{,}  \Gamma_{{\mathrm{2}}}  \ottsym{,}  \Gamma_{{\mathrm{3}}}  \vdash   \bm{ { S } }^{ \mathit{N} } : \bm{ \ottnt{K} }^{ \mathit{N} } $, and
              \item $ \Gamma_{{\mathrm{1}}}  \ottsym{,}  \Gamma_{{\mathrm{2}}}  \ottsym{,}  \Gamma_{{\mathrm{3}}}  \vdash _{ \sigma \,  \! [ {\bm{ { S } } }^{ \mathit{N} } / {\bm{ \alpha } }^{ \mathit{N} } ]  }  \ottnt{h}  :  \ottnt{A'}  ^ { \varepsilon' }  \Rightarrow  ^ { \varepsilon }  \ottnt{A} $.
            \end{itemize}
            Thus, \rname{T}{SHandling} derives
            \begin{align*}
              \Gamma_{{\mathrm{1}}}  \ottsym{,}  \Gamma_{{\mathrm{2}}}  \ottsym{,}  \Gamma_{{\mathrm{3}}}  \vdash   \mathbf{handle}_{ \mathit{l} \,  \bm{ { S } } ^ {  \mathit{N}  }  }  \, \ottnt{e} \, \mathbf{with} \, \ottnt{h}  \ottsym{:}  \ottnt{A}  \mid  \varepsilon.
            \end{align*}

            \item[\rname{SH}{Return}]
            Without loss of generality, we can choose $\mathit{x}$ such that $ \mathit{x}   \notin    \mathrm{dom}   \ottsym{(}   \Gamma_{{\mathrm{2}}}   \ottsym{)}  $.
            For some $\ottnt{e_{\ottmv{r}}}$, the following are given:
            \begin{itemize}
              \item $\ottnt{h} = \ottsym{\{} \, \mathbf{return} \, \mathit{x}  \mapsto  \ottnt{e_{\ottmv{r}}}  \ottsym{\}}$,
              \item $\sigma =  \{\} $,
              \item $\Gamma_{{\mathrm{1}}}  \ottsym{,}  \Gamma_{{\mathrm{3}}}  \ottsym{,}  \mathit{x}  \ottsym{:}  \ottnt{A}  \vdash  \ottnt{e_{\ottmv{r}}}  \ottsym{:}  \ottnt{B}  \mid  \varepsilon$, and
              \item $\Gamma_{{\mathrm{1}}}  \ottsym{,}  \Gamma_{{\mathrm{3}}}  \vdash  \varepsilon'  \ottsym{:}   \mathbf{Eff} $.
            \end{itemize}
            By the induction hypothesis, we have $\Gamma_{{\mathrm{1}}}  \ottsym{,}  \Gamma_{{\mathrm{2}}}  \ottsym{,}  \Gamma_{{\mathrm{3}}}  \ottsym{,}  \mathit{x}  \ottsym{:}  \ottnt{A}  \vdash  \ottnt{e_{\ottmv{r}}}  \ottsym{:}  \ottnt{B}  \mid  \varepsilon$.
            By Lemma~\ref{lem:weakening_shallow}\ref{lem:weakening_shallow:kinding},
            we have $\Gamma_{{\mathrm{1}}}  \ottsym{,}  \Gamma_{{\mathrm{2}}}  \ottsym{,}  \Gamma_{{\mathrm{3}}}  \vdash  \varepsilon'  \ottsym{:}   \mathbf{Eff} $.
            Thus, \rname{SH}{Return} derives $ \Gamma_{{\mathrm{1}}}  \ottsym{,}  \Gamma_{{\mathrm{2}}}  \ottsym{,}  \Gamma_{{\mathrm{3}}}  \vdash _{  \{\}  }  \ottsym{\{} \, \mathbf{return} \, \mathit{x}  \mapsto  \ottnt{e_{\ottmv{r}}}  \ottsym{\}}  :  \ottnt{A}  ^ { \varepsilon' }  \Rightarrow  ^ { \varepsilon }  \ottnt{B} $.

            \item[\rname{SH}{Op}]
            Without loss of generality, we can choose $ \bm{ { \beta } } ^ {  \mathit{J}  } $ and $\mathit{p}$ and $\mathit{k}$ such that:
            \begin{itemize}
              \item $ \{   \bm{ { \beta } } ^ {  \mathit{J}  }   \}   \cap    \mathrm{dom}   \ottsym{(}   \Gamma_{{\mathrm{2}}}   \ottsym{)}    \ottsym{=}  \emptyset$,
              \item $ \mathit{p}   \notin    \mathrm{dom}   \ottsym{(}   \Gamma_{{\mathrm{2}}}   \ottsym{)}  $, and
              \item $ \mathit{k}   \notin    \mathrm{dom}   \ottsym{(}   \Gamma_{{\mathrm{2}}}   \ottsym{)}  $.
            \end{itemize}
            For some $\ottnt{h'}$, $\sigma'$, $\mathsf{op}$, $\ottnt{A'}$, $\ottnt{B'}$, and $\ottnt{e}$,
            the following are given:
            \begin{itemize}
              \item $\ottnt{h} =  \ottnt{h'}   \uplus   \ottsym{\{}  \mathsf{op} \,  {\bm{ \beta } }^{ \mathit{J} } : {\bm{ \ottnt{K} } }^{ \mathit{J} }  \, \mathit{p} \, \mathit{k}  \mapsto  \ottnt{e}  \ottsym{\}} $,
              \item $\sigma =  \sigma'   \uplus   \ottsym{\{}  \mathsf{op}  \ottsym{:}    \forall    {\bm{ \beta } }^{ \mathit{J} } : {\bm{ \ottnt{K} } }^{ \mathit{J} }    \ottsym{.}    \ottnt{A'}   \Rightarrow   \ottnt{B'}   \ottsym{\}} $,
              \item $ \Gamma_{{\mathrm{1}}}  \ottsym{,}  \Gamma_{{\mathrm{3}}}  \vdash _{ \sigma' }  \ottnt{h'}  :  \ottnt{A}  ^ { \varepsilon' }  \Rightarrow  ^ { \varepsilon }  \ottnt{B} $, and
              \item $\Gamma_{{\mathrm{1}}}  \ottsym{,}  \Gamma_{{\mathrm{3}}}  \ottsym{,}   {\bm{ \beta } }^{ \mathit{J} } : {\bm{ \ottnt{K} } }^{ \mathit{J} }   \ottsym{,}  \mathit{p}  \ottsym{:}  \ottnt{A'}  \ottsym{,}  \mathit{k}  \ottsym{:}   \ottnt{B'}    \rightarrow_{ \varepsilon' }    \ottnt{B}   \vdash  \ottnt{e}  \ottsym{:}  \ottnt{B}  \mid  \varepsilon$.
            \end{itemize}
            By the induction hypothesis, we have
            \begin{itemize}
              \item $ \Gamma_{{\mathrm{1}}}  \ottsym{,}  \Gamma_{{\mathrm{2}}}  \ottsym{,}  \Gamma_{{\mathrm{3}}}  \vdash _{ \sigma' }  \ottnt{h'}  :  \ottnt{A}  ^ { \varepsilon' }  \Rightarrow  ^ { \varepsilon }  \ottnt{B} $ and
              \item $\Gamma_{{\mathrm{1}}}  \ottsym{,}  \Gamma_{{\mathrm{2}}}  \ottsym{,}  \Gamma_{{\mathrm{3}}}  \ottsym{,}   {\bm{ \beta } }^{ \mathit{J} } : {\bm{ \ottnt{K} } }^{ \mathit{J} }   \ottsym{,}  \mathit{p}  \ottsym{:}  \ottnt{A'}  \ottsym{,}  \mathit{k}  \ottsym{:}   \ottnt{B'}    \rightarrow_{ \varepsilon' }    \ottnt{B}   \vdash  \ottnt{e}  \ottsym{:}  \ottnt{B}  \mid  \varepsilon$.
            \end{itemize}
            Thus, \rname{SH}{Op} derives $ \Gamma_{{\mathrm{1}}}  \ottsym{,}  \Gamma_{{\mathrm{2}}}  \ottsym{,}  \Gamma_{{\mathrm{3}}}  \vdash _{ \sigma }   \ottnt{h'}   \uplus   \ottsym{\{}  \mathsf{op} \,  {\bm{ \beta } }^{ \mathit{J} } : {\bm{ \ottnt{K} } }^{ \mathit{J} }  \, \mathit{p} \, \mathit{k}  \mapsto  \ottnt{e}  \ottsym{\}}   :  \ottnt{A}  ^ { \varepsilon' }  \Rightarrow  ^ { \varepsilon }  \ottnt{B} $.

            \item[others] Similarly to Lemma~\ref{lem:weakening}\ref{lem:weakening:typing} and \ref{lem:weakening:handling}.
          \end{divcases}
  \end{itemize}
\end{proof}

\begin{lemma}[Substitution of values]\label{lem:subst_value_shallow}
  Suppose that $\Gamma_{{\mathrm{1}}}  \vdash  \ottnt{v}  \ottsym{:}  \ottnt{A}  \mid   \bbZero $.
  \begin{enumerate}
    \item\label{lem:subst_value_shallow:wf} If $\vdash  \Gamma_{{\mathrm{1}}}  \ottsym{,}  \mathit{x}  \ottsym{:}  \ottnt{A}  \ottsym{,}  \Gamma_{{\mathrm{2}}}$, then $\vdash  \Gamma_{{\mathrm{1}}}  \ottsym{,}  \Gamma_{{\mathrm{2}}}$.
    \item\label{lem:subst_value_shallow:kinding} If $\Gamma_{{\mathrm{1}}}  \ottsym{,}  \mathit{x}  \ottsym{:}  \ottnt{A}  \ottsym{,}  \Gamma_{{\mathrm{2}}}  \vdash  S  \ottsym{:}  \ottnt{K}$, then $\Gamma_{{\mathrm{1}}}  \ottsym{,}  \Gamma_{{\mathrm{2}}}  \vdash  S  \ottsym{:}  \ottnt{K}$.
    \item\label{lem:subst_value_shallow:subtyping} If $\Gamma_{{\mathrm{1}}}  \ottsym{,}  \mathit{x}  \ottsym{:}  \ottnt{A}  \ottsym{,}  \Gamma_{{\mathrm{2}}}  \vdash  \ottnt{B}  <:  \ottnt{C}$, then $\Gamma_{{\mathrm{1}}}  \ottsym{,}  \Gamma_{{\mathrm{2}}}  \vdash  \ottnt{B}  <:  \ottnt{C}$.
    \item\label{lem:subst_value_shallow:subtyping_comp} If $\Gamma_{{\mathrm{1}}}  \ottsym{,}  \mathit{x}  \ottsym{:}  \ottnt{A}  \ottsym{,}  \Gamma_{{\mathrm{2}}}  \vdash  \ottnt{B_{{\mathrm{1}}}}  \mid  \varepsilon_{{\mathrm{1}}}  <:  \ottnt{B_{{\mathrm{2}}}}  \mid  \varepsilon_{{\mathrm{2}}}$, then $\Gamma_{{\mathrm{1}}}  \ottsym{,}  \Gamma_{{\mathrm{2}}}  \vdash  \ottnt{B_{{\mathrm{1}}}}  \mid  \varepsilon_{{\mathrm{1}}}  <:  \ottnt{B_{{\mathrm{2}}}}  \mid  \varepsilon_{{\mathrm{2}}}$.
    \item\label{lem:subst_value_shallow:typing} If $\Gamma_{{\mathrm{1}}}  \ottsym{,}  \mathit{x}  \ottsym{:}  \ottnt{A}  \ottsym{,}  \Gamma_{{\mathrm{2}}}  \vdash  \ottnt{e}  \ottsym{:}  \ottnt{B}  \mid  \varepsilon$, then $\Gamma_{{\mathrm{1}}}  \ottsym{,}  \Gamma_{{\mathrm{2}}}  \vdash  \ottnt{e} \,  \! [  \ottnt{v}  /  \mathit{x}  ]   \ottsym{:}  \ottnt{B}  \mid  \varepsilon$.
    \item\label{lem:subst_value_shallow:handling} If $ \Gamma_{{\mathrm{1}}}  \ottsym{,}  \mathit{x}  \ottsym{:}  \ottnt{A}  \ottsym{,}  \Gamma_{{\mathrm{2}}}  \vdash _{ \sigma }  \ottnt{h}  :  \ottnt{B}  ^ { \varepsilon' }  \Rightarrow  ^ { \varepsilon }  \ottnt{C} $, then $ \Gamma_{{\mathrm{1}}}  \ottsym{,}  \Gamma_{{\mathrm{2}}}  \vdash _{ \sigma }  \ottnt{h} \,  \! [  \ottnt{v}  /  \mathit{x}  ]   :  \ottnt{B}  ^ { \varepsilon' }  \Rightarrow  ^ { \varepsilon }  \ottnt{C} $.
  \end{enumerate}
\end{lemma}

\begin{proof}
  \begin{itemize}
    \item[(1)(2)] Similarly to Lemma~\ref{lem:subst_value}\ref{lem:subst_value:wf} and \ref{lem:subst_value:kinding}.

    \item[(3)(4)] Similarly to Lemma~\ref{lem:subst_value}\ref{lem:subst_value:subtyping} and \ref{lem:subst_value:subtyping_comp}.

    \item[(5)(6)]
          By mutual induction on derivations of the judgments.
          We proceed by case analysis on the rule applied lastly to the derivation.
          \begin{divcases}
            \item[\rname{T}{SHandling}]
            For some $\mathit{N}$, $\ottnt{e'}$, $\ottnt{A'}$, $\varepsilon'$, $\mathit{l}$, $ \bm{ { S } } ^ {  \mathit{N}  } $, $ \bm{ { \alpha } } ^ {  \mathit{N}  } $, $ {\bm{ { \ottnt{K} } } }^{ \mathit{N} } $, $\ottnt{h}$, and $\sigma$,
            the following are given:
            \begin{itemize}
              \item $\ottnt{e} =  \mathbf{handle}_{ \mathit{l} \,  \bm{ { S } } ^ {  \mathit{N}  }  }  \, \ottnt{e'} \, \mathbf{with} \, \ottnt{h}$,
              \item $\Gamma_{{\mathrm{1}}}  \ottsym{,}  \mathit{x}  \ottsym{:}  \ottnt{A}  \ottsym{,}  \Gamma_{{\mathrm{2}}}  \vdash  \ottnt{e'}  \ottsym{:}  \ottnt{A'}  \mid  \varepsilon'$,
              \item $ \mathit{l}  ::    \forall    {\bm{ \alpha } }^{ \mathit{N} } : {\bm{ \ottnt{K} } }^{ \mathit{N} }    \ottsym{.}    \sigma    \in   \Xi $,
              \item $\Gamma_{{\mathrm{1}}}  \ottsym{,}  \mathit{x}  \ottsym{:}  \ottnt{A}  \ottsym{,}  \Gamma_{{\mathrm{2}}}  \vdash   \bm{ { S } }^{ \mathit{N} } : \bm{ \ottnt{K} }^{ \mathit{N} } $,
              \item $ \Gamma_{{\mathrm{1}}}  \ottsym{,}  \mathit{x}  \ottsym{:}  \ottnt{A}  \ottsym{,}  \Gamma_{{\mathrm{2}}}  \vdash _{ \sigma \,  \! [ {\bm{ { S } } }^{ \mathit{N} } / {\bm{ \alpha } }^{ \mathit{N} } ]  }  \ottnt{h}  :  \ottnt{A'}  ^ { \varepsilon' }  \Rightarrow  ^ { \varepsilon }  \ottnt{B} $, and
              \item $   \lift{ \mathit{l} \,  \bm{ { S } } ^ {  \mathit{N}  }  }   \mathop{ \odot }  \varepsilon    \sim   \varepsilon' $.
            \end{itemize}
            By the induction hypothesis and case~\ref{lem:subst_value_shallow:kinding}, we have
            \begin{itemize}
              \item $\Gamma_{{\mathrm{1}}}  \ottsym{,}  \Gamma_{{\mathrm{2}}}  \vdash  \ottnt{e'} \,  \! [  \ottnt{v}  /  \mathit{x}  ]   \ottsym{:}  \ottnt{A'}  \mid  \varepsilon'$,
              \item $\Gamma_{{\mathrm{1}}}  \ottsym{,}  \Gamma_{{\mathrm{2}}}  \vdash   \bm{ { S } }^{ \mathit{N} } : \bm{ \ottnt{K} }^{ \mathit{N} } $, and
              \item $ \Gamma_{{\mathrm{1}}}  \ottsym{,}  \Gamma_{{\mathrm{2}}}  \vdash _{ \sigma \,  \! [ {\bm{ { S } } }^{ \mathit{N} } / {\bm{ \alpha } }^{ \mathit{N} } ]  }  \ottnt{h} \,  \! [  \ottnt{v}  /  \mathit{x}  ]   :  \ottnt{A'}  ^ { \varepsilon' }  \Rightarrow  ^ { \varepsilon }  \ottnt{A} $.
            \end{itemize}
            Thus, \rname{T}{SHandling} derives
            \begin{align*}
              \Gamma_{{\mathrm{1}}}  \ottsym{,}  \Gamma_{{\mathrm{2}}}  \vdash   \mathbf{handle}_{ \mathit{l} \,  \bm{ { S } } ^ {  \mathit{N}  }  }  \, \ottnt{e'} \,  \! [  \ottnt{v}  /  \mathit{x}  ]  \, \mathbf{with} \, \ottnt{h} \,  \! [  \ottnt{v}  /  \mathit{x}  ]   \ottsym{:}  \ottnt{B}  \mid  \varepsilon.
            \end{align*}

            \item[\rname{SH}{Return}]
            Without loss of generality, we can choose $\mathit{y}$ such that $\mathit{y} \neq \mathit{x}$ and $ \mathit{y}   \notin    \mathrm{FV}   \ottsym{(}   \ottnt{v}   \ottsym{)}  $.
            For some $\ottnt{e_{\ottmv{r}}}$, the following are given:
            \begin{itemize}
              \item $\ottnt{h} = \ottsym{\{} \, \mathbf{return} \, \mathit{y}  \mapsto  \ottnt{e_{\ottmv{r}}}  \ottsym{\}}$,
              \item $\sigma =  \{\} $,
              \item $\Gamma_{{\mathrm{1}}}  \ottsym{,}  \mathit{x}  \ottsym{:}  \ottnt{A}  \ottsym{,}  \Gamma_{{\mathrm{2}}}  \ottsym{,}  \mathit{y}  \ottsym{:}  \ottnt{B}  \vdash  \ottnt{e_{\ottmv{r}}}  \ottsym{:}  \ottnt{C}  \mid  \varepsilon$, and
              \item $\Gamma_{{\mathrm{1}}}  \ottsym{,}  \mathit{x}  \ottsym{:}  \ottnt{A}  \ottsym{,}  \Gamma_{{\mathrm{2}}}  \vdash  \varepsilon'  \ottsym{:}   \mathbf{Eff} $.
            \end{itemize}
            By the induction hypothesis, we have $\Gamma_{{\mathrm{1}}}  \ottsym{,}  \Gamma_{{\mathrm{2}}}  \ottsym{,}  \mathit{y}  \ottsym{:}  \ottnt{B}  \vdash  \ottnt{e_{\ottmv{r}}} \,  \! [  \ottnt{v}  /  \mathit{x}  ]   \ottsym{:}  \ottnt{C}  \mid  \varepsilon$.
            By Lemma~\ref{lem:subst_value_shallow}\ref{lem:subst_value_shallow:kinding}, we have $\Gamma_{{\mathrm{1}}}  \ottsym{,}  \Gamma_{{\mathrm{2}}}  \vdash  \varepsilon'  \ottsym{:}   \mathbf{Eff} $.
            Thus, \rname{SH}{Return} derives
            \begin{align*}
               \Gamma_{{\mathrm{1}}}  \ottsym{,}  \Gamma_{{\mathrm{2}}}  \vdash _{  \{\}  }  \ottsym{\{} \, \mathbf{return} \, \mathit{y}  \mapsto  \ottnt{e_{\ottmv{r}}} \,  \! [  \ottnt{v}  /  \mathit{x}  ]   \ottsym{\}}  :  \ottnt{B}  ^ { \varepsilon' }  \Rightarrow  ^ { \varepsilon }  \ottnt{C} .
            \end{align*}

            \item[\rname{SH}{Op}]
            Without loss of generality, we can choose $ \bm{ { \beta } } ^ {  \mathit{J}  } $ and $\mathit{p}$ and $\mathit{k}$ such that:
            \begin{itemize}
              \item $\mathit{p} \neq \mathit{x}$,
              \item $\mathit{k} \neq \mathit{x}$,
              \item $ \mathit{p}   \notin    \mathrm{FV}   \ottsym{(}   \ottnt{v}   \ottsym{)}  $,
              \item $ \mathit{k}   \notin    \mathrm{FV}   \ottsym{(}   \mathit{k}   \ottsym{)}  $, and
              \item $ \{   \bm{ { \beta } } ^ {  \mathit{J}  }   \}   \cap    \mathrm{FTV}   \ottsym{(}   \ottnt{v}   \ottsym{)}    \ottsym{=}  \emptyset$.
            \end{itemize}
            For some $\ottnt{h'}$, $\sigma'$, $\mathsf{op}$, $\ottnt{A'}$, $\ottnt{B'}$, and $\ottnt{e}$, the following are given:
            \begin{itemize}
              \item $\ottnt{h} =  \ottnt{h'}   \uplus   \ottsym{\{}  \mathsf{op} \,  {\bm{ \beta } }^{ \mathit{J} } : {\bm{ \ottnt{K} } }^{ \mathit{J} }  \, \mathit{p} \, \mathit{k}  \mapsto  \ottnt{e}  \ottsym{\}} $,
              \item $\sigma =  \sigma'   \uplus   \ottsym{\{}  \mathsf{op}  \ottsym{:}    \forall    {\bm{ \beta } }^{ \mathit{J} } : {\bm{ \ottnt{K} } }^{ \mathit{J} }    \ottsym{.}    \ottnt{A'}   \Rightarrow   \ottnt{B'}   \ottsym{\}} $,
              \item $ \Gamma_{{\mathrm{1}}}  \ottsym{,}  \mathit{x}  \ottsym{:}  \ottnt{A}  \ottsym{,}  \Gamma_{{\mathrm{2}}}  \vdash _{ \sigma' }  \ottnt{h'}  :  \ottnt{B}  ^ { \varepsilon' }  \Rightarrow  ^ { \varepsilon }  \ottnt{C} $, and
              \item $\Gamma_{{\mathrm{1}}}  \ottsym{,}  \mathit{x}  \ottsym{:}  \ottnt{A}  \ottsym{,}  \Gamma_{{\mathrm{2}}}  \ottsym{,}   {\bm{ \beta } }^{ \mathit{J} } : {\bm{ \ottnt{K} } }^{ \mathit{J} }   \ottsym{,}  \mathit{p}  \ottsym{:}  \ottnt{A'}  \ottsym{,}  \mathit{k}  \ottsym{:}   \ottnt{B'}    \rightarrow_{ \varepsilon' }    \ottnt{C}   \vdash  \ottnt{e}  \ottsym{:}  \ottnt{C}  \mid  \varepsilon$.
            \end{itemize}
            By the induction hypothesis, we have
            \begin{itemize}
              \item $ \Gamma_{{\mathrm{1}}}  \ottsym{,}  \Gamma_{{\mathrm{2}}}  \vdash _{ \sigma' }  \ottnt{h'} \,  \! [  \ottnt{v}  /  \mathit{x}  ]   :  \ottnt{A}  ^ { \varepsilon' }  \Rightarrow  ^ { \varepsilon }  \ottnt{B} $ and
              \item $\Gamma_{{\mathrm{1}}}  \ottsym{,}  \Gamma_{{\mathrm{2}}}  \ottsym{,}   {\bm{ \beta } }^{ \mathit{J} } : {\bm{ \ottnt{K} } }^{ \mathit{J} }   \ottsym{,}  \mathit{p}  \ottsym{:}  \ottnt{A'}  \ottsym{,}  \mathit{k}  \ottsym{:}   \ottnt{B'}    \rightarrow_{ \varepsilon' }    \ottnt{B}   \vdash  \ottnt{e} \,  \! [  \ottnt{v}  /  \mathit{x}  ]   \ottsym{:}  \ottnt{B}  \mid  \varepsilon$.
            \end{itemize}
            Thus, \rname{SH}{Op} derives
            \begin{align*}
               \Gamma_{{\mathrm{1}}}  \ottsym{,}  \Gamma_{{\mathrm{2}}}  \vdash _{ \sigma }   \ottnt{h'} \,  \! [  \ottnt{v}  /  \mathit{x}  ]    \uplus   \ottsym{\{}  \mathsf{op} \,  {\bm{ \beta } }^{ \mathit{J} } : {\bm{ \ottnt{K} } }^{ \mathit{J} }  \, \mathit{p} \, \mathit{k}  \mapsto  \ottnt{e} \,  \! [  \ottnt{v}  /  \mathit{x}  ]   \ottsym{\}}   :  \ottnt{B}  ^ { \varepsilon' }  \Rightarrow  ^ { \varepsilon }  \ottnt{C} 
            \end{align*}.

            \item[others] Similarly to Lemma~\ref{lem:subst_value}\ref{lem:subst_value:typing} and \ref{lem:subst_value:handling}.
          \end{divcases}
  \end{itemize}
\end{proof}

\begin{lemma}[Substitution of Typelikes]\label{lem:subst_type_shallow}
  Suppose that $\Gamma_{{\mathrm{1}}}  \vdash   \bm{ { S } }^{ \mathit{I} } : \bm{ \ottnt{K} }^{ \mathit{I} } $.
  \begin{enumerate}
    \item\label{lem:subst_type_shallow:wf} If $\vdash  \Gamma_{{\mathrm{1}}}  \ottsym{,}   {\bm{ \alpha } }^{ \mathit{I} } : {\bm{ \ottnt{K} } }^{ \mathit{I} }   \ottsym{,}  \Gamma_{{\mathrm{2}}}$, then $\vdash  \Gamma_{{\mathrm{1}}}  \ottsym{,}  \Gamma_{{\mathrm{2}}} \,  \! [ {\bm{ { S } } }^{ \mathit{I} } / {\bm{ \alpha } }^{ \mathit{I} } ] $.
    \item\label{lem:subst_type_shallow:kinding} If $\Gamma_{{\mathrm{1}}}  \ottsym{,}   {\bm{ \alpha } }^{ \mathit{I} } : {\bm{ \ottnt{K} } }^{ \mathit{I} }   \ottsym{,}  \Gamma_{{\mathrm{2}}}  \vdash  T  \ottsym{:}  \ottnt{K}$, then $\Gamma_{{\mathrm{1}}}  \ottsym{,}  \Gamma_{{\mathrm{2}}} \,  \! [ {\bm{ { S } } }^{ \mathit{I} } / {\bm{ \alpha } }^{ \mathit{I} } ]   \vdash  T \,  \! [ {\bm{ { S } } }^{ \mathit{I} } / {\bm{ \alpha } }^{ \mathit{I} } ]   \ottsym{:}  \ottnt{K}$.
    \item\label{lem:subst_type_shallow:subtyping} If $\Gamma_{{\mathrm{1}}}  \ottsym{,}   {\bm{ \alpha } }^{ \mathit{I} } : {\bm{ \ottnt{K} } }^{ \mathit{I} }   \ottsym{,}  \Gamma_{{\mathrm{2}}}  \vdash  \ottnt{A}  <:  \ottnt{B}$, then $\Gamma_{{\mathrm{1}}}  \ottsym{,}  \Gamma_{{\mathrm{2}}} \,  \! [ {\bm{ { S } } }^{ \mathit{I} } / {\bm{ \alpha } }^{ \mathit{I} } ]   \vdash  \ottnt{A} \,  \! [ {\bm{ { S } } }^{ \mathit{I} } / {\bm{ \alpha } }^{ \mathit{I} } ]   <:  \ottnt{B} \,  \! [ {\bm{ { S } } }^{ \mathit{I} } / {\bm{ \alpha } }^{ \mathit{I} } ] $.
    \item\label{lem:subst_type_shallow:subtyping_comp} If $\Gamma_{{\mathrm{1}}}  \ottsym{,}   {\bm{ \alpha } }^{ \mathit{I} } : {\bm{ \ottnt{K} } }^{ \mathit{I} }   \ottsym{,}  \Gamma_{{\mathrm{2}}}  \vdash  \ottnt{A_{{\mathrm{1}}}}  \mid  \varepsilon_{{\mathrm{1}}}  <:  \ottnt{A_{{\mathrm{2}}}}  \mid  \varepsilon_{{\mathrm{2}}}$, then $\Gamma_{{\mathrm{1}}}  \ottsym{,}  \Gamma_{{\mathrm{2}}} \,  \! [ {\bm{ { S } } }^{ \mathit{I} } / {\bm{ \alpha } }^{ \mathit{I} } ]   \vdash  \ottnt{A_{{\mathrm{1}}}} \,  \! [ {\bm{ { S } } }^{ \mathit{I} } / {\bm{ \alpha } }^{ \mathit{I} } ]   \mid  \varepsilon_{{\mathrm{1}}} \,  \! [ {\bm{ { S } } }^{ \mathit{I} } / {\bm{ \alpha } }^{ \mathit{I} } ]   <:  \ottnt{A_{{\mathrm{2}}}} \,  \! [ {\bm{ { S } } }^{ \mathit{I} } / {\bm{ \alpha } }^{ \mathit{I} } ]   \mid  \varepsilon_{{\mathrm{2}}} \,  \! [ {\bm{ { S } } }^{ \mathit{I} } / {\bm{ \alpha } }^{ \mathit{I} } ] $.
    \item\label{lem:subst_type_shallow:typing} If $\Gamma_{{\mathrm{1}}}  \ottsym{,}   {\bm{ \alpha } }^{ \mathit{I} } : {\bm{ \ottnt{K} } }^{ \mathit{I} }   \ottsym{,}  \Gamma_{{\mathrm{2}}}  \vdash  \ottnt{e}  \ottsym{:}  \ottnt{A}  \mid  \varepsilon$, then $\Gamma_{{\mathrm{1}}}  \ottsym{,}  \Gamma_{{\mathrm{2}}} \,  \! [ {\bm{ { S } } }^{ \mathit{I} } / {\bm{ \alpha } }^{ \mathit{I} } ]   \vdash  \ottnt{e} \,  \! [ {\bm{ { S } } }^{ \mathit{I} } / {\bm{ \alpha } }^{ \mathit{I} } ]   \ottsym{:}  \ottnt{A} \,  \! [ {\bm{ { S } } }^{ \mathit{I} } / {\bm{ \alpha } }^{ \mathit{I} } ]   \mid  \varepsilon \,  \! [ {\bm{ { S } } }^{ \mathit{I} } / {\bm{ \alpha } }^{ \mathit{I} } ] $.
    \item\label{lem:subst_type_shallow:handling} If $ \Gamma_{{\mathrm{1}}}  \ottsym{,}   {\bm{ \alpha } }^{ \mathit{I} } : {\bm{ \ottnt{K} } }^{ \mathit{I} }   \ottsym{,}  \Gamma_{{\mathrm{2}}}  \vdash _{ \sigma }  \ottnt{h}  :  \ottnt{A}  ^ { \varepsilon' }  \Rightarrow  ^ { \varepsilon }  \ottnt{B} $, then $ \Gamma_{{\mathrm{1}}}  \ottsym{,}  \Gamma_{{\mathrm{2}}} \,  \! [ {\bm{ { S } } }^{ \mathit{I} } / {\bm{ \alpha } }^{ \mathit{I} } ]   \vdash _{ \sigma \,  \! [ \bm{ { S } } / \bm{ \alpha } ]  }  \ottnt{h} \,  \! [ \bm{ { S } } / \bm{ \alpha } ]   :  \ottnt{A} \,  \! [ {\bm{ { S } } }^{ \mathit{I} } / {\bm{ \alpha } }^{ \mathit{I} } ]   ^ { \varepsilon' \,  \! [ {\bm{ { S } } }^{ \mathit{I} } / {\bm{ \alpha } }^{ \mathit{I} } ]  }  \Rightarrow  ^ { \varepsilon \,  \! [ {\bm{ { S } } }^{ \mathit{I} } / {\bm{ \alpha } }^{ \mathit{I} } ]  }  \ottnt{B} \,  \! [ {\bm{ { S } } }^{ \mathit{I} } / {\bm{ \alpha } }^{ \mathit{I} } ]  $.
  \end{enumerate}
\end{lemma}

\begin{proof}
  \begin{itemize}
    \item[(1)(2)] Similarly to Lemma~\ref{lem:subst_type}\ref{lem:subst_type:wf} and \ref{lem:subst_value:kinding}.

    \item[(3)(4)] Similarly to Lemma~\ref{lem:subst_type}\ref{lem:subst_type:subtyping} and \ref{lem:subst_type:subtyping_comp}.

    \item[(5)(6)]
          By mutual induction on derivations of the judgments.
          We proceed by case analysis on the rule applied lastly to the derivation.
          \begin{divcases}
            \item[\rname{T}{SHandling}]
            For some $\mathit{N}$, $\ottnt{e'}$, $\ottnt{A'}$, $\varepsilon'$, $\mathit{l}$, $ \bm{ { S_{{\mathrm{0}}} } } ^ {  \mathit{N}  } $, $ \bm{ { \alpha_{{\mathrm{0}}} } } ^ {  \mathit{N}  } $, $ {\bm{ { \ottnt{K_{{\mathrm{0}}}} } } }^{ \mathit{N} } $, $\ottnt{h}$, and $\sigma$, the following are given:
            \begin{itemize}
              \item $\ottnt{e} =  \mathbf{handle}_{ \mathit{l} \,  \bm{ { S_{{\mathrm{0}}} } } ^ {  \mathit{N}  }  }  \, \ottnt{e'} \, \mathbf{with} \, \ottnt{h}$,
              \item $\Gamma_{{\mathrm{1}}}  \ottsym{,}   {\bm{ \alpha } }^{ \mathit{I} } : {\bm{ \ottnt{K} } }^{ \mathit{I} }   \ottsym{,}  \Gamma_{{\mathrm{2}}}  \vdash  \ottnt{e'}  \ottsym{:}  \ottnt{A'}  \mid  \varepsilon'$,
              \item $ \mathit{l}  ::    \forall    {\bm{ \alpha_{{\mathrm{0}}} } }^{ \mathit{N} } : {\bm{ \ottnt{K_{{\mathrm{0}}}} } }^{ \mathit{N} }    \ottsym{.}    \sigma    \in   \Xi $,
              \item $\Gamma_{{\mathrm{1}}}  \ottsym{,}   {\bm{ \alpha } }^{ \mathit{I} } : {\bm{ \ottnt{K} } }^{ \mathit{I} }   \ottsym{,}  \Gamma_{{\mathrm{2}}}  \vdash   \bm{ { S_{{\mathrm{0}}} } }^{ \mathit{N} } : \bm{ \ottnt{K_{{\mathrm{0}}}} }^{ \mathit{N} } $,
              \item $ \Gamma_{{\mathrm{1}}}  \ottsym{,}   {\bm{ \alpha } }^{ \mathit{I} } : {\bm{ \ottnt{K} } }^{ \mathit{I} }   \ottsym{,}  \Gamma_{{\mathrm{2}}}  \vdash _{ \sigma \,  \! [ {\bm{ { S_{{\mathrm{0}}} } } }^{ \mathit{N} } / {\bm{ \alpha_{{\mathrm{0}}} } }^{ \mathit{N} } ]  }  \ottnt{h}  :  \ottnt{A'}  ^ { \varepsilon' }  \Rightarrow  ^ { \varepsilon }  \ottnt{A} $, and
              \item $   \lift{ \mathit{l} \,  \bm{ { S_{{\mathrm{0}}} } } ^ {  \mathit{N}  }  }   \mathop{ \odot }  \varepsilon    \sim   \varepsilon' $.
            \end{itemize}
            By the induction hypothesis,
            case~\ref{lem:subst_type_shallow:kinding}, and
            that a typelike substitution is homomorphism for $ \odot $ and $ \sim $,
            we have
            \begin{itemize}
              \item $\Gamma_{{\mathrm{1}}}  \ottsym{,}  \Gamma_{{\mathrm{2}}} \,  \! [ {\bm{ { S } } }^{ \mathit{I} } / {\bm{ \alpha } }^{ \mathit{I} } ]   \vdash  \ottnt{e'} \,  \! [ {\bm{ { S } } }^{ \mathit{I} } / {\bm{ \alpha } }^{ \mathit{I} } ]   \ottsym{:}  \ottnt{A'} \,  \! [ {\bm{ { S } } }^{ \mathit{I} } / {\bm{ \alpha } }^{ \mathit{I} } ]   \mid  \varepsilon' \,  \! [ {\bm{ { S } } }^{ \mathit{I} } / {\bm{ \alpha } }^{ \mathit{I} } ] $,
              \item $\Gamma_{{\mathrm{1}}}  \ottsym{,}  \Gamma_{{\mathrm{2}}} \,  \! [ {\bm{ { S } } }^{ \mathit{I} } / {\bm{ \alpha } }^{ \mathit{I} } ]   \vdash   \bm{ { S_{{\mathrm{0}}} \,  \! [ {\bm{ { S } } }^{ \mathit{I} } / {\bm{ \alpha } }^{ \mathit{I} } ]  } }^{ \mathit{N} } : \bm{ \ottnt{K_{{\mathrm{0}}}} }^{ \mathit{N} } $,
              \item $ \Gamma_{{\mathrm{1}}}  \ottsym{,}  \Gamma_{{\mathrm{2}}} \,  \! [ {\bm{ { S } } }^{ \mathit{I} } / {\bm{ \alpha } }^{ \mathit{I} } ]   \vdash _{ \sigma \,  \! [ {\bm{ { S_{{\mathrm{0}}} } } }^{ \mathit{N} } / {\bm{ \alpha_{{\mathrm{0}}} } }^{ \mathit{N} } ]  \,  \! [ {\bm{ { S } } }^{ \mathit{I} } / {\bm{ \alpha } }^{ \mathit{I} } ]  }  \ottnt{h} \,  \! [ {\bm{ { S } } }^{ \mathit{I} } / {\bm{ \alpha } }^{ \mathit{I} } ]   :  \ottnt{A'} \,  \! [ {\bm{ { S } } }^{ \mathit{I} } / {\bm{ \alpha } }^{ \mathit{I} } ]   ^ { \varepsilon' \,  \! [ {\bm{ { S } } }^{ \mathit{I} } / {\bm{ \alpha } }^{ \mathit{I} } ]  }  \Rightarrow  ^ { \varepsilon \,  \! [ {\bm{ { S } } }^{ \mathit{I} } / {\bm{ \alpha } }^{ \mathit{I} } ]  }  \ottnt{A} \,  \! [ {\bm{ { S } } }^{ \mathit{I} } / {\bm{ \alpha } }^{ \mathit{I} } ]  $, and
              \item $   \lift{ \mathit{l} \,  \bm{ { S_{{\mathrm{0}}} \,  \! [ {\bm{ { S } } }^{ \mathit{I} } / {\bm{ \alpha } }^{ \mathit{I} } ]  } } ^ {  \mathit{N}  }  }   \mathop{ \odot }  \varepsilon  \,  \! [ {\bm{ { S } } }^{ \mathit{I} } / {\bm{ \alpha } }^{ \mathit{I} } ]    \sim   \varepsilon' \,  \! [ {\bm{ { S } } }^{ \mathit{I} } / {\bm{ \alpha } }^{ \mathit{I} } ]  $.
            \end{itemize}
            Now, because we can assume that
            \begin{itemize}
              \item $ \{   \bm{ { \alpha } } ^ {  \mathit{I}  }   \}   \cap   \{   \bm{ { \alpha_{{\mathrm{0}}} } } ^ {  \mathit{N}  }   \}   \ottsym{=}  \emptyset$ and
              \item $ \{   \bm{ { \alpha_{{\mathrm{0}}} } } ^ {  \mathit{N}  }   \}   \cap    \mathrm{FTV}   \ottsym{(}    \bm{ { S } } ^ {  \mathit{I}  }    \ottsym{)}    \ottsym{=}  \emptyset$
            \end{itemize}
            without loss of generality, we have
            \begin{align*}
               \Gamma_{{\mathrm{1}}}  \ottsym{,}  \Gamma_{{\mathrm{2}}} \,  \! [ {\bm{ { S } } }^{ \mathit{I} } / {\bm{ \alpha } }^{ \mathit{I} } ]   \vdash _{ \sigma \,  \! [ {\bm{ { S_{{\mathrm{0}}} \,  \! [ {\bm{ { S } } }^{ \mathit{I} } / {\bm{ \alpha } }^{ \mathit{I} } ]  } } }^{ \mathit{N} } / {\bm{ \alpha_{{\mathrm{0}}} } }^{ \mathit{N} } ]  }  \ottnt{h} \,  \! [ {\bm{ { S } } }^{ \mathit{I} } / {\bm{ \alpha } }^{ \mathit{I} } ]   :  \ottnt{A'} \,  \! [ {\bm{ { S } } }^{ \mathit{I} } / {\bm{ \alpha } }^{ \mathit{I} } ]   ^ { \varepsilon' \,  \! [ {\bm{ { S } } }^{ \mathit{I} } / {\bm{ \alpha } }^{ \mathit{I} } ]  }  \Rightarrow  ^ { \varepsilon \,  \! [ {\bm{ { S } } }^{ \mathit{I} } / {\bm{ \alpha } }^{ \mathit{I} } ]  }  \ottnt{A} \,  \! [ {\bm{ { S } } }^{ \mathit{I} } / {\bm{ \alpha } }^{ \mathit{I} } ]  .
            \end{align*}
            Thus, \rname{T}{SHandling} derives
            \begin{align*}
              \Gamma_{{\mathrm{1}}}  \ottsym{,}  \Gamma_{{\mathrm{2}}} \,  \! [ {\bm{ { S } } }^{ \mathit{I} } / {\bm{ \alpha } }^{ \mathit{I} } ]   \vdash   \mathbf{handle}_{ \mathit{l} \,  \bm{ { S_{{\mathrm{0}}} \,  \! [ {\bm{ { S } } }^{ \mathit{I} } / {\bm{ \alpha } }^{ \mathit{I} } ]  } } ^ {  \mathit{N}  }  }  \, \ottnt{e} \,  \! [ {\bm{ { S } } }^{ \mathit{I} } / {\bm{ \alpha } }^{ \mathit{I} } ]  \, \mathbf{with} \, \ottnt{h} \,  \! [ {\bm{ { S } } }^{ \mathit{I} } / {\bm{ \alpha } }^{ \mathit{I} } ]   \ottsym{:}  \ottnt{B} \,  \! [ {\bm{ { S } } }^{ \mathit{I} } / {\bm{ \alpha } }^{ \mathit{I} } ]   \mid  \varepsilon \,  \! [ {\bm{ { S } } }^{ \mathit{I} } / {\bm{ \alpha } }^{ \mathit{I} } ] .
            \end{align*}

            \item[\rname{SH}{Return}]
            For some $\mathit{x}$ and $\ottnt{e_{\ottmv{r}}}$, the following are given:
            \begin{itemize}
              \item $\ottnt{h} = \ottsym{\{} \, \mathbf{return} \, \mathit{y}  \mapsto  \ottnt{e_{\ottmv{r}}}  \ottsym{\}}$,
              \item $\sigma =  \{\} $,
              \item $\Gamma_{{\mathrm{1}}}  \ottsym{,}   {\bm{ \alpha } }^{ \mathit{I} } : {\bm{ \ottnt{K} } }^{ \mathit{I} }   \ottsym{,}  \Gamma_{{\mathrm{2}}}  \ottsym{,}  \mathit{x}  \ottsym{:}  \ottnt{A}  \vdash  \ottnt{e_{\ottmv{r}}}  \ottsym{:}  \ottnt{B}  \mid  \varepsilon$, and
              \item $\Gamma_{{\mathrm{1}}}  \ottsym{,}   {\bm{ \alpha } }^{ \mathit{I} } : {\bm{ \ottnt{K} } }^{ \mathit{I} }   \ottsym{,}  \Gamma_{{\mathrm{2}}}  \vdash  \varepsilon'  \ottsym{:}   \mathbf{Eff} $.
            \end{itemize}
            By the induction hypothesis, we have
            \begin{itemize}
              \item $\Gamma_{{\mathrm{1}}}  \ottsym{,}  \Gamma_{{\mathrm{2}}} \,  \! [ {\bm{ { S } } }^{ \mathit{I} } / {\bm{ \alpha } }^{ \mathit{I} } ]   \ottsym{,}  \mathit{x}  \ottsym{:}  \ottnt{A} \,  \! [ {\bm{ { S } } }^{ \mathit{I} } / {\bm{ \alpha } }^{ \mathit{I} } ]   \vdash  \ottnt{e_{\ottmv{r}}} \,  \! [ {\bm{ { S } } }^{ \mathit{I} } / {\bm{ \alpha } }^{ \mathit{I} } ]   \ottsym{:}  \ottnt{B} \,  \! [ {\bm{ { S } } }^{ \mathit{I} } / {\bm{ \alpha } }^{ \mathit{I} } ]   \mid  \varepsilon \,  \! [ {\bm{ { S } } }^{ \mathit{I} } / {\bm{ \alpha } }^{ \mathit{I} } ] $.
            \end{itemize}
            By Lemma~\ref{lem:subst_type_shallow}\ref{lem:subst_type_shallow:kinding}, we have
            \begin{itemize}
              \item $\Gamma_{{\mathrm{1}}}  \ottsym{,}  \Gamma_{{\mathrm{2}}} \,  \! [ {\bm{ { S } } }^{ \mathit{I} } / {\bm{ \alpha } }^{ \mathit{I} } ]   \vdash  \varepsilon' \,  \! [ {\bm{ { S } } }^{ \mathit{I} } / {\bm{ \alpha } }^{ \mathit{I} } ]   \ottsym{:}   \mathbf{Eff} $.
            \end{itemize}
            Thus, \rname{SH}{Return} derives
            \begin{align*}
               \Gamma_{{\mathrm{1}}}  \ottsym{,}  \Gamma_{{\mathrm{2}}}  \vdash _{  \{\}  }  \ottsym{\{} \, \mathbf{return} \, \mathit{x}  \mapsto  \ottnt{e_{\ottmv{r}}} \,  \! [ {\bm{ { S } } }^{ \mathit{I} } / {\bm{ \alpha } }^{ \mathit{I} } ]   \ottsym{\}}  :  \ottnt{A} \,  \! [ {\bm{ { S } } }^{ \mathit{I} } / {\bm{ \alpha } }^{ \mathit{I} } ]   ^ { \varepsilon' \,  \! [ {\bm{ { S } } }^{ \mathit{I} } / {\bm{ \alpha } }^{ \mathit{I} } ]  }  \Rightarrow  ^ { \varepsilon \,  \! [ {\bm{ { S } } }^{ \mathit{I} } / {\bm{ \alpha } }^{ \mathit{I} } ]  }  \ottnt{B} \,  \! [ {\bm{ { S } } }^{ \mathit{I} } / {\bm{ \alpha } }^{ \mathit{I} } ]  .
            \end{align*}

            \item[\rname{SH}{Op}]
            Without loss of generality, we can choose $ \bm{ { \beta } } ^ {  \mathit{J}  } $ such that:
            \begin{itemize}
              \item $ \{   \bm{ { \beta } } ^ {  \mathit{J}  }   \}   \cap   \{   \bm{ { \alpha } } ^ {  \mathit{I}  }   \}   \ottsym{=}  \emptyset$ and
              \item $ \{   \bm{ { \beta } } ^ {  \mathit{J}  }   \}   \cap    \mathrm{FTV}   \ottsym{(}    \bm{ { S } } ^ {  \mathit{I}  }    \ottsym{)}    \ottsym{=}  \emptyset$.
            \end{itemize}
            For some $\ottnt{h'}$, $\sigma'$, $\mathsf{op}$, $\ottnt{A'}$, $\ottnt{B'}$, and $\ottnt{e}$, the following are given:
            \begin{itemize}
              \item $\ottnt{h} =  \ottnt{h'}   \uplus   \ottsym{\{}  \mathsf{op} \,  {\bm{ \beta } }^{ \mathit{J} } : {\bm{ \ottnt{K} } }^{ \mathit{J} }  \, \mathit{p} \, \mathit{k}  \mapsto  \ottnt{e}  \ottsym{\}} $,
              \item $\sigma =  \sigma'   \uplus   \ottsym{\{}  \mathsf{op}  \ottsym{:}    \forall    {\bm{ \beta } }^{ \mathit{J} } : {\bm{ \ottnt{K} } }^{ \mathit{J} }    \ottsym{.}    \ottnt{A'}   \Rightarrow   \ottnt{B'}   \ottsym{\}} $,
              \item $ \Gamma_{{\mathrm{1}}}  \ottsym{,}   {\bm{ \alpha } }^{ \mathit{I} } : {\bm{ \ottnt{K} } }^{ \mathit{I} }   \ottsym{,}  \Gamma_{{\mathrm{2}}}  \vdash _{ \sigma' }  \ottnt{h'}  :  \ottnt{A}  ^ { \varepsilon' }  \Rightarrow  ^ { \varepsilon }  \ottnt{B} $, and
              \item $\Gamma_{{\mathrm{1}}}  \ottsym{,}   {\bm{ \alpha } }^{ \mathit{I} } : {\bm{ \ottnt{K} } }^{ \mathit{I} }   \ottsym{,}  \Gamma_{{\mathrm{2}}}  \ottsym{,}   {\bm{ \beta } }^{ \mathit{J} } : {\bm{ \ottnt{K} } }^{ \mathit{J} }   \ottsym{,}  \mathit{p}  \ottsym{:}  \ottnt{A'}  \ottsym{,}  \mathit{k}  \ottsym{:}   \ottnt{B'}    \rightarrow_{ \varepsilon' }    \ottnt{B}   \vdash  \ottnt{e}  \ottsym{:}  \ottnt{B}  \mid  \varepsilon$.
            \end{itemize}
            By the induction hypothesis and Definition~\ref{def:subst_typelike}, we have
            \begin{itemize}
              \item $\sigma \,  \! [ {\bm{ { S } } }^{ \mathit{I} } / {\bm{ \alpha } }^{ \mathit{I} } ]  =  \sigma' \,  \! [ {\bm{ { S } } }^{ \mathit{I} } / {\bm{ \alpha } }^{ \mathit{I} } ]    \uplus   \ottsym{\{}  \mathsf{op}  \ottsym{:}    \forall    {\bm{ \beta } }^{ \mathit{J} } : {\bm{ \ottnt{K} } }^{ \mathit{J} }    \ottsym{.}    \ottnt{A'} \,  \! [ {\bm{ { S } } }^{ \mathit{I} } / {\bm{ \alpha } }^{ \mathit{I} } ]    \Rightarrow   \ottnt{B'} \,  \! [ {\bm{ { S } } }^{ \mathit{I} } / {\bm{ \alpha } }^{ \mathit{I} } ]    \ottsym{\}} $,
              \item $ \Gamma_{{\mathrm{1}}}  \ottsym{,}  \Gamma_{{\mathrm{2}}} \,  \! [ {\bm{ { S } } }^{ \mathit{I} } / {\bm{ \alpha } }^{ \mathit{I} } ]   \vdash _{ \sigma' \,  \! [ {\bm{ { S } } }^{ \mathit{I} } / {\bm{ \alpha } }^{ \mathit{I} } ]  }  \ottnt{h'} \,  \! [ {\bm{ { S } } }^{ \mathit{I} } / {\bm{ \alpha } }^{ \mathit{I} } ]   :  \ottnt{A} \,  \! [ {\bm{ { S } } }^{ \mathit{I} } / {\bm{ \alpha } }^{ \mathit{I} } ]   ^ { \varepsilon' \,  \! [ {\bm{ { S } } }^{ \mathit{I} } / {\bm{ \alpha } }^{ \mathit{I} } ]  }  \Rightarrow  ^ { \varepsilon \,  \! [ {\bm{ { S } } }^{ \mathit{I} } / {\bm{ \alpha } }^{ \mathit{I} } ]  }  \ottnt{B} \,  \! [ {\bm{ { S } } }^{ \mathit{I} } / {\bm{ \alpha } }^{ \mathit{I} } ]  $, and
              \item $\Gamma_{{\mathrm{1}}}  \ottsym{,}  \Gamma_{{\mathrm{2}}} \,  \! [ {\bm{ { S } } }^{ \mathit{I} } / {\bm{ \alpha } }^{ \mathit{I} } ]   \ottsym{,}   {\bm{ \beta } }^{ \mathit{J} } : {\bm{ \ottnt{K} } }^{ \mathit{J} }   \ottsym{,}  \mathit{p}  \ottsym{:}  \ottnt{A'} \,  \! [ {\bm{ { S } } }^{ \mathit{I} } / {\bm{ \alpha } }^{ \mathit{I} } ]   \ottsym{,}  \mathit{k}  \ottsym{:}   \ottnt{B'} \,  \! [ {\bm{ { S } } }^{ \mathit{I} } / {\bm{ \alpha } }^{ \mathit{I} } ]     \rightarrow_{ \varepsilon' \,  \! [ {\bm{ { S } } }^{ \mathit{I} } / {\bm{ \alpha } }^{ \mathit{I} } ]  }    \ottnt{B} \,  \! [ {\bm{ { S } } }^{ \mathit{I} } / {\bm{ \alpha } }^{ \mathit{I} } ]    \vdash  \ottnt{e} \,  \! [ {\bm{ { S } } }^{ \mathit{I} } / {\bm{ \alpha } }^{ \mathit{I} } ]   \ottsym{:}  \ottnt{B} \,  \! [ {\bm{ { S } } }^{ \mathit{I} } / {\bm{ \alpha } }^{ \mathit{I} } ]   \mid  \varepsilon \,  \! [ {\bm{ { S } } }^{ \mathit{I} } / {\bm{ \alpha } }^{ \mathit{I} } ] $.
            \end{itemize}
            Thus, \rname{SH}{Op} derives
            \begin{align*}
               \Gamma_{{\mathrm{1}}}  \ottsym{,}  \Gamma_{{\mathrm{2}}} \,  \! [ {\bm{ { S } } }^{ \mathit{I} } / {\bm{ \alpha } }^{ \mathit{I} } ]   \vdash _{ \sigma \,  \! [ {\bm{ { S } } }^{ \mathit{I} } / {\bm{ \alpha } }^{ \mathit{I} } ]  }   \ottnt{h'} \,  \! [ {\bm{ { S } } }^{ \mathit{I} } / {\bm{ \alpha } }^{ \mathit{I} } ]    \uplus   \ottsym{\{}  \mathsf{op} \,  {\bm{ \beta } }^{ \mathit{J} } : {\bm{ \ottnt{K} } }^{ \mathit{J} }  \, \mathit{p} \, \mathit{k}  \mapsto  \ottnt{e} \,  \! [ {\bm{ { S } } }^{ \mathit{I} } / {\bm{ \alpha } }^{ \mathit{I} } ]   \ottsym{\}}   :  \ottnt{A} \,  \! [ {\bm{ { S } } }^{ \mathit{I} } / {\bm{ \alpha } }^{ \mathit{I} } ]   ^ { \varepsilon' \,  \! [ {\bm{ { S } } }^{ \mathit{I} } / {\bm{ \alpha } }^{ \mathit{I} } ]  }  \Rightarrow  ^ { \varepsilon \,  \! [ {\bm{ { S } } }^{ \mathit{I} } / {\bm{ \alpha } }^{ \mathit{I} } ]  }  \ottnt{B} \,  \! [ {\bm{ { S } } }^{ \mathit{I} } / {\bm{ \alpha } }^{ \mathit{I} } ]  .
            \end{align*}

            \item[others] Similarly to Lemma~\ref{lem:subst_type}\ref{lem:subst_type:typing} and \ref{lem:subst_type:handling}.
          \end{divcases}
  \end{itemize}
\end{proof}

\begin{lemma}[Well-formedness of contexts in typing judgments]
  \label{lem:ctx-wf-typing-shallow}
  \phantom{}\\
  \begin{itemize}
    \item If $\Gamma  \vdash  \ottnt{e}  \ottsym{:}  \ottnt{A}  \mid  \varepsilon$, then $\vdash  \Gamma$.
    \item If $ \Gamma  \vdash _{ \sigma }  \ottnt{h}  :  \ottnt{A}  ^ { \varepsilon' }  \Rightarrow  ^ { \varepsilon }  \ottnt{B} $, then $\vdash  \Gamma$.
  \end{itemize}
\end{lemma}
\begin{proof}
  Straightforward by mutual induction on the derivations.
\end{proof}

\begin{lemma}[Well-kinded of Typing]\label{lem:wk_shallow}
  \phantom{}
  \begin{itemize}
    \item\label{lem:wk_shallow:typing} If $\Gamma  \vdash  \ottnt{e}  \ottsym{:}  \ottnt{A}  \mid  \varepsilon$, then $\Gamma  \vdash  \ottnt{A}  \ottsym{:}   \mathbf{Typ} $ and $\Gamma  \vdash  \varepsilon  \ottsym{:}   \mathbf{Eff} $.
    \item\label{lem:wk_shallow:handling} If $ \Gamma  \vdash _{ \sigma }  \ottnt{h}  :  \ottnt{A}  ^ { \varepsilon' }  \Rightarrow  ^ { \varepsilon }  \ottnt{B} $, then $\Gamma  \vdash  \ottnt{A}  \ottsym{:}   \mathbf{Typ} $ and $\Gamma  \vdash  \varepsilon'  \ottsym{:}   \mathbf{Eff} $ and $\Gamma  \vdash  \ottnt{B}  \ottsym{:}   \mathbf{Typ} $ and $\Gamma  \vdash  \varepsilon  \ottsym{:}   \mathbf{Eff} $.
  \end{itemize}
\end{lemma}

\begin{proof}
  By mutual induction on derivations of the judgments.
  We proceed by cases on the typing rule applied lastly to the derivation.
  \begin{divcases}
    \item[\rname{T}{SHandling}]
    For some $\ottnt{A'}$, $\varepsilon'$, $\sigma$, $\mathit{N}$, $ \bm{ { \alpha } } ^ {  \mathit{N}  } $, and $ \bm{ { S } } ^ {  \mathit{N}  } $, we have
    \begin{align*}
       \Gamma  \vdash _{ \sigma \,  \! [ {\bm{ { S } } }^{ \mathit{N} } / {\bm{ \alpha } }^{ \mathit{N} } ]  }  \ottnt{h}  :  \ottnt{A'}  ^ { \varepsilon' }  \Rightarrow  ^ { \varepsilon }  \ottnt{A} .
    \end{align*}
    By the induction hypothesis, we have $\Gamma  \vdash  \ottnt{A}  \ottsym{:}   \mathbf{Typ} $ and $\Gamma  \vdash  \varepsilon  \ottsym{:}   \mathbf{Eff} $.

    \item[\rname{SH}{Return}]
    For some $\mathit{x}$ and $\ottnt{e_{\ottmv{r}}}$, we have
    \begin{itemize}
      \item $\Gamma  \ottsym{,}  \mathit{x}  \ottsym{:}  \ottnt{A}  \vdash  \ottnt{e_{\ottmv{r}}}  \ottsym{:}  \ottnt{B}  \mid  \varepsilon$ and
      \item $\Gamma  \vdash  \varepsilon'  \ottsym{:}   \mathbf{Eff} $.
    \end{itemize}
    By the induction hypothesis, we have
    \begin{itemize}
      \item $\Gamma  \ottsym{,}  \mathit{x}  \ottsym{:}  \ottnt{A}  \vdash  \ottnt{B}  \ottsym{:}   \mathbf{Typ} $ and
      \item $\Gamma  \ottsym{,}  \mathit{x}  \ottsym{:}  \ottnt{A}  \vdash  \varepsilon  \ottsym{:}   \mathbf{Eff} $.
    \end{itemize}
    By Lemma~\ref{lem:delta_context}\ref{lem:delta_context:kinding}, we have
    \begin{itemize}
      \item $ \Delta   \ottsym{(}   \Gamma   \ottsym{)}   \vdash  \ottnt{B}  \ottsym{:}   \mathbf{Typ} $ and
      \item $ \Delta   \ottsym{(}   \Gamma   \ottsym{)}   \vdash  \varepsilon  \ottsym{:}   \mathbf{Eff} $.
    \end{itemize}
    By Lemma~\ref{lem:delta_weakening}, we have
    \begin{itemize}
      \item $\Gamma  \vdash  \ottnt{B}  \ottsym{:}   \mathbf{Typ} $ and
      \item $\Gamma  \vdash  \varepsilon  \ottsym{:}   \mathbf{Eff} $.
    \end{itemize}

    Now, we have $\vdash  \Gamma  \ottsym{,}  \mathit{x}  \ottsym{:}  \ottnt{A}$ by Lemma~\ref{lem:ctx-wf-typing-shallow}.
    Since only \rname{C}{Var} can derive $\vdash  \Gamma  \ottsym{,}  \mathit{x}  \ottsym{:}  \ottnt{A}$, we have $\Gamma  \vdash  \ottnt{A}  \ottsym{:}   \mathbf{Typ} $.

    \item[\rname{SH}{Op}]
    For some $\ottnt{h'}$ and $\sigma'$, we have $ \Gamma  \vdash _{ \sigma' }  \ottnt{h'}  :  \ottnt{A}  ^ { \varepsilon' }  \Rightarrow  ^ { \varepsilon }  \ottnt{B} $.
    By the induction hypothesis, we have $\Gamma  \vdash  \ottnt{A}  \ottsym{:}   \mathbf{Typ} $ and $\Gamma  \vdash  \varepsilon'  \ottsym{:}   \mathbf{Eff} $ and $\Gamma  \vdash  \ottnt{B}  \ottsym{:}   \mathbf{Typ} $ and $\Gamma  \vdash  \varepsilon  \ottsym{:}   \mathbf{Eff} $.

    \item[others] Similarly to Lemma~\ref{lem:wk}\ref{lem:wk:typing} and \ref{lem:wk:handling}.
  \end{divcases}
\end{proof}

\begin{lemma}[Inversion]\label{lem:inversion_shallow}
  \mbox{}
  \begin{enumerate}
    \item\label{lem:inversion_shallow:var} If $\Gamma  \vdash  \ottnt{v}  \ottsym{:}  \ottnt{A}  \mid  \varepsilon$, then $\Gamma  \vdash  \ottnt{v}  \ottsym{:}  \ottnt{A}  \mid   \bbZero $.
    \item\label{lem:inversion_shallow:abs} If $\Gamma  \vdash  \ottkw{fun} \, \ottsym{(}  \mathit{f}  \ottsym{,}  \mathit{x}  \ottsym{,}  \ottnt{e}  \ottsym{)}  \ottsym{:}   \ottnt{A_{{\mathrm{1}}}}    \rightarrow_{ \varepsilon_{{\mathrm{1}}} }    \ottnt{B_{{\mathrm{1}}}}   \mid  \varepsilon$, then $\Gamma  \ottsym{,}  \mathit{f}  \ottsym{:}   \ottnt{A_{{\mathrm{2}}}}    \rightarrow_{ \varepsilon_{{\mathrm{2}}} }    \ottnt{B_{{\mathrm{2}}}}   \ottsym{,}  \mathit{x}  \ottsym{:}  \ottnt{A_{{\mathrm{2}}}}  \vdash  \ottnt{e}  \ottsym{:}  \ottnt{B_{{\mathrm{2}}}}  \mid  \varepsilon_{{\mathrm{2}}}$ for some $\ottnt{A_{{\mathrm{2}}}}$, $\varepsilon_{{\mathrm{2}}}$, and $\ottnt{B_{{\mathrm{2}}}}$ such that $\Gamma  \vdash   \ottnt{A_{{\mathrm{2}}}}    \rightarrow_{ \varepsilon_{{\mathrm{2}}} }    \ottnt{B_{{\mathrm{2}}}}   <:   \ottnt{A_{{\mathrm{1}}}}    \rightarrow_{ \varepsilon_{{\mathrm{1}}} }    \ottnt{B_{{\mathrm{1}}}} $.
    \item\label{lem:inversion_shallow:tabs} If $\Gamma  \vdash  \Lambda  \alpha  \ottsym{:}  \ottnt{K}  \ottsym{.}  \ottnt{e}  \ottsym{:}    \forall   \alpha  \ottsym{:}  \ottnt{K}   \ottsym{.}    \ottnt{A_{{\mathrm{1}}}}    ^{ \varepsilon_{{\mathrm{1}}} }    \mid  \varepsilon$, then $\Gamma  \ottsym{,}  \alpha  \ottsym{:}  \ottnt{K}  \vdash  \ottnt{e}  \ottsym{:}  \ottnt{A_{{\mathrm{1}}}}  \mid  \varepsilon_{{\mathrm{1}}}$.
    \item\label{lem:inversion_shallow:op} If $\Gamma  \vdash   \mathsf{op} _{ \mathit{l} \,  \bm{ { S } } ^ {  \mathit{I}  }  }  \,  \bm{ { T } } ^ {  \mathit{J}  }   \ottsym{:}   \ottnt{A_{{\mathrm{1}}}}    \rightarrow_{ \varepsilon_{{\mathrm{1}}} }    \ottnt{B_{{\mathrm{1}}}}   \mid  \varepsilon$, then the following hold:
          \begin{itemize}
            \item $ \mathit{l}  ::    \forall    {\bm{ \alpha } }^{ \mathit{I} } : {\bm{ \ottnt{K} } }^{ \mathit{I} }    \ottsym{.}    \sigma    \in   \Xi $,
            \item $ \mathsf{op}  \ottsym{:}    \forall    {\bm{ \beta } }^{ \mathit{J} } : {\bm{ \ottnt{K'} } }^{ \mathit{J} }    \ottsym{.}    \ottnt{A}   \Rightarrow   \ottnt{B}    \in   \sigma \,  \! [ {\bm{ { S } } }^{ \mathit{I} } / {\bm{ \alpha } }^{ \mathit{I} } ]  $,
            \item $\vdash  \Gamma$,
            \item $\Gamma  \vdash   \bm{ { S } }^{ \mathit{I} } : \bm{ \ottnt{K} }^{ \mathit{I} } $,
            \item $\Gamma  \vdash   \bm{ { T } }^{ \mathit{J} } : \bm{ \ottnt{K'} }^{ \mathit{J} } $,
            \item $\Gamma  \vdash  \ottnt{A_{{\mathrm{1}}}}  <:  \ottnt{A} \,  \! [ {\bm{ { T } } }^{ \mathit{J} } / {\bm{ \beta } }^{ \mathit{J} } ] $,
            \item $\Gamma  \vdash  \ottnt{B} \,  \! [ {\bm{ { T } } }^{ \mathit{J} } / {\bm{ \beta } }^{ \mathit{J} } ]   <:  \ottnt{B_{{\mathrm{1}}}}$, and
            \item $\Gamma  \vdash    \lift{ \mathit{l} \,  \bm{ { S } } ^ {  \mathit{I}  }  }   \olessthan  \varepsilon_{{\mathrm{1}}} $
          \end{itemize}
          for some $ \bm{ { \alpha } } ^ {  \mathit{I}  } $, $ {\bm{ { \ottnt{K} } } }^{ \mathit{I} } $, $\sigma$, $ \bm{ { \beta } } ^ {  \mathit{J}  } $, $ {\bm{ { \ottnt{K'} } } }^{ \mathit{J} } $, $\ottnt{A}$, and $\ottnt{B}$.
    \item\label{lem:inversion_shallow:app} If $\Gamma  \vdash  \ottnt{v_{{\mathrm{1}}}} \, \ottnt{v_{{\mathrm{2}}}}  \ottsym{:}  \ottnt{B}  \mid  \varepsilon$, then there exists some type $\ottnt{A}$ such that $\Gamma  \vdash  \ottnt{v_{{\mathrm{1}}}}  \ottsym{:}   \ottnt{A}    \rightarrow_{ \varepsilon }    \ottnt{B}   \mid   \bbZero $ and $\Gamma  \vdash  \ottnt{v_{{\mathrm{2}}}}  \ottsym{:}  \ottnt{A}  \mid   \bbZero $.
  \end{enumerate}
\end{lemma}

\begin{proof}
  Similarly to Lemma~\ref{lem:inversion};
  Lemmas~\ref{lem:ctx-wf-typing-shallow} and \ref{lem:wk_shallow} are used
  instead of Lemmas~\ref{lem:ctx-wf-typing} and \ref{lem:wk}, respectively.
\end{proof}

\begin{lemma}[Canonical Form]\label{lem:canonical_shallow}
  \phantom{}
  \begin{enumerate}
    \item\label{lem:canonical_shallow:abs} If $\emptyset  \vdash  \ottnt{v}  \ottsym{:}   \ottnt{A}    \rightarrow_{ \varepsilon }    \ottnt{B}   \mid  \varepsilon'$, then either of the following holds:
          \begin{itemize}
            \item $\ottnt{v} = \ottkw{fun} \, \ottsym{(}  \mathit{f}  \ottsym{,}  \mathit{x}  \ottsym{,}  \ottnt{e}  \ottsym{)}$ for some $\mathit{f}$, $\mathit{x}$, and $\ottnt{e}$, or
            \item $\ottnt{v} =  \mathsf{op} _{ \mathit{l} \,  \bm{ { S } } ^ {  \mathit{I}  }  }  \,  \bm{ { T } } ^ {  \mathit{J}  } $ for some $\mathsf{op}$, $\mathit{l}$, $ \bm{ { S } } ^ {  \mathit{I}  } $, and $ \bm{ { T } } ^ {  \mathit{J}  } $.
          \end{itemize}
    \item\label{lem:canonical_shallow:tabs} If $\emptyset  \vdash  \ottnt{v}  \ottsym{:}    \forall   \alpha  \ottsym{:}  \ottnt{K}   \ottsym{.}    \ottnt{A}    ^{ \varepsilon }    \mid  \varepsilon'$, then $\ottnt{v} = \Lambda  \alpha  \ottsym{:}  \ottnt{K}  \ottsym{.}  \ottnt{e}$ for some $\ottnt{e}$.
  \end{enumerate}
\end{lemma}

\begin{proof}
  Similarly to Lemma~\ref{lem:canonical}.
\end{proof}

\begin{lemma}[Inversion of Handler Typing]\label{lem:inversion_handler_shallow}
  \phantom{}
  \begin{enumerate}
    \item\label{lem:inversion_handler_shallow:return} If $ \Gamma  \vdash _{ \sigma }  \ottnt{h}  :  \ottnt{A}  ^ { \varepsilon' }  \Rightarrow  ^ { \varepsilon }  \ottnt{B} $, then there exist some $\mathit{x}$ and $\ottnt{e_{\ottmv{r}}}$ such that $ \mathbf{return} \, \mathit{x}  \mapsto  \ottnt{e_{\ottmv{r}}}   \in   \ottnt{h} $ and $\Gamma  \ottsym{,}  \mathit{x}  \ottsym{:}  \ottnt{A}  \vdash  \ottnt{e_{\ottmv{r}}}  \ottsym{:}  \ottnt{B}  \mid  \varepsilon$.
    \item\label{lem:inversion_handler_shallow:operation} If $ \Gamma  \vdash _{ \sigma }  \ottnt{h}  :  \ottnt{A}  ^ { \varepsilon' }  \Rightarrow  ^ { \varepsilon }  \ottnt{B} $ and $ \mathsf{op}  \ottsym{:}    \forall    {\bm{ \beta } }^{ \mathit{J} } : {\bm{ \ottnt{K} } }^{ \mathit{J} }    \ottsym{.}    \ottnt{A'}   \Rightarrow   \ottnt{B'}    \in   \sigma $, then
          \begin{itemize}
            \item $ \mathsf{op} \,  {\bm{ \beta } }^{ \mathit{J} } : {\bm{ \ottnt{K} } }^{ \mathit{J} }  \, \mathit{p} \, \mathit{k}  \mapsto  \ottnt{e}   \in   \ottnt{h} $ and
            \item $\Gamma  \ottsym{,}   {\bm{ \beta } }^{ \mathit{J} } : {\bm{ \ottnt{K} } }^{ \mathit{J} }   \ottsym{,}  \mathit{p}  \ottsym{:}  \ottnt{A'}  \ottsym{,}  \mathit{k}  \ottsym{:}   \ottnt{B'}    \rightarrow_{ \varepsilon' }    \ottnt{B}   \vdash  \ottnt{e}  \ottsym{:}  \ottnt{B}  \mid  \varepsilon$
          \end{itemize}
          for some $\mathit{p}$, $\mathit{k}$, and $\ottnt{e}$.
  \end{enumerate}
\end{lemma}

\begin{proof}
  \begin{enumerate}
    \item By induction on a derivation of $ \Gamma  \vdash _{ \sigma }  \ottnt{h}  :  \ottnt{A}  ^ { \varepsilon' }  \Rightarrow  ^ { \varepsilon }  \ottnt{B} $.
          We proceed by cases on the typing rule applied lastly to this derivation.
          \begin{divcases}
            \item[\rname{H}{Return}] Clearly.
            \item[\rname{H}{Op}] Clearly by the induction hypothesis.
          \end{divcases}

    \item By induction on a derivation of $ \Gamma  \vdash _{ \sigma }  \ottnt{h}  :  \ottnt{A}  ^ { \varepsilon' }  \Rightarrow  ^ { \varepsilon }  \ottnt{B} $.
          We proceed by cases on the typing rule applied lastly to this derivation.
          \begin{divcases}
            \item[\rname{H}{Return}]
            Clearly because there is no operation belonging to $ \{\} $.

            \item[\rname{H}{Op}]
            For some $\ottnt{h'}$, $\sigma'$, $\mathsf{op'}$, $ \bm{ { \beta' } } ^ {  \mathit{J'}  } $, $ {\bm{ { \ottnt{K'} } } }^{ \mathit{J'} } $, $\ottnt{A''}$, $\ottnt{B''}$, $\mathit{p'}$, $\mathit{k'}$, $\ottnt{e''}$, and $\varepsilon'$, the following are given:
            \begin{itemize}
              \item $\ottnt{h} =  \ottnt{h'}   \uplus   \ottsym{\{}  \mathsf{op'} \,  {\bm{ \beta' } }^{ \mathit{J'} } : {\bm{ \ottnt{K'} } }^{ \mathit{J'} }  \, \mathit{p'} \, \mathit{k'}  \mapsto  \ottnt{e'}  \ottsym{\}} $,
              \item $\sigma =  \sigma'   \uplus   \ottsym{\{}  \mathsf{op'}  \ottsym{:}    \forall    {\bm{ \beta' } }^{ \mathit{J'} } : {\bm{ \ottnt{K'} } }^{ \mathit{J'} }    \ottsym{.}    \ottnt{A''}   \Rightarrow   \ottnt{B''}   \ottsym{\}} $,
              \item $ \Gamma  \vdash _{ \sigma' }  \ottnt{h'}  :  \ottnt{A}  ^ { \varepsilon' }  \Rightarrow  ^ { \varepsilon }  \ottnt{B} $, and
              \item $\Gamma  \ottsym{,}   {\bm{ \beta' } }^{ \mathit{J'} } : {\bm{ \ottnt{K'} } }^{ \mathit{J'} }   \ottsym{,}  \mathit{p'}  \ottsym{:}  \ottnt{A''}  \ottsym{,}  \mathit{k'}  \ottsym{:}   \ottnt{B''}    \rightarrow_{ \varepsilon' }    \ottnt{B}   \vdash  \ottnt{e}  \ottsym{:}  \ottnt{B}  \mid  \varepsilon$.
            \end{itemize}

            If $\mathsf{op} = \mathsf{op'}$, then clearly.

            If $\mathsf{op} \neq \mathsf{op'}$, then clearly by the induction hypothesis.
          \end{divcases}
  \end{enumerate}
\end{proof}

\begin{lemma}[Independence of Evaluation Contexts]\label{lem:ind_ev_shallow}
  If $\Gamma  \vdash  \ottnt{E}  \ottsym{[}  \ottnt{e}  \ottsym{]}  \ottsym{:}  \ottnt{A}  \mid  \varepsilon$, then
  there exist some $\ottnt{A'}$ and $\varepsilon'$ such that
  \begin{itemize}
    \item $\Gamma  \vdash  \ottnt{e}  \ottsym{:}  \ottnt{A'}  \mid  \varepsilon'$, and
    \item $\Gamma  \ottsym{,}  \Gamma'  \vdash  \ottnt{E}  \ottsym{[}  \ottnt{e'}  \ottsym{]}  \ottsym{:}  \ottnt{A}  \mid  \varepsilon$ holds for any $\ottnt{e'}$ and $\Gamma'$ such that $\Gamma  \ottsym{,}  \Gamma'  \vdash  \ottnt{e'}  \ottsym{:}  \ottnt{A'}  \mid  \varepsilon'$.
  \end{itemize}
\end{lemma}

\begin{proof}
  Similarly to Lemma~\ref{lem:ind_ev};
  Lemmas~\ref{lem:weakening_shallow} and \ref{lem:ctx-wf-typing-shallow} are used
  instead of Lemmas~\ref{lem:weakening} and \ref{lem:ctx-wf-typing}, respectively.
\end{proof}

\begin{lemma}[Progress]\label{lem:progress_shallow}
  If $\emptyset  \vdash  \ottnt{e}  \ottsym{:}  \ottnt{A}  \mid  \varepsilon$, then one of the following holds:
  \begin{itemize}
    \item $\ottnt{e}$ is a value;
    \item There exists some expression $\ottnt{e'}$ such that $\ottnt{e}  \longrightarrow  \ottnt{e'}$; or
    \item There exist some $\mathsf{op}$, $\mathit{l}$, $ \bm{ { S } } ^ {  \mathit{I}  } $, $ \bm{ { T } } ^ {  \mathit{J}  } $, $\ottnt{v}$, $\ottnt{E}$, and $\mathit{n}$
          such that $\ottnt{e} = \ottnt{E}  \ottsym{[}   \mathsf{op} _{ \mathit{l} \,  \bm{ { S } } ^ {  \mathit{I}  }  }  \,  \bm{ { T } } ^ {  \mathit{J}  }  \, \ottnt{v}  \ottsym{]}$ and $ \mathit{n}  \mathrm{-free} ( \mathit{l} \,  \bm{ { S } } ^ {  \mathit{I}  }  ,  \ottnt{E} ) $.
  \end{itemize}
\end{lemma}

\begin{proof}
  Similarly to Lemma~\ref{lem:progress};
  Lemmas~\ref{lem:inversion_shallow}, \ref{lem:canonical_shallow}, and \ref{lem:inversion_handler_shallow} are used
  instead of Lemmas~\ref{lem:inversion}, \ref{lem:canonical}, and \ref{lem:inversion_handler}, respectively.
\end{proof}

\begin{lemma}[Preservation in Reduction]\label{lem:pres_red_shallow}
  If $\emptyset  \vdash  \ottnt{e}  \ottsym{:}  \ottnt{A}  \mid  \varepsilon$ and $\ottnt{e}  \longmapsto  \ottnt{e'}$, then $\emptyset  \vdash  \ottnt{e'}  \ottsym{:}  \ottnt{A}  \mid  \varepsilon$.
\end{lemma}

\begin{proof}
  By induction on a derivation of $\Gamma  \vdash  \ottnt{e}  \ottsym{:}  \ottnt{A}  \mid  \varepsilon$.
  We proceed by cases on the typing rule applied lastly to this derivation.
  \begin{divcases}
    \item[\rname{T}{SHandling}]
    We proceed by cases on the derivation rule which derives $\ottnt{e}  \longmapsto  \ottnt{e'}$.
    \begin{divcases}
      \item[\rname{R}{Handle1}] We have
      \begin{itemize}
        \item $\ottnt{e} =  \mathbf{handle}_{ \mathit{l} \,  \bm{ { S } } ^ {  \mathit{I}  }  }  \, \ottnt{v} \, \mathbf{with} \, \ottnt{h}$,
        \item $ \mathbf{return} \, \mathit{x}  \mapsto  \ottnt{e_{\ottmv{r}}}   \in   \ottnt{h} $,
        \item $\emptyset  \vdash  \ottnt{v}  \ottsym{:}  \ottnt{B}  \mid  \varepsilon'$,
        \item $ \mathit{l}  ::    \forall    {\bm{ \alpha } }^{ \mathit{I} } : {\bm{ \ottnt{K} } }^{ \mathit{I} }    \ottsym{.}    \sigma    \in   \Xi $,
        \item $\emptyset  \vdash   \bm{ { S } }^{ \mathit{I} } : \bm{ \ottnt{K} }^{ \mathit{I} } $,
        \item $ \emptyset  \vdash _{ \sigma \,  \! [ {\bm{ { S } } }^{ \mathit{I} } / {\bm{ \alpha } }^{ \mathit{I} } ]  }  \ottnt{h}  :  \ottnt{B}  ^ { \varepsilon' }  \Rightarrow  ^ { \varepsilon }  \ottnt{A} $,
        \item $   \lift{ \mathit{l} \,  \bm{ { S } } ^ {  \mathit{I}  }  }   \mathop{ \odot }  \varepsilon    \sim   \varepsilon' $, and
        \item $\ottnt{e'} = \ottnt{e_{\ottmv{r}}} \,  \! [  \ottnt{v}  /  \mathit{x}  ] $
      \end{itemize}
      for some $\mathit{l}$, $ \bm{ { S } } ^ {  \mathit{I}  } $, $ \bm{ { \alpha } } ^ {  \mathit{I}  } $, $ {\bm{ { \ottnt{K} } } }^{ \mathit{I} } $, $\sigma$, $\ottnt{v}$, $\ottnt{h}$, $\ottnt{B}$, and $\varepsilon'$.
      By $ \emptyset  \vdash _{ \sigma \,  \! [ {\bm{ { S } } }^{ \mathit{I} } / {\bm{ \alpha } }^{ \mathit{I} } ]  }  \ottnt{h}  :  \ottnt{B}  ^ { \varepsilon' }  \Rightarrow  ^ { \varepsilon }  \ottnt{A} $ and
      $ \mathbf{return} \, \mathit{x}  \mapsto  \ottnt{e_{\ottmv{r}}}   \in   \ottnt{h} $ and
      Lemma~\ref{lem:inversion_handler_shallow}\ref{lem:inversion_handler_shallow:return},
      we have
      \begin{align*}
        \mathit{x}  \ottsym{:}  \ottnt{B}  \vdash  \ottnt{e_{\ottmv{r}}}  \ottsym{:}  \ottnt{A}  \mid  \varepsilon.
      \end{align*}
      By Lemma~\ref{lem:inversion_shallow}\ref{lem:inversion_shallow:var},
      we have $\emptyset  \vdash  \ottnt{v}  \ottsym{:}  \ottnt{B}  \mid   \bbZero $.
      Thus, Lemma~\ref{lem:subst_value_shallow}\ref{lem:subst_value_shallow:typing} makes
      $\emptyset  \vdash  \ottnt{e_{\ottmv{r}}} \,  \! [  \ottnt{v}  /  \mathit{x}  ]   \ottsym{:}  \ottnt{A}  \mid  \varepsilon$ hold as required.

      \item[\rname{R}{Handle2}] We have
      \begin{itemize}
        \item $\ottnt{e} =  \mathbf{handle}_{ \mathit{l} \,  \bm{ { S } } ^ {  \mathit{N}  }  }  \, \ottnt{E}  \ottsym{[}   \mathsf{op_{{\mathrm{0}}}} _{ \mathit{l} \,  \bm{ { S } } ^ {  \mathit{N}  }  }  \,  \bm{ { T } } ^ {  \mathit{J}  }  \, \ottnt{v}  \ottsym{]} \, \mathbf{with} \, \ottnt{h}$,
        \item $ \mathit{l}  ::    \forall    {\bm{ \alpha } }^{ \mathit{N} } : {\bm{ \ottnt{K} } }^{ \mathit{N} }    \ottsym{.}    \sigma    \in   \Xi $,
        \item $\emptyset  \vdash   \bm{ { S } }^{ \mathit{N} } : \bm{ \ottnt{K} }^{ \mathit{N} } $,
        \item $ \mathsf{op_{{\mathrm{0}}}} \,  {\bm{ \beta_{{\mathrm{0}}} } }^{ \mathit{J} } : {\bm{ \ottnt{K_{{\mathrm{0}}}} } }^{ \mathit{J} }  \, \mathit{p_{{\mathrm{0}}}} \, \mathit{k_{{\mathrm{0}}}}  \mapsto  \ottnt{e_{{\mathrm{0}}}}   \in   \ottnt{h} $,
        \item $ 0  \mathrm{-free} ( \mathit{l} \,  \bm{ { S } } ^ {  \mathit{N}  }  ,  \ottnt{E} ) $,
        \item $\emptyset  \vdash  \ottnt{E}  \ottsym{[}   \mathsf{op_{{\mathrm{0}}}} _{ \mathit{l} \,  \bm{ { S } } ^ {  \mathit{N}  }  }  \,  \bm{ { T } } ^ {  \mathit{J}  }  \, \ottnt{v}  \ottsym{]}  \ottsym{:}  \ottnt{B}  \mid  \varepsilon'$,
        \item $ \emptyset  \vdash _{ \sigma \,  \! [ {\bm{ { S } } }^{ \mathit{N} } / {\bm{ \alpha } }^{ \mathit{N} } ]  }  \ottnt{h}  :  \ottnt{B}  ^ { \varepsilon' }  \Rightarrow  ^ { \varepsilon }  \ottnt{A} $,
        \item $   \lift{ \mathit{l} \,  \bm{ { S } } ^ {  \mathit{N}  }  }   \mathop{ \odot }  \varepsilon    \sim   \varepsilon' $, and
        \item $\ottnt{e'} = \ottnt{e_{{\mathrm{0}}}} \,  \! [ {\bm{ { T } } }^{ \mathit{J} } / {\bm{ \beta_{{\mathrm{0}}} } }^{ \mathit{J} } ]  \,  \! [  \ottnt{v}  /  \mathit{p_{{\mathrm{0}}}}  ]  \,  \! [  \lambda  \mathit{z}  \ottsym{.}  \ottnt{E}  \ottsym{[}  \mathit{z}  \ottsym{]}  /  \mathit{k_{{\mathrm{0}}}}  ] $
      \end{itemize}
      for some $\mathit{l}$, $ \bm{ { S } } ^ {  \mathit{N}  } $, $\ottnt{E}$, $\mathsf{op_{{\mathrm{0}}}}$, $ \bm{ { T } } ^ {  \mathit{J}  } $, $\ottnt{v}$, $\ottnt{h}$, $ \bm{ { \alpha } } ^ {  \mathit{N}  } $, $ {\bm{ { \ottnt{K} } } }^{ \mathit{N} } $, $\sigma$, $ \bm{ { \beta_{{\mathrm{0}}} } } ^ {  \mathit{J}  } $, $ {\bm{ { \ottnt{K_{{\mathrm{0}}}} } } }^{ \mathit{J} } $, $\mathit{p_{{\mathrm{0}}}}$, $\mathit{k_{{\mathrm{0}}}}$, $\ottnt{e_{{\mathrm{0}}}}$, $\ottnt{B}$, and $\varepsilon'$.
      By Lemma~\ref{lem:ind_ev_shallow}, there exist some $\ottnt{B_{{\mathrm{1}}}}$ and $\varepsilon_{{\mathrm{1}}}$ such that
      \begin{itemize}
        \item $\emptyset  \vdash   \mathsf{op_{{\mathrm{0}}}} _{ \mathit{l} \,  \bm{ { S } } ^ {  \mathit{N}  }  }  \,  \bm{ { T } } ^ {  \mathit{J}  }  \, \ottnt{v}  \ottsym{:}  \ottnt{B_{{\mathrm{1}}}}  \mid  \varepsilon_{{\mathrm{1}}}$, and
        \item for any $\ottnt{e''}$ and $\Gamma''$,
              if $\Gamma''  \vdash  \ottnt{e''}  \ottsym{:}  \ottnt{B_{{\mathrm{1}}}}  \mid  \varepsilon_{{\mathrm{1}}}$,
              then $\Gamma''  \vdash  \ottnt{E}  \ottsym{[}  \ottnt{e''}  \ottsym{]}  \ottsym{:}  \ottnt{B}  \mid  \varepsilon'$.
      \end{itemize}
      By Lemma~\ref{lem:inversion_shallow}\ref{lem:inversion_shallow:app},
      we have $\emptyset  \vdash   \mathsf{op_{{\mathrm{0}}}} _{ \mathit{l} \,  \bm{ { S } } ^ {  \mathit{N}  }  }  \,  \bm{ { T } } ^ {  \mathit{J}  }   \ottsym{:}   \ottnt{A_{{\mathrm{1}}}}    \rightarrow_{ \varepsilon_{{\mathrm{1}}} }    \ottnt{B_{{\mathrm{1}}}}   \mid   \bbZero $ and
      $\emptyset  \vdash  \ottnt{v}  \ottsym{:}  \ottnt{A_{{\mathrm{1}}}}  \mid   \bbZero $ for some $\ottnt{A_{{\mathrm{1}}}}$.
      By Lemma~\ref{lem:inversion_shallow}\ref{lem:inversion_shallow:op} and
      \ref{lem:inversion_handler_shallow}\ref{lem:inversion_handler_shallow:operation},
      we have
      \begin{itemize}
        \item $ \mathsf{op_{{\mathrm{0}}}}  \ottsym{:}    \forall    {\bm{ \beta_{{\mathrm{0}}} } }^{ \mathit{J} } : {\bm{ \ottnt{K_{{\mathrm{0}}}} } }^{ \mathit{J} }    \ottsym{.}    \ottnt{A_{{\mathrm{0}}}}   \Rightarrow   \ottnt{B_{{\mathrm{0}}}}    \in   \sigma \,  \! [ {\bm{ { S } } }^{ \mathit{N} } / {\bm{ \alpha } }^{ \mathit{N} } ]  $,
        \item $\emptyset  \vdash   \bm{ { S } }^{ \mathit{N} } : \bm{ \ottnt{K} }^{ \mathit{N} } $,
        \item $\emptyset  \vdash   \bm{ { T } }^{ \mathit{J} } : \bm{ \ottnt{K_{{\mathrm{0}}}} }^{ \mathit{J} } $,
        \item $\emptyset  \vdash  \ottnt{A_{{\mathrm{1}}}}  <:  \ottnt{A_{{\mathrm{0}}}} \,  \! [ {\bm{ { T } } }^{ \mathit{J} } / {\bm{ \beta_{{\mathrm{0}}} } }^{ \mathit{J} } ] $,
        \item $\emptyset  \vdash  \ottnt{B_{{\mathrm{0}}}} \,  \! [ {\bm{ { T } } }^{ \mathit{J} } / {\bm{ \beta_{{\mathrm{0}}} } }^{ \mathit{J} } ]   <:  \ottnt{B_{{\mathrm{1}}}}$, and
        \item $\emptyset  \vdash    \lift{ \mathit{l} \,  \bm{ { S } } ^ {  \mathit{N}  }  }   \olessthan  \varepsilon_{{\mathrm{1}}} $.
      \end{itemize}
      for some $\ottnt{A_{{\mathrm{0}}}}$ and $\ottnt{B_{{\mathrm{0}}}}$.
      Thus, \rname{T}{Sub} with $\emptyset  \vdash    \bbZero   \olessthan   \bbZero  $ implied by Lemma~\ref{lem:entailment} derives
      \begin{align*}
        \emptyset  \vdash  \ottnt{v}  \ottsym{:}  \ottnt{A_{{\mathrm{0}}}} \,  \! [ {\bm{ { T } } }^{ \mathit{J} } / {\bm{ \beta_{{\mathrm{0}}} } }^{ \mathit{J} } ]   \mid   \bbZero .
      \end{align*}
      By Lemma~\ref{lem:wk_subtyping}, we have $\emptyset  \vdash  \ottnt{B_{{\mathrm{0}}}} \,  \! [ {\bm{ { T } } }^{ \mathit{J} } / {\bm{ \beta_{{\mathrm{0}}} } }^{ \mathit{J} } ]   \ottsym{:}   \mathbf{Typ} $.
      Thus, \rname{C}{Var} derives $\vdash  \mathit{z}  \ottsym{:}  \ottnt{B_{{\mathrm{0}}}} \,  \! [ {\bm{ { T } } }^{ \mathit{J} } / {\bm{ \beta_{{\mathrm{0}}} } }^{ \mathit{J} } ] $.
      By $\emptyset  \vdash   \bbZero   \ottsym{:}   \mathbf{Eff} $, $\emptyset  \vdash  \varepsilon_{{\mathrm{1}}}  \ottsym{:}   \mathbf{Eff} $ implied by Lemma~\ref{lem:wk}, and $   \bbZero   \mathop{ \odot }  \varepsilon_{{\mathrm{1}}}    \sim   \varepsilon_{{\mathrm{1}}} $,
      we have $\emptyset  \vdash    \bbZero   \olessthan  \varepsilon_{{\mathrm{1}}} $.
      Since \rname{T}{Var} and \rname{T}{Sub} derives $\mathit{z}  \ottsym{:}  \ottnt{B_{{\mathrm{0}}}} \,  \! [ {\bm{ { T } } }^{ \mathit{J} } / {\bm{ \beta_{{\mathrm{0}}} } }^{ \mathit{J} } ]   \vdash  \mathit{z}  \ottsym{:}  \ottnt{B_{{\mathrm{1}}}}  \mid  \varepsilon_{{\mathrm{1}}}$,
      we have
      \begin{align*}
        \mathit{z}  \ottsym{:}  \ottnt{B_{{\mathrm{0}}}} \,  \! [ {\bm{ { T } } }^{ \mathit{J} } / {\bm{ \beta_{{\mathrm{0}}} } }^{ \mathit{J} } ]   \vdash  \ottnt{E}  \ottsym{[}  \mathit{z}  \ottsym{]}  \ottsym{:}  \ottnt{B}  \mid  \varepsilon'
      \end{align*}
      by the result of Lemma~\ref{lem:ind_ev_shallow}.
      Thus, \rname{T}{Abs} derives
      \begin{align*}
        \emptyset  \vdash  \lambda  \mathit{z}  \ottsym{.}  \ottnt{E}  \ottsym{[}  \mathit{z}  \ottsym{]}  \ottsym{:}   \ottnt{B_{{\mathrm{0}}}} \,  \! [ {\bm{ { T } } }^{ \mathit{J} } / {\bm{ \beta_{{\mathrm{0}}} } }^{ \mathit{J} } ]     \rightarrow_{ \varepsilon' }    \ottnt{B}   \mid   \bbZero .
      \end{align*}
      Since
      \[
         {\bm{ \beta_{{\mathrm{0}}} } }^{ \mathit{J} } : {\bm{ \ottnt{K_{{\mathrm{0}}}} } }^{ \mathit{J} }   \ottsym{,}  \mathit{p_{{\mathrm{0}}}}  \ottsym{:}  \ottnt{A_{{\mathrm{0}}}}  \ottsym{,}  \mathit{k_{{\mathrm{0}}}}  \ottsym{:}   \ottnt{B_{{\mathrm{0}}}}    \rightarrow_{ \varepsilon' }    \ottnt{B}   \vdash  \ottnt{e_{{\mathrm{0}}}}  \ottsym{:}  \ottnt{A}  \mid  \varepsilon
      \]
      by $ \emptyset  \vdash _{ \sigma \,  \! [ {\bm{ { S } } }^{ \mathit{N} } / {\bm{ \alpha } }^{ \mathit{N} } ]  }  \ottnt{h}  :  \ottnt{B}  ^ { \varepsilon' }  \Rightarrow  ^ { \varepsilon }  \ottnt{A} $ and
      $ \mathsf{op_{{\mathrm{0}}}}  \ottsym{:}    \forall    {\bm{ \beta_{{\mathrm{0}}} } }^{ \mathit{J} } : {\bm{ \ottnt{K_{{\mathrm{0}}}} } }^{ \mathit{J} }    \ottsym{.}    \ottnt{A_{{\mathrm{0}}}}   \Rightarrow   \ottnt{B_{{\mathrm{0}}}}    \in   \sigma $ and
      Lemma~\ref{lem:inversion_handler_shallow}\ref{lem:inversion_handler_shallow:operation},
      Lemma~\ref{lem:subst_type_shallow}\ref{lem:subst_type_shallow:typing} and
      Lemma~\ref{lem:subst_value_shallow}\ref{lem:subst_value_shallow:typing} imply
      \begin{align*}
        \emptyset  \vdash  \ottnt{e_{{\mathrm{0}}}} \,  \! [ {\bm{ { T } } }^{ \mathit{J} } / {\bm{ \beta_{{\mathrm{0}}} } }^{ \mathit{J} } ]  \,  \! [  \ottnt{v}  /  \mathit{p_{{\mathrm{0}}}}  ]  \,  \! [  \lambda  \mathit{z}  \ottsym{.}  \ottnt{E}  \ottsym{[}  \mathit{z}  \ottsym{]}  /  \mathit{k_{{\mathrm{0}}}}  ]   \ottsym{:}  \ottnt{A}  \mid  \varepsilon
      \end{align*}
      as required.
    \end{divcases}

    \item[others]
    Similarly to Lemma~\ref{lem:pres_red};
    Lemmas~\ref{lem:inversion_shallow}, \ref{lem:weakening_shallow}, \ref{lem:ctx-wf-typing-shallow}, \ref{lem:wk_shallow}, \ref{lem:subst_value_shallow}, and \ref{lem:subst_type_shallow} are used
    instead of Lemmas~\ref{lem:inversion}, \ref{lem:weakening}, \ref{lem:ctx-wf-typing}, \ref{lem:wk}, \ref{lem:subst_value}, and \ref{lem:subst_type} respectively.
  \end{divcases}
\end{proof}

\begin{lemma}[Preservation]\label{lem:preservation_shallow}
  If $\emptyset  \vdash  \ottnt{e}  \ottsym{:}  \ottnt{A}  \mid  \varepsilon$ and $\ottnt{e}  \longrightarrow  \ottnt{e'}$, then $\emptyset  \vdash  \ottnt{e'}  \ottsym{:}  \ottnt{A}  \mid  \varepsilon$.
\end{lemma}

\begin{proof}
  Similarly to Lemma~\ref{lem:preservation};
  Lemmas~\ref{lem:ind_ev_shallow} and \ref{lem:pres_red_shallow} are used
  instead of Lemmas~\ref{lem:ind_ev} and \ref{lem:pres_red}.
\end{proof}

\begin{lemma}\label{lem:effsafe_aux_shallow}
  If $\Gamma  \vdash  \ottnt{E}  \ottsym{[}   \mathsf{op} _{ \mathit{l} \,  \bm{ { S } } ^ {  \mathit{I}  }  }  \,  \bm{ { T } } ^ {  \mathit{J}  }  \, \ottnt{v}  \ottsym{]}  \ottsym{:}  \ottnt{A}  \mid  \varepsilon$ and
  $ \mathit{n}  \mathrm{-free} ( \mathit{l} \,  \bm{ { S } } ^ {  \mathit{I}  }  ,  \ottnt{E} ) $,
  then $\Gamma  \vdash    \lift{ \mathit{l} \,  \bm{ { S } } ^ {  \mathit{I}  }  }   \olessthan  \varepsilon $.
\end{lemma}

\begin{proof}
  Similarly to Lemma~\ref{lem:effsafe_aux};
  Lemma~\ref{lem:inversion_shallow} is used instead of Lemma~\ref{lem:inversion}.
\end{proof}

\begin{lemma}[Effect Safety]\label{lem:effsafe_shallow}
  If $\Gamma  \vdash  \ottnt{E}  \ottsym{[}   \mathsf{op} _{ \mathit{l} \,  \bm{ { S } } ^ {  \mathit{I}  }  }  \,  \bm{ { T } } ^ {  \mathit{J}  }  \, \ottnt{v}  \ottsym{]}  \ottsym{:}  \ottnt{A}  \mid  \varepsilon$ and
  $ \mathit{n}  \mathrm{-free} ( \mathit{l} \,  \bm{ { S } } ^ {  \mathit{I}  }  ,  \ottnt{E} ) $,
  then $ \varepsilon   \nsim    \bbZero  $.
\end{lemma}

\begin{proof}
  Similarly to Lemma~\ref{lem:effsafe};
  Lemma~\ref{lem:effsafe_aux_shallow} is used instead of Lemma\ref{lem:effsafe_aux}.
\end{proof}

\begin{theorem}[Type and Effect Safety]\label{thm:safety_shallow}
  If $\emptyset  \vdash  \ottnt{e}  \ottsym{:}  \ottnt{A}  \mid   \bbZero $ and $\ottnt{e}  \longrightarrow  ^ * \ottnt{e'}$ and $\ottnt{e'} \centernot \longrightarrow $, then $\ottnt{e'}$ is a value.
\end{theorem}

\begin{proof}
  Similarly to Theorem~\ref{thm:safety};
  Lemmas~\ref{lem:preservation_shallow},
  \ref{lem:effsafe_shallow}, and
  \ref{lem:progress_shallow} are used
  instead of Lemmas~\ref{lem:preservation},
  \ref{lem:effsafe}, and
  \ref{lem:progress}, respectively.
\end{proof}
\subsection{Properties with Lift Coercions \TY{Chaneg list (3)}}

This section assumes that the safety conditions in Definition~\ref{def:safe_cond}
and the safety condition for lift coercions in Definition~\ref{def:safe_cond_lift} hold.

\begin{lemma}[Weakening]\label{lem:weakening_lift}
  Suppose that $\vdash  \Gamma_{{\mathrm{1}}}  \ottsym{,}  \Gamma_{{\mathrm{2}}}$ and $  \mathrm{dom}   \ottsym{(}   \Gamma_{{\mathrm{2}}}   \ottsym{)}    \cap    \mathrm{dom}   \ottsym{(}   \Gamma_{{\mathrm{3}}}   \ottsym{)}    \ottsym{=}  \emptyset$.
  \begin{enumerate}
    \item\label{lem:weakening_lift:wf} If $\vdash  \Gamma_{{\mathrm{1}}}  \ottsym{,}  \Gamma_{{\mathrm{3}}}$, then $\vdash  \Gamma_{{\mathrm{1}}}  \ottsym{,}  \Gamma_{{\mathrm{2}}}  \ottsym{,}  \Gamma_{{\mathrm{3}}}$.
    \item\label{lem:weakening_lift:kinding} If $\Gamma_{{\mathrm{1}}}  \ottsym{,}  \Gamma_{{\mathrm{3}}}  \vdash  S  \ottsym{:}  \ottnt{K}$, then $\Gamma_{{\mathrm{1}}}  \ottsym{,}  \Gamma_{{\mathrm{2}}}  \ottsym{,}  \Gamma_{{\mathrm{3}}}  \vdash  S  \ottsym{:}  \ottnt{K}$.
    \item\label{lem:weakening_lift:subtyping} If $\Gamma_{{\mathrm{1}}}  \ottsym{,}  \Gamma_{{\mathrm{3}}}  \vdash  \ottnt{A}  <:  \ottnt{B}$, then $\Gamma_{{\mathrm{1}}}  \ottsym{,}  \Gamma_{{\mathrm{2}}}  \ottsym{,}  \Gamma_{{\mathrm{3}}}  \vdash  \ottnt{A}  <:  \ottnt{B}$.
    \item\label{lem:weakening_lift:subtyping_comp} If $\Gamma_{{\mathrm{1}}}  \ottsym{,}  \Gamma_{{\mathrm{3}}}  \vdash  \ottnt{A_{{\mathrm{1}}}}  \mid  \varepsilon_{{\mathrm{1}}}  <:  \ottnt{A_{{\mathrm{2}}}}  \mid  \varepsilon_{{\mathrm{2}}}$, then $\Gamma_{{\mathrm{1}}}  \ottsym{,}  \Gamma_{{\mathrm{2}}}  \ottsym{,}  \Gamma_{{\mathrm{3}}}  \vdash  \ottnt{A_{{\mathrm{1}}}}  \mid  \varepsilon_{{\mathrm{1}}}  <:  \ottnt{A_{{\mathrm{2}}}}  \mid  \varepsilon_{{\mathrm{2}}}$.
    \item\label{lem:weakening_lift:typing} If $\Gamma_{{\mathrm{1}}}  \ottsym{,}  \Gamma_{{\mathrm{3}}}  \vdash  \ottnt{e}  \ottsym{:}  \ottnt{A}  \mid  \varepsilon$, then $\Gamma_{{\mathrm{1}}}  \ottsym{,}  \Gamma_{{\mathrm{2}}}  \ottsym{,}  \Gamma_{{\mathrm{3}}}  \vdash  \ottnt{e}  \ottsym{:}  \ottnt{A}  \mid  \varepsilon$.
    \item\label{lem:weakening_lift:handling} If $ \Gamma_{{\mathrm{1}}}  \ottsym{,}  \Gamma_{{\mathrm{3}}}  \vdash _{ \sigma }  \ottnt{h}  :  \ottnt{A}   \Rightarrow  ^ { \varepsilon }  \ottnt{B} $, then $ \Gamma_{{\mathrm{1}}}  \ottsym{,}  \Gamma_{{\mathrm{2}}}  \ottsym{,}  \Gamma_{{\mathrm{3}}}  \vdash _{ \sigma }  \ottnt{h}  :  \ottnt{A}   \Rightarrow  ^ { \varepsilon }  \ottnt{B} $.
  \end{enumerate}
\end{lemma}

\begin{proof}
  \begin{itemize}
    \item[(1)(2)] Similarly to Lemma~\ref{lem:weakening}\ref{lem:weakening:wf} and \ref{lem:weakening:kinding}.

    \item[(3)(4)] Similarly to Lemma~\ref{lem:weakening}\ref{lem:weakening:subtyping} and \ref{lem:weakening:subtyping_comp}.

    \item[(5)(6)]
          By mutual induction on derivations of the judgments.
          We proceed by case analysis on the rule applied lastly to the derivation.
          \begin{divcases}
            \item[\rname{T}{Lift}]
            For some $\ottnt{e'}$, $\ottnt{L}$, and $\varepsilon'$, the following are given:
            \begin{itemize}
              \item $\ottnt{e} =  [  \ottnt{e'}  ] _{ \ottnt{L} } $,
              \item $\Gamma_{{\mathrm{1}}}  \ottsym{,}  \Gamma_{{\mathrm{3}}}  \vdash  \ottnt{e'}  \ottsym{:}  \ottnt{A}  \mid  \varepsilon'$,
              \item $\Gamma_{{\mathrm{1}}}  \ottsym{,}  \Gamma_{{\mathrm{3}}}  \vdash  \ottnt{L}  \ottsym{:}   \mathbf{Lab} $, and
              \item $   \lift{ \ottnt{L} }   \mathop{ \odot }  \varepsilon'    \sim   \varepsilon $.
            \end{itemize}
            By the induction hypothesis and case~\ref{lem:weakening_lift:kinding},
            we have
            \begin{itemize}
              \item $\Gamma_{{\mathrm{1}}}  \ottsym{,}  \Gamma_{{\mathrm{2}}}  \ottsym{,}  \Gamma_{{\mathrm{3}}}  \vdash  \ottnt{e'}  \ottsym{:}  \ottnt{A}  \mid  \varepsilon'$ and
              \item $\Gamma_{{\mathrm{1}}}  \ottsym{,}  \Gamma_{{\mathrm{2}}}  \ottsym{,}  \Gamma_{{\mathrm{3}}}  \vdash  \ottnt{L}  \ottsym{:}   \mathbf{Lab} $.
            \end{itemize}
            Thus, \rname{T}{Lift} derives $\Gamma_{{\mathrm{1}}}  \ottsym{,}  \Gamma_{{\mathrm{2}}}  \ottsym{,}  \Gamma_{{\mathrm{3}}}  \vdash   [  \ottnt{e'}  ] _{ \ottnt{L} }   \ottsym{:}  \ottnt{A}  \mid  \varepsilon$.

            \item[others] Similarly to Lemma~\ref{lem:weakening}\ref{lem:weakening:typing} and \ref{lem:weakening:handling}.
          \end{divcases}
  \end{itemize}
\end{proof}

\begin{lemma}[Substitution of values]\label{lem:subst_value_lift}
  Suppose that $\Gamma_{{\mathrm{1}}}  \vdash  \ottnt{v}  \ottsym{:}  \ottnt{A}  \mid   \bbZero $.
  \begin{enumerate}
    \item\label{lem:subst_value_lift:wf} If $\vdash  \Gamma_{{\mathrm{1}}}  \ottsym{,}  \mathit{x}  \ottsym{:}  \ottnt{A}  \ottsym{,}  \Gamma_{{\mathrm{2}}}$, then $\vdash  \Gamma_{{\mathrm{1}}}  \ottsym{,}  \Gamma_{{\mathrm{2}}}$.
    \item\label{lem:subst_value_lift:kinding} If $\Gamma_{{\mathrm{1}}}  \ottsym{,}  \mathit{x}  \ottsym{:}  \ottnt{A}  \ottsym{,}  \Gamma_{{\mathrm{2}}}  \vdash  S  \ottsym{:}  \ottnt{K}$, then $\Gamma_{{\mathrm{1}}}  \ottsym{,}  \Gamma_{{\mathrm{2}}}  \vdash  S  \ottsym{:}  \ottnt{K}$.
    \item\label{lem:subst_value_lift:subtyping} If $\Gamma_{{\mathrm{1}}}  \ottsym{,}  \mathit{x}  \ottsym{:}  \ottnt{A}  \ottsym{,}  \Gamma_{{\mathrm{2}}}  \vdash  \ottnt{B}  <:  \ottnt{C}$, then $\Gamma_{{\mathrm{1}}}  \ottsym{,}  \Gamma_{{\mathrm{2}}}  \vdash  \ottnt{B}  <:  \ottnt{C}$.
    \item\label{lem:subst_value_lift:subtyping_comp} If $\Gamma_{{\mathrm{1}}}  \ottsym{,}  \mathit{x}  \ottsym{:}  \ottnt{A}  \ottsym{,}  \Gamma_{{\mathrm{2}}}  \vdash  \ottnt{B_{{\mathrm{1}}}}  \mid  \varepsilon_{{\mathrm{1}}}  <:  \ottnt{B_{{\mathrm{2}}}}  \mid  \varepsilon_{{\mathrm{2}}}$, then $\Gamma_{{\mathrm{1}}}  \ottsym{,}  \Gamma_{{\mathrm{2}}}  \vdash  \ottnt{B_{{\mathrm{1}}}}  \mid  \varepsilon_{{\mathrm{1}}}  <:  \ottnt{B_{{\mathrm{2}}}}  \mid  \varepsilon_{{\mathrm{2}}}$.
    \item\label{lem:subst_value_lift:typing} If $\Gamma_{{\mathrm{1}}}  \ottsym{,}  \mathit{x}  \ottsym{:}  \ottnt{A}  \ottsym{,}  \Gamma_{{\mathrm{2}}}  \vdash  \ottnt{e}  \ottsym{:}  \ottnt{B}  \mid  \varepsilon$, then $\Gamma_{{\mathrm{1}}}  \ottsym{,}  \Gamma_{{\mathrm{2}}}  \vdash  \ottnt{e} \,  \! [  \ottnt{v}  /  \mathit{x}  ]   \ottsym{:}  \ottnt{B}  \mid  \varepsilon$.
    \item\label{lem:subst_value_lift:handling} If $ \Gamma_{{\mathrm{1}}}  \ottsym{,}  \mathit{x}  \ottsym{:}  \ottnt{A}  \ottsym{,}  \Gamma_{{\mathrm{2}}}  \vdash _{ \sigma }  \ottnt{h}  :  \ottnt{B}  ^ { \varepsilon' }  \Rightarrow  ^ { \varepsilon }  \ottnt{C} $, then $ \Gamma_{{\mathrm{1}}}  \ottsym{,}  \Gamma_{{\mathrm{2}}}  \vdash _{ \sigma }  \ottnt{h} \,  \! [  \ottnt{v}  /  \mathit{x}  ]   :  \ottnt{B}  ^ { \varepsilon' }  \Rightarrow  ^ { \varepsilon }  \ottnt{C} $.
  \end{enumerate}
\end{lemma}

\begin{proof}
  \begin{itemize}
    \item[(1)(2)] Similarly to Lemma~\ref{lem:subst_value}\ref{lem:subst_value:wf} and \ref{lem:subst_value:kinding}.

    \item[(3)(4)] Similarly to Lemma~\ref{lem:subst_value}\ref{lem:subst_value:subtyping} and \ref{lem:subst_value}\ref{lem:subst_value:subtyping_comp}.

    \item[(5)(6)]
          By mutual induction on derivations of the judgments.
          We proceed by case analysis on the rule applied lastly to the derivation.
          \begin{divcases}
            \item[\rname{T}{Lift}]
            For some $\ottnt{e'}$, $\varepsilon'$, and $\ottnt{L}$, the following are given:
            \begin{itemize}
              \item $\ottnt{e} =  [  \ottnt{e'}  ] _{ \ottnt{L} } $,
              \item $\Gamma_{{\mathrm{1}}}  \ottsym{,}  \mathit{x}  \ottsym{:}  \ottnt{A}  \ottsym{,}  \Gamma_{{\mathrm{2}}}  \vdash  \ottnt{e'}  \ottsym{:}  \ottnt{B}  \mid  \varepsilon'$,
              \item $\Gamma_{{\mathrm{1}}}  \ottsym{,}  \mathit{x}  \ottsym{:}  \ottnt{A}  \ottsym{,}  \Gamma_{{\mathrm{2}}}  \vdash  \ottnt{L}  \ottsym{:}   \mathbf{Lab} $, and
              \item $   \lift{ \ottnt{L} }   \mathop{ \odot }  \varepsilon'    \sim   \varepsilon $.
            \end{itemize}
            By the induction hypothesis and case~\ref{lem:subst_value_lift}\ref{lem:subst_value_lift:kinding},
            we have
            \begin{itemize}
              \item $\Gamma_{{\mathrm{1}}}  \ottsym{,}  \Gamma_{{\mathrm{2}}}  \vdash  \ottnt{e'} \,  \! [  \ottnt{v}  /  \mathit{x}  ]   \ottsym{:}  \ottnt{B}  \mid  \varepsilon'$ and
              \item $\Gamma_{{\mathrm{1}}}  \ottsym{,}  \Gamma_{{\mathrm{2}}}  \vdash  \ottnt{L}  \ottsym{:}   \mathbf{Lab} $.
            \end{itemize}
            Thus, \rname{T}{Lift} derives $\Gamma_{{\mathrm{1}}}  \ottsym{,}  \Gamma_{{\mathrm{2}}}  \vdash   [  \ottnt{e'} \,  \! [  \ottnt{v}  /  \mathit{x}  ]   ] _{ \ottnt{L} }   \ottsym{:}  \ottnt{A}  \mid  \varepsilon$.

            \item[others] Similarly to Lemma~\ref{lem:subst_value}\ref{lem:subst_value:typing} and \ref{lem:subst_value:handling}.
          \end{divcases}
  \end{itemize}
\end{proof}

\begin{lemma}[Substitution of Typelikes]\label{lem:subst_type_lift}
  Suppose that $\Gamma_{{\mathrm{1}}}  \vdash   \bm{ { S } }^{ \mathit{I} } : \bm{ \ottnt{K} }^{ \mathit{I} } $.
  \begin{enumerate}
    \item\label{lem:subst_type_lift:wf} If $\vdash  \Gamma_{{\mathrm{1}}}  \ottsym{,}   {\bm{ \alpha } }^{ \mathit{I} } : {\bm{ \ottnt{K} } }^{ \mathit{I} }   \ottsym{,}  \Gamma_{{\mathrm{2}}}$, then $\vdash  \Gamma_{{\mathrm{1}}}  \ottsym{,}  \Gamma_{{\mathrm{2}}} \,  \! [ {\bm{ { S } } }^{ \mathit{I} } / {\bm{ \alpha } }^{ \mathit{I} } ] $.
    \item\label{lem:subst_type_lift:kinding} If $\Gamma_{{\mathrm{1}}}  \ottsym{,}   {\bm{ \alpha } }^{ \mathit{I} } : {\bm{ \ottnt{K} } }^{ \mathit{I} }   \ottsym{,}  \Gamma_{{\mathrm{2}}}  \vdash  T  \ottsym{:}  \ottnt{K}$, then $\Gamma_{{\mathrm{1}}}  \ottsym{,}  \Gamma_{{\mathrm{2}}} \,  \! [ {\bm{ { S } } }^{ \mathit{I} } / {\bm{ \alpha } }^{ \mathit{I} } ]   \vdash  T \,  \! [ {\bm{ { S } } }^{ \mathit{I} } / {\bm{ \alpha } }^{ \mathit{I} } ]   \ottsym{:}  \ottnt{K}$.
    \item\label{lem:subst_type_lift:subtyping} If $\Gamma_{{\mathrm{1}}}  \ottsym{,}   {\bm{ \alpha } }^{ \mathit{I} } : {\bm{ \ottnt{K} } }^{ \mathit{I} }   \ottsym{,}  \Gamma_{{\mathrm{2}}}  \vdash  \ottnt{A}  <:  \ottnt{B}$, then $\Gamma_{{\mathrm{1}}}  \ottsym{,}  \Gamma_{{\mathrm{2}}} \,  \! [ {\bm{ { S } } }^{ \mathit{I} } / {\bm{ \alpha } }^{ \mathit{I} } ]   \vdash  \ottnt{A} \,  \! [ {\bm{ { S } } }^{ \mathit{I} } / {\bm{ \alpha } }^{ \mathit{I} } ]   <:  \ottnt{B} \,  \! [ {\bm{ { S } } }^{ \mathit{I} } / {\bm{ \alpha } }^{ \mathit{I} } ] $.
    \item\label{lem:subst_type_lift:subtyping_comp} If $\Gamma_{{\mathrm{1}}}  \ottsym{,}   {\bm{ \alpha } }^{ \mathit{I} } : {\bm{ \ottnt{K} } }^{ \mathit{I} }   \ottsym{,}  \Gamma_{{\mathrm{2}}}  \vdash  \ottnt{A_{{\mathrm{1}}}}  \mid  \varepsilon_{{\mathrm{1}}}  <:  \ottnt{A_{{\mathrm{2}}}}  \mid  \varepsilon_{{\mathrm{2}}}$, then $\Gamma_{{\mathrm{1}}}  \ottsym{,}  \Gamma_{{\mathrm{2}}} \,  \! [ {\bm{ { S } } }^{ \mathit{I} } / {\bm{ \alpha } }^{ \mathit{I} } ]   \vdash  \ottnt{A_{{\mathrm{1}}}} \,  \! [ {\bm{ { S } } }^{ \mathit{I} } / {\bm{ \alpha } }^{ \mathit{I} } ]   \mid  \varepsilon_{{\mathrm{1}}} \,  \! [ {\bm{ { S } } }^{ \mathit{I} } / {\bm{ \alpha } }^{ \mathit{I} } ]   <:  \ottnt{A_{{\mathrm{2}}}} \,  \! [ {\bm{ { S } } }^{ \mathit{I} } / {\bm{ \alpha } }^{ \mathit{I} } ]   \mid  \varepsilon_{{\mathrm{2}}} \,  \! [ {\bm{ { S } } }^{ \mathit{I} } / {\bm{ \alpha } }^{ \mathit{I} } ] $.
    \item\label{lem:subst_type_lift:typing} If $\Gamma_{{\mathrm{1}}}  \ottsym{,}   {\bm{ \alpha } }^{ \mathit{I} } : {\bm{ \ottnt{K} } }^{ \mathit{I} }   \ottsym{,}  \Gamma_{{\mathrm{2}}}  \vdash  \ottnt{e}  \ottsym{:}  \ottnt{A}  \mid  \varepsilon$, then $\Gamma_{{\mathrm{1}}}  \ottsym{,}  \Gamma_{{\mathrm{2}}} \,  \! [ {\bm{ { S } } }^{ \mathit{I} } / {\bm{ \alpha } }^{ \mathit{I} } ]   \vdash  \ottnt{e} \,  \! [ {\bm{ { S } } }^{ \mathit{I} } / {\bm{ \alpha } }^{ \mathit{I} } ]   \ottsym{:}  \ottnt{A} \,  \! [ {\bm{ { S } } }^{ \mathit{I} } / {\bm{ \alpha } }^{ \mathit{I} } ]   \mid  \varepsilon \,  \! [ {\bm{ { S } } }^{ \mathit{I} } / {\bm{ \alpha } }^{ \mathit{I} } ] $.
    \item\label{lem:subst_type_lift:handling} If $ \Gamma_{{\mathrm{1}}}  \ottsym{,}   {\bm{ \alpha } }^{ \mathit{I} } : {\bm{ \ottnt{K} } }^{ \mathit{I} }   \ottsym{,}  \Gamma_{{\mathrm{2}}}  \vdash _{ \sigma }  \ottnt{h}  :  \ottnt{A}   \Rightarrow  ^ { \varepsilon }  \ottnt{B} $, then $ \Gamma_{{\mathrm{1}}}  \ottsym{,}  \Gamma_{{\mathrm{2}}} \,  \! [ {\bm{ { S } } }^{ \mathit{I} } / {\bm{ \alpha } }^{ \mathit{I} } ]   \vdash _{ \sigma \,  \! [ \bm{ { S } } / \bm{ \alpha } ]  }  \ottnt{h} \,  \! [ \bm{ { S } } / \bm{ \alpha } ]   :  \ottnt{A} \,  \! [ {\bm{ { S } } }^{ \mathit{I} } / {\bm{ \alpha } }^{ \mathit{I} } ]    \Rightarrow  ^ { \varepsilon \,  \! [ {\bm{ { S } } }^{ \mathit{I} } / {\bm{ \alpha } }^{ \mathit{I} } ]  }  \ottnt{B} \,  \! [ {\bm{ { S } } }^{ \mathit{I} } / {\bm{ \alpha } }^{ \mathit{I} } ]  $.
  \end{enumerate}
\end{lemma}

\begin{proof}
  \begin{itemize}
    \item[(1)(2)] Similarly to Lemma~\ref{lem:subst_type}\ref{lem:subst_type:wf} and \ref{lem:subst_type:kinding}.

    \item[(3)(4)] Similarly to Lemma~\ref{lem:subst_type}\ref{lem:subst_type:subtyping} and \ref{lem:subst_type}\ref{lem:subst_type:subtyping}.

    \item[(5)(6)]
          By mutual induction on derivations of the judgments.
          We proceed by case analysis on the rule applied lastly to the derivation.
          \begin{divcases}
            \item[\rname{T}{Lift}]
            For some $\ottnt{e'}$, $\varepsilon'$, and $\ottnt{L}$, the following are given:
            \begin{itemize}
              \item $\ottnt{e} =  [  \ottnt{e'}  ] _{ \ottnt{L} } $,
              \item $\Gamma_{{\mathrm{1}}}  \ottsym{,}   {\bm{ \alpha } }^{ \mathit{I} } : {\bm{ \ottnt{K} } }^{ \mathit{I} }   \ottsym{,}  \Gamma_{{\mathrm{2}}}  \vdash  \ottnt{e'}  \ottsym{:}  \ottnt{A}  \mid  \varepsilon'$,
              \item $\Gamma_{{\mathrm{1}}}  \ottsym{,}   {\bm{ \alpha } }^{ \mathit{I} } : {\bm{ \ottnt{K} } }^{ \mathit{I} }   \ottsym{,}  \Gamma_{{\mathrm{2}}}  \vdash  \ottnt{L}  \ottsym{:}   \mathbf{Lab} $, and
              \item $   \lift{ \ottnt{L} }   \mathop{ \odot }  \varepsilon'    \sim   \varepsilon $.
            \end{itemize}
            By the induction hypothesis,
            case~\ref{lem:subst_type_lift}\ref{lem:subst_type_lift:kinding}, and
            the fact that a typelike substitution is homomorphism for $ \odot $ and $ \sim $,
            we have
            \begin{itemize}
              \item $\Gamma_{{\mathrm{1}}}  \ottsym{,}  \Gamma_{{\mathrm{2}}} \,  \! [ {\bm{ { S } } }^{ \mathit{I} } / {\bm{ \alpha } }^{ \mathit{I} } ]   \vdash  \ottnt{e'} \,  \! [ {\bm{ { S } } }^{ \mathit{I} } / {\bm{ \alpha } }^{ \mathit{I} } ]   \ottsym{:}  \ottnt{A} \,  \! [ {\bm{ { S } } }^{ \mathit{I} } / {\bm{ \alpha } }^{ \mathit{I} } ]   \mid  \varepsilon' \,  \! [ {\bm{ { S } } }^{ \mathit{I} } / {\bm{ \alpha } }^{ \mathit{I} } ] $,
              \item $\Gamma_{{\mathrm{1}}}  \ottsym{,}  \Gamma_{{\mathrm{2}}} \,  \! [ {\bm{ { S } } }^{ \mathit{I} } / {\bm{ \alpha } }^{ \mathit{I} } ]   \vdash  \ottnt{L} \,  \! [ {\bm{ { S } } }^{ \mathit{I} } / {\bm{ \alpha } }^{ \mathit{I} } ]   \ottsym{:}   \mathbf{Lab} $, and
              \item $   \lift{ \ottnt{L} }  \,  \! [ {\bm{ { S } } }^{ \mathit{I} } / {\bm{ \alpha } }^{ \mathit{I} } ]   \mathop{ \odot }  \varepsilon'  \,  \! [ {\bm{ { S } } }^{ \mathit{I} } / {\bm{ \alpha } }^{ \mathit{I} } ]    \sim   \varepsilon \,  \! [ {\bm{ { S } } }^{ \mathit{I} } / {\bm{ \alpha } }^{ \mathit{I} } ]  $.
            \end{itemize}
            Thus, \rname{T}{Lift} derives $\Gamma_{{\mathrm{1}}}  \ottsym{,}  \Gamma_{{\mathrm{2}}} \,  \! [ {\bm{ { S } } }^{ \mathit{I} } / {\bm{ \alpha } }^{ \mathit{I} } ]   \vdash   [  \ottnt{e'} \,  \! [ {\bm{ { S } } }^{ \mathit{I} } / {\bm{ \alpha } }^{ \mathit{I} } ]   ] _{ \ottnt{L} \,  \! [ {\bm{ { S } } }^{ \mathit{I} } / {\bm{ \alpha } }^{ \mathit{I} } ]  }   \ottsym{:}  \ottnt{A} \,  \! [ {\bm{ { S } } }^{ \mathit{I} } / {\bm{ \alpha } }^{ \mathit{I} } ]   \mid  \varepsilon \,  \! [ {\bm{ { S } } }^{ \mathit{I} } / {\bm{ \alpha } }^{ \mathit{I} } ] $.

            \item[others] Similarly to Lemma~\ref{lem:subst_type}\ref{lem:subst_type:typing} and \ref{lem:subst_type:handling}.
          \end{divcases}
  \end{itemize}
\end{proof}

\begin{lemma}[Well-formedness of contexts in typing judgments]
  \label{lem:ctx-wf-typing-lift}
  \phantom{}\\
  \begin{itemize}
    \item If $\Gamma  \vdash  \ottnt{e}  \ottsym{:}  \ottnt{A}  \mid  \varepsilon$, then $\vdash  \Gamma$.
    \item If $ \Gamma  \vdash _{ \sigma }  \ottnt{h}  :  \ottnt{A}   \Rightarrow  ^ { \varepsilon }  \ottnt{B} $, then $\vdash  \Gamma$.
  \end{itemize}
\end{lemma}

\begin{proof}
  Straightforward by mutual induction on the derivations.
\end{proof}

\begin{lemma}[Well-kinded of Typing]\label{lem:wk_lift}
  \phantom{}
  \begin{itemize}
    \item\label{lem:wk_lift:typing} If $\Gamma  \vdash  \ottnt{e}  \ottsym{:}  \ottnt{A}  \mid  \varepsilon$, then $\Gamma  \vdash  \ottnt{A}  \ottsym{:}   \mathbf{Typ} $ and $\Gamma  \vdash  \varepsilon  \ottsym{:}   \mathbf{Eff} $.
    \item\label{lem:wk_lift:handling} If $ \Gamma  \vdash _{ \sigma }  \ottnt{h}  :  \ottnt{A}  ^ { \varepsilon' }  \Rightarrow  ^ { \varepsilon }  \ottnt{B} $, then $\Gamma  \vdash  \ottnt{A}  \ottsym{:}   \mathbf{Typ} $ and $\Gamma  \vdash  \varepsilon'  \ottsym{:}   \mathbf{Eff} $ and $\Gamma  \vdash  \ottnt{B}  \ottsym{:}   \mathbf{Typ} $ and $\Gamma  \vdash  \varepsilon  \ottsym{:}   \mathbf{Eff} $.
  \end{itemize}
\end{lemma}

\begin{proof}
  By mutual induction on derivations of the judgments.
  We proceed by cases on the typing rule applied lastly to the derivation.
  \begin{divcases}
    \item[\rname{T}{Lift}]
    For some $\ottnt{e'}$, $\varepsilon'$, and $\ottnt{L}$, the following are given:
    \begin{itemize}
      \item $\ottnt{e} =  [  \ottnt{e'}  ] _{ \ottnt{L} } $,
      \item $\Gamma  \vdash  \ottnt{e'}  \ottsym{:}  \ottnt{A}  \mid  \varepsilon'$,
      \item $\Gamma  \vdash  \ottnt{L}  \ottsym{:}   \mathbf{Lab} $, and
      \item $   \lift{ \ottnt{L} }   \mathop{ \odot }  \varepsilon'    \sim   \varepsilon $.
    \end{itemize}
    By the induction hypothesis, we have $\Gamma  \vdash  \ottnt{A}  \ottsym{:}   \mathbf{Typ} $ and $\Gamma  \vdash  \varepsilon'  \ottsym{:}   \mathbf{Eff} $.
    $ \lift{ \ottsym{-} } $, $ \odot $, and $ \sim $ preserve well-formedness,
    we have $\Gamma  \vdash  \varepsilon  \ottsym{:}   \mathbf{Eff} $.

    \item[others] Similarly to Lemma~\ref{lem:wk}\ref{lem:wk:typing} and \ref{lem:wk:handling}.
  \end{divcases}
\end{proof}

\begin{lemma}[Inversion]\label{lem:inversion_lift}
  \mbox{}
  \begin{enumerate}
    \item\label{lem:inversion_lift:var} If $\Gamma  \vdash  \ottnt{v}  \ottsym{:}  \ottnt{A}  \mid  \varepsilon$, then $\Gamma  \vdash  \ottnt{v}  \ottsym{:}  \ottnt{A}  \mid   \bbZero $.
    \item\label{lem:inversion_lift:abs} If $\Gamma  \vdash  \ottkw{fun} \, \ottsym{(}  \mathit{g}  \ottsym{,}  \mathit{x}  \ottsym{,}  \ottnt{e}  \ottsym{)}  \ottsym{:}   \ottnt{A_{{\mathrm{1}}}}    \rightarrow_{ \varepsilon_{{\mathrm{1}}} }    \ottnt{B_{{\mathrm{1}}}}   \mid  \varepsilon$, then $\Gamma  \ottsym{,}  \mathit{g}  \ottsym{:}   \ottnt{A_{{\mathrm{2}}}}    \rightarrow_{ \varepsilon_{{\mathrm{2}}} }    \ottnt{B_{{\mathrm{2}}}}   \ottsym{,}  \mathit{x}  \ottsym{:}  \ottnt{A_{{\mathrm{2}}}}  \vdash  \ottnt{e}  \ottsym{:}  \ottnt{B_{{\mathrm{2}}}}  \mid  \varepsilon_{{\mathrm{2}}}$ for some $\ottnt{A_{{\mathrm{2}}}}$, $\varepsilon_{{\mathrm{2}}}$, and $\ottnt{B_{{\mathrm{2}}}}$ such that $\Gamma  \vdash   \ottnt{A_{{\mathrm{2}}}}    \rightarrow_{ \varepsilon_{{\mathrm{2}}} }    \ottnt{B_{{\mathrm{2}}}}   <:   \ottnt{A_{{\mathrm{1}}}}    \rightarrow_{ \varepsilon_{{\mathrm{1}}} }    \ottnt{B_{{\mathrm{1}}}} $.
    \item\label{lem:inversion_lift:tabs} If $\Gamma  \vdash  \Lambda  \alpha  \ottsym{:}  \ottnt{K}  \ottsym{.}  \ottnt{e}  \ottsym{:}    \forall   \alpha  \ottsym{:}  \ottnt{K}   \ottsym{.}    \ottnt{A_{{\mathrm{1}}}}    ^{ \varepsilon_{{\mathrm{1}}} }    \mid  \varepsilon$, then $\Gamma  \ottsym{,}  \alpha  \ottsym{:}  \ottnt{K}  \vdash  \ottnt{e}  \ottsym{:}  \ottnt{A_{{\mathrm{1}}}}  \mid  \varepsilon_{{\mathrm{1}}}$.
    \item\label{lem:inversion_lift:op} If $\Gamma  \vdash   \mathsf{op} _{ \mathit{l} \,  \bm{ { S } } ^ {  \mathit{I}  }  }  \,  \bm{ { T } } ^ {  \mathit{J}  }   \ottsym{:}   \ottnt{A_{{\mathrm{1}}}}    \rightarrow_{ \varepsilon_{{\mathrm{1}}} }    \ottnt{B_{{\mathrm{1}}}}   \mid  \varepsilon$, then the following hold:
          \begin{itemize}
            \item $ \mathit{l}  ::    \forall    {\bm{ \alpha } }^{ \mathit{I} } : {\bm{ \ottnt{K} } }^{ \mathit{I} }    \ottsym{.}    \sigma    \in   \Xi $,
            \item $ \mathsf{op}  \ottsym{:}    \forall    {\bm{ \beta } }^{ \mathit{J} } : {\bm{ \ottnt{K'} } }^{ \mathit{J} }    \ottsym{.}    \ottnt{A}   \Rightarrow   \ottnt{B}    \in   \sigma $,
            \item $\vdash  \Gamma$,
            \item $\Gamma  \vdash   \bm{ { S } }^{ \mathit{I} } : \bm{ \ottnt{K} }^{ \mathit{I} } $,
            \item $\Gamma  \vdash   \bm{ { T } }^{ \mathit{J} } : \bm{ \ottnt{K'} }^{ \mathit{J} } $,
            \item $\Gamma  \vdash  \ottnt{A_{{\mathrm{1}}}}  <:  \ottnt{A} \,  \! [ {\bm{ { S } } }^{ \mathit{I} } / {\bm{ \alpha } }^{ \mathit{I} } ]  \,  \! [ {\bm{ { T } } }^{ \mathit{J} } / {\bm{ \beta } }^{ \mathit{J} } ] $,
            \item $\Gamma  \vdash  \ottnt{B} \,  \! [ {\bm{ { S } } }^{ \mathit{I} } / {\bm{ \alpha } }^{ \mathit{I} } ]  \,  \! [ {\bm{ { T } } }^{ \mathit{J} } / {\bm{ \beta } }^{ \mathit{J} } ]   <:  \ottnt{B_{{\mathrm{1}}}}$, and
            \item $\Gamma  \vdash    \lift{ \mathit{l} \,  \bm{ { S } } ^ {  \mathit{I}  }  }   \olessthan  \varepsilon_{{\mathrm{1}}} $
          \end{itemize}
          for some $ \bm{ { \alpha } } ^ {  \mathit{I}  } $, $ {\bm{ { \ottnt{K} } } }^{ \mathit{I} } $, $\sigma$, $ \bm{ { \beta } } ^ {  \mathit{J}  } $, $ {\bm{ { \ottnt{K'} } } }^{ \mathit{J} } $, $\ottnt{A}$, and $\ottnt{B}$.
    \item\label{lem:inversion_lift:app} If $\Gamma  \vdash  \ottnt{v_{{\mathrm{1}}}} \, \ottnt{v_{{\mathrm{2}}}}  \ottsym{:}  \ottnt{B}  \mid  \varepsilon$, then there exists some type $\ottnt{A}$ such that $\Gamma  \vdash  \ottnt{v_{{\mathrm{1}}}}  \ottsym{:}   \ottnt{A}    \rightarrow_{ \varepsilon }    \ottnt{B}   \mid   \bbZero $ and $\Gamma  \vdash  \ottnt{v_{{\mathrm{2}}}}  \ottsym{:}  \ottnt{A}  \mid   \bbZero $.
  \end{enumerate}
\end{lemma}

\begin{proof}
  Similarly to Lemma~\ref{lem:inversion};
  Lemmas~\ref{lem:ctx-wf-typing-lift} and \ref{lem:wk_lift} are used
  instead of Lemmas~\ref{lem:ctx-wf-typing} and \ref{lem:wk}, respectively.
\end{proof}

\begin{lemma}[Canonical Form]\label{lem:canonical_lift}
  \phantom{}
  \begin{enumerate}
    \item\label{lem:canonical_lift:abs} If $\emptyset  \vdash  \ottnt{v}  \ottsym{:}   \ottnt{A}    \rightarrow_{ \varepsilon }    \ottnt{B}   \mid  \varepsilon'$, then either of the following holds:
          \begin{itemize}
            \item $\ottnt{v} = \ottkw{fun} \, \ottsym{(}  \mathit{g}  \ottsym{,}  \mathit{x}  \ottsym{,}  \ottnt{e}  \ottsym{)}$ for some $\mathit{g}$, $\mathit{x}$, and $\ottnt{e}$, or
            \item $\ottnt{v} =  \mathsf{op} _{ \mathit{l} \,  \bm{ { S } } ^ {  \mathit{I}  }  }  \,  \bm{ { T } } ^ {  \mathit{J}  } $ for some $\mathsf{op}$, $\mathit{l}$, $ \bm{ { S } } ^ {  \mathit{I}  } $, and $ \bm{ { T } } ^ {  \mathit{J}  } $.
          \end{itemize}
    \item\label{lem:canonical_lift:tabs} If $\emptyset  \vdash  \ottnt{v}  \ottsym{:}    \forall   \alpha  \ottsym{:}  \ottnt{K}   \ottsym{.}    \ottnt{A}    ^{ \varepsilon }    \mid  \varepsilon'$, then $\ottnt{v} = \Lambda  \alpha  \ottsym{:}  \ottnt{K}  \ottsym{.}  \ottnt{e}$ for some $\ottnt{e}$.
  \end{enumerate}
\end{lemma}

\begin{proof}
  Similarly to Lemma~\ref{lem:canonical}.
\end{proof}

\begin{lemma}[Independence of Evaluation Contexts]\label{lem:ind_ev_lift}
  If $\Gamma  \vdash  \ottnt{E}  \ottsym{[}  \ottnt{e}  \ottsym{]}  \ottsym{:}  \ottnt{A}  \mid  \varepsilon$, then
  there exist some $\ottnt{A'}$ and $\varepsilon'$ such that
  \begin{itemize}
    \item $\Gamma  \vdash  \ottnt{e}  \ottsym{:}  \ottnt{A'}  \mid  \varepsilon'$, and
    \item $\Gamma  \ottsym{,}  \Gamma'  \vdash  \ottnt{E}  \ottsym{[}  \ottnt{e'}  \ottsym{]}  \ottsym{:}  \ottnt{A}  \mid  \varepsilon$ holds for any $\ottnt{e'}$ and $\Gamma'$ such that $\Gamma  \ottsym{,}  \Gamma'  \vdash  \ottnt{e'}  \ottsym{:}  \ottnt{A'}  \mid  \varepsilon'$.
  \end{itemize}
\end{lemma}

\begin{proof}
  By induction on a derivation of $\Gamma  \vdash  \ottnt{E}  \ottsym{[}  \ottnt{e}  \ottsym{]}  \ottsym{:}  \ottnt{A}  \mid  \varepsilon$.
  We proceed by cases on the typing rule applied lastly to this derivation.
  \begin{divcases}
    \item[\rname{T}{Lift}] If $\ottnt{E} =  \Box $, then the required result is achieved immediately.

    If $\ottnt{E} \neq  \Box $, then we have
    \begin{itemize}
      \item $\ottnt{E} =  [  \ottnt{E'}  ] _{ \ottnt{L} } $,
      \item $\Gamma  \vdash  \ottnt{E'}  \ottsym{[}  \ottnt{e}  \ottsym{]}  \ottsym{:}  \ottnt{A}  \mid  \varepsilon'$,
      \item $\Gamma  \vdash  \ottnt{L}  \ottsym{:}   \mathbf{Lab} $, and
      \item $   \lift{ \ottnt{L} }   \mathop{ \odot }  \varepsilon'    \sim   \varepsilon $,
    \end{itemize}
    for some $\ottnt{E'}$, $\ottnt{L}$, and $\varepsilon'$.
    By the induction hypothesis, there exist some $\ottnt{A'}$ and $\varepsilon''$ such that
    \begin{itemize}
      \item $\Gamma  \vdash  \ottnt{e}  \ottsym{:}  \ottnt{A'}  \mid  \varepsilon''$ and
      \item for any $\ottnt{e'}$ and $\Gamma'$ such that $\Gamma  \ottsym{,}  \Gamma'  \vdash  \ottnt{e'}  \ottsym{:}  \ottnt{A'}  \mid  \varepsilon''$, typing judgment $\Gamma  \ottsym{,}  \Gamma'  \vdash  \ottnt{E'}  \ottsym{[}  \ottnt{e'}  \ottsym{]}  \ottsym{:}  \ottnt{A}  \mid  \varepsilon'$ is derivable.
    \end{itemize}
    Let $\ottnt{e'}$ be an expression and $\Gamma'$ be a typing context
    such that $\Gamma  \ottsym{,}  \Gamma'  \vdash  \ottnt{e'}  \ottsym{:}  \ottnt{A'}  \mid  \varepsilon'$.
    The induction hypothesis gives us $\Gamma  \ottsym{,}  \Gamma'  \vdash  \ottnt{E'}  \ottsym{[}  \ottnt{e'}  \ottsym{]}  \ottsym{:}  \ottnt{A}  \mid  \varepsilon'$.
    By Lemma~\ref{lem:weakening_lift}\ref{lem:weakening_lift:kinding},
    we have $\Gamma  \ottsym{,}  \Gamma'  \vdash  \ottnt{L}  \ottsym{:}   \mathbf{Lab} $.
    Thus, \rname{T}{Lift} derives $\Gamma  \ottsym{,}  \Gamma'  \vdash   [  \ottnt{E'}  \ottsym{[}  \ottnt{e'}  \ottsym{]}  ] _{ \ottnt{L} }   \ottsym{:}  \ottnt{A}  \mid  \varepsilon$ as required.

    \item[others]
    Similarly to Lemma~\ref{lem:ind_ev};
    Lemmas~\ref{lem:weakening_lift} and \ref{lem:ctx-wf-typing-lift} are used
    instead of Lemmas~\ref{lem:weakening} and \ref{lem:ctx-wf-typing}, respectively.
  \end{divcases}
\end{proof}

\begin{lemma}[Progress]\label{lem:progress_lift}
  If $\emptyset  \vdash  \ottnt{e}  \ottsym{:}  \ottnt{A}  \mid  \varepsilon$, then one of the following holds:
  \begin{itemize}
    \item $\ottnt{e}$ is a value;
    \item There exists some expression $\ottnt{e'}$ such that $\ottnt{e}  \longrightarrow  \ottnt{e'}$; or
    \item There exist some $\mathsf{op}$, $\mathit{l}$, $ \bm{ { S } } ^ {  \mathit{I}  } $, $ \bm{ { T } } ^ {  \mathit{J}  } $, $\ottnt{v}$, $\ottnt{E}$, and $\mathit{n}$
          such that $\ottnt{e} = \ottnt{E}  \ottsym{[}   \mathsf{op} _{ \mathit{l} \,  \bm{ { S } } ^ {  \mathit{I}  }  }  \,  \bm{ { T } } ^ {  \mathit{J}  }  \, \ottnt{v}  \ottsym{]}$ and $ \mathit{n}  \mathrm{-free} ( \mathit{l} \,  \bm{ { S } } ^ {  \mathit{I}  }  ,  \ottnt{E} ) $.
  \end{itemize}
\end{lemma}

\begin{proof}
  By induction on a derivation of $\emptyset  \vdash  \ottnt{e}  \ottsym{:}  \ottnt{A}  \mid  \varepsilon$. We proceed by cases on the typing rule applied lastly to this derivation.
  \begin{divcases}
    \item[\rname{T}{Lift}] For some $\ottnt{e_{{\mathrm{1}}}}$, $\ottnt{L}$ and $\varepsilon_{{\mathrm{1}}}$, the following are given:
    \begin{itemize}
      \item $\ottnt{e} =  [  \ottnt{e_{{\mathrm{1}}}}  ] _{ \ottnt{L} } $,
      \item $\emptyset  \vdash  \ottnt{e_{{\mathrm{1}}}}  \ottsym{:}  \ottnt{A}  \mid  \varepsilon_{{\mathrm{1}}}$, and
      \item $\emptyset  \vdash     \lift{ \ottnt{L} }   \mathop{ \odot }  \varepsilon_{{\mathrm{1}}}    \sim   \varepsilon $.
    \end{itemize}
    By the induction hypothesis, we proceed by cases on the following conditions:
    \begin{enumerate}
      \item $\ottnt{e_{{\mathrm{1}}}}$ is a value,
      \item There exists some $\ottnt{e'_{{\mathrm{1}}}}$ such that $\ottnt{e_{{\mathrm{1}}}}  \longrightarrow  \ottnt{e'_{{\mathrm{1}}}}$,
      \item There exist some $\mathsf{op}$, $\mathit{l}$, $ \bm{ { S } } ^ {  \mathit{I}  } $, $ \bm{ { T } } ^ {  \mathit{J}  } $, $\ottnt{v}$, $\ottnt{E}$, and $\mathit{n}$
            such that $\ottnt{e_{{\mathrm{1}}}} = \ottnt{E}  \ottsym{[}   \mathsf{op} _{ \mathit{l} \,  \bm{ { S } } ^ {  \mathit{I}  }  }  \,  \bm{ { T } } ^ {  \mathit{J}  }  \, \ottnt{v}  \ottsym{]}$ and $ \mathit{n}  \mathrm{-free} ( \mathit{l} \,  \bm{ { S } } ^ {  \mathit{I}  }  ,  \ottnt{E} ) $.
    \end{enumerate}
    \begin{divcases}
      \item[(1)] \rname{R}{Lift} derives $ [  \ottnt{e_{{\mathrm{1}}}}  ] _{ \ottnt{L} }   \longmapsto  \ottnt{e_{{\mathrm{1}}}}$ because $\ottnt{e_{{\mathrm{1}}}}$ is a value.

      \item[(2)] Since only \rname{E}{Eval} can derive $\ottnt{e_{{\mathrm{1}}}}  \longrightarrow  \ottnt{e'_{{\mathrm{1}}}}$, we have
      \begin{itemize}
        \item $\ottnt{e_{{\mathrm{1}}}} = \ottnt{E_{{\mathrm{1}}}}  \ottsym{[}  \ottnt{e_{{\mathrm{11}}}}  \ottsym{]}$,
        \item $\ottnt{e'_{{\mathrm{1}}}} = \ottnt{E_{{\mathrm{1}}}}  \ottsym{[}  \ottnt{e_{{\mathrm{12}}}}  \ottsym{]}$, and
        \item $\ottnt{e_{{\mathrm{11}}}}  \longmapsto  \ottnt{e_{{\mathrm{12}}}}$,
      \end{itemize}
      for some $\ottnt{E_{{\mathrm{1}}}}$, $\ottnt{e_{{\mathrm{11}}}}$, and $\ottnt{e_{{\mathrm{12}}}}$.
      Let $\ottnt{E} = ( [  \ottnt{E_{{\mathrm{1}}}}  ] _{ \ottnt{L} } )$.
      \rname{E}{Eval} derives $\ottnt{e}  \longrightarrow  \ottnt{E}  \ottsym{[}  \ottnt{e_{{\mathrm{12}}}}  \ottsym{]}$ because of $\ottnt{e} = \ottnt{E}  \ottsym{[}  \ottnt{e_{{\mathrm{11}}}}  \ottsym{]}$.

      \item[(3)]
      If $\ottnt{L} \neq \mathit{l} \,  \bm{ { S } } ^ {  \mathit{I}  } $, then we have $ \mathit{n}  \mathrm{-free} ( \mathit{l} \,  \bm{ { S } } ^ {  \mathit{I}  }  ,   [  \ottnt{E}  ] _{ \ottnt{L} }  ) $.

      If $\ottnt{L} = \mathit{l} \,  \bm{ { S } } ^ {  \mathit{I}  } $, then we have $ \mathit{n}  \ottsym{+}  1  \mathrm{-free} ( \mathit{l} \,  \bm{ { S } } ^ {  \mathit{I}  }  ,   [  \ottnt{E}  ] _{ \ottnt{L} }  ) $.
    \end{divcases}

    \item[others]
    Similarly to Lemma~\ref{lem:progress};
    Lemmas~\ref{lem:inversion_lift} and \ref{lem:canonical_lift} are used
    instead of Lemmas~\ref{lem:inversion} and \ref{lem:canonical}, respectively.
  \end{divcases}
\end{proof}

\begin{lemma}[Preservation in Reduction]\label{lem:pres_red_lift}
  If $\emptyset  \vdash  \ottnt{e}  \ottsym{:}  \ottnt{A}  \mid  \varepsilon$ and $\ottnt{e}  \longmapsto  \ottnt{e'}$, then $\emptyset  \vdash  \ottnt{e'}  \ottsym{:}  \ottnt{A}  \mid  \varepsilon$.
\end{lemma}

\begin{proof}
  By induction on a derivation of $\Gamma  \vdash  \ottnt{e}  \ottsym{:}  \ottnt{A}  \mid  \varepsilon$.
  We proceed by cases on the typing rule applied lastly to this derivation.
  \begin{divcases}
    \item[\rname{T}{Lift}] Since only \rname{R}{Lift} derives $\ottnt{e}  \longmapsto  \ottnt{e'}$, we have
    \begin{itemize}
      \item $\ottnt{e} =  [  \ottnt{v}  ] _{ \ottnt{L} } $,
      \item $\emptyset  \vdash  \ottnt{v}  \ottsym{:}  \ottnt{A}  \mid  \varepsilon_{{\mathrm{1}}}$,
      \item $\emptyset  \vdash  \ottnt{L}  \ottsym{:}   \mathbf{Lab} $,
      \item $   \lift{ \ottnt{L} }   \mathop{ \odot }  \varepsilon_{{\mathrm{1}}}    \sim   \varepsilon $, and
      \item $\ottnt{e'} = \ottnt{v}$.
    \end{itemize}
    for some $\ottnt{v}$, $\ottnt{L}$, and $\varepsilon_{{\mathrm{1}}}$.
    By Lemma~\ref{lem:inversion_lift}\ref{lem:inversion_lift:var},
    we have $\emptyset  \vdash  \ottnt{v}  \ottsym{:}  \ottnt{A}  \mid   \bbZero $.
    By $   \bbZero   \mathop{ \odot }  \varepsilon    \sim   \varepsilon $,
    we have $\emptyset  \vdash  \ottnt{v}  \ottsym{:}  \ottnt{A}  \mid  \varepsilon$ as required.

    \item[others]
    Similarly to Lemma~\ref{lem:pres_red};
    Lemmas~\ref{lem:inversion_lift}, \ref{lem:weakening_lift}, \ref{lem:ctx-wf-typing-lift}, \ref{lem:wk_lift}, \ref{lem:subst_value_lift}, and \ref{lem:subst_type_lift} are used
    instead of Lemmas~\ref{lem:inversion}, \ref{lem:weakening}, \ref{lem:ctx-wf-typing}, \ref{lem:wk}, \ref{lem:subst_value}, and \ref{lem:subst_type} respectively.
  \end{divcases}
\end{proof}

\begin{lemma}[Preservation]\label{lem:preservation_lift}
  If $\emptyset  \vdash  \ottnt{e}  \ottsym{:}  \ottnt{A}  \mid  \varepsilon$ and $\ottnt{e}  \longrightarrow  \ottnt{e'}$, then $\emptyset  \vdash  \ottnt{e'}  \ottsym{:}  \ottnt{A}  \mid  \varepsilon$.
\end{lemma}

\begin{proof}
  Similarly to Lemma~\ref{lem:preservation};
  Lemmas~\ref{lem:ind_ev_lift} and \ref{lem:pres_red_lift} are used
  instead of Lemmas~\ref{lem:ind_ev} and \ref{lem:pres_red}.
\end{proof}

\begin{definition}[Label Inclusion]
  \phantom{}\\
  \textnormal{\bfseries Label Inclusion}\tquad\fbox{$ \ottnt{L}  \olessthan^{ \mathit{n} }  \varepsilon $}
  \begin{mathpar}
    \inferrule{ }{
       \ottnt{L}  \olessthan^{ 0 }  \varepsilon 
    }\ \rname{LI}{Empty}

    \inferrule{
     \ottnt{L}  \olessthan^{ \mathit{n} }  \varepsilon_{{\mathrm{1}}}  \\    \lift{ \ottnt{L} }   \mathop{ \odot }  \varepsilon_{{\mathrm{1}}}    \sim   \varepsilon_{{\mathrm{2}}} 
    }{
     \ottnt{L}  \olessthan^{ \mathit{n}  \ottsym{+}  1 }  \varepsilon_{{\mathrm{2}}} 
    }\ \rname{LI}{Handling}

    \inferrule{
     \ottnt{L}  \olessthan^{ \mathit{n} }  \varepsilon_{{\mathrm{1}}}  \\    \lift{ \ottnt{L'} }   \mathop{ \odot }  \varepsilon_{{\mathrm{1}}}    \sim   \varepsilon_{{\mathrm{2}}}  \\ \ottnt{L} \neq \ottnt{L'}
    }{
     \ottnt{L}  \olessthan^{ \mathit{n} }  \varepsilon_{{\mathrm{2}}} 
    }\ \rname{LI}{NoHandling}
  \end{mathpar}
\end{definition}

\begin{lemma}\label{lem:label_inclusion_mono}
  If $ \ottnt{L}  \olessthan^{ \mathit{n} }  \varepsilon_{{\mathrm{1}}} $ and $  \varepsilon_{{\mathrm{1}}}  \mathop{ \odot }  \varepsilon_{{\mathrm{2}}}    \sim   \varepsilon_{{\mathrm{3}}} $,
  then $ \ottnt{L}  \olessthan^{ \mathit{n} }  \varepsilon_{{\mathrm{3}}} $.
\end{lemma}

\begin{proof}
  By induction on a derivation of $ \ottnt{L}  \olessthan^{ \mathit{n} }  \varepsilon_{{\mathrm{1}}} $.
  We proceed by case analysis on the rule applied lastly to this derivation.
  \begin{divcases}
    \item[\rname{LI}{Empty}]
    We have $\mathit{n} = 0$.
    \rname{LI}{Empty} derives $ \ottnt{L}  \olessthan^{ 0 }  \varepsilon_{{\mathrm{3}}} $ as required.

    \item[\rname{LI}{Handling}]
    We have
    \begin{itemize}
      \item $\mathit{n} = \mathit{n}'  \ottsym{+}  1$,
      \item $ \ottnt{L}  \olessthan^{ \mathit{n}' }  \varepsilon_{{\mathrm{4}}} $, and
      \item $   \lift{ \ottnt{L} }   \mathop{ \odot }  \varepsilon_{{\mathrm{4}}}    \sim   \varepsilon_{{\mathrm{1}}} $,
    \end{itemize}
    for some $\mathit{n}'$ and $\varepsilon_{{\mathrm{4}}}$.
    By the induction hypothesis, we have $ \ottnt{L}  \olessthan^{ \mathit{n}' }  \varepsilon_{{\mathrm{5}}} $ such that
    $  \varepsilon_{{\mathrm{4}}}  \mathop{ \odot }  \varepsilon_{{\mathrm{2}}}    \sim   \varepsilon_{{\mathrm{5}}} $.
    Thus, \rname{LI}{Handling} derives $ \ottnt{L}  \olessthan^{ \mathit{n}'  \ottsym{+}  1 }  \varepsilon_{{\mathrm{3}}} $ as required.

    \item[\rname{LI}{NoHandling}]
    We have
    \begin{itemize}
      \item $ \ottnt{L}  \olessthan^{ \mathit{n} }  \varepsilon_{{\mathrm{4}}} $,
      \item $   \lift{ \ottnt{L'} }   \mathop{ \odot }  \varepsilon_{{\mathrm{4}}}    \sim   \varepsilon_{{\mathrm{1}}} $, and
      \item $\ottnt{L} \neq \ottnt{L'}$,
    \end{itemize}
    for some $\ottnt{L'}$ and $\varepsilon_{{\mathrm{4}}}$.
    By the induction hypothesis, we have $ \ottnt{L}  \olessthan^{ \mathit{n} }  \varepsilon_{{\mathrm{5}}} $ such that
    $  \varepsilon_{{\mathrm{4}}}  \mathop{ \odot }  \varepsilon_{{\mathrm{2}}}    \sim   \varepsilon_{{\mathrm{5}}} $.
    Thus, \rname{LI}{NoHandling} derives $ \ottnt{L}  \olessthan^{ \mathit{n} }  \varepsilon_{{\mathrm{3}}} $ as required.
  \end{divcases}
\end{proof}

\begin{lemma}\label{lem:label_inclusion_handle}
  If $ \ottnt{L}  \olessthan^{ \mathit{n}  \ottsym{+}  1 }  \varepsilon_{{\mathrm{2}}} $ and $   \lift{ \ottnt{L} }   \mathop{ \odot }  \varepsilon_{{\mathrm{1}}}    \sim   \varepsilon_{{\mathrm{2}}} $,
  then $ \ottnt{L}  \olessthan^{ \mathit{n} }  \varepsilon_{{\mathrm{1}}} $.
\end{lemma}

\begin{proof}
  By induction on a derivation of $ \ottnt{L}  \olessthan^{ \mathit{n}  \ottsym{+}  1 }  \varepsilon_{{\mathrm{2}}} $.
  We proceed by case analysis on the rule lastly applied to this derivation.
  \begin{divcases}
    \item[\rname{LI}{Empty}] Cannot happen.

    \item[\rname{LI}{Handling}]
    We have
    \begin{itemize}
      \item $ \ottnt{L}  \olessthan^{ \mathit{n} }  \varepsilon'_{{\mathrm{1}}} $ and
      \item $   \lift{ \ottnt{L} }   \mathop{ \odot }  \varepsilon'_{{\mathrm{1}}}    \sim   \varepsilon_{{\mathrm{2}}} $
    \end{itemize}
    for some $\varepsilon'_{{\mathrm{1}}}$.
    By safety condition \ref{def:safe_cond_lift:removal}, we have $ \varepsilon_{{\mathrm{1}}}   \sim   \varepsilon'_{{\mathrm{1}}} $.
    By Lemma~\ref{lem:label_inclusion_mono} and $  \varepsilon'_{{\mathrm{1}}}  \mathop{ \odot }   \bbZero     \sim   \varepsilon_{{\mathrm{1}}} $,
    we have $ \ottnt{L}  \olessthan^{ \mathit{n} }  \varepsilon_{{\mathrm{1}}} $ as required.

    \item[\rname{LI}{NoHandling}]
    We have
    \begin{itemize}
      \item $ \ottnt{L}  \olessthan^{ \mathit{n}  \ottsym{+}  1 }  \varepsilon_{{\mathrm{3}}} $,
      \item $   \lift{ \ottnt{L'} }   \mathop{ \odot }  \varepsilon_{{\mathrm{3}}}    \sim   \varepsilon_{{\mathrm{2}}} $, and
      \item $\ottnt{L} \neq \ottnt{L'}$,
    \end{itemize}
    for some $\ottnt{L'}$ and $\varepsilon_{{\mathrm{3}}}$.
    By safety condition \ref{def:safe_cond:pres} and $\ottnt{L} \neq \ottnt{L'}$,
    we have $   \lift{ \ottnt{L} }   \mathop{ \odot }  \varepsilon_{{\mathrm{4}}}    \sim   \varepsilon_{{\mathrm{3}}} $ for some $\varepsilon_{{\mathrm{4}}}$.
    By safety condition \ref{def:safe_cond_lift:removal},
    we have $ \varepsilon_{{\mathrm{1}}}   \sim     \lift{ \ottnt{L'} }   \mathop{ \odot }  \varepsilon_{{\mathrm{4}}}  $.
    By the induction hypothesis, we have $ \ottnt{L}  \olessthan^{ \mathit{n} }  \varepsilon_{{\mathrm{4}}} $.
    Thus, \rname{LI}{NoHandling} derives $ \ottnt{L}  \olessthan^{ \mathit{n} }  \varepsilon_{{\mathrm{1}}} $ as required.
  \end{divcases}
\end{proof}

\begin{lemma}\label{lem:label_inclusion_nohandle}
  If $ \ottnt{L}  \olessthan^{ \mathit{n} }  \varepsilon_{{\mathrm{2}}} $ and $   \lift{ \ottnt{L'} }   \mathop{ \odot }  \varepsilon_{{\mathrm{1}}}    \sim   \varepsilon_{{\mathrm{2}}} $ and $\ottnt{L} \neq \ottnt{L'}$,
  then $ \ottnt{L}  \olessthan^{ \mathit{n} }  \varepsilon_{{\mathrm{1}}} $.
\end{lemma}

\begin{proof}
  By induction on a derivation of $ \ottnt{L}  \olessthan^{ \mathit{n} }  \varepsilon_{{\mathrm{2}}} $.
  We proceed by case analysis on the rule lastly applied to this derivation.
  \begin{divcases}
    \item[\rname{LI}{Empty}]
    We have $\mathit{n} = 0$.
    \rname{LI}{Empty} derives $ \ottnt{L}  \olessthan^{ 0 }  \varepsilon_{{\mathrm{1}}} $ as required.

    \item[\rname{LI}{Handling}]
    We have
    \begin{itemize}
      \item $\mathit{n} = \mathit{n}'  \ottsym{+}  1$,
      \item $ \ottnt{L}  \olessthan^{ \mathit{n}' }  \varepsilon_{{\mathrm{3}}} $, and
      \item $   \lift{ \ottnt{L} }   \mathop{ \odot }  \varepsilon_{{\mathrm{3}}}    \sim   \varepsilon_{{\mathrm{2}}} $,
    \end{itemize}
    for some $\mathit{n}'$ and $\varepsilon_{{\mathrm{3}}}$.
    By safety condition \ref{def:safe_cond:pres} and $\ottnt{L} \neq \ottnt{L'}$,
    we have $   \lift{ \ottnt{L'} }   \mathop{ \odot }  \varepsilon_{{\mathrm{4}}}    \sim   \varepsilon_{{\mathrm{3}}} $ for some $\varepsilon_{{\mathrm{4}}}$.
    By safety condition \ref{def:safe_cond_lift:removal},
    we have $ \varepsilon_{{\mathrm{1}}}   \sim     \lift{ \ottnt{L} }   \mathop{ \odot }  \varepsilon_{{\mathrm{4}}}  $.
    By the induction hypothesis, we have $ \ottnt{L}  \olessthan^{ \mathit{n}' }  \varepsilon_{{\mathrm{4}}} $.
    Thus, \rname{LI}{Handling} derives $ \ottnt{L}  \olessthan^{ \mathit{n}'  \ottsym{+}  1 }  \varepsilon_{{\mathrm{1}}} $ as required.

    \item[\rname{LI}{NoHandling}]
    We have
    \begin{itemize}
      \item $ \ottnt{L}  \olessthan^{ \mathit{n} }  \varepsilon_{{\mathrm{3}}} $,
      \item $   \lift{ \ottnt{L''} }   \mathop{ \odot }  \varepsilon_{{\mathrm{3}}}    \sim   \varepsilon_{{\mathrm{2}}} $, and
      \item $\ottnt{L} \neq \ottnt{L''}$,
    \end{itemize}
    for some $\ottnt{L''}$ and $\varepsilon_{{\mathrm{3}}}$.

    If $\ottnt{L'} = \ottnt{L''}$, then we have $ \varepsilon_{{\mathrm{1}}}   \sim   \varepsilon_{{\mathrm{3}}} $
    by safety condition \ref{def:safe_cond_lift:removal}.
    Thus, Lemma~\ref{lem:label_inclusion_mono} gives us $ \ottnt{L}  \olessthan^{ \mathit{n} }  \varepsilon_{{\mathrm{1}}} $ as required.

    If $\ottnt{L'} \neq \ottnt{L''}$, then we have $   \lift{ \ottnt{L'} }   \mathop{ \odot }  \varepsilon_{{\mathrm{4}}}    \sim   \varepsilon_{{\mathrm{3}}} $ for some $\varepsilon_{{\mathrm{4}}}$
    by safety condition \ref{def:safe_cond:pres} and $\ottnt{L'} \neq \ottnt{L''}$.
    By safety condition \ref{def:safe_cond_lift:removal},
    we have $ \varepsilon_{{\mathrm{1}}}   \sim     \lift{ \ottnt{L''} }   \mathop{ \odot }  \varepsilon_{{\mathrm{4}}}  $.
    By the induction hypothesis, we have $ \ottnt{L}  \olessthan^{ \mathit{n} }  \varepsilon_{{\mathrm{4}}} $.
    Thus, \rname{LI}{NoHandling} derives $ \ottnt{L}  \olessthan^{ \mathit{n} }  \varepsilon_{{\mathrm{1}}} $ as required.
  \end{divcases}
\end{proof}

\begin{lemma}\label{lem:effsafe_aux_lift}
  If $\emptyset  \vdash  \ottnt{E}  \ottsym{[}   \mathsf{op} _{ \mathit{l} \,  \bm{ { S } } ^ {  \mathit{I}  }  }  \,  \bm{ { T } } ^ {  \mathit{J}  }  \, \ottnt{v}  \ottsym{]}  \ottsym{:}  \ottnt{A}  \mid  \varepsilon$ and
  $ \mathit{n}  \mathrm{-free} ( \mathit{l} \,  \bm{ { S } } ^ {  \mathit{I}  }  ,  \ottnt{E} ) $,
  then $ \mathit{l} \,  \bm{ { S } } ^ {  \mathit{I}  }   \olessthan^{ \mathit{n}  \ottsym{+}  1 }  \varepsilon $.
\end{lemma}

\begin{proof}
  By induction on a derivation of $\emptyset  \vdash  \ottnt{E}  \ottsym{[}   \mathsf{op} _{ \mathit{l} \,  \bm{ { S } } ^ {  \mathit{I}  }  }  \,  \bm{ { T } } ^ {  \mathit{J}  }  \, \ottnt{v}  \ottsym{]}  \ottsym{:}  \ottnt{A}  \mid  \varepsilon$. We proceed by case analysis on the typing rule applied lastly to this derivation.
  \begin{divcases}
    \item[\rname{T}{App}]
    For some $\ottnt{B}$, we have
    \begin{itemize}
      \item $\ottnt{E} =  \Box $,
      \item $\emptyset  \vdash   \mathsf{op} _{ \mathit{l} \,  \bm{ { S } } ^ {  \mathit{I}  }  }  \,  \bm{ { T } } ^ {  \mathit{J}  }   \ottsym{:}   \ottnt{B}    \rightarrow_{ \varepsilon }    \ottnt{A}   \mid   \bbZero $, and
      \item $\emptyset  \vdash  \ottnt{v}  \ottsym{:}  \ottnt{B}  \mid   \bbZero $.
    \end{itemize}
    By Lemma~\ref{lem:inversion_lift}\ref{lem:inversion_lift:op},
    we have $\emptyset  \vdash    \lift{ \mathit{l} \,  \bm{ { S } } ^ {  \mathit{I}  }  }   \olessthan  \varepsilon $.
    Thus, the required result is achieved.

    \item[\rname{T}{Let}]
    For some $\mathit{x}$, $\ottnt{E_{{\mathrm{1}}}}$, $\ottnt{e}$, and $\ottnt{B}$, we have
    \begin{itemize}
      \item $\ottnt{E} = (\mathbf{let} \, \mathit{x}  \ottsym{=}  \ottnt{E_{{\mathrm{1}}}} \, \mathbf{in} \, \ottnt{e})$,
      \item $\emptyset  \vdash  \ottnt{E_{{\mathrm{1}}}}  \ottsym{[}   \mathsf{op} _{ \mathit{l} \,  \bm{ { S } } ^ {  \mathit{I}  }  }  \,  \bm{ { T } } ^ {  \mathit{J}  }  \, \ottnt{v}  \ottsym{]}  \ottsym{:}  \ottnt{B}  \mid  \varepsilon$,
      \item $ \mathit{n}  \mathrm{-free} ( \mathit{l} \,  \bm{ { S } } ^ {  \mathit{I}  }  ,  \ottnt{E_{{\mathrm{1}}}} ) $, and
      \item $\mathit{x}  \ottsym{:}  \ottnt{B}  \vdash  \ottnt{e}  \ottsym{:}  \ottnt{A}  \mid  \varepsilon$.
    \end{itemize}
    By the induction hypothesis, we have $ \mathit{l} \,  \bm{ { S } } ^ {  \mathit{I}  }   \olessthan^{ \mathit{n}  \ottsym{+}  1 }  \varepsilon $ as required.

    \item[\rname{T}{Sub}] For some $\ottnt{A''}$ and $\varepsilon''$, we have
    \begin{itemize}
      \item $\emptyset  \vdash  \ottnt{E}  \ottsym{[}   \mathsf{op} _{ \mathit{l} \,  \bm{ { S } } ^ {  \mathit{I}  }  }  \,  \bm{ { T } } ^ {  \mathit{J}  }  \, \ottnt{v}  \ottsym{]}  \ottsym{:}  \ottnt{A'}  \mid  \varepsilon'$ and
      \item $\emptyset  \vdash  \ottnt{A'}  \mid  \varepsilon'  <:  \ottnt{A}  \mid  \varepsilon$.
    \end{itemize}
    By the induction hypothesis, we have $ \mathit{l} \,  \bm{ { S } } ^ {  \mathit{I}  }   \olessthan^{ \mathit{n}  \ottsym{+}  1 }  \varepsilon' $.
    Since only \rname{ST}{Comp} can derive $\emptyset  \vdash  \ottnt{A'}  \mid  \varepsilon'  <:  \ottnt{A}  \mid  \varepsilon$,
    we have $\emptyset  \vdash   \varepsilon'  \olessthan  \varepsilon $.
    Thus, Lemma~\ref{lem:label_inclusion_mono} derives $ \mathit{l} \,  \bm{ { S } } ^ {  \mathit{I}  }   \olessthan^{ \mathit{n}  \ottsym{+}  1 }  \varepsilon $ as required.

    \item[\rname{T}{Lift}]
    For some $\ottnt{L}$, $\varepsilon'$, and $\ottnt{E'}$, we have
    \begin{itemize}
      \item $\ottnt{E} =  [  \ottnt{E'}  ] _{ \ottnt{L} } $,
      \item $\emptyset  \vdash  \ottnt{E'}  \ottsym{[}   \mathsf{op} _{ \mathit{l} \,  \bm{ { S } } ^ {  \mathit{I}  }  }  \,  \bm{ { T } } ^ {  \mathit{J}  }  \, \ottnt{v}  \ottsym{]}  \ottsym{:}  \ottnt{A}  \mid  \varepsilon'$,
      \item $\emptyset  \vdash  \ottnt{L}  \ottsym{:}   \mathbf{Lab} $, and
      \item $   \lift{ \ottnt{L} }   \mathop{ \odot }  \varepsilon'    \sim   \varepsilon $.
    \end{itemize}

    If $\mathit{l} \,  \bm{ { S } } ^ {  \mathit{I}  }  \neq \ottnt{L}$, then $ \mathit{n}  \mathrm{-free} ( \mathit{l} \,  \bm{ { S } } ^ {  \mathit{I}  }  ,  \ottnt{E'} ) $.
    By the induction hypothesis, we have $ \mathit{l} \,  \bm{ { S } } ^ {  \mathit{I}  }   \olessthan^{ \mathit{n}  \ottsym{+}  1 }  \varepsilon' $.
    \rname{LI}{NoHandling} derives $ \mathit{l} \,  \bm{ { S } } ^ {  \mathit{I}  }   \olessthan^{ \mathit{n}  \ottsym{+}  1 }  \varepsilon $ as required.

    If $\mathit{l} \,  \bm{ { S } } ^ {  \mathit{I}  }  = \ottnt{L}$, then there exists some $\mathit{m}$ such that
    $\mathit{n} = \mathit{m}  \ottsym{+}  1$ and $ \mathit{m}  \mathrm{-free} ( \mathit{l} \,  \bm{ { S } } ^ {  \mathit{I}  }  ,  \ottnt{E'} ) $.
    By the induction hypothesis, we have $ \mathit{l} \,  \bm{ { S } } ^ {  \mathit{I}  }   \olessthan^{ \mathit{m}  \ottsym{+}  1 }  \varepsilon' $.
    \rname{LI}{Handling} derives $ \mathit{l} \,  \bm{ { S } } ^ {  \mathit{I}  }   \olessthan^{ \mathit{m}  \ottsym{+}  2 }  \varepsilon $ as required.

    \item[\rname{T}{Handling}]
    For some $\mathit{l'}$, $ \bm{ { S' } } ^ {  \mathit{I'}  } $, $\ottnt{E_{{\mathrm{1}}}}$, $\ottnt{h}$, $\ottnt{B}$, and $\varepsilon'$, we have
    \begin{itemize}
      \item $\ottnt{E} =  \mathbf{handle}_{ \mathit{l'} \,  \bm{ { S' } } ^ {  \mathit{I'}  }  }  \, \ottnt{E_{{\mathrm{1}}}} \, \mathbf{with} \, \ottnt{h}$,
      \item $\emptyset  \vdash  \ottnt{E_{{\mathrm{1}}}}  \ottsym{[}   \mathsf{op} _{ \mathit{l} \,  \bm{ { S } } ^ {  \mathit{I}  }  }  \,  \bm{ { T } } ^ {  \mathit{J}  }  \, \ottnt{v}  \ottsym{]}  \ottsym{:}  \ottnt{B}  \mid  \varepsilon'$, and
      \item $   \lift{ \mathit{l'} \,  \bm{ { S' } } ^ {  \mathit{I'}  }  }   \mathop{ \odot }  \varepsilon    \sim   \varepsilon' $.
    \end{itemize}

    If $\mathit{l} \,  \bm{ { S } } ^ {  \mathit{I}  }  \neq \mathit{l'} \,  \bm{ { S' } } ^ {  \mathit{I'}  } $, then $ \mathit{n}  \mathrm{-free} ( \mathit{l} \,  \bm{ { S } } ^ {  \mathit{I}  }  ,  \ottnt{E_{{\mathrm{1}}}} ) $.
    By the induction hypothesis, we have $ \mathit{l} \,  \bm{ { S } } ^ {  \mathit{I}  }   \olessthan^{ \mathit{n}  \ottsym{+}  1 }  \varepsilon' $.
    By Lemma~\ref{lem:label_inclusion_nohandle}, we have $ \mathit{l} \,  \bm{ { S } } ^ {  \mathit{I}  }   \olessthan^{ \mathit{n}  \ottsym{+}  1 }  \varepsilon $.

    If $\mathit{l} \,  \bm{ { S } } ^ {  \mathit{I}  }  = \mathit{l'} \,  \bm{ { S' } } ^ {  \mathit{I'}  } $, then $ \mathit{n}  \ottsym{+}  1  \mathrm{-free} ( \mathit{l} \,  \bm{ { S } } ^ {  \mathit{I}  }  ,  \ottnt{E_{{\mathrm{1}}}} ) $.
    By the induction hypothesis, we have $ \mathit{l} \,  \bm{ { S } } ^ {  \mathit{I}  }   \olessthan^{ \mathit{n}  \ottsym{+}  2 }  \varepsilon' $.
    By Lemma~\ref{lem:label_inclusion_handle}, we have $ \mathit{l} \,  \bm{ { S } } ^ {  \mathit{I}  }   \olessthan^{ \mathit{n}  \ottsym{+}  1 }  \varepsilon $.

    \item[others] Cannot happen.
  \end{divcases}
\end{proof}

\begin{lemma}[No Inclusion by Empty Effect]\label{lem:no_inclusion}
  If $ \ottnt{L}  \olessthan^{ \mathit{n} }  \varepsilon $ and $ \varepsilon   \sim    \bbZero  $, then $\mathit{n} = 0$.
\end{lemma}

\begin{proof}
  By induction on the derivation of $ \ottnt{L}  \olessthan^{ \mathit{n} }  \varepsilon $.
  We proceed by case analysis on the rule applied lastly to this derivation.
  \begin{divcases}
    \item[\rname{LI}{Empty}] Clearly.

    \item[\rname{LI}{Handling}]
    This case cannot happen.
    If this case happens, we have $   \lift{ \ottnt{L} }   \mathop{ \odot }  \varepsilon'    \sim   \varepsilon $ for some $\mathit{m}$ and $\varepsilon'$.
    Thus, we have $   \lift{ \ottnt{L} }   \mathop{ \odot }  \varepsilon'    \sim    \bbZero  $ by $ \varepsilon   \sim    \bbZero  $.
    However, it is contradictory with safety condition \ref{def:safe_cond:label_notemp}.

    \item[\rname{LI}{NoHandling}]
    This case cannot happen.
    If this case happens, we have $   \lift{ \ottnt{L'} }   \mathop{ \odot }  \varepsilon'    \sim   \varepsilon $ for some $\ottnt{L'}$ and $\varepsilon'$.
    Thus, we have $   \lift{ \ottnt{L'} }   \mathop{ \odot }  \varepsilon'    \sim    \bbZero  $ by $ \varepsilon   \sim    \bbZero  $.
    However, it is contradictory with safety condition \ref{def:safe_cond:label_notemp}.
  \end{divcases}
\end{proof}

\begin{lemma}[Effect Safety]\label{lem:effsafe_lift}
  If $\emptyset  \vdash  \ottnt{E}  \ottsym{[}   \mathsf{op} _{ \mathit{l} \,  \bm{ { S } } ^ {  \mathit{I}  }  }  \,  \bm{ { T } } ^ {  \mathit{J}  }  \, \ottnt{v}  \ottsym{]}  \ottsym{:}  \ottnt{A}  \mid  \varepsilon$ and
  $ \mathit{n}  \mathrm{-free} ( \mathit{l} \,  \bm{ { S } } ^ {  \mathit{I}  }  ,  \ottnt{E} ) $,
  then $ \varepsilon   \nsim    \bbZero  $.
\end{lemma}

\begin{proof}
  Assume that $ \varepsilon   \sim    \bbZero  $.
  By Lemma~\ref{lem:effsafe_aux_lift} and Lemma~\ref{lem:label_inclusion_mono},
  we have $ \mathit{l} \,  \bm{ { S } } ^ {  \mathit{I}  }   \olessthan^{ \mathit{n}  \ottsym{+}  1 }   \bbZero  $.
  However, it is contradictory with Lemma~\ref{lem:no_inclusion}.
\end{proof}

\begin{theorem}[Type and Effect Safety]\label{thm:safety_lift}
  If $\emptyset  \vdash  \ottnt{e}  \ottsym{:}  \ottnt{A}  \mid   \bbZero $ and $\ottnt{e}  \longrightarrow  ^ * \ottnt{e'}$ and $\ottnt{e'} \centernot \longrightarrow $, then $\ottnt{e'}$ is a value.
\end{theorem}

\begin{proof}
  Similarly to Theorem~\ref{thm:safety};
  Lemmas~\ref{lem:preservation_lift}
  , \ref{lem:effsafe_lift},
  and \ref{lem:progress_lift} are used
  instead of Lemmas~\ref{lem:preservation}
  , \ref{lem:effsafe},
  and \ref{lem:progress}, respectively.
\end{proof}
\subsection{Properties with Type-Erasure Semantics}

This section assumes that the safety conditions in Definition~\ref{def:safe_cond}
and the safety condition for type-erasure semantics in Definition~\ref{def:safe_cond_erasure} hold,
and that the semantics adapts \rname{R}{Handle2'}instead of \rname{R}{Handle2}.

\begin{remark}
  The change of semantics only affects Lemma~\ref{lem:progress}, Lemma~\ref{lem:pres_red}, Lemma~\ref{lem:preservation}, Lemma~\ref{lem:effsafe_aux}, Lemma~\ref{lem:effsafe}, and Theorem~\ref{thm:safety}.
  Therefore, we can use other lemmas in this type-erasure setting.
\end{remark}

\begin{lemma}[Progress]\label{lem:progress_erasure}
  If $\emptyset  \vdash  \ottnt{e}  \ottsym{:}  \ottnt{A}  \mid  \varepsilon$, then one of the following holds:
  \begin{itemize}
    \item $\ottnt{e}$ is a value;
    \item There exists some $\ottnt{e'}$ such that $\ottnt{e}  \longrightarrow  \ottnt{e'}$; or
    \item There exist some $\mathsf{op}$, $\mathit{l}$, $ \bm{ { S } } ^ {  \mathit{I}  } $, $ \bm{ { T } } ^ {  \mathit{J}  } $, $\ottnt{v}$, $\ottnt{E}$, and $\mathit{n}$
          such that $\ottnt{e} \ottnt{E}  \ottsym{[}   \mathsf{op} _{ \mathit{l} \,  \bm{ { S } } ^ {  \mathit{I}  }  }  \,  \bm{ { T } } ^ {  \mathit{J}  }  \, \ottnt{v}  \ottsym{]}$ and $ \mathit{n}  \mathrm{-free} ( \mathit{l} ,  \ottnt{E} ) $.
  \end{itemize}
\end{lemma}

\begin{proof}
  By induction on a derivation of $\emptyset  \vdash  \ottnt{e}  \ottsym{:}  \ottnt{A}  \mid  \varepsilon$.
  We proceed by case analysis on the typing rule applied lastly to this derivation.
  \begin{divcases}
    \item[\rname{T}{Handling}]
    For some $\mathit{l}$, $ \bm{ { S } } ^ {  \mathit{N}  } $, $\ottnt{h}$, $\ottnt{e_{{\mathrm{1}}}}$, $\ottnt{A_{{\mathrm{1}}}}$, $\varepsilon_{{\mathrm{1}}}$, $ \bm{ { \alpha } } ^ {  \mathit{N}  } $, $ {\bm{ { \ottnt{K} } } }^{ \mathit{N} } $, $\sigma$, given are the following:
    \begin{itemize}
      \item $\ottnt{e} =  \mathbf{handle}_{ \mathit{l} \,  \bm{ { S } } ^ {  \mathit{N}  }  }  \, \ottnt{e_{{\mathrm{1}}}} \, \mathbf{with} \, \ottnt{h}$,
      \item $\emptyset  \vdash  \ottnt{e_{{\mathrm{1}}}}  \ottsym{:}  \ottnt{A_{{\mathrm{1}}}}  \mid  \varepsilon_{{\mathrm{1}}}$,
      \item $ \mathit{l}  ::    \forall    {\bm{ \alpha } }^{ \mathit{N} } : {\bm{ \ottnt{K} } }^{ \mathit{N} }    \ottsym{.}    \sigma    \in   \Xi $,
      \item $\emptyset  \vdash   \bm{ { S } }^{ \mathit{N} } : \bm{ \ottnt{K} }^{ \mathit{N} } $,
      \item $ \emptyset  \vdash _{ \sigma \,  \! [ {\bm{ { S } } }^{ \mathit{N} } / {\bm{ \alpha } }^{ \mathit{N} } ]  }  \ottnt{h}  :  \ottnt{A_{{\mathrm{1}}}}   \Rightarrow  ^ { \varepsilon }  \ottnt{A} $, and
      \item $   \lift{ \mathit{l} \,  \bm{ { S } } ^ {  \mathit{N}  }  }   \mathop{ \odot }  \varepsilon    \sim   \varepsilon_{{\mathrm{1}}} $.
    \end{itemize}
    By the induction hypothesis, we proceed by case analysis on the following conditions:
    \begin{enumerate}
      \item $\ottnt{e_{{\mathrm{1}}}}$ is a value,
      \item There exists some $\ottnt{e'_{{\mathrm{1}}}}$ such that $\ottnt{e_{{\mathrm{1}}}}  \longrightarrow  \ottnt{e'_{{\mathrm{1}}}}$, and
      \item There exist some $\mathsf{op'}$, $\mathit{l'}$, $ \bm{ { S' } } ^ {  \mathit{N'}  } $, $ \bm{ { T } } ^ {  \mathit{J}  } $, $\ottnt{v}$, $\ottnt{E}$, and $\mathit{n}$
            such that $\ottnt{e_{{\mathrm{1}}}} = \ottnt{E}  \ottsym{[}   \mathsf{op'} _{ \mathit{l'} \,  \bm{ { S' } } ^ {  \mathit{N'}  }  }  \,  \bm{ { T } } ^ {  \mathit{J}  }  \, \ottnt{v}  \ottsym{]}$ and $ \mathit{n}  \mathrm{-free} ( \mathit{l'} ,  \ottnt{E} ) $.
    \end{enumerate}
    \begin{divcases}
      \item[(1)]
      By Lemma~\ref{lem:inversion_handler}\ref{lem:inversion_handler:return},
      there exists some $\mathit{x}$ and $\ottnt{e_{\ottmv{r}}}$ such that $ \mathbf{return} \, \mathit{x}  \mapsto  \ottnt{e_{\ottmv{r}}}   \in   \ottnt{h} $.
      Thus, \rname{R}{Handle1} derives $\ottnt{e}  \longmapsto  \ottnt{e_{\ottmv{r}}} \,  \! [  \ottnt{v_{{\mathrm{1}}}}  /  \mathit{x}  ] $ because $\ottnt{e_{{\mathrm{1}}}}$ is a value $\ottnt{v_{{\mathrm{1}}}}$.

      \item[(2)] Since only \rname{E}{Eval} can derive $\ottnt{e_{{\mathrm{1}}}}  \longrightarrow  \ottnt{e'_{{\mathrm{1}}}}$, we have
      \begin{itemize}
        \item $\ottnt{e_{{\mathrm{1}}}} = \ottnt{E_{{\mathrm{1}}}}  \ottsym{[}  \ottnt{e_{{\mathrm{11}}}}  \ottsym{]}$,
        \item $\ottnt{e'_{{\mathrm{1}}}} = \ottnt{E_{{\mathrm{1}}}}  \ottsym{[}  \ottnt{e_{{\mathrm{12}}}}  \ottsym{]}$, and
        \item $\ottnt{e_{{\mathrm{11}}}}  \longmapsto  \ottnt{e_{{\mathrm{12}}}}$,
      \end{itemize}
      for some $\ottnt{E_{{\mathrm{1}}}}$, $\ottnt{e_{{\mathrm{11}}}}$, and $\ottnt{e_{{\mathrm{12}}}}$.
      Let $\ottnt{E} =  \mathbf{handle}_{ \mathit{l} \,  \bm{ { S } } ^ {  \mathit{N}  }  }  \, \ottnt{E_{{\mathrm{1}}}} \, \mathbf{with} \, \ottnt{h}$,
      \rname{E}{Eval} derives $\ottnt{e}  \longrightarrow  \ottnt{E}  \ottsym{[}  \ottnt{e_{{\mathrm{12}}}}  \ottsym{]}$ because $\ottnt{e} = \ottnt{E}  \ottsym{[}  \ottnt{e_{{\mathrm{11}}}}  \ottsym{]}$.

      \item[(3)] If $\mathit{l} \neq \mathit{l'}$, then $\ottnt{e} = \ottsym{(}   \mathbf{handle}_{ \mathit{l} \,  \bm{ { S } } ^ {  \mathit{N}  }  }  \, \ottnt{E} \, \mathbf{with} \, \ottnt{h}  \ottsym{)}  \ottsym{[}   \mathsf{op'} _{ \mathit{l'} \,  \bm{ { S' } } ^ {  \mathit{N'}  }  }  \,  \bm{ { T } } ^ {  \mathit{J}  }  \, \ottnt{v}  \ottsym{]}$ and $ \mathit{n}  \mathrm{-free} ( \mathit{l'} ,   \mathbf{handle}_{ \mathit{l} \,  \bm{ { S } } ^ {  \mathit{N}  }  }  \, \ottnt{E} \, \mathbf{with} \, \ottnt{h} ) $.

      If $\mathit{l} = \mathit{l'}$, then by Lemma~\ref{lem:ind_ev} and \ref{lem:inversion}\ref{lem:inversion:op}, we have
      \begin{itemize}
        \item $ \mathit{l'}  ::    \forall    {\bm{ \alpha' } }^{ \mathit{N'} } : {\bm{ \ottnt{K'} } }^{ \mathit{N'} }    \ottsym{.}    \sigma'    \in   \Xi $ and
        \item $ \mathsf{op'}  \ottsym{:}    \forall    {\bm{ \beta' } }^{ \mathit{J} } : {\bm{ \ottnt{K'_{{\mathrm{0}}}} } }^{ \mathit{J} }    \ottsym{.}    \ottnt{A'}   \Rightarrow   \ottnt{B'}    \in   \sigma' \,  \! [ {\bm{ { S' } } }^{ \mathit{N'} } / {\bm{ \alpha' } }^{ \mathit{N'} } ]  $,
      \end{itemize}
      for some $ \bm{ { \alpha' } } ^ {  \mathit{N'}  } $, $ {\bm{ { \ottnt{K'} } } }^{ \mathit{N'} } $, $\sigma'$, $ \bm{ { \beta' } } ^ {  \mathit{J}  } $, $\ottnt{A'}$, and $\ottnt{B'}$.
      Therefore, since $\mathit{l} = \mathit{l'}$, we have
      \begin{itemize}
        \item $ \bm{ { \alpha } } ^ {  \mathit{N}  }  =  \bm{ { \alpha' } } ^ {  \mathit{N'}  } $,
        \item $ {\bm{ { \ottnt{K} } } }^{ \mathit{N} }  =  {\bm{ { \ottnt{K'} } } }^{ \mathit{N'} } $, and
        \item $\sigma$ = $\sigma'$.
      \end{itemize}
      By $ \emptyset  \vdash _{ \sigma \,  \! [ {\bm{ { S } } }^{ \mathit{N} } / {\bm{ \alpha } }^{ \mathit{N} } ]  }  \ottnt{h}  :  \ottnt{A_{{\mathrm{1}}}}   \Rightarrow  ^ { \varepsilon }  \ottnt{A} $,
      $ \mathsf{op'}  \ottsym{:}    \forall    {\bm{ \beta' } }^{ \mathit{J} } : {\bm{ \ottnt{K'_{{\mathrm{0}}}} } }^{ \mathit{J} }    \ottsym{.}    \ottnt{A''}   \Rightarrow   \ottnt{B''}    \in   \sigma \,  \! [ {\bm{ { S } } }^{ \mathit{N} } / {\bm{ \alpha } }^{ \mathit{N} } ]  $ for some $\ottnt{A''}$ and $\ottnt{B''}$, and
      Lemma~\ref{lem:inversion_handler}\ref{lem:inversion_handler:operation},
      we have
      \begin{align*}
         \mathsf{op'} \,  {\bm{ \beta' } }^{ \mathit{J} } : {\bm{ \ottnt{K'_{{\mathrm{0}}}} } }^{ \mathit{J} }  \, \mathit{p} \, \mathit{k}  \mapsto  \ottnt{e'}   \in   \ottnt{h} 
      \end{align*}
      for some $\mathit{p}$, $\mathit{k}$, and $\ottnt{e'}$.
      If $\mathit{n} = 0$, the evaluation of $\ottnt{e}$ proceeds by \rname{R}{Handle2'}.
      Otherwise, there exists some $\mathit{m}$ such that
      $\mathit{n} = \mathit{m}  \ottsym{+}  1$ and $ \mathit{m}  \mathrm{-free} ( \mathit{l} ,   \mathbf{handle}_{ \mathit{l} \,  \bm{ { S } } ^ {  \mathit{N}  }  }  \, \ottnt{E} \, \mathbf{with} \, \ottnt{h} ) $.
    \end{divcases}
    \item[others] Similarly to Lemma~\ref{lem:progress}.
  \end{divcases}
\end{proof}

\begin{lemma}\label{lem:zero_freeness_erasure}
  If $ \mathit{n}  \mathrm{-free} ( \mathit{l} ,  \ottnt{E} ) $, then $\mathit{n} = \mathtt{0}$.
\end{lemma}

\begin{proof}
  Straightforward by the induction on the derivation of $ \mathit{n}  \mathrm{-free} ( \mathit{l} ,  \ottnt{E} ) $.
\end{proof}

\begin{lemma}\label{lem:effsafe_aux_erasure}
  If $\Gamma  \vdash  \ottnt{E}  \ottsym{[}   \mathsf{op} _{ \mathit{l} \,  \bm{ { S } } ^ {  \mathit{I}  }  }  \,  \bm{ { T } } ^ {  \mathit{J}  }  \, \ottnt{v}  \ottsym{]}  \ottsym{:}  \ottnt{A}  \mid  \varepsilon$ and
  $ \mathit{n}  \mathrm{-free} ( \mathit{l} ,  \ottnt{E} ) $,
  then $  \lift{ \mathit{l} \,  \bm{ { S } } ^ {  \mathit{I}  }  }   \olessthan  \varepsilon $.
\end{lemma}

\begin{proof}
  By induction on a derivation of $\Gamma  \vdash  \ottnt{E}  \ottsym{[}   \mathsf{op} _{ \mathit{l} \,  \bm{ { S } } ^ {  \mathit{I}  }  }  \,  \bm{ { T } } ^ {  \mathit{J}  }  \, \ottnt{v}  \ottsym{]}  \ottsym{:}  \ottnt{A}  \mid  \varepsilon$.
  We proceed by case analysis on the typing rule applied lastly to this derivation.
  \begin{divcases}
    \item[\rname{T}{App}]
    For some $\ottnt{B}$, we have
    \begin{itemize}
      \item $\ottnt{E} =  \Box $,
      \item $\Gamma  \vdash   \mathsf{op} _{ \mathit{l} \,  \bm{ { S } } ^ {  \mathit{I}  }  }  \,  \bm{ { T } } ^ {  \mathit{J}  }   \ottsym{:}   \ottnt{B}    \rightarrow_{ \varepsilon }    \ottnt{A}   \mid   \bbZero $, and
      \item $\Gamma  \vdash  \ottnt{v}  \ottsym{:}  \ottnt{B}  \mid   \bbZero $.
    \end{itemize}
    By Lemma~\ref{lem:inversion}\ref{lem:inversion:op},
    we have $\Gamma  \vdash    \lift{ \mathit{l} \,  \bm{ { S } } ^ {  \mathit{I}  }  }   \olessthan  \varepsilon $.
    Thus, the required result is achieved.

    \item[\rname{T}{Let}]
    For some $\mathit{x}$, $\ottnt{E_{{\mathrm{1}}}}$, $\ottnt{e}$, and $\ottnt{B}$, we have
    \begin{itemize}
      \item $\ottnt{E} = (\mathbf{let} \, \mathit{x}  \ottsym{=}  \ottnt{E_{{\mathrm{1}}}} \, \mathbf{in} \, \ottnt{e})$,
      \item $\Gamma  \vdash  \ottnt{E_{{\mathrm{1}}}}  \ottsym{[}   \mathsf{op} _{ \mathit{l} \,  \bm{ { S } } ^ {  \mathit{I}  }  }  \,  \bm{ { T } } ^ {  \mathit{J}  }  \, \ottnt{v}  \ottsym{]}  \ottsym{:}  \ottnt{B}  \mid  \varepsilon$, and
      \item $\Gamma  \ottsym{,}  \mathit{x}  \ottsym{:}  \ottnt{B}  \vdash  \ottnt{e}  \ottsym{:}  \ottnt{A}  \mid  \varepsilon$.
    \end{itemize}
    By the induction hypothesis, we have $  \lift{ \mathit{l} \,  \bm{ { S } } ^ {  \mathit{I}  }  }   \olessthan  \varepsilon $ as required.

    \item[\rname{T}{Sub}]
    For some $\ottnt{A'}$ and $\varepsilon'$, we have
    \begin{itemize}
      \item $\Gamma  \vdash  \ottnt{E}  \ottsym{[}   \mathsf{op} _{ \mathit{l} \,  \bm{ { S } } ^ {  \mathit{I}  }  }  \,  \bm{ { T } } ^ {  \mathit{J}  }  \, \ottnt{v}  \ottsym{]}  \ottsym{:}  \ottnt{A'}  \mid  \varepsilon'$ and
      \item $\Gamma  \vdash  \ottnt{A'}  \mid  \varepsilon'  <:  \ottnt{A}  \mid  \varepsilon$.
    \end{itemize}
    Since only \rname{ST}{Comp} can derive $\Gamma  \vdash  \ottnt{A'}  \mid  \varepsilon'  <:  \ottnt{A}  \mid  \varepsilon$,
    we have $\Gamma  \vdash   \varepsilon'  \olessthan  \varepsilon $.
    By the induction hypothesis, we have $  \lift{ \mathit{l} \,  \bm{ { S } } ^ {  \mathit{I}  }  }   \olessthan  \varepsilon' $.
    By the associativity of $ \odot $, we have $  \lift{ \mathit{l} \,  \bm{ { S } } ^ {  \mathit{I}  }  }   \olessthan  \varepsilon $ as required.

    \item[\rname{T}{Handling}]
    For some $\mathit{l'}$, $ \bm{ { S' } } ^ {  \mathit{I'}  } $, $\ottnt{E_{{\mathrm{1}}}}$, $\ottnt{h}$, $\ottnt{B}$, and $\varepsilon'$,
    we have
    \begin{itemize}
      \item $\ottnt{E} =  \mathbf{handle}_{ \mathit{l'} \,  \bm{ { S' } } ^ {  \mathit{I'}  }  }  \, \ottnt{E_{{\mathrm{1}}}} \, \mathbf{with} \, \ottnt{h}$,
      \item $\Gamma  \vdash  \ottnt{E_{{\mathrm{1}}}}  \ottsym{[}   \mathsf{op} _{ \mathit{l} \,  \bm{ { S } } ^ {  \mathit{I}  }  }  \,  \bm{ { T } } ^ {  \mathit{J}  }  \, \ottnt{v}  \ottsym{]}  \ottsym{:}  \ottnt{B}  \mid  \varepsilon'$, and
      \item $   \lift{ \mathit{l'} \,  \bm{ { S' } } ^ {  \mathit{I'}  }  }   \mathop{ \odot }  \varepsilon    \sim   \varepsilon' $.
    \end{itemize}
    By Lemma~\ref{lem:zero_freeness_erasure}, we have $ \mathit{l}   \neq   \mathit{l'} $ and $ 0  \mathrm{-free} ( \mathit{l} ,  \ottnt{E_{{\mathrm{1}}}} ) $.
    By the induction hypothesis, we have $  \lift{ \mathit{l} \,  \bm{ { S } } ^ {  \mathit{I}  }  }   \olessthan  \varepsilon' $.
    Thus, safety condition \ref{def:safe_cond:pres} makes $  \lift{ \mathit{l} \,  \bm{ { S } } ^ {  \mathit{I}  }  }   \olessthan  \varepsilon $ hold as required.
    \item[others] Cannot happen.
  \end{divcases}
\end{proof}

\begin{lemma}[Preservation in Reduction]\label{lem:pres_red_erasure}
  If $\emptyset  \vdash  \ottnt{e}  \ottsym{:}  \ottnt{A}  \mid  \varepsilon$ and $\ottnt{e}  \longmapsto  \ottnt{e'}$, then $\emptyset  \vdash  \ottnt{e'}  \ottsym{:}  \ottnt{A}  \mid  \varepsilon$.
\end{lemma}

\begin{proof}
  By induction on a derivation of $\Gamma  \vdash  \ottnt{e}  \ottsym{:}  \ottnt{A}  \mid  \varepsilon$. We proceed by case analysis on the typing rule applied lastly to this derivation.
  \begin{divcases}
    \item[\rname{T}{Handling}] We proceed by case analysis on the derivation rule that derives $\ottnt{e}  \longmapsto  \ottnt{e'}$.
    \begin{divcases}
      \item[\rname{R}{Handle1}] Similarly to Lemma~\ref{lem:pres_red}.

      \item[\rname{R}{Handle2'}] For some $\mathit{l}$, $ \bm{ { S } } ^ {  \mathit{N}  } $, $\ottnt{E}$, $\mathsf{op_{{\mathrm{0}}}}$, $ \bm{ { S' } } ^ {  \mathit{N}  } $, $ \bm{ { T } } ^ {  \mathit{J}  } $, $\ottnt{v}$, $\ottnt{h}$, $ \bm{ { \alpha } } ^ {  \mathit{N}  } $, $ {\bm{ { \ottnt{K} } } }^{ \mathit{N} } $, $\sigma$, $ \bm{ { \beta_{{\mathrm{0}}} } } ^ {  \mathit{J}  } $, $ {\bm{ { \ottnt{K_{{\mathrm{0}}}} } } }^{ \mathit{J} } $, $\ottnt{A_{{\mathrm{0}}}}$, $\ottnt{B_{{\mathrm{0}}}}$, $\mathit{p_{{\mathrm{0}}}}$, $\mathit{k_{{\mathrm{0}}}}$, $\ottnt{e_{{\mathrm{0}}}}$, $\ottnt{B}$, and $\varepsilon'$, we have
      \begin{itemize}
        \item $\ottnt{e} =  \mathbf{handle}_{ \mathit{l} \,  \bm{ { S } } ^ {  \mathit{N}  }  }  \, \ottnt{E}  \ottsym{[}   \mathsf{op_{{\mathrm{0}}}} _{ \mathit{l} \,  \bm{ { S' } } ^ {  \mathit{N}  }  }  \,  \bm{ { T } } ^ {  \mathit{J}  }  \, \ottnt{v}  \ottsym{]} \, \mathbf{with} \, \ottnt{h}$,
        \item $ \mathit{l}  ::    \forall    {\bm{ \alpha } }^{ \mathit{N} } : {\bm{ \ottnt{K} } }^{ \mathit{N} }    \ottsym{.}    \sigma    \in   \Xi $,
        \item $\emptyset  \vdash   \bm{ { S } }^{ \mathit{N} } : \bm{ \ottnt{K} }^{ \mathit{N} } $,
        \item $ \mathsf{op_{{\mathrm{0}}}} \,  {\bm{ \beta_{{\mathrm{0}}} } }^{ \mathit{J} } : {\bm{ \ottnt{K_{{\mathrm{0}}}} } }^{ \mathit{J} }  \, \mathit{p_{{\mathrm{0}}}} \, \mathit{k_{{\mathrm{0}}}}  \mapsto  \ottnt{e_{{\mathrm{0}}}}   \in   \ottnt{h} $,
        \item $ 0  \mathrm{-free} ( \mathit{l} ,  \ottnt{E} ) $,
        \item $\emptyset  \vdash  \ottnt{E}  \ottsym{[}   \mathsf{op_{{\mathrm{0}}}} _{ \mathit{l} \,  \bm{ { S' } } ^ {  \mathit{N}  }  }  \,  \bm{ { T } } ^ {  \mathit{J}  }  \, \ottnt{v}  \ottsym{]}  \ottsym{:}  \ottnt{B}  \mid  \varepsilon'$,
        \item $ \emptyset  \vdash _{ \sigma \,  \! [ {\bm{ { S } } }^{ \mathit{N} } / {\bm{ \alpha } }^{ \mathit{N} } ]  }  \ottnt{h}  :  \ottnt{B}   \Rightarrow  ^ { \varepsilon }  \ottnt{A} $,
        \item $   \lift{ \mathit{l} \,  \bm{ { S } } ^ {  \mathit{N}  }  }   \mathop{ \odot }  \varepsilon    \sim   \varepsilon' $, and
        \item $\ottnt{e'} = \ottnt{e_{{\mathrm{0}}}} \,  \! [ {\bm{ { T } } }^{ \mathit{J} } / {\bm{ \beta_{{\mathrm{0}}} } }^{ \mathit{J} } ]  \,  \! [  \ottnt{v}  /  \mathit{p_{{\mathrm{0}}}}  ]  \,  \! [  \lambda  \mathit{z}  \ottsym{.}   \mathbf{handle}_{ \mathit{l} \,  \bm{ { S } } ^ {  \mathit{N}  }  }  \, \ottnt{E}  \ottsym{[}  \mathit{z}  \ottsym{]} \, \mathbf{with} \, \ottnt{h}  /  \mathit{k_{{\mathrm{0}}}}  ] $.
      \end{itemize}
      By Lemma~\ref{lem:effsafe_aux_erasure}, we have $  \lift{ \mathit{l} \,  \bm{ { S' } } ^ {  \mathit{N}  }  }   \olessthan  \varepsilon' $.
      Thus, we get $ \bm{ { S' } } ^ {  \mathit{N}  }  =  \bm{ { S } } ^ {  \mathit{N}  } $ by $   \lift{ \mathit{l} \,  \bm{ { S } } ^ {  \mathit{N}  }  }   \mathop{ \odot }  \varepsilon    \sim   \varepsilon' $ and safety condition \ref{def:safe_cond_erasure:uniq}.
      By Lemma~\ref{lem:ind_ev}, there exist some $\ottnt{B_{{\mathrm{1}}}}$ and $\varepsilon_{{\mathrm{1}}}$ such that
      \begin{itemize}
        \item $\emptyset  \vdash   \mathsf{op_{{\mathrm{0}}}} _{ \mathit{l} \,  \bm{ { S } } ^ {  \mathit{N}  }  }  \,  \bm{ { T } } ^ {  \mathit{J}  }  \, \ottnt{v}  \ottsym{:}  \ottnt{B_{{\mathrm{1}}}}  \mid  \varepsilon_{{\mathrm{1}}}$, and
        \item for any $\ottnt{e'}$ and $\Gamma'$, if $\Gamma'  \vdash  \ottnt{e'}  \ottsym{:}  \ottnt{B_{{\mathrm{1}}}}  \mid  \varepsilon_{{\mathrm{1}}}$, then $\Gamma'  \vdash  \ottnt{E}  \ottsym{[}  \ottnt{e'}  \ottsym{]}  \ottsym{:}  \ottnt{B}  \mid  \varepsilon'$.
      \end{itemize}
      By Lemma~\ref{lem:inversion}\ref{lem:inversion:app}, we have
      $\emptyset  \vdash   \mathsf{op_{{\mathrm{0}}}} _{ \mathit{l} \,  \bm{ { S } } ^ {  \mathit{N}  }  }  \,  \bm{ { T } } ^ {  \mathit{J}  }   \ottsym{:}   \ottnt{A_{{\mathrm{1}}}}    \rightarrow_{ \varepsilon_{{\mathrm{1}}} }    \ottnt{B_{{\mathrm{1}}}}   \mid   \bbZero $ and
      $\emptyset  \vdash  \ottnt{v}  \ottsym{:}  \ottnt{A_{{\mathrm{1}}}}  \mid   \bbZero $ for some $\ottnt{A_{{\mathrm{1}}}}$.
      By Lemma~\ref{lem:inversion}\ref{lem:inversion:op} and
      \ref{lem:inversion_handler}\ref{lem:inversion_handler:operation}, we have
      \begin{itemize}
        \item $ \mathsf{op_{{\mathrm{0}}}}  \ottsym{:}    \forall    {\bm{ \beta_{{\mathrm{0}}} } }^{ \mathit{J} } : {\bm{ \ottnt{K_{{\mathrm{0}}}} } }^{ \mathit{J} }    \ottsym{.}    \ottnt{A_{{\mathrm{0}}}}   \Rightarrow   \ottnt{B_{{\mathrm{0}}}}    \in   \sigma \,  \! [ {\bm{ { S } } }^{ \mathit{N} } / {\bm{ \alpha } }^{ \mathit{N} } ]  $,
        \item $\emptyset  \vdash   \bm{ { S } }^{ \mathit{N} } : \bm{ \ottnt{K} }^{ \mathit{N} } $,
        \item $\emptyset  \vdash   \bm{ { T } }^{ \mathit{J} } : \bm{ \ottnt{K_{{\mathrm{0}}}} }^{ \mathit{J} } $,
        \item $\emptyset  \vdash  \ottnt{A_{{\mathrm{1}}}}  <:  \ottnt{A_{{\mathrm{0}}}} \,  \! [ {\bm{ { T } } }^{ \mathit{J} } / {\bm{ \beta_{{\mathrm{0}}} } }^{ \mathit{J} } ] $,
        \item $\emptyset  \vdash  \ottnt{B_{{\mathrm{0}}}} \,  \! [ {\bm{ { T } } }^{ \mathit{J} } / {\bm{ \beta_{{\mathrm{0}}} } }^{ \mathit{J} } ]   <:  \ottnt{B_{{\mathrm{1}}}}$, and
        \item $\emptyset  \vdash    \lift{ \mathit{l} \,  \bm{ { S } } ^ {  \mathit{N}  }  }   \olessthan  \varepsilon_{{\mathrm{1}}} $,
      \end{itemize}
      for some $\ottnt{A_{{\mathrm{0}}}}$ and $\ottnt{B_{{\mathrm{0}}}}$.
      Thus, \rname{T}{Sub} with $\emptyset  \vdash    \bbZero   \olessthan   \bbZero  $ implied by Lemma~\ref{lem:entailment} derives
      \begin{align*}
        \emptyset  \vdash  \ottnt{v}  \ottsym{:}  \ottnt{A_{{\mathrm{0}}}} \,  \! [ {\bm{ { T } } }^{ \mathit{J} } / {\bm{ \beta_{{\mathrm{0}}} } }^{ \mathit{J} } ]   \mid   \bbZero .
      \end{align*}
      By Lemma~\ref{lem:wk_subtyping}, we have $\emptyset  \vdash  \ottnt{B_{{\mathrm{0}}}} \,  \! [ {\bm{ { T } } }^{ \mathit{J} } / {\bm{ \beta_{{\mathrm{0}}} } }^{ \mathit{J} } ]   \ottsym{:}   \mathbf{Typ} $.
      Thus, \rname{C}{Var} derives $\vdash  \mathit{z}  \ottsym{:}  \ottnt{B_{{\mathrm{0}}}} \,  \! [ {\bm{ { T } } }^{ \mathit{J} } / {\bm{ \beta_{{\mathrm{0}}} } }^{ \mathit{J} } ] $.
      By $\emptyset  \vdash   \bbZero   \ottsym{:}   \mathbf{Eff} $,
      $\emptyset  \vdash  \varepsilon_{{\mathrm{1}}}  \ottsym{:}   \mathbf{Eff} $ implied by Lemma~\ref{lem:wk}, and
      $   \bbZero   \mathop{ \odot }  \varepsilon_{{\mathrm{1}}}    \sim   \varepsilon_{{\mathrm{1}}} $,
      we have $\emptyset  \vdash    \bbZero   \olessthan  \varepsilon_{{\mathrm{1}}} $.
      Since \rname{T}{Var} and \rname{T}{Sub} derives $\mathit{z}  \ottsym{:}  \ottnt{B_{{\mathrm{0}}}} \,  \! [ {\bm{ { T } } }^{ \mathit{J} } / {\bm{ \beta_{{\mathrm{0}}} } }^{ \mathit{J} } ]   \vdash  \mathit{z}  \ottsym{:}  \ottnt{B_{{\mathrm{1}}}}  \mid  \varepsilon_{{\mathrm{1}}}$,
      we have
      \begin{align*}
        \mathit{z}  \ottsym{:}  \ottnt{B_{{\mathrm{0}}}} \,  \! [ {\bm{ { T } } }^{ \mathit{J} } / {\bm{ \beta_{{\mathrm{0}}} } }^{ \mathit{J} } ]   \vdash   \mathbf{handle}_{ \mathit{l} \,  \bm{ { S } } ^ {  \mathit{N}  }  }  \, \ottnt{E}  \ottsym{[}  \mathit{z}  \ottsym{]} \, \mathbf{with} \, \ottnt{h}  \ottsym{:}  \ottnt{A}  \mid  \varepsilon
      \end{align*}
      by the result of Lemma~\ref{lem:ind_ev}, Lemma~\ref{lem:weakening}, and \rname{T}{Handling}.
      Thus, \rname{T}{Abs} derives
      \begin{align*}
        \emptyset  \vdash  \lambda  \mathit{z}  \ottsym{.}   \mathbf{handle}_{ \mathit{l} \,  \bm{ { S } } ^ {  \mathit{N}  }  }  \, \ottnt{E}  \ottsym{[}  \mathit{z}  \ottsym{]} \, \mathbf{with} \, \ottnt{h}  \ottsym{:}   \ottnt{B_{{\mathrm{0}}}} \,  \! [ {\bm{ { T } } }^{ \mathit{J} } / {\bm{ \beta_{{\mathrm{0}}} } }^{ \mathit{J} } ]     \rightarrow_{ \varepsilon }    \ottnt{A}   \mid   \bbZero .
      \end{align*}
      Since
      \[
         {\bm{ \beta_{{\mathrm{0}}} } }^{ \mathit{J} } : {\bm{ \ottnt{K_{{\mathrm{0}}}} } }^{ \mathit{J} }   \ottsym{,}  \mathit{p_{{\mathrm{0}}}}  \ottsym{:}  \ottnt{A_{{\mathrm{0}}}}  \ottsym{,}  \mathit{k_{{\mathrm{0}}}}  \ottsym{:}   \ottnt{B_{{\mathrm{0}}}}    \rightarrow_{ \varepsilon }    \ottnt{A}   \vdash  \ottnt{e_{{\mathrm{0}}}}  \ottsym{:}  \ottnt{A}  \mid  \varepsilon
      \]
      by $ \emptyset  \vdash _{ \sigma \,  \! [ {\bm{ { S } } }^{ \mathit{N} } / {\bm{ \alpha } }^{ \mathit{N} } ]  }  \ottnt{h}  :  \ottnt{B}   \Rightarrow  ^ { \varepsilon }  \ottnt{A} $ and
      $ \mathsf{op_{{\mathrm{0}}}}  \ottsym{:}    \forall    {\bm{ \beta_{{\mathrm{0}}} } }^{ \mathit{J} } : {\bm{ \ottnt{K_{{\mathrm{0}}}} } }^{ \mathit{J} }    \ottsym{.}    \ottnt{A_{{\mathrm{0}}}}   \Rightarrow   \ottnt{B_{{\mathrm{0}}}}    \in   \sigma \,  \! [ {\bm{ { S } } }^{ \mathit{N} } / {\bm{ \alpha } }^{ \mathit{N} } ]  $ and
      Lemma~\ref{lem:inversion_handler}\ref{lem:inversion_handler:operation},
      Lemma~\ref{lem:subst_type}\ref{lem:subst_type:typing} and Lemma~\ref{lem:subst_value}\ref{lem:subst_value:typing} imply
      \begin{align*}
        \emptyset  \vdash  \ottnt{e_{{\mathrm{0}}}} \,  \! [ {\bm{ { T } } }^{ \mathit{J} } / {\bm{ \beta_{{\mathrm{0}}} } }^{ \mathit{J} } ]  \,  \! [  \ottnt{v}  /  \mathit{p_{{\mathrm{0}}}}  ]  \,  \! [  \lambda  \mathit{z}  \ottsym{.}   \mathbf{handle}_{ \mathit{l} \,  \bm{ { S } } ^ {  \mathit{N}  }  }  \, \ottnt{E}  \ottsym{[}  \mathit{z}  \ottsym{]} \, \mathbf{with} \, \ottnt{h}  /  \mathit{k_{{\mathrm{0}}}}  ]   \ottsym{:}  \ottnt{A}  \mid  \varepsilon
      \end{align*}
      as required.
    \end{divcases}

    \item[others] Similarly to Lemma~\ref{lem:pres_red}.
  \end{divcases}
\end{proof}

\begin{lemma}[Preservation]\label{lem:preservation_erasure}
  If $\emptyset  \vdash  \ottnt{e}  \ottsym{:}  \ottnt{A}  \mid  \varepsilon$ and $\ottnt{e}  \longrightarrow  \ottnt{e'}$, then $\emptyset  \vdash  \ottnt{e'}  \ottsym{:}  \ottnt{A}  \mid  \varepsilon$.
\end{lemma}

\begin{proof}
  Similarly to Lemma~\ref{lem:preservation};
  Lemma~\ref{lem:pres_red_erasure} is used instead of Lemma~\ref{lem:pres_red}.
\end{proof}

\begin{lemma}[Effect Safety]\label{lem:effsafe_erasure}
  If $\Gamma  \vdash  \ottnt{E}  \ottsym{[}   \mathsf{op} _{ \mathit{l} \,  \bm{ { S } } ^ {  \mathit{I}  }  }  \,  \bm{ { T } } ^ {  \mathit{J}  }  \, \ottnt{v}  \ottsym{]}  \ottsym{:}  \ottnt{A}  \mid  \varepsilon$ and
  $ \mathit{n}  \mathrm{-free} ( \mathit{l} ,  \ottnt{E} ) $,
  then $ \varepsilon   \nsim    \bbZero  $.
\end{lemma}

\begin{proof}
  Similarly to Lemma~\ref{lem:effsafe};
  Lemma~\ref{lem:effsafe_aux_erasure} is used instead of Lemma~\ref{lem:effsafe_aux}.
\end{proof}

\begin{theorem}[Type and Effect Safety]\label{thm:safety_erasure}
  If $\emptyset  \vdash  \ottnt{e}  \ottsym{:}  \ottnt{A}  \mid   \bbZero $ and $\ottnt{e}  \longrightarrow  ^ * \ottnt{e'}$ and $\ottnt{e'} \centernot \longrightarrow $, then $\ottnt{e'}$ is a value.
\end{theorem}

\begin{proof}
  Similarly to Theorem~\ref{thm:safety};
  Lemmas~\ref{lem:preservation_erasure}
  , \ref{lem:effsafe_erasure},
  and \ref{lem:progress_erasure} are used
  instead of Lemmas~\ref{lem:preservation}
  , \ref{lem:effsafe},
  and \ref{lem:progress}, respectively.
\end{proof}

\subsection{Properties with Lift Coercions and Type-Erasure Semantics \TY{Change list (4)}}

This section assumes that the safety conditions in Definition~\ref{def:safe_cond}
and the safety conditions for type-erasure semantics and lift coercions in Definition~\ref{def:safe_cond_erasure} and \ref{def:safe_cond_lift} hold,
and that the semantics adapts \rname{R}{Handle2'} instead of \rname{R}{Handle2}.

\begin{lemma}[Progress]\label{lem:progress_lift_erasure}
  If $\emptyset  \vdash  \ottnt{e}  \ottsym{:}  \ottnt{A}  \mid  \varepsilon$, then one of the following holds:
  \begin{itemize}
    \item $\ottnt{e}$ is a value;
    \item There exists some expression $\ottnt{e'}$ such that $\ottnt{e}  \longrightarrow  \ottnt{e'}$; or
    \item There exist some $\mathsf{op}$, $\mathit{l}$, $ \bm{ { S } } ^ {  \mathit{I}  } $, $ \bm{ { T } } ^ {  \mathit{J}  } $, $\ottnt{v}$, $\ottnt{E}$, and $\mathit{n}$
          such that $\ottnt{e} = \ottnt{E}  \ottsym{[}   \mathsf{op} _{ \mathit{l} \,  \bm{ { S } } ^ {  \mathit{I}  }  }  \,  \bm{ { T } } ^ {  \mathit{J}  }  \, \ottnt{v}  \ottsym{]}$ and $ \mathit{n}  \mathrm{-free} ( \mathit{l} ,  \ottnt{E} ) $.
  \end{itemize}
\end{lemma}

\begin{proof}
  Similarly to Lemma~\ref{lem:progress_lift}.
\end{proof}

\begin{definition}[Label Inclusion with Type-Erasure]
  \phantom{}\\
  \textnormal{\bfseries Label Inclusion with Type-Erasure}\tquad\fbox{$ \mathit{l}  \olessthan^{ \mathcal{P} }  \varepsilon $} where $\mathcal{P} \Coloneqq  \bullet   \mid    \bm{ { S } } ^ {  \mathit{I}  }   \blacktriangleright  \mathcal{P} $ 
  \begin{mathpar}
    \inferrule{ }{
       \mathit{l}  \olessthan^{  \bullet  }  \varepsilon 
    }\ \rname{LITE}{Empty}

    \inferrule{
     \mathit{l}  \olessthan^{ \mathcal{P} }  \varepsilon_{{\mathrm{1}}}  \\    \lift{ \mathit{l} \,  \bm{ { S_{{\mathrm{0}}} } } ^ {  \mathit{I_{{\mathrm{0}}}}  }  }   \mathop{ \odot }  \varepsilon_{{\mathrm{1}}}    \sim   \varepsilon_{{\mathrm{2}}} 
    }{
     \mathit{l}  \olessthan^{   \bm{ { S_{{\mathrm{0}}} } } ^ {  \mathit{I_{{\mathrm{0}}}}  }   \blacktriangleright  \mathcal{P}  }  \varepsilon_{{\mathrm{2}}} 
    }\ \rname{LITE}{Handling}

    \inferrule{
     \mathit{l}  \olessthan^{ \mathcal{P} }  \varepsilon_{{\mathrm{1}}}  \\    \lift{ \ottnt{L} }   \mathop{ \odot }  \varepsilon_{{\mathrm{1}}}    \sim   \varepsilon_{{\mathrm{2}}}  \\
    \forall  \bm{ { S_{{\mathrm{0}}} } } ^ {  \mathit{I_{{\mathrm{0}}}}  }  . ( \ottnt{L} \neq \mathit{l} \,  \bm{ { S_{{\mathrm{0}}} } } ^ {  \mathit{I_{{\mathrm{0}}}}  } )
    }{
     \mathit{l}  \olessthan^{ \mathcal{P} }  \varepsilon_{{\mathrm{2}}} 
    }\ \rname{LITE}{NoHandling}
  \end{mathpar}
  If $n = 0$, then $  \bm{ { S_{{\mathrm{1}}} } } ^ {  \mathit{I_{{\mathrm{1}}}}  }   \blacktriangleright \cdots \blacktriangleright   \bm{ { S_{\ottmv{n}} } } ^ {  \mathit{I_{\ottmv{n}}}  }   \blacktriangleright  \mathcal{P} $ means $\mathcal{P}$.
\end{definition}

\begin{lemma}\label{lem:label_inclusion_mono_erasure}
  If $ \mathit{l}  \olessthan^{ \mathcal{P} }  \varepsilon_{{\mathrm{1}}} $ and $  \varepsilon_{{\mathrm{1}}}  \mathop{ \odot }  \varepsilon_{{\mathrm{2}}}    \sim   \varepsilon_{{\mathrm{3}}} $,
  then $ \mathit{l}  \olessthan^{ \mathcal{P} }  \varepsilon_{{\mathrm{3}}} $.
\end{lemma}

\begin{proof}
  By induction on a derivation of $ \mathit{l}  \olessthan^{ \mathcal{P} }  \varepsilon_{{\mathrm{1}}} $.
  We proceed by case analysis on the rule applied lastly to this derivation.
  \begin{divcases}
    \item[\rname{LITE}{Empty}]
    We have $\mathcal{P} =  \bullet $.
    \rname{LITE}{Empty} derives $ \mathit{l}  \olessthan^{  \bullet  }  \varepsilon_{{\mathrm{2}}} $ as required.

    \item[\rname{LITE}{Handling}]
    We have
    \begin{itemize}
      \item $\mathcal{P} =   \bm{ { S } } ^ {  \mathit{I}  }   \blacktriangleright  \mathcal{P}' $,
      \item $ \mathit{l}  \olessthan^{ \mathcal{P}' }  \varepsilon_{{\mathrm{4}}} $, and
      \item $   \lift{ \mathit{l} \,  \bm{ { S } } ^ {  \mathit{I}  }  }   \mathop{ \odot }  \varepsilon_{{\mathrm{4}}}    \sim   \varepsilon_{{\mathrm{1}}} $,
    \end{itemize}
    for some $\mathcal{P}'$, $\varepsilon_{{\mathrm{4}}}$, and $ \bm{ { S } } ^ {  \mathit{I}  } $.
    By the induction hypothesis, we have $ \mathit{l}  \olessthan^{ \mathcal{P}' }  \varepsilon_{{\mathrm{5}}} $ such that
    $  \varepsilon_{{\mathrm{4}}}  \mathop{ \odot }  \varepsilon_{{\mathrm{2}}}    \sim   \varepsilon_{{\mathrm{5}}} $.
    Thus, \rname{LITE}{Handling} derives $ \mathit{l}  \olessthan^{   \bm{ { S } } ^ {  \mathit{I}  }   \blacktriangleright  \mathcal{P}'  }  \varepsilon_{{\mathrm{2}}} $ as required.

    \item[\rname{LITE}{NoHandling}]
    We have
    \begin{itemize}
      \item $ \mathit{l}  \olessthan^{ \mathcal{P} }  \varepsilon_{{\mathrm{4}}} $,
      \item $   \lift{ \ottnt{L} }   \mathop{ \odot }  \varepsilon_{{\mathrm{4}}}    \sim   \varepsilon_{{\mathrm{1}}} $, and
      \item $\forall  \bm{ { S } } ^ {  \mathit{I}  }  . (\ottnt{L} \neq \mathit{l} \,  \bm{ { S } } ^ {  \mathit{I}  } )$,
    \end{itemize}
    for some $\ottnt{L}$ and $\varepsilon_{{\mathrm{4}}}$.
    By the induction hypothesis, we have $ \mathit{l}  \olessthan^{ \mathcal{P} }  \varepsilon_{{\mathrm{5}}} $ such that
    $  \varepsilon_{{\mathrm{4}}}  \mathop{ \odot }  \varepsilon_{{\mathrm{2}}}    \sim   \varepsilon_{{\mathrm{5}}} $.
    Thus, \rname{LITE}{NoHandling} derives $ \mathit{l}  \olessthan^{ \mathcal{P} }  \varepsilon_{{\mathrm{3}}} $ as required.
  \end{divcases}
\end{proof}

\begin{lemma}\label{lem:label_inclusion_handle_erasure}
  If $ \mathit{l}  \olessthan^{   \bm{ { S } } ^ {  \mathit{I}  }   \blacktriangleright  \mathcal{P}  }  \varepsilon_{{\mathrm{2}}} $ and $   \lift{ \mathit{l} \,  \bm{ { S } } ^ {  \mathit{I}  }  }   \mathop{ \odot }  \varepsilon_{{\mathrm{1}}}    \sim   \varepsilon_{{\mathrm{2}}} $,
  then $ \mathit{l}  \olessthan^{ \mathcal{P} }  \varepsilon_{{\mathrm{1}}} $.
\end{lemma}

\begin{proof}
  By induction on a derivation of $ \mathit{l}  \olessthan^{   \bm{ { S } } ^ {  \mathit{I}  }   \blacktriangleright  \mathcal{P}  }  \varepsilon_{{\mathrm{2}}} $.
  We proceed by case analysis on the rule lastly applied to this derivation.
  \begin{divcases}
    \item[\rname{LITE}{Empty}] Cannot happen.

    \item[\rname{LITE}{Handling}]
    We have
    \begin{itemize}
      \item $ \mathit{l}  \olessthan^{ \mathcal{P} }  \varepsilon'_{{\mathrm{1}}} $ and
      \item $   \lift{ \mathit{l} \,  \bm{ { S } } ^ {  \mathit{I}  }  }   \mathop{ \odot }  \varepsilon'_{{\mathrm{1}}}    \sim   \varepsilon_{{\mathrm{2}}} $
    \end{itemize}
    for some $\varepsilon'_{{\mathrm{1}}}$.
    By safety condition \ref{def:safe_cond_lift:removal}, we have $ \varepsilon_{{\mathrm{1}}}   \sim   \varepsilon'_{{\mathrm{1}}} $.
    By Lemma~\ref{lem:label_inclusion_mono_erasure} and $  \varepsilon'_{{\mathrm{1}}}  \mathop{ \odot }   \bbZero     \sim   \varepsilon_{{\mathrm{1}}} $,
    we have $ \mathit{l}  \olessthan^{ \mathcal{P} }  \varepsilon_{{\mathrm{1}}} $ as required.

    \item[\rname{LITE}{NoHandling}]
    We have
    \begin{itemize}
      \item $ \mathit{l}  \olessthan^{   \bm{ { S } } ^ {  \mathit{I}  }   \blacktriangleright  \mathcal{P}  }  \varepsilon_{{\mathrm{3}}} $,
      \item $   \lift{ \ottnt{L} }   \mathop{ \odot }  \varepsilon_{{\mathrm{3}}}    \sim   \varepsilon_{{\mathrm{2}}} $, and
      \item $\forall  \bm{ { S_{{\mathrm{0}}} } } ^ {  \mathit{I_{{\mathrm{0}}}}  } . (\ottnt{L} \neq \mathit{l} \,  \bm{ { S_{{\mathrm{0}}} } } ^ {  \mathit{I_{{\mathrm{0}}}}  } )$,
    \end{itemize}
    for some $\ottnt{L}$ and $\varepsilon_{{\mathrm{3}}}$.
    By safety condition \ref{def:safe_cond:pres} and $\ottnt{L} \neq \mathit{l} \,  \bm{ { S } } ^ {  \mathit{I}  } $,
    we have $   \lift{ \mathit{l} \,  \bm{ { S } } ^ {  \mathit{I}  }  }   \mathop{ \odot }  \varepsilon_{{\mathrm{4}}}    \sim   \varepsilon_{{\mathrm{3}}} $ for some $\varepsilon_{{\mathrm{4}}}$.
    By safety condition \ref{def:safe_cond_lift:removal},
    we have $ \varepsilon_{{\mathrm{1}}}   \sim     \lift{ \ottnt{L} }   \mathop{ \odot }  \varepsilon_{{\mathrm{4}}}  $.
    By the induction hypothesis, we have $ \mathit{l}  \olessthan^{ \mathcal{P} }  \varepsilon_{{\mathrm{4}}} $.
    Thus, \rname{LITE}{NoHandling} derives $ \mathit{l}  \olessthan^{ \mathcal{P} }  \varepsilon_{{\mathrm{1}}} $ as required.
  \end{divcases}
\end{proof}

\begin{lemma}\label{lem:label_inclusion_nohandle_erasure}
  If $ \mathit{l}  \olessthan^{ \mathcal{P} }  \varepsilon_{{\mathrm{2}}} $ and $   \lift{ \ottnt{L} }   \mathop{ \odot }  \varepsilon_{{\mathrm{1}}}    \sim   \varepsilon_{{\mathrm{2}}} $ and $\forall  \bm{ { S } } ^ {  \mathit{I}  } . ( \ottnt{L} \neq \mathit{l} \,  \bm{ { S } } ^ {  \mathit{I}  }  )$,
  then $ \mathit{l}  \olessthan^{ \mathcal{P} }  \varepsilon_{{\mathrm{1}}} $.
\end{lemma}

\begin{proof}
  By induction on a derivation of $ \mathit{l}  \olessthan^{ \mathcal{P} }  \varepsilon_{{\mathrm{2}}} $.
  We proceed by case analysis on the rule lastly applied to this derivation.
  \begin{divcases}
    \item[\rname{LITE}{Empty}]
    We have $\mathcal{P} =  \bullet $.
    \rname{LITE}{Empty} derives $ \mathit{l}  \olessthan^{  \bullet  }  \varepsilon_{{\mathrm{1}}} $ as required.

    \item[\rname{LITE}{Handling}]
    We have
    \begin{itemize}
      \item $\mathcal{P} =   \bm{ { S } } ^ {  \mathit{I}  }   \blacktriangleright  \mathcal{P}' $,
      \item $ \mathit{l}  \olessthan^{ \mathcal{P}' }  \varepsilon_{{\mathrm{3}}} $, and
      \item $   \lift{ \mathit{l} \,  \bm{ { S } } ^ {  \mathit{I}  }  }   \mathop{ \odot }  \varepsilon_{{\mathrm{3}}}    \sim   \varepsilon_{{\mathrm{2}}} $,
    \end{itemize}
    for some $\mathcal{P}'$, $\varepsilon_{{\mathrm{3}}}$, and $ \bm{ { S } } ^ {  \mathit{I}  } $.
    By safety condition \ref{def:safe_cond:pres} and $\ottnt{L} \neq \mathit{l} \,  \bm{ { S } } ^ {  \mathit{I}  } $,
    we have $   \lift{ \ottnt{L} }   \mathop{ \odot }  \varepsilon_{{\mathrm{4}}}    \sim   \varepsilon_{{\mathrm{3}}} $ for some $\varepsilon_{{\mathrm{4}}}$.
    By safety condition \ref{def:safe_cond_lift:removal},
    we have $ \varepsilon_{{\mathrm{1}}}   \sim     \lift{ \mathit{l} \,  \bm{ { S } } ^ {  \mathit{I}  }  }   \mathop{ \odot }  \varepsilon_{{\mathrm{4}}}  $.
    By the induction hypothesis, we have $ \mathit{l}  \olessthan^{ \mathcal{P}' }  \varepsilon_{{\mathrm{4}}} $.
    Thus, \rname{LITE}{Handling} derives $ \mathit{l}  \olessthan^{   \bm{ { S } } ^ {  \mathit{I}  }   \blacktriangleright  \mathcal{P}'  }  \varepsilon_{{\mathrm{1}}} $ as required.

    \item[\rname{LITE}{NoHandling}]
    We have
    \begin{itemize}
      \item $ \mathit{l}  \olessthan^{ \mathcal{P} }  \varepsilon_{{\mathrm{3}}} $,
      \item $   \lift{ \ottnt{L'} }   \mathop{ \odot }  \varepsilon_{{\mathrm{3}}}    \sim   \varepsilon_{{\mathrm{2}}} $, and
      \item $\forall  \bm{ { S } } ^ {  \mathit{I}  } . (\ottnt{L'} \neq \mathit{l} \,  \bm{ { S } } ^ {  \mathit{I}  } )$,
    \end{itemize}
    for some $\ottnt{L'}$ and $\varepsilon_{{\mathrm{3}}}$.

    If $\ottnt{L} = \ottnt{L'}$, then we have $ \varepsilon_{{\mathrm{1}}}   \sim   \varepsilon_{{\mathrm{3}}} $
    by safety condition \ref{def:safe_cond_lift:removal}.
    Thus, Lemma~\ref{lem:label_inclusion_mono_erasure} gives us $ \mathit{l}  \olessthan^{ \mathcal{P} }  \varepsilon_{{\mathrm{1}}} $ as required.

    If $\ottnt{L} \neq \ottnt{L'}$, then we have $   \lift{ \ottnt{L} }   \mathop{ \odot }  \varepsilon_{{\mathrm{4}}}    \sim   \varepsilon_{{\mathrm{3}}} $ for some $\varepsilon_{{\mathrm{4}}}$ by safety condition \ref{def:safe_cond:pres} and $\ottnt{L} \neq \ottnt{L'}$.
    By safety condition \ref{def:safe_cond_lift:removal},
    we have $ \varepsilon_{{\mathrm{1}}}   \sim     \lift{ \ottnt{L'} }   \mathop{ \odot }  \varepsilon_{{\mathrm{4}}}  $.
    By the induction hypothesis, we have $ \mathit{l}  \olessthan^{ \mathcal{P}' }  \varepsilon_{{\mathrm{4}}} $.
    Thus, \rname{LITE}{NoHandling} derives $ \mathit{l}  \olessthan^{ \mathcal{P} }  \varepsilon_{{\mathrm{1}}} $ as required.
  \end{divcases}
\end{proof}

\begin{lemma}\label{lem:handle_label_uniq}
  If $ \mathit{l}  \olessthan^{   \bm{ { S_{{\mathrm{0}}} } } ^ {  \mathit{I_{{\mathrm{0}}}}  }   \blacktriangleright  \mathcal{P}  }  \varepsilon $ and $  \lift{ \mathit{l} \,  \bm{ { S } } ^ {  \mathit{I}  }  }   \olessthan  \varepsilon $,
  then $ \bm{ { S } } ^ {  \mathit{I}  }  =  \bm{ { S_{{\mathrm{0}}} } } ^ {  \mathit{I_{{\mathrm{0}}}}  } $.
\end{lemma}

\begin{proof}
  By induction on a derivation of $ \mathit{l}  \olessthan^{   \bm{ { S_{{\mathrm{0}}} } } ^ {  \mathit{I_{{\mathrm{0}}}}  }   \blacktriangleright  \mathcal{P}  }  \varepsilon $.
  We proceed by case analysis on the rule lastly applied to this derivation.
  \begin{divcases}
    \item[\rname{LITE}{Empty}] Cannot happen.

    \item[\rname{LITE}{Handling}]
    We have
    \begin{itemize}
      \item $ \mathit{l}  \olessthan^{ \mathcal{P} }  \varepsilon_{{\mathrm{1}}} $ and
      \item $   \lift{ \mathit{l} \,  \bm{ { S_{{\mathrm{0}}} } } ^ {  \mathit{I_{{\mathrm{0}}}}  }  }   \mathop{ \odot }  \varepsilon_{{\mathrm{1}}}    \sim   \varepsilon $
    \end{itemize}
    for some $\varepsilon_{{\mathrm{1}}}$.
    By safety condition \ref{def:safe_cond_erasure:uniq},
    we have $ \bm{ { S } } ^ {  \mathit{I}  }  =  \bm{ { S_{{\mathrm{0}}} } } ^ {  \mathit{I_{{\mathrm{0}}}}  } $ as required.

    \item[\rname{LITE}{NoHandling}]
    We have
    \begin{itemize}
      \item $ \mathit{l}  \olessthan^{   \bm{ { S_{{\mathrm{0}}} } } ^ {  \mathit{I_{{\mathrm{0}}}}  }   \blacktriangleright  \mathcal{P}  }  \varepsilon_{{\mathrm{1}}} $,
      \item $   \lift{ \ottnt{L} }   \mathop{ \odot }  \varepsilon_{{\mathrm{1}}}    \sim   \varepsilon $, and
      \item $\forall  \bm{ { S' } } ^ {  \mathit{I'}  } . (\ottnt{L} \neq  \bm{ { S' } } ^ {  \mathit{I'}  } )$
    \end{itemize}
    for some $\ottnt{L}$ and $\varepsilon_{{\mathrm{1}}}$.
    By safety condition \ref{def:safe_cond:pres} and $\ottnt{L} \neq \mathit{l} \,  \bm{ { S } } ^ {  \mathit{I}  } $,
    we have $  \lift{ \mathit{l} \,  \bm{ { S } } ^ {  \mathit{I}  }  }   \olessthan  \varepsilon_{{\mathrm{1}}} $.
    Thus, by the induction hypothesis, we have $ \bm{ { S } } ^ {  \mathit{I}  }  =  \bm{ { S_{{\mathrm{0}}} } } ^ {  \mathit{I_{{\mathrm{0}}}}  } $ as required.
  \end{divcases}
\end{proof}

\begin{lemma}\label{lem:handling_order}
  If $\emptyset  \vdash  \ottnt{E}  \ottsym{[}   \mathsf{op} _{ \mathit{l} \,  \bm{ { S } } ^ {  \mathit{I}  }  }  \,  \bm{ { T } } ^ {  \mathit{J}  }  \, \ottnt{v}  \ottsym{]}  \ottsym{:}  \ottnt{A}  \mid  \varepsilon$ and
  $ \mathit{n}  \mathrm{-free} ( \mathit{l} ,  \ottnt{E} ) $,
  then $ \mathit{l}  \olessthan^{   \bm{ { S_{{\mathrm{1}}} } } ^ {  \mathit{I_{{\mathrm{1}}}}  }   \blacktriangleright \cdots \blacktriangleright   \bm{ { S_{\ottmv{n}} } } ^ {  \mathit{I_{\ottmv{n}}}  }   \ottsym{,}   \bm{ { S } } ^ {  \mathit{I}  }   \blacktriangleright   \bullet   }  \varepsilon $.
\end{lemma}

\begin{proof}
  By induction on a derivation of $\emptyset  \vdash  \ottnt{E}  \ottsym{[}   \mathsf{op} _{ \mathit{l} \,  \bm{ { S } } ^ {  \mathit{I}  }  }  \,  \bm{ { T } } ^ {  \mathit{J}  }  \, \ottnt{v}  \ottsym{]}  \ottsym{:}  \ottnt{A}  \mid  \varepsilon$. We proceed by case analysis on the typing rule applied lastly to this derivation.
  \begin{divcases}
    \item[\rname{T}{App}]
    For some $\ottnt{B}$, we have
    \begin{itemize}
      \item $\ottnt{E} =  \Box $,
      \item $\emptyset  \vdash   \mathsf{op} _{ \mathit{l} \,  \bm{ { S } } ^ {  \mathit{I}  }  }  \,  \bm{ { T } } ^ {  \mathit{J}  }   \ottsym{:}   \ottnt{B}    \rightarrow_{ \varepsilon }    \ottnt{A}   \mid   \bbZero $, and
      \item $\emptyset  \vdash  \ottnt{v}  \ottsym{:}  \ottnt{B}  \mid   \bbZero $.
    \end{itemize}
    By Lemma~\ref{lem:inversion_lift}\ref{lem:inversion_lift:op},
    we have $\emptyset  \vdash    \lift{ \mathit{l} \,  \bm{ { S } } ^ {  \mathit{I}  }  }   \olessthan  \varepsilon $.
    Thus, \rname{LITE}{Empty} and \rname{LITE}{Handling} derive $ \mathit{l}  \olessthan^{   \bm{ { S } } ^ {  \mathit{I}  }   \blacktriangleright   \bullet   }  \varepsilon $.

    \item[\rname{T}{Let}]
    For some $\mathit{x}$, $\ottnt{E_{{\mathrm{1}}}}$, $\ottnt{e}$, and $\ottnt{B}$, we have
    \begin{itemize}
      \item $\ottnt{E} = (\mathbf{let} \, \mathit{x}  \ottsym{=}  \ottnt{E_{{\mathrm{1}}}} \, \mathbf{in} \, \ottnt{e})$,
      \item $\emptyset  \vdash  \ottnt{E_{{\mathrm{1}}}}  \ottsym{[}   \mathsf{op} _{ \mathit{l} \,  \bm{ { S } } ^ {  \mathit{I}  }  }  \,  \bm{ { T } } ^ {  \mathit{J}  }  \, \ottnt{v}  \ottsym{]}  \ottsym{:}  \ottnt{B}  \mid  \varepsilon$,
      \item $ \mathit{n}  \mathrm{-free} ( \mathit{l} ,  \ottnt{E_{{\mathrm{1}}}} ) $, and
      \item $\mathit{x}  \ottsym{:}  \ottnt{B}  \vdash  \ottnt{e}  \ottsym{:}  \ottnt{A}  \mid  \varepsilon$.
    \end{itemize}
    By the induction hypothesis,
    we have $ \mathit{l}  \olessthan^{   \bm{ { S_{{\mathrm{1}}} } } ^ {  \mathit{I_{{\mathrm{1}}}}  }   \blacktriangleright \cdots \blacktriangleright   \bm{ { S_{\ottmv{n}} } } ^ {  \mathit{I_{\ottmv{n}}}  }   \ottsym{,}   \bm{ { S } } ^ {  \mathit{I}  }   \blacktriangleright   \bullet   }  \varepsilon $ as required.

    \item[\rname{T}{Sub}]
    For some $\ottnt{A'}$ and $\varepsilon'$, we have
    \begin{itemize}
      \item $\emptyset  \vdash  \ottnt{E}  \ottsym{[}   \mathsf{op} _{ \mathit{l} \,  \bm{ { S } } ^ {  \mathit{I}  }  }  \,  \bm{ { T } } ^ {  \mathit{J}  }  \, \ottnt{v}  \ottsym{]}  \ottsym{:}  \ottnt{A'}  \mid  \varepsilon'$ and
      \item $\emptyset  \vdash  \ottnt{A'}  \mid  \varepsilon'  <:  \ottnt{A}  \mid  \varepsilon$.
    \end{itemize}
    By the induction hypothesis, we have $ \mathit{l}  \olessthan^{   \bm{ { S_{{\mathrm{1}}} } } ^ {  \mathit{I_{{\mathrm{1}}}}  }   \blacktriangleright \cdots \blacktriangleright   \bm{ { S_{\ottmv{n}} } } ^ {  \mathit{I_{\ottmv{n}}}  }   \ottsym{,}   \bm{ { S } } ^ {  \mathit{I}  }   \blacktriangleright   \bullet   }  \varepsilon' $.
    Since only \rname{ST}{Comp} can derives $\emptyset  \vdash  \ottnt{A'}  \mid  \varepsilon'  <:  \ottnt{A}  \mid  \varepsilon$,
    we have $\emptyset  \vdash   \varepsilon'  \olessthan  \varepsilon $.
    Thus, Lemma~\ref{lem:label_inclusion_mono_erasure} derives
    $ \mathit{l}  \olessthan^{   \bm{ { S_{{\mathrm{1}}} } } ^ {  \mathit{I_{{\mathrm{1}}}}  }   \blacktriangleright \cdots \blacktriangleright   \bm{ { S_{\ottmv{n}} } } ^ {  \mathit{I_{\ottmv{n}}}  }   \ottsym{,}   \bm{ { S } } ^ {  \mathit{I}  }   \blacktriangleright   \bullet   }  \varepsilon $ as required.

    \item[\rname{T}{Lift}]
    For some $\ottnt{L}$, $\varepsilon'$, and $\ottnt{E'}$, we have
    \begin{itemize}
      \item $\ottnt{E} =  [  \ottnt{E'}  ] _{ \ottnt{L} } $,
      \item $\emptyset  \vdash  \ottnt{E'}  \ottsym{[}   \mathsf{op} _{ \mathit{l} \,  \bm{ { S } } ^ {  \mathit{I}  }  }  \,  \bm{ { T } } ^ {  \mathit{J}  }  \, \ottnt{v}  \ottsym{]}  \ottsym{:}  \ottnt{A}  \mid  \varepsilon'$,
      \item $\emptyset  \vdash  \ottnt{L}  \ottsym{:}   \mathbf{Lab} $, and
      \item $   \lift{ \ottnt{L} }   \mathop{ \odot }  \varepsilon'    \sim   \varepsilon $.
    \end{itemize}

    If $\ottnt{L} \neq \mathit{l} \,  \bm{ { S' } } ^ {  \mathit{I'}  } $ for any $ \bm{ { S' } } ^ {  \mathit{I'}  } $, then we have $ \mathit{n}  \mathrm{-free} ( \mathit{l} ,  \ottnt{E'} ) $.
    By the induction hypothesis, we have $ \mathit{l}  \olessthan^{   \bm{ { S_{{\mathrm{1}}} } } ^ {  \mathit{I_{{\mathrm{1}}}}  }   \blacktriangleright \cdots \blacktriangleright   \bm{ { S_{\ottmv{n}} } } ^ {  \mathit{I_{\ottmv{n}}}  }   \ottsym{,}   \bm{ { S } } ^ {  \mathit{I}  }   \blacktriangleright   \bullet   }  \varepsilon' $.
    Thus, \rname{LITE}{NoHandling} derives $ \mathit{l}  \olessthan^{   \bm{ { S_{{\mathrm{1}}} } } ^ {  \mathit{I_{{\mathrm{1}}}}  }   \blacktriangleright \cdots \blacktriangleright   \bm{ { S_{\ottmv{n}} } } ^ {  \mathit{I_{\ottmv{n}}}  }   \ottsym{,}   \bm{ { S } } ^ {  \mathit{I}  }   \blacktriangleright   \bullet   }  \varepsilon $ as required.

    If $\ottnt{L} = \mathit{l} \,  \bm{ { S' } } ^ {  \mathit{I'}  } $ for some $ \bm{ { S' } } ^ {  \mathit{I'}  } $,
    then there exists some $\mathit{m}$ such that $\mathit{n} = \mathit{m}  \ottsym{+}  1$ and $ \mathit{m}  \mathrm{-free} ( \mathit{l} ,  \ottnt{E'} ) $.
    By the induction hypothesis, we have $ \mathit{l}  \olessthan^{   \bm{ { S_{{\mathrm{1}}} } } ^ {  \mathit{I_{{\mathrm{1}}}}  }   \blacktriangleright \cdots \blacktriangleright   \bm{ { S_{\ottmv{m}} } } ^ {  \mathit{I_{\ottmv{m}}}  }   \ottsym{,}   \bm{ { S } } ^ {  \mathit{I}  }   \blacktriangleright   \bullet   }  \varepsilon' $.
    Thus, \rname{LITE}{Handling} derives $ \mathit{l}  \olessthan^{   \bm{ { S' } } ^ {  \mathit{I'}  }   \ottsym{,}   \bm{ { S_{{\mathrm{1}}} } } ^ {  \mathit{I_{{\mathrm{1}}}}  }   \blacktriangleright \cdots \blacktriangleright   \bm{ { S_{\ottmv{m}} } } ^ {  \mathit{I_{\ottmv{m}}}  }   \ottsym{,}   \bm{ { S } } ^ {  \mathit{I}  }   \blacktriangleright   \bullet   }  \varepsilon $ as required.

    \item[\rname{T}{Handling}]
    For some $\mathit{l'}$, $ \bm{ { S' } } ^ {  \mathit{I'}  } $, $\ottnt{E_{{\mathrm{1}}}}$, $\ottnt{h}$, $\ottnt{B}$, and $\varepsilon'$, we have
    \begin{itemize}
      \item $\ottnt{E} =  \mathbf{handle}_{ \mathit{l'} \,  \bm{ { S' } } ^ {  \mathit{I'}  }  }  \, \ottnt{E_{{\mathrm{1}}}} \, \mathbf{with} \, \ottnt{h}$,
      \item $\emptyset  \vdash  \ottnt{E_{{\mathrm{1}}}}  \ottsym{[}   \mathsf{op} _{ \mathit{l} \,  \bm{ { S } } ^ {  \mathit{I}  }  }  \,  \bm{ { T } } ^ {  \mathit{J}  }  \, \ottnt{v}  \ottsym{]}  \ottsym{:}  \ottnt{B}  \mid  \varepsilon'$, and
      \item $   \lift{ \mathit{l'} \,  \bm{ { S' } } ^ {  \mathit{I'}  }  }   \mathop{ \odot }  \varepsilon    \sim   \varepsilon' $.
    \end{itemize}

    If $\mathit{l} \neq \mathit{l'}$, then $ \mathit{n}  \mathrm{-free} ( \mathit{l} ,  \ottnt{E_{{\mathrm{1}}}} ) $.
    By the induction hypothesis, we have $ \mathit{l}  \olessthan^{   \bm{ { S_{{\mathrm{1}}} } } ^ {  \mathit{I_{{\mathrm{1}}}}  }   \blacktriangleright \cdots \blacktriangleright   \bm{ { S_{\ottmv{n}} } } ^ {  \mathit{I_{\ottmv{n}}}  }   \ottsym{,}   \bm{ { S } } ^ {  \mathit{I}  }   \blacktriangleright   \bullet   }  \varepsilon' $.
    By Lemma~\ref{lem:label_inclusion_nohandle_erasure},
    we have $ \mathit{l}  \olessthan^{   \bm{ { S_{{\mathrm{1}}} } } ^ {  \mathit{I_{{\mathrm{1}}}}  }   \blacktriangleright \cdots \blacktriangleright   \bm{ { S_{\ottmv{n}} } } ^ {  \mathit{I_{\ottmv{n}}}  }   \ottsym{,}   \bm{ { S } } ^ {  \mathit{I}  }   \blacktriangleright   \bullet   }  \varepsilon $ as required.

    If $\mathit{l} = \mathit{l'}$, then $ \mathit{n}  \ottsym{+}  1  \mathrm{-free} ( \mathit{l} ,  \ottnt{E_{{\mathrm{1}}}} ) $.
    By the induction hypothesis, we have $ \mathit{l}  \olessthan^{   \bm{ { S_{{\mathrm{0}}} } } ^ {  \mathit{I_{{\mathrm{0}}}}  }   \ottsym{,}   \bm{ { S_{{\mathrm{1}}} } } ^ {  \mathit{I_{{\mathrm{1}}}}  }   \blacktriangleright \cdots \blacktriangleright   \bm{ { S_{\ottmv{n}} } } ^ {  \mathit{I_{\ottmv{n}}}  }   \ottsym{,}   \bm{ { S } } ^ {  \mathit{I}  }   \blacktriangleright   \bullet   }  \varepsilon' $.
    By Lemma~\ref{lem:handle_label_uniq}, we have $ \bm{ { S_{{\mathrm{0}}} } } ^ {  \mathit{I_{{\mathrm{0}}}}  }  =  \bm{ { S' } } ^ {  \mathit{I'}  } $.
    By Lemma~\ref{lem:label_inclusion_handle_erasure},
    we have $ \mathit{l}  \olessthan^{   \bm{ { S_{{\mathrm{1}}} } } ^ {  \mathit{I_{{\mathrm{1}}}}  }   \blacktriangleright \cdots \blacktriangleright   \bm{ { S_{\ottmv{n}} } } ^ {  \mathit{I_{\ottmv{n}}}  }   \ottsym{,}   \bm{ { S } } ^ {  \mathit{I}  }   \blacktriangleright   \bullet   }  \varepsilon $ as required.

    \item[others] Cannot happen.
  \end{divcases}
\end{proof}

\begin{lemma}[Preservation in Reduction]\label{lem:pres_red_lift_erasure}
  If $\emptyset  \vdash  \ottnt{e}  \ottsym{:}  \ottnt{A}  \mid  \varepsilon$ and $\ottnt{e}  \longmapsto  \ottnt{e'}$, then $\emptyset  \vdash  \ottnt{e'}  \ottsym{:}  \ottnt{A}  \mid  \varepsilon$.
\end{lemma}

\begin{proof}
  By induction on a derivation of $\Gamma  \vdash  \ottnt{e}  \ottsym{:}  \ottnt{A}  \mid  \varepsilon$.
  We proceed by cases on the typing rule applied lastly to this derivation.
  \begin{divcases}
    \item[\rname{T}{Handling}]
    We proceed by cases on the derivation rule which derives $\ottnt{e}  \longmapsto  \ottnt{e'}$.
    \begin{divcases}
      \item[\rname{R}{Handle1}] Similarly to Lemma~\ref{lem:pres_red_lift}.

      \item[\rname{R}{Handle2}] We have
      \begin{itemize}
        \item $\ottnt{e} =  \mathbf{handle}_{ \mathit{l} \,  \bm{ { S } } ^ {  \mathit{N}  }  }  \, \ottnt{E}  \ottsym{[}   \mathsf{op_{{\mathrm{0}}}} _{ \mathit{l} \,  \bm{ { S' } } ^ {  \mathit{N}  }  }  \,  \bm{ { T } } ^ {  \mathit{J}  }  \, \ottnt{v}  \ottsym{]} \, \mathbf{with} \, \ottnt{h}$,
        \item $ \mathit{l}  ::    \forall    {\bm{ \alpha } }^{ \mathit{N} } : {\bm{ \ottnt{K} } }^{ \mathit{N} }    \ottsym{.}    \sigma    \in   \Xi $,
        \item $\emptyset  \vdash   \bm{ { S } }^{ \mathit{N} } : \bm{ \ottnt{K} }^{ \mathit{N} } $,
        \item $ \mathsf{op_{{\mathrm{0}}}} \,  {\bm{ \beta_{{\mathrm{0}}} } }^{ \mathit{J} } : {\bm{ \ottnt{K_{{\mathrm{0}}}} } }^{ \mathit{J} }  \, \mathit{p_{{\mathrm{0}}}} \, \mathit{k_{{\mathrm{0}}}}  \mapsto  \ottnt{e_{{\mathrm{0}}}}   \in   \ottnt{h} $,
        \item $\emptyset  \vdash  \ottnt{E}  \ottsym{[}   \mathsf{op_{{\mathrm{0}}}} _{ \mathit{l} \,  \bm{ { S' } } ^ {  \mathit{N}  }  }  \,  \bm{ { T } } ^ {  \mathit{J}  }  \, \ottnt{v}  \ottsym{]}  \ottsym{:}  \ottnt{B}  \mid  \varepsilon'$,
        \item $ \emptyset  \vdash _{ \sigma \,  \! [ {\bm{ { S } } }^{ \mathit{N} } / {\bm{ \alpha } }^{ \mathit{N} } ]  }  \ottnt{h}  :  \ottnt{B}   \Rightarrow  ^ { \varepsilon }  \ottnt{A} $,
        \item $   \lift{ \mathit{l} \,  \bm{ { S } } ^ {  \mathit{N}  }  }   \mathop{ \odot }  \varepsilon    \sim   \varepsilon' $,
        \item $ 0  \mathrm{-free} ( \mathit{l} ,  \ottnt{E} ) $, and
        \item $\ottnt{e'} = \ottnt{e_{{\mathrm{0}}}} \,  \! [ {\bm{ { T } } }^{ \mathit{J} } / {\bm{ \beta_{{\mathrm{0}}} } }^{ \mathit{J} } ]  \,  \! [  \ottnt{v}  /  \mathit{p_{{\mathrm{0}}}}  ]  \,  \! [  \lambda  \mathit{z}  \ottsym{.}   \mathbf{handle}_{ \mathit{l} \,  \bm{ { S } } ^ {  \mathit{N}  }  }  \, \ottnt{E}  \ottsym{[}  \mathit{z}  \ottsym{]} \, \mathbf{with} \, \ottnt{h}  /  \mathit{k_{{\mathrm{0}}}}  ] $
      \end{itemize}
      for some $\mathit{l}$, $ \bm{ { S } } ^ {  \mathit{N}  } $, $\ottnt{E}$, $\mathsf{op_{{\mathrm{0}}}}$, $ \bm{ { S' } } ^ {  \mathit{N}  } $, $ \bm{ { T } } ^ {  \mathit{J}  } $, $\ottnt{v}$, $\ottnt{h}$, $ \bm{ { \alpha } } ^ {  \mathit{N}  } $, $ {\bm{ { \ottnt{K} } } }^{ \mathit{N} } $, $\sigma$, $ \bm{ { \beta_{{\mathrm{0}}} } } ^ {  \mathit{J}  } $, $ {\bm{ { \ottnt{K_{{\mathrm{0}}}} } } }^{ \mathit{J} } $, $\mathit{p_{{\mathrm{0}}}}$, $\mathit{k_{{\mathrm{0}}}}$, $\ottnt{e_{{\mathrm{0}}}}$, $\ottnt{B}$, and $\varepsilon'$.
      By Lemma~\ref{lem:handling_order}, we have $ \mathit{l}  \olessthan^{   \bm{ { S' } } ^ {  \mathit{N}  }   \blacktriangleright   \bullet   }  \varepsilon' $.
      By Lemma~\ref{lem:handle_label_uniq} and $   \lift{ \mathit{l} \,  \bm{ { S } } ^ {  \mathit{N}  }  }   \mathop{ \odot }  \varepsilon    \sim   \varepsilon' $,
      we have $ \bm{ { S } } ^ {  \mathit{N}  }  =  \bm{ { S' } } ^ {  \mathit{N}  } $.
      By Lemma~\ref{lem:ind_ev_lift}, there exist some $\ottnt{B_{{\mathrm{1}}}}$ and $\varepsilon_{{\mathrm{1}}}$ such that
      \begin{itemize}
        \item $\emptyset  \vdash   \mathsf{op_{{\mathrm{0}}}} _{ \mathit{l} \,  \bm{ { S } } ^ {  \mathit{N}  }  }  \,  \bm{ { T } } ^ {  \mathit{J}  }  \, \ottnt{v}  \ottsym{:}  \ottnt{B_{{\mathrm{1}}}}  \mid  \varepsilon_{{\mathrm{1}}}$, and
        \item for any $\ottnt{e''}$ and $\Gamma''$,
              if $\Gamma''  \vdash  \ottnt{e''}  \ottsym{:}  \ottnt{B_{{\mathrm{1}}}}  \mid  \varepsilon_{{\mathrm{1}}}$,
              then $\Gamma''  \vdash  \ottnt{E}  \ottsym{[}  \ottnt{e''}  \ottsym{]}  \ottsym{:}  \ottnt{B}  \mid  \varepsilon'$.
      \end{itemize}
      By Lemma~\ref{lem:inversion_lift}\ref{lem:inversion_lift:app},
      we have $\emptyset  \vdash   \mathsf{op_{{\mathrm{0}}}} _{ \mathit{l} \,  \bm{ { S } } ^ {  \mathit{N}  }  }  \,  \bm{ { T } } ^ {  \mathit{J}  }   \ottsym{:}   \ottnt{A_{{\mathrm{1}}}}    \rightarrow_{ \varepsilon_{{\mathrm{1}}} }    \ottnt{B_{{\mathrm{1}}}}   \mid   \bbZero $ and $\emptyset  \vdash  \ottnt{v}  \ottsym{:}  \ottnt{A_{{\mathrm{1}}}}  \mid   \bbZero $ for some $\ottnt{A_{{\mathrm{1}}}}$.
      By Lemma~\ref{lem:inversion_lift}\ref{lem:inversion_lift:op} and
      \ref{lem:inversion_handler}\ref{lem:inversion_handler:operation}, we have
      \begin{itemize}
        \item $ \mathsf{op_{{\mathrm{0}}}}  \ottsym{:}    \forall    {\bm{ \beta_{{\mathrm{0}}} } }^{ \mathit{J} } : {\bm{ \ottnt{K_{{\mathrm{0}}}} } }^{ \mathit{J} }    \ottsym{.}    \ottnt{A_{{\mathrm{0}}}}   \Rightarrow   \ottnt{B_{{\mathrm{0}}}}    \in   \sigma \,  \! [ {\bm{ { S } } }^{ \mathit{N} } / {\bm{ \alpha } }^{ \mathit{N} } ]  $,
        \item $\emptyset  \vdash   \bm{ { S } }^{ \mathit{N} } : \bm{ \ottnt{K} }^{ \mathit{N} } $,
        \item $\emptyset  \vdash   \bm{ { T } }^{ \mathit{J} } : \bm{ \ottnt{K_{{\mathrm{0}}}} }^{ \mathit{J} } $,
        \item $\emptyset  \vdash  \ottnt{A_{{\mathrm{1}}}}  <:  \ottnt{A_{{\mathrm{0}}}} \,  \! [ {\bm{ { T } } }^{ \mathit{J} } / {\bm{ \beta_{{\mathrm{0}}} } }^{ \mathit{J} } ] $,
        \item $\emptyset  \vdash  \ottnt{B_{{\mathrm{0}}}} \,  \! [ {\bm{ { T } } }^{ \mathit{J} } / {\bm{ \beta_{{\mathrm{0}}} } }^{ \mathit{J} } ]   <:  \ottnt{B_{{\mathrm{1}}}}$, and
        \item $\emptyset  \vdash    \lift{ \mathit{l} \,  \bm{ { S } } ^ {  \mathit{N}  }  }   \olessthan  \varepsilon_{{\mathrm{1}}} $,
      \end{itemize}
      for some $\ottnt{A_{{\mathrm{0}}}}$ and $\ottnt{B_{{\mathrm{0}}}}$.
      Thus, \rname{T}{Sub} with $\emptyset  \vdash    \bbZero   \olessthan   \bbZero  $ implied by Lemma~\ref{lem:entailment} derives
      \begin{align*}
        \emptyset  \vdash  \ottnt{v}  \ottsym{:}  \ottnt{A_{{\mathrm{0}}}} \,  \! [ {\bm{ { T } } }^{ \mathit{J} } / {\bm{ \beta_{{\mathrm{0}}} } }^{ \mathit{J} } ]   \mid   \bbZero .
      \end{align*}
      By Lemma~\ref{lem:wk_subtyping}, we have $\emptyset  \vdash  \ottnt{B_{{\mathrm{0}}}} \,  \! [ {\bm{ { T } } }^{ \mathit{J} } / {\bm{ \beta_{{\mathrm{0}}} } }^{ \mathit{J} } ]   \ottsym{:}   \mathbf{Typ} $.
      Thus, \rname{C}{Var} derives $\vdash  \mathit{z}  \ottsym{:}  \ottnt{B_{{\mathrm{0}}}} \,  \! [ {\bm{ { T } } }^{ \mathit{J} } / {\bm{ \beta_{{\mathrm{0}}} } }^{ \mathit{J} } ] $.
      By $\emptyset  \vdash   \bbZero   \ottsym{:}   \mathbf{Eff} $,
      $\emptyset  \vdash  \varepsilon_{{\mathrm{1}}}  \ottsym{:}   \mathbf{Eff} $ implied by Lemma~\ref{lem:wk}, and
      $   \bbZero   \mathop{ \odot }  \varepsilon_{{\mathrm{1}}}    \sim   \varepsilon_{{\mathrm{1}}} $,
      we have $\emptyset  \vdash    \bbZero   \olessthan  \varepsilon_{{\mathrm{1}}} $.
      Since \rname{T}{Var} and \rname{T}{Sub} derives $\mathit{z}  \ottsym{:}  \ottnt{B_{{\mathrm{0}}}} \,  \! [ {\bm{ { T } } }^{ \mathit{J} } / {\bm{ \beta_{{\mathrm{0}}} } }^{ \mathit{J} } ]   \vdash  \mathit{z}  \ottsym{:}  \ottnt{B_{{\mathrm{1}}}}  \mid  \varepsilon_{{\mathrm{1}}}$,
      we have
      \begin{align*}
        \mathit{z}  \ottsym{:}  \ottnt{B_{{\mathrm{0}}}} \,  \! [ {\bm{ { T } } }^{ \mathit{J} } / {\bm{ \beta_{{\mathrm{0}}} } }^{ \mathit{J} } ]   \vdash   \mathbf{handle}_{ \mathit{l} \,  \bm{ { S } } ^ {  \mathit{N}  }  }  \, \ottnt{E}  \ottsym{[}  \mathit{z}  \ottsym{]} \, \mathbf{with} \, \ottnt{h}  \ottsym{:}  \ottnt{A}  \mid  \varepsilon
      \end{align*}
      by the result of Lemma~\ref{lem:ind_ev}, Lemma~\ref{lem:weakening}, and \rname{T}{Handling}.
      Thus, \rname{T}{Abs} derives
      \begin{align*}
        \emptyset  \vdash  \lambda  \mathit{z}  \ottsym{.}   \mathbf{handle}_{ \mathit{l} \,  \bm{ { S } } ^ {  \mathit{N}  }  }  \, \ottnt{E}  \ottsym{[}  \mathit{z}  \ottsym{]} \, \mathbf{with} \, \ottnt{h}  \ottsym{:}   \ottnt{B_{{\mathrm{0}}}} \,  \! [ {\bm{ { T } } }^{ \mathit{J} } / {\bm{ \beta_{{\mathrm{0}}} } }^{ \mathit{J} } ]     \rightarrow_{ \varepsilon }    \ottnt{A}   \mid   \bbZero .
      \end{align*}
      Since
      \[
         {\bm{ \beta_{{\mathrm{0}}} } }^{ \mathit{J} } : {\bm{ \ottnt{K_{{\mathrm{0}}}} } }^{ \mathit{J} }   \ottsym{,}  \mathit{p_{{\mathrm{0}}}}  \ottsym{:}  \ottnt{A_{{\mathrm{0}}}}  \ottsym{,}  \mathit{k_{{\mathrm{0}}}}  \ottsym{:}   \ottnt{B_{{\mathrm{0}}}}    \rightarrow_{ \varepsilon }    \ottnt{A}   \vdash  \ottnt{e_{{\mathrm{0}}}}  \ottsym{:}  \ottnt{A}  \mid  \varepsilon
      \]
      by $ \emptyset  \vdash _{ \sigma \,  \! [ {\bm{ { S } } }^{ \mathit{N} } / {\bm{ \alpha } }^{ \mathit{N} } ]  }  \ottnt{h}  :  \ottnt{B}   \Rightarrow  ^ { \varepsilon }  \ottnt{A} $ and
      $ \mathsf{op_{{\mathrm{0}}}}  \ottsym{:}    \forall    {\bm{ \beta_{{\mathrm{0}}} } }^{ \mathit{J} } : {\bm{ \ottnt{K_{{\mathrm{0}}}} } }^{ \mathit{J} }    \ottsym{.}    \ottnt{A_{{\mathrm{0}}}}   \Rightarrow   \ottnt{B_{{\mathrm{0}}}}    \in   \sigma \,  \! [ {\bm{ { S } } }^{ \mathit{N} } / {\bm{ \alpha } }^{ \mathit{N} } ]  $ and
      Lemma~\ref{lem:inversion_handler}\ref{lem:inversion_handler:operation},
      Lemma~\ref{lem:subst_type}\ref{lem:subst_type:typing} and Lemma~\ref{lem:subst_value}\ref{lem:subst_value:typing} imply
      \begin{align*}
        \emptyset  \vdash  \ottnt{e_{{\mathrm{0}}}} \,  \! [ {\bm{ { T } } }^{ \mathit{J} } / {\bm{ \beta_{{\mathrm{0}}} } }^{ \mathit{J} } ]  \,  \! [  \ottnt{v}  /  \mathit{p_{{\mathrm{0}}}}  ]  \,  \! [  \lambda  \mathit{z}  \ottsym{.}   \mathbf{handle}_{ \mathit{l} \,  \bm{ { S } } ^ {  \mathit{N}  }  }  \, \ottnt{E}  \ottsym{[}  \mathit{z}  \ottsym{]} \, \mathbf{with} \, \ottnt{h}  /  \mathit{k_{{\mathrm{0}}}}  ]   \ottsym{:}  \ottnt{A}  \mid  \varepsilon
      \end{align*}
      as required.
    \end{divcases}

    \item[others]
    Similarly to Lemma~\ref{lem:pres_red_lift}.
  \end{divcases}
\end{proof}

\begin{lemma}[Preservation]\label{lem:preservation_lift_erasure}
  If $\emptyset  \vdash  \ottnt{e}  \ottsym{:}  \ottnt{A}  \mid  \varepsilon$ and $\ottnt{e}  \longrightarrow  \ottnt{e'}$, then $\emptyset  \vdash  \ottnt{e'}  \ottsym{:}  \ottnt{A}  \mid  \varepsilon$.
\end{lemma}

\begin{proof}
  Similarly to Lemma~\ref{lem:preservation_lift};
  Lemma~\ref{lem:pres_red_lift_erasure} is used instead of Lemma~\ref{lem:pres_red_lift}.
\end{proof}

\begin{lemma}[No Inclusion by Empty Effect]\label{lem:no_inclusion_erasure}
  If $ \mathit{l}  \olessthan^{ \mathcal{P} }  \varepsilon $ and $ \varepsilon   \sim    \bbZero  $, then $\mathcal{P} =  \bullet $.
\end{lemma}

\begin{proof}
  By induction on the derivation of $ \mathit{l}  \olessthan^{ \mathcal{P} }  \varepsilon $.
  We proceed by case analysis on the rule applied lastly to this derivation.
  \begin{divcases}
    \item[\rname{LITE}{Empty}] Clearly.

    \item[\rname{LITE}{Handling}]
    This case cannot happen.
    If this case happens, we have $   \lift{ \mathit{l} \,  \bm{ { S_{{\mathrm{0}}} } } ^ {  \mathit{I_{{\mathrm{0}}}}  }  }   \mathop{ \odot }  \varepsilon'    \sim   \varepsilon $ for some $\varepsilon'$ and $ \bm{ { S_{{\mathrm{0}}} } } ^ {  \mathit{I_{{\mathrm{0}}}}  } $.
    Thus, we have $   \lift{ \mathit{l} \,  \bm{ { S_{{\mathrm{0}}} } } ^ {  \mathit{I_{{\mathrm{0}}}}  }  }   \mathop{ \odot }  \varepsilon'    \sim    \bbZero  $ by $ \varepsilon   \sim    \bbZero  $.
    However, it is contradictory with safety condition \ref{def:safe_cond:label_notemp}.

    \item[\rname{LITE}{NoHandling}]
    This case cannot happen.
    If this case happens, we have $   \lift{ \ottnt{L} }   \mathop{ \odot }  \varepsilon'    \sim   \varepsilon $ for some $\ottnt{L}$ and $\varepsilon'$.
    Thus, we have $   \lift{ \ottnt{L} }   \mathop{ \odot }  \varepsilon'    \sim    \bbZero  $ by $ \varepsilon   \sim    \bbZero  $.
    However, it is contradictory with safety condition \ref{def:safe_cond:label_notemp}.
  \end{divcases}
\end{proof}

\begin{lemma}[Effect Safety]\label{lem:effsafe_lift_erasure}
  If $\emptyset  \vdash  \ottnt{E}  \ottsym{[}   \mathsf{op} _{ \mathit{l} \,  \bm{ { S } } ^ {  \mathit{I}  }  }  \,  \bm{ { T } } ^ {  \mathit{J}  }  \, \ottnt{v}  \ottsym{]}  \ottsym{:}  \ottnt{A}  \mid  \varepsilon$ and
  $ \mathit{n}  \mathrm{-free} ( \mathit{l} \,  \bm{ { S } } ^ {  \mathit{I}  }  ,  \ottnt{E} ) $,
  then $ \varepsilon   \nsim    \bbZero  $.
\end{lemma}

\begin{proof}
  Assume that $ \varepsilon   \sim    \bbZero  $.
  By Lemma~\ref{lem:handling_order} and Lemma~\ref{lem:label_inclusion_mono_erasure},
  we have $ \mathit{l}  \olessthan^{   \bm{ { S_{{\mathrm{1}}} } } ^ {  \mathit{I_{{\mathrm{1}}}}  }   \blacktriangleright \cdots \blacktriangleright   \bm{ { S_{\ottmv{n}} } } ^ {  \mathit{I_{\ottmv{n}}}  }   \ottsym{,}   \bm{ { S } } ^ {  \mathit{I}  }   \blacktriangleright   \bullet   }  \varepsilon $.
  However, it is contradictory with Lemma~\ref{lem:no_inclusion_erasure}.
\end{proof}

\begin{theorem}[Type and Effect Safety]\label{thm:safety_lift_erasure}
  If $\emptyset  \vdash  \ottnt{e}  \ottsym{:}  \ottnt{A}  \mid   \bbZero $ and $\ottnt{e}  \longrightarrow  ^ * \ottnt{e'}$ and $\ottnt{e'} \centernot \longrightarrow $, then $\ottnt{e'}$ is a value.
\end{theorem}

\begin{proof}
  Similarly to Theorem~\ref{thm:safety_lift};
  Lemmas~\ref{lem:preservation_lift_erasure}
  , \ref{lem:effsafe_lift_erasure},
  and \ref{lem:progress_lift_erasure} are used
  instead of Lemmas~\ref{lem:preservation_lift}
  , \ref{lem:effsafe_lift},
  and \ref{lem:progress_lift}, respectively.
\end{proof}
\documentclass[comment]{paper}

\usepackage{xr}
\externaldocument{defns}

\begin{document}

\subsection{Safety Conditions about Instances}

\begin{lemma}\label{lem:effset_aux}
  In Example~\ref{exa:effset}, we write $a$ and $b$ to denote $\{  \} \tor \rho \tor \{  \ottnt{L}  \}$.
  If $  a_{{\mathrm{1}}}  \,\underline{ \cup }\, \cdots \,\underline{ \cup }\,  a_{\ottmv{m}}     \sim_{\eanameSet}     b_{{\mathrm{1}}}  \,\underline{ \cup }\, \cdots \,\underline{ \cup }\,  b_{\ottmv{n}}  $, then
  \begin{itemize}
    \item for any $i \in \{1, \ldots, m\}$,
          $a_{\ottmv{i}} = \{  \}$ or there exists some $\ottmv{j}$ such that $a_{\ottmv{i}} = b_{\ottmv{j}}$, and
    \item for any $j \in \{1, \ldots, n\}$,
          $b_{\ottmv{j}} = \{  \}$ or there exists some $\ottmv{i}$ such that $a_{\ottmv{i}} = b_{\ottmv{j}}$.
  \end{itemize}
\end{lemma}

\begin{proof}
  By induction on the derivation of $  a_{{\mathrm{1}}}  \,\underline{ \cup }\, \cdots \,\underline{ \cup }\,  a_{\ottmv{m}}     \sim_{\eanameSet}     b_{{\mathrm{1}}}  \,\underline{ \cup }\, \cdots \,\underline{ \cup }\,  b_{\ottmv{n}}  $.
\end{proof}

\begin{theorem}\label{thm:effset_safe}
  Example~\ref{exa:effset} meets safety conditions.
\end{theorem}

\begin{proof}
  \phantom{}
  \begin{itemize}
    \item[\ref{def:safe_cond:label_notemp}]
          Clearly by Lemma~\ref{lem:effset_aux}.
    \item[\ref{def:safe_cond:pres}]
          Clearly by Lemma~\ref{lem:effset_aux}.
  \end{itemize}
\end{proof}

\begin{lemma}\label{lem:effmultiset_aux}
  In Example~\ref{exa:effmultiset},
  we write $a$ and $b$ to denote $\{  \} \tor \rho \tor \{  \ottnt{L}  \}$.
  If $  a_{{\mathrm{1}}}  \,\underline{ \sqcup }\, \cdots \,\underline{ \sqcup }\,  a_{\ottmv{m}}     \sim_{\eanameMSet}     b_{{\mathrm{1}}}  \,\underline{ \sqcup }\, \cdots \,\underline{ \sqcup }\,  b_{\ottmv{n}}  $, then
  \begin{itemize}
    \item for any $a$ such that $a \neq \{  \}$,
          the number of $a_{\ottmv{i}}$ such that $a_{\ottmv{i}} = a$ is equal to
          the number of $b_{\ottmv{j}}$ such that $b_{\ottmv{i}} = a$.
  \end{itemize}
\end{lemma}

\begin{proof}
  By induction on the derivation of $  a_{{\mathrm{1}}}  \,\underline{ \sqcup }\, \cdots \,\underline{ \sqcup }\,  a_{\ottmv{m}}     \sim_{\eanameMSet}     b_{{\mathrm{1}}}  \,\underline{ \sqcup }\, \cdots \,\underline{ \sqcup }\,  b_{\ottmv{n}}  $.
\end{proof}

\begin{theorem}\label{thm:effmultiset_safe}
  Example~\ref{exa:effmultiset} meets safety conditions (for lift coercions).
\end{theorem}

\begin{proof}
  \phantom{}
  \begin{itemize}
    \item[\ref{def:safe_cond:label_notemp}]
          Clearly by Lemma~\ref{lem:effmultiset_aux}.
    \item[\ref{def:safe_cond:pres}]
          Clearly by Lemma~\ref{lem:effmultiset_aux}.
    \item[\ref{def:safe_cond_lift:removal}]
          Clearly by Lemma~\ref{lem:effmultiset_aux}.
  \end{itemize}
\end{proof}

\begin{lemma}\label{lem:eff_simplerow_aux}
  In Example~\ref{exa:eff_simple_row}, we write $a$ and $b$ to denote $\langle  \rangle \tor \rho$.
  If $ \langle  \ottnt{L_{{\mathrm{1}}}}  \mid  \langle  \cdots  \langle  \ottnt{L_{\ottmv{m}}}  \mid  a  \rangle  \cdots  \rangle  \rangle    \sim_{\eanameSimpleRow}    \langle  \ottnt{L'_{{\mathrm{1}}}}  \mid  \langle  \cdots  \langle  \ottnt{L'_{\ottmv{m}}}  \mid  b  \rangle  \cdots  \rangle  \rangle $, then
  \begin{itemize}
    \item $a = b$,
    \item for any $i \in \{1, \ldots, m\}$, there exists some $j$ such that $\ottnt{L_{\ottmv{i}}} = \ottnt{L'_{\ottmv{j}}}$, and
    \item for any $j \in \{1, \ldots, n\}$, there exists some $i$ such that $\ottnt{L_{\ottmv{i}}} = \ottnt{L'_{\ottmv{j}}}$.
  \end{itemize}
\end{lemma}

\begin{proof}
  By induction on the derivation of
  $ \langle  \ottnt{L_{{\mathrm{1}}}}  \mid  \langle  \cdots  \langle  \ottnt{L_{\ottmv{m}}}  \mid  a  \rangle  \cdots  \rangle  \rangle    \sim_{\eanameSimpleRow}    \langle  \ottnt{L'_{{\mathrm{1}}}}  \mid  \langle  \cdots  \langle  \ottnt{L'_{\ottmv{m}}}  \mid  b  \rangle  \cdots  \rangle  \rangle $.
\end{proof}

\begin{theorem}\label{thm:eff_simplerow_safe}
  Example~\ref{exa:eff_simple_row} meets safety conditions.
\end{theorem}

\begin{proof}
  \phantom{}
  \begin{itemize}
    \item[\ref{def:safe_cond:label_notemp}]
          Clearly by Lemma~\ref{lem:eff_simplerow_aux}.
    \item[\ref{def:safe_cond:pres}]
          Clearly by Lemma~\ref{lem:eff_simplerow_aux}.
  \end{itemize}
\end{proof}

\begin{lemma}\label{lem:effrow_aux}
  In Example~\ref{exa:effrow},
  we write $a$ and $b$ to denote $\langle  \rangle \tor \rho$.
  If $ \langle  \ottnt{L_{{\mathrm{1}}}}  \mid  \langle  \cdots  \langle  \ottnt{L_{\ottmv{m}}}  \mid  a  \rangle  \cdots  \rangle  \rangle    \sim_{\eanameScopedRow}    \langle  \ottnt{L'_{{\mathrm{1}}}}  \mid  \langle  \cdots  \langle  \ottnt{L'_{\ottmv{m}}}  \mid  b  \rangle  \cdots  \rangle  \rangle $, then
  \begin{itemize}
    \item $a = b$ and
    \item for any $\ottnt{L}$,
          the number of $\ottnt{L_{\ottmv{i}}}$ such that $\ottnt{L_{\ottmv{i}}} = \ottnt{L}$ is equal to
          the number of $\ottnt{L'_{\ottmv{j}}}$ such that $\ottnt{L'_{\ottmv{j}}} = \ottnt{L}$.
  \end{itemize}
\end{lemma}

\begin{proof}
  By induction on the derivation of
  $ \langle  \ottnt{L_{{\mathrm{1}}}}  \mid  \langle  \cdots  \langle  \ottnt{L_{\ottmv{m}}}  \mid  a  \rangle  \cdots  \rangle  \rangle    \sim_{\eanameScopedRow}    \langle  \ottnt{L'_{{\mathrm{1}}}}  \mid  \langle  \cdots  \langle  \ottnt{L'_{\ottmv{m}}}  \mid  b  \rangle  \cdots  \rangle  \rangle $.
\end{proof}

\begin{theorem}\label{thm:effrow_safe}
  Example~\ref{exa:effrow} meets safety conditions (for lift coercions).
\end{theorem}

\begin{proof}
  \item[\ref{def:safe_cond:label_notemp}]
  Clearly by Lemma~\ref{lem:effrow_aux}.
  \item[\ref{def:safe_cond:pres}]
  Clearly by Lemma~\ref{lem:effrow_aux}.
  \item[\ref{def:safe_cond_lift:removal}]
  Clearly by Lemma~\ref{lem:effrow_aux}.
\end{proof}

\begin{lemma}\label{lem:effset_erasure_aux}
  In Example~\ref{exa:effset_erasure}, we write $a$ and $b$ to denote $\{  \} \tor \rho \tor \{  \ottnt{L}  \}$.
  If $  a_{{\mathrm{1}}}  \,\underline{ \cup }\, \cdots \,\underline{ \cup }\,  a_{\ottmv{m}}     \sim_\mathrm{ESet}     b_{{\mathrm{1}}}  \,\underline{ \cup }\, \cdots \,\underline{ \cup }\,  b_{\ottmv{n}}  $, then
  \begin{itemize}
    \item for any $i \in \{1, \ldots, m\}$,
          $a_{\ottmv{i}} = \{  \}$ or there exists some $\ottmv{j}$ such that $a_{\ottmv{i}} = b_{\ottmv{j}}$ or label names of them are the same, and
    \item for any $j \in \{1, \ldots, n\}$,
          $b_{\ottmv{j}} = \{  \}$ or there exists some $\ottmv{i}$ such that $a_{\ottmv{i}} = b_{\ottmv{j}}$ or label names of them are the same.
  \end{itemize}
\end{lemma}

\begin{proof}
  By induction on the derivation of $  a_{{\mathrm{1}}}  \,\underline{ \cup }\, \cdots \,\underline{ \cup }\,  a_{\ottmv{m}}     \sim_\mathrm{ESet}     b_{{\mathrm{1}}}  \,\underline{ \cup }\, \cdots \,\underline{ \cup }\,  b_{\ottmv{n}}  $.
\end{proof}

\begin{lemma}\label{lem:effset_erasure_fo_label}
  In Example~\ref{exa:effset_erasure}, we define the function $\mathit{FO}$ as follows:
  \begin{mathpar}
    \mathit{FO}(\mathit{l}, \{  \}) = \bot

    \mathit{FO}(\mathit{l}, \{  \iota  \}) = \bot

    \mathit{FO}(\mathit{l}, \rho) = \bot

    \mathit{FO}(\mathit{l}, \{  \mathit{l} \,  \bm{ { S } } ^ {  \mathit{I}  }   \}) =  \bm{ { S } } ^ {  \mathit{I}  } 

    \mathit{FO}(\mathit{l}, \{  \mathit{l'} \,  \bm{ { S } } ^ {  \mathit{I}  }   \}) = \bot \quad (\twhere \mathit{l} \neq \mathit{l'})

    \mathit{FO}(\mathit{l},  \varepsilon_{{\mathrm{1}}}  \,\underline{ \cup }\,  \varepsilon_{{\mathrm{2}}} ) = \begin{dcases}
      \mathit{FO}(\mathit{l}, \varepsilon_{{\mathrm{2}}}) & (\tif \mathit{FO}(\mathit{l}, \varepsilon_{{\mathrm{1}}}) = \bot) \\
      \mathit{FO}(\mathit{l}, \varepsilon_{{\mathrm{1}}}) & (\mathit{otherwise})
    \end{dcases}
  \end{mathpar}
  If $ \varepsilon_{{\mathrm{1}}}    \sim_\mathrm{ESet}    \varepsilon_{{\mathrm{2}}} $, then for any $\mathit{l}$, $\mathit{FO}(\mathit{l}, \varepsilon_{{\mathrm{1}}}) = \mathit{FO}(\mathit{l}, \varepsilon_{{\mathrm{2}}})$.
\end{lemma}

\begin{proof}
  By induction on the derivation of $ \varepsilon_{{\mathrm{1}}}    \sim_\mathrm{ESet}    \varepsilon_{{\mathrm{2}}} $.
\end{proof}

\begin{theorem}\label{thm:effset_erasure_safe}
  Example~\ref{exa:effset_erasure} meets safety conditions.
\end{theorem}

\begin{proof}
  \phantom{}
  \begin{itemize}
    \item[\ref{def:safe_cond:label_notemp}]
          Clearly by Lemma~\ref{lem:effset_erasure_aux}.
    \item[\ref{def:safe_cond:pres}]
          Clearly by Lemma~\ref{lem:effset_erasure_aux} and \ref{lem:effset_erasure_fo_label}.
    \item[\ref{def:safe_cond_erasure:uniq}]
          Clearly by Lemma~\ref{lem:effset_erasure_fo_label}.
  \end{itemize}
\end{proof}

\begin{lemma}\label{lem:effmultiset_erasure_aux}
  In Example~\ref{exa:effmultiset_erasure}, we write $a$ and $b$ to denote $\{  \} \tor \rho \tor \{  \ottnt{L}  \}$.
  If $  a_{{\mathrm{1}}}  \,\underline{ \sqcup }\, \cdots \,\underline{ \sqcup }\,  a_{\ottmv{m}}     \sim_\mathrm{EMSet}     b_{{\mathrm{1}}}  \,\underline{ \sqcup }\, \cdots \,\underline{ \sqcup }\,  b_{\ottmv{n}}  $, then
  \begin{itemize}
    \item for any $a$ such that $a \neq \{  \}$,
          the number of $a_{\ottmv{i}}$ such that $a_{\ottmv{i}} = a$ is equal to
          the number of $b_{\ottmv{j}}$ such that $b_{\ottmv{i}} = a$.
  \end{itemize}
\end{lemma}

\begin{proof}
  By induction on the derivation of $  a_{{\mathrm{1}}}  \,\underline{ \sqcup }\, \cdots \,\underline{ \sqcup }\,  a_{\ottmv{m}}     \sim_\mathrm{EMSet}     b_{{\mathrm{1}}}  \,\underline{ \sqcup }\, \cdots \,\underline{ \sqcup }\,  b_{\ottmv{n}}  $.
\end{proof}

\begin{lemma}\label{lem:effmultiset_erasure_fo_label}
  In Example~\ref{exa:effmultiset_erasure}, we define the function $\mathit{FO}$ as follows:
  \begin{mathpar}
    \mathit{FO}(\mathit{l}, \{  \}) = \bot

    \mathit{FO}(\mathit{l}, \{  \iota  \}) = \bot

    \mathit{FO}(\mathit{l}, \rho) = \bot

    \mathit{FO}(\mathit{l}, \{  \mathit{l} \,  \bm{ { S } } ^ {  \mathit{I}  }   \}) =  \bm{ { S } } ^ {  \mathit{I}  } 

    \mathit{FO}(\mathit{l}, \{  \mathit{l'} \,  \bm{ { S } } ^ {  \mathit{I}  }   \}) = \bot \quad (\twhere \mathit{l} \neq \mathit{l'})

    \mathit{FO}(\mathit{l},  \varepsilon_{{\mathrm{1}}}  \,\underline{ \sqcup }\,  \varepsilon_{{\mathrm{2}}} ) = \begin{dcases}
      \mathit{FO}(\mathit{l}, \varepsilon_{{\mathrm{2}}}) & (\tif \mathit{FO}(\mathit{l}, \varepsilon_{{\mathrm{1}}}) = \bot) \\
      \mathit{FO}(\mathit{l}, \varepsilon_{{\mathrm{1}}}) & (\mathit{otherwise})
    \end{dcases}
  \end{mathpar}
  If $ \varepsilon_{{\mathrm{1}}}    \sim_\mathrm{EMSet}    \varepsilon_{{\mathrm{2}}} $, then for any $\mathit{l}$, $\mathit{FO}(\mathit{l}, \varepsilon_{{\mathrm{1}}}) = \mathit{FO}(\mathit{l}, \varepsilon_{{\mathrm{2}}})$.
\end{lemma}

\begin{proof}
  By induction on the derivation of $ \varepsilon_{{\mathrm{1}}}    \sim_\mathrm{EMSet}    \varepsilon_{{\mathrm{2}}} $.
\end{proof}

\begin{theorem}\label{thm:effmultiset_erasure_safe}
  Example~\ref{exa:effmultiset_erasure} meets safety conditions.
\end{theorem}

\begin{proof}
  \phantom{}
  \begin{itemize}
    \item[\ref{def:safe_cond:label_notemp}]
          Clearly by Lemma~\ref{lem:effmultiset_erasure_aux}.
    \item[\ref{def:safe_cond:pres}]
          Clearly by Lemma~\ref{lem:effmultiset_erasure_aux}.
    \item[\ref{def:safe_cond_erasure:uniq}]
          Clearly by Lemma~\ref{lem:effmultiset_erasure_fo_label}.
  \end{itemize}
\end{proof}

\begin{lemma}\label{lem:eff_simplerow_erasure_aux}
  In Example~\ref{exa:eff_simple_row_erasure}, we write $a$ and $b$ to denote $\langle  \rangle \tor \rho$.
  If $ \langle  \ottnt{L_{{\mathrm{1}}}}  \mid  \langle  \cdots  \langle  \ottnt{L_{\ottmv{m}}}  \mid  a  \rangle  \cdots  \rangle  \rangle    \sim_\mathrm{ESimpR}    \langle  \ottnt{L'_{{\mathrm{1}}}}  \mid  \langle  \cdots  \langle  \ottnt{L'_{\ottmv{m}}}  \mid  b  \rangle  \cdots  \rangle  \rangle $, then
  \begin{itemize}
    \item $a = b$,
    \item for any $i \in \{1, \ldots, m\}$, there exists some $j$ such that $\ottnt{L_{\ottmv{i}}} = \ottnt{L'_{\ottmv{j}}}$ or label names of them are the same, and
    \item for any $j \in \{1, \ldots, n\}$, there exists some $i$ such that $\ottnt{L_{\ottmv{i}}} = \ottnt{L'_{\ottmv{j}}}$ or label names of them are the same.
  \end{itemize}
\end{lemma}

\begin{proof}
  By induction on the derivation of
  $ \langle  \ottnt{L_{{\mathrm{1}}}}  \mid  \langle  \cdots  \langle  \ottnt{L_{\ottmv{m}}}  \mid  a  \rangle  \cdots  \rangle  \rangle    \sim_\mathrm{ESimpR}    \langle  \ottnt{L'_{{\mathrm{1}}}}  \mid  \langle  \cdots  \langle  \ottnt{L'_{\ottmv{m}}}  \mid  b  \rangle  \cdots  \rangle  \rangle $.
\end{proof}

\begin{lemma}\label{lem:eff_simplerow_fo_label}
  In Example~\ref{exa:eff_simple_row_erasure}, we define the function $\mathit{FO}$ as follows:
  \begin{mathpar}
    \mathit{FO}(\mathit{l}, \langle  \rangle) = \bot

    \mathit{FO}(\mathit{l}, \rho) = \bot

    \mathit{FO}(\mathit{l}, \langle  \mathit{l} \,  \bm{ { S } } ^ {  \mathit{I}  }   \mid  \varepsilon  \rangle) =  \bm{ { S } } ^ {  \mathit{I}  } 

    \mathit{FO}(\mathit{l}, \langle  \mathit{l'} \,  \bm{ { S } } ^ {  \mathit{I}  }   \mid  \varepsilon  \rangle) = \mathit{FO}(\mathit{l}, \varepsilon) \quad (\text{where } \mathit{l} \neq \mathit{l'})

    \mathit{FO}(\mathit{l}, \langle  \iota  \mid  \varepsilon  \rangle) = \mathit{FO}(\mathit{l}, \varepsilon)
  \end{mathpar}
  If $ \varepsilon_{{\mathrm{1}}}    \sim_\mathrm{ESimpR}    \varepsilon_{{\mathrm{2}}} $, then for any $\mathit{l}$, $\mathit{FO}(\mathit{l}, \varepsilon_{{\mathrm{1}}}) = \mathit{FO}(\mathit{l}, \varepsilon_{{\mathrm{2}}})$.
\end{lemma}

\begin{proof}
  By induction on the derivation of $ \varepsilon_{{\mathrm{1}}}    \sim_\mathrm{ESimpR}    \varepsilon_{{\mathrm{2}}} $.
\end{proof}

\begin{theorem}\label{thm:eff_simplerow_erasure_safe}
  Example~\ref{exa:eff_simple_row_erasure} meets safety conditions (for type-erasure).
\end{theorem}

\begin{proof}
  \phantom{}
  \begin{itemize}
    \item[\ref{def:safe_cond:label_notemp}]
          Clearly by Lemma~\ref{lem:eff_simplerow_erasure_aux}.
    \item[\ref{def:safe_cond:pres}]
          Clearly by Lemma~\ref{lem:eff_simplerow_erasure_aux} and \ref{lem:eff_simplerow_fo_label}.
    \item[\ref{def:safe_cond_erasure:uniq}]
          Clearly by Lemma~\ref{lem:eff_simplerow_fo_label}.
  \end{itemize}
\end{proof}

\begin{lemma}\label{lem:effrow_erasure_aux}
  In Example~\ref{exa:effrow_erasure},
  we write $a$ and $b$ to denote $\langle  \rangle \tor \rho$.
  If $ \langle  \ottnt{L_{{\mathrm{1}}}}  \mid  \langle  \cdots  \langle  \ottnt{L_{\ottmv{m}}}  \mid  a  \rangle  \cdots  \rangle  \rangle    \sim_\mathrm{EScpR}    \langle  \ottnt{L'_{{\mathrm{1}}}}  \mid  \langle  \cdots  \langle  \ottnt{L'_{\ottmv{m}}}  \mid  b  \rangle  \cdots  \rangle  \rangle $, then
  \begin{itemize}
    \item $a = b$ and
    \item for any $\ottnt{L}$,
          the number of $\ottnt{L_{\ottmv{i}}}$ such that $\ottnt{L_{\ottmv{i}}} = \ottnt{L}$ is equal to
          the number of $\ottnt{L'_{\ottmv{j}}}$ such that $\ottnt{L'_{\ottmv{j}}} = \ottnt{L}$.
  \end{itemize}
\end{lemma}

\begin{proof}
  By induction on the derivation of
  $ \langle  \ottnt{L_{{\mathrm{1}}}}  \mid  \langle  \cdots  \langle  \ottnt{L_{\ottmv{m}}}  \mid  a  \rangle  \cdots  \rangle  \rangle    \sim_\mathrm{EScpR}    \langle  \ottnt{L'_{{\mathrm{1}}}}  \mid  \langle  \cdots  \langle  \ottnt{L'_{\ottmv{m}}}  \mid  b  \rangle  \cdots  \rangle  \rangle $.
\end{proof}

\begin{lemma}\label{lem:effrow_fo_label}
  In Example~\ref{exa:effrow_erasure}, we define the function $\mathit{FO}$ as follows:
  \begin{mathpar}
    \mathit{FO}(\mathit{l}, \langle  \rangle) = \bot

    \mathit{FO}(\mathit{l}, \rho) = \bot

    \mathit{FO}(\mathit{l}, \langle  \mathit{l} \,  \bm{ { S } } ^ {  \mathit{I}  }   \mid  \varepsilon  \rangle) =  \bm{ { S } } ^ {  \mathit{I}  } 

    \mathit{FO}(\mathit{l}, \langle  \mathit{l'} \,  \bm{ { S } } ^ {  \mathit{I}  }   \mid  \varepsilon  \rangle) = \mathit{FO}(\mathit{l}, \varepsilon) \quad (\text{where } \mathit{l} \neq \mathit{l'})

    \mathit{FO}(\mathit{l}, \langle  \iota  \mid  \varepsilon  \rangle) = \mathit{FO}(\mathit{l}, \varepsilon)
  \end{mathpar}
  If $ \varepsilon_{{\mathrm{1}}}    \sim_\mathrm{EScpR}    \varepsilon_{{\mathrm{2}}} $, then for any $\mathit{l}$, $\mathit{FO}(\mathit{l}, \varepsilon_{{\mathrm{1}}}) = \mathit{FO}(\mathit{l}, \varepsilon_{{\mathrm{2}}})$.
\end{lemma}

\begin{proof}
  By induction on the derivation of $ \varepsilon_{{\mathrm{1}}}    \sim_\mathrm{EScpR}    \varepsilon_{{\mathrm{2}}} $.
\end{proof}

\begin{theorem}
  Example~\ref{exa:effrow_erasure} meets safety conditions (for lift coercions and type-erasure).
\end{theorem}

\begin{proof}
  \phantom{}
  \begin{itemize}
    \item[\ref{def:safe_cond:label_notemp}]
          Clearly by Lemma~\ref{lem:effrow_erasure_aux}.
    \item[\ref{def:safe_cond:pres}]
          Clearly by Lemma~\ref{lem:effrow_erasure_aux}.
    \item[\ref{def:safe_cond_lift:removal}]
          Clearly by Lemma~\ref{lem:effrow_erasure_aux}.
    \item[\ref{def:safe_cond_erasure:uniq}]
          Clearly by Lemma~\ref{lem:effrow_fo_label}.
  \end{itemize}
\end{proof}

\begin{theorem}[Unsafe Effect Algebras with Lift Coercions]
  The effect algebras {\eaSet} and {\eaSimpleRow} do not meet safety condition~\ref{def:safe_cond_lift:removal}.
  Furthermore, there exists an expression such that it is well typed under {\eaSet} and {\eaSimpleRow}, but its evaluation gets stuck.
\end{theorem}

\begin{proof}
  We consider only {\eaSet} here; a similar discussion can be applied to {\eaSimpleRow}.
  Recall that the operation $ \odot $ in {\eaSet} is implemented by the
  set union, so it meets idempotence: $  \{  \ottnt{L}  \}  \,\underline{ \cup }\,  \{  \ottnt{L}  \}    \sim   \{  \ottnt{L}  \} $.
  Furthermore, we can use the empty set as the identity element, so
  $  \{  \ottnt{L}  \}  \,\underline{ \cup }\,  \{  \ottnt{L}  \}    \sim    \{  \ottnt{L}  \}  \,\underline{ \cup }\,  \{  \}  $.
  If safety condition~\ref{def:safe_cond_lift:removal} was met,
  $ \{  \ottnt{L}  \}   \sim   \{  \} $ (where $\{  \ottnt{L}  \}$, $\{  \}$, and $0$ are
  taken as $\varepsilon_{{\mathrm{1}}}$, $\varepsilon_{{\mathrm{2}}}$, and $\ottmv{n}$, respectively, in Definition~\ref{def:safe_cond_lift}).
  However, the equivalence does not hold.

  As a program that is typeable under {\eaSet}, consider $ \mathbf{handle}_{ \mathsf{Exc} }  \,  [   \mathsf{raise} _{ \mathsf{Exc} }  \,  \mathsf{Unit}  \,  ()   ] _{ \mathsf{Exc} }  \, \mathbf{with} \, \ottnt{h}$
  where
  $ \mathsf{Exc}  ::  \ottsym{\{}  \mathsf{raise}  \ottsym{:}    \forall   \alpha  \ottsym{:}   \mathbf{Typ}    \ottsym{.}     \mathsf{Unit}    \Rightarrow   \alpha   \ottsym{\}} $.
  This program can be typechecked under an appropriate assumption as illustrated by the following typing derivation:
  \[
    \hspace{-5em}
    \inferrule* [Right=T\_Handling] {
    \cdots \\
      \{  \mathsf{Exc}  \}  \,\underline{ \cup }\,  \{  \}    \sim   \{  \mathsf{Exc}  \} \\
    \inferrule* [Right=T\_Lift] {
    \emptyset  \vdash   \mathsf{raise} _{ \mathsf{Exc} }  \,  \mathsf{Unit}  \,  ()   \ottsym{:}  \ottnt{A}  \mid  \{  \mathsf{Exc}  \} \\
      \{  \mathsf{Exc}  \}  \,\underline{ \cup }\,  \{  \mathsf{Exc}  \}    \sim   \{  \mathsf{Exc}  \} 
    }{
    \emptyset  \vdash   [   \mathsf{raise} _{ \mathsf{Exc} }  \,  \mathsf{Unit}  \,  ()   ] _{ \mathsf{Exc} }   \ottsym{:}  \ottnt{A}  \mid  \{  \mathsf{Exc}  \}
    }
    }{
    \emptyset  \vdash   \mathbf{handle}_{ \mathsf{Exc} }  \,  [   \mathsf{raise} _{ \mathsf{Exc} }  \,  \mathsf{Unit}  \,  ()   ] _{ \mathsf{Exc} }  \, \mathbf{with} \, \ottnt{h}  \ottsym{:}  \ottnt{B}  \mid  \{  \}
    }
  \]
  However, the call to $\mathsf{raise}$ is not handled because it needs to be handled by the \emph{second} closest effect handler.
\end{proof}

\begin{theorem}[Unsafe Effect Algebras in Type-Erasure Semantics]
  \label{thm:unsafe_instances_erasure}
  The effect algebras {\eaSet}, {\eaMSet}, {\eaSimpleRow}, and {\eaScopedRow} do not meet safety condition~\ref{def:safe_cond_erasure:uniq}.
  Furthermore, there exists an expression that is well typed under these algebras and gets stuck.
\end{theorem}
\begin{proof}
  Here we focus on the effect algebra {\eaSet}, but a similar discussions can be applied to
  the other algebras.
  Recall that $ \odot $ in {\eaSet} is implemented by the union operation for sets, and therefore it is commutative
  (i.e., it allows exchanging labels in a set no matter what label names and what type arguments are in the labels).
  Hence, for example, $  \{  \mathit{l} \,  \mathsf{Int}   \}  \,\underline{ \cup }\,  \{  \mathit{l} \,  \mathsf{Bool}   \}     \sim_{\eanameSet}     \{  \mathit{l} \,  \mathsf{Bool}   \}  \,\underline{ \cup }\,  \{  \mathit{l} \,  \mathsf{Int}   \}  $
  for a label name $\mathit{l}$ taking one type parameter.
  It means that {\eaSet} violates safety condition~\ref{def:safe_cond_erasure:uniq}.

  To give a program that is typeable under {\eaSet} but unsafe in the type-erasure semantics,
  consider the following which uses an effect label $ \mathsf{Writer}  ::    \forall   \alpha  \ottsym{:}   \mathbf{Typ}    \ottsym{.}    \ottsym{\{}  \mathsf{tell}  \ottsym{:}   \alpha   \Rightarrow    \mathsf{Unit}    \ottsym{\}}  $:
  \begin{flalign*}
     &  \mathbf{handle}_{ \mathsf{Writer} \,  \mathsf{Int}  }  \,   \mathbf{handle}_{ \mathsf{Writer} \,  \mathsf{Bool}  }  \,  & \\ &    \quad    \quad        \mathsf{tell} _{ \mathsf{Writer} \,  \mathsf{Int}  }  \,  {}  \, 42    & \\ &    \quad      \, \mathbf{with} \,  \ottsym{\{} \, \mathbf{return} \, \mathit{x}  \mapsto  0  \ottsym{\}}   \uplus   \ottsym{\{}  \mathsf{tell} \,  {}  \, \mathit{p} \, \mathit{k}  \mapsto  \mathbf{if} \, \mathit{p} \, \mathbf{then} \, 0 \, \mathbf{else} \, 42  \ottsym{\}}    & \\ &  \, \mathbf{with} \,  \ottsym{\{} \, \mathbf{return} \, \mathit{x}  \mapsto  \mathit{x}  \ottsym{\}}   \uplus   \ottsym{\{}  \mathsf{tell} \,  {}  \, \mathit{p} \, \mathit{k}  \mapsto  \mathit{p}  \ottsym{\}} 
  \end{flalign*}
  This program is well typed because
  \begin{itemize}
    \item the operation call $ \mathsf{tell} _{ \mathsf{Writer} \,  \mathsf{Int}  }  \,  {}  \, 42$ can have effect $ \{  \mathsf{Writer} \,  \mathsf{Bool}   \}  \,\underline{ \cup }\,  \{  \mathsf{Writer} \,  \mathsf{Int}   \} $ via subeffecting $ \{  \mathsf{Writer} \,  \mathsf{Int}   \}  \olessthan   \{  \mathsf{Writer} \,  \mathsf{Bool}   \}  \,\underline{ \cup }\,  \{  \mathsf{Writer} \,  \mathsf{Int}   \}  $ (which holds because $\mathsf{Writer} \,  \mathsf{Int} $ and $\mathsf{Writer} \,  \mathsf{Bool} $ are exchangeable),
    \item the inner handling expression is well typed and its effect is $\{  \mathsf{Writer} \,  \mathsf{Int}   \}$, and
    \item the outer one is well typed and its effect is $\{  \}$.
  \end{itemize}
  Note that this typing rests on the fact that the inner handler assumes that the argument variable $\mathit{p}$ of its $\mathsf{tell}$ clause will be replaced by Boolean values as indicated by the type argument $ \mathsf{Bool} $ to $\mathsf{Writer}$.
  However, this program reaches the stuck state:
  because the operation call is handled by the innermost handler for the label \emph{name} $\mathsf{Writer}$,
  the inner handler is chosen and then the Boolean parameter
  $\mathit{p}$ of the $\mathsf{tell}$ clause in it will be replaced by integer
  $42$.
\end{proof}

\end{document}

\clearpage
\section{Comparison of Instances and Previous Work}

\subsection{Comparison to \cite{pretnar_introduction_2015}}

We define the targets of comparison:
one is an instance of {\lang} (Example~\ref{exa:effset}), and another is a minor changed language of \cite{pretnar_introduction_2015}.

\renewcommand{\arraystretch}{1.2}

{
  %
  \newcommand{\typA}[1][]{A_{#1}}
  \newcommand{\typB}[1][]{B_{#1}}
  \newcommand{\typFun}[2]{#1 \rightarrow #2}
  \newcommand{\typH}[2]{#1 \Rightarrow #2}
  \newcommand{\compC}[1][]{\underline{C}_{#1}}
  \newcommand{\compD}[1][]{\underline{D}_{#1}}
  \newcommand{\dirt}[1][]{\Delta_{#1}}
  \newcommand{\val}[1][]{v_{#1}}
  \newcommand{\comp}[1][]{c_{#1}}
  \newcommand{\handler}[1]{\mathbf{handler}\, \{#1\}}
  \newcommand{\op}[1][]{\mathsf{op}_{#1}}
  \newcommand{\fun}[2]{\mathbf{fun} \, #1 \mapsto #2}
  \newcommand{\ret}[1]{\mathbf{return} \, #1}
  \newcommand{\opcall}[4]{#1 (#2 ; #3 . #4)}
  \newcommand{\doin}[3]{\mathbf{do} \, #1 \leftarrow #2 \,\mathbf{in}\, #3}
  \newcommand{\withhandle}[2]{\mathbf{with}\, #1 \,\mathbf{handle}\, #2}
  \newcommand{\retc}[2]{\mathbf{return}\, #1 \mapsto #2}
  \newcommand{\opc}[4]{#1 (#2; #3) \mapsto #4}
  \newcommand{\sig}[1][]{\Sigma_{#1}}
  \newcommand{\opsig}[3]{#1 : #2 \rightarrow #3}
  \newcommand{\empG}{\emptyset}
  \newcommand{\ctx}[1][]{\Gamma_{#1}}
  \newcommand{\bind}[2]{#1 : #2}
  %
  \newcommand{\ctxwf}[1]{\vdash #1}
  \newcommand{\kinding}[2]{#1 \vdash #2}
  \newcommand{\typing}[3]{#1 \vdash #2 : #3}

  \begin{definition}[Minor Changed Version of \cite{pretnar_introduction_2015}]
    Change list:
    \begin{itemize}
      \item removing Boolean and if expressions,
      \item removing handlers from values and handler types from types,
      \item adding well-formedness of contexts and type, and
      \item adding well-formedness of dirt to the return rule.
    \end{itemize}

    The syntax of a minor changed version of \cite{pretnar_introduction_2015} is as follows.
    \[
      \begin{array}{rclr}
        \typA, \typB   & \Coloneqq
                       & \typFun{\typA}{\compC}
                       & \text{(value types)}                                                                                            \\
        \compC, \compD & \Coloneqq
                       & \typA ! \dirt
                       & \text{(computation types)}                                                                                      \\
        \dirt          & \Coloneqq
                       & \{\op[1], \ldots, \op[n]\}
                       & \text{(dirt)}                                                                                                   \\
        \val           & \Coloneqq
                       & x \mid \fun{x}{\comp}
                       & \text{(values)}                                                                                                 \\
        \comp          & \Coloneqq
                       & \ret{\val} \mid \opcall{\op}{\val}{y}{\comp} \mid \doin{x}{\comp[1]}{\comp[2]} \mid \app{\val[1]}{\val[2]} \mid
                       & \text{(computation)}                                                                                            \\
                       &
                       & \withhandle{h}{\comp}
                       &                                                                                                                 \\
        h              & \Coloneqq
                       & \handler{\retc{x}{\comp[r]}, \opc{\op[1]}{x_1}{k_1}{\comp[1]}, \ldots, \opc{\op[n]}{x_n}{k_n}{\comp[n]}}
                       & \text{(handlers)}                                                                                               \\
        \sig           & \Coloneqq
                       & \{\opsig{\op[1]}{\typA[1]}{\typB[1]}, \ldots, \opsig{\op[n]}{\typA[n]}{\typB[n]}\}
                       & \text{(signature)}                                                                                              \\
        \ctx           & \Coloneqq
                       & \empG \mid \ctx, \bind{x}{\typA}
                       & \text{(typing contexts)}
      \end{array}
    \]

    Well-formedness rules consist of the following.\\
    \textnormal{\bfseries Contexts Well-formedness}\tquad\fbox{$\ctxwf{\ctx}$}
    \begin{mathpar}
      \inferrule{ }{\ctxwf{\empG}}\quad\rname{Cp}{Empty}

      \inferrule{x \notin \dom(\ctx) \\ \kinding{\ctx}{\typA}}{\ctxwf{\ctx, \bind{x}{\typA}}}\quad\rname{Cp}{Var}
    \end{mathpar}
    \textnormal{\bfseries Kinding}\tquad\fbox{$\kinding{\ctx}{\typA}$}
    \begin{mathpar}
      \inferrule{
        \kinding{\ctx}{\typA} \\ \dirt \subseteq \dom(\sig) \\ \kinding{\ctx}{\typB}
      }{
        \kinding{\ctx}{\typFun{\typA}{\typB ! \dirt}}
      }\quad\rname{Kp}{Fun}
    \end{mathpar}
    \textnormal{\bfseries Typing}\tquad\fbox{$\typing{\ctx}{\val}{\typA}$}\tquad\fbox{$\typing{\ctx}{\comp}{\compC}$}
    \begin{mathpar}
      \inferrule{
        \ctxwf{\ctx} \\ \bind{x}{\typA} \in \ctx
      }{
        \typing{\ctx}{x}{\typA}
      }\quad\rname{Tp}{Var}

      \inferrule{
        \typing{\ctx, \bind{x}{\typA}}{\comp}{\compC}
      }{
        \typing{\ctx}{\fun{x}{\comp}}{\typFun{\typA}{\compC}}
      }\quad\rname{Tp}{Abs}

      \inferrule{
        \typing{\ctx}{\val}{\typA} \\ \dirt \subseteq \dom(\sig)
      }{
        \typing{\ctx}{\ret{\val}}{\typA ! \dirt}
      }\quad\rname{Tp}{Return}

      \inferrule{
        \typing{\ctx}{\val[1]}{\typFun{\typA}{\compC}} \\ \typing{\ctx}{\val[2]}{\typA}
      }{
        \typing{\ctx}{\app{\val[1]}{\val[2]}}{\compC}
      }\quad\rname{Tp}{App}

      \inferrule{
        \opsig{\op}{\typA}{\typB} \in \sig \\ \typing{\ctx}{\val}{\typA} \\
        \typing{\ctx, \bind{y}{\typB}}{\comp}{\typA[0] ! \dirt} \\ \op \in \dirt
      }{
        \typing{\ctx}{\opcall{\op}{\val}{y}{\comp}}{\typA[0] ! \dirt}
      }\quad\rname{Tp}{OpApp}

      \inferrule{
        \typing{\ctx}{\comp[1]}{\typA ! \dirt} \\ \typing{\ctx, \bind{x}{\typA}}{\comp[2]}{\typB ! \dirt}
      }{
        \typing{\ctx}{\doin{x}{\comp[1]}{\comp[2]}}{\typB ! \dirt}
      }\quad\rname{Tp}{Do}

      \inferrule{
        \typing{\ctx}{\comp}{\compC} \\
        \typing{\ctx}{h}{\typH{\compC}{\compD}}
      }{
        \typing{\ctx}{\withhandle{h}{\comp}}{\compD}
      }\quad\rname{Tp}{Handle}
    \end{mathpar}
    \textnormal{\bfseries Handler Typing}\tquad\fbox{$\typing{\ctx}{h}{\typH{\compC}{\compD}}$}
    \begin{mathpar}
      \inferrule{
        \typing{\ctx, \bind{x}{\typA}}{\comp[r]}{\typB ! \dirt'} \\
        \dirt \setminus \{\op[1], \ldots, \op[n]\} \subseteq \dirt' \\\\
        \forall i \in \{1, \ldots, n\} .
        (
        \opsig{\op[i]}{\typA[i]}{\typB[i]} \in \sig \\
        \typing{\ctx, \bind{x_i}{\typA[i]}, \bind{k_i}{\typFun{\typB[i]}{\typB ! \dirt'}}}{\comp[i]}{\typB ! \dirt'}
        )
      }{
        \typing
        {\ctx}
        {\handler{\retc{x}{\comp[r]}, \opc{\op[1]}{x_1}{k_1}{\comp[1]}, \ldots, \opc{\op[n]}{x_n}{k_n}{\comp[n]}}}
        {\typH{\typA ! \dirt}{\typB ! \dirt'}}
      }\quad\rname{Hp}{Handler}
    \end{mathpar}
  \end{definition}

  \newcommand{\ucup}{\,\underline{\cup}\,}
  \newcommand{\ptoi}{\textnormal{\texttt{P2I}}}
  \newcommand{\stos}{\textnormal{\texttt{S2s}}}
  \newcommand{\dtol}{\textnormal{\texttt{d2l}}}
  %
  \newcommand{\typFuni}[3]{#1 \rightarrow_{#2} #3}
  \newcommand{\typHi}[3]{#1 \Rightarrow ^ {#3} #2}
  \newcommand{\expi}[1][]{e_{#1}}
  \newcommand{\lam}[2]{\lambda #1 . #2}
  \newcommand{\funi}[3]{\mathbf{fun}(#1, #2, #3)}
  \newcommand{\letin}[3]{\mathbf{let}\, #1 = #2 \,\mathbf{in}\, #3}
  \newcommand{\opi}[1][]{\mathsf{op}_{#1}}
  \newcommand{\handlewith}[3]{\mathbf{handle}_{#1}\, #2 \,\mathbf{with}\, #3}
  \newcommand{\lDecl}[2]{#1 :: #2}
  \newcommand{\opsigi}[3]{#1 : #2 \Rightarrow #3}
  \newcommand{\retci}[2]{\{\mathbf{return}\, #1 \mapsto #2\}}
  \newcommand{\opci}[4]{\{#1\, #2\, #3 \mapsto #4\}}
  %
  \newcommand{\kTyp}{\mathbf{Typ}}
  \newcommand{\kLab}{\mathbf{Lab}}
  \newcommand{\kEff}{\mathbf{Eff}}
  \newcommand{\eps}[1][]{\varepsilon_{#1}}
  \newcommand{\emptyE}{\{  \}}
  \newcommand{\typingE}[4]{#1 \vdash #2 : #3 \mid #4}
  \newcommand{\typingH}[4]{#1 \vdash_{#2} #3 : #4}

  \begin{definition}[Translation from Pretnars to An Instance]
    We assume that\footnote{
      These assumptions arise from our formalization of labels and operations.
      They are easily removed if we omit labels.}:
    \begin{itemize}
      \item there exists a unique partition of $\sig$,
      \item any dirt is a disjoint union of the partition results of $\sig$, and
      \item target operations of any handlers must be one of the partition results of $\sig$.
    \end{itemize}
    We write $\stos(\sig)$ to denote the set of the partition results of $\sig$.
    We write $\dtol$ to denote the function that assigns unique label $l$ such that $ \mathit{l}   \ottsym{:}    \mathbf{Lab}    \in   \Slabel $ to $s \in \stos(\sig)$.

    We define $\dtol(\dirt)$ as the labels whose label is $\dtol(s)$ where $\dom(s) \subseteq \dirt$ and $s \in \stos(\Sigma)$.
    We define $\dtol (h)$ as $\dtol(s)$ where $h = \handler{\retc{x}{\comp[r]}, \opc{\op[1]}{x_1}{k_1}{\comp[1]}, \ldots, \opc{\op[n]}{x_n}{k_n}{\comp[n]}}$ and $s = \{\opsig{\op[1]}{\typA[1]}{\typB[1]}, \ldots, \opsig{\op[n]}{\typA[n]}{\typB[n]}\}$.

    We define $\ptoi$ as follows.
    \newcommand{\arrayspacing}{\hspace{2em}}
    \\\textnormal{\bfseries Types}\hfill\phantom{}
    \[
      \ptoi(\typFun{\typA}{\typB ! \dirt}) = \typFuni{\ptoi(\typA)}{\ptoi(\dirt)}{\ptoi(\typB)}
    \]
    \textnormal{\bfseries Dirts}\hfill\phantom{}
    \[
      \begin{array}{rcl@{\arrayspacing}rcl}
        \ptoi(\emptyset)            & = & \{\}                                                         &
        \ptoi(\dirt \uplus \dom(s)) & = & \ptoi(\dirt) \cup \{\dtol(s)\} \quad(\tif s \in \stos(\sig))
      \end{array}
    \]
    \textnormal{\bfseries Values}\hfill\phantom{}
    \[
      \begin{array}{rcl@{\arrayspacing}rcl}
        \ptoi(x)                          & = & x                                                             &
        \ptoi(\mathbf{fun}\, x \mapsto c) & = & \funi{f}{x}{\ptoi(\comp)} \quad (\twhere \text{$f$ is fresh})
      \end{array}
    \]
    \textnormal{\bfseries Computations}\hfill\phantom{}
    \[
      \begin{array}{rcl}
        \ptoi(\ret{\val})                   & = & \ptoi(\val)                                                                                \\
        \ptoi(\app{\val[1]}{\val[2]})       & = & \app{\ptoi(\val[1])}{\ptoi(\val[2])}                                                       \\
        \ptoi(\doin{x}{\comp[1]}{\comp[2]}) & = & \letin{x}{\ptoi(\comp[1])}{\ptoi(\comp[2])}                                                \\
        \ptoi(\opcall{\op}{\val}{y}{\comp}) & = & \letin{y}{\app{\opi[\dtol(s)]}{\ptoi(\val)}}{\ptoi(\comp)} \quad (\twhere \op \in \dom(s)) \\
        \ptoi(\withhandle{h}{\comp})        & = & \handlewith{\dtol(h)}{\ptoi(c)}{\ptoi(h)}
      \end{array}
    \]
    \textnormal{\bfseries Handlers}\hfill\phantom{}
    \[
      \begin{array}{rcl}
        \ptoi(h) & = & \retci{x}{\ptoi(\comp[r])} \uplus \opci{\op[1]}{x_1}{k_1}{\ptoi(\comp[1])} \uplus \cdots \uplus \opci{\op[n]}{x_n}{k_n}{\ptoi(\comp[n])} \\
                 &   & (\twhere h = \handler{\retc{x}{\comp[r]}, \opc{\op[1]}{x_1}{k_1}{\comp[1]}, \ldots, \opc{\op[n]}{x_n}{k_n}{\comp[n]}})
      \end{array}
    \]
    \textnormal{\bfseries Effect contexts}\hfill\phantom{}
    \[
      \begin{array}{rcl}
        \ptoi(\sig) & = & \bigcup_{s \in \stos(\sig)} \{\lDecl{\dtol(s)}{\{\opsigi{\op[1]}{\ptoi(\typA[1])}{\ptoi(\typB[1])}, \ldots, \opsigi{\op[n]}{\ptoi(\typA[n])}{\ptoi(\typB[n])} \}} \} \\
                    &   & (\twhere s = \{\opsig{\op[1]}{\typA[1]}{\typB[1]}, \ldots, \opsig{\op[n]}{\typA[n]}{\typB[n]} \})
      \end{array}
    \]
    \textnormal{\bfseries Typing Contexts}\hfill\phantom{}
    \[
      \begin{array}{rcl@{\arrayspacing}rcl}
        \ptoi(\emptyset)             & = & \emptyset                           &
        \ptoi(\ctx, \bind{x}{\typA}) & = & \ptoi(\ctx), \bind{x}{\ptoi(\typA)}
      \end{array}
    \]
  \end{definition}

  \begin{lemma}\label{lem:ctx_dom_ptoi}
    $\dom(\ctx) = \dom(\ptoi(\ctx))$.
  \end{lemma}

  \begin{proof}
    Clearly by definition of $\ptoi$.
  \end{proof}

  \begin{lemma}\label{lem:eff_dom_ptoi}
    If $\dirt \subseteq \dom(\sig)$, then $\typing{\ctx}{\ptoi(\dirt)}{\kEff}$ for any $\ctx$ such that $\ctxwf{\ctx}$.
  \end{lemma}

  \begin{proof}
    By induction on the size of $\ctx$.

    If $\dirt = \emptyset$, then clearly because $\ptoi(\emptyset) = \{\}$.

    If $\dirt = \dirt' \uplus \dom(s)$ for some $\dirt'$ and $s \in \stos(\Sigma)$,
    then $\ptoi(\dirt) = \ptoi(\dirt') \ucup \{\dtol(s)\}$ where $\dtol(s) : \mathbf{Lab} \in \Slabel$.
    Let $\ctx$ be a typing context such that $\ctxwf{\ctx}$.
    By the induction hypothesis, we have $\typing{\ctx}{\ptoi(\dirt')}{\kEff}$.
    Thus, \rname{K}{Cons} derives $\typing{\ctx}{\ptoi(\dirt') \ucup \{\dtol(s)\}}{\kEff}$
    because we have $\typing{\ctx}{\{\dtol(s)\}}{\kEff}$.
  \end{proof}

  \begin{lemma}\label{lem:ctx_var_ptoi}
    If $\ctxwf{\ctx}$ and $\bind{x}{\typA} \in \ctx$, then $\bind{x}{\ptoi(\typA)} \in \ptoi(\ctx)$.
  \end{lemma}

  \begin{proof}
    By structual induction on $\ctx$.

    If $\ctx = \empG$, then $\bind{x}{\typA} \in \ctx$ cannot happen.

    If $\ctx = \ctx', \bind{y}{\typB}$ for some $y$, $\typB$, and $\ctx'$,
    then we have $\ptoi(\ctx) = \ptoi(\ctx'), y : \ptoi(\typB)$.
    In this case, if $x = y$, then we have $\typA = \typB$ and $\bind{y}{\ptoi(\typB)} \in \ptoi(\ctx)$ as required.
    If $x \neq y$, then we have $\bind{x}{\typA} \in \ctx'$.
    By the induction hypothesis, we have $\bind{x}{\ptoi(\typA)} \in \ptoi(\ctx')$.
    Thus, we have $\bind{x}{\ptoi(A)} \in \ptoi(\ctx)$ as required.
  \end{proof}

  \begin{theorem}\label{thm:typability_ptoi}
    \phantom{}\\
    \begin{enumerate}
      \item\label{thm:typability_ptoi_ctx}
            If $\ctxwf{\ctx}$, then $\ctxwf{\ptoi(\ctx)}$.

      \item\label{thm:typability_ptoi_typ}
            If $\kinding{\ctx}{\typA}$, then $\typing{\ptoi(\ctx)}{\ptoi(\typA)}{\kTyp}$.

      \item\label{thm:typability_ptoi_value}
            If $\typing{\ctx}{\val}{\typA}$, then $\typingE{\ptoi(\ctx)}{\ptoi(\val)}{\ptoi(\typA)}{\emptyE}$.

      \item\label{thm:typability_ptoi_exp}
            If $\typing{\ctx}{\comp}{\typA ! \dirt}$, then $\typingE{\ptoi(\ctx)}{\ptoi(\comp)}{\ptoi(\typA)}{\ptoi(\dirt)}$.
      \item\label{thm:typability_ptoi_handler}
            If $\typing{\ctx}{h}{\typH{\typA ! \dirt}{\typB ! \dirt'}}$,
            then $\ptoi(\dirt) \ucup \varepsilon  \sim_{\eanameSet}  \dtol(h) \ucup \ptoi(\dirt')$ for some $\varepsilon$ and
            there exists some $\sigma$ such that
            $\typingH{\ptoi(\ctx)}{\sigma}{\ptoi(h)}{\typHi{\ptoi(\typA)}{\ptoi(\typB)}{\ptoi(\dirt')}}$ and
            $\lDecl{\dtol(h)}{\sigma} \in \ptoi(\sig)$.
    \end{enumerate}
  \end{theorem}

  \begin{proof}
    \begin{itemize}
      \item[(1)(2)]
            By mutual induction on derivations of the judgments.
            We proceed by case analysis on the rule applied lastly to the derivation.
            \begin{divcases}
              \item[\rname{Cp}{Empty}] Clearly.

              \item[\rname{Cp}{Var}]
              We have
              \begin{itemize}
                \item $\ctx = \ctx', \bind{x}{\typA}$,
                \item $x \notin \dom(\ctx')$, and
                \item $\kinding{\ctx'}{\typA}$,
              \end{itemize}
              for some $x$, $\typA$, and $\ctx'$.
              By the induction hypothesis and Lemma~\ref{lem:ctx_dom_ptoi},
              we have $x \notin \dom(\ptoi(\ctx'))$ and $\typing{\ptoi(\ctx')}{\ptoi(\typA)}{\kTyp}$.
              Thus, \rname{C}{Var} derives $\ctxwf{\ptoi(\ctx'), \bind{x}{\ptoi(\typA)}}$ as required.

              \item[\rname{Kp}{Fun}]
              We have
              \begin{itemize}
                \item $\typA = \typFun{\typA[1]}{\typB[1] ! \dirt}$,
                \item $\kinding{\ctx}{\typA[1]}$,
                \item $\dirt \subseteq \dom(\sig)$, and
                \item $\kinding{\ctx}{\typB[1]}$,
              \end{itemize}
              for some $\typA[1]$, $\typB[1]$, and $\dirt$.
              By the induction hypothesis and Lemma~\ref{lem:eff_dom_ptoi},
              we have
              \begin{itemize}
                \item $\typing{\ptoi(\ctx)}{\ptoi(\typA[1])}{\kTyp}$,
                \item $\typing{\ptoi(\ctx)}{\ptoi(\dirt)}{\kEff}$, and
                \item $\typing{\ptoi(\ctx)}{\ptoi(\typB[1])}{\kTyp}$.
              \end{itemize}
              Thus, \rname{K}{Fun} derives
              \[\typing{\ptoi(\ctx)}{\typFuni{\ptoi(\typA[1])}{\ptoi(\dirt)}{\ptoi(\typB[1])}}{\kTyp}\]
              as required.
            \end{divcases}

      \item[(3)(4)(5)]
            By mutual induction on derivations of the judgments.
            We proceed by case analysis on the rule applied lastly to the derivation.
            \begin{divcases}
              \item[\rname{Tp}{Var}]
              We have
              \begin{itemize}
                \item $\val = x$,
                \item $\ctxwf{\ctx}$, and
                \item $\bind{x}{\typA} \in \ctx$,
              \end{itemize}
              for some $x$.
              By Lemma~\ref{lem:ctx_var_ptoi} and Theorem~\ref{thm:typability_ptoi}\ref{thm:typability_ptoi_ctx},
              we have
              \begin{itemize}
                \item $\ctxwf{\ptoi(\ctx)}$ and
                \item $\bind{x}{\ptoi(\typA)} \in \ptoi(\ctx)$.
              \end{itemize}
              Thus, \rname{T}{Var} derives
              \[\typingE{\ptoi(\ctx)}{x}{\ptoi(A)}{\emptyE}\] as required.

              \item[\rname{Tp}{Abs}]
              We have
              \begin{itemize}
                \item $\val = \fun{x}{\comp}$ and
                \item $\typing{\ctx, \bind{x}{\typA}}{\comp}{\typB ! \dirt}$
              \end{itemize}
              for some $x$, $\comp$, $\typA$, $\typB$, and $\dirt$.
              By the induction hypothesis, we have
              \[\typingE{\ptoi(\ctx), \bind{x}{\ptoi(\typA)}}{\ptoi(\comp)}{\ptoi(\typB)}{\ptoi(\dirt)}.\]
              Without loss of generality, we can choose $f$ such that
              \begin{itemize}
                \item $f \notin \mathit{FV}(\ptoi(\comp))$,
                \item $f \neq x$,
                \item $f \notin \dom(\ctx)$, and
                \item $\ptoi(\fun{x}{\comp}) = \funi{f}{x}{\ptoi(\comp)}$.
              \end{itemize}
              By Lemma~\ref{lem:wk} and
              Lemma~\ref{lem:delta_context}\ref{lem:delta_context:kinding} and
              Lemma~\ref{lem:delta_weakening}, we have
              \begin{itemize}
                \item $\typing{\ptoi(\ctx)}{\ptoi(\typB)}{\kTyp}$ and
                \item $\typing{\ptoi(\ctx)}{\ptoi(\dirt)}{\kEff}$.
              \end{itemize}
              By Lemma~\ref{lem:ctx-wf-typing}, we have $\ctxwf{\ptoi(\ctx), \bind{x}{\ptoi(\typA)}}$.
              Since only \rname{C}{Var} can derive $\ctxwf{\ptoi(\ctx), \bind{x}{\ptoi(\typA)}}$,
              we have $\typing{\ptoi(\ctx)}{\ptoi(\typA)}{\kTyp}$.
              Thus, \rname{C}{Var} derives
              \[\ctxwf{\ptoi(\ctx), \bind{f}{\typFuni{\ptoi(\typA)}{\ptoi(\dirt)}{\ptoi(\typB)}}}.\]
              Thus, Lemma~\ref{lem:weakening} and \rname{T}{Abs} derives
              \[\typingE{\ptoi(\ctx)}{\funi{f}{x}{\ptoi(\comp)}}{\typFuni{\ptoi(\typA)}{\ptoi(\dirt)}{\ptoi(\typB)}}{\emptyE}\]
              as required.

              \item[\rname{Tp}{Return}]
              We have
              \begin{itemize}
                \item $\comp = \ret{\val}$,
                \item $\typing{\ctx}{\val}{\typA}$, and
                \item $\dirt \subseteq \dom(\sig)$,
              \end{itemize}
              for some $v$.
              By the induction hypothesis and Lemma~\ref{lem:eff_dom_ptoi}, we have
              \begin{itemize}
                \item $\typingE{\ptoi(\ctx)}{\ptoi(\val)}{\ptoi(A)}{\emptyE}$ and
                \item $\typing{\ptoi(\ctx)}{\ptoi(\dirt)}{\kEff}$.
              \end{itemize}
              Thus, \rname{T}{Sub} derives
              \[\typingE{\ptoi(\ctx)}{\ptoi(\val)}{\ptoi(A)}{\ptoi(\dirt)}\] as required.

              \item[\rname{Tp}{App}]
              We have
              \begin{itemize}
                \item $\comp = \app{\val[1]}{\val[2]}$,
                \item $\typing{\ctx}{\val[1]}{\typFun{\typB}{\typA ! \dirt}}$, and
                \item $\typing{\ctx}{\val[2]}{\typB}$,
              \end{itemize}
              for some $\val[1]$, $\val[2]$, and $\typB$.
              By the induction hypothesis, we have
              \begin{itemize}
                \item $\typingE{\ptoi(\ctx)}{\ptoi(\val[1])}{\typFuni{\ptoi(\typB)}{\ptoi(\dirt)}{\ptoi(\typA)}}{\emptyE}$ and
                \item $\typingE{\ptoi(\ctx)}{\ptoi(\val[2])}{\ptoi(\typB)}{\emptyE}$.
              \end{itemize}
              Thus, \rname{T}{App} derives
              \[\typingE{\ptoi(\ctx)}{\app{\ptoi(\val[1])}{\ptoi(\val[2])}}{\ptoi(\typA)}{\ptoi(\dirt)}\] as required.

              \item[\rname{Tp}{OpApp}]
              We have
              \begin{itemize}
                \item $\comp = \opcall{\op}{\val}{y}{\comp'}$,
                \item $\opsig{\op}{\typA'}{\typB'} \in \sig$,
                \item $\typing{\ctx}{\val}{\typA'}$,
                \item $\typing{\ctx, \bind{y}{\typB'}}{\comp'}{\typA ! \dirt}$, and
                \item $\op \in \dirt$,
              \end{itemize}
              for some $\op$, $\val$, $y$, $\comp'$, $\typA'$, $\typB'$, and $\dirt$.
              By $\op \in \dirt$,
              there uniquely exists some $s$ such that
              \begin{itemize}
                \item $s \in \stos(\sig)$,
                \item $\opsig{\op}{\typA'}{\typB'} \in s$,
                \item $\dom(s) \subseteq \dirt$.
              \end{itemize}
              Thus, we have
              \begin{itemize}
                \item $l :: \sigma \in \ptoi(\sig)$,
                \item $\opsigi{\op}{\ptoi(\typA')}{\ptoi(\typB')} \in \sigma$, and
                \item $\{\dtol(s)\} \ucup \varepsilon  \sim_{\eanameSet}  \ptoi(\dirt)$,
              \end{itemize}
              for some $\sigma$.
              By the induction hypothesis, we have
              \begin{itemize}
                \item $\typingE{\ptoi(\ctx)}{\ptoi(\val)}{\ptoi(\typA')}{\emptyE}$ and
                \item $\typingE{\ptoi(\ctx), \bind{y}{\ptoi(\typB')}}{\ptoi(\comp')}{\ptoi(\typA)}{\ptoi(\dirt)}$.
              \end{itemize}
              Thus, \rname{T}{Op} and \rname{T}{App} derives
              \[\typingE{\ptoi(\ctx)}{\app{\opi[\dtol(s)]}{\ptoi(\val)}}{\ptoi(\typB')}{\{\dtol(s)\}}.\]
              Thus, \rname{T}{Sub} and \rname{T}{Let} derives
              \[\typingE{\ptoi(\ctx)}{\letin{y}{\app{\opi[\dtol(s)]}{\ptoi(\val)}}{\ptoi(\comp')}}{\ptoi(A)}{\ptoi(\dirt)}\]
              as required.

              \item[\rname{Tp}{Do}]
              We have
              \begin{itemize}
                \item $\comp = \doin{x}{\comp[1]}{\comp[2]}$,
                \item $\typing{\ctx}{\comp[1]}{\typB ! \dirt}$, and
                \item $\typing{\ctx, \bind{x}{\typB}}{\comp[2]}{\typA ! \dirt}$,
              \end{itemize}
              for some $x$, $\comp[1]$, $\comp[2]$, and $\dirt$.
              By the induction hypothesis, we have
              \begin{itemize}
                \item $\typingE{\ptoi(\ctx)}{\ptoi(\comp[1])}{\ptoi(\typB)}{\ptoi(\dirt)}$ and
                \item $\typingE{\ptoi(\ctx), \bind{x}{\ptoi(\typB)}}{\ptoi(\comp[2])}{\ptoi(\typA)}{\ptoi(\dirt)}$.
              \end{itemize}
              Thus, \rname{T}{Let} derives
              \[\typingE{\ptoi(\ctx)}{\letin{x}{\ptoi(\comp[1])}{\ptoi(\comp[2])}}{\ptoi(\typA)}{\ptoi(\dirt)}\]
              as required.

              \item[\rname{Tp}{Handle}]
              We have
              \begin{itemize}
                \item $\comp = \withhandle{h}{\comp'}$,
                \item $\typing{\ctx}{\comp'}{\typA' ! \dirt'}$, and
                \item $\typing{\ctx}{h}{\typH{\typA' ! \dirt'}{\typA ! \dirt}}$.
              \end{itemize}
              for some $\comp'$, $h$, $\typA'$, and $\dirt'$.
              By the induction hypothesis, we have
              \begin{itemize}
                \item $\typingE{\ptoi(\ctx)}{\ptoi(\comp')}{\ptoi(\typA')}{\ptoi(\dirt')}$,
                \item $\lDecl{\dtol(h)}{\sigma} \in \ptoi(\sig)$,
                \item $\typingH{\ptoi(\ctx)}{\sigma}{\ptoi(h)}{\typHi{\ptoi(\typA')}{\ptoi(\typA)}{\ptoi(\dirt)}}$, and
                \item $\ptoi(\dirt') \ucup \varepsilon  \sim_{\eanameSet}  \dtol(h) \ucup \ptoi(\dirt)$,
              \end{itemize}
              for some $\varepsilon$ and $\sigma$.
              Thus, \rname{T}{Sub} and \rname{T}{Handling} derive
              \[\typingE{\ptoi(\ctx)}{\handlewith{\ptoi(h)}{\ptoi(\comp')}{\ptoi(h)}}{\ptoi(\typA)}{\ptoi(\dirt)}\]
              as required.

              \item[\rname{Hp}{Handler}]
              We have
              \begin{itemize}
                \item $h = \handler{\retc{x}{\comp[r]}, \opc{\op[1]}{x_1}{k_1}{\comp[1]}, \ldots, \opc{\op[n]}{x_n}{k_n}{\comp[n]}}$,
                \item $\typing{\ctx, \bind{x}{\typA}}{\comp[r]}{\typB ! \dirt'}$,
                \item $\opsig{\op[i]}{\typA[i]}{\typB[i]} \in \sig$ for any $i \in \{1, \ldots, n\}$,
                \item $\typing{\ctx, \bind{x_i}{\typA[i]}, \bind{k_i}{\typFun{\typB[i]}{\typB ! \dirt'}}}{\comp[i]}{\typB ! \dirt'}$ for any $i \in \{1, \ldots, n\}$, and
                \item $\dirt \setminus \{\op[1], \ldots, \op[n]\} \subseteq \dirt'$,
              \end{itemize}
              for some $n$, $x$, $\comp[r]$, $\op[i]$, $x_i$, $k_i$, $\comp[i]$, $\typA[i]$, and $\typB[i]$,
              where $i \in \{1, \ldots, n\}$.

              By the assumptions, we have
              \begin{itemize}
                \item $s \in \stos(\sig)$ and
                \item $\dtol(h) = \dtol(s)$
              \end{itemize}
              where $s = \{\opsig{\op[1]}{\typA[1]}{\typB[1]}, \ldots, \opsig{\op[n]}{\typA[n]}{\typB[n]}\}$.
              Thus, we have $\lDecl{\dtol(h)}{\sigma} \in \ptoi(\sig)$
              where $\sigma = \{\opsigi{\op[1]}{\ptoi(\typA[1])}{\ptoi(\typB[1])}, \ldots, \opsigi{\op[n]}{\ptoi(\typA[n])}{\ptoi(\typB[n])}\}$.

              By $\dirt \setminus \{\op[1], \ldots, \op[n]\} \subseteq \dirt'$,
              we have $\dirt \subseteq \dom(s) \cup \dirt'$.
              By the assumptions, we have either $\dom(s) \subseteq \dirt'$ or $\op[i] \notin \dirt'$ for any $i$.
              In any case, we have $\ptoi(\dirt) \ucup \varepsilon  \sim_{\eanameSet}  \dtol(h) \ucup \ptoi(\dirt')$ for some $\varepsilon$.

              By the induction hypothesis, we have
              \begin{itemize}
                \item $\typingE{\ptoi(\ctx), \bind{x}{\ptoi(\typA)}}{\ptoi(\comp[r])}{\ptoi(\typB)}{\ptoi(\dirt')}$ and
                \item $\typingE{\ptoi(\ctx), \bind{x_i}{\ptoi(\typA[i])}, \bind{k_i}{\typFuni{\ptoi(\typB[i])}{\ptoi(\dirt')}{\ptoi(\typB)}}}{\ptoi(\comp[i])}{\ptoi(\typB)}{\ptoi(\dirt')}$ for any $i \in \{1, \ldots, n\}$.
              \end{itemize}
              Therefore, \rname{H}{Return} and \rname{H}{Op} derive
              $\typingH{\ptoi(\ctx)}{\sigma}{\ptoi(h)}{\typHi{\ptoi(\typA)}{\ptoi(\typB)}{\ptoi(\dirt')}}$.

              Thus, the required result is achieved.
            \end{divcases}
    \end{itemize}
  \end{proof}

}
\subsection{Comparison to \cite{hillerstrom_continuation_2017}}

{
  We give the targets of comparison:
  one is an instance of {\lang} (Example~\ref{exa:eff_simple_row}), and
  another is a minorly changed language of \cite{hillerstrom_continuation_2017}.

  \newcommand{\lam}[3]{\lambda #1 ^ {#2} . #3}
  \newcommand{\tlam}[3]{\Lambda #1 ^ {#2} . #3}
  \newcommand{\ret}[1]{\mathbf{return}\, #1}
  \newcommand{\letinh}[3]{\mathbf{let}\, #1 \leftarrow #2 \,\mathbf{in}\, #3}
  \newcommand{\doop}[3]{(\mathbf{do}\, #1\, #2)^{#3}}
  \newcommand{\handlewithh}[2]{\mathbf{handle}\, #1 \,\mathbf{with}\, #2}
  \newcommand{\retc}[2]{\{\mathbf{return}\, #1 \mapsto #2\}}
  \newcommand{\opc}[4]{\{#1\, #2\, #3 \mapsto #4\}}
  \newcommand{\tfunh}[2]{#1 \rightarrow #2}
  \newcommand{\tabsh}[3]{\forall #1 ^ {#2} . #3}
  \newcommand{\pre}[1]{\mathsf{Pre}(#1)}
  \newcommand{\abs}{\mathsf{Abs}}
  \newcommand{\kType}{\mathsf{Type}}
  \newcommand{\kRow}[1][\emptyset]{\mathsf{Row}_{#1}}
  \newcommand{\labelset}{\mathcal{L}}
  \newcommand{\kComp}{\mathsf{Comp}}
  \newcommand{\kEffect}{\mathsf{Effect}}
  \newcommand{\ctx}{\Gamma}
  \newcommand{\kctx}{\Delta}
  \newcommand{\wfctx}[1]{\vdash #1}
  \newcommand{\wkctx}[2]{#1 \vdash #2}
  \newcommand{\kinding}[3]{#1 \vdash #2 : #3}
  \newcommand{\typingh}[4]{#1; #2 \vdash #3 : #4}
  \newcommand{\htypingh}[5]{#1; #2 \vdash #3 : #4 \Rightarrow #5}
  \begin{definition}[Minor Changed Version of \cite{hillerstrom_continuation_2017}]
    Change list:
    \begin{itemize}
      \item removing variants and records,
      \item removing presence and handler types,
      \item removing computation kinds, and
      \item adding well-formedness rules of contexts.
    \end{itemize}

    The syntax of a minor changed version of \cite{hillerstrom_continuation_2017} is as follows.
    \[
      \begin{array}{rclr}
        V, W      & \Coloneqq & x \mid \lam{x}{A}{M} \mid \tlam{\alpha}{K}{M}       & \text{(values)}             \\
        M, N      & \Coloneqq & V\, W \mid V\, T \mid \ret{M} \mid \letinh{x}{M}{N} & \text{(computations)}       \\
                  &           & \mid \doop{l}{V}{E} \mid \handlewithh{M}{H}         &                             \\
        H         & \Coloneqq & \retc{x}{M} \mid H \uplus \opc{l}{p}{r}{M}          & \text{(handlers)}           \\
        A, B      & \Coloneqq & \tfunh{A}{C} \mid \tabsh{\alpha}{K}{C} \mid \alpha  & \text{(value types)}        \\
        C, D      & \Coloneqq & A ! E                                               & \text{(computations types)} \\
        E         & \Coloneqq & \{ R \}                                             & \text{(effect types)}       \\
        R         & \Coloneqq & l : P ; R \mid \rho \mid \cdot                      & \text{(row types)}          \\
        P         & \Coloneqq & \pre{\tfunh{A}{B}} \mid \abs                        & \text{(presence types)}     \\
        T         & \Coloneqq & A \mid C \mid E \mid R                              & \text{(types)}              \\
        K         & \Coloneqq & \kType \mid \kRow[\labelset] \mid \kEffect          & \text{(kinds)}              \\
        \labelset & \Coloneqq & \emptyset \mid \{l\} \uplus \labelset               & \text{(label sets)}         \\
        \ctx      & \Coloneqq & \cdot \mid \ctx, x : A                              & \text{(type environments)}  \\
        \kctx     & \Coloneqq & \cdot \mid \kctx, \alpha : K                        & \text{(kind environments)}
      \end{array}
    \]

    Well-formedness, kinding, and typing rules consist of the following.\\
    \normalfont{\bfseries Kinding Contexts Well-formedness}\tquad\fbox{$\wfctx{\kctx}$}
    \begin{mathpar}
      \inferrule{ }{
        \wfctx{\cdot}
      }\quad\rname{KCh}{Empty}

      \inferrule{
        \wfctx{\kctx} \\ \alpha \notin \dom(\kctx)
      }{
        \wfctx{\kctx, \alpha : K}
      }\quad\rname{KCh}{TVar}
    \end{mathpar}
    \normalfont{\bfseries Contexts Well-formedness}\tquad\fbox{$\wfctx{\kctx; \ctx}$}
    \begin{mathpar}
      \inferrule{
        \wfctx{\kctx}
      }{
        \wkctx{\kctx}{\cdot}
      }\quad\rname{Ch}{Empty}

      \inferrule{
        \wkctx{\kctx}{\ctx} \\ x \notin \dom(\ctx) \\ \kinding{\kctx}{A}{\kType}
      }{
        \wkctx{\kctx}{\ctx, x : A}
      }\quad\rname{Ch}{Var}
    \end{mathpar}
    \normalfont{\bfseries Kinding}\tquad\fbox{$\kinding{\kctx}{T}{K}$}
    \begin{mathpar}
      \inferrule{
        \wfctx{\kctx, \alpha : K}
      }{
        \kinding{\kctx, \alpha : K}{\alpha}{K}
      }\quad\rname{Kh}{Var}

      \inferrule{
        \kinding{\kctx}{A}{\kType} \\
        \kinding{\kctx}{B}{\kType} \\ \kinding{\kctx}{E}{\kEffect}
      }{
        \kinding{\kctx}{\tfunh{A}{B ! E}}{\kType}
      }\quad\rname{Kh}{Fun}

      \inferrule{
        \kinding{\kctx, \alpha : K}{A}{\kType} \\
        \kinding{\kctx, \alpha : K}{E}{\kEffect}
      }{
        \kinding{\kctx}{\tabsh{\alpha}{K}{A ! E}}{\kType}
      }\quad\rname{Kh}{Forall}

      \inferrule{
        \kinding{\kctx}{R}{\kRow}
      }{
        \kinding{\kctx}{\{R\}}{\kEffect}
      }\quad\rname{Kh}{Effect}

      \inferrule{
        \forall i \in \{1, \ldots, n\} .
        (P_i = \abs \tor
        (P_i = \pre{\tfunh{A_i}{B_i}} \tand \kinding{\kctx}{A_i}{\kType} \tand \kinding{\kctx}{B_i}{\kType})) \\
        \wfctx{\kctx}
      }{
        \kinding{\kctx}{l_1 : P_1; \cdots; l_n : P_n; \cdot}{\kRow}
      }\quad\rname{Kh}{CloseRow}

      \inferrule{
        \forall i \in \{1, \ldots, n\} .
        (P_i = \abs \tor
        (P_i = \pre{\tfunh{A_i}{B_i}} \tand \kinding{\kctx}{A_i}{\kType} \tand \kinding{\kctx}{B_i}{\kType})) \\
        \kinding{\kctx}{\rho}{\kRow[\labelset]} \\ \labelset = \{l_1, \ldots, l_n\}
      }{
        \kinding{\kctx}{l_1 : P_1; \cdots; l_n : P_n; \rho}{\kRow}
      }\quad\rname{Kh}{OpenRow}
    \end{mathpar}
    \normalfont{\bfseries Typing}\tquad\fbox{$\typingh{\kctx}{\ctx}{V}{A}$}\tquad\fbox{$\typingh{\kctx}{\ctx}{M}{C}$}
    \begin{mathpar}
      \inferrule{
        \wkctx{\kctx}{\ctx} \\ x : A \in \ctx
      }{
        \typingh{\kctx}{\ctx}{x}{A}
      }\quad\rname{Th}{Var}

      \inferrule{
        \typingh{\kctx}{\ctx, x : A}{M}{C}
      }{
        \typingh{\kctx}{\ctx}{\lam{x}{A}{M}}{\tfunh{A}{C}}
      }\quad\rname{Th}{Lam}

      \inferrule{
        \typingh{\kctx, \alpha : K}{\ctx}{M}{C}
        \\ \wkctx{\kctx}{\ctx}
      }{
        \typingh{\kctx}{\ctx}{\tlam{\alpha}{K}{M}}{\tabsh{\alpha}{K}{C}}
      }\quad\rname{Th}{PolyLam}

      \inferrule{
        \typingh{\kctx}{\ctx}{V}{\tfunh{A}{C}} \\ \typingh{\kctx}{\ctx}{W}{A}
      }{
        \typingh{\kctx}{\ctx}{V\, W}{C}
      }\quad\rname{Th}{App}

      \inferrule{
        \typingh{\kctx}{\ctx}{V}{\tabsh{\alpha}{K}{C}} \\ \kinding{\kctx}{T}{K}
      }{
        \typingh{\kctx}{\ctx}{V\, T}{C[T/\alpha]}
      }\quad\rname{Th}{PolyApp}

      \inferrule{
        \typingh{\kctx}{\ctx}{V}{A} \\ \kinding{\kctx}{E}{\kEffect}
      }{
        \typingh{\kctx}{\ctx}{\ret{V}}{A ! E}
      }\quad\rname{Th}{Return}

      \inferrule{
        \typingh{\kctx}{\ctx}{M}{A ! E} \\ \typingh{\kctx}{\ctx, x : A}{N}{B ! E}
      }{
        \typingh{\kctx}{\ctx}{\letinh{x}{M}{N}}{B ! E}
      }\quad\rname{Th}{Let}

      \inferrule{
        \typingh{\kctx}{\ctx}{V}{A} \\ E = \{l : \pre{\tfunh{A}{B}}; R\} \\ \kinding{\kctx}{E}{\kEffect}
      }{
        \typingh{\kctx}{\ctx}{\doop{l}{V}{E}}{B ! E}
      }\quad\rname{Th}{Do}

      \inferrule{
        \typingh{\kctx}{\ctx}{M}{C} \\ \htypingh{\kctx}{\ctx}{H}{C}{D}
      }{
        \typingh{\kctx}{\ctx}{\handlewithh{M}{H}}{D}
      }\quad\rname{Th}{Handle}
    \end{mathpar}
    \normalfont{\bfseries Handler Typing}\tquad\fbox{$\htypingh{\kctx}{\ctx}{H}{C}{D}$}
    \begin{mathpar}
      \inferrule{
        C = A ! \{l_1: \pre{\tfunh{A_1}{B_1}}; \cdots; l_n: \pre{\tfunh{A_n}{B_n}}; R \} \\
        D = B ! \{l_1: P_1; \cdots; l_n: P_n; R \} \\
        H = \retc{x}{M} \uplus \opc{l_1}{y_1}{r_1}{N_1} \uplus \cdots \uplus \opc{l_n}{y_n}{r_n}{N_n} \\
        \typingh{\kctx}{\ctx, x : A}{M}{D} \\
        \forall i \in \{1, \ldots, n\} . (\typingh{\kctx}{\ctx, y_i : A_i, r_i : \tfunh{B_i}{D}}{N_i}{D})
      }{
        \htypingh{\kctx}{\ctx}{H}{C}{D}
      }\quad\rname{Hh}{Handler}
    \end{mathpar}
  \end{definition}

  \newcommand{\bigL}{\mathbb{L}}
  \newcommand{\ltoS}{\mathtt{L2S}}
  \newcommand{\rtol}{\mathtt{r2l}}
  \newcommand{\ltoT}{\mathtt{l2T}}
  \newcommand{\ltoOp}{\mathtt{l2Op}}
  \newcommand{\htol}{\mathtt{h2l}}
  \newcommand{\htoi}{\mathtt{H2I}}
  \newcommand{\fun}[3]{\mathbf{fun}(#1, #2, #3)}
  \newcommand{\ffun}[3]{\Lambda #1 : #2 . #3}
  \newcommand{\letin}[3]{\mathbf{let}\, #1 = #2 \,\mathbf{in}\, #3}
  \newcommand{\handlewith}[3]{\mathbf{handle}_{#1}\, #2 \,\mathbf{with}\, #3}
  \newcommand{\tfun}[3]{#1 \rightarrow_{#2} #3}
  \newcommand{\tabs}[4]{\forall #1 : #2 . {#3} ^ {#4}}
  \newcommand{\kTyp}{\mathbf{Typ}}
  \newcommand{\kEff}{\mathbf{Eff}}
  \newcommand{\kLab}{\mathbf{Lab}}
  \newcommand{\emptyE}{\emptyset_E}
  \newcommand{\erow}{\langle\rangle}
  \newcommand{\row}[2]{\langle #1 \mid #2 \rangle}
  \begin{definition}[Translation from Hillerström's to An Instance]
    We assume that:
    \begin{itemize}
      \item there exists a unique set that has any label (we call it $\bigL$),
      \item there exists a unique partition of $\bigL$,
      \item for any row, a set of presence labels in that row is a disjoint union of the partition result of $\bigL$,
      \item for any handler, target labels of that handler is one of the partition result of $\bigL$, and
      \item a unique closed type can be attached to $l$ as presence.
    \end{itemize}
    We write $\ltoS(\bigL)$ to denote the set of partition results of $\bigL$,
    $\rtol$ to denote the function that assigns a unique label $l$ such that $ \mathit{l}   \ottsym{:}    \mathbf{Lab}    \in   \Sbase $
    to $\labelset \in \ltoS(\bigL)$.
    We write $\rtol(H)$ to denote $l$ such that
    $\rtol(\{l_1, \ldots, l_n\}) = l$ where
    $H = \retc{x}{M} \uplus \opc{l_1}{p_1}{r_1}{N_1} \uplus \cdots \opc{l_n}{p_n}{r_n}{N_n}$.
    We define $\ltoT$ as the function that takes a label $l$ and returns the type that corresponds to the unique presence type of $l$.
    We define $\ltoOp$ as the function that takes a label $l$ and returns a unique operation name.
    We also assume that
    \[\rtol(\{l_1, \ldots, l_n\}) :: \{\ltoOp(l_1) : \ltoT(l_1), \ldots, \ltoOp(l_n) : \ltoT(l_n)\} \in \Sigma.\]
    We define $\htoi$ as follows.
    \newcommand{\arrayspacing}{\hspace{2em}}
    \\\textnormal{\bfseries Kinds}\hfill\phantom{}
    \[
      \begin{array}{rcl@{\arrayspacing}rcl}
        \htoi(\kType)           & = & \kTyp                   &
        \htoi(\kRow[\labelset]) & = & \htoi(\kEffect) = \kEff
      \end{array}
    \]
    \textnormal{\bfseries Types}\hfill\phantom{}
    \[
      \begin{array}{rcl@{\arrayspacing}rcl}
        \htoi(\tfunh{A}{B ! E})         & = & \tfun{\htoi(A)}{\htoi(E)}{\htoi(B)}         &
        \htoi(\tabsh{\alpha}{K}{A ! E}) & = & \tabs{\alpha}{\htoi(K)}{\htoi(A)}{\htoi(E)}
      \end{array}
    \]
    \textnormal{\bfseries Effects}\hfill\phantom{}
    \[
      \begin{array}{rcl}
        \htoi(\{R\})                             & = & \htoi(R)                                                      \\
        \htoi(l_1: P_1; \cdots; l_n: P_n; \cdot) & = &
        \row{l'_1}{\row{\cdots}{\row{l'_m}{\erow}}} \quad
        (\twhere l'_i = \rtol(\labelset_i) \tand \labelset_1 \uplus \cdots \labelset_m = \{l_j \mid P_j \neq \abs\}) \\
        \htoi(l_1: P_1; \cdots; l_n: P_n; \rho)  & = &
        \row{l'_1}{\row{\cdots}{\row{l'_m}{\rho}}} \quad
        (\twhere l'_i = \rtol(\labelset_i) \tand \labelset_1 \uplus \cdots \labelset_m = \{l_j \mid P_j \neq \abs\})
      \end{array}
    \]
    \textnormal{\bfseries Values}\hfill\phantom{}
    \[
      \begin{array}{rcl@{\arrayspacing}rcl}
        \htoi(x)                   & = & x                                 &
        \htoi(\tlam{\alpha}{K}{M}) & = & \ffun{\alpha}{\htoi(K)}{\htoi(M)}   \\
        \htoi(\lam{x}{A}{M})       & = & \multicolumn{4}{l}{
          \fun{z}{x}{\htoi(M)} \quad (\twhere z \text{ is fresh})
        }
      \end{array}
    \]
    \textnormal{\bfseries Computations}\hfill\phantom{}
    \[
      \begin{array}{rcl@{\arrayspacing}rcl}
        \htoi(V\, W)              & = & \htoi(V)\, \htoi(W)           &
        \htoi(V\, T)              & = & \htoi(V)\, \htoi(T)             \\
        \htoi(\ret{M})            & = & \htoi(M)                      &
        \htoi(\letinh{x}{M}{N})   & = & \letin{x}{\htoi(M)}{\htoi(N)}   \\
        \htoi(\doop{l}{V}{E})     & = & \multicolumn{4}{l}{
          \ltoOp(l)_{\rtol(\labelset)}\, \htoi(V) \quad (\twhere l \in \labelset \in \ltoS(\bigL))
        }                                                               \\
        \htoi(\handlewithh{M}{H}) & = & \multicolumn{4}{l}{
          \handlewith{\rtol(H)}{\htoi(M)}{\htoi(H)}
        }
      \end{array}
    \]
    \textnormal{\bfseries Handlers}\hfill\phantom{}
    \[
      \begin{array}{rclrcl}
        \htoi(\retc{x}{M})               & = & \retc{x}{\htoi(M)}                              &
        \htoi(\opc{l}{p}{r}{M} \uplus H) & = & \htoi(H) \uplus \opc{\ltoOp(l)}{p}{r}{\htoi(M)}
      \end{array}
    \]
    \textnormal{\bfseries Contexts}\hfill\phantom{}
    \[
      \begin{array}{rcl@{\arrayspacing}rcl}
        \htoi(\cdot)             & = & \emptyset                       &
        \htoi(\ctx, x : A)       & = & \htoi(\ctx), x : \htoi(A)         \\
        \htoi(\kctx, \alpha : K) & = & \htoi(\kctx), \alpha : \htoi(K) &
                                 &   &
      \end{array}
    \]
  \end{definition}

  \newcommand{\typing}[4]{#1 \vdash #2 : #3 \mid #4}
  \newcommand{\htyping}[6]{#1 \vdash_{#2} #3 : #4 \Rightarrow ^ {#5} #6}

  \begin{lemma}\label{lem:exchange_tvar}
    \phantom{}\\
    \begin{enumerate}
      \item If $\vdash  \Gamma_{{\mathrm{1}}}  \ottsym{,}  \alpha  \ottsym{:}  \ottnt{K}  \ottsym{,}  \mathit{x}  \ottsym{:}  \ottnt{A}  \ottsym{,}  \Gamma_{{\mathrm{3}}}$ and $\vdash  \Gamma_{{\mathrm{1}}}  \ottsym{,}  \mathit{x}  \ottsym{:}  \ottnt{A}$, then $\vdash  \Gamma_{{\mathrm{1}}}  \ottsym{,}  \mathit{x}  \ottsym{:}  \ottnt{A}  \ottsym{,}  \alpha  \ottsym{:}  \ottnt{K}  \ottsym{,}  \Gamma_{{\mathrm{3}}}$.

      \item If $\Gamma_{{\mathrm{1}}}  \ottsym{,}  \alpha  \ottsym{:}  \ottnt{K}  \ottsym{,}  \mathit{x}  \ottsym{:}  \ottnt{A}  \ottsym{,}  \Gamma_{{\mathrm{3}}}  \vdash  S  \ottsym{:}  \ottnt{K'}$ and $\vdash  \Gamma_{{\mathrm{1}}}  \ottsym{,}  \mathit{x}  \ottsym{:}  \ottnt{A}$,
            then $\Gamma_{{\mathrm{1}}}  \ottsym{,}  \mathit{x}  \ottsym{:}  \ottnt{A}  \ottsym{,}  \alpha  \ottsym{:}  \ottnt{K}  \ottsym{,}  \Gamma_{{\mathrm{3}}}  \vdash  S  \ottsym{:}  \ottnt{K'}$.

      \item If $\Gamma_{{\mathrm{1}}}  \ottsym{,}  \alpha  \ottsym{:}  \ottnt{K}  \ottsym{,}  \mathit{x}  \ottsym{:}  \ottnt{A}  \ottsym{,}  \Gamma_{{\mathrm{3}}}  \vdash  \ottnt{B}  <:  \ottnt{C}$ and $\vdash  \Gamma_{{\mathrm{1}}}  \ottsym{,}  \mathit{x}  \ottsym{:}  \ottnt{A}$,
            then $\Gamma_{{\mathrm{1}}}  \ottsym{,}  \mathit{x}  \ottsym{:}  \ottnt{A}  \ottsym{,}  \alpha  \ottsym{:}  \ottnt{K}  \ottsym{,}  \Gamma_{{\mathrm{3}}}  \vdash  \ottnt{B}  <:  \ottnt{C}$.

      \item If $\Gamma_{{\mathrm{1}}}  \ottsym{,}  \alpha  \ottsym{:}  \ottnt{K}  \ottsym{,}  \mathit{x}  \ottsym{:}  \ottnt{A}  \ottsym{,}  \Gamma_{{\mathrm{3}}}  \vdash  \ottnt{B_{{\mathrm{1}}}}  \mid  \varepsilon_{{\mathrm{1}}}  <:  \ottnt{B_{{\mathrm{2}}}}  \mid  \varepsilon_{{\mathrm{2}}}$ and $\vdash  \Gamma_{{\mathrm{1}}}  \ottsym{,}  \mathit{x}  \ottsym{:}  \ottnt{A}$,
            then $\Gamma_{{\mathrm{1}}}  \ottsym{,}  \mathit{x}  \ottsym{:}  \ottnt{A}  \ottsym{,}  \alpha  \ottsym{:}  \ottnt{K}  \ottsym{,}  \Gamma_{{\mathrm{3}}}  \vdash  \ottnt{B_{{\mathrm{1}}}}  \mid  \varepsilon_{{\mathrm{1}}}  <:  \ottnt{B_{{\mathrm{2}}}}  \mid  \varepsilon_{{\mathrm{2}}}$.

      \item If $\Gamma_{{\mathrm{1}}}  \ottsym{,}  \alpha  \ottsym{:}  \ottnt{K}  \ottsym{,}  \mathit{x}  \ottsym{:}  \ottnt{A}  \ottsym{,}  \Gamma_{{\mathrm{3}}}  \vdash  \ottnt{e}  \ottsym{:}  \ottnt{B}  \mid  \varepsilon$ and $\vdash  \Gamma_{{\mathrm{1}}}  \ottsym{,}  \mathit{x}  \ottsym{:}  \ottnt{A}$,
            then $\Gamma_{{\mathrm{1}}}  \ottsym{,}  \mathit{x}  \ottsym{:}  \ottnt{A}  \ottsym{,}  \alpha  \ottsym{:}  \ottnt{K}  \ottsym{,}  \Gamma_{{\mathrm{3}}}  \vdash  \ottnt{e}  \ottsym{:}  \ottnt{B}  \mid  \varepsilon$.

      \item If $ \Gamma_{{\mathrm{1}}}  \ottsym{,}  \alpha  \ottsym{:}  \ottnt{K}  \ottsym{,}  \mathit{x}  \ottsym{:}  \ottnt{A}  \ottsym{,}  \Gamma_{{\mathrm{3}}}  \vdash _{ \sigma }  \ottnt{h}  :  \ottnt{B}   \Rightarrow  ^ { \varepsilon }  \ottnt{C} $ and $\vdash  \Gamma_{{\mathrm{1}}}  \ottsym{,}  \mathit{x}  \ottsym{:}  \ottnt{A}$,
            then $ \Gamma_{{\mathrm{1}}}  \ottsym{,}  \mathit{x}  \ottsym{:}  \ottnt{A}  \ottsym{,}  \alpha  \ottsym{:}  \ottnt{K}  \ottsym{,}  \Gamma_{{\mathrm{3}}}  \vdash _{ \sigma }  \ottnt{h}  :  \ottnt{B}   \Rightarrow  ^ { \varepsilon }  \ottnt{C} $.
    \end{enumerate}
  \end{lemma}

  \begin{proof}
    Straightforward by mutual induction on the derivations.
  \end{proof}

  \begin{theorem}\label{thm:typability_htoi}
    \phantom{}\\
    \begin{enumerate}
      \item\label{thm:typability_htoi:kctx}
            If $\wfctx{\kctx}$, then $\wfctx{\htoi(\kctx)}$.

      \item\label{thm:typability_htoi:kinding}
            If $\kinding{\kctx}{T}{K}$, then $\kinding{\htoi(\kctx)}{\htoi(T)}{\htoi(K)}$.

      \item\label{thm:typability_htoi:ctx}
            If $\wkctx{\kctx}{\ctx}$, then $\wfctx{\htoi(\kctx), \htoi(\ctx)}$.

      \item\label{thm:typability_htoi:typing_val}
            If $\typingh{\kctx}{\ctx}{V}{A}$, then $\typing{\htoi(\kctx), \htoi(\ctx)}{\htoi(V)}{\htoi(A)}{\emptyE}$.

      \item\label{thm:typability_htoi:typing_comp}
            If $\typingh{\kctx}{\ctx}{M}{A ! E}$, then $\typing{\htoi(\kctx), \htoi(\ctx)}{\htoi(M)}{\htoi(A)}{\htoi(E)}$.

      \item\label{thm:typability_htoi:typing_handler}
            If $\htypingh{\kctx}{\ctx}{
                \retc{x}{M} \uplus \opc{l_1}{p_1}{r_1}{N_1} \uplus \cdots \uplus \opc{l_n}{p_n}{r_n}{N_n}
              }{A ! E}{B ! E'}$, then
            \begin{itemize}
              \item $\htyping{\htoi(\kctx), \htoi(\ctx)}{\sigma}{\htoi(H)}{\htoi(A)}{\htoi(E')}{\htoi(B)}$,
              \item $\rtol(H) : \sigma \in \Sigma$, and
              \item $\row{\rtol(H)}{\htoi(E')}  \sim_{\eanameSimpleRow}  \htoi(E)$,
            \end{itemize}
            where $\sigma = \{\ltoOp(l_1) : \ltoT(l_1), \ldots, \ltoOp(l_n) : \ltoT(l_n)\}$.
    \end{enumerate}
  \end{theorem}

  \begin{proof}
    \phantom{}\\
    \begin{itemize}
      \item[(1)]
            Straightforward by induction on the derivation.

      \item[(2)]
            By induction on a derivation of the judgment.
            We proceed by case analysis on the rule applied lastly to the derivation.
            \begin{divcases}
              \item[\rname{Kh}{Var}]
              We have
              \begin{itemize}
                \item $T = \alpha$,
                \item $\wfctx{\kctx', \alpha : K}$, and
                \item $\kctx = \kctx', \alpha : K$,
              \end{itemize}
              for some $\kctx'$.

              By definition of $\htoi$, we have $\alpha : \htoi(K) \in \htoi(\kctx', \alpha : K)$.
              By case~\ref{thm:typability_htoi:kctx}, \rname{K}{Var} derives $\kinding{\htoi(\ctx)}{\alpha}{\htoi(K)}$

              \item[\rname{Kh}{Fun}]
              We have
              \begin{itemize}
                \item $T = \tfunh{A}{B!E}$,
                \item $K = \kType$,
                \item $\kinding{\kctx}{A}{\kType}$,
                \item $\kinding{\kctx}{B}{\kType}$,
                \item $\kinding{\kctx}{E}{\kEffect}$,
              \end{itemize}
              for some $A$, $B$, and $E$.
              By the induction hypothesis, we have
              \begin{itemize}
                \item $\kinding{\htoi(\kctx)}{\htoi(A)}{\kTyp}$,
                \item $\kinding{\htoi(\kctx)}{\htoi(B)}{\kTyp}$, and
                \item $\kinding{\htoi(\kctx)}{\htoi(E)}{\kEff}$.
              \end{itemize}
              Thus, \rname{K}{Fun} derives
              \[\kinding{\htoi(\kctx)}{\tfun{\htoi(A)}{\htoi(E)}{\htoi(B)}}{\kTyp}\]
              as required.

              \item[\rname{Kh}{Forall}]
              We have
              \begin{itemize}
                \item $T = \tabsh{\alpha}{K'}{A ! E}$,
                \item $K = \kType$,
                \item $\kinding{\kctx, \alpha : K'}{A}{\kType}$, and
                \item $\kinding{\kctx, \alpha : K'}{E}{\kEffect}$,
              \end{itemize}
              for some $\alpha$, $K'$, $A$, and $E$.
              By the induction hypothesis, we have
              \begin{itemize}
                \item $\kinding{\htoi(\kctx, \alpha : K')}{\htoi(A)}{\kTyp}$ and
                \item $\kinding{\htoi(\kctx, \alpha : K')}{\htoi(E)}{\kEff}$.
              \end{itemize}
              Thus, \rname{K}{Poly} derives
              \[\kinding{\htoi(\kctx)}{\tabs{\alpha}{\htoi(K')}{\htoi(A)}{\htoi(E)}}{\kTyp}\]
              as required.

              \item[\rname{Kh}{Effect}]
              Clearly by the induction hypothesis.

              \item[\rname{Kh}{CloseRow}]
              Clearly by the assumptions and \rname{K}{Cons}.

              \item[\rname{Kh}{OpenRow}]
              Clearly by the assumptions and \rname{K}{Cons}.
            \end{divcases}

      \item[(3)]
            By induction on a derivation of the judgment.
            We proceed by case analysis on the rule applied lastly to the derivation.
            \begin{divcases}
              \item[\rname{Ch}{Empty}]
              Clearly because case~\ref{thm:typability_htoi:kctx}.

              \item[\rname{Ch}{Var}]
              We have
              \begin{itemize}
                \item $\ctx = \ctx', x : A$,
                \item $\wkctx{\kctx}{\ctx'}$,
                \item $x \notin \dom(\ctx')$, and
                \item $\kinding{\kctx}{A}{\kType}$,
              \end{itemize}
              for some $\ctx'$, $x$, and $A$.
              By the induction hypothesis and case~\ref{thm:typability_htoi:kinding}, we have
              \begin{itemize}
                \item $\wfctx{\htoi(\kctx), \htoi(\ctx')}$ and
                \item $\kinding{\htoi(\kctx)}{\htoi(A)}{\kTyp}$.
              \end{itemize}
              By $\wfctx{\htoi(\kctx), \htoi(\ctx')}$ and Lemma~\ref{lem:weakening}\ref{lem:weakening:kinding},
              we have $\kinding{\htoi(\kctx), \htoi(\ctx')}{\htoi(A)}{\kTyp}$.
              By definition of $\htoi$, we have $x \notin \dom(\htoi(\kctx), \htoi(\ctx'))$.
              Thus, \rname{C}{Var} derives
              \[\wfctx{\htoi(\kctx), \htoi(\ctx'), x : \htoi(A)}\]
              as required.
            \end{divcases}

      \item[(4)(5)(6)]
            By mutual induction on derivations of the judgments.
            We proceed by case analysis on the rule applied lastly to the derivations.
            \begin{divcases}
              \item[\rname{Th}{Var}]
              We have
              \begin{itemize}
                \item $V = x$,
                \item $\wkctx{\kctx}{\ctx}$, and
                \item $x : A \in \ctx$,
              \end{itemize}
              for some $x$.
              By Theorem~\ref{thm:typability_htoi:ctx}, we have $\wfctx{\htoi(\kctx), \htoi(\ctx)}$.
              By definition of $\htoi$, we have $x : \htoi(A) \in \htoi(\ctx)$.
              Thus, \rname{T}{Var} derives
              \[\typing{\htoi(\kctx), \htoi(\ctx)}{x}{\htoi(A)}{\emptyE}\]
              as required.

              \item[\rname{Th}{Lam}]
              We have
              \begin{itemize}
                \item $V = \lam{x}{A_0}{M}$,
                \item $A = \tfunh{A_0}{A_1 ! E}$, and
                \item $\typingh{\kctx}{\ctx, x : A_0}{M}{A_1 ! E}$,
              \end{itemize}
              for some $x$, $A_0$, $A_1$, $E$, and $M$.
              By the induction hypothesis, we have
              \[\typing{\htoi(\kctx), \htoi(\ctx), x : \htoi(A_0)}{\htoi(M)}{\htoi(A_1)}{\htoi(E)}.\]
              Without loss of generality, we can choose $z$ such that
              \begin{itemize}
                \item $z \notin \mathit{FV}(\htoi(M))$,
                \item $z \neq x$,
                \item $z \notin \dom(\htoi(\kctx), \htoi(\ctx))$, and
                \item $\htoi(\lam{x}{A_0}{M}) = \fun{z}{x}{\htoi(M)}$.
              \end{itemize}
              By Lemma~\ref{lem:wk} and Lemma~\ref{lem:delta_context}\ref{lem:delta_context:kinding} and
              Lemma~\ref{lem:delta_weakening}, we have
              \begin{itemize}
                \item $\kinding{\htoi(\kctx), \htoi(\ctx)}{\htoi(A_1)}{\kTyp}$ and
                \item $\kinding{\htoi(\kctx), \htoi(\ctx)}{\htoi(E)}{\kEff}$.
              \end{itemize}
              By Lemma~\ref{lem:ctx-wf-typing}, we have $\wfctx{\htoi(\kctx), \htoi(\ctx), x : \htoi(A_0)}$.
              Since only \rname{C}{Var} can derive this judgment,
              we have $\kinding{\htoi(\kctx), \htoi(\ctx)}{\htoi(A_0)}{\kTyp}$.
              Thus, \rname{C}{Var} derives
              \[\wfctx{\htoi(\kctx), \htoi(\ctx), z : \tfun{\htoi(A_0)}{\htoi(E)}{\htoi(A_1)}}.\]
              Thus, Lemma~\ref{lem:weakening}\ref{lem:weakening:typing} and \rname{T}{Abs} derives
              \[\typing{\htoi(\kctx), \htoi(\ctx)}{\fun{z}{x}{\htoi(M)}}{\tfun{\htoi(A_0)}{\htoi(E)}{\htoi(A_1)}}{\emptyE}\]
              as required.

              \item[\rname{Th}{PolyLam}]
              We have
              \begin{itemize}
                \item $V = \tlam{\alpha}{K}{M}$,
                \item $A = \tabsh{\alpha}{K}{B ! E}$,
                \item $\typingh{\kctx, \alpha : K}{\ctx}{M}{B ! E}$, and
                \item $\wkctx{\kctx}{\ctx}$,
              \end{itemize}
              for some $\alpha$, $K$, $M$, $B$, and $E$.
              By the induction hypothesis and case~\ref{thm:typability_htoi:ctx}, we have
              \begin{itemize}
                \item $\wfctx{\htoi(\kctx), \htoi(\ctx)}$ and
                \item $\typing{\htoi(\kctx), \alpha : \htoi(K), \htoi(\ctx)}{\htoi(M)}{\htoi(B)}{\htoi(E)}$.
              \end{itemize}
              By applying Lemma~\ref{lem:exchange_tvar} repeatedly, we have
              \[\typing{\htoi(\kctx), \htoi(\ctx), \alpha : \htoi(K)}{\htoi(M)}{\htoi(B)}{\htoi(E)}.\]
              Thus, \rname{T}{TAbs} derives
              \[\typing{\htoi(\kctx), \htoi(\ctx)}{\ffun{\alpha}{\htoi(K)}{\htoi(M)}}
                {\tabs{\alpha}{\htoi(K)}{\htoi(B)}{\htoi(E)}}{\emptyE}\]
              as required.

              \item[\rname{Th}{App}]
              We have
              \begin{itemize}
                \item $M = V\, W$,
                \item $\typingh{\kctx}{\ctx}{V}{\tfunh{B}{A ! E}}$, and
                \item $\typingh{\kctx}{\ctx}{W}{B}$,
              \end{itemize}
              for some $V$, $W$, and $B$.
              By the induction hypothesis, we have
              \begin{itemize}
                \item $\typing{\htoi(\kctx), \htoi(\ctx)}{\htoi(V)}{\tfun{\htoi(B)}{\htoi(E)}{\htoi(A)}}{\emptyE}$ and
                \item $\typing{\htoi(\kctx), \htoi(\ctx)}{\htoi(W)}{\htoi(B)}{\emptyE}$.
              \end{itemize}
              Thus, \rname{T}{App} derives
              \[\typing{\htoi(\kctx), \htoi(\ctx)}{\htoi(V)\, \htoi(W)}{\htoi(A)}{\htoi(E)}\]
              as required.

              \item[\rname{Th}{PolyApp}]
              We have
              \begin{itemize}
                \item $M = V\, T$,
                \item $A = (B ! E)[T/\alpha]$,
                \item $\typingh{\kctx}{\ctx}{V}{\tabsh{\alpha}{K}{B ! E}}$, and
                \item $\kinding{\kctx}{T}{K}$
              \end{itemize}
              for some $V$, $T$, $\alpha$, $K$, $B$, and $E$.
              By the induction hypothesis and
              case~\ref{thm:typability_htoi:kinding}, we have
              \begin{itemize}
                \item $\typing{\htoi(\kctx), \htoi(\ctx)}{\htoi(V)}{\tabs{\alpha}{\htoi(K)}{\htoi(B)}{\htoi(E)}}{\emptyE}$ and
                \item $\kinding{\htoi(\kctx)}{\htoi(T)}{\htoi(K)}$.
              \end{itemize}
              By Lemma~\ref{lem:ctx-wf-typing} and Lemma~\ref{lem:weakening}\ref{lem:weakening:kinding},
              we have $\kinding{\htoi(\kctx), \htoi(\ctx)}{\htoi(T)}{\htoi(K)}$.
              Thus, \rname{T}{TApp} derives
              \[\typing{\htoi(\kctx), \htoi(\ctx)}{\htoi(V)\, \htoi(T)}{\htoi(B[T/\alpha])}{\htoi(E[T/\alpha])}\]
              as required.

              \item[\rname{Th}{Return}]
              We have
              \begin{itemize}
                \item $M = \ret{V}$,
                \item $\typingh{\kctx}{\ctx}{V}{A}$, and
                \item $\kinding{\kctx}{E}{\kEffect}$,
              \end{itemize}
              for some $V$.
              By the induction hypothesis and case~\ref{thm:typability_htoi:kinding}, we have
              \begin{itemize}
                \item $\typing{\htoi(\kctx), \htoi(\ctx)}{\htoi(V)}{\htoi(A)}{\emptyE}$ and
                \item $\kinding{\htoi(\kctx)}{\htoi(E)}{\kEff}$.
              \end{itemize}
              Thus, \rname{T}{Sub} derives
              \[\typing{\htoi(\kctx), \htoi(\ctx)}{\htoi(V)}{\htoi(A)}{\htoi(E)}\]
              as required.

              \item[\rname{Th}{Let}]
              We have
              \begin{itemize}
                \item $M = \letinh{x}{M_0}{M_1}$,
                \item $\typingh{\kctx}{\ctx}{M_0}{B ! E}$, and
                \item $\typingh{\kctx}{\ctx, x : B}{M_1}{A ! E}$,
              \end{itemize}
              for some $x$, $M_0$, $M_1$, and $B$.
              By the induction hypothesis, we have
              \begin{itemize}
                \item $\typing{\htoi(\kctx), \htoi(\ctx)}{\htoi(M_0)}{\htoi(B)}{\htoi(E)}$ and
                \item $\typing{\htoi(\kctx), \htoi(\ctx), x : \htoi(B)}{\htoi(M_1)}{\htoi(A)}{\htoi(E)}$.
              \end{itemize}
              Thus, \rname{T}{Let} derives
              \[\typing{\htoi(\kctx), \htoi(\ctx)}{\letin{x}{\htoi(M_0)}{\htoi(M_1)}}{\htoi(A)}{\htoi(E)}\]
              as required.

              \item[\rname{Th}{Do}]
              We have
              \begin{itemize}
                \item $M = \doop{l}{V}{E}$,
                \item $\typingh{\kctx}{\ctx}{V}{B}$,
                \item $\{l : \pre{\tfunh{B}{A}}; R\}$, and
                \item $\kinding{\kctx}{E}{\kEffect}$,
              \end{itemize}
              for some $l$, $V$, $E$, $B$, and $R$.
              By the induction hypothesis and case~\ref{thm:typability_htoi:kinding}, we have
              \begin{itemize}
                \item $\typing{\htoi(\kctx), \htoi(\ctx)}{\htoi(V)}{\htoi(B)}{\emptyE}$ and
                \item $\kinding{\htoi(\kctx)}{\htoi(E)}{\kEff}$.
              \end{itemize}
              There uniquely exists some $\labelset$ such that
              \begin{itemize}
                \item $\labelset \in \ltoS(\bigL)$,
                \item $l \in \labelset$, and
                \item $\labelset \subseteq \dom(E)$.
              \end{itemize}
              Thus, we have
              \begin{itemize}
                \item $\rtol(\labelset) :: \sigma \in \Sigma$,
                \item $\ltoOp(l) : \ltoT(l) \in \sigma$, and
                \item $\row{\rtol(\labelset)}{\varepsilon}  \sim_{\eanameSimpleRow}  \htoi(E)$,
              \end{itemize}
              for some $\sigma$ and $\varepsilon$.
              Because Lemma~\ref{lem:ctx-wf-typing} gives us $\wfctx{\htoi(\kctx), \htoi(\ctx)}$,
              \rname{T}{Op} and \rname{T}{App} and \rname{T}{Sub} derive
              \[\typing{\htoi(\kctx), \htoi(\ctx)}{\ltoOp(l)_{\rtol(\labelset)}\, \htoi(V)}{\htoi(B)}{\htoi(E)}\]
              as required.

              \item[\rname{Th}{Handle}]
              We have
              \begin{itemize}
                \item $M = \handlewithh{N}{H}$,
                \item $\typingh{\kctx}{\ctx}{N}{B ! E'}$, and
                \item $\htypingh{\kctx}{\ctx}{H}{B ! E'}{A ! E}$,
              \end{itemize}
              for some $N$, $H$, $B$, and $E'$.
              By the induction hypothesis, we have
              \begin{itemize}
                \item $\typing{\htoi(\kctx), \htoi(\ctx)}{\htoi(N)}{\htoi(B)}{\htoi(E')}$,
                \item $\htyping{\htoi(\kctx), \htoi(\ctx)}{\sigma}{\htoi(H)}{\htoi(B)}{\htoi(E)}{\htoi(A)}$,
                \item $\rtol(H) :: \sigma \in \Sigma$, and
                \item $\row{\rtol(H)}{\htoi(E)}  \sim_{\eanameSimpleRow}  \htoi(E')$.
              \end{itemize}
              for some $\sigma$.
              Thus, \rname{T}{Handling} derives
              \[\typing{\htoi(\kctx), \htoi(\ctx)}{\handlewith{\rtol(H)}{\htoi(N)}{\htoi(H)}}{\htoi(A)}{\htoi(E)}\]
              as required.

              \item[\rname{Hh}{Handler}]
              We have
              \begin{itemize}
                \item $\typingh{\kctx}{\ctx, x : A}{M}{B ! E'}$,
                \item $\typingh{\kctx}{\ctx, y_i : A_i, r_i : \tfunh{B_i}{B ! E'}}{N_i}{B ! E'}$ for any $i \in \{1, \ldots, n\}$,
                \item $E = \{l_1 : \pre{\tfunh{A_1}{B_1}}; \cdots; l_n : \pre{\tfunh{A_n}{B_n}}; R\}$, and
                \item $E' = \{l_1 : P_1; \cdots; l_n : P_n; R\}$,
              \end{itemize}
              for some $A_i$, $B_i$, and $P_i$, where $i \in \{1, \ldots, n\}$.
              By the assumptions, we have
              \begin{itemize}
                \item $\{l_1, \ldots, l_n\} \in \ltoS(\bigL)$,
                \item $\ltoT(l_i) = {\htoi(A_i)} \Rightarrow {\htoi(B_i)}$ for any $i \in \{1, \ldots, n\}$,
                \item $\rtol(\{l_1, \ldots, l_n\}) :: \{\ltoOp(l_1) : \ltoT(l_1), \ldots, \ltoOp(l_n) : \ltoT(l_n)\} \in \Sigma$, and
                \item $\forall i \in \{1, \ldots, n\} . (P_i = \abs)$ or
                      $\forall i \in \{1, \ldots, n\} . (P_i = \pre{\tfunh{A_i}{B_i}})$.
              \end{itemize}
              Thus, we have $\row{\rtol(H)}{\htoi(E')}  \sim_{\eanameSimpleRow}  \htoi(E)$.

              By the induction hypothesis, we have
              \begin{itemize}
                \item $\typing{\htoi(\kctx), \htoi(\ctx), x : \htoi(A)}{\htoi(M)}{\htoi(B)}{\htoi(E')}$ and
                \item $\typing
                        {\htoi(\kctx), \htoi(\ctx), y_i : \htoi(A_i), r_i : \tfun{\htoi(B_i)}{\htoi(E')}{\htoi(B)}}
                        {\htoi(N_i)}{\htoi(B)}{\htoi(E')}$ for any $i \in \{1, \ldots, n\}$.
              \end{itemize}
              Therefore, \rname{H}{Return} and \rname{H}{Op} derive
              \[\htyping{\htoi(\kctx), \htoi(\ctx)}{\{\ltoOp(l_1) : \ltoT(l_1), \ldots, \ltoOp(l_n) : \ltoT(l_n)\}}
                {\htoi(H)}{\htoi(A)}{\htoi(E')}{\htoi(B)}.\]

              Thus, the required result is achieved.
            \end{divcases}
    \end{itemize}
  \end{proof}
}
\subsection{Comparison to \cite{leijen_type_2017}}

{
  We give the targets of comparison:
  one is an instance of {\lang} (Example~\ref{exa:effrow}), and
  another is a minorly changed language of \cite{leijen_type_2017}.

  \renewcommand{\arraystretch}{1.2}

  \newcommand{\val}[3]{\mathsf{val}\, #1 = #2 ; #3}
  \newcommand{\handlel}[2]{\mathsf{handle}\{#1\}(#2)}
  \newcommand{\lam}[2]{\lambda #1 . #2}
  \newcommand{\tlaml}[2]{\Lambda #1 . #2}
  \newcommand{\opl}{\mathit{op}}
  \newcommand{\retcl}[2]{\mathsf{return}\, #1 \rightarrow #2}
  \newcommand{\opcl}[3]{#1 (#2) \rightarrow #3}
  \newcommand{\resume}{\mathit{resume}}
  \newcommand{\typ}[1][]{\tau^{#1}}
  \renewcommand{\tvar}[1][]{\alpha^{#1}}
  \newcommand{\tconst}[2]{c^{#1} \langle #2 \rangle}
  \newcommand{\keffl}{\mathsf{e}}
  \newcommand{\klabl}{\mathsf{k}}
  \newcommand{\tsch}{\sigma}
  \newcommand{\tabs}[2]{\forall #1 . #2}
  \newcommand{\eps}{\epsilon}
  \newcommand{\erow}{\langle\rangle}
  \newcommand{\row}[2]{\langle #1 \mid #2 \rangle}
  \newcommand{\tfunl}[3]{#1 \rightarrow #2\, #3}
  \newcommand{\tinst}[2]{[#1 \mapsto #2]}
  \renewcommand{\ftv}{\mathsf{ftv}}
  \newcommand{\bra}[1]{\left\llbracket #1 \right\rrbracket}
  \newcommand{\deri}{\mathcal{D}}
  \newcommand{\derit}[2]{#1 :: #2}
  \newcommand{\otyping}[5]{#1 \vdash #2 : #3 \mid #4\, \colorbox{syntaxhighlight}{$\!\dashv #5$}}
  \newcommand{\typing}[4]{#1 \vdash #2 : #3 \mid #4}
  \newcommand{\ctx}{\Gamma}
  \newcommand{\wfctx}[1]{\vdash #1}
  \newcommand{\tfuni}[3]{#1 \rightarrow_{#2} #3}
  \newcommand{\tabsi}[3]{\forall #1 : #2 . #3}
  \newcommand{\letin}[3]{\mathbf{let}\, #1 = #2 \,\mathbf{in}\, #3}
  \newcommand{\funi}[3]{\mathbf{fun}(#1, #2, #3)}
  \newcommand{\tlami}[3]{\Lambda #1 : #2 . #3}
  \newcommand{\opi}[1][]{\mathsf{op}_{#1}}
  \newcommand{\handlewith}[3]{\mathbf{handle}_{#1}\, #2 \,\mathbf{with}\, #3}
  \newcommand{\lDecl}[2]{#1 :: #2}
  \newcommand{\opsigi}[3]{#1 : #2 \Rightarrow #3}
  \newcommand{\retci}[2]{\{\mathbf{return}\, #1 \mapsto #2\}}
  \newcommand{\opci}[4]{\{#1\, #2\, #3 \mapsto #4\}}
  \newcommand{\kTyp}{\mathbf{Typ}}
  \newcommand{\kLab}{\mathbf{Lab}}
  \newcommand{\kEff}{\mathbf{Eff}}
  \newcommand{\emptyE}{\langle  \rangle}
  \newcommand{\kinding}[3]{#1 \vdash #2 : #3}
  \newcommand{\typingH}[4]{#1 \vdash_{#2} #3 : #4}
  \renewcommand{\bar}[1]{\overline{#1}}
  \newcommand{\ctol}{\mathtt{c2l}}

  \begin{definition}[Minor Changed Version of \cite{leijen_type_2017}]
    Change list:
    \begin{itemize}
      \item changing implicit polymorphism to explicit polymorphism,
      \item removing constants from values,
      \item removing the assumption that the initial environment has effect declarations, and adding such declarations to $\Sigma$,
      \item adding type variables to contexts, and
      \item adding well-formedness of contexts.
    \end{itemize}

    The syntax of a minor changed version of \cite{leijen_type_2017} is as follows.
    \[
      \begin{array}{rclr}
        e               & \Coloneqq & v \mid e (e) \mid e (\typ[k]) \mid \val{x}{e_1}{e_2} \mid \handlel{h}{e}                  & \text{(expressions)}           \\
        v               & \Coloneqq & x \mid \opl \mid \lam{x}{e} \mid \tlaml{\tvar[k]}{e}                                      & \text{(values)}                \\
        h               & \Coloneqq & \retcl{x}{e} \mid \opcl{\opl}{x}{e}; h                                                    & \text{(clauses)}               \\
        \typ[k]         & \Coloneqq & \tvar[k] \mid \tconst{(k_1, \ldots, k_n) \rightarrow k}{\typ[k_1]_1, \ldots, \typ[k_n]_n} & \text{(types)}                 \\
        k               & \Coloneqq & * \mid \keffl \mid \klabl \mid (k_1, \ldots, k_n) \rightarrow k                           & \text{(kinds)}                 \\
        \tsch           & \Coloneqq & \tabs{\tvar[k]}{\tsch} \mid \typ[*]                                                       & \text{(type scheme)}           \\
        \ctx            & \Coloneqq & \emptyset \mid \ctx, x : \tsch \mid \ctx, \tvar[k]                                        & \text{(typing contexts)}       \\
        \Sigma          & \Coloneqq & \emptyset \mid \Sigma, l : \{\opl_1, \ldots, \opl_n\}                                     & \text{(signature environment)} \\
        \tfunl{-}{-}{-} & ::        & (*, \keffl, *) \rightarrow *                                                              & \text{(functions)}             \\
        \erow           & ::        & \keffl                                                                                    & \text{(empty effect)}          \\
        \row{-}{-}      & ::        & (\klabl, \keffl) \rightarrow \keffl                                                       & \text{(effect extension)}      \\
        l               & \coloneqq & \tconst{(k_1, \ldots, k_n) \rightarrow \klabl}{\typ[k_1]_1, \ldots, \typ[k_n]_n}          & \text{(effect labels)}
      \end{array}
    \]

    Well-formedness rules,
    free type variable,
    and typing rules consist of the following.\\
    \textnormal{\bfseries Contexts Well-formedness}\tquad\fbox{$\wfctx{\ctx}$} \\
    \begin{mathpar}
      \inferrule{ }{
        \wfctx{\emptyset}
      }\quad\rname{Cl}{Empty}

      \inferrule{
        \wfctx{\ctx} \\
        x \notin \dom(\ctx) \\
        \ftv(\typ[*]) \setminus \{\bar{\tvar[k']}\} \subseteq \ctx \\
        \forall \bar{k} . (\{\bar{\tvar[k]}\} \cap \ctx = \emptyset)
      }{
        \wfctx{\ctx, x : \tabs{\bar{\tvar[k']}}{\typ[*]}}
      }\quad\rname{Cl}{Var}

      \inferrule{
        \wfctx{\ctx} \\ \forall k . (\tvar[k] \notin \ctx)
      }{
        \wfctx{\ctx, \tvar[k']}
      }\quad\rname{Cl}{TVar}
    \end{mathpar}
    \textnormal{\bfseries Free Type Variable}\tquad\fbox{$\ftv(\typ[k])$}\tquad\fbox{$\ftv(\tsch)$}
    \begin{mathpar}
      \ftv(\tvar[k]) = \{\tvar[k]\}

      \ftv(\tfunl{\typ[*]_1}{\typ[\keffl]_2}{\typ[*]_3}) = \ftv(\typ[*]_1) \cup \ftv(\typ[\keffl]_2) \cup \ftv(\typ[*]_3)

      \ftv(\erow) = \emptyset

      \ftv(\row{\typ[\klabl]_1}{\typ[\keffl]_2}) = \ftv(\typ[\klabl]_1) \cup \ftv(\typ[\keffl]_2)

      \ftv(\tconst{(k_1, \ldots, k_n) \rightarrow \klabl}{\typ[k_1]_1, \ldots, \typ[k_n]_n})
      = \bigcup_{i \in \{1, \ldots, n\}} \ftv(\typ[k_i]_i)

      \ftv(\tabs{\tvar[k]}{\tsch}) = \ftv(\tsch) \setminus \{\tvar[k]\}
    \end{mathpar}
    \textnormal{\bfseries Typing}\tquad\fbox{$\typing{\ctx}{e}{\tsch}{\eps}$}
    \begin{mathpar}
      \inferrule{
        \wfctx{\ctx} \\
        \ctx(x) = \tsch \\ \ftv(\eps) \subseteq \ctx
      }{
        \typing{\ctx}{x}{\tsch}{\eps}
      }\quad\rname{Tl}{Var}

      \inferrule{
        \typing{\ctx, x : \typ_1}{e}{\typ_2}{\eps'} \\ \ftv(\eps) \subseteq \ctx
      }{
        \typing{\ctx}{\lam{x}{e}}{\tfunl{\typ_1}{\eps'}{\typ_2}}{\eps}
      }\quad\rname{Tl}{Lam}

      \inferrule{
        \typing{\ctx}{e_1}{\tsch_1}{\eps} \\ \typing{\ctx, x : \tsch}{e_2}{\typ}{\eps}
      }{
        \typing{\ctx}{\val{x}{e_1}{e_2}}{\typ}{\eps}
      }\quad\rname{Tl}{Let}

      \inferrule{
        \typing{\ctx}{e_1}{\tfunl{\typ_2}{\eps}{\typ}}{\eps} \\
        \typing{\ctx}{e_2}{\typ_2}{\eps}
      }{
        \typing{\ctx}{e_1 (e_2)}{\typ}{\eps}
      }\quad\rname{Tl}{App}

      \inferrule{
        \typing{\ctx, \bar{\tvar[k]}}{e}{\typ}{\erow} \\
        \ftv(\eps) \subseteq \ctx
      }{
        \typing{\ctx}{\tlaml{\bar{\tvar[k]}}{e}}{\tabs{\bar{\tvar[k]}}{\typ}}{\eps}
      }\quad\rname{Tl}{TAbs}

      \inferrule{
        \typing{\ctx}{e}{\tabs{\bar{\tvar[k]}}{\typ}}{\eps} \\
        \ftv(\bar{\typ[k]_0}) \subseteq \ctx
      }{
        \typing{\ctx}{e (\bar{\typ[k]_0})}{\typ\tinst{\bar{\tvar[k]}}{\bar{\typ[k]_0}}}{\eps}
      }\quad\rname{Tl}{TApp}

      \inferrule{
        \typing{\ctx}{e}{\typ}{\row{l}{\eps}} \\
        \typing{\ctx, x : \typ}{e_r}{\typ_r}{\eps} \\\\
        \Sigma(l) = \{\opl_1, \ldots, \opl_n\} \\
        \typing{\ctx}{\opl_i}{\tfunl{\typ_i}{\row{l}{\erow}}{{\typ}'_i}}{\erow} \\\\
        \typing{\ctx, x_i : \typ_i, \resume : \tfunl{{\typ}'_i}{\eps}{\typ_r}}{e_i}{\typ_r}{\eps}
      }{
        \typing{\ctx}{\handlel{\opcl{\opl_1}{x_1}{e_1}; \cdots; \opcl{\opl_n}{x_n}{e_n}; \retcl{x}{e_r}}{e}}{\typ_r}{\eps}
      }\quad\rname{Tl}{Handle}

      \inferrule{
        \typing{\ctx}{e}{\tfunl{\typ_1}{\row{l_1, \ldots, l_n}{\erow}}{\typ_2}}{\eps} \\
        \ftv(\eps') \subseteq \ctx
      }{
        \typing{\ctx}{e}{\tfunl{\typ_1}{\row{l_1, \ldots, l_n}{\eps'}}{\typ_2}}{\eps}
      }\quad\rname{Tl}{Open}
    \end{mathpar}
  \end{definition}

  \newcommand{\ltoi}{\mathtt{L2I}}
  \renewcommand{\ctol}{\mathtt{c2l}}
  \newcommand{\htol}{\mathtt{h2l}}
  \begin{definition}[Translation from Leijen's to An Instance]
    We assume that there is no constants other than $\tfunl{-}{-}{-}$, $\erow$, $\row{-}{-}$ and
    $c^{(k_1, \ldots, k_n) \rightarrow \klabl}$.
    We define $\ctol$ as the injective function that assigns a label name $l$ such that $l : $ (belonging to an instance) to
    $c^{(k_1, \ldots, k_n) \rightarrow \klabl}$ (belonging to Leijen's).
    We define $\ltoi$, $\htol$, and $\ctol$ as follows. We require $\ctol$ to be injective.
    \newcommand{\arrayspacing}{\hspace{2em}}
    \\\textnormal{\bfseries Kinds}
    \[
      \begin{array}{rcl@{\arrayspacing}rcl@{\arrayspacing}rcl}
        \ltoi(*)
         & = & \kTyp &
        \ltoi(\keffl)
         & = & \kEff &
        \ltoi(\klabl)
         & = & \kLab
      \end{array}
    \]
    \textnormal{\bfseries Types}
    \[
      \begin{array}{rcl@{\arrayspacing}rcl}
        \ltoi(\tvar[k])
         & = & \tvar                                                             &
        \ltoi(\tfunl{\typ[*]_1}{\typ[\keffl]_2}{\typ[*]_3})
         & = & \tfuni{\ltoi(\typ[*]_1)}{\ltoi(\typ[\keffl]_2)}{\ltoi(\typ[*]_3)}   \\
        \ltoi(\erow)
         & = & \erow                                                             &
        \ltoi(\row{l}{\eps})
         & = & \row{\ltoi(l)}{\ltoi(\eps)}                                         \\
        \ltoi(\tabs{\tvar[k]}{\tsch})
         & = & \tabsi{\tvar}{\ltoi(k)}{\ltoi(\tsch)}^{\emptyE}                   &
         &   &                                                                     \\
        \ltoi(\tconst{(k_1, \ldots, k_n) \rightarrow \klabl}{\typ[k_1]_1, \ldots, \typ[k_n]_n})
         & = &
        \multicolumn{4}{l}{
          \ctol(c^{(k_1, \ldots, k_n) \rightarrow \klabl})\, \ltoi(\typ[k_1]_1)\, \cdots \, \ltoi(\typ[k_n]_n)
        }                                                                          \\
        \ctol (c^{(k_1, \ldots, k_n) \rightarrow \klabl})
         & = & \multicolumn{4}{l}{
          l \quad (\twhere l : \ltoi(k_1) \times \ldots \times \ltoi(k_n) \rightarrow \kLab \in \Sbase)
        }
      \end{array}
    \]
    \textnormal{\bfseries Expressions}
    \[
      \begin{array}{rcl}
        \ltoi(x)
         & = & x                                                        \\
        \ltoi(\opl)
         & = & \opi[\ctol(c)\, \bar{\ltoi(\typ[k])}] \quad
        (\twhere \opl \in \Sigma(\tconst{}{\bar{\typ[k]}}))             \\
        \ltoi(\lam{x}{e})
         & = & \funi{z}{x}{\ltoi(e)} \quad (\twhere z \text{ is fresh}) \\
        \ltoi(\tlaml{\tvar[k]}{e})
         & = & \tlami{\tvar}{\ltoi(k)}{\ltoi(e)}                        \\
        \ltoi(e_1(e_2))
         & = & \letin{x}{\ltoi(e_1)}{\letin{y}{\ltoi(e_2)}{x\, y}}      \\
        \ltoi(e(\typ[k]))
         & = & \letin{x}{\ltoi(e)}{x\, \ltoi(\typ[k])}                  \\
        \ltoi(\val{x}{e_1}{e_2})
         & = & \letin{x}{\ltoi(e_1)}{\ltoi(e_2)}                        \\
        \ltoi(\handlel{h}{e})
         & = & \handlewith{\htol(h)}{\ltoi(e)}{\ltoi(h)}
      \end{array}
    \]
    \textnormal{\bfseries Handlers}
    \[
      \begin{array}{rcl}
        \ltoi(\retcl{x}{e})
         & = & \retci{x}{\ltoi(e)}                               \\
        \ltoi(\opcl{\opl}{x}{e}; h)
         & = & \ltoi(h) \uplus \opci{\opi}{x}{\resume}{\ltoi(e)} \\[2ex]
      \end{array}
    \]
    \textnormal{\bfseries Contexts}
    \[
      \begin{array}{rcl@{\arrayspacing}rcl@{\arrayspacing}rcl}
        \ltoi(\emptyset)
         & = & \emptyset                     &
        \ltoi(\ctx, x : \tsch)
         & = & \ltoi(\ctx), x : \ltoi(\tsch) &
        \ltoi(\ctx, \tvar[k])
         & = & \ltoi(\ctx), \tvar : \ltoi(k)
      \end{array}
    \]
    \textnormal{\bfseries Effect Contexts}
    \[
      \begin{array}{rcl}
        \ltoi(\emptyset)
         & = & \emptyset                                                                \\
        \ltoi(\Sigma, \tconst{}{\bar{\typ[k]}} : \{\opl_1, \ldots, \opl_n\})
         & = & \ltoi(\Sigma), \ctol(c) :: \forall \bar{\tvar} : \bar{\ltoi(k)} . \sigma \\
         &   & (\twhere \ctx_0 \ni \opl_i :
        \tfunl{\typ_i}{\row{\tconst{}{\bar{\typ[k]}}}{\erow}}{{\typ}'_i}                \\
         &   & \tand \sigma[\bar{\ltoi(\typ[k])} / \bar{\tvar}] =
        \{\opsigi{{\opi}_1}{\typ_1}{{\typ_1}'}, \ldots, \opsigi{{\opi}_n}{\typ_n}{{\typ_n}'}\} )
      \end{array}
    \]
    \textnormal{\bfseries Translation from Handlers to Labels}
    \begin{align*}
       & \htol(\opcl{\opl_1}{x_1}{e_1}; \cdots ; \opcl{\opl_n}{x_n}{e_n}; \retcl{x}{e})                          \\
       & = \begin{dcases}
             l                  & (\tif l = \ctol(c) \tand \{\opl_1, \ldots, \opl_n\} = \Sigma(\tconst{}{\cdots})) \\
             \mathit{undefined} & (\text{otherwise})
           \end{dcases}
    \end{align*}
  \end{definition}

  \begin{lemma}\label{lem:ctx_dom_var_ltoi}
    If $x \notin \dom(\ctx)$, then $x \notin \dom(\ltoi(\ctx))$.
  \end{lemma}

  \begin{proof}
    Straightforward by structual induction on $\ctx$ and the definition of $\ltoi$.
  \end{proof}

  \begin{lemma}\label{lem:ctx_dom_tvar_ltoi}
    $\tvar[k] \in \ctx$ iff $\tvar : \ltoi(k) \in \ltoi(\ctx)$.
  \end{lemma}

  \begin{proof}
    Straightforward by structual induction on $\ctx$ and the definition of $\ltoi$.
  \end{proof}

  \begin{lemma}\label{lem:ctx_binding_ltoi}
    If $x : \tsch \in \ctx$, then $x : \ltoi(\tsch) \in \ltoi(\ctx)$.
  \end{lemma}

  \begin{proof}
    Straightforward by structual induction on $\ctx$ and the definition of $\ltoi$.
  \end{proof}

  \begin{lemma}\label{lem:wfctx_typing}
    If $\typing{\ctx}{e}{\tsch}{\eps}$, then $\wfctx{\ctx}$.
  \end{lemma}

  \begin{proof}
    Straightforward by induction on a derivation of $\typing{\ctx}{e}{\tsch}{\eps}$.
  \end{proof}

  \begin{theorem}\label{thm:typability_ltoi}
    \phantom{}\\
    \begin{enumerate}
      \item\label{thm:typability_ltoi_kinding}
            If $\ftv(\typ[k]) \subseteq \ctx$ and $\wfctx{\ltoi(\ctx)}$, then $\kinding{\ltoi(\ctx)}{\ltoi(\typ[k])}{\ltoi(k)}$.

      \item\label{thm:typability_ltoi_ctx}
            If $\wfctx{\ctx}$, then $\wfctx{\ltoi(\ctx)}$.

      \item\label{thm:typability_ltoi_typing}
            If $\typing{\ctx}{e}{\tsch}{\eps}$,
            then $\typing{\ltoi(\ctx)}{\ltoi(e)}{\ltoi(\tsch)}{\ltoi(\eps)}$.
    \end{enumerate}
  \end{theorem}

  \begin{proof}
    \phantom{}\\
    \begin{enumerate}
      \item
            By structual induction on $\typ[k]$.
            \begin{divcases}
              \item[${\typ[k]} = {\tvar[k]}$]
              We have $\tvar[k] \in \ctx$.
              By Lemma~\ref{lem:ctx_dom_tvar_ltoi}, we have $\tvar : \ltoi(k) \in \ltoi(\ctx)$.
              Thus, \rname{K}{Var} derives
              \[\kinding{\ltoi(\ctx)}{\tvar}{\ltoi(k)}\]
              as required.

              \item[${\typ[k]} = \tfunl{\typ[*]_1}{\typ[\keffl]_2}{\typ[*]_3}$]
              We have
              \begin{itemize}
                \item $k = *$,
                \item $\ftv(\typ[*]_1) \subseteq \ctx$,
                \item $\ftv(\typ[\keffl]_2) \subseteq \ctx$, and
                \item $\ftv(\typ[*]_3) \subseteq \ctx$.
              \end{itemize}
              By the induction hypothesis, we have
              \begin{itemize}
                \item $\kinding{\ltoi(\ctx)}{\ltoi(\typ[*]_1)}{\kTyp}$,
                \item $\kinding{\ltoi(\ctx)}{\ltoi(\typ[\keffl]_2)}{\kEff}$, and
                \item $\kinding{\ltoi(\ctx)}{\ltoi(\typ[*]_3)}{\kTyp}$.
              \end{itemize}
              Thus, \rname{K}{Fun} derives
              \[\kinding{\ltoi(\ctx)}{\tfuni{\ltoi(\typ[*]_1)}{\ltoi(\typ[\keffl]_2)}{\ltoi(\typ[*]_3)}}{\kTyp}\]
              as required.

              \item[${\typ[k]} = \erow$]
              We have $k = \keffl$.
              Thus, by $\wfctx{\ltoi(\ctx)}$, we have
              \[\kinding{\ltoi(\ctx)}{\erow}{\kEff}\]
              as required.

              \item[${\typ[k]} = \row{\typ[\klabl]_1}{\typ[\keffl]_2}$]
              We have
              \begin{itemize}
                \item $k = \keffl$,
                \item $\ftv(\typ[\klabl]_1) \subseteq \ctx$, and
                \item $\ftv(\typ[\keffl]_2) \subseteq \ctx$.
              \end{itemize}
              By the induction hypothesis, we have
              \begin{itemize}
                \item $\kinding{\ltoi(\ctx)}{\ltoi(\typ[\klabl]_1)}{\kLab}$ and
                \item $\kinding{\ltoi(\ctx)}{\ltoi(\typ[\keffl]_2)}{\kEff}$.
              \end{itemize}
              Thus, \rname{K}{Cons} derives
              \[\kinding{\ltoi(\ctx)}{\row{\ltoi(\typ[\klabl]_1)}{\ltoi(\typ[\keffl]_2)}}{\kEff}\]
              as required.

              \item[${\typ[k]} = \tconst{(k_1, \ldots, k_n) \rightarrow \klabl}{\typ[k_1]_1, \ldots, \typ[k_n]_n}$]
              We have
              $\ftv(\typ[i]_i) \subseteq \ctx$
              for any $i \in \{1, \ldots, n\}$.
              By the induction hypothesis and definition of $\ctol$, we have
              \begin{itemize}
                \item $\kinding{\ltoi(\ctx)}{\ltoi(\typ[k_i]_i)}{\ltoi(k_i)}$ for any $i \in \{1, \ldots, n\}$,
                \item $\ctol(c^{(k_1, \ldots, k_n) \rightarrow \klabl}) : \ltoi(k_1) \times \ldots \times \ltoi(k_n) \rightarrow \kLab \in \Sbase$.
              \end{itemize}
              Thus, \rname{K}{Cons} derives
              \[\kinding{\ltoi(\ctx)}{\ctol(c^{(k_1, \ldots, k_n) \rightarrow \klabl})\, \ltoi(\typ[k_1]_1) \,\cdots\, \ltoi(\typ[k_n]_n)}{\kLab}\]
              as required.
            \end{divcases}

      \item
            By induction on a derivation of the judgment.
            We proceed by case analysis on the rule applied lastly to the derivation.
            \begin{divcases}
              \item[\rname{Cl}{Empty}] Clearly by \rname{C}{Empty} and the definition of $\ltoi$.

              \item[\rname{Cl}{Var}]
              We have
              \begin{itemize}
                \item $\ctx = \ctx', x : \tabs{\bar{\tvar[k']}}{\typ[*]}$,
                \item $\wfctx{\ctx'}$,
                \item $x \notin \dom(\ctx')$,
                \item $\ftv(\typ[*]) \setminus \{\bar{\tvar[k']}\} \subseteq \ctx'$, and
                \item $\forall \bar{k} . (\{\bar{\tvar[k]}\} \cap \ctx' = \emptyset)$,
              \end{itemize}
              for some $\ctx'$, $x$, $\bar{\tvar[k']}$, and $\typ[*]$.
              By the induction hypothesis and Lemma~\ref{lem:ctx_dom_var_ltoi}, we have
              \begin{itemize}
                \item $\wfctx{\ltoi(\ctx')}$ and
                \item $x \notin \dom(\ltoi(\ctx'))$.
              \end{itemize}
              By Lemma~\ref{lem:ctx_dom_tvar_ltoi} and \rname{C}{TVar}, we have
              \[\wfctx{\ltoi(\ctx'), \bar{\tvar} : \bar{\ltoi(k')}}.\]
              By $\ftv(\typ[*]) \subseteq \ctx', \bar{\tvar[k']}$ and case~\ref{thm:typability_ltoi_kinding}, we have
              \[\kinding{\ltoi(\ctx'), \bar{\tvar} : \bar{\ltoi(k')}}{\ltoi(\typ[*])}{\kTyp}.\]
              Thus, \rname{K}{Poly} and \rname{C}{Var} derives
              \[\wfctx{\ltoi(\ctx'), x : \ltoi(\tabs{\bar{\tvar[k']}}{\typ[*]})}\]
              as required.

              \item[\rname{Cl}{TVar}]
              We have
              \begin{itemize}
                \item $\ctx = \ctx', \tvar[k]$,
                \item $\wfctx{\ctx'}$, and
                \item $\forall k . (\tvar[k] \notin \ctx')$,
              \end{itemize}
              for some $\ctx'$ and $\tvar[k]$.
              By the induction hypothesis and Lemma~\ref{lem:ctx_dom_tvar_ltoi}, we have
              \begin{itemize}
                \item $\wfctx{\ltoi(\ctx')}$ and
                \item $\tvar \notin \dom(\ltoi(\ctx'))$.
              \end{itemize}
              Thus, \rname{C}{TVar} derives
              \[\wfctx{\ltoi(\ctx'), \tvar : \ltoi(k)}\]
              as required.
            \end{divcases}

      \item
            By induction on a derivation of the judgement.
            We proceed by case analysis on the rule applied lastly to the derivation.
            \begin{divcases}
              \item[\rname{Tl}{Var}]
              We have
              \begin{itemize}
                \item $e = x$,
                \item $\ctx(x) = \tsch$, and
                \item $\ftv(\eps) \subseteq \ctx$
              \end{itemize}
              for some $x$.
              By Lemma~\ref{lem:wfctx_typing} and
              case~\ref{thm:typability_ltoi_ctx},
              we have $\wfctx{\ltoi(\ctx)}$.
              By case~\ref{thm:typability_ltoi_kinding}, we have
              \[\kinding{\ltoi(\ctx)}{\ltoi(\eps)}{\kEff}.\]
              By Lemma~\ref{lem:ctx_binding_ltoi}, we have $x : \ltoi(\tsch) \in \ltoi(\ctx)$.
              Thus, \rname{T}{Var} derives
              \[\typing{\ltoi(\ctx)}{x}{\ltoi(\tsch)}{\emptyE}.\]
              By Lemma~\ref{lem:wk}, we have $\kinding{\ltoi(\ctx)}{\ltoi(\tsch)}{\kTyp}$.
              Thus, \rname{ST}{Refl} and Lemma~\ref{lem:entailment}\ref{lem:entailment:refl} and \rname{T}{Sub} derive
              \[\typing{\ltoi(\ctx)}{x}{\ltoi(\tsch)}{\ltoi(\eps)}\]
              as required.

              \item[\rname{Tl}{Lam}]
              We have
              \begin{itemize}
                \item $e = \lam{x}{e'}$,
                \item $\tsch = \tfunl{\typ_1}{\eps'}{\typ_2}$,
                \item $\typing{\ctx, x : \typ_1}{e'}{\typ_2}{\eps'}$, and
                \item $\ftv(\eps) \subseteq \ctx$
              \end{itemize}
              for some $x$, $e'$, $\typ_1$, $\eps'$, and $\typ_2$.
              By Lemma~\ref{lem:wfctx_typing}, we have $\wfctx{\ctx, x : \typ_1}$.
              Since only \rname{Cl}{Var} can derive $\wfctx{\ctx, x : \typ_1}$,
              we have $\wfctx{\ctx}$.
              By case~\ref{thm:typability_ltoi_ctx},
              we have $\wfctx{\ltoi(\ctx)}$.
              By the induction hypothesis and case~\ref{thm:typability_ltoi_kinding}, we have
              \begin{itemize}
                \item $\typing{\ltoi(\ctx, x : \typ_1)}{\ltoi(e')}{\ltoi(\typ_2)}{\ltoi(\eps')}$ and
                \item $\kinding{\ltoi(\ctx)}{\ltoi(\eps)}{\kEff}$.
              \end{itemize}
              Without loss of generality, we can choose $z$ such that
              \begin{itemize}
                \item $z \notin \mathit{FV}(\ltoi(e'))$,
                \item $z \neq x$,
                \item $z \notin \dom(\ltoi(\ctx))$, and
                \item $\ltoi(\lam{x}{e'}) = \funi{z}{x}{\ltoi(e')}$.
              \end{itemize}
              By Lemma~\ref{lem:wk} and
              Lemma~\ref{lem:delta_context}\ref{lem:delta_context:kinding} and
              Lemma~\ref{lem:delta_weakening}, we have
              \begin{itemize}
                \item $\kinding{\ltoi(\ctx)}{\ltoi(\typ_2)}{\kTyp}$ and
                \item $\kinding{\ltoi(\ctx)}{\ltoi(\eps')}{\kEff}$.
              \end{itemize}
              By Lemma~\ref{lem:ctx-wf-typing}, we have $\wfctx{\ltoi(\ctx), x : \ltoi(\typ_1)}$.
              Since only \rname{C}{Var} can derive $\wfctx{\ltoi(\ctx), x : \ltoi(\typ_1)}$,
              we have $\kinding{\ltoi(\ctx)}{\ltoi(\typ_1)}{\kTyp}$.
              Thus, \rname{K}{Fun} derives
              \[\kinding{\ltoi(\ctx)}{\tfuni{\ltoi(\typ_1)}{\ltoi(\eps')}{\ltoi(\typ_2)}}{\kTyp}.\]
              Thus, \rname{C}{Var} derives
              \[\wfctx{\ltoi(\ctx), z : \tfuni{\ltoi(\typ_1)}{\ltoi(\eps')}{\ltoi(\typ_2)}}.\]
              Thus, Lemma~\ref{lem:weakening} and \rname{T}{Abs} derives
              \[\typing{\ltoi(\ctx)}{\funi{z}{x}{\ltoi(e')}}{\tfuni{\ltoi(\typ_1)}{\ltoi(\eps')}{\ltoi(\typ_2)}}{\emptyE}.\]
              Thus, \rname{ST}{Refl} and \rname{T}{Sub} derive
              \[\typing{\ltoi(\ctx)}{\funi{z}{x}{\ltoi(e')}}{\tfuni{\ltoi(\typ_1)}{\ltoi(\eps')}{\ltoi(\typ_2)}}{\ltoi(\eps)}.\]
              as required.

              \item[\rname{Tl}{Let}]
              We have
              \begin{itemize}
                \item $e = \val{x}{e_1}{e_2}$,
                \item $\tsch = \typ$,
                \item $\typing{\ctx}{e_1}{\tsch'}{\eps}$, and
                \item $\typing{\ctx, x : \tsch'}{e_2}{\typ}{\eps}$,
              \end{itemize}
              for some $x$, $e_1$, $e_2$, $\typ$, and $\tsch'$.
              By the induction hypothesis, we have
              \begin{itemize}
                \item $\typing{\ltoi(\ctx)}{\ltoi(e_1)}{\ltoi(\tsch')}{\ltoi(\eps)}$ and
                \item $\typing{\ltoi(\ctx, x : \tsch')}{\ltoi(e_2)}{\ltoi(\typ)}{\ltoi(\eps)}$.
              \end{itemize}
              By definition of $\ltoi$ and \rname{T}{Let}, we have
              \[\typing{\ltoi(\ctx)}{\letin{x}{\ltoi(e_1)}{\ltoi(e_2)}}{\ltoi(\typ)}{\ltoi(\eps)}\]
              as required.

              \item[\rname{Tl}{App}]
              We have
              \begin{itemize}
                \item $e = e_1 (e_2)$,
                \item $\tsch = \typ$,
                \item $\typing{\ctx}{e_1}{\tfunl{\typ_2}{\eps}{\typ}}{\eps}$, and
                \item $\typing{\ctx}{e_2}{\typ_2}{\eps}$,
              \end{itemize}
              for some $e_1$, $e_2$, $\typ$, and $\typ_2$.
              By the induction hypothesis, we have
              \begin{itemize}
                \item $\typing{\ltoi(\ctx)}{\ltoi(e_1)}{\tfuni{\ltoi(\typ_2)}{\ltoi(\eps)}{\ltoi(\typ)}}{\ltoi(\eps)}$ and
                \item $\typing{\ltoi(\ctx)}{\ltoi(e_2)}{\ltoi(\typ_2)}{\ltoi(\eps)}$.
              \end{itemize}
              Without loss of generality, we can choose $x$ and $y$ such that
              \begin{itemize}
                \item $x \neq y$,
                \item $x \notin \dom(\ltoi(\ctx))$, and
                \item $y \notin \dom(\ltoi(\ctx))$.
              \end{itemize}
              Because Lemma~\ref{lem:wk}\ref{lem:wk:typing} and \rname{C}{Var} give us
              \begin{itemize}
                \item $\wfctx{\ltoi(\ctx), x : \tfuni{\ltoi(\typ_2)}{\ltoi(\eps)}{\ltoi(\typ)}}$ and
                \item $\wfctx{\ltoi(\ctx), x : \tfuni{\ltoi(\typ_2)}{\ltoi(\eps)}{\ltoi(\typ)}}, y : \ltoi(\typ_2)$.
              \end{itemize}
              Thus, \rname{T}{Var} and \rname{T}{App} derive
              \[\typing
                {\ltoi(\ctx), x : \tfuni{\ltoi(\typ_2)}{\ltoi(\eps)}{\ltoi(\typ)}, y : \ltoi(\typ_2)}
                {x \, y}{\ltoi(\typ)}{\ltoi(\eps)}.\]
              By Lemma~\ref{lem:weakening}\ref{lem:weakening:typing}, \rname{T}{Let} derive
              \[
                \typing
                {\ltoi(\ctx), x : \tfuni{\ltoi(\typ_2)}{\ltoi(\eps)}{\ltoi(\typ)}}
                {\letin{y}{\ltoi(e_2)}{x\, y}}{\ltoi(\typ)}{\ltoi(\eps)}.
              \]
              Thus, \rname{T}{Let} derives
              \[\typing{\ltoi(\ctx)}{\letin{x}{e_1}{\letin{y}{\ltoi(e_2)}{x\, y}}}{\ltoi(\typ)}{\ltoi(\eps)}\]
              as required.

              \item[\rname{Tl}{TAbs}]
              We have
              \begin{itemize}
                \item $e = \tlaml{\bar{\tvar[k]}}{e'}$,
                \item $\tsch = \tabs{\bar{\tvar[k]}}{\typ}$,
                \item $\typing{\ctx, \bar{\tvar[k]}}{e'}{\typ}{\erow}$, and
                \item $\ftv(\eps) \subseteq \ctx$,
              \end{itemize}
              for some $\bar{\tvar[k]}$, $e'$, and $\typ$.
              By Lemma~\ref{lem:wfctx_typing}, we have $\wfctx{\ctx, \bar{\tvar[k]}}$.
              Since only \rname{Cl}{TVar} derive $\wfctx{\ctx, \bar{\tvar[k]}}$,
              we have $\wfctx{\ctx}$.
              By case~\ref{thm:typability_ltoi_ctx},
              we have $\wfctx{\ltoi(\ctx)}$.
              By the induction hypothesis and case~\ref{thm:typability_ltoi_kinding}, we have
              \begin{itemize}
                \item $\typing{\ltoi(\ctx), \bar{\tvar} : \bar{\ltoi(k)}}{\ltoi(e')}{\ltoi(\typ)}{\erow}$ and
                \item $\kinding{\ltoi(\ctx)}{\ltoi(\eps)}{\kEff}$.
              \end{itemize}
              By applying \rname{T}{TAbs} repeatedly, we have
              \[\typing{\ltoi(\ctx)}
                {\tlami{\bar{\tvar}}{\bar{\ltoi(k)}}{\ltoi(e')}}
                {\tabsi{\tvar_0}{\ltoi(k_0)}{(\cdots (\tabsi{\tvar_n}{\ltoi(k_n)}{\ltoi(\typ)^{\emptyE}}) \cdots)^{\emptyE}} }
                {\emptyE}.
              \]
              Thus, \rname{T}{Sub} derives
              \[\typing{\ltoi(\ctx)}
                {\tlami{\bar{\tvar}}{\bar{\ltoi(k)}}{\ltoi(e')}}
                {\tabsi{\tvar_0}{\ltoi(k_0)}{(\cdots (\tabsi{\tvar_n}{\ltoi(k_n)}{\ltoi(\typ)^{\emptyE}}) \cdots)^{\emptyE}} }
                {\ltoi(\eps)}.
              \]
              as required.

              \item[\rname{Tl}{TApp}]
              We have
              \begin{itemize}
                \item $e = e' (\bar{\typ[k]_0})$,
                \item $\tsch = \typ \tinst{\bar{\tvar[k]}}{\bar{\typ[k]_0}}$,
                \item $\typing{\ctx}{e'}{\tabs{\bar{\tvar[k]}}{\typ}}{\eps}$, and
                \item $\ftv(\bar{\typ[k]_0}) \subseteq \ctx$,
              \end{itemize}
              for some $e'$, $\bar{\typ[k]_0}$, and $\bar{\tvar[k]}$.
              By Lemma~\ref{lem:wfctx_typing}, we have $\wfctx{\ctx}$.
              By case~\ref{thm:typability_ltoi_ctx},
              we have $\wfctx{\ltoi(\ctx)}$.
              By the induction hypothesis and case~\ref{thm:typability_ltoi_kinding}, we have
              \begin{itemize}
                \item $\typing
                        {\ltoi(\ctx)}{\ltoi(e')}
                        {\tabsi{\tvar_0}{\ltoi(k_0)}{(\cdots (\tabsi{\tvar_n}{\ltoi(k_n)}{\ltoi(\typ)^{\emptyE}}) \cdots)^{\emptyE}} }
                        {\ltoi(\eps)}$ and
                \item $\kinding{\ltoi(\ctx)}{\ltoi(\bar{\typ[k]_0})}{\bar{\ltoi(k)}}$.
              \end{itemize}
              Without loss of generality, we can choose $x$ such that $x \notin \dom(\ltoi(\ctx))$.
              By Lemma~\ref{lem:wk}\ref{lem:wk:typing} and \rname{C}{Var}, we have
              \[\wfctx{\ltoi(\ctx), x : \tabsi{\tvar_0}{\ltoi(k_0)}{(\cdots (\tabsi{\tvar_n}{\ltoi(k_n)}{\ltoi(\typ)^{\emptyE}}) \cdots)^{\emptyE}} }.\]
              By Lemma~\ref{lem:weakening}\ref{lem:weakening:kinding}, we have
              \[\kinding{
                  \ltoi(\ctx), x : \tabsi{\tvar_0}{\ltoi(k_0)}{(\cdots (\tabsi{\tvar_n}{\ltoi(k_n)}{\ltoi(\typ)^{\emptyE}}) \cdots)^{\emptyE}}
                }{\bar{\typ[k]_0}}{\bar{\ltoi(k)}}.\]
              Thus, \rname{T}{Var} and applying \rname{T}{TApp} repeatedly derive
              \[\typing
                {\ltoi(\ctx), x : \tabsi{\tvar_0}{\ltoi(k_0)}{(\cdots (\tabsi{\tvar_n}{\ltoi(k_n)}{\ltoi(\typ)^{\emptyE}}) \cdots)^{\emptyE}} }
                {x\, \ltoi(\bar{\typ[k]_0})}{\ltoi(\typ)[\bar{\typ[k]_0}/\bar{\tvar[k]}] }{\emptyE}.\]
              Because Lemma~\ref{lem:wk}\ref{lem:wk:typing} and Lemma~\ref{lem:weakening}\ref{lem:weakening:kinding} give us
              \begin{itemize}
                \item $\kinding{\ltoi(\ctx), x : \tabsi{\tvar_0}{\ltoi(k_0)}{(\cdots (\tabsi{\tvar_n}{\ltoi(k_n)}{\ltoi(\typ)^{\emptyE}}) \cdots)^{\emptyE}} }{\ltoi(\typ)[\bar{\typ[k]_0}/\bar{\tvar[k]}]}{\kEff}$ and
                \item $\kinding{\ltoi(\ctx), x : \tabsi{\tvar_0}{\ltoi(k_0)}{(\cdots (\tabsi{\tvar_n}{\ltoi(k_n)}{\ltoi(\typ)^{\emptyE}}) \cdots)^{\emptyE}} }{\ltoi(\eps)}{\kEff}$,
              \end{itemize}
              \rname{ST}{Refl} and \rname{T}{Sub} derives
              \[\typing
                {\ltoi(\ctx), x : \tabsi{\tvar_0}{\ltoi(k_0)}{(\cdots (\tabsi{\tvar_n}{\ltoi(k_n)}{\ltoi(\typ)^{\emptyE}}) \cdots)^{\emptyE}} }
                {x\, \ltoi(\bar{\typ[k]_0})}{\ltoi(\typ)[\bar{\typ[k]_0}/\bar{\tvar[k]}] }{\ltoi(\eps)}.\]
              Thus, \rname{T}{Let} derives
              \[
                \typing{\ltoi(\ctx)}{\letin{x}{\ltoi(e')}{x\, \ltoi(\typ[k]_0)}}{\ltoi(\typ)[\bar{\typ[k]_0}/\bar{\tvar[k]}]}{\ltoi(\eps)}
              \]
              as required.

              \item[\rname{Tl}{Handle}]
              We have
              \begin{itemize}
                \item $h = \opcl{\opl_1}{x_1}{e_1}; \cdots; \opcl{\opl_n}{x_n}{e_n}; \retcl{x}{e_r}$,
                \item $e = \handlel{h}{e'}$,
                \item $\tsch = \typ_r$,
                \item $\typing{\ctx}{e'}{\typ}{\row{\tconst{}{\bar{\typ[k]}}}{\eps}}$,
                \item $\typing{\ctx, x : \typ}{e_r}{\typ_r}{\eps}$,
                \item $\Sigma(\tconst{}{\bar{\typ[k]}}) = \{\opl_1, \ldots, \opl_n\}$,
                \item $\typing{\ctx}{\opl_i}{\tfunl{\typ_i}{\row{\tconst{}{\bar{\typ[k]}}}{\erow}}{{\typ}'_i}}{\erow}$, and
                \item $\typing{\ctx, \resume : \tfunl{{\typ}'_i}{\eps}{\typ_r}, x_i : \typ_i}{e_i}{\typ_r}{\eps}$ for any $i \in \{1, \ldots, n\}$,
              \end{itemize}
              for some $h$, $e'$, $x$, $e_r$, $\opl_i$, $x_i$, $e_i$, $\typ_i$, ${\typ}'_i$, and $\tconst{}{\bar{\typ[k]}}$
              where $i \in \{1, \ldots, n\}$.
              By the induction hypothesis and definition of $\ltoi$, we have
              \begin{itemize}
                \item $\typing{\ltoi(\ctx)}{\ltoi(e')}{\ltoi(\typ)}{\row{\ctol(c)\, \bar{\ltoi(\typ[k])}}{\ltoi(\eps)}}$,
                \item $\typing{\ltoi(\ctx), x : \ltoi(\typ)}{\ltoi(e_r)}{\ltoi(\typ_r)}{\ltoi(\eps)}$,
                \item $
                        \typing
                        {\ltoi(\ctx), x_i : \ltoi(\typ_i), \resume : \tfuni{\ltoi({\typ}'_i)}{\ltoi(\eps)}{\ltoi(\typ_r)}}
                        {\ltoi(e_i)}{\ltoi(\typ_r)}{\ltoi(\eps)}
                      $ for any $i \in \{1, \ldots, n\}$,
                \item $\ctol(c) :: \forall \bar{\tvar} : \bar{\ltoi(k)} . \sigma \in \ltoi(\Sigma)$, and
                \item $\sigma[\bar{\ltoi(\typ[k])} / \bar{\tvar}] = \{\opsigi{{\opi}_1}{\typ_1}{{\typ}'_1}, \ldots, \opsigi{{\opi}_n}{\typ_n}{{\typ}'_n}\}$.
              \end{itemize}
              Because \rname{H}{Return} and \rname{H}{Op} derive
              \[\typingH{\ltoi(\ctx)}{\sigma[\bar{\ltoi(\typ[k])} / \bar{\tvar}]}
                {\ltoi(h)}
                {\ltoi(\typ) \Rightarrow ^ {\ltoi(\eps)} \ltoi(\typ_r) },\]
              \rname{T}{Handling} derives
              \[\typing{\ltoi(\ctx)}
                {\handlewith{\htol(h)}
                  {\ltoi(e')}{\ltoi(h)}}
                {\ltoi(\typ_r)}{\ltoi(\eps)}\]
              as required.

              \item[\rname{Tl}{Open}]
              We have
              \begin{itemize}
                \item $\tsch = \tfunl{\typ_1}{\row{l_1, \ldots, l_n}{\eps'}}{\typ_2}$,
                \item $\typing{\ctx}{e}{\tfunl{\typ_1}{\row{l_1, \ldots, l_n}{\erow}}{\typ_2}}{\eps}$, and
                \item $\ftv(\eps') \subseteq \ctx$,
              \end{itemize}
              for some $\typ_1$, $\typ_2$, $l_1, \ldots, l_n$, and $\eps'$.
              By Lemma~\ref{lem:wfctx_typing}, we have $\wfctx{\ctx}$.
              By case~\ref{thm:typability_ltoi_ctx}, we have $\wfctx{\ltoi(\ctx)}$.
              By the induction hypothesis and case~\ref{thm:typability_ltoi_kinding}, we have
              \begin{itemize}
                \item $\typing{\ltoi(\ctx)}{\ltoi(e)}
                        {\tfuni{\ltoi(\typ_1)}{\row{\ltoi(l_1), \ldots, \ltoi(l_n)}{\erow}}{\ltoi(\typ_2)}}{\ltoi(\eps)}$ and
                \item $\kinding{\ltoi(\ctx)}{\ltoi(\eps')}{\kEff}$.
              \end{itemize}
              By Lemma~\ref{lem:wk}\ref{lem:wk:typing}, we have
              \[\kinding
                {\ltoi(\ctx)}
                {\tfuni{\ltoi(\typ_1)}{\row{\ltoi(l_1), \ldots, \ltoi(l_n)}{\erow}}{\ltoi(\typ_2)}}
                {\kTyp}.\]
              Since only \rname{K}{Fun} can derive this judgment, we have
              \begin{itemize}
                \item $\kinding{\ltoi(\ctx)}{\ltoi(\typ_1)}{\kTyp}$,
                \item $\kinding{\ltoi(\ctx)}{\ltoi(\row{\ltoi(l_1), \ldots, \ltoi(l_n)}{\erow})}{\kEff}$, and
                \item $\kinding{\ltoi(\ctx)}{\ltoi(\typ_2)}{\kTyp}$.
              \end{itemize}
              Thus, by \rname{ST}{Refl} and \rname{ST}{Fun} and Lemma~\ref{lem:entailment}\ref{lem:entailment:refl} and \rname{T}{Sub}, we have
              \[\typing{\ltoi(\ctx)}{\ltoi(e)}
                {\tfuni{\ltoi(\typ_1)}{\row{\ltoi(l_1), \ldots, \ltoi(l_n)}{\eps'}}{\ltoi(\typ_2)}}{\ltoi(\eps)}
              \]
              as required.
            \end{divcases}
    \end{enumerate}
  \end{proof}
}

\bibliographystyle{ACM-Reference-Format}
\bibliography{paper}